\pdfoutput=1
\documentclass[11pt,twoside]{book}
\usepackage{fancyhdr}
\usepackage{time}
\fancypagestyle{plain}{%
\fancyhf{} %clear all head and foot styles
\fancyfoot[R]{\textsf{ILC Reference Design Report\hspace{0.5cm}III-\thepage}}}
\usepackage{epsfig}
\usepackage{rotate}
\usepackage[figuresright]{rotating}
\usepackage{lscape}
\usepackage{longtable}
\usepackage{hyperref}
\usepackage{url}
\usepackage{graphicx}
\usepackage{color}
\usepackage{epstopdf}
\usepackage{epsf}
\usepackage{hhline}
\usepackage{mytrc_style}
\usepackage{ulem}
\usepackage{verbatim}

\usepackage{eso-pic}
\usepackage{type1cm}
\newcommand{\vbdlspacing}{\\[-15pt]}
\newcommand{\vbabovecaption}{\vspace{-24pt}}

\newcommand{\vbabove}{\vspace{-12pt}}
\newcommand{\vbbelow}{\vspace{-12pt}}
\newcommand{\itemspace}{\vspace{-6pt}}

\newcounter{figlcl}[section]
\newcounter{tablcl}[section]

\renewcommand{\thepage}{\arabic{chapter}.\arabic{section}-\arabic{page}}
\renewcommand{\thefigure}{\arabic{chapter}.\arabic{section}-\arabic{figlcl}}
\renewcommand{\thetable}{\arabic{chapter}.\arabic{section}-\arabic{tablcl}}

\begin{document}
\renewcommand{\thepage}{\roman{page}}
%\frontmatter
%\begin{titlepage}

%\begin{center}

{\sffamily\bfseries
\begin{titlepage}
\begin{center}
~
 ~ \vskip 4cm

    {\Huge I}{\huge NTERNATIONAL} 
    {\Huge L}{\huge INEAR} 
    {\Huge C}{\huge OLLIDER}
    
  \vskip 1.2cm

    {\Huge R}{\huge EFERENCE}
    {\Huge D}{\huge ESIGN}
    {\Huge R}{\huge EPORT}

  \vskip 1.2cm

    %{\Huge 2007}

%\vskip 3cm

    %{\huge Volume 1:~~~EXECUTIVE SUMMARY}

\vskip 3cm

{\huge ILC Global Design Effort and} \\
    
  \vskip 0.5cm

{\huge World Wide Study }

  \vskip 3cm

    {\huge AUGUST, 2007}

\end{center}
\end{titlepage}

\newpage\thispagestyle{empty}
~
 ~ \vskip 2cm

{\LARGE Volume 1:~~~EXECUTIVE SUMMARY}
 \vskip 0.5cm
{\Large Editors:} 
 \vskip 0.25cm
{\Large James~Brau, Yasuhiro~Okada, Nicholas~Walker}

  \vskip 1.5cm

{\LARGE Volume 2:~~~PHYSICS AT THE ILC}
 \vskip 0.5cm
{\Large Editors:} 
 \vskip 0.25cm
{\Large Abdelhak~Djouadi, Joseph~Lykken, Klaus~M{\"o}nig} 
 \vskip 0.25cm
{\Large Yasuhiro~Okada, Mark~Oreglia, Satoru~Yamashita}

  \vskip 1.5cm

{\LARGE Volume 3:~~~ACCELERATOR}
 \vskip 0.5cm
{\Large Editors:} 
 \vskip 0.25cm
{\Large Nan~Phinney, Nobukasu~Toge, Nicholas~Walker}

  \vskip 1.5cm

{\LARGE Volume 4:~~~DETECTORS}
 \vskip 0.5cm
{\Large Editors:} 
 \vskip 0.25cm
{\Large Ties~Behnke, Chris~Damerell, John~Jaros, Akiya~Miyamoto}

\newpage\thispagestyle{empty}
%\begin{center}
~
 ~ \vskip 5cm
~~~{\Huge Volume 3:~~~ACCELERATOR}
 \vskip 1cm
~~~{\LARGE Editors:} 
 \vskip 0.5cm
~~~{\LARGE Nan~Phinney, Nobukazu~Toge, Nicholas~Walker}

%\end{center}
\newpage\thispagestyle{empty}

}

\cleardoublepage\setcounter{page}{1}
\chapter*{List of Contributors} 

\begin{center}
%============== This is Author list ==========================

\begin{center}

Gerald~Aarons$^{203}$,
Toshinori~Abe$^{290}$,
Jason~Abernathy$^{293}$,
Medina~Ablikim$^{87}$,
Halina~Abramowicz$^{216}$,
David~Adey$^{236}$,
Catherine~Adloff$^{128}$,
Chris~Adolphsen$^{203}$,
Konstantin~Afanaciev$^{11,47}$,
Ilya~Agapov$^{192,35}$
% ...

\end{center}

\clearpage

\chapter*{List of Institutions}

\begin{center}

{\sl $^{1}$ Abdus Salam International Centre for Theoretical Physics, Strada Costriera 11, 34014 Trieste, Italy}

{\sl $^{2}$ Academy, RPR, National Institute of Physics and Nuclear Engineering `Horia Hulubei' (IFIN-HH), Str. Atomistilor no. 407, P.O. Box MG-6, R-76900 Bucharest - Magurele, Romania}

{\sl $^{3}$ AGH University of Science and Technology Akademia Gorniczo-Hutnicza im. Stanislawa Staszica w Krakowie al. Mickiewicza 30 PL-30-059 Cracow, Poland}

{\sl $^{4}$ Albert-Ludwigs Universit{\"a}t Freiburg, Physikalisches Institut, Hermann-Herder Str. 3, D-79104 Freiburg, Germany}

{\sl $^{5}$ Aligarh Muslim University, Aligarh, Uttar Pradesh 202002, India}

{\sl $^{6}$ Amberg Engineering AG, Trockenloostr. 21, P.O.Box 27, 8105 Regensdorf-Watt, Switzerland}

{\sl $^{7}$ Angstromquelle Karlsruhe (ANKA), Forschungszentrum Karlsruhe, Hermann-von-Helmholtz-Platz 1, D-76344 Eggenstein-Leopoldshafen, Germany}

{\sl $^{8}$ Argonne National Laboratory (ANL), 9700 S. Cass Avenue, Argonne, IL 60439, USA}

{\sl $^{9}$ Baylor University, Department of Physics, 101 Bagby Avenue, Waco, TX 76706, USA}

{\sl $^{10}$ Beijing University, Department of Physics, Beijing, China 100871}

{\sl $^{11}$ Belarusian State University, National Scientific \& Educational Center, Particle \& HEP Physics, M. Bogdanovich St., 153, 240040 Minsk, Belarus}

{\sl $^{12}$ Benares Hindu University, Benares, Varanasi 221005, India}

{\sl $^{13}$ Bhabha Atomic Research Centre, Trombay, Mumbai 400085, India}

{\sl $^{14}$ Birla Institute of Technology and Science, EEE Dept., Pilani, Rajasthan, India}

{\sl $^{15}$ Bogazici University, Physics Department, 34342 Bebek / Istanbul, 80820 Istanbul, Turkey}

{\sl $^{16}$ Boston University, Department of Physics, 590 Commonwealth Avenue, Boston, MA 02215, USA}

{\sl $^{17}$ Brandenburg University of Technology, Postfach 101344, D-03013 Cottbus, Germany}

{\sl $^{18}$ Brno University of Technology, Anton\'insk\'a; 548/1, CZ 601 90 Brno, Czech Republic}

{\sl $^{19}$ Brookhaven National Laboratory (BNL), P.O.Box 5000, Upton, NY 11973-5000, USA}

{\sl $^{20}$ Brown University, Department of Physics, Box 1843, Providence, RI 02912, USA}

{\sl $^{21}$ Budkar Institute for Nuclear Physics (BINP), 630090 Novosibirsk, Russia}

{\sl $^{22}$ Calcutta University, Department of Physics, 92 A.P.C. Road, Kolkata 700009, India}

{\sl $^{23}$ California Institute of Technology, Physics, Mathematics and Astronomy (PMA), 1200 East California Blvd, Pasadena, CA 91125, USA}

{\sl $^{24}$ Carleton University, Department of Physics, 1125 Colonel By Drive, Ottawa, Ontario, Canada K1S 5B6}

{\sl $^{25}$ Carnegie Mellon University, Department of Physics, Wean Hall 7235, Pittsburgh, PA 15213, USA}

{\sl $^{26}$ CCLRC Daresbury Laboratory, Daresbury, Warrington, Cheshire WA4 4AD, UK }

{\sl $^{27}$ CCLRC Rutherford Appleton Laboratory, Chilton, Didcot, Oxton OX11 0QX, UK }

{\sl $^{28}$ CEA Saclay, DAPNIA, F-91191 Gif-sur-Yvette, France}

{\sl $^{29}$ CEA Saclay, Service de Physique Th{\'e}orique, CEA/DSM/SPhT, F-91191 Gif-sur-Yvette Cedex, France}

{\sl $^{30}$ Center for High Energy Physics (CHEP) / Kyungpook National University, 1370 Sankyuk-dong, Buk-gu, Daegu 702-701, Korea}

{\sl $^{31}$ Center for High Energy Physics (TUHEP), Tsinghua University, Beijing, China 100084}

{\sl $^{32}$ Centre de Physique Theorique, CNRS - Luminy, Universiti d'Aix - Marseille II, Campus of Luminy, Case 907, 13288 Marseille Cedex 9, France}

{\sl $^{33}$ Centro de Investigaciones Energ\'eticas, Medioambientales y Technol\'ogicas, CIEMAT, Avenia Complutense 22, E-28040 Madrid, Spain}

{\sl $^{34}$ Centro Nacional de Microelectr\'onica (CNM), Instituto de Microelectr\'onica de Barcelona (IMB), Campus UAB, 08193 Cerdanyola del Vall\`es (Bellaterra), Barcelona, Spain}

{\sl $^{35}$ CERN, CH-1211 Gen\`eve 23, Switzerland}

{\sl $^{36}$ Charles University, Institute of Particle \& Nuclear Physics, Faculty of Mathematics and Physics, V Holesovickach 2, CZ-18000 Praque 8, Czech Republic}

{\sl $^{37}$ Chonbuk National University, Physics Department, Chonju 561-756, Korea}

{\sl $^{38}$ Cockcroft Institute, Daresbury, Warrington WA4 4AD, UK }

{\sl $^{39}$ College of William and Mary, Department of Physics, Williamsburg, VA, 23187, USA}

{\sl $^{40}$ Colorado State University, Department of Physics, Fort Collins, CO 80523, USA}

{\sl $^{41}$ Columbia University, Department of Physics, New York, NY 10027-6902, USA}

{\sl $^{42}$ Concordia University, Department of Physics, 1455 De Maisonneuve Blvd. West, Montreal, Quebec, Canada H3G 1M8}

{\sl $^{43}$ Cornell University, Laboratory for Elementary-Particle Physics (LEPP), Ithaca, NY 14853, USA}

{\sl $^{44}$ Cukurova University, Department of Physics, Fen-Ed. Fakultesi 01330, Balcali, Turkey}

{\sl $^{45}$ D.~V. Efremov Research Institute, SINTEZ, 196641 St. Petersburg, Russia}

{\sl $^{46}$ Dartmouth College, Department of Physics and Astronomy, 6127 Wilder Laboratory, Hanover, NH 03755, USA}

{\sl $^{47}$ DESY-Hamburg site, Deutsches Elektronen-Synchrotoron in der Helmholtz-Gemeinschaft, Notkestrasse 85, 22607 Hamburg, Germany}

{\sl $^{48}$ DESY-Zeuthen site, Deutsches Elektronen-Synchrotoron in der Helmholtz-Gemeinschaft, Platanenallee 6, D-15738 Zeuthen, Germany}

{\sl $^{49}$ Durham University,  Department of Physics, Ogen Center for Fundamental Physics, South Rd., Durham DH1 3LE, UK}

{\sl $^{50}$ Ecole Polytechnique, Laboratoire Leprince-Ringuet (LLR), Route de Saclay, F-91128 Palaiseau Cedex, France}

{\sl $^{51}$ Ege University, Department of Physics, Faculty of Science, 35100 Izmir, Turkey}

{\sl $^{52}$ Enrico Fermi Institute, University of Chicago, 5640 S. Ellis Avenue, RI-183, Chicago, IL 60637, USA}

{\sl $^{53}$ Ewha Womans University, 11-1 Daehyun-Dong, Seodaemun-Gu, Seoul, 120-750, Korea}

{\sl $^{54}$ Fermi National Accelerator Laboratory (FNAL), P.O.Box 500, Batavia, IL 60510-0500, USA}

{\sl $^{55}$ Fujita Gakuen Health University, Department of Physics, Toyoake, Aichi 470-1192, Japan}

{\sl $^{56}$ Fukui University of Technology, 3-6-1 Gakuen, Fukui-shi, Fukui 910-8505, Japan}

{\sl $^{57}$ Fukui University, Department of Physics, 3-9-1 Bunkyo, Fukui-shi, Fukui 910-8507, Japan}

{\sl $^{58}$ Georg-August-Universit{\"a}t G{\"o}ttingen, II. Physikalisches Institut, Friedrich-Hund-Platz 1, 37077 G{\"o}ttingen, Germany}

{\sl $^{59}$ Global Design Effort}

{\sl $^{60}$ Gomel State University, Department of Physics, Ul. Sovietskaya 104, 246699 Gomel, Belarus}

{\sl $^{61}$ Guangxi University, College of Physics science and Engineering Technology, Nanning, China 530004}

{\sl $^{62}$ Hanoi University of Technology, 1 Dai Co Viet road, Hanoi, Vietnam}

{\sl $^{63}$ Hanson Professional Services, Inc., 1525 S. Sixth St., Springfield, IL 62703, USA}

{\sl $^{64}$ Harish-Chandra Research Institute, Chhatnag Road, Jhusi, Allahabad 211019, India}

{\sl $^{65}$ Helsinki Institute of Physics (HIP), P.O. Box 64, FIN-00014 University of Helsinki, Finland}

{\sl $^{66}$ Henan Normal University, College of Physics and Information Engineering, Xinxiang, China 453007}

{\sl $^{67}$ High Energy Accelerator Research Organization, KEK, 1-1 Oho, Tsukuba, Ibaraki 305-0801, Japan}

{\sl $^{68}$ Hiroshima University, Department of Physics, 1-3-1 Kagamiyama, Higashi-Hiroshima, Hiroshima 739-8526, Japan}

{\sl $^{69}$ Humboldt Universit{\"a}t zu Berlin, Fachbereich Physik, Institut f\"ur Elementarteilchenphysik, Newtonstr. 15, D-12489 Berlin, Germany}

{\sl $^{70}$ Hungarian Academy of Sciences, KFKI Research Institute for Particle and Nuclear Physics, P.O. Box 49, H-1525 Budapest, Hungary}

{\sl $^{71}$ Ibaraki University, College of Technology, Department of Physics, Nakanarusawa 4-12-1, Hitachi, Ibaraki 316-8511, Japan}

{\sl $^{72}$ Imperial College, Blackett Laboratory, Department of Physics, Prince Consort Road, London, SW7 2BW, UK}

{\sl $^{73}$ Indian Association for the Cultivation of Science, Department of Theoretical Physics and Centre for Theoretical Sciences, Kolkata 700032, India}

{\sl $^{74}$ Indian Institute of Science, Centre for High Energy Physics, Bangalore 560012, Karnataka, India}

{\sl $^{75}$ Indian Institute of Technology, Bombay, Powai, Mumbai 400076, India}

{\sl $^{76}$ Indian Institute of Technology, Guwahati, Guwahati, Assam 781039, India}

{\sl $^{77}$ Indian Institute of Technology, Kanpur, Department of Physics,  IIT Post Office, Kanpur 208016, India}

{\sl $^{78}$ Indiana University - Purdue University, Indianapolis, Department of Physics, 402 N. Blackford St., LD 154, Indianapolis, IN 46202, USA}

{\sl $^{79}$ Indiana University, Department of Physics, Swain Hall West 117, 727 E. 3rd St., Bloomington, IN 47405-7105, USA}

{\sl $^{80}$ Institucio Catalana de Recerca i Estudis, ICREA,  Passeig Lluis Companys, 23, Barcelona 08010, Spain}

{\sl $^{81}$ Institut de Physique Nucl\'eaire, F-91406 Orsay, France }

{\sl $^{82}$ Institut f\"ur Theorie Elektromagnetischer Felder (TEMF), Technische Universit\"at Darmstadt, Schlo{\ss}gartenstr. 8, D-64289 Darmstadt, Germany}

{\sl $^{83}$ Institut National de Physique Nucleaire et de Physique des Particules, 3, Rue Michel- Ange, 75794 Paris Cedex 16, France}

{\sl $^{84}$ Institut Pluridisciplinaire Hubert Curien, 23 Rue du Loess - BP28, 67037 Strasbourg Cedex 2, France}

{\sl $^{85}$ Institute for Chemical Research, Kyoto University, Gokasho, Uji, Kyoto 611-0011, Japan}

{\sl $^{86}$ Institute for Cosmic Ray Research, University of Tokyo, 5-1-5 Kashiwa-no-Ha, Kashiwa, Chiba 277-8582, Japan}

{\sl $^{87}$ Institute of High Energy Physics - IHEP, Chinese Academy of Sciences, P.O. Box 918, Beijing, China 100049}

{\sl $^{88}$ Institute of Mathematical Sciences, Taramani, C.I.T. Campus, Chennai 600113, India}

{\sl $^{89}$ Institute of Physics and Electronics, Vietnamese Academy of Science and Technology (VAST), 10 Dao-Tan, Ba-Dinh, Hanoi 10000, Vietnam}

{\sl $^{90}$ Institute of Physics, ASCR, Academy of Science of the Czech Republic, Division of Elementary Particle Physics, Na Slovance 2, CS-18221 Prague 8, Czech Republic}

{\sl $^{91}$ Institute of Physics, Pomorska 149/153, PL-90-236 Lodz, Poland}

{\sl $^{92}$ Institute of Theoretical and Experimetal Physics, B. Cheremushkinskawa, 25, RU-117259, Moscow, Russia}

{\sl $^{93}$ Institute of Theoretical Physics, Chinese Academy of Sciences, P.O.Box 2735, Beijing, China 100080}

{\sl $^{94}$ Instituto de Fisica Corpuscular (IFIC), Centro Mixto CSIC-UVEG, Edificio Investigacion Paterna, Apartado 22085, 46071 Valencia, Spain}

{\sl $^{95}$ Instituto de Fisica de Cantabria, (IFCA, CSIC-UC), Facultad de Ciencias, Avda. Los Castros s/n, 39005 Santander, Spain}

{\sl $^{96}$ Instituto Nazionale di Fisica Nucleare (INFN), Laboratorio LASA, Via Fratelli Cervi 201, 20090 Segrate, Italy}

{\sl $^{97}$ Instituto Nazionale di Fisica Nucleare (INFN), Sezione di Ferrara, via Paradiso 12, I-44100 Ferrara, Italy}

{\sl $^{98}$ Instituto Nazionale di Fisica Nucleare (INFN), Sezione di Firenze, Via G. Sansone 1, I-50019 Sesto Fiorentino (Firenze), Italy}

{\sl $^{99}$ Instituto Nazionale di Fisica Nucleare (INFN), Sezione di Lecce, via Arnesano, I-73100 Lecce, Italy}

{\sl $^{100}$ Instituto Nazionale di Fisica Nucleare (INFN), Sezione di Napoli, Complesso Universit{\'a} di Monte Sant'Angelo,via, I-80126 Naples, Italy}

{\sl $^{101}$ Instituto Nazionale di Fisica Nucleare (INFN), Sezione di Pavia, Via Bassi 6, I-27100 Pavia, Italy}

{\sl $^{102}$ Instituto Nazionale di Fisica Nucleare (INFN), Sezione di Pisa, Edificio C - Polo Fibonacci Largo B. Pontecorvo, 3, I-56127 Pisa, Italy}

{\sl $^{103}$ Instituto Nazionale di Fisica Nucleare (INFN), Sezione di Torino, c/o Universit{\'a}' di Torino facolt{\'a}' di Fisica, via P Giuria 1, 10125 Torino, Italy}

{\sl $^{104}$ Instituto Nazionale di Fisica Nucleare (INFN), Sezione di Trieste, Padriciano 99, I-34012 Trieste (Padriciano), Italy}

{\sl $^{105}$ Inter-University Accelerator Centre, Aruna Asaf Ali Marg, Post Box 10502, New Delhi 110067, India}

{\sl $^{106}$ International Center for Elementary Particle Physics, University of Tokyo, Hongo 7-3-1, Bunkyo District, Tokyo 113-0033, Japan}

{\sl $^{107}$ Iowa State University, Department of Physics, High Energy Physics Group, Ames, IA 50011, USA}

{\sl $^{108}$ Jagiellonian University, Institute of Physics, Ul. Reymonta 4, PL-30-059 Cracow, Poland}

{\sl $^{109}$ Jamia Millia Islamia, Centre for Theoretical Physics, Jamia Nagar, New Delhi 110025, India}

{\sl $^{110}$ Jamia Millia Islamia, Department of Physics, Jamia Nagar, New Delhi 110025, India}

{\sl $^{111}$ Japan Aerospace Exploration Agency, Sagamihara Campus, 3-1-1 Yoshinodai, Sagamihara, Kanagawa 220-8510 , Japan}

{\sl $^{112}$ Japan Atomic Energy Agency, 4-49 Muramatsu, Tokai-mura, Naka-gun, Ibaraki 319-1184, Japan}

{\sl $^{113}$ Johannes Gutenberg Universit{\"a}t Mainz, Institut f{\"u}r Physik, 55099 Mainz, Germany}

{\sl $^{114}$ Johns Hopkins University, Applied Physics Laboratory, 11100 Johns Hopkins RD., Laurel, MD 20723-6099, USA}

{\sl $^{115}$ Joint Institute for Nuclear Research (JINR), Joliot-Curie 6, 141980, Dubna, Moscow Region, Russia}

{\sl $^{116}$ Kansas State University, Department of Physics, 116 Cardwell Hall, Manhattan, KS 66506, USA}

{\sl $^{117}$ KCS Corp., 2-7-25 Muramatsukita, Tokai, Ibaraki 319-1108, Japan}

{\sl $^{118}$ Kharkov Institute of Physics and Technology, National Science Center, 1, Akademicheskaya St., Kharkov, 61108, Ukraine}

{\sl $^{119}$ Kinki University, Department of Physics, 3-4-1 Kowakae, Higashi-Osaka, Osaka 577-8502, Japan}

{\sl $^{120}$ Kobe University, Faculty of Science, 1-1 Rokkodai-cho, Nada-ku, Kobe, Hyogo 657-8501, Japan}

{\sl $^{121}$ Kogakuin University, Department of Physics, Shinjuku Campus, 1-24-2 Nishi-Shinjuku, Shinjuku-ku, Tokyo 163-8677, Japan}

{\sl $^{122}$ Konkuk University, 93-1 Mojin-dong, Kwanglin-gu, Seoul 143-701, Korea}

{\sl $^{123}$ Korea Advanced Institute of Science \& Technology, Department of Physics, 373-1 Kusong-dong, Yusong-gu, Taejon 305-701, Korea}

{\sl $^{124}$ Korea Institute for Advanced Study (KIAS), School of Physics, 207-43 Cheongryangri-dong, Dongdaemun-gu, Seoul 130-012, Korea}

{\sl $^{125}$ Korea University, Department of Physics, Seoul 136-701, Korea}

{\sl $^{126}$ Kyoto University, Department of Physics, Kitashirakawa-Oiwakecho, Sakyo-ku, Kyoto 606-8502, Japan}

{\sl $^{127}$ L.P.T.A., UMR 5207 CNRS-UM2, Universit{\'e} Montpellier II, Case Courrier 070, B{\^a}t. 13, place Eug{\`e}ne Bataillon, 34095 Montpellier Cedex 5, France}

{\sl $^{128}$ Laboratoire d'Annecy-le-Vieux de Physique des Particules (LAPP), Chemin du Bellevue, BP 110, F-74941 Annecy-le-Vieux Cedex, France}

{\sl $^{129}$ Laboratoire d'Annecy-le-Vieux de Physique Theorique (LAPTH), Chemin de Bellevue, BP 110, F-74941 Annecy-le-Vieux Cedex, France}

{\sl $^{130}$ Laboratoire de l'Acc\'el\'erateur Lin\'eaire (LAL), Universit\'e Paris-Sud 11, B\^atiment 200, 91898 Orsay, France}

{\sl $^{131}$ Laboratoire de Physique Corpusculaire de Clermont-Ferrand (LPC), Universit\'e Blaise Pascal, I.N.2.P.3./C.N.R.S., 24 avenue des Landais, 63177 Aubi\`ere Cedex, France}

{\sl $^{132}$ Laboratoire de Physique Subatomique et de Cosmologie (LPSC), Universit\'e Joseph Fourier (Grenoble 1), 53, ave. des Marthyrs, F-38026 Grenoble Cedex, France}

{\sl $^{133}$ Laboratoire de Physique Theorique, Universit\'e de Paris-Sud XI, Batiment 210, F-91405 Orsay Cedex, France}

{\sl $^{134}$ Laboratori Nazionali di Frascati, via E. Fermi, 40, C.P. 13, I-00044 Frascati, Italy}

{\sl $^{135}$ Laboratory of High Energy Physics and Cosmology, Department of Physics, Hanoi National University, 334 Nguyen Trai, Hanoi, Vietnam}

{\sl $^{136}$ Lancaster University, Physics Department, Lancaster LA1 4YB, UK}

{\sl $^{137}$ Lawrence Berkeley National Laboratory (LBNL), 1 Cyclotron Rd, Berkeley, CA 94720, USA}

{\sl $^{138}$ Lawrence Livermore National Laboratory (LLNL), Livermore, CA 94551, USA}

{\sl $^{139}$ Lebedev Physical Institute, Leninsky Prospect 53, RU-117924 Moscow, Russia}

{\sl $^{140}$ Liaoning Normal University, Department of Physics, Dalian, China 116029}

{\sl $^{141}$ Lomonosov Moscow State University, Skobeltsyn Institute of Nuclear Physics (MSU SINP), 1(2), Leninskie gory, GSP-1, Moscow 119991, Russia}

{\sl $^{142}$ Los Alamos National Laboratory (LANL), P.O.Box 1663, Los Alamos, NM 87545, USA}

{\sl $^{143}$ Louisiana Technical University, Department of Physics, Ruston, LA 71272, USA}

{\sl $^{144}$ Ludwig-Maximilians-Universit{\"a}t M{\"u}nchen, Department f{\"u}r Physik, Schellingstr. 4, D-80799 Munich, Germany}

{\sl $^{145}$ Lunds Universitet, Fysiska Institutionen, Avdelningen f{\"o}r Experimentell H{\"o}genergifysik, Box 118, 221 00 Lund, Sweden}

{\sl $^{146}$ Massachusetts Institute of Technology, Laboratory for Nuclear Science \& Center for Theoretical Physics, 77 Massachusetts Ave., NW16, Cambridge, MA 02139, USA}

{\sl $^{147}$ Max-Planck-Institut f{\"u}r Physik (Werner-Heisenberg-Institut), F{\"o}hringer Ring 6, 80805 M{\"u}nchen, Germany}

{\sl $^{148}$ McGill University, Department of Physics, Ernest Rutherford Physics Bldg., 3600 University Ave., Montreal, Quebec, H3A 2T8 Canada}

{\sl $^{149}$ Meiji Gakuin University, Department of Physics, 2-37 Shirokanedai 1-chome, Minato-ku, Tokyo 244-8539, Japan}

{\sl $^{150}$ Michigan State University, Department of Physics and Astronomy, East Lansing, MI 48824, USA}

{\sl $^{151}$ Middle East Technical University, Department of Physics, TR-06531 Ankara, Turkey}

{\sl $^{152}$ Mindanao Polytechnic State College, Lapasan, Cagayan de Oro City 9000, Phillipines}

{\sl $^{153}$ MSU-Iligan Institute of Technology, Department of Physics, Andres Bonifacio Avenue, 9200 Iligan City, Phillipines}

{\sl $^{154}$ Nagasaki Institute of Applied Science, 536 Abamachi, Nagasaki-Shi, Nagasaki 851-0193, Japan}

{\sl $^{155}$ Nagoya University, Fundamental Particle Physics Laboratory, Division of Particle and Astrophysical Sciences, Furo-cho, Chikusa-ku, Nagoya, Aichi 464-8602, Japan}

{\sl $^{156}$ Nanchang University, Department of Physics, Nanchang, China 330031}

{\sl $^{157}$ Nanjing University, Department of Physics, Nanjing, China 210093}

{\sl $^{158}$ Nankai University, Department of Physics, Tianjin, China 300071}

{\sl $^{159}$ National Central University, High Energy Group, Department of Physics, Chung-li, Taiwan 32001}

{\sl $^{160}$ National Institute for Nuclear \& High Energy Physics, PO Box 41882, 1009 DB Amsterdam, Netherlands}

{\sl $^{161}$ National Institute of Radiological Sciences, 4-9-1 Anagawa, Inaga, Chiba 263-8555, Japan}

{\sl $^{162}$ National Synchrotron Radiation Laboratory, University of Science and Technology of china, Hefei, Anhui, China 230029}

{\sl $^{163}$ National Synchrotron Research Center, 101 Hsin-Ann Rd., Hsinchu Science Part, Hsinchu, Taiwan 30076}

{\sl $^{164}$ National Taiwan University, Physics Department, Taipei, Taiwan 106}

{\sl $^{165}$ Niels Bohr Institute (NBI), University of Copenhagen, Blegdamsvej 17, DK-2100 Copenhagen, Denmark}

{\sl $^{166}$ Niigata University, Department of Physics, Ikarashi, Niigata 950-218, Japan}

{\sl $^{167}$ Nikken Sekkai Ltd., 2-18-3 Iidabashi, Chiyoda-Ku, Tokyo 102-8117, Japan}

{\sl $^{168}$ Nippon Dental University, 1-9-20 Fujimi, Chiyoda-Ku, Tokyo 102-8159, Japan}

{\sl $^{169}$ North Asia University, Akita 010-8515, Japan}

{\sl $^{170}$ North Eastern Hill University, Department of Physics, Shillong 793022, India}

{\sl $^{171}$ Northern Illinois University, Department of Physics, DeKalb, Illinois 60115-2825, USA}

{\sl $^{172}$ Northwestern University, Department of Physics and Astronomy, 2145 Sheridan Road., Evanston, IL 60208, USA}

{\sl $^{173}$ Novosibirsk State University (NGU), Department of Physics, Pirogov st. 2, 630090 Novosibirsk, Russia}

{\sl $^{174}$ Obninsk State Technical University for Nuclear Engineering (IATE), Obninsk, Russia}

{\sl $^{175}$ Ochanomizu University, Department of Physics, Faculty of Science, 1-1 Otsuka 2, Bunkyo-ku, Tokyo 112-8610, Japan}

{\sl $^{176}$ Osaka University, Laboratory of Nuclear Studies, 1-1 Machikaneyama, Toyonaka, Osaka 560-0043, Japan}

{\sl $^{177}$ {\"O}sterreichische Akademie der Wissenschaften, Institut f{\"u}r Hochenergiephysik, Nikolsdorfergasse 18, A-1050 Vienna, Austria}

{\sl $^{178}$ Panjab University, Chandigarh 160014, India}

{\sl $^{179}$ Pavel Sukhoi Gomel State Technical University, ICTP Affiliated Centre \& Laboratory for Physical Studies, October Avenue, 48, 246746, Gomel, Belarus}

{\sl $^{180}$ Pavel Sukhoi Gomel State Technical University, Physics Department, October Ave. 48, 246746 Gomel, Belarus}

{\sl $^{181}$ Physical Research Laboratory, Navrangpura, Ahmedabad 380 009, Gujarat, India}

{\sl $^{182}$ Pohang Accelerator Laboratory (PAL), San-31 Hyoja-dong, Nam-gu, Pohang, Gyeongbuk 790-784, Korea}

{\sl $^{183}$ Polish Academy of Sciences (PAS), Institute of Physics, Al. Lotnikow 32/46, PL-02-668 Warsaw, Poland}

{\sl $^{184}$ Primera Engineers Ltd., 100 S Wacker Drive, Suite 700, Chicago, IL 60606, USA}

{\sl $^{185}$ Princeton University, Department of Physics, P.O. Box 708, Princeton, NJ 08542-0708, USA}

{\sl $^{186}$ Purdue University, Department of Physics, West Lafayette, IN 47907, USA}

{\sl $^{187}$ Pusan National University, Department of Physics, Busan 609-735, Korea}

{\sl $^{188}$ R. W. Downing Inc., 6590 W. Box Canyon Dr., Tucson, AZ 85745, USA}

{\sl $^{189}$ Raja Ramanna Center for Advanced Technology, Indore 452013, India}

{\sl $^{190}$ Rheinisch-Westf{\"a}lische Technische Hochschule (RWTH), Physikalisches Institut, Physikzentrum, Sommerfeldstrasse 14, D-52056 Aachen, Germany}

{\sl $^{191}$ RIKEN, 2-1 Hirosawa, Wako, Saitama 351-0198, Japan}

{\sl $^{192}$ Royal Holloway, University of London (RHUL), Department of Physics, Egham, Surrey TW20 0EX, UK }

{\sl $^{193}$ Saga University, Department of Physics, 1 Honjo-machi, Saga-shi, Saga 840-8502, Japan}

{\sl $^{194}$ Saha Institute of Nuclear Physics, 1/AF Bidhan Nagar, Kolkata 700064, India}

{\sl $^{195}$ Salalah College of Technology (SCOT), Engineering Department, Post Box No. 608, Postal Code 211, Salalah, Sultanate of Oman}

{\sl $^{196}$ Saube Co., Hanabatake, Tsukuba, Ibaraki 300-3261, Japan}

{\sl $^{197}$ Seoul National University, San 56-1, Shinrim-dong, Kwanak-gu, Seoul 151-742, Korea}

{\sl $^{198}$ Shandong University, 27 Shanda Nanlu, Jinan, China 250100}

{\sl $^{199}$ Shanghai Institute of Applied Physics, Chinese Academy of Sciences, 2019 Jiaruo Rd., Jiading, Shanghai, China 201800}

{\sl $^{200}$ Shinshu University, 3-1-1, Asahi, Matsumoto, Nagano 390-8621, Japan}

{\sl $^{201}$ Sobolev Institute of Mathematics, Siberian Branch of the Russian Academy of Sciences, 4 Acad. Koptyug Avenue, 630090 Novosibirsk, Russia}

{\sl $^{202}$ Sokendai, The Graduate University for Advanced Studies, Shonan Village, Hayama, Kanagawa 240-0193, Japan}

{\sl $^{203}$ Stanford Linear Accelerator Center (SLAC), 2575 Sand Hill Road, Menlo Park, CA 94025, USA}

{\sl $^{204}$ State University of New York at Binghamton, Department of Physics, PO Box 6016, Binghamton, NY 13902, USA}

{\sl $^{205}$ State University of New York at Buffalo, Department of Physics \& Astronomy, 239 Franczak Hall, Buffalo, NY 14260, USA}

{\sl $^{206}$ State University of New York at Stony Brook, Department of Physics and Astronomy, Stony Brook, NY 11794-3800, USA}

{\sl $^{207}$ Sumitomo Heavy Industries, Ltd., Natsushima-cho, Yokosuka, Kanagawa 237-8555, Japan}

{\sl $^{208}$ Sungkyunkwan University (SKKU), Natural Science Campus 300, Physics Research Division, Chunchun-dong, Jangan-gu, Suwon, Kyunggi-do 440-746, Korea}

{\sl $^{209}$ Swiss Light Source (SLS), Paul Scherrer Institut (PSI), PSI West, CH-5232 Villigen PSI, Switzerland}

{\sl $^{210}$ Syracuse University, Department of Physics, 201 Physics Building, Syracuse, NY 13244-1130, USA}

{\sl $^{211}$ Tata Institute of Fundamental Research, School of Natural Sciences, Homi Bhabha Rd., Mumbai 400005, India}

{\sl $^{212}$ Technical Institute of Physics and Chemistry, Chinese Academy of Sciences, 2 North 1st St., Zhongguancun, Beijing, China 100080}

{\sl $^{213}$ Technical University of Lodz, Department of Microelectronics and Computer Science, al. Politechniki 11, 90-924 Lodz, Poland}

{\sl $^{214}$ Technische Universit{\"a}t Dresden, Institut f{\"u}r Kern- und Teilchenphysik, D-01069 Dresden, Germany}

{\sl $^{215}$ Technische Universit{\"a}t Dresden, Institut f{\"u}r Theoretische Physik,D-01062 Dresden, Germany}

{\sl $^{216}$ Tel-Aviv University, School of Physics and Astronomy, Ramat Aviv, Tel Aviv 69978, Israel}

{\sl $^{217}$ Texas A\&M University, Physics Department, College Station, 77843-4242 TX, USA}

{\sl $^{218}$ Texas Tech University, Department of Physics, Campus Box 41051, Lubbock, TX 79409-1051, USA}

{\sl $^{219}$ The Henryk Niewodniczanski Institute of Nuclear Physics (NINP), High Energy Physics Lab, ul. Radzikowskiego 152, PL-31342 Cracow, Poland}

{\sl $^{220}$ Thomas Jefferson National Accelerator Facility (TJNAF), 12000 Jefferson Avenue, Newport News, VA 23606, USA}

{\sl $^{221}$ Tohoku Gakuin University, Faculty of Technology, 1-13-1 Chuo, Tagajo, Miyagi 985-8537, Japan}

{\sl $^{222}$ Tohoku University, Department of Physics, Aoba District, Sendai, Miyagi 980-8578, Japan}

{\sl $^{223}$ Tokyo Management College, Computer Science Lab, Ichikawa, Chiba 272-0001, Japan}

{\sl $^{224}$ Tokyo University of Agriculture Technology, Department of Applied Physics, Naka-machi, Koganei, Tokyo 183-8488, Japan}

{\sl $^{225}$ Toyama University, Department of Physics, 3190 Gofuku, Toyama-shi 930-8588, Japan}

{\sl $^{226}$ TRIUMF, 4004 Wesbrook Mall, Vancouver, BC V6T 2A3, Canada}

{\sl $^{227}$ Tufts University, Department of Physics and Astronomy, Robinson Hall, Medford, MA 02155, USA}

{\sl $^{228}$ Universidad Aut\`onoma de Madrid (UAM), Facultad de Ciencias C-XI, Departamento de Fisica Teorica, Cantoblanco, Madrid 28049, Spain}

{\sl $^{229}$ Universitat Aut\`onoma de Barcelona, Institut de Fisica d'Altes Energies (IFAE), Campus UAB, Edifici Cn, E-08193 Bellaterra, Barcelona, Spain}

{\sl $^{230}$ University College of London (UCL), High Energy Physics Group, Physics and Astronomy Department, Gower Street, London WC1E 6BT, UK }

{\sl $^{231}$ University College, National University of Ireland (Dublin), Department of Experimental Physics, Science Buildings, Belfield, Dublin 4, Ireland}

{\sl $^{232}$ University de Barcelona, Facultat de F\'isica, Av. Diagonal, 647, Barcelona 08028, Spain}

{\sl $^{233}$ University of Abertay Dundee, Department of Physics, Bell St, Dundee, DD1 1HG, UK}

{\sl $^{234}$ University of Auckland, Department of Physics, Private Bag, Auckland 1, New Zealand}

{\sl $^{235}$ University of Bergen, Institute of Physics, Allegaten 55, N-5007 Bergen, Norway}

{\sl $^{236}$ University of Birmingham, School of Physics and Astronomy, Particle Physics Group, Edgbaston, Birmingham B15 2TT, UK}

{\sl $^{237}$ University of Bristol, H. H. Wills Physics Lab, Tyndall Ave., Bristol BS8 1TL, UK}

{\sl $^{238}$ University of British Columbia, Department of Physics and Astronomy, 6224 Agricultural Rd., Vancouver, BC V6T 1Z1, Canada}

{\sl $^{239}$ University of California Berkeley, Department of Physics, 366 Le Conte Hall, \#7300, Berkeley, CA 94720, USA}

{\sl $^{240}$ University of California Davis, Department of Physics, One Shields Avenue, Davis, CA 95616-8677, USA}

{\sl $^{241}$ University of California Irvine, Department of Physics and Astronomy, High Energy Group, 4129 Frederick Reines Hall, Irvine, CA 92697-4575 USA}

{\sl $^{242}$ University of California Riverside, Department of Physics, Riverside, CA 92521, USA}

{\sl $^{243}$ University of California Santa Barbara, Department of Physics, Broida Hall, Mail Code 9530, Santa Barbara, CA 93106-9530, USA}

{\sl $^{244}$ University of California Santa Cruz, Department of Astronomy and Astrophysics, 1156 High Street, Santa Cruz, CA 05060, USA}

{\sl $^{245}$ University of California Santa Cruz, Institute for Particle Physics, 1156 High Street, Santa Cruz, CA 95064, USA}

{\sl $^{246}$ University of Cambridge, Cavendish Laboratory, J J Thomson Avenue, Cambridge CB3 0HE, UK}

{\sl $^{247}$ University of Colorado at Boulder, Department of Physics, 390 UCB, University of Colorado, Boulder, CO 80309-0390, USA}

{\sl $^{248}$ University of Delhi, Department of Physics and Astrophysics, Delhi 110007, India}

{\sl $^{249}$ University of Delhi, S.G.T.B. Khalsa College, Delhi 110007, India}

{\sl $^{250}$ University of Dundee, Department of Physics, Nethergate, Dundee, DD1 4HN,  Scotland, UK}

{\sl $^{251}$ University of Edinburgh, School of Physics, James Clerk Maxwell Building, The King's Buildings, Mayfield Road, Edinburgh EH9 3JZ, UK}

{\sl $^{252}$ University of Essex, Department of Physics, Wivenhoe Park, Colchester CO4 3SQ, UK}

{\sl $^{253}$ University of Florida, Department of Physics, Gainesville, FL 32611, USA}

{\sl $^{254}$ University of Glasgow, Department of Physics \& Astronomy, University Avenue, Glasgow G12 8QQ, Scotland, UK}

{\sl $^{255}$ University of Hamburg, Physics Department, Institut f{\"u}r Experimentalphysik, Luruper Chaussee 149, 22761 Hamburg, Germany}

{\sl $^{256}$ University of Hawaii, Department of Physics and Astronomy, HEP, 2505 Correa Rd., WAT 232, Honolulu, HI 96822-2219, USA}

{\sl $^{257}$ University of Heidelberg, Kirchhoff Institute of Physics, Albert {\"U}berle Strasse 3-5, DE-69120 Heidelberg, Germany}

{\sl $^{258}$ University of Helsinki, Department of Physical Sciences, P.O. Box 64 (Vaino Auerin katu 11), FIN-00014, Helsinki, Finland}

{\sl $^{259}$ University of Hyogo, School of Science, Kouto 3-2-1, Kamigori, Ako, Hyogo 678-1297,  Japan}

{\sl $^{260}$ University of Illinois at Urbana-Champaign, Department of Phys., High Energy Physics, 441 Loomis Lab. of Physics1110 W. Green St., Urbana, IL 61801-3080, USA}

{\sl $^{261}$ University of Iowa, Department of Physics and Astronomy, 203 Van Allen Hall, Iowa City, IA 52242-1479, USA}

{\sl $^{262}$ University of Kansas, Department of Physics and Astronomy, Malott Hall, 1251 Wescoe Hall Drive, Room 1082, Lawrence, KS 66045-7582, USA}

{\sl $^{263}$ University of Liverpool, Department of Physics, Oliver Lodge Lab, Oxford St., Liverpool L69 7ZE, UK}

{\sl $^{264}$ University of Louisville, Department of Physics, Louisville, KY 40292, USA}

{\sl $^{265}$ University of Manchester, School of Physics and Astronomy, Schuster Lab, Manchester M13 9PL, UK}

{\sl $^{266}$ University of Maryland, Department of Physics and Astronomy, Physics Building (Bldg. 082), College Park, MD 20742, USA}

{\sl $^{267}$ University of Melbourne, School of Physics, Victoria 3010, Australia}

{\sl $^{268}$ University of Michigan, Department of Physics, 500 E. University Ave., Ann Arbor, MI 48109-1120, USA}

{\sl $^{269}$ University of Minnesota, 148 Tate Laboratory Of Physics, 116 Church St. S.E., Minneapolis, MN 55455, USA}

{\sl $^{270}$ University of Mississippi, Department of Physics and Astronomy, 108 Lewis Hall, PO Box 1848, Oxford, Mississippi 38677-1848, USA}

{\sl $^{271}$ University of Montenegro, Faculty of Sciences and Math., Department of Phys., P.O. Box 211, 81001 Podgorica, Serbia and Montenegro}

{\sl $^{272}$ University of New Mexico, New Mexico Center for Particle Physics, Department of Physics and Astronomy, 800 Yale Boulevard N.E., Albuquerque, NM 87131, USA}

{\sl $^{273}$ University of Notre Dame, Department of Physics, 225 Nieuwland Science Hall, Notre Dame, IN 46556, USA}

{\sl $^{274}$ University of Oklahoma, Department of Physics and Astronomy, Norman, OK 73071, USA}

{\sl $^{275}$ University of Oregon, Department of Physics, 1371 E. 13th Ave., Eugene, OR 97403, USA}

{\sl $^{276}$ University of Oxford, Particle Physics Department, Denys Wilkinson Bldg., Keble Road, Oxford OX1 3RH England, UK }

{\sl $^{277}$ University of Patras, Department of Physics, GR-26100 Patras, Greece}

{\sl $^{278}$ University of Pavia, Department of Nuclear and Theoretical Physics, via Bassi 6, I-27100 Pavia, Italy}

{\sl $^{279}$ University of Pennsylvania, Department of Physics and Astronomy, 209 South 33rd Street, Philadelphia, PA 19104-6396, USA}

{\sl $^{280}$ University of Puerto Rico at Mayaguez, Department of Physics, P.O. Box 9016, Mayaguez, 00681-9016 Puerto Rico}

{\sl $^{281}$ University of Regina, Department of Physics, Regina, Saskatchewan, S4S 0A2 Canada}

{\sl $^{282}$ University of Rochester, Department of Physics and Astronomy, Bausch \& Lomb Hall, P.O. Box 270171, 600 Wilson Boulevard, Rochester, NY 14627-0171 USA}

{\sl $^{283}$ University of Science and Technology of China, Department of Modern Physics (DMP), Jin Zhai Road 96, Hefei, China 230026}

{\sl $^{284}$ University of Silesia, Institute of Physics, Ul. Uniwersytecka 4, PL-40007 Katowice, Poland}

{\sl $^{285}$ University of Southampton, School of Physics and Astronomy, Highfield, Southampton S017 1BJ, England, UK}

{\sl $^{286}$ University of Strathclyde, Physics Department, John Anderson Building, 107 Rottenrow, Glasgow, G4 0NG, Scotland, UK}

{\sl $^{287}$ University of Sydney, Falkiner High Energy Physics Group, School of Physics, A28, Sydney, NSW 2006, Australia}

{\sl $^{288}$ University of Texas, Center for Accelerator Science and Technology, Arlington, TX 76019, USA}

{\sl $^{289}$ University of Tokushima, Institute of Theoretical Physics, Tokushima-shi 770-8502, Japan}

{\sl $^{290}$ University of Tokyo, Department of Physics, 7-3-1 Hongo, Bunkyo District, Tokyo 113-0033, Japan}

{\sl $^{291}$ University of Toronto, Department of Physics, 60 St. George St., Toronto M5S 1A7, Ontario, Canada}

{\sl $^{292}$ University of Tsukuba, Institute of Physics, 1-1-1 Ten'nodai, Tsukuba, Ibaraki 305-8571, Japan}

{\sl $^{293}$ University of Victoria, Department of Physics and Astronomy, P.O.Box 3055 Stn Csc, Victoria, BC V8W 3P6, Canada}

{\sl $^{294}$ University of Warsaw, Institute of Physics, Ul. Hoza 69, PL-00 681 Warsaw, Poland}

{\sl $^{295}$ University of Warsaw, Institute of Theoretical Physics, Ul. Hoza 69, PL-00 681 Warsaw, Poland}

{\sl $^{296}$ University of Washington, Department of Physics, PO Box 351560, Seattle, WA 98195-1560, USA}

{\sl $^{297}$ University of Wisconsin, Physics Department, Madison, WI 53706-1390, USA}

{\sl $^{298}$ University of Wuppertal, Gau{\ss}stra{\ss}e 20, D-42119 Wuppertal, Germany}

{\sl $^{299}$ Universit\'e Claude Bernard Lyon-I, Institut de Physique Nucl\'eaire de Lyon (IPNL), 4, rue Enrico Fermi, F-69622 Villeurbanne Cedex, France}

{\sl $^{300}$ Universit\'e de Gen\`eve, Section de Physique, 24, quai E. Ansermet, 1211 Gen\`eve 4, Switzerland}

{\sl $^{301}$ Universit\'e Louis Pasteur (Strasbourg I), UFR de Sciences Physiques, 3-5 Rue de l'Universit\'e, F-67084 Strasbourg Cedex, France}

{\sl $^{302}$ Universit\'e Pierre et Marie Curie (Paris VI-VII) (6-7) (UPMC), Laboratoire de Physique Nucl\'eaire et de Hautes Energies (LPNHE), 4 place Jussieu, Tour 33, Rez de chausse, 75252 Paris Cedex 05, France}

{\sl $^{303}$ Universit{\"a}t Bonn, Physikalisches Institut, Nu{\ss}allee 12, 53115 Bonn, Germany}

{\sl $^{304}$ Universit{\"a}t Karlsruhe, Institut f{\"u}r Physik, Postfach 6980, Kaiserstrasse 12, D-76128 Karlsruhe, Germany}

{\sl $^{305}$ Universit{\"a}t Rostock, Fachbereich Physik, Universit{\"a}tsplatz 3, D-18051 Rostock, Germany}

{\sl $^{306}$ Universit{\"a}t Siegen, Fachbereich f{\"u}r Physik, Emmy Noether Campus, Walter-Flex-Str.3, D-57068 Siegen, Germany}

{\sl $^{307}$ Universit{\`a} de Bergamo, Dipartimento di Fisica, via Salvecchio, 19, I-24100 Bergamo, Italy}

{\sl $^{308}$ Universit{\`a} degli Studi di Roma La Sapienza, Dipartimento di Fisica, Istituto Nazionale di Fisica Nucleare, Piazzale Aldo Moro 2, I-00185 Rome, Italy}

{\sl $^{309}$ Universit{\`a} degli Studi di Trieste, Dipartimento di Fisica, via A. Valerio 2, I-34127 Trieste, Italy}

{\sl $^{310}$ Universit{\`a} degli Studi di ``Roma Tre'', Dipartimento di Fisica ``Edoardo Amaldi'', Istituto Nazionale di Fisica Nucleare, Via della Vasca Navale 84, 00146 Roma, Italy}

{\sl $^{311}$ Universit{\`a} dell'Insubria in Como, Dipartimento di Scienze CC.FF.MM., via Vallegio 11, I-22100 Como, Italy}

{\sl $^{312}$ Universit{\`a} di Pisa, Departimento di Fisica 'Enrico Fermi', Largo Bruno Pontecorvo 3, I-56127 Pisa, Italy}

{\sl $^{313}$ Universit{\`a} di Salento, Dipartimento di Fisica, via Arnesano, C.P. 193, I-73100 Lecce, Italy}

{\sl $^{314}$ Universit{\`a} di Udine, Dipartimento di Fisica, via delle Scienze, 208, I-33100 Udine, Italy}

{\sl $^{315}$ Variable Energy Cyclotron Centre, 1/AF, Bidhan Nagar, Kolkata 700064, India}

{\sl $^{316}$ VINCA Institute of Nuclear Sciences, Laboratory of Physics, PO Box 522, YU-11001 Belgrade, Serbia and Montenegro}

{\sl $^{317}$ Vinh University, 182 Le Duan, Vinh City, Nghe An Province, Vietnam}

{\sl $^{318}$ Virginia Polytechnic Institute and State University, Physics Department, Blacksburg, VA 2406, USA}

{\sl $^{319}$ Visva-Bharati University, Department of Physics, Santiniketan 731235, India}

{\sl $^{320}$ Waseda University, Advanced Research Institute for Science and Engineering, Shinjuku, Tokyo 169-8555, Japan}

{\sl $^{321}$ Wayne State University, Department of Physics, Detroit, MI 48202, USA}

{\sl $^{322}$ Weizmann Institute of Science, Department of Particle Physics, P.O. Box 26, Rehovot 76100, Israel}

{\sl $^{323}$ Yale University, Department of Physics, New Haven, CT 06520, USA}

{\sl $^{324}$ Yonsei University, Department of Physics, 134 Sinchon-dong, Sudaemoon-gu, Seoul 120-749, Korea}

{\sl $^{325}$ Zhejiang University, College of Science, Department of Physics, Hangzhou, China 310027}

{\sl * deceased } 

\end{center}

\end{center}

\chapter*{Acknowledgements} 
We would like to acknowledge the support and guidance of the International Committee on Future Accelerators (ICFA), chaired by A. Wagner of DESY, and the International Linear Collider Steering Committee (ILCSC), chaired by S. Kurokawa of KEK, who established the ILC Global Design Effort, as well as the World Wide Study of the Physics and Detectors.
      
\medskip
We are grateful to the ILC Machine Advisory Committee (MAC), chaired by F. Willeke of DESY and the International ILC Cost Review Committee, chaired by L. Evans of CERN, for their advice on the ILC Reference Design. We also thank the consultants who particpated in the Conventional Facilities Review at CalTech and in the RDR Cost Review at SLAC. 

\medskip 
We would like to thank the directors of the institutions who have hosted ILC meetings: KEK, ANL/FNAL/SLAC/U. Colorado (Snowmass), INFN/Frascati, IIT/Bangalore, TRIUMF/U. British Columbia, U. Valencia, IHEP/Beijing and DESY.

\medskip 
We are grateful for the support of the Funding Agencies for Large Colliders (FALC), chaired by R. Petronzio of INFN, and we thank all of the international, regional and national funding agencies whose generous support has made the ILC 
Reference Design possible.

\medskip 
Each of the GDE regional teams in the Americas, Asia and Europe are grateful for the support of their local scientific societies, industrial forums, advisory committees and reviewers.

%\begin{comment}
%\setcounter{tocdepth}{2)
\tableofcontents %
%\listoffigures %
%\listoftables

%\end{comment}
% Want to set counter for section numbering to 0 for preface (\textit{i.e.}, Executive Summary)
% This means that the sections will not be numbered, but will appear in the TOC and
% on the running headers (odd pages only).
\setcounter{secnumdepth}{0}

%\setlength{\parskip}{1ex plus0.2ex minus0.2ex} %
%\raggedright

%\chapter{\textsf{Executive Summary}}
%\renewcommand{\arraystretch}{1.0}
%\input{exec}

%\mainmatter %

% Want to reset counter for section numbering to 4 for rest of report
%\setcounter{secnumdepth}{4}
% Want to show section number in running head for rest of report
%\renewcommand{\sectionmark}[1]{\markright{\thesection.\ #1}}
%\chapter{\textsf{Steering Committee, Charge, Working Groups, Milestones, and Methodology}\label{chap1}}
%\renewcommand{\arraystretch}{1.0}
%\input{chap1}

%\begin{comment}
%\renewcommand{\thepage}{\arabic{chapter}.\arabic{section}-\arabic{page}}

\renewcommand{\thefigure}{\arabic{chapter}.\arabic{section}-\arabic{figlcl}}
\renewcommand{\thetable}{\arabic{chapter}.\arabic{section}-\arabic{tablcl}}
%\end{comment}

\setcounter{chapter}{0}
\newcommand{\picturefolder}{}

\setcounter{figlcl}{0}
\setcounter{tablcl}{0}
\setcounter{secnumdepth}{4} \setcounter{subsection}{0}
\setcounter{subsubsection}{0}
\usecounter{subsubsection}

\setcounter{section}{0} \renewcommand{\picturefolder}{./Overview/}

\chapter{\textsf{Overview}\label{chapIntro}}
\setcounter{page}{1}  \renewcommand{\thepage}{\arabic{page}}
\setcounter{figlcl}{0}
\setcounter{tablcl}{0}
\setcounter{secnumdepth}{4} \setcounter{subsection}{0}
\setcounter{subsubsection}{0}
\usecounter{subsubsection}

\setcounter{section}{0} \renewcommand{\picturefolder}{./Overview/}

\section{Introduction}

The International Linear Collider (ILC) is a 200-500 GeV
center-of-mass high-luminosity linear electron-positron collider,
based on 1.3 GHz superconducting radio-frequency (SCRF) accelerating
cavities. The use of the SCRF technology was recommended by
the International Technology Recommendation Panel (ITRP) in August
2004 \cite{itrp}, and shortly thereafter endorsed by the International
Committee for Future Accelerators (ICFA). In an unprecedented
milestone in high-energy physics, the many institutes around the world
involved in linear collider R\&D united in a common effort to produce a
global design for the ILC.  In November 2004, the 1st International
Linear Collider Workshop was held at KEK, Tsukuba, Japan. The workshop
was attended by some 200 physicists and engineers from around the world,
and paved the way for the 2nd ILC Workshop in August 2005, held at
Snowmass, Colorado, USA, where the ILC Global Design Effort (GDE) was
officially formed. The GDE membership reflects the global nature of
the collaboration, with accelerator experts from all three regions
(Americas, Asia and Europe). The first major goal of the GDE was to
define the basic parameters and layout of the machine -- the Baseline
Configuration. This was achieved at the first GDE meeting held at
INFN, Frascati, Italy in December 2005 with the creation of the
Baseline Configuration Document (BCD). During the next 14 months, the
BCD was used as the basis for the detailed design work and value
estimate (as described in Section~\ref{sec:ValueEstimate}) culminating in the completion
of the second major milestone, the publication of the draft ILC
Reference Design Report (RDR).

The technical design and cost estimate for the ILC is based on two
decades of world-wide Linear Collider R\&D, beginning with the
construction and operation of the SLAC Linear Collider (SLC). The SLC
is acknowledged as a proof-of-principle machine for the linear
collider concept. The ILC SCRF linac technology was pioneered by the
TESLA collaboration\footnote{Now known as the TESLA Technology
  Collaboration (TTC); see http://tesla.desy.de}, culminating in a
proposal for a 500 GeV center-of-mass linear collider in
2001~\cite{tdr}. The concurrent (competing) design work on a normal
conducting collider (NLC with X-band~\cite{nlc} and GLC with X- or
C-Band~\cite{glc}), has advanced the design concepts for the ILC
injectors, Damping Rings (DR) and Beam Delivery System (BDS), as well
as addressing overall operations, machine protection and availability
issues. The X- and C-band R\&D has led to concepts for RF power
sources that may eventually produce either cost and/or performance
benefits. Finally, the European XFEL \cite{euro-xfel} to be constructed at DESY,
Hamburg, Germany, will make use of the TESLA linac technology, and
represents a significant on-going R\&D effort of great
benefit for the ILC.

The ILC design has been developed to achieve the following physics
performance goals\cite{ilcsc-param}:

\begin{itemize}
  \item a continuous center-of-mass energy range between 200 GeV and 500 GeV;
\itemspace
  \item a peak luminosity of $\sim2\times10~{34}$ cm$^{-2}$s$^{-1}$,
    and an availability (75\%) consistent with producing 500 fb$^{-1}$
    in the first four years of operation\footnote{This assumes one
      additional year for commissioning, followed by a ramp up to the
      peak design performance over the four year operation period.} ;\itemspace

  \item $>80$\% electron polarization at the Interaction Point (IP);\itemspace

  \item an energy stability and precision of $\leq 0.1$\%;\itemspace

  \item an option for $\sim$60\% positron polarization;\itemspace

  \item options for $e^-$-$e^-$ and $\gamma$-$\gamma$ collisions.\itemspace

\end{itemize}

In addition, the machine must be upgradeable to a center-of-mass
energy of 1 TeV.

These goals guarantee a rich and varied program of physics. The energy
of the ILC will be sufficient to produce a very large number of $t
\overline{t}$ pairs, which will allow top-quark physics to be studied
with unprecedented precision. The energy range of the ILC spans all
predictions for the mass of a Standard Model Higgs boson based on the
precision electroweak data. Any supersymmetric particles found by LHC
will lead to a rich harvest of new phenomena at ILC; in addition, the
ILC has its own unique discovery capabilities which will be the only
way to produce a full picture of any of the new physics that might
exist at the Terascale. The ILC physics case has been endorsed by
recent major reviews conducted by distinguished scientists -- some
outside the field of particle physics -- in all three regions. The ILC
has established its place as the next major project on the world
particle physics roadmap.

\stepcounter{tablcl}\begin{table}[htb] \vbabove \caption[Basic
design parameters for the ILC.]{ Basic design parameters for the ILC
($^{a)}$ values at 500 GeV
  center-of-mass energy). \label{tab:BaseParams}}
\begin{center}
    \begin{tabular}{| l | l | l |} \hline
    Parameter                   & Unit              &          \\ \hline
    & & \vbdlspacing \hline
    Center-of-mass energy range & GeV               & 200 - 500         \\ \hline
    Peak luminosity$^{a)}$      & cm$^{-2}$s$^{-1}$ & $2\times 10^{34}$ \\ \hline
%                                &                   &                    \vbdlspacing  \hline
    Average beam current in pulse& mA                & 9.0               \\ \hline
    Pulse rate                  & Hz                & 5.0               \\ \hline
    Pulse length (beam)         & ms                & $\sim1$           \\ \hline
    Number of bunches per pulse &                   & 1000 - 5400       \\ \hline
    Charge per bunch            & nC                & 1.6 - 3.2         \\ \hline
    Accelerating gradient$^{a)}$      & MV/m        & 31.5              \\ \hline
    RF pulse length             & ms                & 1.6               \\ \hline
    Beam power (per beam)$^{a)}$      & MW          & 10.8              \\ \hline
%                                &                   &                    \vbdlspacing \hline
    Typical beam size at IP$^{a)}$ ($h\times v$) & nm & 640 $\times$ 5.7\\ \hline
%                               &                   &                    \vbdlspacing \hline
    Total AC Power consumption$^{a)}$ &MW           & 230               \\
    \hline
    \end{tabular}
\end{center}
\vbbelow
\end{table}

The current ILC baseline assumes an average accelerating gradient of
31.5 MV/m in the cavities to achieve a center-of-mass energy of 500
GeV. The high luminosity requires the use of high power and small
emittance beams. The choice of 1.3 GHz SCRF is well suited to the
requirements, primarily because the very low power loss in the SCRF
cavity walls allows the use of long RF pulses, relaxing the
requirements on the peak-power generation, and ultimately leading to
high wall-plug to beam transfer efficiency.

The primary cost drivers are the SCRF Main Linac technology and the
Conventional Facilities (including civil engineering). The choice of
gradient is a key cost and performance parameter, since it dictates
the length of the linacs, while the cavity quality factor %{\it }
($Q_0$) relates to the required cryogenic cooling power. The
achievement of 31.5 MV/m as the baseline average operational
accelerating gradient -- requiring a minimum performance of 35 MV/m
during cavity mass-production acceptance testing -- represents the
primary challenge to the global ILC R\&D

With the completion of the RDR, the GDE will begin an
engineering design study, closely coupled with a prioritized R\&D
program. The goal is to produce an Engineering Design Report (EDR) by
2010, presenting the matured technology, design and construction plan
for the ILC, allowing the world High Energy Physics community to seek
government-level project approvals, followed by start of construction
in 2012. When combined with the seven-year construction phase that is
assumed in studies presented in RDR, this timeline will allow
operations to begin in 2019. This is consistent with a technically
driven schedule for this international project.

\clearpage

\setcounter{figlcl}{0}
\setcounter{tablcl}{0}
\setcounter{secnumdepth}{4} \setcounter{subsection}{0}
\setcounter{subsubsection}{0}
\usecounter{subsubsection}

\section{Superconducting RF}

The primary cost driver for the ILC is the superconducting RF
technology used for the Main Linacs, bunch compressors and injector
linacs. In 1992, the TESLA Collaboration began R\&D on 1.3 GHz
technology with a goal of reducing the cost per MeV by a factor of 20
over the then state-of-the-art SCRF installation (CEBAF). This was
achieved by increasing the operating accelerating gradient by a factor
of five from ~5 MV/m to ~25 MV/m, and reducing the cost per meter of
the complete accelerating module by a factor of four for large-scale
production.

\stepcounter{figlcl}\begin{figure}[htb] \vbabove
\begin{center}
  \includegraphics[width=0.95\textwidth]{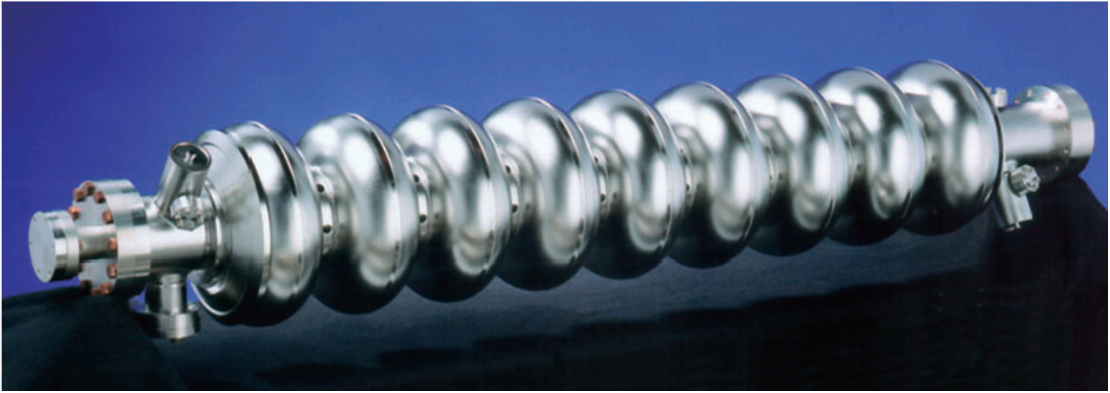}
  \vbabovecaption \caption{A TESLA nine-cell 1.3 GHz superconducting niobium cavity.}
  \label{fig:OVtesla9cell}
\end{center} \vbbelow
\end{figure}

The TESLA cavity R\&D was based on extensive existing experience from
CEBAF (Jefferson Lab), CERN, Cornell University, KEK, Saclay and
Wuppertal. The basic element of the technology is a nine-cell 1.3 GHz
niobium cavity, shown in Figure~\ref{fig:OVtesla9cell}. Approximately
160 of these cavities have been fabricated by industry as part of the
on-going R\&D program at DESY; some 17,000 are needed for the ILC.

\begin{figure}[htb] \vbabove
\begin{center}
  \includegraphics[width=0.95\textwidth]{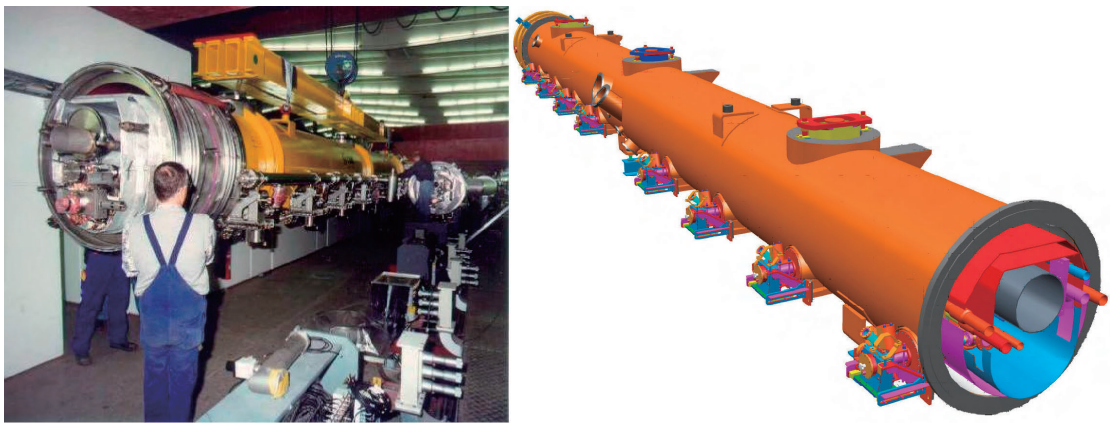}
 \vbabovecaption  \caption[SCRF Cryomodules.] {SCRF Cryomodules. Left: an 8     cavity TESLA cryomodule is
    installed into the FLASH linac at DESY. Right: design for the 4th
    generation ILC prototype cryomodule, due to be constructed at
    Fermilab National Laboratory.}
  \label{fig:OVcryomodule}
\end{center} \vbbelow
\end{figure}

A single cavity is approximately 1 m long. The cavities must be
operated at 2 K to achieve their performance. Eight or nine cavities are
mounted together in a string and assembled into a common
low-temperature cryostat or {\it cryomodule}
(Figure~\ref{fig:OVcryomodule}), the design of which is already in the
third generation. Ten cryomodules have been produced to-date, five of
which are currently installed in the in the VUV free-electron laser
(FLASH)\footnote{Originally known as the TESLA Test Facility (TTF).}
at DESY, where they are routinely operated. DESY is currently
preparing for the construction of the European XFEL facility, which
will have a $\sim20$ GeV superconducting linac containing 116
cryomodules.

\stepcounter{figlcl}

\begin{figure}[htb] \vbabove
\begin{center}
  \includegraphics[width=0.95\textwidth]{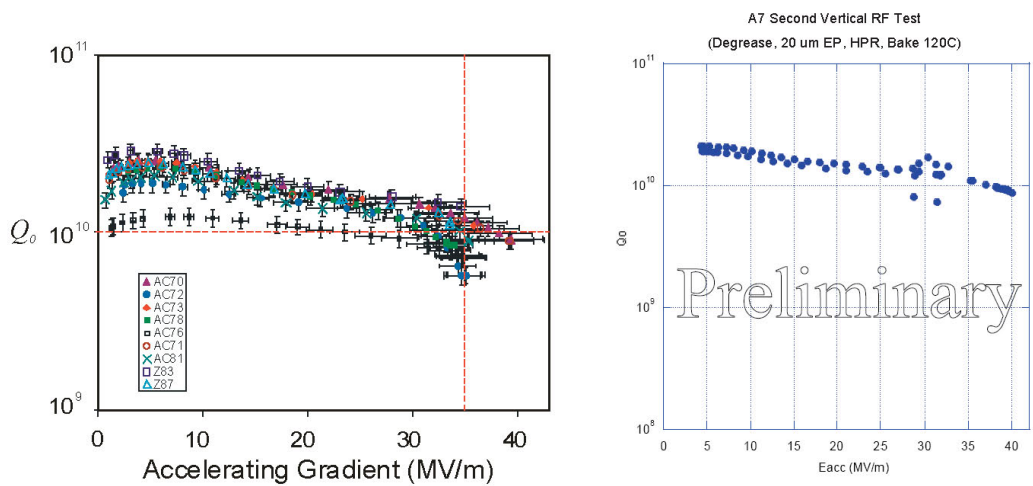}
  \vbabovecaption \caption[High-performance nine-cell cavities.]
    {High-performance nine-cell cavities. Left: Examples of DESY
    nine-cell cavities achieving $\ge 35$ MV/m. Right: Recent result
    from Jefferson Lab of nine-cell cavity achieving ~40 MV/m.  }
  \label{fig:OV9cellperf}
\end{center} \vbbelow
\end{figure}

The ILC community has set an aggressive goal of routinely
achieving\footnote{Acceptance test.} 35 MV/m in nine-cell cavities,
with a minimum production yield of 80\%. Several cavities have already
achieved these and higher gradients (see
Figure~\ref{fig:OV9cellperf}), demonstrating proof of
principle. Records of over 50 MV/m have been achieved in single-cell
cavities at KEK and Cornell\cite{high-g}. However, it is still a challenge
to achieve the desired production yield for nine-cell cavities
at the mass-production levels ($\sim$17,000 cavities) required.

\begin{figure}[htb] \vbabove
\begin{center}
  \includegraphics[width=0.95\textwidth]{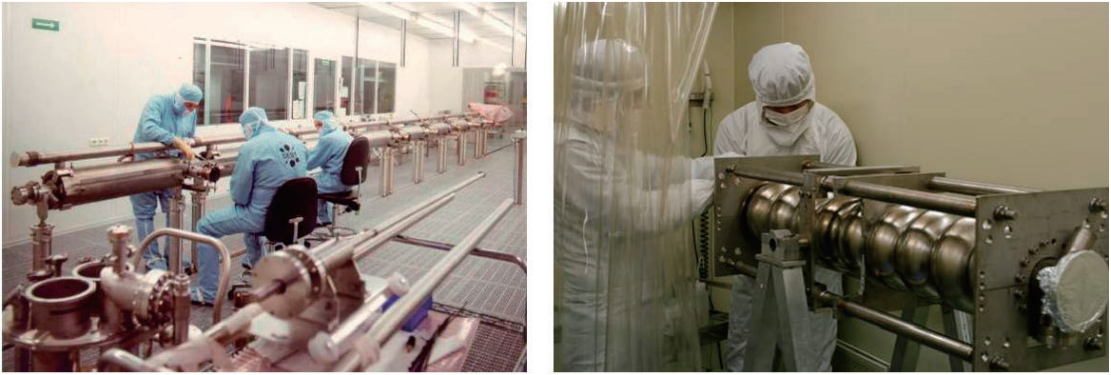}
  \vbabovecaption \caption[Clean room environments are mandatory.]
    {Clean room environments are mandatory. Left: the assembly
    of eight nine-cell TESLA cavities into a cryomodule string at
    DESY. Right: an ICHIRO nine-cell cavity is prepared for initial
    tests at the Superconducting RF Test Facility (STF) at KEK. }
  \label{fig:OVcleanrm}
\end{center} \vbbelow
\end{figure}

%\begin{comment}
\stepcounter{figlcl}
%\end{comment}
The key to high-gradient performance is the ultra-clean and
defect-free inner surface of the cavity. Both cavity preparation and
assembly into cavity strings for the cryomodules must be performed in
clean-room environments (Figure~\ref{fig:OVcleanrm}).

\stepcounter{figlcl}

\begin{figure}[htb] \vbabove
\begin{center}
  \includegraphics[width=0.95\textwidth]{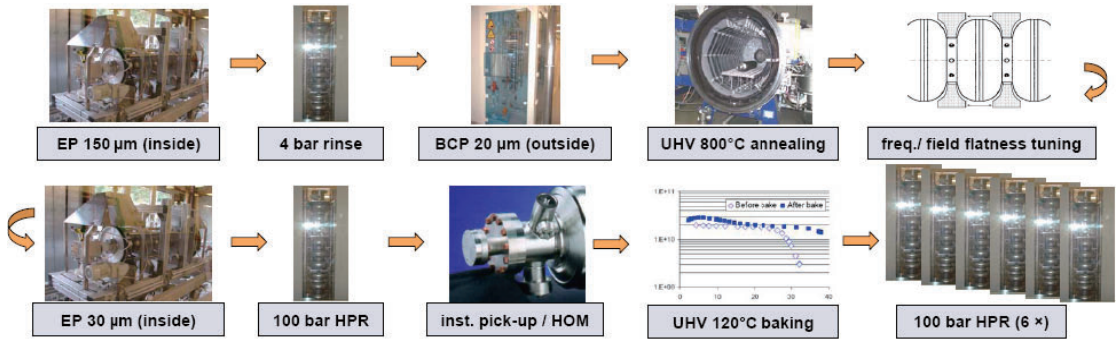}
  \vbabovecaption \caption[Birth of a nine-cell cavity.]
    {Birth of a nine-cell cavity: basic steps in surface
    treatment needed to achieve high-performance superconducting
    cavities. (EP = electropolishing; HPR = high-pressure rinsing.) }
  \label{fig:OVcavproc}
\end{center} \vbbelow
\end{figure}

The best
cavities have been achieved using electropolishing, a common industry
practice which was first developed for use with superconducting
cavities by CERN and KEK. Over the last few years, research at
Cornell, DESY, KEK and Jefferson Lab has led to an agreed standard
procedure for cavity preparation, depicted in
Figure~\ref{fig:OVcavproc}. The focus of the R\&D is now to optimize
the process to guarantee the required yield. The ILC SCRF community
has developed an internationally agreed-upon plan to address the
priority issues.

\stepcounter{figlcl}

The high-gradient SCRF R\&D required for ILC is expected to ramp-up
world-wide over the next years. The U.S. is currently investing in new
infrastructure for nine-cell cavity preparation and string and
cryomodule assembly. These efforts are centered at Fermilab (ILC Test
Accelerator, or ILCTA), together with ANL, Cornell University, SLAC
and Jefferson Lab. In Japan, KEK is developing the
Superconducting RF Test Facility (STF). In Europe, the focus of R\&D at
DESY has shifted to industrial preparation for construction of the
XFEL. There is continued R\&D to support the high-gradient program, as
well as other critical ILC-related R\&D such as high-power RF couplers
(LAL, Orsay, France) and cavity tuners (CEA Saclay, France; INFN
Milan, Italy).

The quest for high-gradient and affordable SCRF technology for
high-energy physics has revolutionized accelerator applications. In addition
to the recently completed Spallation Neutron Source (SNS) in Oak Ridge, Tennessee and the European XFEL under construction, many linac-based
projects utilizing
SCRF technology are being developed, including 4th-generation light
sources such as single-pass FELs and energy-recovery linacs. For the large
majority of new accelerator-based projects, SCRF has become the
technology of choice.

\clearpage

\setcounter{figlcl}{0}
\setcounter{tablcl}{0}
\setcounter{secnumdepth}{4} \setcounter{subsection}{0}
\setcounter{subsubsection}{0}
\usecounter{subsubsection}

\section{The ILC Baseline Design}\label{Ovr:baseline}

\begin{figure}[htb] \vbabove
\begin{center}
  \includegraphics[width=8.5cm]{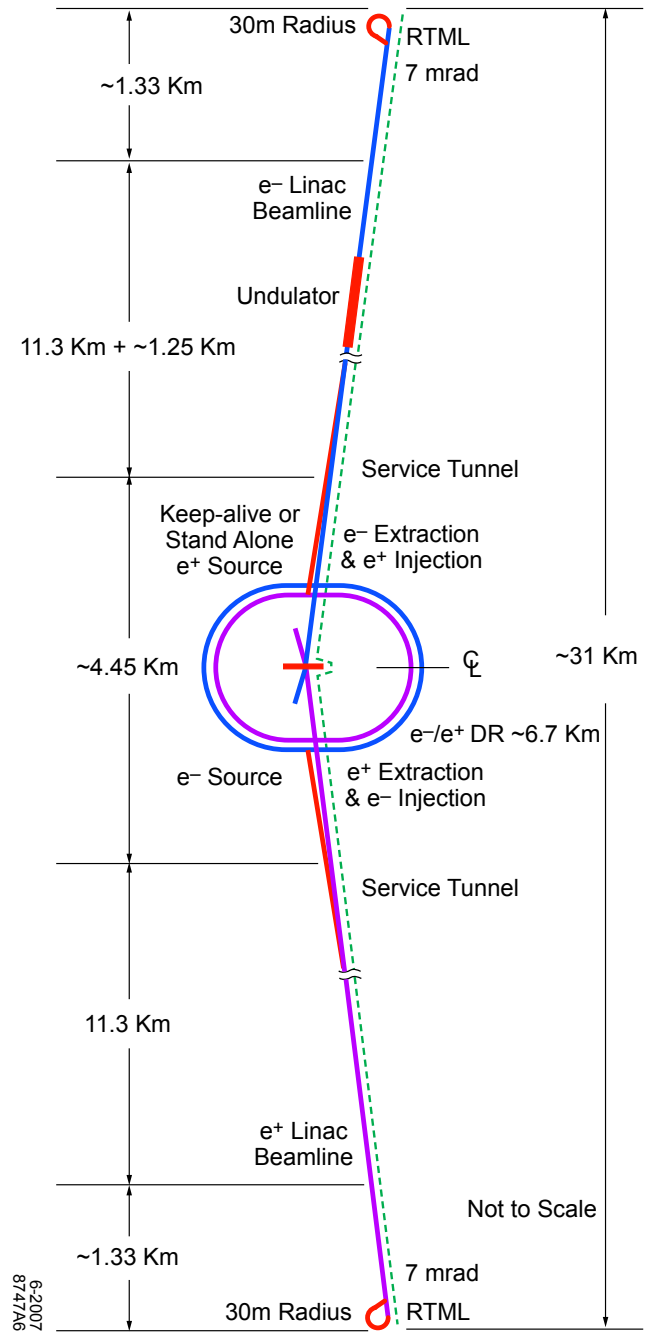}
 \vbabovecaption  \caption{Schematic layout of the ILC complex for 500 GeV CM.}
  \label{fig:OVlayout}
\end{center} \vbbelow
\end{figure}

The overall system design has been chosen to realize the physics
requirements with a maximum CM energy of 500 GeV and a peak luminosity
of $2\times10^{34}$ cm$^{-2}$s$^{-1}$. Figure~\ref{fig:OVlayout} shows
a schematic view of the overall layout of the ILC, indicating the
location of the major sub-systems:

\begin{itemize}

  \item a polarized electron source based on a photocathode DC gun;\itemspace
  \item an undulator-based positron source, driven by a the 150 GeV
    main electron beam;\itemspace
  \item 5 GeV electron and positron damping rings (DR) with a
    circumference of 6.7 km, housed in a common tunnel at the center
    of the ILC complex;\itemspace
  \item beam transport from the damping rings to the main linacs,
    followed by a two-stage bunch compressor system prior to injection into the
    main linac;\itemspace
  \item two 11 km long main linacs, utilizing 1.3 GHz SCRF cavities,
    operating at an average gradient of 31.5 MV/m, with a pulse length
    of 1.6 ms;\itemspace
  \item a 4.5 km long beam delivery system, which brings the two beams
    into collision with a 14 mrad crossing angle, at a single
    interaction point which can be shared by two detectors.\itemspace

\end{itemize}

\stepcounter{figlcl}

The total footprint is $\sim$31 km. The electron source, the damping
rings, and the positron auxiliary (`keep-alive') source are centrally
located around the interaction region (IR). The plane of the damping
rings is elevated by $\sim$10 m above that of the BDS to avoid
interference.

To upgrade the machine to $E_{\rm cms} =1$ TeV, the linacs and the beam
transport lines from the damping rings would be extended by another
$\sim 11$ km each. Certain components in the beam delivery system
would also need to be augmented or replaced.

\subsection{Beam Parameters}

The nominal beam parameter set, corresponding to the design luminosity
of $2\times10^{34}$ cm$^{-2}$s$^{-1}$ at $E_{\rm cms} = 500$ GeV is given in Table~\ref{tab:BaseParams}. These parameters have been chosen to optimize between known accelerator physics and technology
challenges throughout the whole accelerator complex. Examples of such
challenges are:

\begin{itemize}

\item beam instability and kicker hardware constraints in the
    damping rings; \itemspace
\item beam current, beam power, and pulse length limitations in the
    main linacs; \itemspace
\item emittance preservation requirements, in the main linacs and in
    the beam delivery system; \itemspace
\item background control and kink instability issues in the interaction region.
\itemspace

\end{itemize}

Nearly all high-energy physics accelerators have shown unanticipated
difficulties in reaching their design luminosity. The ILC design
specifies that each subsystem support a range of beam parameters. The
resulting flexibility in operating parameters will allow identified
problems in one area to be compensated for in another. The nominal IP
beam parameters and design ranges are presented in
Table~\ref{tab:OVparamrange}.

\stepcounter{tablcl}\begin{table}[htb] \vbabove \caption[Nominal and
design range of beam parameters at the IP.]
  {Nominal and design range of beam parameters at the IP. The
  min. and max. columns do not represent consistent sets of
  parameters, but only indicate the span of the design range for each
  parameter. (Nominal vertical emittance assumes a 100\% emittance
  dilution budget from the damping ring to the IP.)}
\begin{center}
\begin{tabular}{| l | r | r | r | l |} \hline
             &  min  & nominal. & max. & unit \\ \hline & & & & \vbdlspacing \hline
Bunch population                            & 1       & 2      & 2    & $\times10^{10}$ \\ \hline
Number of bunches                           & 1260    & 2625   & 5340 &                 \\ \hline
Linac bunch interval                        & 180     & 369    & 500  & ns              \\ \hline
RMS bunch length                            & 200     & 300    & 500  & $\mu$m          \\ \hline
Normalized horizontal emittance at IP       & 10      & 10     & 12   & mm$\cdot$mrad   \\ \hline
Normalized vertical emittance at IP         & 0.02    & 0.04   & 0.08 & mm$\cdot$mrad   \\ \hline
Horizontal beta function at IP              & 10      & 20     & 20   & mm              \\ \hline
Vertical beta function at IP                & 0.2     & 0.4    & 0.6  & mm              \\ \hline
RMS horizontal beam size at IP              & 474     & 640    & 640  & nm              \\ \hline
RMS vertical beam size at IP                & 3.5     & 5.7    & 9.9  & nm              \\ \hline
Vertical disruption parameter               & 14      & 19.4   & 26.1 &                 \\ \hline
Fractional RMS energy loss to beamstrahlung &   1.7   &  2.4  &   5.5 &  \% \\
\hline
\end{tabular}
\label{tab:OVparamrange}
\end{center} \vbbelow
\end{table}

\subsection{Electron Source}

\subparagraph{Functional Requirements} ~\\
The ILC polarized electron source must:
\begin{itemize}
  \item generate the required bunch train of polarized electrons
       ($>80\%$ polarization); \itemspace
  \item capture and accelerate the beam to 5 GeV; \itemspace
  \item transport the beam to the electron damping ring with minimal
       beam loss, and perform an energy compression and spin rotation
       prior to injection. \itemspace
\end{itemize}

\begin{figure}[htb] \vbabove
\begin{center}
  \includegraphics[width=0.95\textwidth]{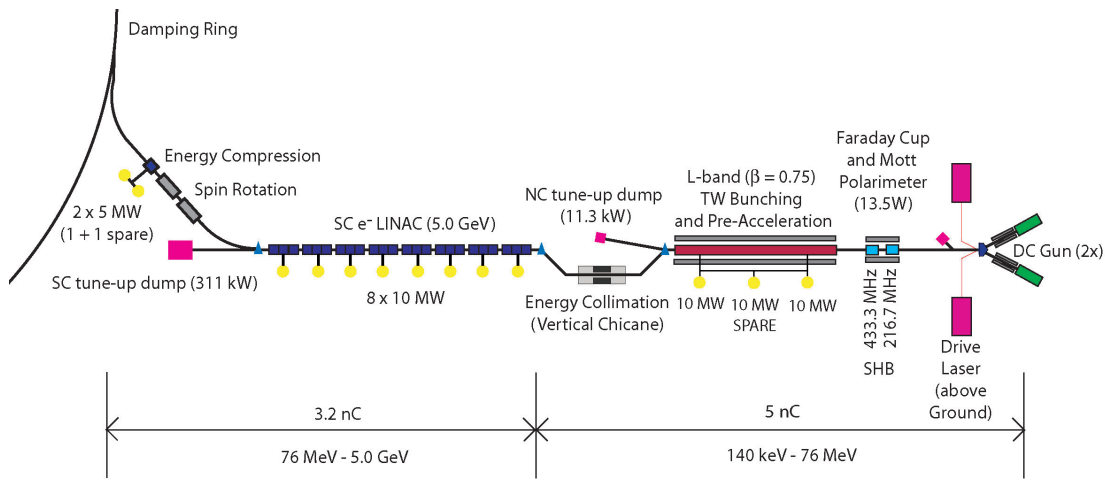}
  \vbabovecaption \caption{Schematic View of the Polarized Electron Source.}
  \label{fig:OVPES}
\end{center} \vbbelow
\end{figure}

\subparagraph{System Description} ~\\
The polarized electron source is located on the positron linac side of
the damping rings. The beam is produced by a laser illuminating a
photocathode in a DC gun. Two independent laser and gun systems
provide redundancy. Normal-conducting structures are used for bunching
and pre-acceleration to 76 MeV, after which the beam is accelerated to
5 GeV in a superconducting linac. Before injection into the damping
ring, superconducting solenoids rotate the spin vector into the
vertical, and a separate superconducting RF structure is used for
energy compression. The layout of the polarized electron source is
shown in Figure~\ref{fig:OVPES}.

\stepcounter{figlcl}

\subparagraph{Challenges} ~\\
The SLC polarized electron source already meets the requirements for
polarization, charge and lifetime. The primary challenge for the ILC
electron source is the ~1 ms long bunch train, which demands a laser
system beyond that used at any existing accelerator.

\subsection{Positron Source}

\subparagraph{Functional requirements} ~\\
The positron source must perform several critical functions:

\begin{itemize}

  \item generate a high-power multi-MeV photon production drive beam
    in a suitably short-period, high K-value helical undulator; \itemspace

  \item produce the needed positron bunches in a metal target that can
    reliably deal with the beam power and induced radioactivity; \itemspace

  \item capture and accelerate the beam to 5 GeV; \itemspace

  \item transport the beam to the positron damping ring with minimal
    beam loss, and perform an energy compression and spin rotation
    prior to injection. \itemspace
\end{itemize}

\begin{figure}[htb] \vbabove
\begin{center}
  \includegraphics[width=0.95\textwidth]{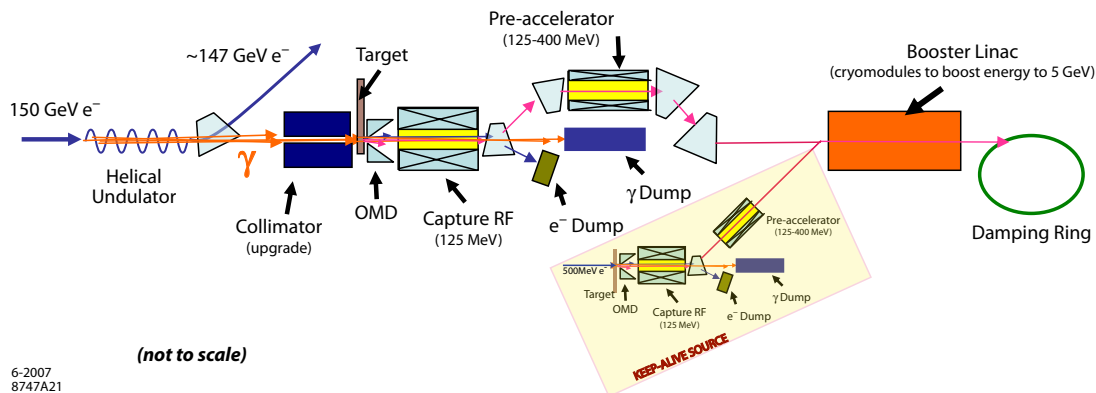}
 \vbabovecaption  \caption{Overall Layout of the Positron Source.}
  \label{fig:OVposi}
\end{center} \vbbelow
\end{figure}

\subparagraph{System Description} ~\\
The major elements of the ILC positron source are shown in
Figure~\ref{fig:OVposi}. The source uses photoproduction to generate
positrons. After acceleration to 150 GeV, the electron beam is
diverted into an offset beamline, transported through a 150-meter
helical undulator, and returned to the electron linac. The high-energy
($\sim$10 MeV) photons from the undulator are directed onto a rotating
0.4 radiation-length Ti-alloy target $\sim$500 meters downstream,
producing a beam of electron and positron pairs. This beam is then
matched using an optical-matching device into a normal conducting (NC)
L-band RF and solenoidal-focusing capture system and accelerated to
125 MeV. The electrons and remaining photons are separated from the
positrons and dumped. The positrons are accelerated to 400 MeV in a NC
L-band linac with solenoidal focusing. The beam is transported ~5 km
through the rest of the electron main linac tunnel, brought to the
central injector complex, and accelerated to 5 GeV using
superconducting L-band RF. Before injection into the damping ring,
superconducting solenoids rotate the spin vector into the vertical,
and a separate superconducting RF structure is used for energy
compression.

\stepcounter{figlcl}

The baseline design is for unpolarized positrons, although the beam
has a polarization of 30\%, and beamline space has been reserved for
an eventual upgrade to 60\% polarization.

To allow commissioning and tuning of the positron systems while the
high-energy electron beam is not available, a low-intensity auxiliary
(or ``keep-alive'') positron source is provided. This is a
conventional positron source, which uses a 500 MeV electron beam
impinging on a heavy-metal target to produce $\sim$10\% of the nominal
positron beam.  The keep-alive and primary sources use the same linac
to accelerate from 400 MeV to 5 GeV.

\subparagraph{Challenges} ~\\
The most challenging elements of the positron source are:

\begin{itemize}

  \item the 150 m long superconducting helical undulator, which has a
    period of 1.15 cm and a K-value of 0.92, and a 6 mm inner diameter
    vacuum chamber; \itemspace

  \item the Ti-alloy target, which is a cylindrical wheel 1.4 cm thick
    and 1 m in diameter, which must rotate at 100 m/s in vacuum to
    limit damage by the photon beam; \itemspace

  \item the normal-conducting RF system which captures the positron
    beam, which must sustain high accelerator gradients during
    millisecond-long pulses in a strong magnetic field, while
    providing adequate cooling in spite of high RF and particle-loss
    heating. \itemspace

\end{itemize}

The target and capture sections are also high-radiation areas which
present remote handing challenges.

\subsection{Damping Rings}

\subparagraph{Functional requirements} ~\\
The damping rings must perform four critical functions:

\begin{itemize}

  \item accept $e^-$ and $e^+$ beams with large transverse and
    longitudinal emittances and damp to the low emittance beam
    required for luminosity production (by five orders of magnitude
    for the positron vertical emittance), within the 200 ms between
    machine pulses; \itemspace

  \item inject and extract individual bunches without affecting the
    emittance or stability of the remaining stored bunches; \itemspace

  \item damp incoming beam jitter (transverse and longitudinal) and
    provide highly stable beams for downstream systems; \itemspace

  \item delay bunches from the source to allow feed-forward systems to
    compensate for pulse-to-pulse variations in parameters such as the
    bunch charge. \itemspace

\end{itemize}

\subparagraph{System Description} ~\\
The ILC damping rings include one electron and one positron ring, each
6.7 km long, operating at a beam energy of 5 GeV. The two rings are
housed in a single tunnel near the center of the site, with one ring
positioned directly above the other. The plane of the DR tunnel is
located $\sim$10 m higher than that of the beam delivery system.  This
elevation difference gives adequate shielding to allow operation of
the injector system while other systems are open to human access.

The damping ring lattice is divided into six arcs and six straight
sections. The arcs are composed of TME cells; the straight sections
use a FODO lattice. Four of the straight sections contain the RF
systems and the superconducting wigglers. The remaining two sections
are used for beam injection and extraction. Except for the wigglers,
all of the magnets in the ring, are normal-conducting. Approximately
200 m of superferric wigglers are used in each damping ring. The
wigglers are 2.5 m long devices, operating at 4.5K, with a peak field
of 1.67 T.

The superconducting RF system is operated CW at 650 MHz, and provides
24 MV for each ring. The frequency is chosen to be half the linac RF frequency to easily accommodate different bunch patterns.  The single-cell cavities
operate at 4.5 K and are housed in eighteen 3.5 m long
cryomodules. Although a number of 500 MHz CW RF systems are currently
in operation, development work is required for this 650 MHz system,
both for cavities and power sources.

The momentum compaction of the lattice is relatively large, which
helps to maintain single bunch stability, but requires a relatively
high RF voltage to achieve the design RMS bunch length (9 mm). The
dynamic aperture of the lattice is sufficient to allow the large
emittance injected beam to be captured with minimal loss.

\subparagraph{Challenges} ~\\
The principal challenges in the damping ring are:
\begin{itemize}

  \item control of the electron cloud effect in the positron damping
    ring. This effect, which can cause instability, tune spread, and
    emittance growth, has been seen in a number of other rings and is
    relatively well understood. Simulations indicate that it can be
    controlled by proper surface treatment of the vacuum chamber to
    suppress secondary emission, and by the use of solenoids and
    clearing electrodes to suppress the buildup of the cloud. \itemspace

  \item control of the fast ion instability in the electron damping
    ring. This effect can be controlled by limiting the pressure in
    the electron damping ring to below 1 nTorr, and by the use of
    short gaps in the ring fill pattern. \itemspace

  \item development of a very fast rise and fall time kicker for single
    bunch injection and extraction in the ring. For the most demanding
    region of the beam parameter range, the bunch spacing in the
    damping ring is $\sim$3 ns, and the kicker must have a rise plus fall
    time no more than twice this. Short stripline kicker structures
    can achieve this, but the drive pulser technology still needs
    development. \itemspace

\end{itemize}

\subsection{Ring to Main Linac (RTML)}
\subparagraph{Functional requirements} ~\\
The RTML must perform several critical functions for each beam:

\begin{itemize}

  \item transport the beam from the damping ring to the upstream end
    of the linac; \itemspace

  \item collimate the beam halo generated in the damping ring;

  \item rotate the polarization from the vertical to any arbitrary
    angle required at the IP; \itemspace

  \item compress the long Damping Ring bunch length by a factor of
    $30\sim45$ to provide the short bunches required by the Main Linac
    and the IP; \itemspace

\end{itemize}

\stepcounter{figlcl}\begin{figure}[htb] \vbabove
\begin{center}
  \includegraphics[width=0.95\textwidth]{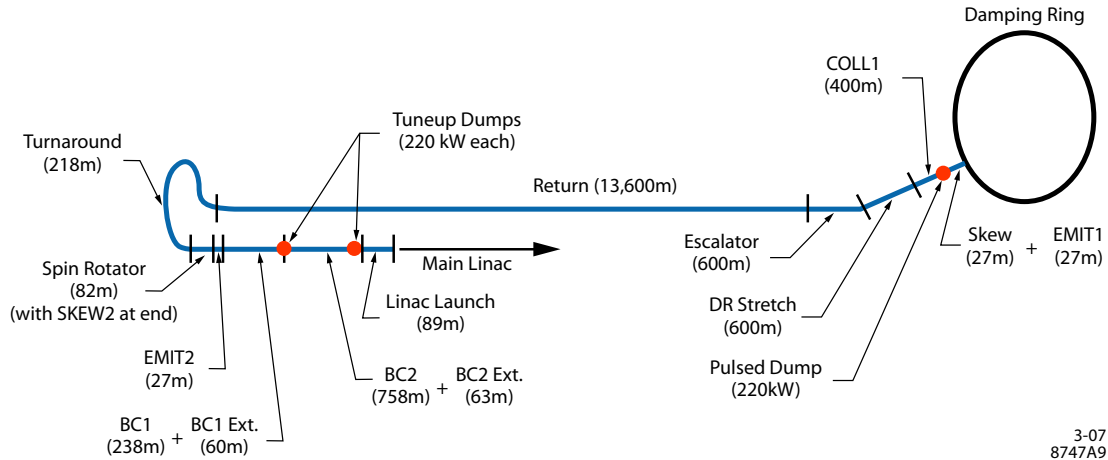}
  \vbabovecaption \caption{Schematic of the RTML.}
  \label{fig:OVRTML}
\end{center} \vbbelow
\end{figure}

\subparagraph{System Description} ~\\
The layout of the RTML is identical for both electrons and positrons,
and is shown in Figure~\ref{fig:OVRTML}. The RTML consists of the
following subsystems:

\begin{itemize}
  \item an $\sim$15 km long 5 GeV transport line; \itemspace

  \item betatron and energy collimation systems; \itemspace

  \item a $180^{\circ}$ turn-around, which enables feed-forward beam
    stabilization; \itemspace

  \item spin rotators to orient the beam polarization to the desired
    direction; \itemspace

  \item a 2-stage bunch compressor to compress the beam bunch length
    from several millimeters to a few hundred microns as required at
    the IP. \itemspace

\end{itemize}

The bunch compressor includes acceleration from 5 GeV to 13-15 GeV in
order to limit the increase in fractional energy spread associated
with bunch compression.

\subparagraph{Challenges} ~\\
The principal challenges in the RTML are:
\begin{itemize}

  \item control of emittance growth due to static misalignments,
    resulting in dispersion and coupling. Simulations indicate that
    the baseline design for beam-based alignment can limit the
    emittance growth to tolerable levels. \itemspace

  \item suppression of phase and amplitude jitter in the bunch
    compressor RF, which can lead to timing errors at the IP. RMS
    phase jitter of 0.24$^{\circ}$ between the electron and positron
    RF systems results in a 2\% loss of luminosity. Feedback loops in
    the bunch compressor low-level RF system should be able to limit
    the phase jitter to this level. \itemspace
\end{itemize}

\subsection{Main Linacs}

\subparagraph{Functional requirements} ~\\
The two main linacs accelerate the electron and positron beams from
their injected energy of 15 GeV to the final beam energy of 250 GeV,
over a combined length of 23 km. The main linacs must:

\begin{itemize}

  \item accelerate the beam while preserving the small bunch
    emittances, which requires precise orbit control based on data
    from high resolution beam position monitors, and also requires
    control of higher-order modes in the accelerating cavities; \itemspace

  \item maintain the beam energy spread within the design requirement
    of $\sim$0.1~\% at the IP; \itemspace

  \item not introduce significant transverse or longitudinal jitter,
    which could cause the beams to miss at the collision point. \itemspace

\end{itemize}

\subparagraph{System description} ~\\
The ILC Main Linacs accelerate the beam from 15 GeV to a maximum
energy of 250 GeV at an average accelerating gradient of 31.5
MV/m. The linacs are composed of RF units, each of which are formed by
three contiguous SCRF cryomodules containing 26 nine-cell
cavities. The layout of one unit is illustrated in
Figure~\ref{fig:OVRFunit}. The positron linac contains 278 RF units,
and the electron linac has 282 RF units\footnote{Approximately 3 GeV of
  extra energy is required in the electron linac to compensate for
  positron production.}.

\stepcounter{figlcl}\begin{figure}[htb] \vbabove
\begin{center}
  \includegraphics[width=0.95\textwidth]{\picturefolder RFunit.pdf}
  \vbabovecaption \caption{RF unit layout.}
  \label{fig:OVRFunit}
\end{center} \vbbelow
\end{figure}

Each RF unit has a stand-alone RF source, which includes a
conventional pulse-transformer type high-voltage (120 kV) modulator, a
10 MW multi-beam klystron, and a waveguide system that distributes the
RF power to the cavities (see Figure~\ref{fig:OVRFunit}). It also
includes the low-level RF (LLRF) system to regulate the cavity field
levels, interlock systems to protect the source components, and the
power supplies and support electronics associated with the operation
of the source.

The cryomodule design is a modification of the Type-3 version
(Figure~\ref{fig:OVcryomodule}) developed and used at DESY. Within the
cryomodules, a 300 mm diameter helium gas return pipe serves as a
strongback to support the cavities and other beam line components.
The middle cryomodule in each RF unit contains a quad package that
includes a superconducting quadrupole magnet at the center, a cavity
BPM, and superconducting horizontal and vertical corrector
magnets. The quadrupoles establish the main linac magnetic lattice,
which is a weak focusing FODO optics with an average beta function of
$\sim$80 m. All cryomodules are 12.652 m long, so the active-length to
actual-length ratio in a nine-cavity cryomodule is 73.8\%. Every
cryomodule also contains a 300 mm long high-order mode beam absorber
assembly that removes energy through the 40-80 K cooling system from
beam-induced higher-order modes above the cavity cutoff frequency.

To operate the cavities at 2 K, they are immersed in a saturated He II
bath, and helium gas-cooled shields intercept thermal radiation and
thermal conduction at 5-8 K and at 40-80 K. The estimated static and
dynamic cryogenic heat loads per RF unit at 2 K are 5.1 W and 29 W,
respectively. Liquid helium for the main linacs and the RTML is
supplied from 10 large cryogenic plants, each of which has an
installed equivalent cooling power of $\sim$20 kW at 4.5 K. The main
linacs follow the average Earth's curvature to simplify the liquid
helium transport.

\begin{figure}[htb] \vbabove
\begin{center}
  \includegraphics[width=0.95\textwidth]{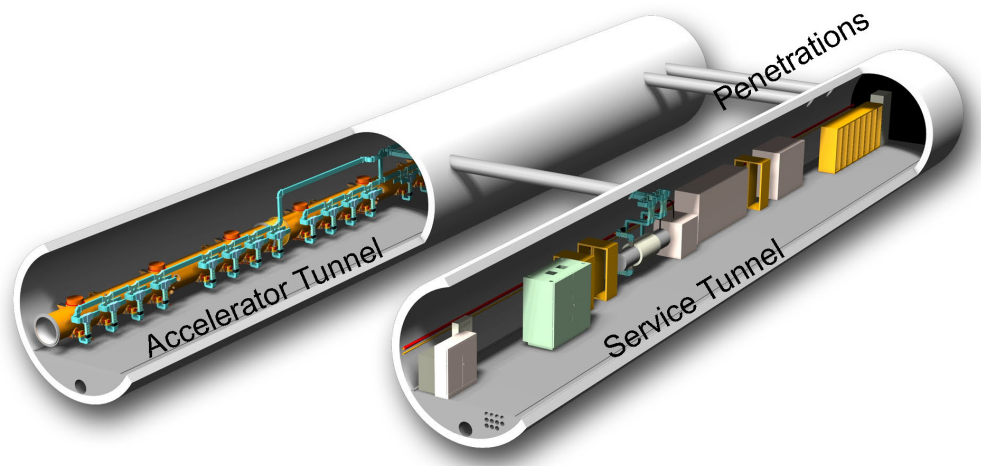}
  \vbabovecaption \caption{Cutaway view of the linac dual-tunnel configuration.}
  \label{fig:OVtunnel}
\end{center} \vbbelow
\end{figure}

The Main Linac components are housed in two tunnels, an accelerator
tunnel and a service tunnel, each of which has an interior diameter of
4.5 meters. To facilitate maintenance and limit radiation exposure,
the RF source is housed mainly in the service tunnel as illustrated in
Figure~\ref{fig:OVtunnel}.

\stepcounter{figlcl}

The tunnels are typically hundreds of meters underground and are
connected to the surface through vertical shafts\footnote{Except for
the Asian sample site: see Section~\ref{sect:OVRss}.}. Each of the main linacs
includes three shafts, roughly 5 km apart as dictated by the cryogenic
system. The upstream shafts in each linac have diameters of 14 m to
accommodate lowering cryomodules horizontally, and the downstream
shaft in each linac is 9 m in diameter, which is the minimum size
required to accommodate tunnel boring machines. At the base of each
shaft is a 14,100 cubic meter cavern for staging installation; it also houses utilities and parts of the cryoplant, most of which are
located on the surface.

\subparagraph{Challenges} ~\\
The principal challenges in the main linac are:
\begin{itemize}

  \item achieving the design average accelerating gradient of 31.5
    MV/m. This operating gradient is higher than that typically
    achievable today and assumes further progress will be made during
    the next few years in the aggressive program that is being pursued
    to improve cavity performance. \itemspace

  \item control of emittance growth due to static misalignments,
    resulting in dispersion and coupling. Beam-based alignment
    techniques should be able to limit the single-bunch emittance
    growth. Long-range multibunch effects are mitigated via HOM
    damping ports on the cavities, HOM absorbers at the quadrupoles,
    and HOM detuning. Coupling from mode-rotation HOMs is limited by
    splitting the horizontal and vertical betatron tunes. \itemspace

  \item control of the beam energy spread. The LLRF system monitors
    the vector sum of the fields in the 26 cavities of each RF unit
    and makes adjustments to flatten the energy gain along the bunch
    train and maintain the beam-to-RF phase constant. Experience from
    FLASH and simulations indicate that the baseline system should
    perform to specifications. \itemspace

\end{itemize}

%-----------------------------------------------------

\subsection{Beam Delivery System}

\subparagraph{Functional requirements} ~\\
The ILC Beam Delivery System (BDS) is responsible for transporting the
$e^+e^-$ beams from the exit of the high energy linacs, focusing them to
the sizes required to meet the ILC luminosity goals, bringing them
into collision, and then transporting the spent beams to the main beam
dumps. In addition, the BDS must perform several other critical
functions:

\begin{itemize}

  \item measure the linac beam and match it into the final focus; \itemspace

  \item protect the beamline and detector against mis-steered beams
    from the main linacs; \itemspace

  \item remove any large amplitude particles (beam-halo) from the
    linac to minimize background in the detectors; \itemspace

  \item measure and monitor the key physics parameters such as energy
    and polarization before and after the collisions. \itemspace

\end{itemize}

\begin{figure}[htb] \vbabove
\begin{center}
  \includegraphics[width=0.95\textwidth]{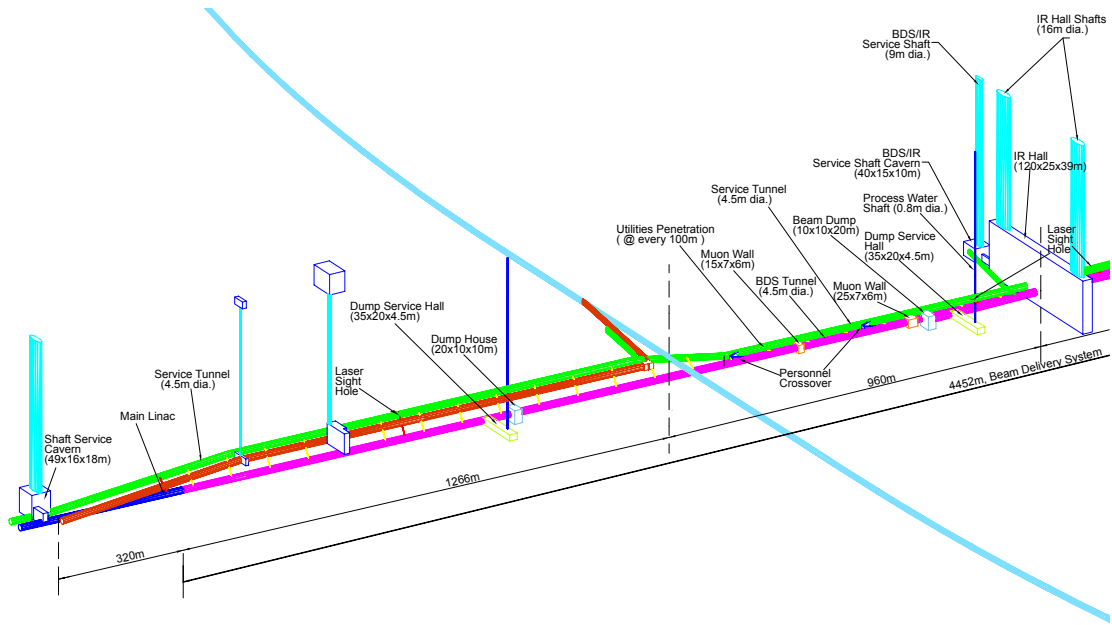}
 \vbabovecaption  \caption[BDS layout, beam and service tunnels.]
    {BDS layout, beam and service tunnels (shown in magenta and
    green), shafts, experimental hall.  The line crossing the BDS
    beamline at right angles is the damping ring, located 10 m above
    the BDS tunnels.}
  \label{fig:OVBDS}
\end{center} \vbbelow
\end{figure}

\subparagraph{System Description} ~\\
The layout of the beam delivery system is shown in
Figure~\ref{fig:OVBDS}. There is a single collision point with a 14
mrad total crossing angle. The 14 mrad geometry provides space for
separate extraction lines but requires crab cavities to rotate the
bunches in the horizontal plane for effective head-on
collisions. There are two detectors in a common interaction region
(IR) hall in a so-called ``push-pull'' configuration. The detectors
are pre-assembled on the surface and then lowered into the IR hall
when the hall is ready for occupancy.

The BDS is designed for 500 GeV center-of-mass energy but can be
upgraded to 1 TeV with additional magnets.

The main subsystems of the beam delivery, starting from the exit of
the main linacs, are:

\stepcounter{figlcl}

The BDS is designed for 500 GeV center-of-mass energy but can be
upgraded to 1 TeV with additional magnets.

The main subsystems of the beam delivery, starting from the exit of
the main linacs, are:

\begin{itemize}

  \item a section containing post-linac emittance measurement and
    matching (correction) sections, trajectory feedback, polarimetry
    and energy diagnostics; \itemspace

  \item a fast pulsed extraction system used to extract beams in case
    of a fault, or to dump the beam when not needed at the IP; \itemspace

  \item a collimation section which removes beam halo particles that
    would otherwise generate unacceptable background in the detector,
    and also contains magnetized iron shielding to deflect muons; \itemspace

  \item the final focus (FF) which uses strong compact superconducting
    quadrupoles to focus the beam at the IP, with sextupoles providing
    local chromaticity correction; \itemspace

  \item the interaction region, containing the experimental
    detectors. The final focus quadrupoles closest to the IP are
    integrated into the detector to facilitate detector ``push-pull''; \itemspace

   \item the extraction line, which has a large enough bandwidth to
     cleanly transport the heavily disrupted beam to a high-powered
     water-cooled dump. The extraction line also contains important
     polarization and energy diagnostics. \itemspace

\end{itemize}

\subparagraph{Challenges} ~\\
The principal challenges in the beam delivery system are:
\begin{itemize}
  \item tight tolerances on magnet motion (down to tens of
    nanometers), which make the use of fast beam-based feedback
    systems mandatory, and may well require mechanical stabilization
    of critical components (e.g. final doublets). \itemspace

  \item uncorrelated relative phase jitter between the crab cavity
    systems, which must be limited to the level of tens of
    femtoseconds. \itemspace

  \item control of emittance growth due to static misalignments, which
    requires beam-based alignment and tuning techniques similar to the
    RTML. \itemspace

  \item control of backgrounds at the IP via careful tuning and
    optimization of the collimation systems and the use of the
    tail-folding octupoles. \itemspace

  \item clean extraction of the high-powered disrupted beam to the
    dump. Simulations indicate that the current design is adequate
    over the full range of beam parameters. \itemspace
\end{itemize}
\clearpage

\setcounter{figlcl}{0}
\setcounter{tablcl}{0}
\setcounter{secnumdepth}{4} \setcounter{subsection}{0}
\setcounter{subsubsection}{0}
\usecounter{subsubsection}

\section{Sample Sites}\label{sect:OVRss}

Conventional Facilities and Siting (CFS) is responsible for civil
engineering, power distribution, water cooling and air conditioning
systems. The value estimate (see Section~\ref{sec:ValueEstimate})
for the CFS is approximately 38\% of the total estimated
project value.

In the absence of a single agreed-upon location for the ILC, a sample
site in each region was developed. Each site was designed to support
the baseline design described in Section~\ref{Ovr:baseline}. Although many of the basic requirements are identical, differences in geology, topography
and local standards and regulations lead to different construction
approaches, resulting in a slight variance in value estimates across
the three regions. Although many aspects of the CFS (and indeed
machine design) will ultimately depend on the specific host site
chosen, the approach taken here is considered sufficient for the
current design phase, while giving a good indication of the influence
of site-specific issues on the project as a whole.

Early in the RDR process, the regional CFS groups agreed upon a matrix of criteria for any sample site. All three sites satisfied these criteria, including the mandatory requirement that the site can support the extension to the 1 TeV
center-of-mass machine.

\stepcounter{figlcl}\begin{figure}[htb] \vbabove
\begin{center}
  \includegraphics[width=0.95\textwidth]{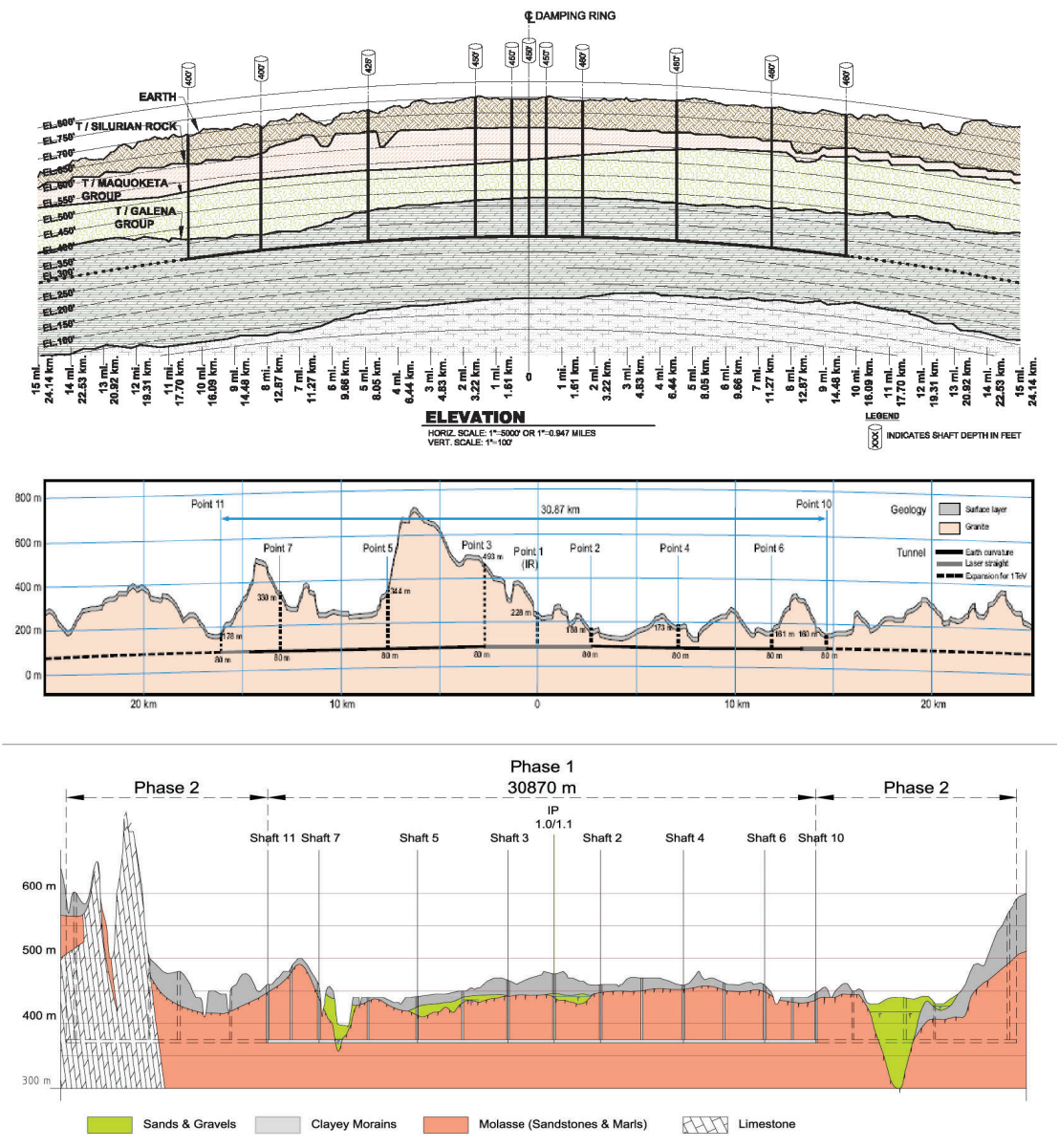}
  \vbabovecaption \caption[Geology and tunnel profiles for the three regional sites.]{Geology and tunnel profiles for the three regional sites,
    showing the location of the major access shafts (tunnels for the
    Asian site). Top: the Americas site close to Fermilab. Middle: the
    Asian site in Japan. Bottom: the European site close to CERN.}
  \label{fig:OVsitesec}
\end{center} \vbbelow
\end{figure}

The three sample sites have the following characteristics:

\begin{itemize}

  \item The Americas sample site lies in Northern Illinois near Fermilab.
    The site provides a range of locations to
    position the ILC in a north-south orientation. The site chosen has
    approximately one-quarter of the machine on the Fermilab site. The
    surface is primarily flat. The long tunnels are bored in a
    contiguous dolomite rock strata (¡ÆGalena Platteville¡Ç), at a
    typical depth of 30-100 m below the surface. \itemspace

  \item The Asian site has been chosen from several possible ILC
    candidate sites in Japan. The sample site has a uniform terrain
    located along a mountain range, with a tunnel depth ranging from
    40 m to 600 m. The chosen geology is uniform granite highly suited
    to modern tunneling methods. One specific difference for the Asian
    site is the use of long sloping access tunnels instead of vertical
    shafts, the exception being the experimental hall at the
    Interaction Region, which is accessed via two 112 m deep vertical
    shafts. The sloping access tunnels take advantage of the
    mountainous location. \itemspace

  \item The European site is located at CERN, Geneva, Switzerland, and
    runs parallel to the Jura mountain range, close to the CERN
    site. The majority of the machine is located in the `Molasse' (a
    local impermeable sedimentary rock), at a typical depth of 370 m. \itemspace

\end{itemize}

The elevations of the three sample sites are shown in
Figure~\ref{fig:OVsitesec}. The tunnels for all three sites would be
predominantly constructed using Tunnel Boring Machines (TBM), at
typical rates of 20--30 m per day. The Molasse of the European site
near CERN requires a reinforced concrete lining for the entire tunnel
length. The Asian site (granite) requires rock bolts and a 5 cm
`shotcrete' lining. The US site is expected to require a concrete
lining for only approximately 20\% of its length, with rock-bolts
being sufficient for permanent structural support.

A second European sample site near DESY, Hamburg, Germany, has also
been developed. This site is significantly different from the three
reported sites, both in geology and depth (25~m deep), and requires
further study.

In addition, the Joint Institute for Nuclear Research has submitted a
proposal to site the ILC in the neighborhood of Dubna, Russian
Federation.

The three sites reported in detail here are all `deep-tunnel'
solutions. The DESY and Dubna sites are examples of `shallow' sites. A
more complete study of shallow sites -- shallow tunnel or
cut-and-cover -- will be made in the future as part of the Engineering
Design phase.

%------------------------------------------
\clearpage

\setcounter{figlcl}{0}
\setcounter{tablcl}{0}
\setcounter{secnumdepth}{4} \setcounter{subsection}{0}
\setcounter{subsubsection}{0}
\usecounter{subsubsection}

\section{The RDR Process}

\begin{figure}[htb] \vbabove
\begin{center}
  \includegraphics[width=0.95\textwidth]{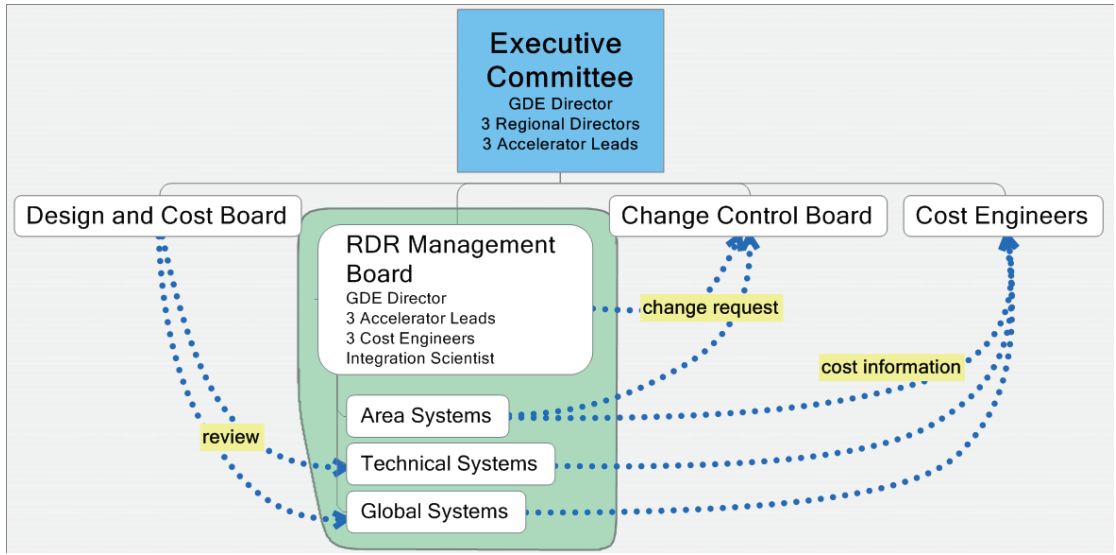}
  \vbabovecaption \caption{GDE structure for producing the ILC Reference Design and
    Cost.}
  \label{fig:OVGDEorg}
\end{center} \vbbelow
\end{figure}

Figure~\ref{fig:OVGDEorg} shows those GDE entities directly
responsible for producing the RDR:

\stepcounter{figlcl}

\begin{itemize}

  \item An {\bf Executive Committee} (EC), chaired by the GDE
    Director, responsible for all major decisions and overall GDE
    management. The committee membership included the three Regional
    GDE Directors and the three Accelerator Leads (one from each
    region). \itemspace
  \item Three {\bf Cost Engineers}, one from each region, who were
    responsible for coordinating the cost effort, defining and
    maintaining the Work Breakdown Structure (WBS) and its associated
    dictionary, and ultimately assembling and reviewing the cost
    estimate. \itemspace
  \item The {\bf RDR Management Board}, responsible for the day-to-day
    technical management of the RDR process. Membership included the
    GDE Director, the three Cost Engineers, the three Accelerator
    Leads, and an Integration Scientist. \itemspace
  \item The {\bf Area, Technical and Global Systems}, who were
    directly responsible for developing the accelerator design and
    producing the value estimate (described in detail below). \itemspace
  \item A {\bf Design and Cost Board} (DCB), charged with defining the
    costing methodology and reviewing the progress of the ILC design
    and costs. The board membership was made up of the three Cost
    Engineers and additional GDE members. \itemspace
  \item A {\bf Change Control Board} (CCB), responsible for
    implementing Change Control for the BCD as the design
    developed. Membership was drawn from the GDE. \itemspace
\end{itemize}

The important concept of Change Control was implemented early in the
ILC design effort, as a mechanism of maintaining a history of the
baseline design, and reviewing the cost/performance trade-off of any
proposed modification. Change Control was formally implemented via the
GDE Change Control Board (CCB), whose regionally-balanced membership
was taken from accelerator expertise within the GDE.

The tasks of producing the technical design and cost estimation were
the primary function of the Area, Technical and Global System groups,
under the leadership of the RDR Management Board. These groups were
arranged in the matrix structure shown in Figure~\ref{fig:OVRDRmatrix}.

\stepcounter{figlcl}\begin{figure}[htb] \vbabove
\begin{center}
  \includegraphics[width=0.95\textwidth]{\picturefolder RDRmatrix.pdf}
  \vbabovecaption \caption{Organizational structures for the Reference Design
    technical design and costing.}
  \label{fig:OVRDRmatrix}
\end{center} \vbbelow
\end{figure}

The design of the machine was geographically broken down into Area
Systems (Electron Source, Positron Source, Damping Rings (DR), Ring to
Main Linac (RTML), Main Linac (ML) and Beam Delivery System (BDS)). At
least two coordinators were assigned to each Area System from
different regions. Critical systems such as the Main Linac, Damping
Rings and Beam Delivery System were assigned coordinators from all
three regions. In all cases, a lead coordinator was identified.

The Area Systems coordinators were given the following
responsibilities:

\begin{itemize}

  \item produce the detailed design and requirements for the layout
    and components of their sub-systems; \itemspace

  \item coordinate cost- and performance-driven design modifications,
    and submit the associated formal Change Requests to the Change
    Control Board; \itemspace

  \item roll-up and maintain the cost estimates for their specific
    Area System, and supply that information to the Cost Engineers. \itemspace

\end{itemize}

The Technical and Global systems were responsible for component design
and producing the unit cost estimates:

\begin{itemize}
  \item {\bf Technical Systems}, are generally associated with
    specific accelerator components found in nearly all the Area
    Systems: Magnets (conventional and superconducting) included power
    supplies and supports; Vacuum systems included insulating vacuum
    for the cryogenic systems as well as beamline UHV; Instrumentation
    covered beam position, profile, length and loss monitoring; Dumps
    and collimators were responsible for low- and high-powered beam
    dumps, and numerous collimator systems throughout the machine; RF
    power sources supplied estimates for klystrons, modulators and
    waveguide distribution systems (dominated by the Main Linac RF
    unit); Cryomodule and Cavity Package were special cases, both
    being focused on the Main Linac superconducting RF. Warm RF
    sections in the source capture sections, as well as the
    superconducting RF for the Damping Rings, were directly estimated
    by experts in the respective Area Systems. \itemspace

  \item {\bf Global Systems} represent more global aspects of the
    machine design which are not directly related to specific
    areas. Of these, the Civil Construction and Siting (CFS) system is
    by far the largest cost driver. Others include cryogenics,
    controls, availability and operations (including machine
    protection) and installation. \itemspace

\end{itemize}

The Technical/Global Systems were responsible for:

\begin{itemize}

  \item obtaining and consolidating lists of components and their
    requirements from the Area System Coordinators; \itemspace

  \item producing cost estimates of the components/systems, using a
    suitably justifiable method (e.g. comparison to existing machines,
    bottoms-up approximate designs, in-house estimate or direct
    industrial quotes); \itemspace

  \item iteration of the designs, where either the technical
    feasibility of the requirements was not practical, or an
    alternative more cost effective solution was identified; \itemspace

  \item supplying the cost information to the relevant Area Systems,
    and to the Cost Engineers for review. \itemspace
\end{itemize}

Each Technical/Global system was assigned a coordinator from each
region (considered important for maintaining cost input information
from all regions). Points of contact between Technical/Global and Area
systems were identified to enable exchange of information between the
two.

The detailed design work and cost estimation began shortly after the
Baseline Configuration was agreed upon at the Frascati GDE meeting
(November 2005). The effort that followed can be loosely split into
two half-year periods:

\begin{itemize}
  \item Frascati GDE Meeting (Dec. 2005) -- Vancouver GDE Meeting
    (July 2006) Consolidation of the detailed Baseline Design;
    production of component specifications and requirements for
    Technical/Global Systems; Area/Technical/Global Systems
    preparation of a first estimate of total project cost for review
    at the Vancouver meeting. \itemspace

  \item Vancouver GDE Meeting (July 2006) -- Valencia GDE Meeting
    (Nov. 2006) Cost-driven iteration of Baseline Design (Area
    Systems) and technical component costs. This phase saw a
    re-evaluation of the Frascati Baseline Design, resulting in
    several significant cost-driven machine layout modifications. \itemspace

\end{itemize}

Figure~\ref{fig:OVGDEskd} shows a more detailed schedule, identifying
the critical interim milestones in the process.

\stepcounter{figlcl}\begin{figure}[htb] \vbabove
\begin{center}
  \includegraphics[width=0.95\textwidth]{\picturefolder GDEskd.pdf}
 \vbabovecaption  \caption{Milestones in producing the Reference Design Report,
    including costs.}
 \label{fig:OVGDEskd}
\end{center} \vbbelow
\end{figure}

The lack of a `geographically centralized' design group has required
additional formality and discipline in the way the work has been
organized. Significant use has been made of teleconferencing
facilities and web-based conferencing tools (e.g. WebEx) wherever
possible. Several ¡Æglobal¡Ç teleconferences including all three
regions were scheduled every week. Use of a wiki site for all
technical information (on http://www.linearcollider.org/wiki/ ) also
facilitated the distribution of key information between the RDR
groups.

\clearpage

\setcounter{figlcl}{0}
\setcounter{tablcl}{0}
\setcounter{secnumdepth}{4} \setcounter{subsection}{0}
\setcounter{subsubsection}{0}
\usecounter{subsubsection}

\section{Value Estimate}\label{sec:ValueEstimate}

A preliminary cost analysis has been performed for the ILC Reference
Design.  A primary goal of the estimate was to allow
cost-to-performance optimization in the Reference Design, before
entering into the engineering design phase.  Over the past year, the
component costs were estimated, various options compared and the
design evolved through about ten significant cost-driven changes,
resulting in a cost reduction of about 25\%, while still maintaining
the physics performance goals.

The ILC cost estimates have been performed using a ``value'' costing
system, which provides basic agreed-to value costs for components in
ILC Units\footnote{For this value estimate, 1 ILC Unit = 1 US 2007\$
  (= 0.83 Euro = 117 Yen).}, and an estimate of the explicit labor (in
person hours) that is required to support the project.  The estimates
are based on making world-wide tenders (major industrialized nations),
using the lowest reasonable price for the required quality.  There are
three classes of costs:

\begin{itemize}
  \item site-specific costs, where a separate estimate was made in
    each of the three regions; \itemspace
  \item conventional costs for items where there is global capability
    -- here a single cost was determined; \itemspace
  \item costs for specialized high-tech components (e.g. the SCRF linac
    technology), where industrial studies and engineering estimates
    were used. \itemspace
\end{itemize}

The total estimated value for the shared ILC costs for the Reference
Design is 4.79 Billion (ILC Units). An important outcome of the value
costing has been to provide a sound basis for determining the relative
value of the various components or work packages. This will enable
equitable division of the commitments of the world-wide collaboration.

In addition, the site specific costs, which are related to the direct
costs to provide the infrastructure required to site the machine, are
estimated to be 1.83 Billion (ILC Units). These costs include the
underground civil facilities, water and electricity distribution and
buildings directly supporting ILC operations and construction on the
surface.  The costs were determined to be almost identical for the
Americas, Asian, and European sample sites. It should be noted that
the actual site-specific costs will depend on where the machine is
constructed, and the facilities that already exist at that location.

Finally, the explicit labor required to support the construction
project is estimated at 24 million person-hours; this includes
administration and project management, installation and testing.  This
labor may be provided in different ways, with some being contracted
and some coming from existing labor in collaborating institutions.

The ILC Reference Design cost estimates and the tools that have been
developed will play a crucial role in the engineering design effort,
both in terms of studying options for reducing costs or improving
performance, and in guiding value engineering studies, as well as
supporting the continued development of a prioritized R\&D program.

The total estimated value cost for the ILC, defined by the Reference
Design, including shared value costs, site specific costs and explicit
labor, is comparable to other recent major international projects,
e.g. ITER, and the CERN LHC when the cost of pre-existing facilities
are taken into account. The GDE is confident that the overall scale of
the project has been reliably estimated and that cost growth can be
contained in the engineering phase, leading to a final project cost
consistent with that determined at this early stage in the design.

\clearpage

\setcounter{figlcl}{0}
\setcounter{tablcl}{0}
\setcounter{secnumdepth}{4} \setcounter{subsection}{0}
\setcounter{subsubsection}{0}
\usecounter{subsubsection}

\section{R\&D and the Engineering Design Phase}

For the last year, the focus of the core GDE activity has been on
producing the RDR and value estimate. In parallel, ILC R\&D programs
around the world have been ramping up to face the considerable
challenges ahead. The GDE Global R\&D Board -- a group of twelve GDE
members from the three regions -- has evaluated existing programs, and
has convened task forces of relevant experts to produce an
internationally agreed-upon prioritized R\&D plan for the critical
items. The highest-priority task force (S0/S1) addresses the SCRF
accelerating gradient:

\begin{itemize}
  \item S0: high-gradient cavity -- aiming to achieve 35 MV/m
    nine-cell cavity performance with an 80\% production yield; \itemspace
  \item S1: high-gradient cryomodule -- the development of one or more
    high-gradient ILC cryomodules with an average operational gradient
    of 31.5 MV/m. \itemspace

\end{itemize}

The S0/S1 task force has already produced focused and comprehensive
R\&D plans. Other task forces (S2: test linac; S3: Damping Ring; S4:
Beam Delivery System, etc.) are in the process of either completing
their reports, or just beginning their work.

\begin{figure}[htb] \vbabove
\begin{center}
  \includegraphics[width=\textwidth]{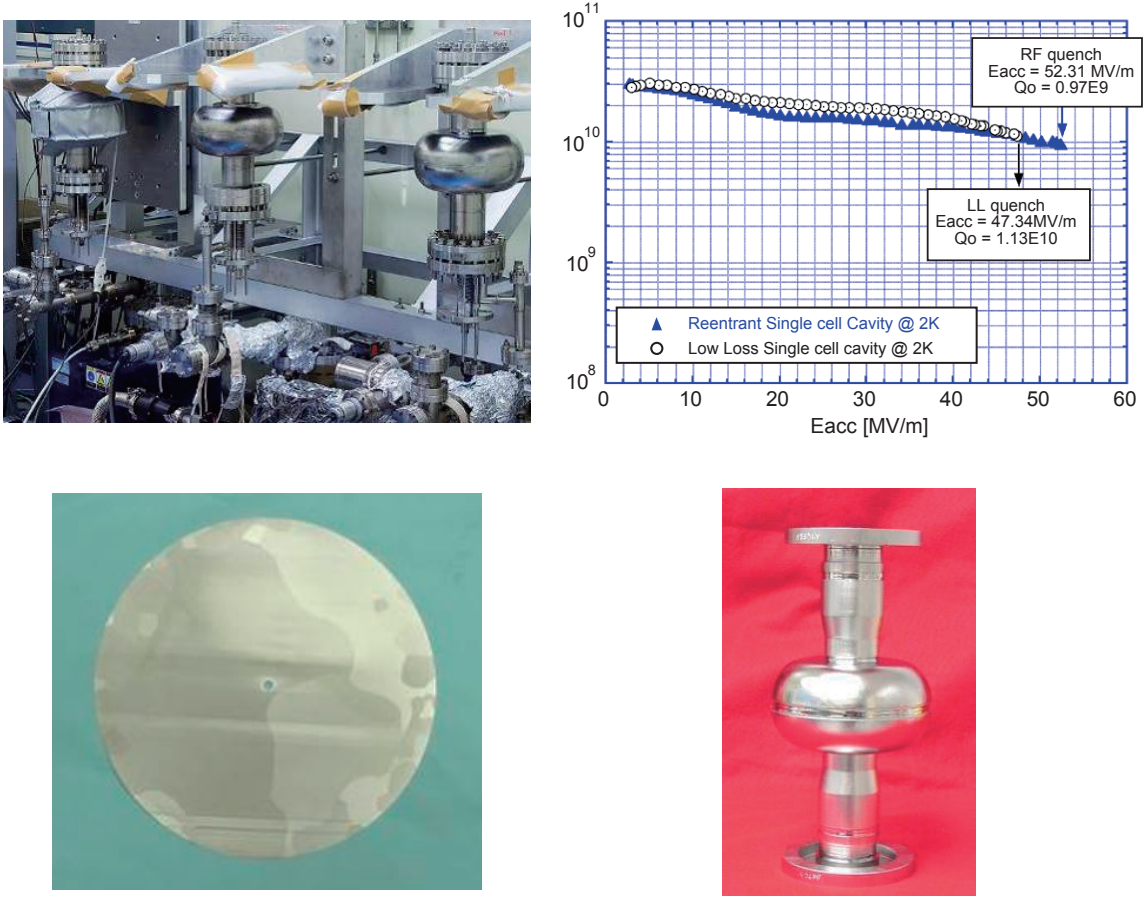}
 \vbabovecaption  \caption[Cutting-edge SCRF R\&D.]
    {Cutting-edge SCRF R\&D. Top-left: ICHIRO single-cells being
    prepared for testing at KEK. Top-right: world-record performance
    from novel shape single-cells (ICHIRO and Cornell's reentrant
    cavity). Bottom-left: large-grain niobium disk (Jefferson
    Lab). Bottom-right: single-cell cavity produced from large-grain
    niobium material (Jefferson Lab).}
  \label{fig:OVcavRD}
\end{center} \vbbelow
\end{figure}

For the cost- and performance-critical SCRF, the primary focus of
S0/S1 remains the baseline choice, the relatively mature TESLA
nine-cell elliptical cavity. However, additional research into
alternative cavity shapes and materials continues in parallel. One
promising technique is the use of `large-grain'
niobium~\cite{large-grain}, as opposed to the small-grain material
that has been used in the past (Figure~\ref{fig:OVcavRD}).  Use of
large grain material may remove the need for electropolishing, since
the same surface finish can potentially be achieved with Buffered
Chemical Polishing (BCP) -- a possible cost saving. Several
single-cells have achieved gradients in excess of 35 MV/m (without
electropolishing) and more recent nine-cell cavity tests have shown
very promising results.

Various new and promising cavity shapes are also being investigated,
primarily at KEK and Cornell. While the basic nine-cell form remains,
the exact shape of the `cells' is modified to reduce the peak magnetic
field at the niobium surface. In principle these new shapes can
achieve higher gradients, or higher quality factors
($Q_0$). Single-cells at KEK (ICHIRO) and Cornell (reentrant) have
achieved the highest gradients to date ($\sim$50 MV/m, see
Figure~\ref{fig:OVcavRD}). R\&D towards making high-performance
nine-cell cavities using these designs continues as future possible
alternatives to the ILC baseline cavity.

The GDE formally supports R\&D on alternative designs for components other than the cavities, where the new designs promise potential cost and/or performance benefits. Some key examples are alternative RF power
source components, of which the Marx modulator is currently the most
promising.  In addition, R\&D on critical technologies will
continue through the EDR. Topics include items such as the damping
ring kickers and electron-cloud mitigation techniques, the positron
target and undulator, the magnets around the beam interaction point,
and global issues that require very high availability such as the
control system, the low-level RF, and the magnet power supplies.

\stepcounter{figlcl}

While investment into the critical R\&D remains a priority, a
significant ramping-up of global engineering resources will be
required to produce an engineered technical design by 2010. An
important aspect of this work will be the refinement and control of
the published cost estimate by value engineering. The EDR phase will
also require a restructuring of the GDE to support the expanded
scope. A more traditional project structure will be adopted based on
the definition of a discrete set of Work Packages. The responsibility
for achieving the milestones and deliverables of each Work Package
will be assigned to either a single institute, or consortium of
institutes, under the overall coordination of a central project
management team. The Work Packages need to be carefully constructed to
accommodate the direct needs of the Engineering Design phase,
while at the same time reflecting the global nature of the project. An
important goal of the current planning is to integrate the engineering
design and fundamental R\&D efforts, since these two aspects of the
project are clearly not independent. The new project structure will be
in place by mid 2007.

\setcounter{chapter}{1}

\chapter{\textsf{Accelerator Description}\label{chapACC}}

\setcounter{figlcl}{0}
\setcounter{tablcl}{0}
\setcounter{secnumdepth}{4} \setcounter{subsection}{0}
\setcounter{subsubsection}{0}
\usecounter{subsubsection}

\setcounter{section}{0} \renewcommand{\picturefolder}{./AccOverview/}

\section{Beam Parameters } \label{sect:ACCparams}

The International Linear Collider (ILC) is designed to achieve the specifications listed in the ILCSC Parameter Subcommittee
Report \cite{bib:ACCilcsc}.  The three most important requirements are: (1) an initial center-of-mass (cms) energy up to 500 GeV with the ability to upgrade to 1 TeV, (2) an integrated luminosity in the first four years of 500 fb$^{-1}$
 at 500 GeV cms or equivalent at lower energies, and (3) the ability to scan in energy between 200 and 500 GeV cms.

The ILC Reference Design Report describes a collider that is designed to meet these requirements.  The installed RF system is capable of accelerating beams for collisions at 500 GeV cms.  The peak luminosity of $2\times10^{34}$\,cm$^{-2}$s$^{-1}$ at 500 GeV and a collider availability of 75\% should enable the delivery of 500 fb$^{-1}$ in the first four years of physics operation assuming an annual physics run of 9 months, and a gradual ramp up of luminosity over the four years.  The energy flexibility has been a consideration throughout the design and essential items to facilitate a future upgrade to 1 TeV, such as the length of the beam delivery system and the power rating of the main beam dumps, have been incorporated.

%\subsubsection{Choice of the accelerating gradient and the RF system}
%31.5 MV/m is adopted as the average accelerating gradient in the cavities.
%This is somewhat higher than the presently available value
%\textcolor{blue}{but is considered
%a reasonable goal in the R\&D works in the next couple of years. The packing
%factor of the main linac is assumed to be 70\% , by taking into account
%other linac components such as the magnets and the spaces between cavities.}
%
%One unit of the main linac consists of 26 accelerating cavities
%\textcolor{blue}{powered by Ene klystron whose output power is 10MW in $\sim 1.4$ms
%pulse length. Availability of such RF power sources has been nearly established
%through efforts at the European XFEL project. The beam current to be accelerate
%in this system is chosen to be 9~mA. The beam pulse length is around
%1~ms.}

\subsection{Collider and Beam Parameters}

\stepcounter{tablcl}\begin{table}[htb] \vbabove \caption{Global
Accelerator Parameters for 500 GeV cms.}
\label{tab:GlobalColliderParam}
\begin{center}
\begin{tabular}{| l | r | l |}
\hline
Parameter  &  Value &  Units  \\ \hline & & \vbdlspacing  \hline
Center-of-mass energy   &  500 & GeV \\ \hline
Peak luminosity   &  $2\times10^{34}$   & cm$^{-2}$s$^{-1}$ \\ \hline
Availability   &  75 & \% \\ \hline
Repetition rate         &  5  & Hz  \\ \hline
Duty cycle         &  0.5 & \%       \\ \hline
Main Linacs           &      &       \\ \hline
\ \ \ \ \ Average accelerating gradient in cavities   &   31.5 & MV/m  \\ \hline
\ \ \ \ \ Length of each Main Linac    &   11 & km   \\ \hline
\ \ \ \ \ Beam pulse length    &   1  & ms   \\ \hline
\ \ \ \ \ Average beam current in pulse    &   9.0 & mA   \\ \hline
Damping Rings          &           &     \\ \hline
\ \ \ \ \ Beam energy     &  5 & GeV   \\ \hline
\ \ \ \ \ Circumference   &  6.7 & km   \\ \hline
Length of Beam Delivery section (2 beams)  &  4.5 & km \\ \hline
Total site length       &  31 & km  \\ \hline
Total site power consumption  &   230 & MW   \\ \hline
Total installed power  &   $\sim$300 & MW   \\
\hline
\end{tabular}
\end{center} \vbbelow
\end{table}

The ILC is based on 1.3~GHz superconducting RF cavities operating at a gradient of 31.5 MV/m.  The collider operates at a repetition rate of 5 Hz with a beam pulse length of roughly 1~msec.  The site length is 31~km for a cms energy of 500 GeV; the site would have to be extended to reach 1~TeV.  The beams are prepared in low energy damping rings that operate at 5~GeV and are 6.7~km in circumference. They are then accelerated in the main linacs which are $\sim$11~km per side. Finally, they are focused down to very small spot sizes at the collision point with a beam delivery system that is $\sim$2.2~km per side.  To attain a peak luminosity of $2\times10^{34}$\,cm$^{-2}$s$^{-1}$,  the collider requires $\sim$230~MW of electrical power.  A summary of the overall collider parameters appears in Table~\ref{tab:GlobalColliderParam}.
%and the collider subsystems are introduced in Section~\ref{sec:layout} before being described in detail in the body of this report.

The beam parameters to reach a peak luminosity of $2\times10^{34}$\,cm$^{-2}$s$^{-1}$ are listed in Table~\ref{tab:GlobalBeamParam}.  The table lists a set of nominal parameters and three other sets that define a \lq parameter plane\rq.  The collider has been designed to the nominal parameter set which was optimized considering aspects of the whole accelerator system such as: the beam instabilities and kicker hardware in the damping rings, the beam current and the pulse length in the linacs, and the kink instability and background in the final focus system. The parameter plane establishes a range of operating parameters that represent slightly different tradeoffs between these considerations. Experience with past accelerators indicates that there will be operational difficulties, which will be eased by modifying the beam parameters. The parameter plane provides flexibility to cope with such problems without sacrificing performance. It can also be useful during collider commissioning and when tuning the luminosity characteristics for different measurements and particle physics detectors.

\subsection{The Nominal Parameter Set}

The main linac RF system is designed to accelerate beam at a gradient of 31.5~MV/m.  The nominal beam current is 9.0~mA and the beam pulse length is 970~$\mu$s so that the RF pulse length (including the fill time of the cavities) is 1.56~ms.  The optimal single bunch charge is a balance between effects at the IP and in the damping ring; the choice of $2\times10^{10}$ is similar to that specified in the TESLA TDR \cite{tdr} and the US Technical Options Study \cite{bib:ACCustos}.

The normalized vertical emittance at the IP is chosen to be $4\times10^{-8}$\,m$\cdot$rad. This corresponds to a geometric emittance of $\sim$2\,pm from the damping rings (5 GeV) and assumes 100\% emittance growth during the transport to the IP. This damping ring emittance is slighty lower than what has already been achieved but is thought to be well within the present technology.  The 100\% emittance growth estimate is based on calculations made during the ILC Technical Review Report \cite{bib:ACCiltrc}.

\stepcounter{tablcl}\begin{table}[htb] \vbabove \caption{Beam and IP
Parameters for 500 GeV cms.} \label{tab:GlobalBeamParam}
\begin{center}
\setlength{\tabcolsep}{4pt}
\begin{tabular}{| l l | r | r r r|}
\hline
 Parameter   & Symbol/Units & Nominal & Low~N & Large~Y & Low~P  \\  \hline & & & & & \vbdlspacing  \hline
Repetition rate  &  $f_{rep}$  (Hz)  &  5   &  5  &  5  &  5    \\  \hline
Number of particles per bunch  &  $N$ $(10^{10})$  & 2 & 1 & 2 & 2  \\  \hline
Number of bunches per pulse &  $n_b$  & 2625 & 5120 & 2625 & 1320  \\  \hline
Bunch interval in the Main Linac  & $t_b$ (ns) & 369.2 & 189.2 & 369.2 & 480.0 \\  \hline
\ \ \ \ \ \ \ in units of RF buckets  &  & 480 & 246  & 480 &  624 \\  \hline
Average beam current in pulse   & $I_{ave}$ (mA) & 9.0 & 9.0 & 9.0 & 6.8  \\  \hline
Normalized emittance at IP  & $\gamma\epsilon_x^*$ (mm$\cdot$mrad) & 10 & 10 & 10 & 10  \\  \hline
Normalized emittance at IP  & $\gamma\epsilon_y^*$ (mm$\cdot$mrad) & 0.04 & 0.03 & 0.08 & 0.036  \\  \hline
Beta function at IP  & $\beta_x^*$  (mm) & 20 & 11 & 11 & 11  \\  \hline
Beta function at IP  & $\beta_y^*$ (mm) & 0.4 & 0.2 & 0.6 & 0.2  \\  \hline
R.m.s. beam size at IP  & $\sigma_x^*$ (nm)  & 639 & 474 & 474 & 474  \\  \hline
R.m.s. beam size at IP    & $\sigma_y^*$ (nm)  & 5.7 & 3.5 & 9.9 & 3.8  \\  \hline
R.m.s. bunch length  & $\sigma_z$ ($\mu$m)  & 300 & 200 & 500 & 200  \\  \hline
Disruption parameter  & $D_x$   & 0.17 & 0.11 & 0.52 & 0.21  \\  \hline
Disruption parameter  & $D_y$   & 19.4 & 14.6 & 24.9 & 26.1  \\  \hline
Beamstrahlung parameter  & $\Upsilon_{ave}$  & 0.048 & 0.050 & 0.038 & 0.097  \\  \hline
Energy loss by beamstrahlung  & $\delta_{BS}$    & 0.024 & 0.017 & 0.027 & 0.055  \\  \hline
Number of beamstrahlung photons & $n_\gamma$    & 1.32 & 0.91 & 1.77 & 1.72  \\  \hline
Luminosity enhancement factor   & $H_D$   & 1.71  & 1.48  & 2.18  & 1.64    \\  \hline
Geometric luminosity   & ${\cal L}_{geo}$ $10^{34}/$cm$^2/$s  & 1.20  & 1.35  & 0.94  & 1.21    \\  \hline
Luminosity   & ${\cal L}$ $10^{34}/$cm$^2/$s    & 2  & 2  & 2  & 2    \\  \hline
\end{tabular}
\end{center} \vbbelow
\end{table}

\subsection{Parameter Plane}
The parameter sets labeled `Low~N' (low number of particles per bunch),
`Large~Y' (large vertical emittance) and `Low~P' (low beam power) in Table \ref{tab:GlobalBeamParam} are representative points in the parameter plane.
These parameter sets deliver essentially the same luminosity $2\times10^{34}$\,cm$^{-2}$s$^{-1}$ at 500 GeV but with different values for the specific beam parameters.  The collider subsystems have been designed such that any point in the parameter plane is attainable.  At present, it is not believed that there is a large cost impact of maintaining the parameter plane and there is a significant gain in operational flexibility; this will need to be examined again during the next phase of design optimization.

\medskip
\noindent\underline{Low~N}

The bunch population of $2\times 10^{10}$ may lead to problems such as
microwave instabilities in the damping rings, single bunch wakefield emittance dilutions, or a large disruption parameter
at the IP which can cause a kink instability and may make the IP feedback difficult.
In such cases, it could be desirable to reduce the bunch population.

The Low~N parameter set addressed these possible difficulties with a reduced single bunch charge and reduced bunch length.  Halving the bunch population with fixed current (twice the number of bunches and half the bunch interval) reduces the luminosity but this is compensated by focusing more tightly at the IP.  These parameters also have lower beamstrahlung and possibly lower backgrounds in the particle physics detectors at the IP which may be desirable for some measurements.  All these changes are beneficial, however, the Low~N parameter set is more demanding in terms of the damping ring kicker, the bunch compressor, and the multi-bunch collective effects in the damping rings.

\medskip
\noindent\underline{Large~Y}

The vertical emittance at the IP of $4\times 10^{-8}$\,m$\cdot$rad may not be achieved due to tuning difficulties in
the damping rings and beam delivery system or wakefield effects in the linac. The Large~Y parameters assume a vertical emittance that is twice the design and the luminosity is recovered by focusing more tightly in the horizontal at the IP and using a longer bunch to reduce the increased beamstrahlung. Unfortunately, the disruption parameter at the interaction point is increased and kink instability may be more pronounced.

\medskip
\noindent\underline{Low~P}

Another condition that may arise are limitations due to the beam current or beam power.  These may arise in the injector systems, damping rings, main linacs or beam delivery system.  In this case, the collider could be optimized in the direction of the Low~P parameters where the beam current is reduced by 30\% and the beam power is reduced by a factor of two.  Again, the luminosity is recovered with increased focusing at the IP in the horizontal plane.   In this case, the beamstrahlung cannot be reduced by increasing the bunch length because of the tight focusing in the vertical plane.  This results in a beamstrahlung that is roughly double that in the nominal parameters and this may limit the performance of the particle physics detector and the beam delivery extraction line.

\subsection{Range of Parameters}
The parameter plane described above defines a range of parameters as shown
in Table~\ref{tab:ParamRange}. Note, however, the parameters, when
they are varied, are correlated. For example,
the shortest bunch length is required only when the bunch population is low.

\stepcounter{tablcl}\begin{table}[htb]
\begin{center}
\vbabove \caption{Range of parameters.}
\label{tab:ParamRange}
\begin{tabular}{| l | c | r c r c r | l |}
\hline
Parameter   &  Symbol &  min  &  \multicolumn{3}{c}{ nominal}  &   max &  Units \\ \hline & & & & & & & \vbdlspacing  \hline
Bunch population   &  $N$    &  1  &-&  2  &-&  2  & $\times 10^{10}$ \\ \hline
Number of bunches  &  $n_b$  &  1320  &-&  2625  &-& 5120  &     \\ \hline
Linac bunch interval  &  $t_b$  &  189  &-&  369  &-&  480  &  ns  \\ \hline
Bunch length         &  $\sigma_z$  &  200  &-&  300  &-&  500  &  $\mu$m \\ \hline
Vertical emittance  &  $\gamma\epsilon_y^*$  &  0.03  &-&  0.04  &-&  0.08  &  mm$\cdot$mrad \\ \hline
Beta function at IP &  $\beta_x^*$   &  11  &-&  20  &-&  20  &  mm  \\  [-4pt]
                    &  $\beta_y^*$   &  0.2  &-&  0.4  &-&  0.4  &  mm  \\ \hline
\end{tabular}
\end{center} \vbbelow
\end{table}

\subsection{Bunch Spacing and Path Length Considerations} \label{sssect:OVRtiming}
In order to extract the bunches in the damping ring one by one and inject into the
main linac there are certain constraints to satisfy among
 the DR circumference, number of
bunches, RF frequencies and bunch distances in the DR and main linac.
The present beam parameters do not meet all of the constraints needed
to best facilitate injection and extraction from the damping rings \cite{bib:ACCtimingdoc}.
The parameters will continue to be optimized during the next design phase to
better satisfy the constraints, and it is expected that the damping ring circumference
and linac bunch spacing will change by small amounts.

In addition, there is another constraint due to the fact
that the positrons are generated by electrons on the previous pulse.
For flexible operation, it is highly desireable that the sum of certain
beamline lengths such as the main linac and the transport lines be a
multiple of the DR circumference. Because of this constraint, the exact location
of the injector complex and the layout of the transport lines is a subject that
can be fixed only after the final component lengths and the site are decided.

\clearpage %\setcounter{page}{1}
\setcounter{figlcl}{0}
\setcounter{tablcl}{0}
\setcounter{secnumdepth}{4} \setcounter{subsection}{0}
\setcounter{subsubsection}{0}
\usecounter{subsubsection}

\setcounter{section}{1} \renewcommand{\picturefolder}{./electron/}

\section{Electron Source}\label{sect:ELEes}

\subsection{Overview}\label{ssect:ELEo}

The ILC polarized electron source must produce the required
train of polarized electron bunches and transport them to the
Damping Ring. The nominal train is 2625 bunches of 2.0$\times$\(10^{10}\)
electrons at 5~Hz with polarization greater than 80\%. The
beam is produced by a laser illuminating a photocathode in a
DC gun. Two independent laser and gun systems provide redundancy.
Normal-conducting structures are used for bunching and
pre-acceleration to 76~MeV, after which the beam is accelerated to
5~GeV in a superconducting linac. Before injection into the damping ring,
superconducting solenoids rotate the spin vector into the vertical,
and a separate superconducting RF structure is used for energy
compression. A third polarized electron source (500~MeV)
drives the Positron Keep Alive Source (KAS). Polarization is not required in the baseline, but will be required for either the $e^-$-$e^-$ or $\gamma$-$\gamma$ options.

The SLC polarized electron
source already meets the requirements for polarization, charge
and lifetime.  The primary challenge for the ILC source is the
long bunch train, which demands a laser system beyond that
used at any existing accelerator, and normal conducting structures which can handle high RF power. Both R\&D developments are considered manageable.

\subsection{Beam Parameters}\label{ssect:ELEbp}

The key beam parameters for the electron source are listed in
Table~\ref{tab:ELEparam}.

\stepcounter{tablcl}\begin{table} [htb] \vbabove \caption{Electron
Source system parameters.}
   \begin{center}
   \label{tab:ELEparam}
      \begin{tabular}{| l | c | c | c |}
         \hline
         Parameter & Symbol & Value & Units \\
         \hline & & & \vbdlspacing \hline
         Electrons per bunch (at gun exit) & $n_{e}$ & 3$\times$\(10^{10}\)  & Number   \\ \hline
         Electrons per bunch (at DR injection) & $n_{e}$ & 2$\times$\(10^{10}\) & Number   \\ \hline
         Number of bunches & $N_{e}$ & 2625 & Number   \\ \hline
         Bunch repetition rate & $F_{ \mu b}$ & 3 & MHz   \\ \hline
         Bunch train repetition rate & $F_{mb}$ & 5 & Hz   \\ \hline
         Bunch length at source & $\Delta t$ & 1 & ns   \\ \hline
         Peak current in bunch at source & $I_{avg}$ & 3.2 & A   \\ \hline
         Energy stability & S & $<$5 & \% rms   \\ \hline
         Polarization & $P_{e}$ & 80 (min) & \%   \\ \hline
         Photocathode Quantum Efficiency & QE & 0.5 & \%   \\ \hline
         Drive laser wavelength & $\Lambda$ & 790$\pm$20 (tunable) & nm   \\ \hline
         Single bunch laser energy & E & 5 & $\mu J$   \\ \hline
      \end{tabular}
   \vbbelow
   \end{center}
\end{table}

\subsection{System Description}\label{ssect:ELEsd}

\begin{figure}[htb]
   \begin{center} \vbabove
      \includegraphics[width=0.99\textwidth]{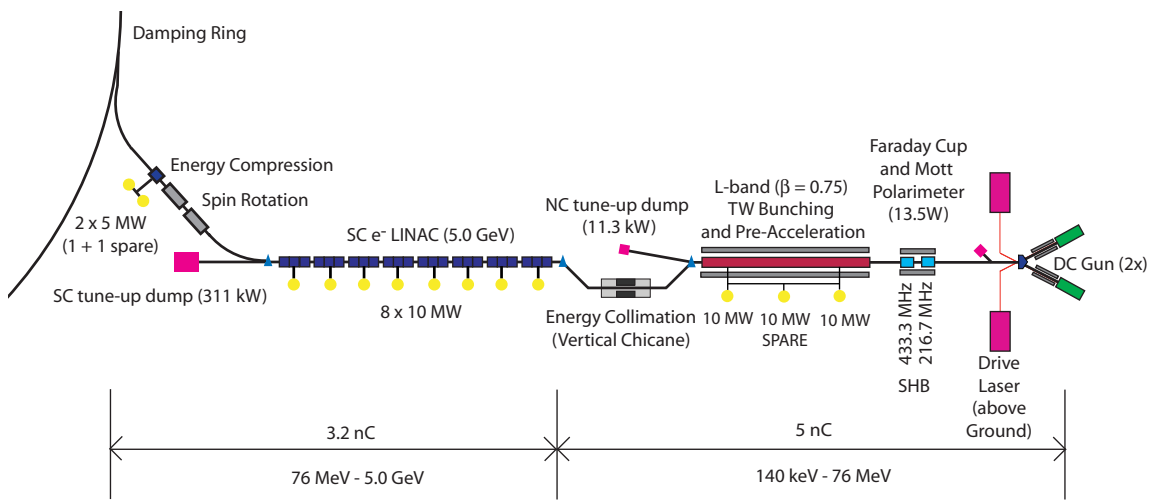}
      \vbabovecaption \caption{Schematic view of the polarized Electron Source.}
      \label{fig:ELElayout}
   \end{center} \vbbelow
\end{figure}

Figure~\ref{fig:ELElayout} depicts schematically the layout of the
polarized electron source. Two independent laser systems are
located in a surface building. The light is transported down an
evacuated light pipe to the DC guns. The beam from either gun
is deflected on line by a magnet system which includes a
spectrometer, and it then passes through the normal-conducting
subharmonic bunchers, traveling wave bunchers and pre-accelerating
sections. This is followed by the 5~GeV superconducting linac.
The Linac-to-Ring transfer line that brings the beam
to the damping rings contains the spin rotators and energy compression.

\stepcounter{figlcl}

\subsubsection{Photocathodes for Polarized Beams}\label{sssect:ELEppb}

\begin{figure}[htb]
   \begin{center} \vbabove
      \includegraphics[width=0.8\textwidth]{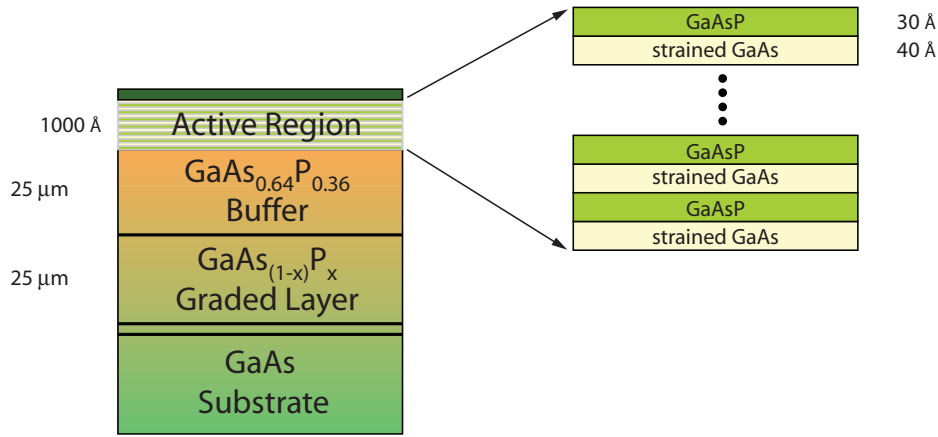}
     \vbabovecaption
      \caption{Structure of a strained GaAs/GaAsP superlattice photocathode for polarized electrons.}
      \label{fig:ELEphotocathode}
   \vbbelow
   \end{center}
\end{figure}

Photocathode materials have been the subject of intense R\&D efforts
for more than 15 years. The most promising candidates for the ILC
polarized electron source are strained GaAs/GaAsP superlattice
structures (see Figure~\ref{fig:ELEphotocathode}). GaAs/GaAsP
superlattice photocathodes routinely yield at least 85\%
polarization with a maximum QE of ~1\% (routinely 0.3 to 0.5\%)
\cite{bib:ELSrefr1, bib:ELSrefr2, bib:ELSJ1}. The present cathodes consist of
very thin quantum well layers (GaAs) alternating with
lattice-mismatched barrier layers (GaAsP). Each layer of the
superlattice (typically 4~nm) is considerably thinner than the
critical thickness ($\sim$10~nm) for the onset of strain relaxation,
while the transport efficiency for elec relaxation, while the
transport efficiency for electrons in the conduction band still can
be high \cite{bib:ELSJ2}. The structures are p-doped using a high-gradient
doping technique, consisting of a thin (10~nm), very highly doped                           % 3 x  \(10^{10}\)
(5$\times10^{19}$~cm$^{-3}$) surface layer with a lower density
doping (5$\times10^{17}$~cm$^{-3}$) in the remaining active
layer(s). A high surface doping density is necessary to achieve high
QE while reducing the surface-charge-limit problem \cite{bib:ELSJ3, bib:ELSJ4}. A lower doping
density is used to maximize the polarization \cite{bib:ELSJ5}. With bunch spacing of
$\sim$300~ns, the surface-charge-limit problem for the ILC is not
expected to be a major issue. The optimum doping level remains to be
determined. An alternative under study is the InAlGaAs/GaAs strained
superlattice with minimum conduction band offset where a peak
polarization of 91\% has been observed \cite{bib:ELSrefr3}. Research
continues on various cleaning and surface preparation techniques.
Atomic hydrogen cleaning (AHC) is a well-known technique for
removing oxides and carbon-related contaminants at relatively low
temperatures \cite{bib:ELSrefr4} and will be further explored in the
near future.

\stepcounter{figlcl}

\subsubsection{Polarized Electron Gun}\label{sssect:ELEpeg}

The ILC polarized electron gun is a DC gun based on the design of
the gun used for the SLC~\cite{bib:ELSrefr5}. However, DC gun
technology for polarized sources has evolved considerably,
\cite{bib:ELSrefr6} and technological advances will be incorporated
into the ILC gun design. The ILC gun will be optimized for a space
charge limited peak current of 4.5-5~A (4.5-5~nC/1ns). This provides
overhead to compensate for losses that occur primarily through the
bunching system. The gun power supply provides a cathode bias of
-140 to -160~kV. An ultrahigh vacuum system with a total pressure
$\leq$ 10$^{-12}$ Torr (excluding H$_{2}$) is required to maintain
the negative electron affinity (NEA) of the cathode. An SF$_{6}$/dry
air gas system is used to maintain a high dielectric gun environment
to avoid HV breakdown between ground and HV components. During HV
operation the electric field on the cathode surface must be kept
below 7~MeV/m to ensure low dark current ($<$ 25~nC). Excessive dark
current will lead to field emission resulting in molecular
desorption from nearby surfaces. This process leads to deterioration
of the gun vacuum and is destructive to the cathode's NEA surface.

The gun area will be equipped with a Mott polarimeter to measure polarization
and a Faraday cup to measure the charge. Several Residual
Gas Analyzers (RGAs) characterize
the vacuum near the gun. Other special diagnostics  for the DC gun
include means to measure the quantum efficiency of the cathode
(a cw diode laser integrated into the gun) and a nano-ammeter for
dark current monitoring.

An NEA cathode requires periodic cesiation. Cesiator channels
are located near the cathode to allow in situ cesiation of the photocathode.
An improvement of the current SLC gun design will be to locate the cesiation
channels behind a retractable photocathode. This will eliminate the deposition
of Cesium on electrode surfaces, thereby reducing the dark current of the gun.
The SLC and subsequent polarized beam experiments at SLAC have
demonstrated the operation of an efficient and highly automated cesiation
system with minimal source downtime. The gun will have an integrated cathode
preparation chamber and load-lock system. The activation chamber will be
semi-permanently attached to the gun and both volumes will be semi-permanently
maintained under high vacuum. The preparation chamber will allow the option
of local cathode cleaning and activation as well as storage of spare cathodes. Cathodes
may be rapidly exchanged between the gun and preparation chamber. The load-lock consists
of a small rapidly-pumped vacuum chamber for transferring cathodes from an external
atmospheric source into or out of the preparation chamber without affecting the latter's vacuum.

The dominant source of intensity variations and timing jitter is
the laser system. A secondary source for intensity variations is
the gun power supply and beam dynamics influenced by space
charge forces within the gun and the low energy sections of the
injector.

\subsubsection{ILC Source Laser System}\label{sssect:ELEisls}

\begin{figure}[htb]
   \begin{center} \vbabove
      \includegraphics[width=0.99\textwidth]{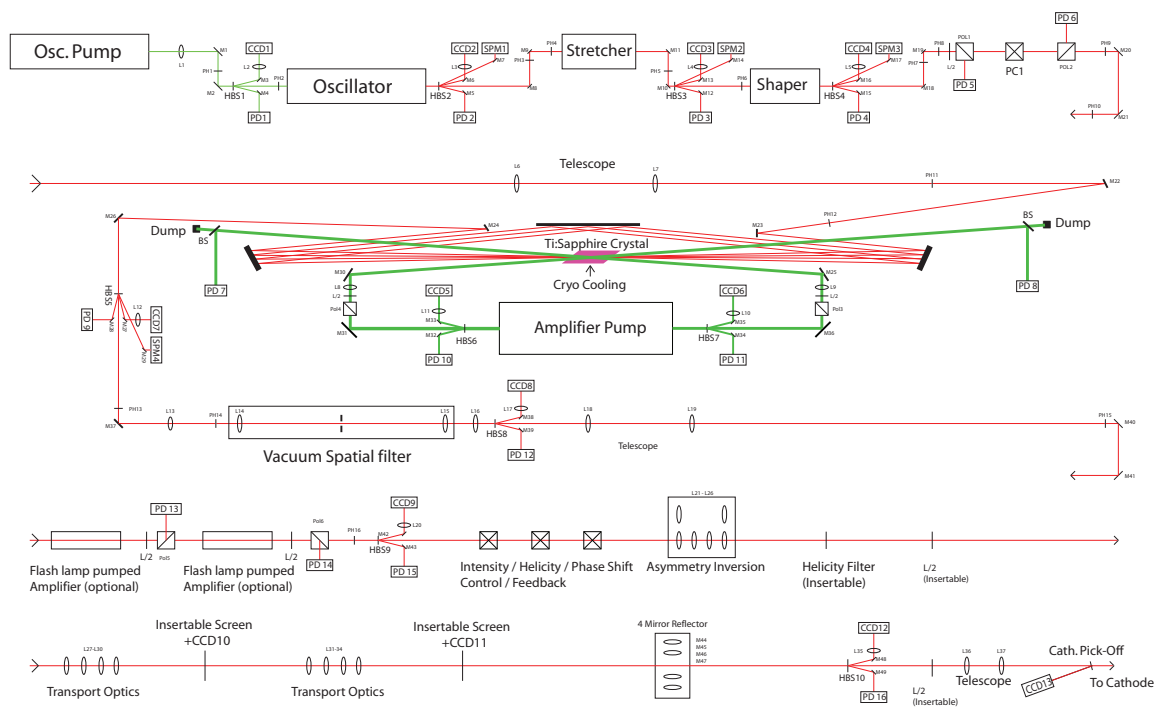}
      \vbabovecaption
      \caption{Schematic view of source drive laser system.}
      \label{fig:ELElaser}
      \vbbelow
   \end{center}
\end{figure}

A conceptual layout schematic of the laser system is depicted in
Figure~\ref{fig:ELElaser}. To match the bandgap energy
of GaAs photocathodes, the wavelength of the laser system
must be 790~nm and provide tunability ($\pm$20~nm)  to optimize
conditions for a specific photocathode. Therefore, the  laser
system is based on Ti:sapphire  technology.

\stepcounter{figlcl}

The 3~MHz pulse train is generated by a cavity-dumped mode-locked
oscillator. After diffractive pulse stretching to 1~ns and temporal
pulse shaping, the bunch train is amplified using a multi-pass
Ti:sapphire amplifier. The amplifier crystal must be cryogenically
cooled to facilitate power dissipation and minimize instabilities
caused by thermal lensing induced by the high power amplifier
pump. A cw frequency-doubled Nd:YAG (or similar such as Nd:vanadate)
diode pumped solid state (DPSS) laser provides the pump power
for the Ti:sapphire amplifier. Additional amplification can be supplied
by one or multiple flashlamp pumped Ti:sapphire stages. Final laser
pulse energy and helicity control is achieved by electro-optical techniques.
This system can also be used as a feed-back device to compensate
for the QE decay of the photocathode between cesiations, to remove
slow intensity drifts of laser and/or electron beam, and to maintain
the circular polarization state of the laser beam. Various optical
techniques are used to cancel systematic effects caused by an
asymmetric laser beam profile or effects associated with the sign
of the helicity of the laser light.

\subsubsection{Bunching and Pre-Acceleration}\label{sssect:ELEbpa}

The bunching system compresses the 1~ns micro-bunches generated
by the gun down to $\sim$20~ps FWHM. It includes two subharmonic bunchers
(SHBs) and a 5 cell traveling wave  $\beta$=0.75  L-band buncher.
The SHB cavities operate at 216.7~MHz and 433.3~MHz, respectively.
Together they compress the bunch to $\sim$200~ps
FWHM. The L-band bunching system is a modification of the TESLA Test
Facility \cite{bib:ELSrefr7} design with a traveling wave buncher to maximize
capture efficiency. The buncher has 5 cells with $\beta$=0.75 and a gradient of
5.5~MV/m and compresses the bunch to 20~ps FWHM. The buncher and
the first few cells of the following TW pre-accelerator are immersed in a
660~G solenoidal field to focus the beam. Two 50 cell $\beta$=1 normal
conducting (NC) TW accelerating sections at a gradient of 8.5~MV/m
increase the beam energy to 76~MeV.  These structures must withstand very high RF power for the duration of the very long pulse but they are identical to those being developed for the positron
source. Further details of the bunching system
are summarized in reference \cite{bib:ELSrefr8}.

\subsubsection{Chicane, Emittance Measurement and Matching Sections}\label{sssect:ELEcemms}

Immediately downstream of the NC pre-acceleration a vertical chicane
provides energy collimation before injection into the SC booster  linac.
The chicane consists of four bending magnets and several 90$^{\circ}$ FODO
cells. The initial dipole at the chicane entrance can be used as a
spectrometer magnet (see Figure~\ref{fig:ELElayout}). A short
beam line leads to a diagnostic section that includes a spectrometer
screen. The injector beam emittance is measured by conventional wire
scanners downstream of the chicane. Two matching sections combine the
chicane and emittance measurement station with the downstream SC booster linac.

\subsubsection{The 5 GeV Superconducting Pre-Acceleration (Booster) Linac}\label{sssect:ELEspabl}

Twenty-one standard ILC-type SC cryomodules accelerate the beam to
5~GeV,and typical FODO cells integrated into the cryomodules
transversely focus the beam. An additional string of three
cryomodules is added to provide redundancy (total of 24
cryomodules). The booster linac consists of two sections. In the
first section, the e$^-$ beam is accelerated from 76~MeV to 1.7~GeV
in cryomodules with one quadrupole per module. In the second
section, the e$^{-}$ beam is accelerated to the final 5~GeV in
cryomodules with one quadrupole every other module.

\subsubsection{Linac to Damping Ring Beamline and Main e$^{-}$ Source Beam Dump}\label{sssect:ldrbmsbd}

The Linac To Ring (LTR) beam line transports the beam to the
damping ring injection point and performs spin rotation and energy
compression. The 5~GeV longitudinally polarized electron beam is
first bent through an arc. At 5~GeV, the spin component in the
plane normal to the magnetic field precesses 90$^{\circ}$ in that plane
for every n $\times$ 7.9$^{\circ}$ (n: odd integer) of rotation of momentum vector.
An axial solenoid field integral of 26.2~T-m rotates the spin direction
into the vertical \cite{bib:ELSrefr9}. A 5~GeV beam dump is installed near the LTR.
To dump the 5~GeV beam, the first bend of the LTR is turned off, and
the dump bend downstream energized. The dump drift is $\sim$12~m.

\subsection{Accelerator Physics Issues}\label{ssect:ELEap}

Simulations indicate that $>$95\% of the electrons produced by the
DC gun are captured within the 6-D damping ring acceptance:
$\gamma(A_{x}$+$A_{y})$ $\leq$0.09 m and $\Delta$E x $\Delta$z $\leq$($\pm$25 MeV) x ($\pm$3.46~cm).
The starting beam size diameter at the gun is 2~cm, and this is
focused to a few mm diameter before it is injected into the DR.
Calculations in the low energy regions of the injector ($\leq$76 MeV)
include space charge effects and use PARMELA \cite{bib:ELSrefr10}.
The beam propagation through the superconducting booster
linac and LTR beam line has been optimized using MAD \cite{bib:ELSrefr11} and
tracked by the ELEGANT code \cite{bib:ELSrefr12}.

\stepcounter{figlcl}\begin{figure}[htb]
   \begin{center}
     \vbabove
     \includegraphics[width=0.99\textwidth]{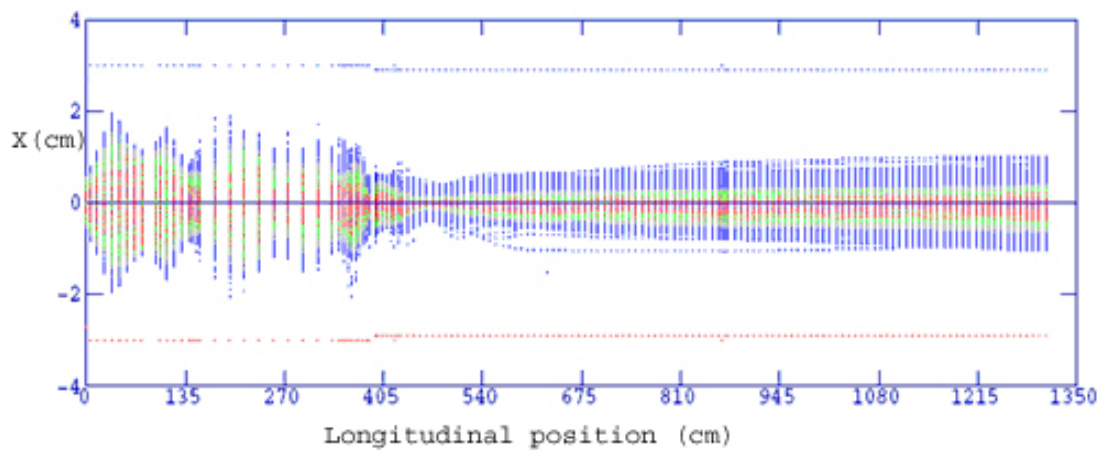}
      \vbabovecaption
      \caption{Beam envelope along the 76 MeV injector.}
        \label{fig:ELEibenvelope}
    \vbbelow
   \end{center}
\end{figure}

\subsubsection{DC Gun and Bunchers}\label{sssect:ELEdgb}

The DC gun creates a 140-160~keV electron beam with a bunch
charge of 4.5-5~nC with a bunch length of 1~ns and an
unnormalized transverse edge emittance at the gun exit of 70~mm-mrad.
To minimize longitudinal growth of the bunch it is desirable to
locate the first subharmonic buncher as close to the gun as possible.
However, the beam lines needed to combine both guns require a
distance of $\sim$1-1.5~m between gun and first SHB. The SHBs
capture almost 100\% of the electrons generated at the gun.
The beam parameters at 76~MeV are summarized in Table~\ref{tab:ELEinjparam}.
A plot of the beam envelope from gun up through the bunching
system is given in Figure~\ref{fig:ELEibenvelope}.
%The transverse emittance through
%bunching and pre-acceleration is shown in Figure~\ref{fig:ELEinjemittance}.

\begin{comment}
\stepcounter{figlcl}\begin{figure}[htb]
   \begin{center} \vbabove
      \includegraphics[width=0.6\textwidth]{\picturefolder injemittance.jpg}
     \vbabovecaption
      \caption{Normalized rms emittance as a function of z up to 76 MeV.}
      \label{fig:ELEinjemittance}
      \vbbelow
   \end{center}
\end{figure}
\end{comment}

\stepcounter{tablcl}\begin{table} [htb] \vbabove
\caption{76 MeV beam parameters after NC bunching and pre-acceleration.} %(end of NC acceleration): }
   \begin{center}
   \label{tab:ELEinjparam}
      \begin{tabular}{| l | c |}
         \hline
         Parameters & $\beta$ = 0.75 TW Buncher Design  \\
         \hline & \vbdlspacing \hline
          Initial charge & 4.5 - 5 nC   \\ \hline
          Transmitted charge & 92\%   \\ \hline
          Phase extension FWHM & 9 deg L-band   \\ \hline
          %Phase extension FW & 25 deg L-band   \\ \hline
          Energy spread FWHM & $<$100 keV   \\ \hline
          %Energy spread FW & $<$ 1.2 MeV   \\ \hline
          Normalized rms emittance & 70 $\mu$m-rad   \\ \hline
      \end{tabular}
   \end{center} \vbbelow
\end{table}

\subsubsection{The 5 GeV Booster  Linac and  Linac to Damping Ring Line (eLTR)}\label{sssect:ELEblldrtl}

\begin{figure}[htb]
   \begin{center}
    \vbabove
      \includegraphics[width=0.99\textwidth]{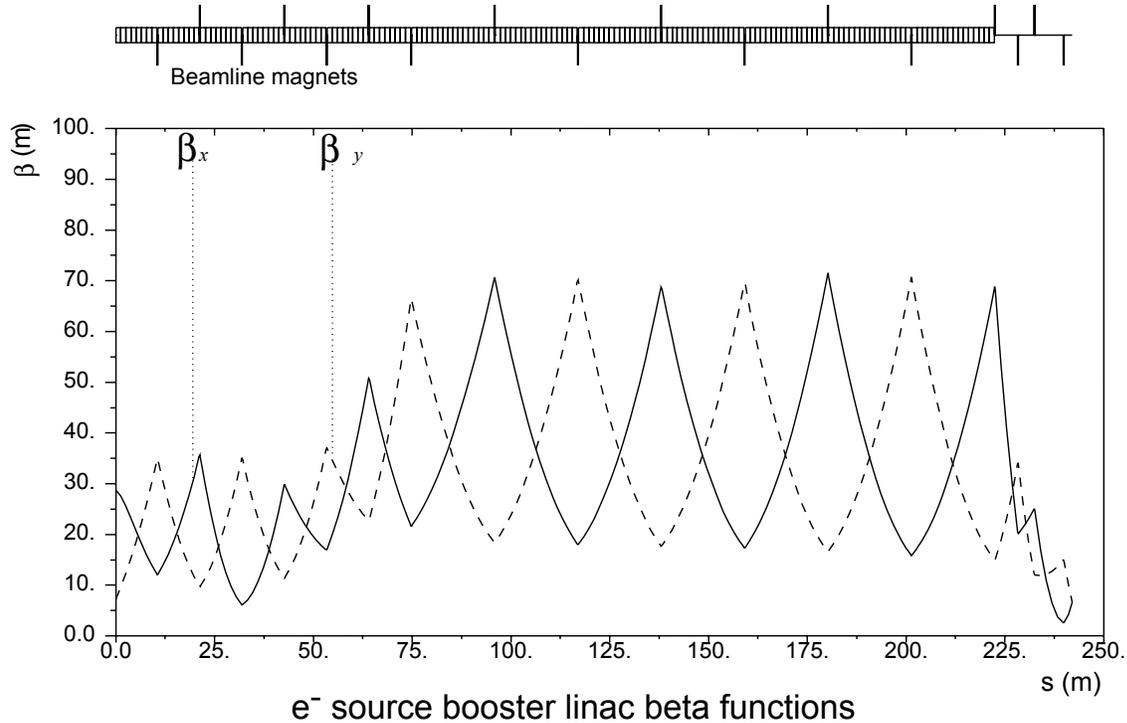}
      \vbabovecaption
      \caption{Optics of the SC electron booster linac.}
        \label{fig:ELEbloptics}
    \vbbelow
   \end{center}
\end{figure}

The optics of the superconducting booster linac are
shown in Figure~\ref{fig:ELEbloptics}.
%and the optics of the LTR in Figure~\ref{fig:ELEdumpline}.

\stepcounter{figlcl}

\stepcounter{figlcl}\begin{figure}[htb]
   \begin{center} \vbabove
      \includegraphics[width=0.99\textwidth]{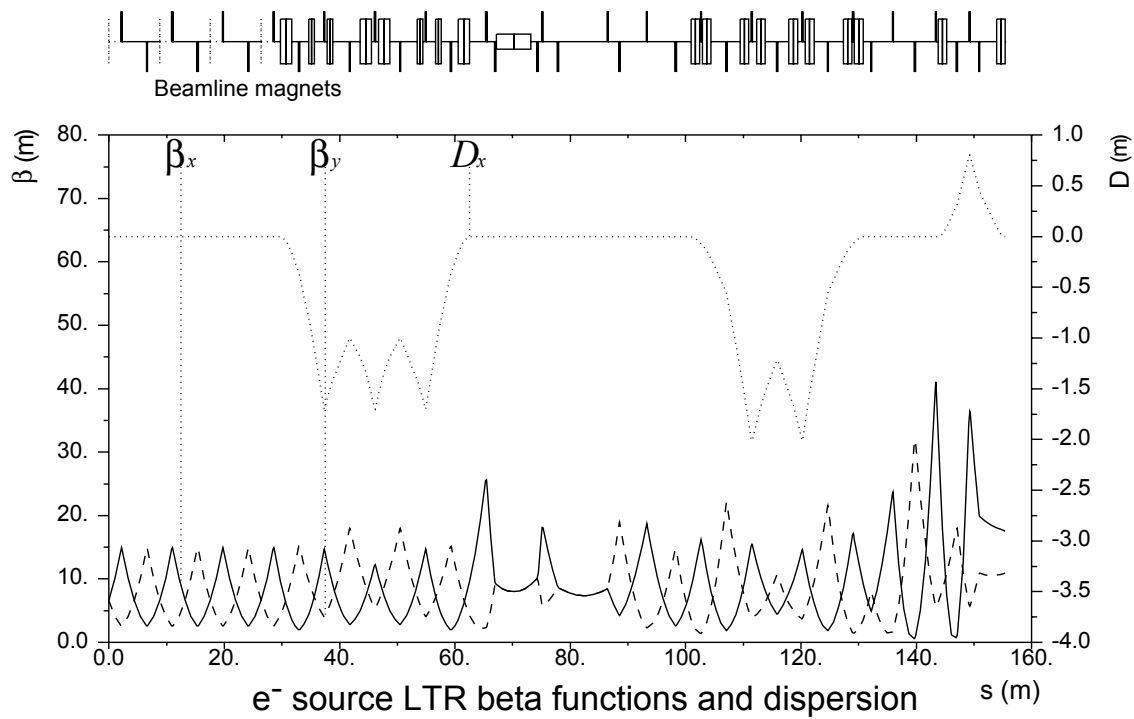}
     \vbabovecaption
      \caption{Optics of the LTR.}
       \label{fig:ELEltroptics}
     \vbbelow
   \end{center}
\end{figure}

At the dump window, the e-edge beam size $\sigma_{x} $/$\sigma_{y}$ is
0.72~cm/1.4~cm and 13.9~cm/1.4~cm for 0\% and $\pm$10\%
energy spread, respectively. These beam sizes are within
the dump window specifications. At the monitor location
the dispersion dominates the beam size and thus the
dump also serves as an energy spectrometer with 0.1\% resolution.

\begin{comment}
\stepcounter{figlcl}\begin{figure}[htb]
   \begin{center} \vbabove
      \includegraphics[width=0.9\textwidth]{\picturefolder dumpline.jpg}
     \vbabovecaption
      \caption{5~GeV beam dump line downstream of the booster linac system.}
      \label{fig:ELEdumpline}
      \vbbelow
   \end{center}
\end{figure}
\end{comment}

The LTR arc consists of four FODO cells with eight bends. The total
arc bending angle is 7 $\times$ 7.9$^{\circ}$. The $R_{56}$ (path
length – energy correlation) is adjustable (86~$\pm$40 cm). The arc
is followed by the solenoid sections and RF unit, which occupy 5.5~m
and 8.32~m, respectively. There are three PPS stoppers with 1~m
space in the LTR arc. Two FODO cells upstream of the LTR arc have
laser wire emittance measurement stations. The optics of the LTR
system are shown in Figure~\ref{fig:ELEltroptics}.
%Figure~\ref{fig:ELElongemitDR} illustrates the result of the energy %compression.

\begin{comment}
\stepcounter{figlcl}\begin{figure}[htb]
   \begin{center} \vbabove
      \includegraphics[width=0.9\textwidth]{\picturefolder longemitDR.jpg}
      \vbabovecaption
      \caption[Longitudinal phase space at the entrance of the DR injection     line.]{Longitudinal phase space at the entrance of the DR injection line after the energy compression; all electrons are within 6-D DR acceptances}
      \label{fig:ELElongemitDR}
    \vbbelow
   \end{center}
\end{figure}
\end{comment}

The arc of the eLTR is designed to rotate the spin vector by 90
degrees from longitudinal into a horizontal position before injection
into the damping ring and to provide the R56 necessary for energy
compression. For a n $\times$ 90$^{\circ}$ of spin rotation, an arc angle
of n $\times$ 7.9$^{\circ}$ is required. A 8.3-m-long superconducting solenoid
with 3.16~T magnetic field solenoid rotates the spin vector into a
vertical orientation. After the bunch is decompressed by the arc,
an RF voltage of 180~MV provided by a 9-m-long 6-cavity superconducting
linac, rotates the electrons in longitudinal phase space to match
with longitudinal DR acceptance. The LTR also includes an
additional 34.5$^{\circ}$ horizontal bend, a matching section with 4
quadrupoles and a double bend achromat to match Twiss
parameters at the DR injection line \cite{bib:ELSrefr13}.

\subsection{Accelerator Components}\label{ssect:ELEc}

\subsubsection{Table of Parts Count}\label{sssect:ELEtpc}

Table~\ref{tab:ELStot} lists the major components of the ILC electron source and
Table~\ref{tab:ELSlen} the lengths of the various electron source beamlines.

\stepcounter{tablcl}\begin{table} [htb]
    \begin{center} \vbabove
     \caption{Total number of components for the polarized electron source.}
    \setlength{\tabcolsep}{6pt}
     \label{tab:ELStot}
    \begin{tabular}{| l | r || l | r || l | r |}
        \hline
        \multicolumn{2}{| c ||}{Magnets} &
        \multicolumn{2}{| c ||}{Instrumentation} &
        \multicolumn{2}{| c |}{ RF} \\
        \hline & & & & & \vbdlspacing \hline
       Bends     & 25 & BPMs           & 100 & 216.7 SHB Cavity             & 1 \\ \hline
       Quads (NC)     & 76 & Wirescanners          &  4 & 433.3 SHB Cavity          &  1 \\ \hline
       Quads (SC)   & 16 & Laserwires           &   1 & 5 Cell L-band buncher & 1 \\ \hline
       Solenoids(NC)   &  12 & BLMs           &   5 & L-band TW structure  &   2 \\ \hline
       Solenoids(SC) &   2 & OTRs &   2 & 1.3 GHz cryomodules  &   25 \\ \hline
       Correctors(SC)     &  32 & Phase monitors      &   2 &  L-band klystrons/modulators  &   13 \\ \hline
     \end{tabular}
    \vbbelow
    \end{center}
\end{table}

\stepcounter{tablcl}\begin{table} [t]
   \begin{center} \vbabove
   \caption{System lengths for the e- source beamlines.}
   \label{tab:ELSlen}
      \begin{tabular}{| l | l |}
          \hline
          Beam Line Section &  Length \\
          \hline & \vbdlspacing \hline
          Gun area & 7 m \\ \hline
          NC beam lines & 14 m \\ \hline
          Chicane + emittance station & 54 m \\  \hline
          SC beam lines & 245 m \\  \hline
          eLTR & 157 m \\ \hline
          Dumplines & 12 m \\
          \hline\hline
          Total beam line length & 489 m \\ \hline
          Total tunnel length & 505 m \\  \hline
      \end{tabular}
   \end{center}
   \vbbelow
\end{table}

\clearpage %\setcounter{page}{1}
\setcounter{figlcl}{0}
\setcounter{tablcl}{0}
\setcounter{secnumdepth}{4} \setcounter{subsection}{0}
\setcounter{subsubsection}{0}
\usecounter{subsubsection}

\setcounter{section}{2} \renewcommand{\picturefolder}{./positron/}

\section{Positron Source}\label{sect:POSps}

\subsection{Overview}\label{ssect:POSo}

The ILC Positron Source uses photoproduction to generate positrons.
The electron main linac beam passes through a long helical undulator
to generate a multi-MeV photon beam which then strikes a thin metal target
to generate positrons in an electromagnetic shower. The positrons are
captured, accelerated, separated from the shower constituents and unused
photon beam and then are transported to the Damping Ring.
Although the baseline design only requires unpolarized positrons,
the positron beam produced by the baseline source has a
polarization of $\sim$30\%, and beamline space has been reserved for an
eventual upgrade to $\sim$60\% polarization.

The positron source must perform three critical functions:

\begin{itemize}

    \item generate a high power multi-MeV photon production drive beam in a suitable short
            period, high K-value helical undulator; \itemspace
    \item produce the needed positron bunches in a metal target that can reliably
            deal with the beam power and induced radioactivity; \itemspace
    \item capture and transport the positron bunch to the ILC Damping Rings
             with minimal beam loss. \itemspace

\end{itemize}
In addition, the Positron Source requires sufficient instrumentation, diagnostics
and feedback (feedforward) systems to ensure optimal operation of the source and ILC.

\subsection{Beam Parameters}\label{ssect:POSbp}

The key parameters of the Positron Source are listed in Tables~\ref{tab:POSparambeam},
\ref{tab:POSparamund}, \ref{tab:POSparamtar}. The source is
required to deliver 2$\times$\(10^{10}\) positrons per bunch at the IP with the nominal ILC
bunch structure and pulse repetition rate. The source target system is designed with a 50\%
overhead and can deliver up to 3$\times$\(10^{10}\) positrons per bunch to the 400~MeV point. There is sufficient RF power to accelerate
2.5$\times$\(10^{10}\) to the damping ring within the 0.09 m-rad transverse dynamic aperture.

\stepcounter{tablcl}\begin{table} [t] \vbabove
 \caption{Nominal Positron Source parameters ($^{\dagger}$ upgrade values).}
   \begin{center}
   \label{tab:POSparambeam}
      \begin{tabular}{| l |c |c | c |}
         \hline
         Beam Parameters & Symbol & Value & Units \\
         \hline & & &  \vbdlspacing  \hline
         Positrons per bunch at IP & $n_{b}$ & 2$\times$\(10^{10}\) & number   \\  \hline
         Bunches per pulse &  $N_{b}$ & 2625 & number \\  \hline
         Pulse repetition rate & $f_{rep}$ & 5 & Hz \\  \hline
         Positron energy (DR injection) & $E_{0}$ & 5 & GeV \\  \hline
         DR transverse acceptance & $\gamma(A_{x}$+$A_{y})$ & 0.09 & m-rad \\  \hline
         DR energy acceptance & $\delta$ &  $\pm$ 0.5 &  \% \\ \hline
         DR longitudinal acceptance & $A_{l}$ & $\pm3.4 \times \pm$25 & cm-MeV \\  \hline
         Electron drive beam energy & $E_{e}$ & 150 & GeV \\  \hline
         Electron beam energy loss in undulator & $ \Delta E_{e} $  & 3.01 & GeV \\ \hline
         Positron polarization $^{\dagger}$ & \it{P} & $\sim$60 & \% \\  \hline
      \end{tabular}
   \end{center}
\vbbelow
\end{table}

\stepcounter{figlcl}\begin{figure} [b] \vbabove
   \begin{center}
      \includegraphics[width=0.95\textwidth]{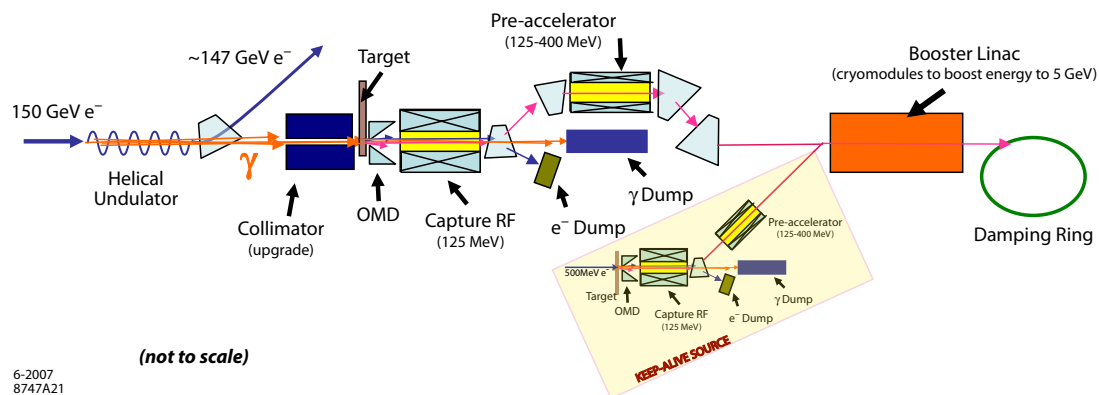}
\vbabovecaption
      \caption{Overall layout of the Positron Source.}
      \label{fig:POSscheme}
   \end{center}
\vbbelow
\end{figure}

\subsection{System Description}\label{ssect:POSsd}

\begin{figure} [hb!]
\vbabove\vbabove\vbabove\vbabove\vbabove\vbabove\vbabove\vbabove
   \begin{center}
      \includegraphics[width=0.95\textwidth]{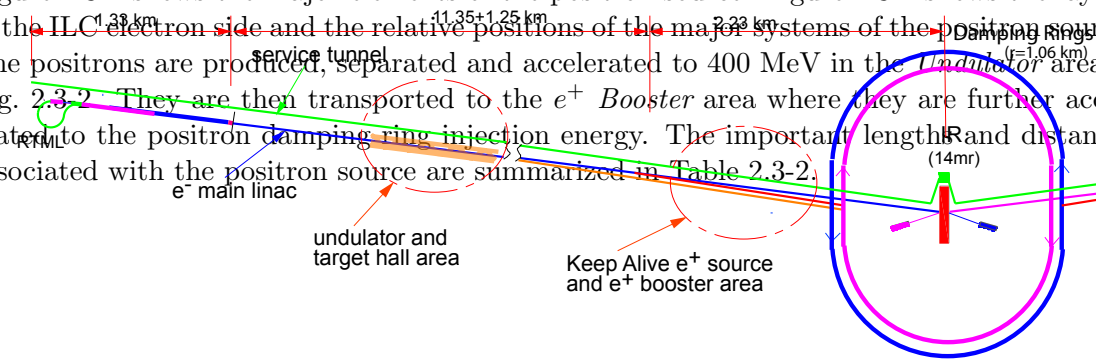}
\vbabovecaption
      \caption{Positron Source locations within the ILC complex.}
      \label{fig:POSlayout}
   \end{center}
\vbbelow\vbbelow
\end{figure}

Figure~\ref{fig:POSscheme} shows the major elements
 of the positron source. Figure~\ref{fig:POSlayout} shows the layout of the ILC electron side and the relative positions of the
 major systems of the positron source. The positrons are produced, separated and accelerated to 400 MeV in
 the {\it Undulator} area of Fig.~\ref{fig:POSlayout}. They are then transported to the {\it $e^{+}$ Booster} area
 where they are further accelerated to the positron damping ring injection energy.  The important
 lengths and distances associated with the positron source are summarized in Table~\ref{tab:POSdistances}.

\stepcounter{figlcl}
Positrons are produced in electromagnetic showers when a multi-MeV photon beam impinges
 on a metal target. The photon beam is produced by passing the main electron linac beam
 through a long undulator. This photon beam is transported $\sim$ 500 meters to the positron source
 target hall where it hits a 0.4 radiation length thick Ti-alloy target producing showers of
 electrons and positrons. The resulting beam is captured using an optical matching device
 (OMD) and normal conducting (NC) L-band RF with solenoidal focusing and
 accelerated to 125 MeV. The electrons and remaining photons are separated from the positrons
 and dumped. The positrons are accelerated to 400 MeV in a NC L-band linac with solenoidal
focusing. They are transported $\sim$5~km to the central damping ring complex, where they are
 boosted to 5 GeV in a linac using superconducting (SC) L-band RF
and injected into the positron damping ring.

The positron source system also includes a Keep Alive Source to generate a low intensity
 positron beam that can be injected into the SC L-band linac. This allows various beam
 feedbacks to remain active if the main electron beam, and hence the undulator based positrons,
is lost. ILC availability studies (see Section 2.9.1) show that the Keep Alive Source makes a
significant improvement in accelerator uptime and delivered luminosity.
This source uses a 500~MeV electron drive beam impinging on a tungsten-rhenium
 target to produce positrons which are then captured and accelerated to 400 MeV similar to the main positron source. The Keep Alive Source is designed to produce 10\%
bunch intensity for the full 2625 bunch ILC pulse train at 5~Hz.

\stepcounter{tablcl}\begin{table} [htb] \vbabove
 \caption{Positron Source beamline lengths.}
   \begin{center}
    \label{tab:POSdistances}
     \begin{tabular}{| l | c |}
         \hline
         Area & Length (meters) \\
         \hline & \vbdlspacing \hline
         Undulator chicane insert & 1257  \\  \hline
         Undulator center to target & 500   \\  \hline
         Undulator insert length &  200  \\  \hline
         Target Hall length & 150 \\  \hline
         400 MeV long transport line & 5032 \\  \hline
         Total RF acceleration length & 350 \\  \hline
         Damping Ring injection line &  431 \\
         \hline
      \end{tabular}
   \end{center}
\vbbelow
\end{table}

\subsubsection{Photon Production}\label{sssect:POSpp}

The Positron Source relies upon an intense beam of high energy photons impinging upon
 a metal target. The photons must be of sufficient energy, typically of order 10 MeV, to
 generate electron-positron pairs that can escape from the target material and be captured
 and accelerated. The photons are generated by the radiation from relativistic electrons as they pass through the periodic, helical, magnetic
 field of the undulator. Details of the undulator are provided in  Section~\ref{sssect:POSu}. To generate the required
 photon energy, very high energy electrons are required. To avoid the expense of a
 dedicated electron beam, the undulator is installed part way along the electron main
 linac, where the electron energy has reached 150~GeV. After passing through the
 undulator the electrons continue through the remainder of the main electron linac, gaining
 energy up to 250~GeV. The first harmonic cut-off energy for the photon spectrum is 10~MeV.

A helical undulator generates twice the synchrotron radiation power per period than the
equivalent planar undulator, enabling the overall undulator to be shorter for the
same number of positrons. The helical device also produces circularly polarized light
which in turn generates longitudinally polarized positrons.

%This can be increased
% by lengthening the undulator to produce more photons. Photons with the wrong
% polarization state are then rejected by careful spatial collimation of the photon beam
% close to the target. In such an upgrade, positron polarization of the order of 60\% should
% readily be achieved with no reduction in positron bunch intensity delivered to the damping ring.

\subsubsection{Positron Production and Capture}\label{sssect:POSppc}

\stepcounter{figlcl}\begin{figure} [htb] \vbabove
   \begin{center}
      \includegraphics[width=0.9\textwidth]{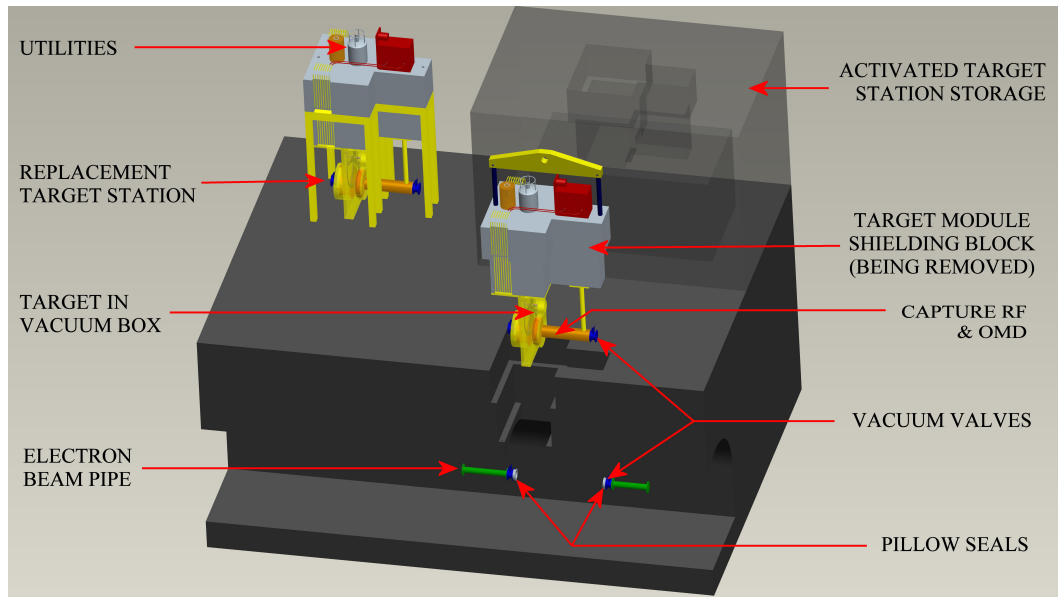}
\vbabovecaption
      \caption[Target removal scheme.]{Target removal scheme (5-spoke target wheel, OMD and first RF section is seen being
       removed from the beamline).}
      \label{fig:POSremotehand}
   \end{center}
\vbbelow
\end{figure}

The positron production, capture and transport to the damping rings are shown in
 Figure~\ref{fig:POSscheme}.
The photon beam generated by the helical undulator is incident on the rim of a
 0.4 $X_0$ thick rotating target (see Section~\ref{sssect:POSt} )
%of thickness 0.4 radiation lengths
contained in a
 vacuum vessel. The photon beam has a transverse size of $\sim$1 mm rms and deposits
 $\sim$10.5 kW of power in the target. Photons up to the 8$^{\rm th}$ harmonic contribute to the positron generation.
The particles emerging from the downstream side of the target are captured in the 0.09 m-rad transverse
 dynamic aperture defined by the positron damping ring. The energy of the beam coming
 out of the target is 3 - 55 MeV. The target is followed by the tapering magnetic field of an
 Optical Matching Device (OMD) (see Section~\ref{sssect:POSomd} ) which has a field which decays
 from 5 - 0.5 T over 20 cm. The OMD matches the beam phase space from the target into the capture L-band RF
%. The capture RF is placed after the target and OMD and
 which accelerates the beam to 125 MeV.  The RF cavities have an average
 gradient of 8~MV/m and are located inside 0.5 Tesla solenoids which provide beam
 focusing. Details of the RF are given in Section~\ref{sssect:POSncras}.

The target and equipment immediately downstream are expected to become highly
activated. A remote-handling system is used to replace the target, OMD and 1.3 meter
 NC RF cavities. Due to the underground location, the activated equipment
 needs to be removed vertically from the target vault. Figure~\ref{fig:POSremotehand} shows the
 conceptual design of such a system, where the target wheel, OMD and the first 1.3 meter NC
 RF cavity is shown being removed. This design does require
 special vacuum seals to enable speedy removal from the beam line, as the power
 deposition from the beam does not allow for windows in these lines.

\subsubsection{Low Energy Positron Transport}\label{sssect:POSlept}

Downstream of the capture RF, the positrons are separated from electrons and
 photons in an achromatic dogleg which horizontally deflects the positron line by 2.5 m.
 A set of collimators remove positrons with large diverging angles and
 large energy offsets. Normal conducting
 L-band RF structures embedded in a constant solenoid field of 0.5 T accelerate the
 positrons from 125 MeV to 400 MeV. The accelerating gradient is $\sim$8 MV/m and the
 length is 34.6 m.

A dogleg deflects the beam 5 m horizontally to the electron main linac tunnel and 2 m vertically to position the long positron transport line
 above the electron main linac. This beamline carries the positrons
 4.09 km to the end of the main linac tunnel, then 941 m to the positron booster linac in a separate tunnel.

\subsubsection{Keep Alive Source}\label{sssect:POSkas}

The Keep Alive Source (KAS) is designed to deliver a low intensity ( $\sim$10\%) beam of
 positrons at 400 MeV to the positron booster linac in case the primary positron beam
 is unavailable. It occupies $\sim$500 meters of tunnel just before the booster linac. A 500 MeV electron beam impinges on a tungsten-rhenium target to produce positrons. The electron drive beam is similar to the main electron source.
 The KAS positron target has a simpler design because of
 the lower incoming beam power, but still requires remote handling. The positrons are captured, separated and accelerated to 400 MeV using the
 same system as for the primary positron beam.

\subsubsection{5-GeV SC Booster Linac}\label{sssect:POSfsbl}

The SC booster linac accelerates the beam from 400 MeV to 5 GeV in three sections
of periodic FODO lattice. The first section up to 1083 MeV has four non-standard
 cryomodules, each containing six 9-cell cavities and six quadrupoles. The quad field
 strength ranges from 0.88-2.0 T. The second section up to 2626 MeV has six
 non-standard cryomodules, each containing eight 9-cell cavities and two quadrupoles
. The quad strength ranges from 0.62-1.3 T. Finally, the positrons are accelerated to 5
 GeV using twelve standard ILC-type cryomodules, each with eight 9-cell cavities and
 one quadrupole with strength ranging from 0.95-1.63 T.

\subsubsection{Linac to Damping Ring Beam Line}\label{sssect:POSldrbl}

The Linac to Ring (LTR) brings the positrons from the booster linac
to the Damping Ring (DR) injection line. In addition, the LTR
orients the beam polarization and compresses the beam energy to
improve acceptance into the DRs. The LTR design is the same as for
the electron source LTR described in \ref{sssect:ELEblldrtl}.
 The longitudinal
polarization of the positrons from the target is preserved to the LTR. If polarization is needed
at the IP it must be preserved through the DR. This is achieved by
rotating the spin to vertical before injection into the DR. The LTR contains bending
magnets which rotate the spin vector from longitudinal to horizontal, followed by
solenoids, if turned on, that rotate from horizontal to vertical. At 5 GeV, the total bending angle
must be an odd integer multiple of 7.9$^{\circ}$  to produce a net 90$^{\circ}$  of spin rotation.
26.23 T-m of solenoidal field is required to produce the horizontal-to-vertical spin rotation
which is provided by two 2.5 meter 5.2 T solenoids.

\subsection{Accelerator Physics Issues}\label{ssect:POSapi}

\subsubsection{Photon Drive Beam}\label{sssect:POSpdb}

The photon drive beam is generated by passing the main electron beam through a long,
 small-aperture undulator which sits in the middle of a magnetic chicane. The design
of this system has to ensure that this does not compromise the main electron beam
 quality, and hence the ILC luminosity. In addition the undulator system and the main
 linac downstream of the undulator need to be protected from any beam failures.

The electron beam transport through the complete undulator system is based upon
a simple FODO arrangement with quadrupole spacing of $\sim$12 m (in the room
 temperature section). There are beam position monitors (BPM) at every
 quadrupole and horizontal and vertical corrector magnets
in each cryomodule.
Preliminary studies \cite{bib:POSdscott} indicate that the total emittance growth in this insertion is small compared to the overall main linac budget.
%A recent study \cite{bib:POSdscott} indicated that a quadrupole to
%BPM misalignment of 10 $\mu$m
%leads to a vertical emittance growth of less than 2\%, due to the
%quadrupole mislignment rather than the undulator.
%Electron beam transport
% calculations have shown that excellent relative alignment between the quadrupoles
%and neighboring BPMs is required. In this simple model, a quadrupole to BPM
% misalignment of $\sim$5 $\mu$m leads to an emittance growth of $\sim$2\%. This is not
%however due to the undulator but to the quadrupoles and may be mitigated by
% dispersion free steering correction algorithms.
The undulator increases the energy spread of the electron beam from
0.16\% to 0.23\%.
%, corresponding to an 0.17\% in quadrature contribution at the IP.

The baseline pressure requirement of \(10^{-8}\) torr has been set to avoid fast ion
instability problems. Vacuum calculations confirm that the cryopumping will be
 adequate provided that photons with energy $>$10 eV
%striking the
%vessel surface is kept low enough. Extensive calculations of the undulator %photon
%output down to these very low energies have been carried out and these indicate
%that low power photon absorbers should be
are intercepted by absorbers spaced approximately every 12 m to
shadow the cold vessel surfaces. These absorbers are
in room temperature sections.

To protect the undulator and downstream linac from beam failure,
there is a fast extraction system before the chicane that can dump the main
electron beam into a full power beam dump. A collimator in front
of the undulator can intercept a few bunches before the dump system fires.

\subsubsection{Positron Generation}\label{sssect:POSpg}

\begin{comment}
The major issues for positron production concern the extremely
 high beam power, the activation of the target and beamline components and the
 need to match the positrons from the target into the phase space of the
 capture RF system. Also it is not possible to place a beam window after the target,
 adding to the complexity of the target system.

The photons used to make positrons have $\sim$ 131 kW beam power in $\sim$ 1 mm spot size.
 Approximately 10.5 kW of power is deposited in the target. The deposited energy
causes a temperature rise and the near instantaneous nature of the beam bunches
 cause shock/stress effects that will destroy any stationary metal target.
The ILC positron target design uses a 1-meter diameter wheel spinning with a rim
speed of 100 m/s causing the deposited energy to be smeared out into a 1 mm by 10 cm
stripe thereby mitigating the shock/stress effects. Because it is not possible to
place a window after the target material, the target wheel is in vacuum necessitating
 special vacuum feedthroughs to allow for cooling and for the target to spin.
\end{comment}
The primary issue for positron production is to efficiently capture the positrons which are produced
with a small spatial extent and large angles. Point-to-parallel focusing immediately
after the target increases the positron capture. An optical matching device (OMD)
 placed immediately after the target produces a longitudinal field that decays
 from 5 Tesla to 0.5 Tesla in $\sim$ 20 cm. Calculations show a factor of two
 improvement in positron capture from the OMD.

\subsubsection{Beam Transport}\label{sssect:POSbt}

The positron beam transport must
 efficiently bring the large emittance beam from the target through several km of beamline. The beam at damping ring injection must match the damping ring
phase space to avoid beam losses in the damping ring. Beam outside the acceptance must be absorbed on collimators to localize radioactivation.

\begin{comment}
The optics of the positron source system starting from target to DR injection is
shown in Figure~\ref{fig:POSoptics}.

\stepcounter{figlcl}\begin{figure}[bhtp] \vbabove
   \begin{center}
      \includegraphics[width=0.6\textwidth]{\picturefolder optics.pdf}
\vbabovecaption
      \caption{Positron Source optical layout (from 125 MeV to DR Injection).}
      \label{fig:POSoptics}
   \end{center}
\vbbelow
\end{figure}
\end{comment}

The linac transfer line that takes the 400~MeV positron beam from the target hall
to the booster linac has 16.8 m long FODO cells with 90 degree phase advance
per cell and $\sim$28.5 m maximum $\beta$-function. It follows the earth's curvature as
does the linac tunnel. The
vertical dogleg which brings the positron beam 8~m vertically from the linac
tunnel to the booster tunnel, has at each end a double bend achromat to
provide 17.1~mrad of bending angle. Four quads are inserted in between two
bends to create 180$^{\circ}$  phase advance between the two bends and cancel the
dispersion. The last section connects to the positron booster linac.

\begin{figure} [htb] \vbabove
   \begin{center}
      \includegraphics[width=0.75\textwidth]{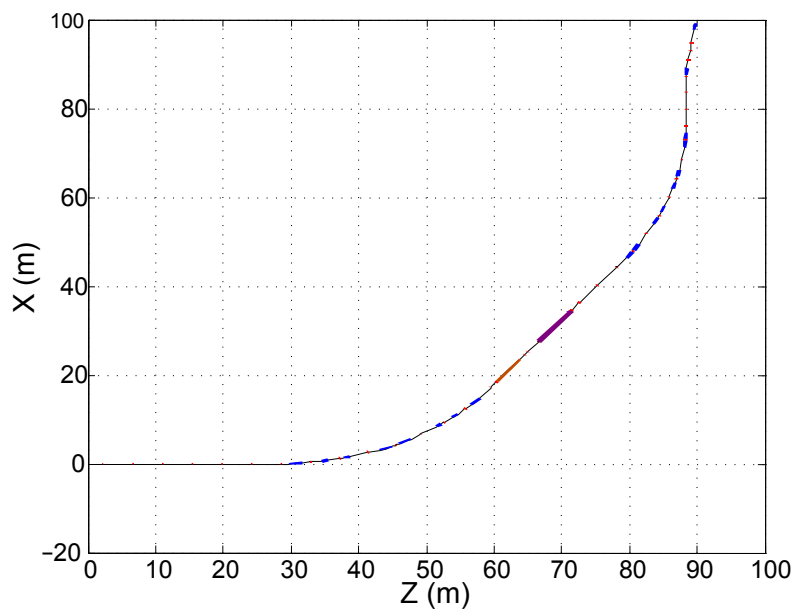}
\vbabovecaption
      \caption[Plan view of the LTR beamline.]{Plan view of the LTR beamline (matching happens from 1-25 meters and
                   DR injection is at z=90meters).}
      \label{fig:POSltr}
   \end{center}
\vbbelow
\end{figure}

In order to match the positron beam into the longitudinal acceptance of the damping
 ring, the beam energy spread is reduced from $\pm 2.8\%$ to $\pm 0.5\%$.
%before injection into the damping
% ring. This energy compression is accomplished by a combination of booster
% linac RF phase and beamline.
 The energy compression and spin rotation take place in
 four FODO-like cells with 8 bends in the first arc of the LTR. The total bending angle is  55.5$^{\circ}$.
The nominal momentum compaction, $R_{56}$,  is 86~cm but it is adjustable. After
 the bunch decompression, a standard 12~m superconducting cryomodule at an RF voltage of
 180 MV rotates the positrons in longitudinal phase space to match the DR acceptance. The rest of the LTR includes a section with an
 additional 34.5$^{\circ}$  horizontal bend, a matching section with 4 quadrupoles and a
 double bend achromat used to match into the DR injection line.
The geometry is shown in Figure~\ref{fig:POSltr}.

\stepcounter{figlcl}

\begin{figure} [htb] \vbabove
   \begin{center}
      \includegraphics[width=0.8\textwidth]{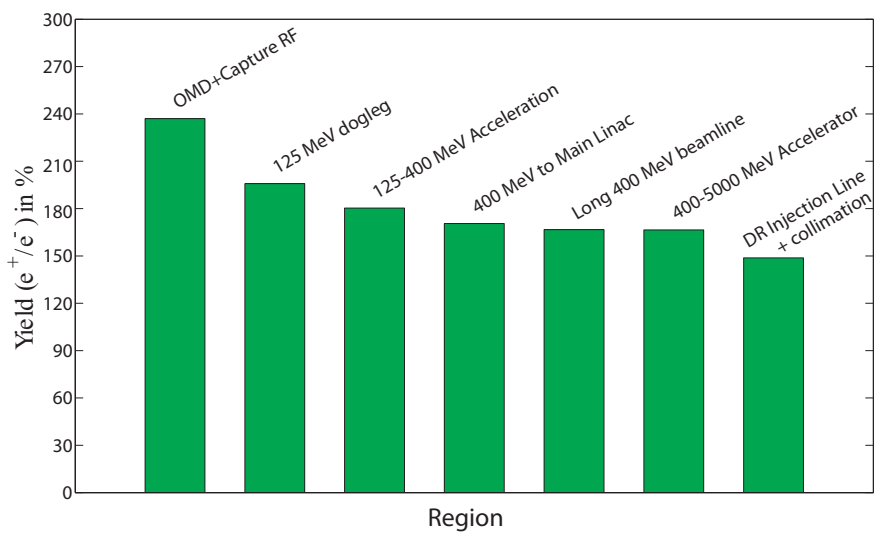}
\vbabovecaption
      \caption{Positron yield in various parts of the Positron Source.}
      \label{fig:POSbeamloss}
   \end{center}
\vbbelow
\end{figure}

Multi-particle tracking has been performed from the target to the DR injection.
%The ELEGANT code, \cite{bib:ELSrefr12} was chosen to track the unique positron beam with
The ELEGANT code \cite{bib:ELSrefr12} was chosen to track the
positron beam with large angular divergence and long low-energy
tails. The LTR energy compression
 was optimized to maximize the positron beam within the 6-dimensional acceptance
in the DR equal to $\gamma(A_{x}$+$A_{y})$  $<$0.09  m and ($\pm$25
MeV)$\times$($\pm$3.46 cm). Of the positrons from the target, 55\%
survive the transport through the complete beamline based on the
physical apertures of the beam pipes \cite{bib:POSoptics} and
 $\sim$50\% of the positrons are within the DR 6-dimensional acceptance.
An energy collimator in the LTR second arc reduces the number of unwanted
 particles reaching the DR from 5.6\% to 1.1\%. Additional betatron and energy
 collimators may be required to collimate the rest of the unwanted 1.1\% of
 particles, 0.8\% of which are outside of the transverse DR acceptance.
Tracking with realistic magnet errors shows similar results after orbit correction.
 Figure~\ref{fig:POSbeamloss} shows the positron yield in various parts of the ILC Positron Source.

\stepcounter{figlcl}

\subsection{Accelerator Components}\label{ssect:POSac}

Table~\ref{tab:POScomponents} lists the components for the positron source. In
addition to this there are two target stations, the first of which is the main
production target and the second used in the Keep Alive Source, and their
 associated instrumentation. Except for the undulator, target, remote handling and the
OMD, costing for the positron source system was done by the global systems
 groups. The undulator, target and OMD costs were estimated by the design engineers
and the remote handling system costs were projected from costs associated
with remote handling in other accelerator facilities.

\stepcounter{tablcl}\begin{table} [htb] \vbabove \caption{Total
number of components in the Positron Source.}
     \label{tab:POScomponents}
    \begin{center}
    \begin{tabular}{| l | c || l | c |}
        \hline
       Magnets & \# & Instrumentation & \#  \\
       \hline & & & \vbdlspacing \hline
       Dipoles & 157 & BPM x,y pairs  & 922  \\ \hline
       NC quads & 871 & BPM readout channels & 922 \\ \hline
       SC quads     & 51 & Wire scanners          &  29  \\ \hline
       Sextupoles   & 32 & Beam length monitors  & 2   \\ \hline
       NC solenoids & 38 & Profile monitors & 7 \\ \hline
       SC solenoids & 2 & Photon profile monitors & 3 \\ \hline
       NC correctors & 871 & & \\ \hline
       SC correctors & 102 & RF & \#   \\ \hline
       Kickers   &  15 & NC L-band structures  &   30  \\ \hline
       Septa &   4 & 1.3 GHz SC cavities & 200 \\ \hline
       SC undulator cryomodules & 42 & 1.3 GHz cryomodules & 26  \\ \hline
       OMD & 2 & 1.3 GHz klystrons/modulators & 37 \\ \hline
     \end{tabular}
    \end{center}
\vbbelow
\end{table}

\subsubsection{Undulator}\label{sssect:POSu}

\stepcounter{figlcl}\begin{figure} [htb] \vbabove
   \begin{center}
      \rotatebox{-90} {\includegraphics[width=0.5\textwidth]{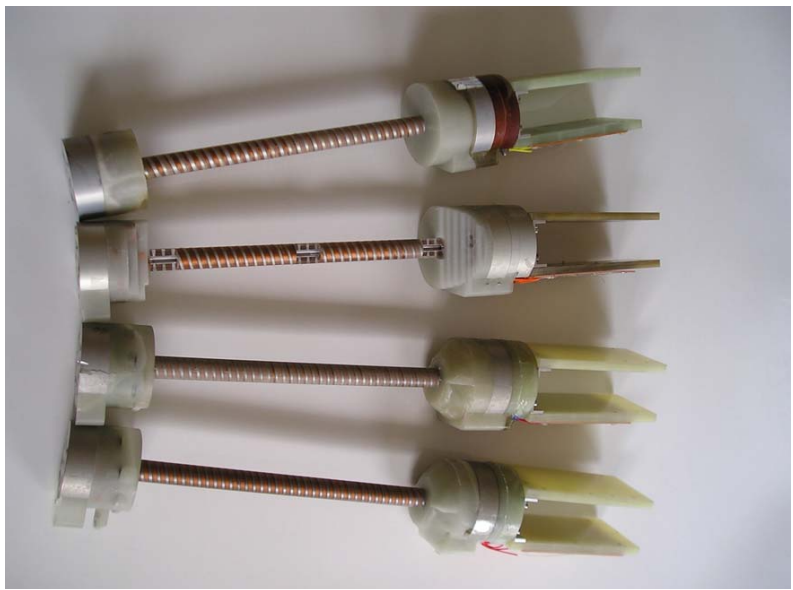}}
\vbabovecaption
      \caption{Short sample undulator prototypes.}
      \label{fig:POSundshort}
   \end{center}
\vbbelow
\end{figure}

\begin{figure} %[bhtp]
\vbabove
   \begin{center}
      \includegraphics[width=0.95\textwidth]{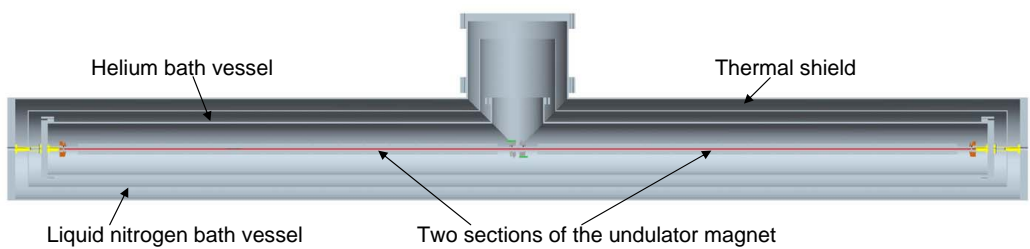}
\vbabovecaption
      \caption{4-meter undulator cryomodule.}
      \label{fig:POScryomod}
   \end{center}
\vbbelow
\end{figure}

The undulator must be superconducting to achieve the required parameters of high
 field and short period. The present baseline parameters are given in Table~\ref{tab:POSparamund}.
 Two interleaved helical windings of NbTi spaced half a period apart generate the
 transverse helical field. Figure~\ref{fig:POSundshort} is a picture of some short sample
undulator prototypes showing the forms for the helical windings. The 147 m
of undulator is supplied by forty-two 4 m long cryomodules containing
two separate undulators with an active undulator length per cryomodule of
3.5 m. Figure~\ref{fig:POScryomod} shows the cryomodules with the two undulators running
along the center.

The undulator vacuum chamber has a nominal inner diameter of 5.85 mm and is made of
 copper. The extremely high conductivity of copper at cryogenic temperatures
 mitigates resistive wall effects. The material between the superconducting
windings is soft magnetic iron which also serves as an outer yoke to increase
the field and provides additional support. Each cryomodule contains a liquid
helium bath and in-situ cryocoolers are used to achieve zero liquid boil off.

Since the electron vacuum vessel is at cryogenic temperatures, each module
 effectively acts as a long cryopump. Roughing pumps are installed in room
 temperature sections between cryomodules (approximately every 12 m) but
 achieving UHV conditions relies upon cryopumping. To achieve the baseline
 pressure requirement of  \(10^{-8}\)  torr absorbers to prevent photons striking the
cold vessel surfaces are placed every 12 meter in room temperature section.

\stepcounter{figlcl}

\stepcounter{tablcl}\begin{table} \vbabove \caption{Nominal
undulator parameters.}
    \label{tab:POSparamund}
   \begin{center}
      \begin{tabular}{| l |c |c | c |}
         \hline
         Undulator Parameters & Symbol & Value & Units \\
         \hline & & & \vbdlspacing \hline
         Undulator period & $\lambda$ & 1.15 & cm \\  \hline
         Undulator strength & K  &  0.92 &   \\  \hline
         Undulator type &  & helical &  \\  \hline
         Active undulator length & $L_{u}$ & 147 & m \\  \hline
         Field on axis & B & 0.86 & T \\  \hline
         Beam aperture & & 5.85 & mm \\  \hline
         Photon energy ($1^{\rm st}$ harmonic cutoff) & $E_{c10}$ & 10.06 & MeV \\  \hline
         Photon beam power & $P_{\gamma}$ & 131 & kW \\  \hline
      \end{tabular}
   \end{center}
\vbbelow
\end{table}

\subsubsection{Target}\label{sssect:POSt}

\begin{figure} [htb] \vbabove
   \begin{center}
      \includegraphics[width=0.8\textwidth] {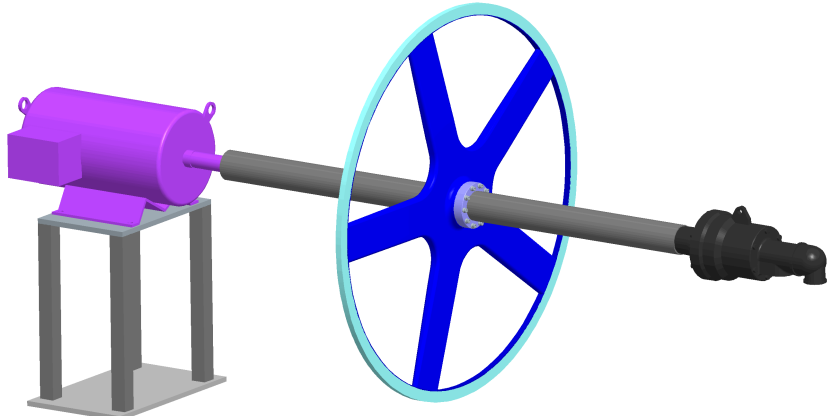}
\vbabovecaption
      \caption{Target station layout.}
      \label{fig:POStarget}
   \end{center}
\vbbelow
\end{figure}

The ILC positron target parameters are shown in Table~\ref{tab:POSparamtar}.
The positron production target is a rotating wheel made of titanium alloy (Ti-6\%Al-4\%V).
 The photon beam is incident on the rim of the spinning wheel, whose diameter
is 1 m and thickness is 0.4 radiation lengths (1.4 cm). During operation the
outer edge of the rim moves at 100 m/s. This combination of wheel size and
speed offsets radiation damage, heating and the shock-stress in the wheel
from the $\sim$131 kW photon beam. A picture of the conceptual target layout is
shown in Figure~\ref{fig:POStarget}. A shaft that extends on both sides of the wheel with
the motor mounted on one shaft end, and a rotating water union on the other
end to feed cooling water. The target wheels sit in a vacuum enclosure
at  \(10^{-8}\)  torr (needed for NC RF operation), which requires vacuum seals to
enable access to the chamber. The rotating shaft penetrates the enclosure
using two vacuum pass-throughs, one on each end. The optical matching
device (OMD - see Section~\ref{sssect:POSomd} ), is mounted on the target assembly,
and requires an additional liquid nitrogen cooling plant. The motor
driving the target wheel is sized to overcome forces due to eddy currents
induced in the wheel by the OMD.

\stepcounter{tablcl}\begin{table} \vbabove
 \caption{Nominal target parameters.}
   \label{tab:POSparamtar}
   \begin{center}
      \begin{tabular}{| l |c |c | c |}
         \hline
         Target Parameters & Symbol & Value & Units \\
         \hline & & & \vbdlspacing \hline
         Target material &  & Ti-6\%Al-4\%V &  \\  \hline
         Target thickness & $L_{t}$ & 0.4 / 1.4 & r.l. / cm \\  \hline
         Target power adsorption &   &  8 & \% \\  \hline
         Incident spot size on target & $\sigma_{i}$ & $>1.7$ & mm, rms \\
         \hline
      \end{tabular}
   \end{center}
\end{table}

\stepcounter{figlcl}

The target wheel assembly is designed for an operational life of two years. In the
 event that the target fails during a run, the assembly can be replaced by a new
 assembly in about a day using vertical remote handling.

A series of sensors provide information on the target behavior. An infrared camera
 tracks temperatures on the wheel, to allow for quick shutdown in the case of a
 cooling failure.  Flowmeters monitor cooling water flow in and out of the wheel
(to watch for leaks), along with thermocouples to check ingoing and outgoing flow
 temperature. A torque sensor is placed on the shaft, with vibration sensors on
the wheel to report mechanical behavior.  Finally, the wheel's rotational speed
is monitored.

\subsubsection{Optical Matching Device}\label{sssect:POSomd}

The OMD generates a solenoidal magnetic field which peaks in strength at
5 Tesla close to the target and falls off to 0.5 Tesla to match the solenoidal
field at the entrance of the capture section. The OMD increases the capture
 efficiency by a factor of 2. The OMD is a normal conducting pulsed flux
 concentrator based on an extrapolation of a magnet created for a hyperon
 experiment \cite{bib:POSbrechna}.

The magnetic field of the OMD interacts with the spinning metal of the target
to create eddy currents.  The target design must accommodate this drag force
 which increases the average heat load and requires a stronger target drive
motor.  The OMD may possibly induce 5 Hz resonance effects in the target
that will need to be mitigated.

\subsubsection{Normal Conducting RF Accelerator System}\label{sssect:POSncras}

\begin{figure} [htb] \vbabove
   \begin{center}
      \includegraphics[width=0.95\textwidth]{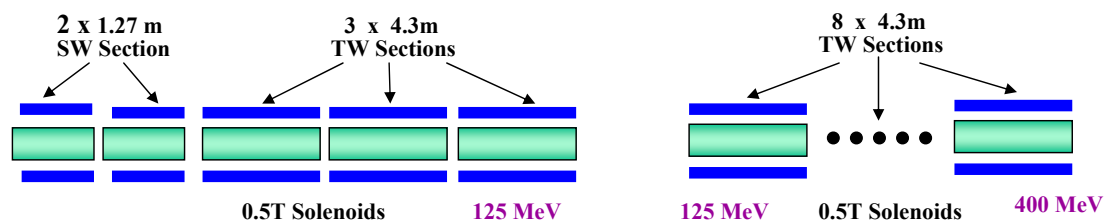}
\vbabovecaption
      \caption{Layout of the capture region (left) and pre-accelerator region (right).}
      \label{fig:POSNCRF}
   \end{center}
\vbbelow
\end{figure}

Due to the extremely high energy deposition from positrons, electrons, photons
and neutrons downstream of the positron target, normal conducting structures must
be used up to an energy of 125 MeV. This normal-conducting section is
challenging but feasible, and a prototype test structure is under construction. It must sustain high accelerator gradients during
 millisecond-long pulses in a strong magnetic field, provide adequate cooling
in spite of high RF and particle loss heating, and produce a high positron
yield with the required emittance. The design contains both
standing-wave (SW) and traveling-wave (TW) L-band accelerator structures
\cite{bib:POSrf}. The capture region has two 1.27 m SW accelerator sections at 15 MV/m
and three 4.3 m TW accelerator sections at 8.5 MV/m accelerating gradient.
All accelerator sections are surrounded with 0.5 T solenoids. Figure~\ref{fig:POSNCRF}
shows the schematic layout.

\stepcounter{figlcl}

\stepcounter{figlcl}\begin{figure} [htb] \vbabove
   \begin{center}
      \includegraphics[width=0.95\textwidth]{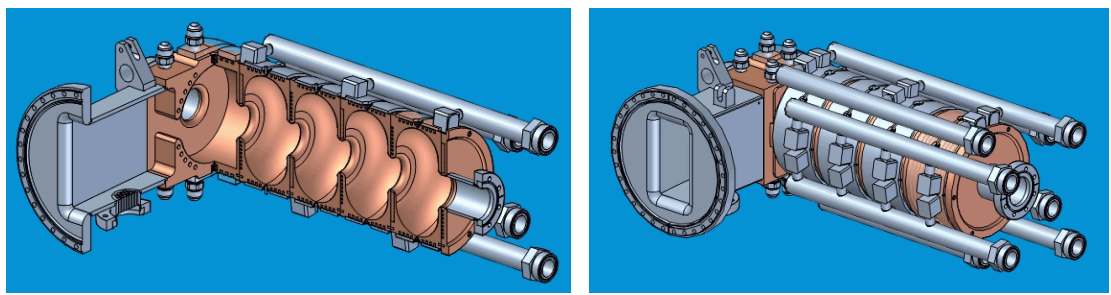}
\vbabovecaption
      \caption{SW structures - cut-away and external views.}
      \label{fig:POSNCRFpic}
   \end{center}
\vbbelow
\end{figure}

The high gradient (15 MV/m) positron capture sections are 11-cell
$\pi$ mode SW
 accelerator structures. The SW structures have a more effective cooling system and higher shunt
 impedance with larger aperture (60 mm) and require RF circulators to protect the klystrons from reflected power. The mode and amplitude stability under various
cooling conditions for this type of structure have been theoretically verified.
Figure~\ref{fig:POSNCRFpic} shows engineering drawings of the SW structures.

The TW sections are 4.3 m long, 3$\pi$/4 mode constant gradient
accelerator
 structures. The phase advance per cell has been chosen to optimize
RF efficiency for a large aperture TW structure. The TW structures allows easy installation for long solenoids and do not need circulators.
Each accelerator section has an individual  1.3 GHz RF power source.

\subsubsection{Magnets}\label{sssect:POSm}

The Positron Source has more than 2000 magnets, see Table~\ref{tab:POScomponents}. Most of the magnet designs are quite
 straightforward.
The large aperture DC solenoids
 surrounding the L-band capture RF must be normal
conducting because of the high beam losses in the target region and as
such use a large amount of electrical power. In addition, there are two long
high field SC solenoids for spin rotation in the LTR.

\subsubsection{Diagnostics}\label{sssect:POSd}

The Positron Source has the normal complement of beamline instrumentation to
 measure orbit, emittance, charge and energy spread. Specialized diagnostics
are designed into the unique positron systems, e.g. target. The major cost is
in the BPM system because of the large channel count coming from the long
 beamlines. The number of readout channels is halved by processing only one
 transverse plane of the BPM x,y pair at each quadrupole. Performance
 specifications for the diagnostics are in most cases equal to or less than the
Main Linac or RTML.

\subsubsection{Electron and Photon Beam Dumps}\label{sssect:POSepbd}

There are 9 beam dumps, 16 variable aperture collimators, 1 fixed aperture
 collimator and 5 stoppers with burn through monitors planned for the positron
 source system.  Three of the beam dumps must absorb sufficiently large beam
 power that they require dump designs with water in the path of the beam.
The plumbing to cool and treat the resulting radioactive water is the
dominant cost.

There is a tune-up dump at the 150 GeV pre-undulator extraction
point of the electron linac. For tune-up, the number of
bunches per train is limited to 100; with absorbed power of
 240 kW at nominal beam parameters.
This dump, roughly in line with the linac, also serves as the abort
dump for up to a full train of electrons (1.35 MJ) to protect the undulator.
The dump consists of a 40 cm diameter by 250 cm long stainless
vessel filled with 10 mm diameter aluminum balls through which flows
 $\sim$30 gallons per minute of water; it is backed by a short
length of peripherally cooled solid copper.  The dump is shielded
from the access passageway by 10 cm of steel and 40 cm of concrete.  A service
 cavern houses a heat exchanger, pumps and a system to treat the water for
 hydrogen, oxygen and  $^{7}$Be.
A second dump, technically identical (225 kW at nominal beam parameters), is
 located near the damping ring to tune the 5 GeV positrons before injection.

The most challenging dump in the positron production system
absorbs the non-interacting undulator photons from the target.
 This dump must absorb up to 300 kW (upgrade value) continuously (1.9 x \(10^{17}\)
 photons/sec of 10 MeV average energy). The primary absorber in this case must be
water, contained in a vessel with a thin window.
%Preliminary calculations indicate that, at the nominal mid-undulator to %positron target distance of 500m
%and the nominal target to dump separation of 150m,
For a dump located 150 m downstream of the target, calculations indicate that the power density on a
1 mm Ti window is 0.5 kW/cm$^{2}$, the resultant temperature rise after the
passage of one bunch train is 425$^{\circ}$C, and in the core of the beam the
rise in the water temperature is 190$^{\circ}$C.   With this geometry,
a compact (10 cm diameter by 100 cm long) pressurized (12 bar) water vessel
and Ti window, with a radioactive water processing system, is required.
Lengthening the target to dump distance by several hundred meters would
ease requirements on the dump, but incur the expense of a longer transport.

The remaining dumps and collimators in the positron system all are based on
 peripherally cooled solid metal construction, with the cooling water supplied
directly from the accelerator low conductivity water (LCW) system and do not
 present technical or cost challenges.

\clearpage %\setcounter{page}{1}
\setcounter{figlcl}{0}
\setcounter{tablcl}{0}
\setcounter{secnumdepth}{4} \setcounter{subsection}{0}
\setcounter{subsubsection}{0}
\usecounter{subsubsection}

\setcounter{section}{3} \renewcommand{\picturefolder}{./DR/}
\section{Damping Rings}\label{sectDR}

\subsection{Overview}

The ILC damping rings include one electron and one positron ring,
each 6.7 km long, operating at a beam energy of 5 GeV. The two rings
are housed in a single tunnel near the center of the site, with one
ring positioned directly above the other. The damping rings must
perform three critical functions:
\begin{itemize}
\item Accept e$ ^{-} $ and e$ ^{+} $ beams with large transverse and longitudinal
  emittances and produce the low-emittance beams required for
  luminosity production \itemspace
\item Damp incoming beam jitter (transverse and longitudinal)
  and provide highly stable beams for downstream systems \itemspace
\item Delay bunches from the source to allow feed-forward
  systems to compensate for pulse-to-pulse variations in
  parameters such as the bunch charge. \itemspace
\end{itemize}

The damping ring system includes the injection and extraction
systems, which themselves include sections of transport lines
matching to the sources (upstream of the damping rings) and the RTML
system (downstream of the damping rings).

\subsection{Beam Parameters}

The key parameters for both the electron and positron damping rings are listed in Table
\ref{tab:pdrparams}.

\subsection{System Description}

The configuration of the damping rings is constrained by the timing
scheme of the main linac. In particular, each damping ring must be
capable of storing a full bunch train (roughly 3000-6000 bunches)
and reducing the emittances to the required level within the 200 ms
interval between machine pulses. In addition, the relatively large
bunch separation in the main linacs means that the damping rings
must be capable of injecting and extracting individual bunches
without affecting the emittance or stability of the remaining stored
bunches.

Several configuration options capable of satisfying the various
constraints were evaluated in 2005 on the basis of cost and
technical risk, and the 6.7 km ring was selected \cite{dr1}. The
exact circumference has been chosen to provide flexibility in the
operational timing scheme, allowing variation in the bunch charge
and number of bunches per pulse, for a fixed total number of
particles per pulse and constant pulse length in the linac
\cite{bib:ACCtimingdoc}. The superconducting RF system is operated at 650 MHz. To achieve the short damping times necessary to reduce the emittances
(by roughly six orders of magnitude in the case of the positron vertical
emittance) within the allowed 200 ms interval, superconducting
wigglers of total length roughly 200 m are used in each damping
ring.

\stepcounter{tablcl}\begin{table} \vbabove \caption[Positron damping
ring parameters.] {Positron damping ring parameters. The electron
damping ring is identical except for a smaller injected emittance.}
\begin{center} \label{tab:pdrparams}
\begin{tabular}{| l | c | r |} \hline
Parameter & Units & Value \\  \hline  & & \vbdlspacing  \hline
  Energy &  GeV & 5.0 \\ \hline
  Circumference &  km & 6.695 \\ \hline
  Nominal \# of bunches &  & 2625 \\ \hline
  Nominal bunch population &  & $2.0\times10^{10}$ \\ \hline
  Maximum \# of bunches & & 5534 \\ \hline
  Bunch population at max \# of bunches  & & $1.0\times10^{10}$ \\
   \hline
  Average current & A & 0.40 \\ \hline
  Energy loss per turn & MeV & 8.7 \\ \hline
  Beam power & MW & 3.5 \\ \hline
  Nominal bunch current & mA & 0.14 \\ \hline
  RF Frequency & MHz & 650 \\ \hline
  Total RF voltage & MV & 24 \\ \hline
  RF bucket height & \% & 1.5 \\ \hline
  Injected betatron amplitude, $A_x+A_y$ & m.rad & 0.09 \\ \hline
  Equilibrium $\gamma\epsilon_x$ & $\mu$m.rad & 5.0 \\ \hline
  Chromaticity, $\Xi_x/\Xi_y$ & & -63/-62 \\ \hline
  Partition Numbers, $J_x/J_y/J_E$ & & 0.9998/1.0000/2.0002 \\ \hline
  $h$ & & 14,516 \\ \hline
  $\nu_s$ & & 0.067 \\ \hline
  $f_s$ & kHz & 3.0 \\ \hline
  $\alpha_c$ & & $4.2\times10^{-4}$ \\ \hline
  $\nu_x/\nu_y$ & & 52.40/49.31 \\ \hline
  $\sigma_z$ & mm & 9.0 \\ \hline
  $\sigma_p/p$ & & $1.28\times10^{-3}$ \\ \hline
  $\tau_x$ & ms & 25.7 \\ \hline
  $\tau_s$ & ms & 12.9 \\ \hline
\end{tabular}
\end{center} \vbbelow
\end{table}

As noted in Section \ref{sssect:OVRtiming}, there are significant constraints on the DR circumference, the fill patterns and the bunch spacing in the main linac. These issues will need to be optimized during the next design hase and it is likely that small changes will be made to the DR circumference and the bunch spacing.

\subsubsection{Lattice Design Considerations}

The ring lattice satisfies the basic requirements of damping time,
normalized horizontal emittance and bunch length, has sufficient
aperture for injecting a large emittance positron beam, and has a
sufficiently large momentum compaction factor to maintain single
bunch stability. However, there remains design work to be done on
the lattice, for example to optimize the dynamic aperture to ensure
efficient acceptance of the large emittance beam from the positron
source, and to minimize sensitivity to tuning and alignment errors
that could degrade the emittance.

\stepcounter{figlcl}\begin{figure}[t]
   \begin{center} \vbabove
      \includegraphics[width=13cm]{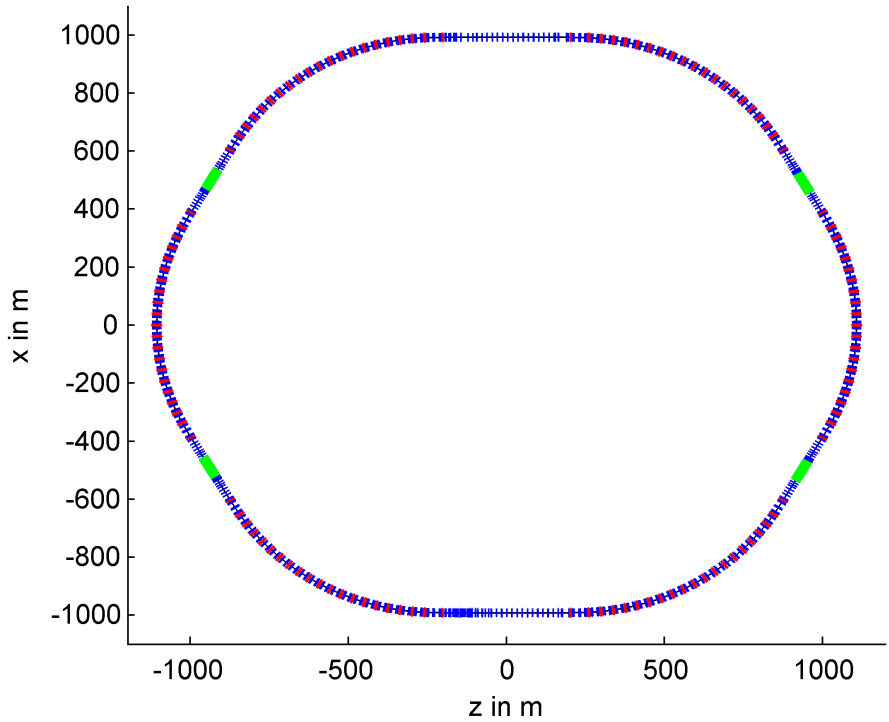}
\vbabovecaption
      \caption{Layout of the ILC Damping Ring.}
      \label{fig:drlayout}
   \end{center} \vbbelow
\end{figure}

\stepcounter{figlcl}\begin{figure}[b]
   \begin{center} \vbabove
      \includegraphics[width=13cm]{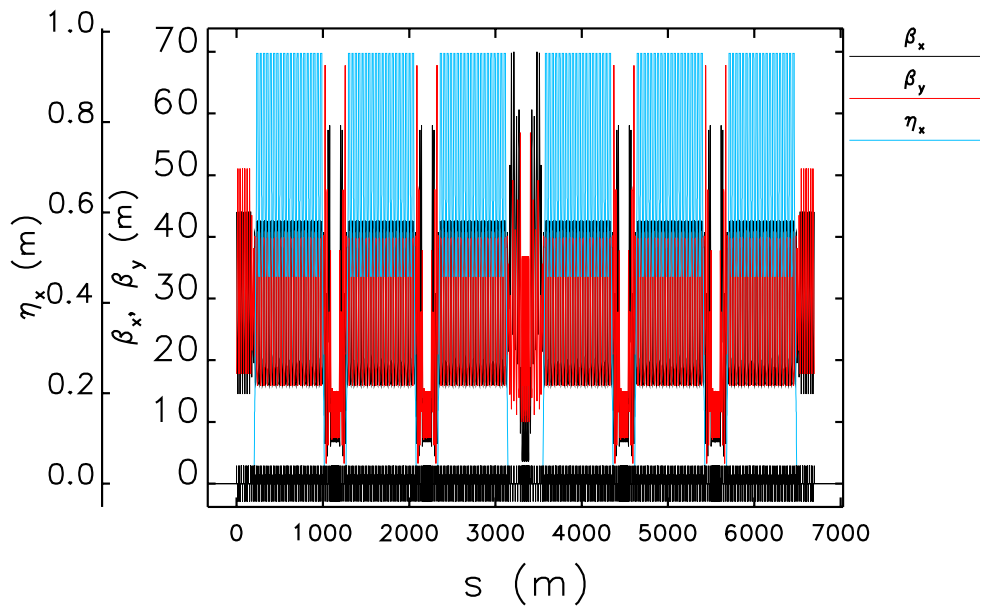}
\vbabovecaption
      \caption{Optical functions of the ILC Damping Ring.}
      \label{fig:drtwiss}
   \end{center} \vbbelow
\end{figure}

The ring is divided into six arcs and six straight sections (see
Figure \ref{fig:drlayout}). The arcs are composed of
Theoretical-Minimum-Emittance (TME) cells to give low quantum
excitation, and the straight sections are composed of FODO cells for
the damping wigglers, RF cavities, and injection and extraction
sections. Optical parameters are shown in Figure \ref{fig:drtwiss}. The
parameters of the TME cells and the wigglers (peak field of 1.67 T)
were selected to obtain the required damping time, momentum spread,
and normalized horizontal emittance.

\begin{figure}[htb]
   \begin{center} \vbabove
      \includegraphics[width=13cm]{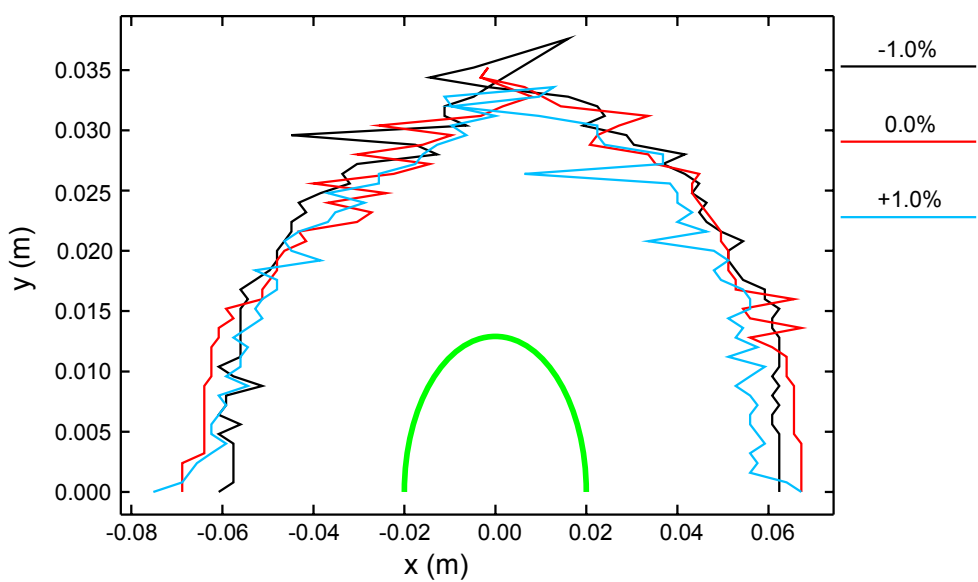}
\vbabovecaption
      \caption[Dynamic aperture of the ILC Damping Ring.]
    {Dynamic aperture of the ILC Damping Ring (without
      field and alignment errors) for relative momentum errors of
      -1\%, 0\% and 1\% at  x = 44 m and  y = 18 m. The thick green
      line represents the size of the injected positron beam.}
      \label{fig:drdynap}
   \end{center} \vbbelow
\end{figure}

Two families of sextupole magnets are inserted in the TME cells for
correcting the first-order chromatic effects of the linear optics.
To reduce nonlinear effects of the sextupoles, the betatron phase
advance of the TME cells was set to 90$^{\circ}$ in each plane. The
resulting dynamic and momentum apertures (see Figure
\ref{fig:drdynap}) were found to depend on the number of straight
sections (i.e., the symmetry of the lattice) and on the betatron
phase advances of the straight sections. The straight section
betatron phase advances were adjusted for maximum dynamic aperture.
While a larger number of straight sections was found to improve the
nonlinearities, this comes at a higher cost for subsystems. A
compromise configuration of six straight sections was eventually
chosen for the baseline lattice.

\stepcounter{figlcl}

The choice of momentum compaction factor, controlled chiefly by the
total number of TME cells, results from a balance between competing
requirements: single-bunch stability against the impedance of the
vacuum chamber (favoring a high value of $\alpha_c$) and a lower
cost RF system (favoring a low value of $\alpha_c$) . The value $4.2
\times10^{-4}$ is somewhat on the high side to reduce the risk of
single-bunch instability. Unfortunately, a high momentum compaction
factor makes it difficult to achieve a low equilibrium emittance and
strong damping wigglers are required. The Twiss parameters in the
wiggler region were adjusted to produce the required equilibrium
emittance.

The injection/extraction sections accommodate a large number of fast
stripline kickers (their large number being due to their inherent
weakness). Optical functions were designed to ensure that the beam
goes through the stripline kickers without hitting their apertures.
For the injection section, the beam traverses the kickers at an
angle but with a small trajectory offset.

\subsubsection{Fast Ion Instability}

Of significant concern to the electron damping ring is the fast ion
instability. As opposed to the more familiar ion-trapping effect,
where ions oscillate stably for long periods in the potential well
of the stored beam, the fast ion instability is associated with ions
created in the beam path by interaction with the circulating beam
during a single turn. Ions created at the head of the bunch train
move slowly, and remain in the beam path, influencing the motion of
subsequent bunches. The resultant ion-induced beam instabilities and
tune shifts are critical issues for the electron damping ring due to
its ultra-low vertical emittance. A low base vacuum pressure at the
1 nTorr level is essential to reduce the number of ions formed. To
mitigate bunch motion, we also employ bunch-by-bunch feedback
systems with a damping time of $\approx$0.1 ms.

\begin{figure}[htb]
   \begin{center} \vbabove
      \includegraphics[width=12cm]{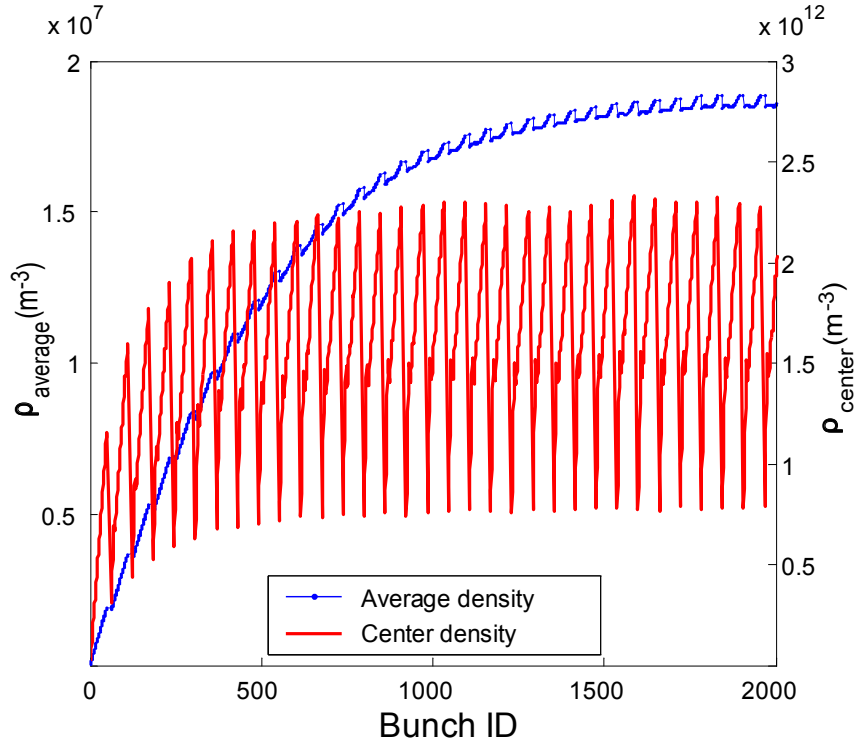}
\vbabovecaption
      \caption[Buildup of CO$^+$ ion cloud at extraction.]
    {Buildup of CO$^+$ ion cloud at extraction.
      The total number of bunches is 5782 (118 trains with 49
      bunches per train). The beam has a bunch separation of two
      RF bucket spacings, and a train gap of 25 RF bucket spacings.
      There are $0.97\times10^{10}$ particles per bunch, and the
      partial vacuum pressure is 1 nTorr.}
      \label{fig:fastion}
   \end{center} \vbbelow
\end{figure}

To further reduce the core ion density, short gaps are introduced in
the electron beam bunch train by omitting a number of successive
bunches. The use of such ``mini-gaps'' in the train significantly
mitigates the fast ion instability by reducing the core ion density
and by inducing tune variation along the train. Figure
\ref{fig:fastion} shows the buildup of the ion cloud in the case of
a particular multi-train pattern with 118 bunch trains and 49
bunches per train. In this case, the reduction in the core ion
density is a factor of 60 compared with a fill consisting of a
single long bunch train. It is worth pointing out that the effect of
train gaps is a function of beam size, so they are less effective
early in the damping cycle. The simulated growth time for the beam
pattern corresponding to Fig. \ref{fig:fastion} is 280 $\mu$s.

\stepcounter{figlcl}

The tune spread due to both linear and nonlinear tune shifts
provides Landau damping that helps control ion-induced instabilities
\cite{dr3}. With a multi-train fill pattern, the size of the ion
cloud is much larger than the vertical beam size, so there is a
larger tune spread. When the oscillation amplitude of the beam
reaches the beam size, the nonlinearity effectively saturates the
instability.
% The presence of multiple gas species in the vacuum will
% also provide some Landau damping due to the variation of oscillation
% frequencies for different masses \cite{dr4}.

\subsubsection{Electron Cloud Instability}

Electron cloud instabilities and tune shifts are critical issues for
the positron damping ring. The electron cloud develops quickly as
photons striking the vacuum chamber wall knock out electrons that
are then accelerated by the beam, gain energy, and strike the
chamber again, producing more electrons. The peak secondary electron
yield (SEY) of typical vacuum chamber materials is $>$1.5 even after
surface treatment, leading to amplification of the cascade. Once the
cloud is present, coupling between the electrons and the circulating
beam can cause a single-bunch (head-tail) instability and incoherent
tune spreads that may lead to increased emittance, beam
oscillations, or even beam losses. Because the electron cloud is
difficult to suppress in the dipole and wiggler regions of the ring,
this is where its effects are expected to be most severe. A large
synchrotron tune is beneficial, as it raises the threshold for the
electron cloud head-tail driven instability.

\begin{figure}[b]
   \begin{center} \vbabove
      \includegraphics[width=13cm]{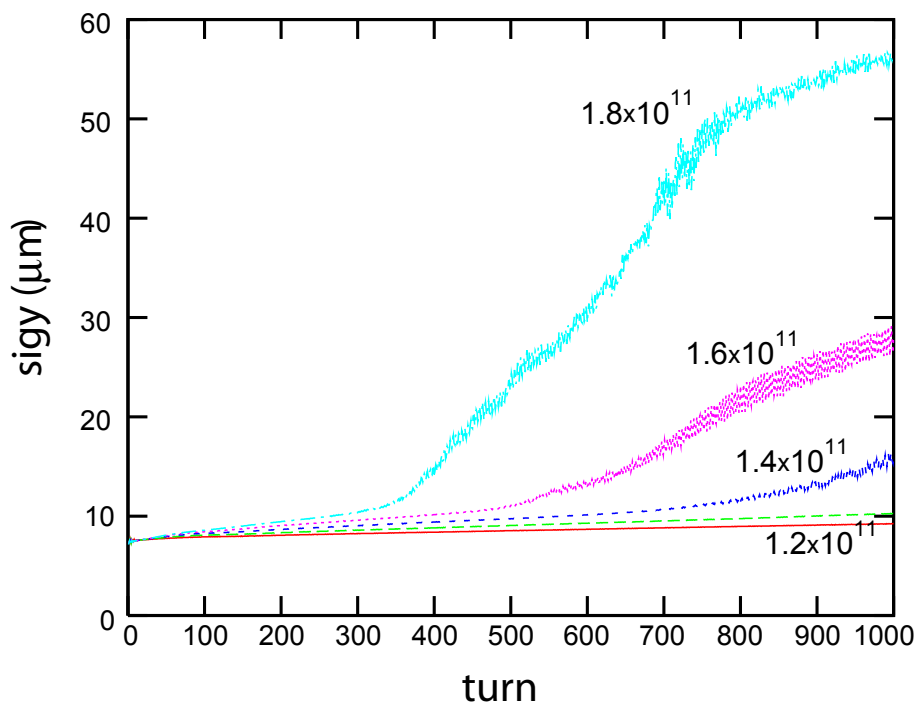}
\vbabovecaption
      \caption[Emittance growth from single-bunch instability
      driven by electron cloud.] {Emittance growth from single-bunch instability
      driven by electron cloud in the 6.7 km OCS ring.
      Electron cloud densities in e$ ^{-} $/m$^3$ are indicated.}
      \label{fig:ecloud}
   \end{center} \vbbelow
\end{figure}

Single-bunch instability simulations for the 6.7 km damping ring
lattice show that the instability sets in above an average cloud
density of $1.4\times10^{11}$ e$^-$/m$^3$, where an incoherent
emittance growth is observed, see Figure \ref{fig:ecloud}. Analytic
calculations give higher density thresholds by roughly a factor of 4
\cite{dr5,dr6}. Tune shifts on the order of 0.01 are expected near
threshold.

\stepcounter{figlcl}

\begin{figure}[htb]
   \begin{center} \vbabove
      \includegraphics[width=13cm]{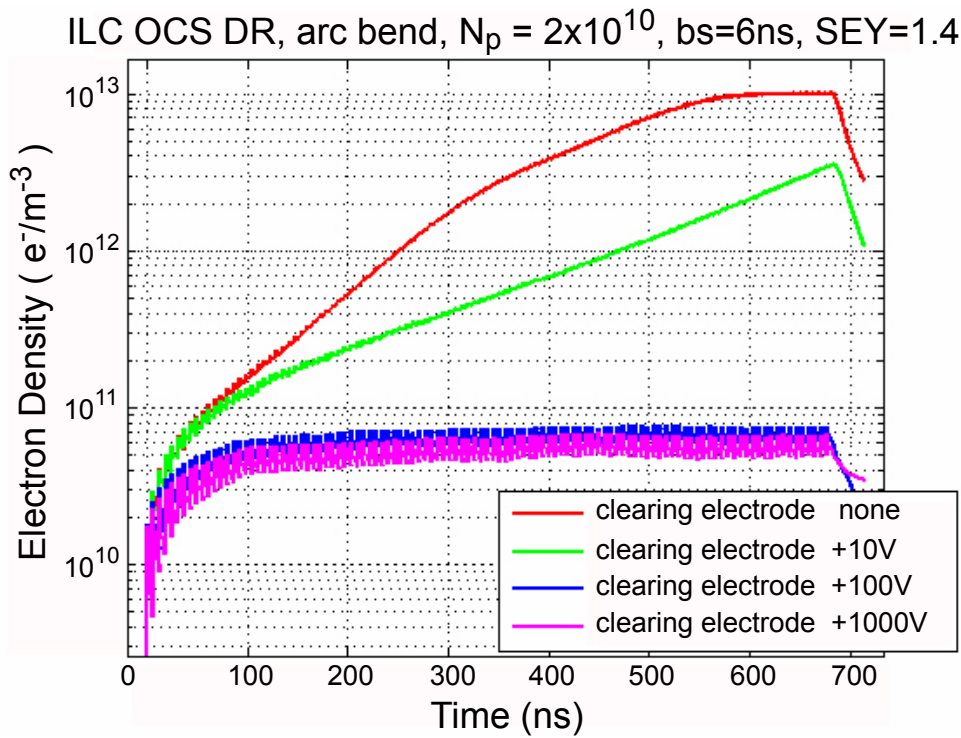}
\vbabovecaption
      \caption[Electron cloud buildup in an arc bend of the 6.7
      km ring.] {Electron cloud buildup in an arc bend of the 6.7
      km ring and suppression effect of clearing electrodes
      biased at the indicated voltages.}
      \label{fig:ecloud2}
   \end{center} \vbbelow
\end{figure}

Simulations indicate that a peak secondary electron yield of 1.2
results in a cloud density close to the instability threshold. Based
on this, the aim of ongoing experimental studies is to obtain a
surface secondary electron yield of 1.1. Simulations also indicate
that techniques such as grooves in the chamber walls or clearing
electrodes will be effective at suppressing the development of an
electron cloud \cite{dr7,dr8}. Figure \ref{fig:ecloud2} shows the
buildup of the electron cloud and the suppression effect of clearing
electrodes in an arc bend of the 6.7 km ring. A clearing electrode
bias potential of +100 V is sufficient to suppress the average (and
central) cloud density by more than two orders of magnitude.
Techniques such as triangular or rectangular fins or clearing
electrodes need further R\&D studies and a full demonstration before
being adopted. Nonetheless, mitigation techniques appear to be
sufficient to adopt a single 6.7 km ring as the baseline design for
the positron damping ring.

\stepcounter{figlcl}

\subsubsection{Injection and Extraction}

The bunch separation in the main linacs is much longer than in the
damping rings, so individual bunches must be injected and extracted
without affecting the emittance or stability of the remaining stored
bunches. For this to be the case, the kicker field must be
negligible for any stored bunch upstream or downstream of the
injected or extracted bunch, requiring that the effective kicker
pulse width be less than twice the bunch spacing. Injection is
interleaved with extraction, to minimize excursions in beam loading
of the RF system.

Extraction is located near the center of one long straight section.
A set of kickers deflects a single damped bunch horizontally. A
horizontally defocusing quadrupole increases the deflection. About
90$^{\circ}$ of betatron phase downstream of the kickers, the bunch
passes through the bending fields of a pair of septum magnets. These
deflect the trajectory further horizontally, so it passes outside of
the next focusing quadrupole and into the extraction line optics.
The stored orbit is located in the nominally field-free region of
the septum magnets and is not significantly affected. The extraction
straight section also includes an abort dump.

Injection is located near the center of the opposite long straight
section. The injection line grazes the outside of a quadrupole, and
is deflected horizontally by a pair of septum bend magnets so the
trajectory passes inside the aperture of the next quadrupole. This
horizontally defocusing quadrupole makes the trajectory nearly
parallel to the stored orbit. About 90$^{\circ}$ of betatron phase
downstream from the septa, where the injection trajectory crosses
the stored orbit, a set of kickers deflects the single injected
bunch onto the stored orbit. As mentioned previously, the kicker is
distributed into several modules at the axis-crossings of the
injected trajectory, so the module aperture can be minimized.

The kicker modules are 50 $\Omega$ stripline structures inside the
vacuum pipe, each 30 cm long, operating at a voltage of 22 kV,
provided as +11 and -11 kV pulses on opposite electrodes. Twenty-one
modules are required for injection in each ring and eleven for extraction. The 30 cm stripline gives a 2 ns contribution
to the kicker pulse width, leaving less than 10 ns for the
electrical pulse width at the nominal ring bunch spacing of 6 ns.
The kickers pulse about every 300 ns during the linac pulse of about
1 ms. For the low bunch charge parameters, the ring bunch spacing is
3 ns, requiring an electrical pulse width of less than 4 ns and a
pulse about every 150 ns. The electrical pulser requirements are
challenging, and the subject of an extensive R\&D program.

\begin{figure}[htb]
   \begin{center} \vbabove
      \includegraphics[width=10cm]{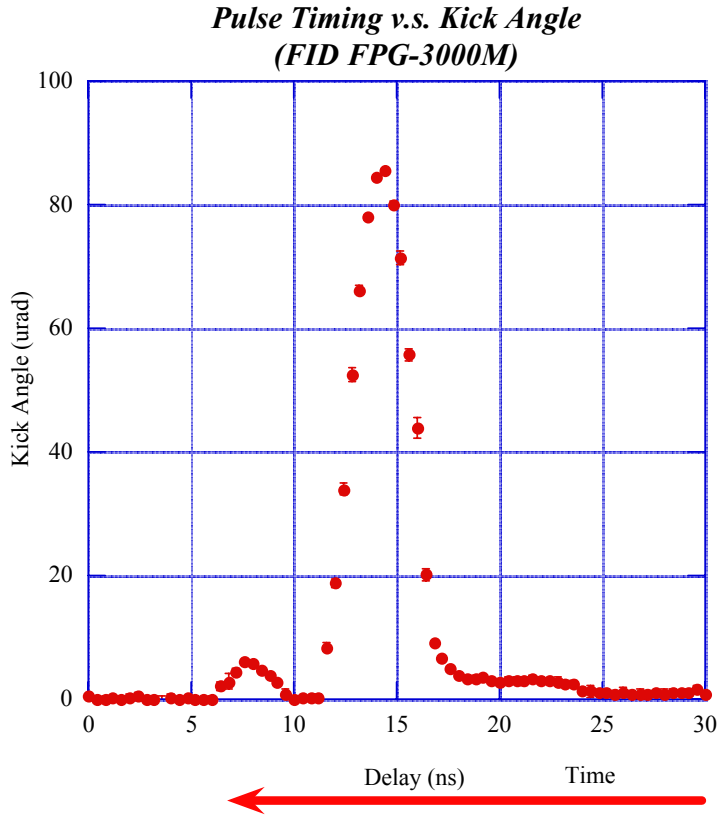}
\vbabovecaption
      \caption[Kick angle vs. time.] {Kick angle vs. time. Note that time   increases to the left here.}
      \label{fig:kicker}
   \end{center} \vbbelow
\end{figure}

Figure \ref{fig:kicker} shows beam deflection vs. kicker time
measured at the ATF storage ring at KEK with 33 cm striplines and a
5 kV, 3 MHz pulser built by FID GmbH. The main pulse easily meets
the width and rate requirement for 6 ns bunch spacing, although at
half the desired amplitude and with undesireable structure before
and after the main pulse.

\stepcounter{figlcl}

\begin{comment}
The positron injection kick amplitude is $\approx$20$\sigma_x$ of
the damped beam, so if the extraction kick amplitude were the same, a
0.5\% amplitude jitter would give a 1 $\sigma$ horizontal beam
jitter. This results in an extraction kicker amplitude jitter
tolerance in the $10^{-3}$ range.
\end{comment}
The tolerance on horizontal beam jitter of the extracted beam is
0.1-0.2 $\sigma$, which requires the extraction kicker amplitude stability
to be 0.1\% or better. A similar tolerance applies to the
kicker amplitude for bunches before and after the target bunch in
the ring bunch train. As a tolerance on the absolute kicker field
before and after the pulse, this is very difficult to achieve and is
the subject of ongoing R\&D.

The septum magnets are modeled after the Argonne APS injection
septa. The thin (2 mm) septum magnet has a 0.73 T field, and the
thick (30 mm) septum magnet has a 1.08 T field. Each magnet has an
effective length of 1 m. Both magnets are pulsed once per linac
cycle to reduce power dissipation, with eddy currents in the septum
shielding the circulating beam. A half-sine pulse of about 10 ms
width is used, and post-regulation is required to produce a 1 ms
plateau flat to $10^{-4}$.

\subsection{Accelerator Components}

The damping ring has conventional electromagnets for the dipole,
quadrupole, sextupole, and corrector magnets. This technology choice
offers flexibility for tuning and optimizing the rings as well as
for adjusting the operating beam energy by a few percent around the
nominal value of 5 GeV. Superconducting wigglers based on the CESR-c
design \cite{dr9} provide sufficient field quality that the wigglers
impose no limitation on the damping ring dynamic aperture. The large
wiggler aperture improves the ring acceptance for the large injected
positron beam and reduces the growth of the electron cloud in the
wiggler region. Power supplies and controllers are located in
alcoves at the centers of the RF-wiggler straights. Magnet counts
are shown in Table \ref{tab:drmags}. Table \ref{tab:drmags2} gives
the key magnet parameters and maximum higher-order harmonic content
specifications.

\stepcounter{tablcl}\begin{table} [b] \vbabove \caption[Magnet types
and counts
  for a single ILC Damping Ring.] {Magnet types and counts
  for a single ILC Damping Ring using the OCS6 lattice.
  These counts do not include injection and extraction
  line magnets nor magnets, kickers, and septa associated
  with the damping ring abort beam dump.  Wiggler magnets are
  superconducting, all others are room-temperature.}
\label{tab:drmags}
\begin{center}
\begin{tabular}{| l | c | c |} \hline
 Type & Number & Power Method \\ \hline & &  \vbdlspacing \hline
  Dipoles (6 m) & 114 & 6 strings, 1 per arc \\ \hline
  Dipoles (3 m) & 12 & 6 strings, 1 per arc \\ \hline
  Quadrupoles & 747 & Individual \\ \hline
  Sextupoles & 504 & Individual \\ \hline
  Horizontal correctors & 150  & Individual \\ \hline
  Vertical correctors & 150 & Individual \\ \hline
  Skew quadrupoles & 240 & Individual \\ \hline
  Wigglers & 80 & Individual \\ \hline
  Kickers & 64 & Individual \\ \hline
  Septa & 4 & Individual \\ \hline
 \end{tabular}
\end{center} \vbbelow
 \end{table}

\stepcounter{tablcl}\begin{table}\label{tab:drmags2} \vbabove
\caption[Target field tolerances.] {Target
%\begin{table}[htb] \caption{Target
  field tolerances at a reference radius of 20 mm for
  damping ring magnets. Magnet aperture radii are 30
  mm except for the wigglers. For the wigglers,
  the operating field is 1.67 T and the field quality
  is specified by the observed roll-off for a horizontal
  displacement from the beam axis by the indicated distance.
  The maximum KL-value specifies the nominal strength of
  the strongest magnet of each magnet type.{\it (Wigglers have reference radius 20 mm (H); aperture radius 37.5 mm (H), 45 mm (V))}}
\begin{center}
\begin{tabular}{| l | c | c | c | c |} \hline
Type & Max $KL$ & L [m] & Max field & \# of \\ [-6pt]
     &          &       & error & types \\ \hline & & & & \vbdlspacing \hline
 Dipoles & 0.0524 & 6 ; 3 & $2\times10^{-4}$ & 2 \\ \hline
 Quadrupoles & 0.31 m$^{-1}$ & 0.3 & $2\times10^{-4}$ & 4 \\ \hline
 Sextupoles & 0.24 m$^{-2}$ & 0.25 &  $2\times10^{-3}$ & 1 \\ \hline
 H correctors & 0.002 & 0.25 & $5\times10^{-3}$ & 1 \\ \hline
 V correctors & 0.002 & 0.25 & $5\times10^{-3}$ & 1 \\ \hline
 Skew quads  & 0.03 m$^{-1}$ & 0.25 & $3\times10^{-3}$ & 1 \\ \hline
 Wigglers & -- & 2.5 & $3\times10^{-3}$ & 1 \\ \hline
\end{tabular}
\end{center} \vbbelow
\end{table}

The superconducting damping wigglers are 4.5 K devices with static
heat loads of 2 W/m or less, based on CESR-c experience. To avoid a
significant dynamic heat load, care must be taken to ensure that
only tiny amounts of scattered synchrotron radiation reach the cold
mass. Two of the wigglers are co-located in the damping ring
straight sections with the superconducting RF cavities, where the
necessary crygenic infrastructure is readily available. The other
wigglers are fed by transfer lines, with a single transfer line
infrastructure for both rings.

All quadrupoles, sextupoles, wigglers and corrector magnets (dipole,
skew quadrupole, and possibly other multipoles yet to be specified)
have individual power supplies. Individual control of the quadrupole
and sextupole magnets significantly enhances the ability to tune and
locally correct the machine optics in a ring with very aggressive
operating parameters. Individual power supplies for the wigglers
offer simplified control in the event of a magnet quench by
eliminating the power system coupling between magnets. Because of
the long distances between individually powered magnets and the
alcoves, the power supply system uses bulk supplies located in the
main alcoves that power a master ``bus'' from which DC-to-DC
converters supply power to individual magnets. This design minimizes
cable heat loads in the ring and provides a more efficient power
system. For the dipole magnets, each arc section is powered
separately. The pulsed power supplies for the stripline kickers
require short cable runs to preserve the necessary timing
synchronization, and must be located in the tunnel or in small
secondary alcoves near each group of kickers.

\subsubsection{RF System}

The damping ring RF frequency of 650 MHz has a simple relationship
with the main linac RF (1.3 GHz) to accommodate varying bunch
patterns.
%Table \ref{tab:drrf} lists the DR parameters of importance
%for the RF system design.
While high power 650 MHz RF sources are
not commercially available, several major klystron manufacturers can
develop them by modifying 500 MHz klystrons of equivalent power
level. Similarly, the RF cavity units can be designed by scaling
from existing 500 MHz superconducting module designs currently in
operation at CESR, KEK, and elsewhere. The RF cryomodule dimensions
are 3.5 m in length and 1.5 m in diameter \cite{dr9}.

For either ring, the beam power and the total RF voltage is shared
among 18 superconducting cavities. These are located in two RF
straight sections roughly 40 m long. Operating 18 SC-cavity modules
per ring ensures adequate energy and beam power margin in case of an
RF station fault, and permits continued operation with 14 cavity
modules at full performance by increasing the RF field in the
remaining units. Table \ref{tab:drrf} summarizes the RF system main
features and compares the parameters for the nominal case with that
when one RF station is off. Parameters are scaled from the 500 MHz
units developed by industry and being operated in various
laboratories.

\stepcounter{tablcl}\begin{table} \vbabove \caption[Estimated 650
MHz SC cavity parameters.] {Estimated 650 MHz SC
%\begin{table}[htb]\label{tab:drrf}\caption{Estimated 650 MHz SC
cavity parameters (scaled from 500 MHz model) for both electron
and positron damping rings.}
\label{tab:drrf}
\begin{center}
\begin{tabular}{| l | c | c | c |} \hline
  Parameter & Units & All Stations & One Station \\ [-6pt]
   & & On & Off \\ \hline & & & \vbdlspacing \hline
  Frequency & MHz & \multicolumn{2}{c|}{650} \\ \hline
  Active cavity length & m & \multicolumn{2}{c|}{0.23} \\ \hline
  R/Q [$\Omega$] & & \multicolumn{2}{c|}{89} \\ \hline
  Operating Temperature & K  & \multicolumn{2}{c|}{4.5} \\ \hline
  Standby losses at 4.5 K & W & \multicolumn{2}{c|}{30} \\ \hline
  Operating SC modules per ring & & 18 & 14 \\ \hline
  Accelerating gradient & MV/m  & 5.8 & 7.5 \\ \hline
  Accelerating voltage & MV  & 1.33 & 1.72 \\ \hline
  Q$_0$ at operating gradient & $10^9$ & 1.0 & 0.9 \\ \hline
  RF cyrogenic losses per cavity & W  & 20 & 33 \\ \hline
  Total cryo losses per ring & W & 900 & 925 \\ \hline
  Beam power per cavity & kW & 194 & 250 \\ \hline
  Klystrons per ring & & 5 & 4 \\ \hline
  Klystron emitted power & kW & 780 & 1000 \\ \hline
\end{tabular}
\end{center} \vbbelow
\end{table}

\begin{figure}[htb]
   \begin{center} \vbabove
      \includegraphics[width=13cm]{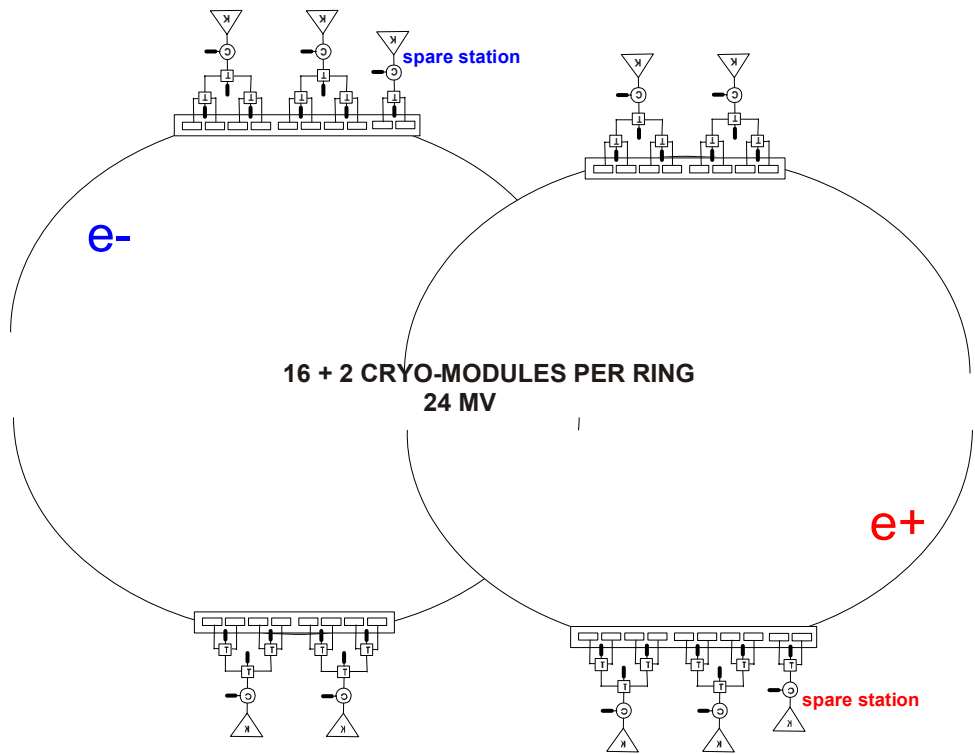}
\vbabovecaption
      \caption[Schematic layout of DR RF systems.]
    {Schematic layout of DR RF systems. Each of the two
      RF-wiggler sections accommodates three stations from one ring,
      and two from the other. All stations are situated upstream of
      the wiggler in that ring.}
      \label{fig:drrfstation}
   \end{center} \vbbelow
\end{figure}

Two or three RF stations are located in each RF-wiggler straight
section, as indicated schematically in Figure \ref{fig:drrfstation}.
Each klystron can feed 4 SC cavities by means of a distribution
system having magic-tees for power splitting and 3-port circulators
for protecting the klystron. To guarantee sufficient power margin in
case of a klystron fault, the power sources are 1.2 MW CW. One
``hot-spare'' station in each ring is operated with only two
cavities, rather than four. The stations are upstream of the
wigglers at opposite ends of the straight section tunnel, with
waveguides connecting them to the klystrons housed in centrally
located alcoves having access shafts to the surface.

\stepcounter{figlcl}

The selection of 650 MHz requires a redesign of existing 500-MHz
sources, cavities and cryomodules. A critical element is the input
coupler because the power handling capability must be kept at a
level of about 260 kW CW, comparable to presently operating 500 MHz
systems. The HOM dampers and cryostat mechanical details must also
be revised.

\subsubsection{Cryogenic Plant}

The DR cavities operate at 4.5 K and the total cryogenic losses per
ring are approximately 900 W with 14 modules operating in case of
one klystron fault. The design has one cryogenic plant in each RF
straight section. With this choice, the helium transfer lines to the
RF are not very long and do not impact the cryogenic plant cost. The
cryogenic plant capability must be sufficient to handle the
worst-case scenario of one klystron fault, where the cryogenic power
in the other straight section could increase to a total of 925 W.
With the standard refrigerator efficiency of 0.3\% at 4.5 K, the
total wall-plug power for each straight-section refrigerator is
about 300 kW. Table \ref{tab:drcryo} summarizes the specifications
of the cryogenic system.

\stepcounter{tablcl}\begin{table} [hb] \vbabove \caption{Main
specifications of
  the RF cryogenic system, with 18 modules per ring.}
\label{tab:drcryo}
\begin{center}
\begin{tabular}{| l | c | r |} \hline
  Parameter & Units & Value \\ \hline & & \vbdlspacing \hline
  Nominal cryogenic losses per straight section & W & 900 \\ \hline
  Design cryogenic losses per straight section & W & 925 \\ \hline
  Wall plug power per cryogenic plant & kW & 300 \\ \hline
  Total number of cryogenic plants & & 2 \\ \hline
\end{tabular}
\end{center} \vbbelow
\end{table}

\subsubsection{Fast Feedback System}

With thousands of bunches circulating in the ring, wakefields
induced in vacuum chamber components can give rise to coupled-bunch
instabilities that cause bunch jitter and/or emittance growth. To
combat this, the rings have fast bunch-by-bunch feedback systems in
all three oscillation planes (longitudinal, horizontal and
vertical) \cite{dr15}. Modern commercial FPGAs (Field Programmable Gate Arrays)
can easily manage the requirements of the feedback systems in terms
of speed and number of bunches. The bandwidth of the fast feedback
system must be at least $f_{\rm RF}$ (that is, 650 MHz). This means
that every block of the system must have the capability to manage
the full bandwidth except for the power section (amplifiers and
kickers), where half bandwidth is sufficient. The main elements of
each system are the analog front end, digital processing unit,
analog back end, amplifier and kicker.

The pickups are 4-button monitors (two or three for each beam line)
with at least full bandwidth and adequate dynamic range. The analog
front ends must be capable of extracting the oscillation signals
from the monitors in each of the three planes (L, H, V) and giving
them to the digital sections with a swing in the range of $\sim$0.5
V (typical of many analog-to-digital converters).

To minimize the quantization noise and have an adequate dynamic
range, the digital units are based on a 16-bit signal processing
system. The processing is able to compute the correction signal for
all buckets (including the empty ones) to decouple the feedback
behavior from the fill pattern. This means that all feedback systems
must have the capability to process, in real time, 14,516
input/output channels, although the real bunches are in, at most,
5,534 buckets. The digital unit sampling frequency is 650 MHz. A
real time FIR (finite impulse response) filter (with $\ge$50 taps)
provides the correction synchrotron or betatron phase advance using
only one pickup for each system. The feedback setup should be easily
configurable using software tools. A down-sampling feature is also
needed to manage very low oscillation frequencies.

The analog back-end systems adapt the output correction signals to
the power section. The longitudinal kicker (a cavity) works at a
frequency between 800 and 1600 MHz, whereas the transverse kickers
(striplines) operate at baseband (from 10 kHz up to half the
bandwidth of the fast feedback system). Each power section has four
250 W amplifiers (1 kW total), with the bandwidth required by the
kicker.

\subsubsection{Vacuum System}

The vacuum design for the damping rings is similar to those for
modern storage rings and synchrotron radiation sources. The need to
avoid the fast ion instability leads to very demanding
specifications for the vacuum in the electron damping ring: $<$0.5
nTorr CO-equivalent in the arc cells, $<$2 nTorr CO-equivalent in
the wiggler cells, and $<$0.1 nTorr CO-equivalent in the straight
sections \cite{dr10}.  A combination of coatings, grooved chamber
profiles, and clearing electrodes is required to suppress the
electron cloud in the positron damping ring. The baseline design
uses a non-evaporable getter (NEG) coated aluminum tubular vacuum
chamber. With NEG coating, fewer pumps with lower pumping speed are
required. Issues associated with the ultimate lifetime of the NEG
material, its regeneration, and the synchrotron radiation power
density on the chamber walls need further study. Each of the 120 arc
cells requires one sputter ion pump with an effective pumping speed
of 20 L/s installed immediately downstream of the dipole. In the
long straight sections, similar sputter ion pumps are required every
10 m for $0 < z < 80$ m, every 20 m for $80 < z < 160$ m, and every
40 m for $160 < z < 400$ m.

\begin{figure}[htb] \vbabove
   \begin{center}
      \includegraphics[width=13cm]{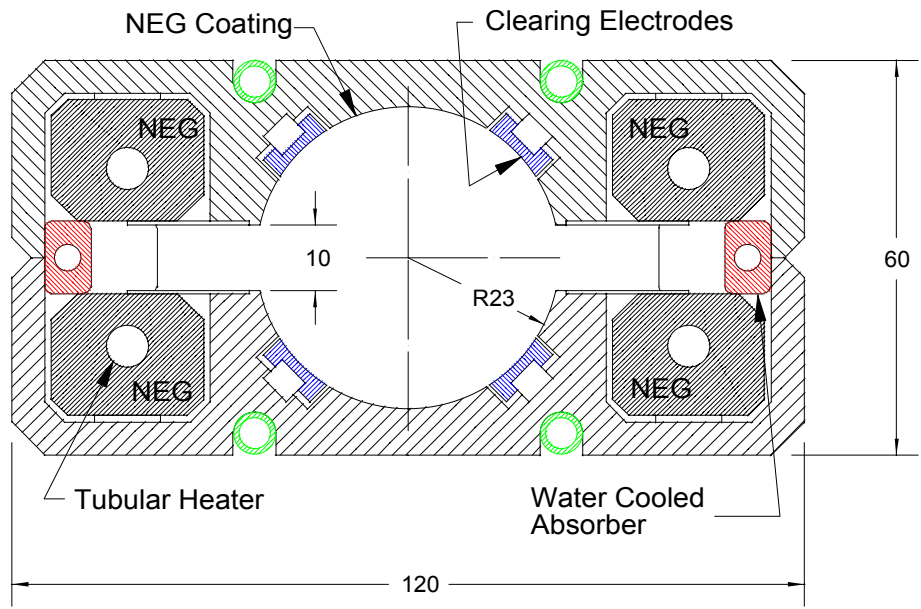}
\vbabovecaption
      \caption{ILC damping ring wiggler chamber; dimensions in mm.}
      \label{fig:wigglervacchamber}
   \end{center}
\end{figure}

The wiggler straight section vacuum system consists of separate
chambers for the wiggler and quadrupole sections. A cross section of
the wiggler chamber is shown in Figure \ref{fig:wigglervacchamber}.
The chamber is a machined and welded aluminum unit, designed as a
warm-bore insert, mechanically decoupled from the wiggler and
cryogenic system. A NEG pumping system and photon absorber are
incorporated in antechambers. Integral cooling is incorporated to
minimize distortion of the chamber and thermal load on the wiggler
cryostat during NEG regeneration. A TiZrV NEG surface coating is
used on the main chamber bore to minimize secondary electron yield
\cite{dr11}. Clearing electrodes are also incorporated to reduce the
electron cloud.

\stepcounter{figlcl} \vbbelow

The photon absorbers are hollow water-cooled copper conductor
designed to absorb photon power from upstream wigglers in the same
straight section. Power radiated from the first wiggler in the
straight section is intercepted initially by wiggler number three.
Intercepted power increases for successive wigglers up to number
nine; thereafter, a constant 3 W/mm$^2$ peak power density is
reached. The total power absorbed per wiggler is 26 kW, that is, 13
kW per absorber.

The NEG pumping system is similar to that designed for the PEP-II B
Factory. The assembly consists of NEG-coated fins and an integral
heating rod for regeneration. The estimated pumping speed for CO is
1000 L/s/m. With a total incident photon flux of $2\times10^{18}$
photons/s/m, the estimated yield of CO is $2\times10^{13}$
molecules/s/m. This results in an equilibrium CO partial pressure of
$7\times10^{-10}$ Torr.

Between each wiggler chamber is a separate chamber for the
quadrupole section. This chamber is welded aluminum, incorporating
TiZrV NEG coating for secondary electron yield reduction. Bellows, a
BPM assembly, and an ion pump for pumping non-reactive gases are
included. The ion pump also serves as a vacuum gauge. The quadrupole
chamber is completely shadowed by the wiggler chamber photon
absorbers and does not absorb any of the photon power from upstream
wigglers.

\subsubsection{Cost Methodology}

Several of the technical subsystems in the damping rings have
specific requirements that distinguish them from corresponding
subsystems in other parts of the ILC; generally, this is because of
the relatively high average current and synchrotron radiation power
(for the vacuum system, wigglers, and fast feedback systems), CW
operation (for the RF system), or unique functionality (for the
injection/extraction kickers). The cost estimates for these
damping-ring-specific designs were developed by the damping ring
group.

The RF system is CW and operates at 650 MHz, a different frequency
from the RF systems used elsewhere in the ILC. The designs of
high-power RF components, such as klystrons and circulators, were
scaled from commercially available 500 MHz devices. Estimates from
klystron manufacturers indicated that development costs would
increase the total cost by roughly the cost of one additional unit
at the standard catalog price. Manufacturing costs for the cavity cryomodules were assumed to be
the same as for commercial versions of 500 MHz systems developed at
Cornell and KEKB, with increased engineering effort to account for
the rescaling, or in some cases redesign, of the existing
subcomponents.

A preliminary design for the vacuum system was based on estimates
for required vacuum levels (set by ion instabilities in the case of
the electron damping ring), handling of synchrotron radiation,
aperture requirements, and conditioning rates. Standard commercial
component costs were used for extruded aluminum vacuum chambers,
bellows, pumps, valves, and bake-out systems. Coating the chambers
with NEG was assumed to be done with in-house labor. The cost
estimate for the complex damping wiggler vacuum chamber  was based
on fabrication of similar systems for other projects.

The engineering and fabrication experience for the CESR-c wigglers
were used to provide reliable cost estimates for the ILC damping
wigglers, taking proper account of the well-defined differences in
specification. Costs for the kicker pulser were based on a
commercially available pulser (a fast ionization dynistor, or FID,
device) that comes close to meeting the specifications for the
damping ring injection/extraction kickers; this cost dominates the
total cost of the injection/extraction system. Other components,
including the stripline electrodes and the septa, are relatively
conventional, and costs were based on similar existing devices.

Costs of the ILC damping ring fast feedback systems were taken
directly from comparable systems in existing machines. Power
amplifiers dominate the cost of the fast feedback systems.
Amplifiers operating in the appropriate parameter regime are
available commercially, and costs for these were obtained from an
experienced manufacturer.

\clearpage %\setcounter{page}{1}
\setcounter{figlcl}{0}
\setcounter{tablcl}{0}
\setcounter{secnumdepth}{4} \setcounter{subsection}{0}
\setcounter{subsubsection}{0}
\usecounter{subsubsection}

\setcounter{section}{4} \renewcommand{\picturefolder}{./RTML/}
\section{Ring to Main Linac}\label{sectRTML}

%\chapter{Ring to Main Linac (RTML)}
%\minitoc
\subsection{Overview}\label{sect:RTMLOverview}

The ILC Ring to Main Linac (RTML) is responsible for transporting
and matching the beam from the Damping Ring to the entrance of the
Main Linac. The RTML must perform several critical functions:

\begin{itemize}
    \item transport of the electron and positron beams from the damping rings,
    at the center of the ILC accelerator complex, to the upstream ends of their
    respective linacs (``geometry matching''); \itemspace
    \item collimation of the beam halo generated in the damping ring; \itemspace
    \item rotation of the spin polarization vector from the vertical to any
    arbitrary angle required at the IP; \itemspace
    \item compression of the long Damping Ring bunch length by a factor of
          \(30\sim45\) to provide the short bunches required by the Main Linac
          and the IP; \itemspace
\end{itemize}

In addition, the RTML must provide sufficient instrumentation, diagnostics and feedback
(feedforward) systems to preserve and tune the beam quality.

\subsection{Beam Parameters}

The key beam parameters of the RTML are listed in Table~\ref{tab:RTMLparam}.
Parameters are shown for the nominal configuration and for the ``low
charge'' configuration (which requires a shorter bunch at the IP).

\stepcounter{tablcl}\begin{table} [htb] \vbabove
   \caption{Basic beam parameters for the RTML.}
   \label{tab:RTMLparam}
   \begin{center}
      \begin{tabular}{| l | r | r |}
         \hline
         Parameter & Nominal Value & Low Charge Value  \\
         \hline & \multicolumn{2}{| c |}{ } \vbdlspacing \hline
         Initial energy & \multicolumn{2}{c |}{5.0 GeV} \\   \hline
         Initial energy spread & \multicolumn{2}{c |}{0.15\%}   \\   \hline
         Initial emittances & \multicolumn{2}{c |}{8.0$\mu$m $\times$ 20 nm}   \\   \hline
         Initial horizontal beam jitter & \multicolumn{2}{c |}{1 $\sigma$}   \\   \hline
         Initial bunch length & \multicolumn{2}{c |}{9.0 mm }  \\   \hline
         Final bunch length & 0.3 mm & 0.2 mm  \\   \hline
         Final energy & 15.0 GeV & 13.7 GeV  \\   \hline
         Final energy spread & 1.5\% & 2.7 \%  \\   \hline
         Final horizontal beam jitter & \multicolumn{2}{c |}{0.1$\sigma$}   \\   \hline
         ISR emittance growth & 0.9 $\mu$m & 0.8 $\mu$m  \\   \hline
         Emittance budget & \multicolumn{2}{c |}{1$\mu$m $\times$ 4 nm }  \\   \hline
      \end{tabular}

   \end{center} \vbbelow
\end{table}

\subsection{System Description}

\subsubsection{ Layout}

\begin{figure}[htb]
   \begin{center} \vbabove
      \includegraphics[width=\textwidth]{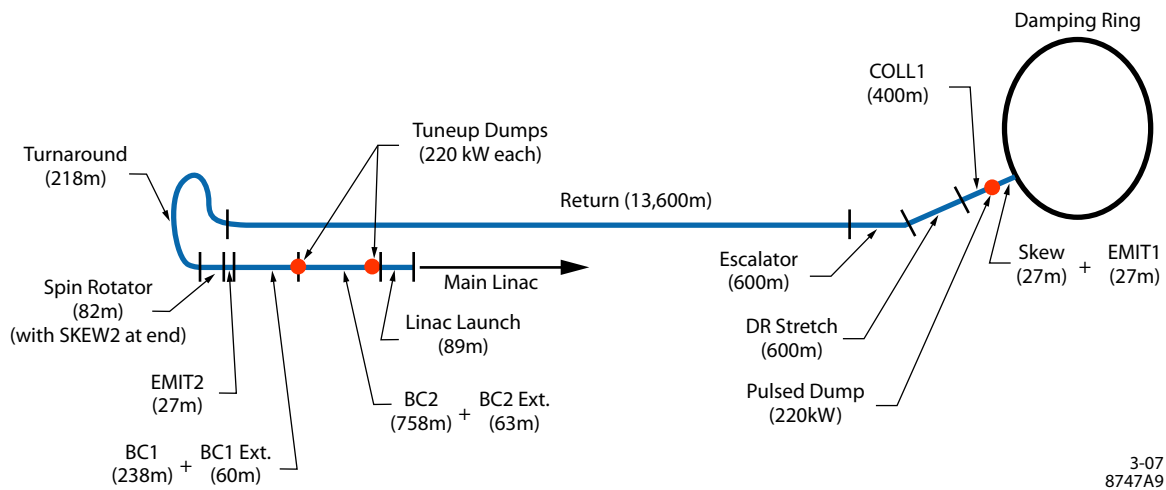}
\vbabovecaption
      \caption{Schematic of RTML, indicating the various functions
      described in the text.}
      \label{fig:RTMLschema}
   \end{center} \vbbelow
\end{figure}

\begin{figure}[htb]
   \begin{center} \vbabove
      \includegraphics[width=13cm]{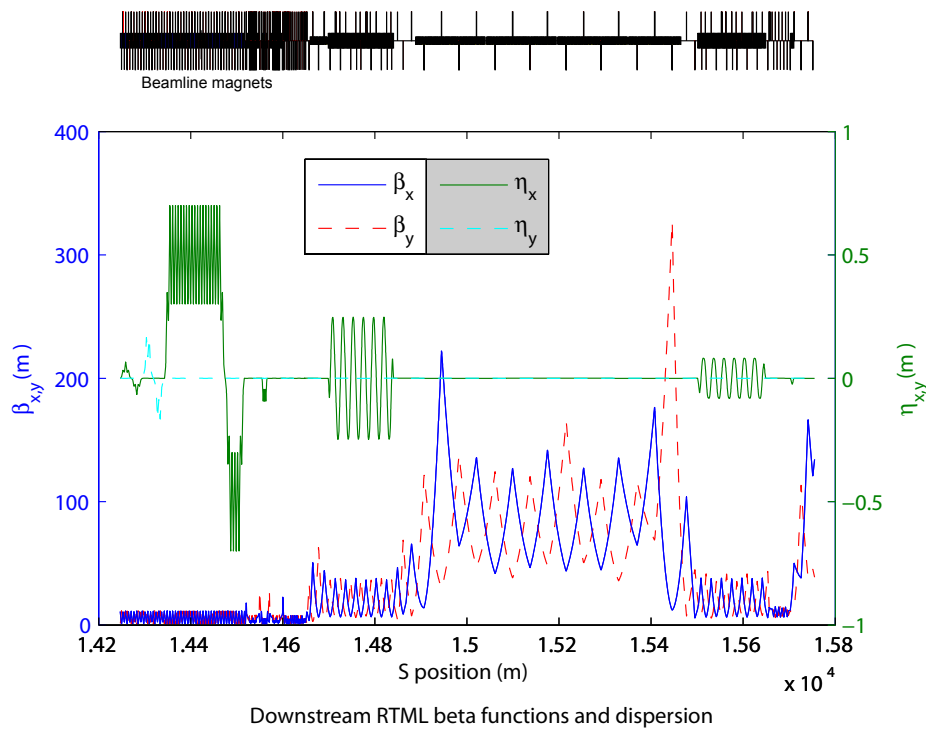}
\vbabovecaption
      \caption{Twiss functions of the downstream RTML, from the start of the
      turnaround to the match into the main linac.}
      \label{fig:RTMLDownstreamTwiss}
   \end{center} \vbbelow
\end{figure}

Figure~\ref{fig:RTMLschema} depicts schematically the layout and
location of the various sub-beamlines of the RTML.
%, while
%Figure~\ref{fig:RTMLgeom} shows the actual geometry. As shown in
%Figure~\ref{fig:RTMLschema},
The RTML includes the long
low-emittance transport from the Damping Ring, followed by a
\(180^{\circ}\) turn-around, the spin-rotation and
two-stage bunch compression sections.
%are located, before injection into the Main Linac.
The beamlines upstream of the turnaround are collectively known as
the ``upstream RTML,'' while those from the turnaround to the launch
into the main linac are collectively known as the ``downstream
RTML.''  Figure \ref{fig:RTMLDownstreamTwiss} shows the Twiss
functions of the downstream RTML.
In order to accommodate the different damping ring elevations and
linac lengths, the electron and positron RTMLs have slight
differences in their long transport sections, but are otherwise
identical.
%
%{\bf NOTE:}  I need the schematic in EPS format from Nobu, and need
%to generate some sort of layout of the whole mess...

\stepcounter{figlcl}

\stepcounter{figlcl}

%I'd skip this one - Nan
%\stepcounter{figlcl}\begin{figure}[bhtp]
%   \begin{center}
%      \includegraphics[width=13cm]{RTMLgeom.jpg}
%      \caption{{\bf PLACEHOLDER} ``Footprint'' of the RTML transfer line, including pulsed
%          extraction lines (``BC1DL'' and ``BC2DL'')  used for machine
%          protection and tune-up purposes.}
%      \label{fig:RTMLgeom}
%   \end{center}
%\end{figure}

Each of the key functions of the RTML listed in
\ref{sect:RTMLOverview} is supported by several of the sub-beamlines
shown in Figure~\ref{fig:RTMLschema}.

\subsubsection{Geometry Match}

%The electron and positron beams are extracted from the damping rings
%near the center of the ILC complex; at this location the beams are
%traveling perpendicular to the long axis of the ILC and are at the
%damping ring elevation, which is different from the linac elevation
%by about 10 m. The RTML must transport these beams such that they
%are traveling parallel to the long axis and are injected into
%upstream ends of the main linacs, which are about 15 km away from
%the injection points.

%This is accomplished by first bringing the beams
Following extraction from the damping rings, the beams are brought
parallel to the long axis via the 90$^{\circ}$ arcs in the Arc
sections; transported from the the damping ring elevation to the
main linac tunnel elevation via the vertical doglegs in the
Escalator sections; transported out to their respective ends of the
site via the Return lines, which are suspended from the ceiling of
the main linac tunnel; and reversed in direction by the Turnaround
sections. In addition, small vertical and horizontal doglegs at the
upstream end of the Turnaround change the beam elevation from the
ceiling of the linac tunnel to the nominal linac elevation, and
adjust the horizontal position between the Return line axis and the
main linac axis.

\subsubsection{Collimation}

%The Stanford Linear Collider (SLC) observed a beam halo that was on
%the order of $10^{-3}$ of the beam population \cite{SLCHalo}, and it
%was believed that this halo was primarily generated in the SLC
%damping rings.  In the ILC, the desired halo specification at the
%entry to the BDS is at the level of $10^{-5}$ of the beam population
%\cite{ILCHalo}.  Thus, it is essential that any significant halo
%generated in the damping ring be promptly removed.  In addition, the
%ILC requires limiting apertures which can provide some segmentation
%of the machine protection system (MPS), {\it i.e.}, ensure that
%errors which occur upstream of a given segment boundary cannot
%result in damage to devices which are downstream of that boundary.
%Both of these functions are generally provided by collimators.

%The RTML has two main betatron collimation sections which are
%functionally identical:  one shortly downstream of the damping ring
%extraction point (between the Arc and the Escalator), and one
%immediately upstream of the Turnaround.  Each section is constructed
The RTML's betatron collimation section is downstream of the damping
ring extraction arc.  It is constructed
from two sets of thin spoiler and thick absorber pairs, placed
\(90^{\circ}\) apart in betatron phase. This is considered
sufficient to reduce the halo density by 3-4 orders of magnitude. The thin
spoilers are needed
to protect the absorbers from a direct hit from an errant beam in
the event of some machine error \cite{bib:RTMLzdrcoll}.  The spoilers in
the upstream section are protected by their proximity to the damping
ring, which permits extraction to be halted prior to spoiler damage
if the beam begins to hit the spoiler.
%
%The spoilers in the downstream section are protected by the main
%intra-train MPS algorithm of the ILC (see \ref{sect:MPS}).
%
There are additional
collimators for energy collimation placed in %the Escalator,
the Turnaround, and in the wigglers of the Bunch Compressor.

\subsubsection{Spin Rotation}

%The polarization vector of the beam at extraction from the damping
%ring is vertical.  This orientation is preserved by the Arc, which
%bends only in the horizontal plane; it is also preserved by the
%Escalator, which produces no net change in the beam direction. There
%is a slight vertical bending necessary in the Return line, since it
%is in the linac tunnel and the linac tunnel is vertically curved to
%follow a gravitational equipotential; the net bend is approximately
%2.4 mrad, which at 5 GeV beam energy leads to a precession of the
%spin vector of 1.5$^{\circ}$.  The Turnaround, like the Arc,
%contains only horizontal bends, which means that the vertical
%component of the spin vector is unaffected by the Turnaround while
%the small longitudinal component introduced by the bending in the
%Return line precesses about the vertical axis approximately 5.6
%times; since the energy spread in the Turnaround is small, the loss
%in net polarization of the beam from this effect is negligible.

The beam polarization in the damping rings is vertical, and this
polarization is transported with negligible loss or precession to
the end of the Turnaround.  At that point
%
%Once past the Turnaround,
it is necessary to be able to reorient the spin vector to any
direction required by the experimental physicists.  To achieve this,
both the electron and positron RTMLs have a complete spin rotation
system.  Each system includes a pair of superconducting solenoids,
followed by an arc with a net 7.9$^{\circ}$ bend angle, which is in
turn followed by another pair of solenoids.  By adjusting the
excitation in the solenoid pairs, the spin vector at the end of the
spin rotator can be oriented in any desired direction.  In order to
rotate the spin without introducing undesired x-y coupling, the
solenoid-based rotators each use a pair of identical solenoids
separated by a quadrupole lattice which introduces a $+I$
transformation in the horizontal plane and a $-I$ transformation in
the vertical plane \cite{bib:RTMLemmarotator}, the net effect of
which is to cancel the cross-plane coupling.

\subsubsection{Bunch Compression}

In order to achieve the required bunch compression factor of 30-45,
a two-stage system is adopted. A single-stage compressor would
produce a beam with a relative energy-spread that is unacceptably
high, leading to unachievable alignment tolerances in both the RTML
and the early stages of the Main Linac.

Table~\ref{tab:BCparam} summarizes the important parameters for both
the first-stage (BC1) and second-stage (BC2) compressor.

\stepcounter{tablcl}\begin{table} \vbabove
   \caption[Key parameters for the two-stage bunch compressor.]{Key parameters for the two-stage bunch compressor in the nominal configuration,
   when compression to 0.3 mm RMS length is desired.}
   \label{tab:BCparam}
   \begin{center}
      \begin{tabular}{| l | r | r |}
         \hline
         Parameter & Nominal BC1 Value & Nominal BC2 Value \\
         \hline & & \vbdlspacing \hline
         Initial energy & 5 GeV & 4.88 GeV \\   \hline
         Initial energy spread &  0.15\% & 2.5\% \\   \hline
         Initial bunch length & 9 mm & 1.0 mm \\   \hline
         RF voltage & 448 MV & 11.4 GV \\   \hline
         RF phase & -105$^{\circ}$ & -27.6$^{\circ}$ \\   \hline
         Wiggler $R_{56}$ & -376 mm & -54 mm \\   \hline
         Final energy & 4.88 GeV & 15.0 GeV \\   \hline
         Final energy spread & 2.5\% & 1.5\% \\   \hline
         Final bunch length & 1.0 mm & 0.3 mm \\   \hline
      \end{tabular}
   \end{center} \vbbelow
\end{table}

In addition to flexibility in the final bunch length, the two-stage
bunch compressor allows some flexibility to balance longitudinal and
transverse tolerances by adjustment of the wiggler magnet strengths,
RF voltages, and RF phases. The nominal compressor configurations
ease tolerances on damping ring extraction phase, damping ring bunch
length, and bunch compressor phase stability, at the expense of
tightening the tolerances on transverse alignment of accelerator
components.  There are also alternate configurations which loosen
transverse alignment tolerances but tighten the longitudinal ({\it
i.e.} phase) tolerances.

The linacs in both compressor stages use standard SCRF cryomodules
and an RF power unit configuration similar to that of the Main Linac
({\it i.e.} one klystron driving three cryomodules). The first-stage
compressor has a single RF unit with 8 cavities and one quadrupole
in each of its 3 cryomodules; the second-stage compressor uses 14 RF
units (plus one redundant spare) which are identical to the main
linac configuration ({\it i.e.}, 26 cavities and 1 quad per unit,
arranged in 3 cryomodules). The stronger focusing in the first stage
is necessary to mitigate the higher wakefields and cavity-tilt
effects resulting from the longer bunch length in the compressors.
The first-stage has no spare RF unit; instead, a spare klystron and
modulator are connected via a waveguide switch to provide some
degree of redundancy.

Each bunch compressor stage includes a 150~m lattice of bend
magnets(``wiggler'') which provides the momentum compaction required
for bunch compression.  As implied by the name, the wigglers
introduce no net offset or angle to the beam.
%The momentum compaction of each wiggler can be
%adjusted over a considerable range.  This flexibility, when coupled
%with the voltage and phase adjustability of the RF sections, allows
%the bunch compressors to be tuned to produce the necessary range of
%final bunch lengths.

\begin{figure}[htb]
   \begin{center} \vbabove
      \includegraphics[width=13cm]{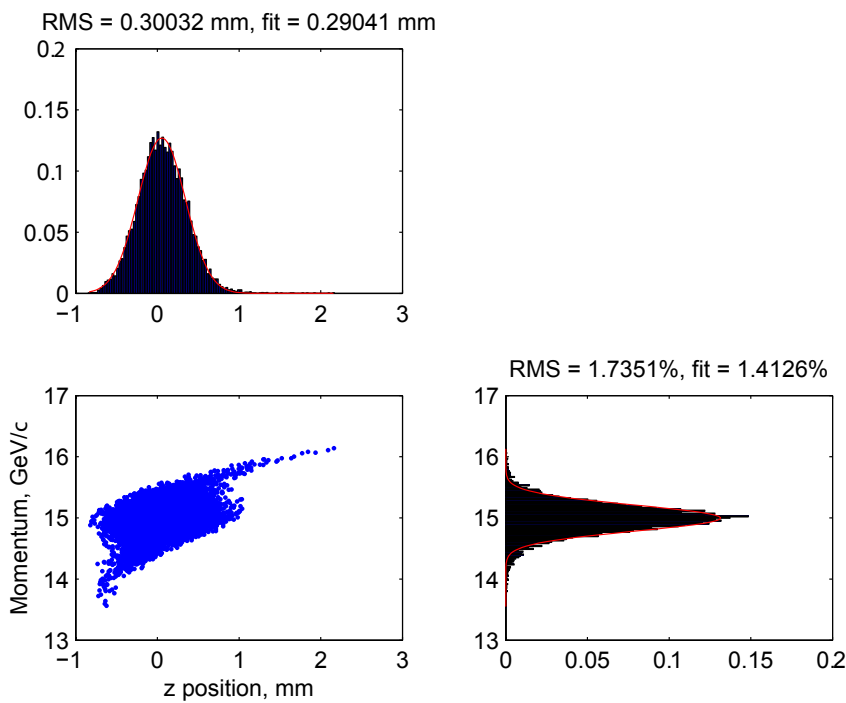}
\vbabovecaption
      \caption{Longitudinal phase space of the compressed bunch.}
      \label{fig:RTMLPZPhaseSpace}
   \end{center} \vbbelow
\end{figure}

Figure \ref{fig:RTMLPZPhaseSpace} shows the longitudinal phase space
after compression from 9 mm to 0.3 mm RMS length.

\stepcounter{figlcl}

\subsubsection{Tuning, Correction, and Operations}

The diagnostic, correction, and operational requirements of the RTML
have been carefully integrated into the design of the entire
beamline.

{\bf Global Dispersion Correction:} The Arc, the BC1 wiggler, and
the BC2 wiggler contain normal and skew quads in regions of
horizontal dispersion which are used to tune any residual dispersion
due to misalignments and errors.  The quads are arranged in pairs,
with an optical $-I$ transform between the two quads in a pair; this
permits the tuning of the dispersion without introducing any
betatron coupling or beta beats.  The dispersions in the Turnaround
are adjusted by tuning normal quads in the horizontal and vertical
doglegs at the upstream end of the Turnaround; similarly, tuning the
normal quads in the Escalator allows its vertical dispersion to be
tuned.

{\bf Global Coupling Correction:}  There are two decoupling regions:
the first is immediately downstream of the Arc, and the second is
immediately downstream of the Spin Rotator.  Each decoupling region
contains 4 orthonormal skew quads in regions of zero dispersion,
which allow complete and independent control of the 4 betatron
coupling terms.  The first station is conceptually intended to
correct the coupling introduced by the damping ring extraction
system, while the second corrects coupling generated by errors in
%the Emma rotators of
the spin rotation system, as well as the
remaining betatron coupling from small rotation errors on the RTML
quads.

{\bf Emittance Measurement:}  There are three emittance measurement
stations:  the first is between the first decoupling section and the
first collimation section, the second is between the second
decoupling station and the bunch compressor, and the third is
between the bunch compressor and the linac.  Each of these stations
contains 4 laser wire scanners embedded in a FODO lattice with
45$^{\circ}$ betatron phase; each station can therefore measure the
projected x and y emittances of the beam.  The first station can be
used to tune the Arc dispersion and the skew quads in the first
decoupler; the second station can be used to tune the Turnaround
dispersion and the skew quads in the second decoupler; the third
station can be used to tune the dispersion correction in the Bunch
Compressor wigglers.  Although none of the systems have the
capability to directly measure normal-mode emittances and coupling
parameters, the optics of the first two stations
are compatible with a later upgrade if needed.
%would permit such a
%measurement with an upgrade to the laser wires if such an upgrade
%was found to be beneficial.

{\bf Beam Position Monitors:}  There are cavity-type beam position
monitors (BPMs) with horizontal and vertical readout at each
quadrupole, with additional units close to the laser wires, at
high-dispersion regions in the Bunch Compressor wigglers, and at
other critical locations. The BPMs in the room-temperature sections
of the RTML almost all operate in the 6 GHz frequency band
(``C-band''), while the BPMs in the cryomodules and at a handful of
other locations use the 1 GHz frequency band (``L-band'').  At the
nominal bunch charge of 3.2 nC, these BPMs can achieve
sub-micron single-bunch resolution.  The standard RTML BPM
requires high precision and stability of the BPM's offset with
respect to the device's mechanical center; a few of the BPMs have
other requirements, such as high bandwidth or low latency.

{\bf Longitudinal Diagnostics:}  Each stage of the Bunch Compressor
contains arrival-time (phase) monitors, beam position monitors at
dispersive locations, X-ray synchrotron light monitors, and two
types of bunch length monitors (a passive device based on measuring
the RF spectrum of the bunch, and an active device based on
transverse deflecting cavities \cite{bib:RTMLlola}).  The active
bunch length monitor can also measure the correlation between energy
and longitudinal position within a bunch.

{\bf Feedback and Feed-Forward:}  The RTML is not expected to
require any intra-train trajectory feedback systems, although there
will be a number of train-to-train (5 Hz) trajectory feedbacks.  In
addition, the beam energy at BC1 and BC2 will be controlled by a 5
Hz feedback, as will the electron-positron path length difference
through their respective bunch compressors (see
\ref{sect:RTMLAccPhys}). There is also a trajectory feed-forward that uses
BPMs at the end of the Return line to make bunch-by-bunch
orbit measurements, which are fed forward to a set of fast
correctors downstream of the Turnaround.  The speed-of-light travel
time between these two points is about 600 nanoseconds, and the
actual distance between them is on the order of a few tens of
meters; the resulting delay of the beam relative to the propagated
signal is more than adequate for a digital low-latency orbit
correction system \cite{bib:RTMLfont}.

{\bf Intermediate Extraction Points:}  There are 3 locations where
the beam in the RTML may be directed to a beam dump:  downstream of
the first collimation section, downstream of BC1, and downstream of
BC2. At each of these locations, there are both pulsed kickers and
pulsed bends for beam extraction. The kickers are used when an
intra-train extraction is required, for example during a machine
protection fault, while the bends are used to send entire trains to
their beam dumps.  The pulsed bends can also be energized by DC
power supplies if a long period of continual dump running is
foreseen.  All 3 dumps are capable of absorbing 220 kW of beam
power.  This implies that the first 2 dumps, which are at 5 GeV, can
absorb the full beam power, while the third dump, at 15 GeV, can
absorb only about 1/3 of the nominal beam power.  Full trains can be
run to this dump at reduced repetition rate, or short trains at full
rate.

{\bf Access Segmentation:}  During personnel access to the main
linac or downstream RTML beam tunnels, the beam can be sent to the
first RTML dump.  For additional safety, the bend magnets in the
Escalator are switched off and additional personnel protection
stoppers are inserted into the beamline.  This allows the damping
ring complex, the Arc dispersion tuning, the first decoupler, and
the first emittance measurement station to be used at full beam
power during linac access.

\subsection{Accelerator Physics Issues}\label{sect:RTMLAccPhys}

A number of beam dynamics issues were considered in the design and
specifications of the RTML.

{\bf Incoherent (ISR) and Coherent (CSR) Synchrotron Radiation:}
%In order to limit
%emittance growth from this effect, all bending systems require a
%combination of weak bends and strong focusing; furthermore, since
%the vertical emittance is much smaller than the horizontal it was
%necessary to observe these requirements to a greater degree in the
%Escalator than in the Arc or the Turnaround.
Current estimates indicate that the horizontal emittance growth from
ISR will be around 90 nm (1.1\%) in the Arc, 380 nm (4.8\%) in the
Turnaround, and 430 nm (5.4\%) in the Bunch Compressor in its
nominal configuration. Vertical emittance growth from ISR in the
Escalator is negligible.

%{\bf Synchrotron Radiation:}
Studies of the ILC
Bunch Compressor indicate that there are no important effects of
coherent synchrotron radiation, primarily because the longitudinal
emittance of the beam extracted from the damping ring is so large
\cite{bib:RTMLjuhao}.

{\bf Stray Fields:}
%Stray fields in the Return line will lead to
%beam jitter at the injection point of the Turnaround.
Studies have found that fields at the level of 2.0 nTesla will lead
to beam jitter at the level of 0.2 $\sigma_y$ \cite{bib:RTMLkubo}.  This is
considered acceptable since the orbit feed-forward will correct most
of this beam motion. Measurements at existing
laboratories \cite{bib:RTMLfrisch} indicate that 2 nTesla is a reasonable estimate
for the stray
field magnitude in the ILC. Emittance growth considerations also place limits on the acceptable stray fields, but these are significantly higher.

{\bf Beam-Ion Instabilities:}  Because of its length and its weak
focusing, the electron Return line will have potential issues with
ion instabilities.  To limit these to acceptable levels, the base
pressure in the Return line must be limited to 20 nTorr
\cite{bib:RTMLlanfa}.

{\bf Static Misalignments:}  The main issues for emittance growth
are: betatron coupling introduced by the Spin Rotator or by rotated
quads; dispersion introduced by rotated bends, rotated quads in
dispersive regions, or misaligned components; wakefields from
misaligned RF cavities; and time-varying transverse kicks from
pitched RF cavities.

Studies of emittance growth and control in the region from the start
of the Turnaround to the end of the second emittance region have
shown that a combination of beam steering, global dispersion
correction, and global decoupling can reduce emittance growth from
magnetostatic sources to negligible levels, subject to the
resolution limits of the measurements performed by the laser wires
\cite{bib:RTMLtuning1a,bib:RTMLtuning1b}. Although the upstream RTML
is much longer than the downstream RTML, its focusing is relatively
weak and as a result its alignment tolerances are actually looser.
Studies have shown that the same tuning techniques can be used in
the upstream RTML with the desired effectiveness
\cite{bib:RTMLtuning2}. The tolerances for RF cavity misalignment in
the RTML are large (0.5 mm RMS would be acceptable) because the
number of cavities is small and the wakefields are relatively weak
\cite{bib:RTMLtuning3}. Although in principle the RF pitch effect is
difficult to manage, in practice it leads to a position-energy
correlation which can be addressed by the Bunch Compressor global
dispersion correction \cite{bib:RTMLtuning4}. A full and complete
set of tuning simulations have not yet been performed, but it is
expected that the baseline design for the RTML can satisfy the
emittance preservation requirements.

{\bf Phase Jitter:}  Phase and amplitude errors in the bunch
compressor RF will lead to energy and timing jitter at the IP, the
latter directly resulting in a loss of luminosity.  Table
\ref{tab:BCtols} shows the RMS tolerances required to limit the
integrated luminosity loss to 2\%, and to limit growth in IP energy
spread to 10\% of the nominal energy spread \cite{bib:RTMLchurch}.
\stepcounter{tablcl}\begin{table} \vbabove
   \caption{Key tolerances for the two-stage bunch compressor.
%   This table may need some
%   updating when we have more of a lattice than we do now.
   }
   \label{tab:BCtols}
\setlength{\tabcolsep}{4pt}
   \begin{center}
      \begin{tabular}{| l | r | r |}
         \hline
         Parameter & Arrival Time Tolerance & Energy Spread Tolerance \\
         \hline & & \vbdlspacing \hline
         Correlated BC phase errors & 0.24$^{\circ}$ &
         0.35$^{\circ}$ \\ \hline
         Uncorrelated BC phase errors & 0.48$^{\circ}$ &
         0.59$^{\circ}$ \\ \hline
         Correlated BC amplitude errors & 0.5\% & 1.8\% \\ \hline
         Uncorrelated BC amplitude errors & 1.6\% & 2.8\% \\  \hline
      \end{tabular}
   \end{center} \vbbelow
\end{table}
The tightest tolerance which influences the arrival time is the
relative phase of the RF systems on the two sides:  in the nominal
configuration, a phase jitter of the electron and positron RF
systems of \(0.24^{\circ}\) RMS, relative to a common master
oscillator, results in 2\% luminosity loss. The tight tolerances
will be met through a three-level system:
\begin{itemize}
\item Over short time scales, such as 1 second, the low-level RF system will be
required to keep the two RF systems phase-locked to the level of
0.24~degrees of 1.3~GHz.  See Section \ref{sect:LLRF} for a fuller
description of the low-level RF system. \itemspace

\item Over longer time periods, the arrival times of the two beams will be directly
measured at the IP and a feedback loop will adjust the low-level RF
system to synchronize the beams. This system is
required to compensate for drifts in the low-level RF phase-locking
system which occur over time scales long compared to a second. \itemspace

\item Over a period of many minutes to a few hours, the arrival time of one beam will
be ``dithered'' with respect to the arrival time of the other beam,
and the relative offset which maximizes the luminosity will be
determined.  This offset will be used as a new set-point for the IP
arrival-time feedback loop, and serve to eliminate drifts which
arise over time scales long compared to a minute. \itemspace
\end{itemize}

{\bf Halo Formation from Scattering:}
%The formation of beam halo in
%the downstream RTML due to scattering from beam gas and thermal
%photons has been studied \cite{sergei}.
Halo formation is dominated by Coulomb scattering from the nuclei of
residual gas atoms, and it is estimated that 100 nTorr base pressure
in the downstream RTML will cause approximately $9\times10^{-7}$ of
the beam population to enter the halo \cite{bib:RTMLsergei}.  A
similar calculation was performed for the upstream RTML, which
indicates that 20 nTorr base pressure will cause approximately
$2\times10^{-6}$ of the beam population to enter the halo.  This is
well below the budget of $10^{-5}$ which has been set for all
beamlines between the damping ring and the BDS collimators (see
\ref{sect:BDSHaloPower}).

{\bf Space Charge:}  In the long, low-energy, low-emittance transfer
line from the damping ring to the bunch compressor, the incoherent
space-charge tune shift will be on the order of 0.15 in the
vertical.  The implications of such large values in a single-pass
beamline have not been studied.

{\bf Collimator Wakefields:}  Assuming collimation of the beam
extracted from the damping ring at $10\sigma_x$, $60\sigma_y$, and
$\pm1.5\%$ ($10\sigma_{\delta}$) in momentum, the worst-case jitter
amplification for untapered, ``razor-blade'' spoilers is expected to
be around 10\% in x, around 75\% in y, and the contribution to x
jitter from energy jitter is expected to be negligible
\cite{bib:RTMLcollwake,bib:RTMLcollwakejitter}.  The vertical jitter amplification
figure is marginal, but can be substantially improved through use of
spoilers with modest longitudinal tapers.  The other collimator
wakefield ``figures of merit'' are acceptable even assuming
untapered spoilers.

\subsection{Accelerator Components}

Table~\ref{tab:RTMLtot} shows the total number of components of each
type in each RTML.  The number of quadrupoles, dipole correctors,
and BPMs is larger in the electron RTML than in the positron RTML
due to the longer electron Return line; for these 3 component
classes, the different totals for each side are shown in
Table~\ref{tab:RTMLtot}.  Each quadrupole and dipole has its own
power supply, while other magnets are generally powered in series
with one power supply supporting many magnets. The cost estimate for
the S-band dipole-mode structures was developed by the RTML Area
Systems group based on recent experience with accelerator structure
construction at IHEP; all other component cost estimates were
developed by the ILC Technical and Global Systems groups.

\stepcounter{tablcl}\begin{table}[htb] \vbabove
     \caption[Total number of components in each RTML.] {Total number of components in each RTML.  Where 2 totals are shown, the larger
     number refers to the longer electron-side RTML, the smaller number refers to the shorter
     positron-side RTML.}
     \label{tab:RTMLtot}
    \begin{center}
     \setlength{\tabcolsep}{4pt}
    \begin{tabular}{| l | r || l | r || l | r |}
        \hline
        \multicolumn{2}{|  c || }{Magnets} &
        \multicolumn{2}{   c || }{Instrumentation} &
        \multicolumn{2}{   c  |}{RF} \\
        \hline & & & & & \vbdlspacing \hline
       Bends     & 362  & BPMs           & 772/740 & 1.3~GHz cavities             & 414 \\   \hline
       Quads     & 789/752  & Wires          &  12 & 1.3~GHz cryomodules          &  48 \\   \hline
       Dipoles   & 1185/1137 & BLMs           &   2 & 1.3~GHz sources & $16+1$ \\   \hline
       Kickers   &  17 & OTRs           &   5 & S-band structures            &   2 \\   \hline
       Septa     &   7 & Phase monitors &   3 & S-band sources  &   2 \\   \hline
       Rasters   &  6 & Xray SLMs      &   2 &  &   \\   \hline
       Solenoids   &  4  &  &  &  & \\
%       Solenoids &   4 & Phase Monitors &   3 & S-band Klystrons/Modulators  &   2 \\
%       Septa     &  7 & Xray SLMs      &   2 &  &   \\
%       Rasters   &  6  &  &  &  & \\
        \hline
     \end{tabular}

    \end{center} \vbbelow
\end{table}

Table \ref{tab:RTMLlen} shows the system lengths for the RTML
beamlines.

\stepcounter{tablcl}\begin{table}[htb] \vbabove
   \caption[System lengths for each RTML beamline.] {System lengths for each RTML beamline.  Where 2 values are shown, the larger
     number refers to the longer electron-side RTML, the smaller number refers to the shorter
     positron-side RTML.}
   \label{tab:RTMLlen}
   \begin{center}
      \begin{tabular}{| c | c | c | c | c | c |}
          \hline
          Upstream RTML & Turn & Spin & Emit & BC  & Dumplines  \\
          \hline & & & & & \vbdlspacing \hline
          15,447 m / 14,247 m & 275 m &  82 m &  47 m & 1,105 m &  180 m  \\   \hline
          \multicolumn{4}{| l}{Total} & \multicolumn{2}{| c |}{17,136 m / 15,936 m} \\   \hline
          \multicolumn{4}{| l}{Total, excluding extraction lines} & \multicolumn{2}{| c |}{16,956 m / 15,756 m} \\   \hline
          \multicolumn{4}{| l}{Footprint length} & \multicolumn{2}{| c |}{1,301 m} \\   \hline
      \end{tabular}
   \end{center} \vbbelow

\end{table}

\subsubsection{Vacuum Systems}

The base pressure requirement for the downstream RTML is set by
limiting the generation of beam halo to tolerable levels, while in
the upstream RTML it is set by the necessity of avoiding beam-ion
instabilities.  As described in \ref{sect:RTMLAccPhys}, the base
pressure requirement for the downstream RTML is 100 nTorr, while in
the upstream RTML it is 20 nTorr.  Both upstream and downstream RTML
vacuum systems will be stainless steel with 2 cm OD; the upstream
RTML vacuum system will be installed with heaters to allow {\it in
situ} baking, while the downstream RTML vacuum system will not. The
bending sections of the turnaround and bunch compressors are not
expected to need photon stops or other sophisticated vacuum systems,
as the average beam current is low, and the fractional power loss of
the beam in the bending regions is already small to limit emittance growth from ISR.
%The vacuum system in the cryomodules will be an exact duplicate of the
%main linac cryomodule vacuum system.

\subsubsection{Service Tunnel}

There is a service tunnel that runs parallel to the beam tunnel for the full length of the RTML and is shared with other systems.
All of the power supplies, RF sources, and
rack-mounted instrumentation and controls equipment and computers are installed in the service tunnel
This configuration allows repairs and maintenance to be performed
while minimizing disruption to the accelerator itself.

\clearpage %\setcounter{page}{1}
\setcounter{figlcl}{0}
\setcounter{tablcl}{0}
\setcounter{secnumdepth}{4} \setcounter{subsection}{0}
\setcounter{subsubsection}{0}
\usecounter{subsubsection}

\setcounter{section}{5} \renewcommand{\picturefolder}{./linac/}
\section{Main Linacs}\label{sect:ML}

\subsection{Overview}

The two main linacs accelerate the electron and positron beams from
their injected energy of 15~GeV to the final beam energy of 250~GeV
over a combined length of 23~km. This must be accomplished while
preserving the small bunch emittances, which requires precise orbit
control based on data from high resolution beam position monitors.
The linacs utilize L-band (1.3~GHz) superconducting technology, with
nine-cell standing-wave niobium cavities operating at an average
gradient of 31.5~MV/m in a 2K superfluid helium bath. The choice of
operating frequency is a balance between the high cavity cost due to
size at lower frequency and the lower sustainable gradient due to
increased surface resistivity at higher frequency. The accelerator
gradient is somewhat higher than that typically achievable today and
assumes that further progress will be made during the next few years in
the aggressive program that is being pursued to improve cavity
performance.

\subsection{Beam Parameters}

Table \ref{tab:MLBeamPars} lists the key beam parameters in the main
linac.  A description of the tradeoffs which led to the selection of
the parameters can be found in Section~\ref{sect:ACCparams}.

\stepcounter{tablcl}\begin{table}[htb]   \vbabove \caption{Nominal
beam parameters in the ILC Main Linacs.} \label{tab:MLBeamPars}
\begin{center}
\setlength{\tabcolsep}{4pt}
\begin{tabular}{| l | r | l || l | r | l |} \hline
Parameter & Value & Units & Parameter & Value & Units \\ \hline & & & & \vbdlspacing \hline
  Initial beam energy & 15 & GeV & Initial $\gamma\epsilon_x$ & 8.4 &
    $\mu$m \\ \hline
  Final beam energy & 250 & GeV & Final $\gamma\epsilon_x$ & 9.4 &
    $\mu$m \\ \hline
  Particles per Bunch & $2\times10^{10}$ & & Initial
    $\gamma\epsilon_y$ & 24 & nm \\ \hline
  Beam current & 9.0 & mA & Final $\gamma\epsilon_y$ & 34 & nm \\ \hline
  Bunch spacing & 369  & ns & $\sigma_z$ & 0.3  & mm \\ \hline
  Bunch train length & 969 & $\mu$s & Initial $\sigma_E/E$ & 1.5 & \% \\
    \hline
  Number of bunches & & 2625 & Final $\sigma_E/E$ (e$^-$,e$^+$)& 0.14,0.10 & \% \\ \hline
  Pulse repetition rate & 5 & Hz & Beam phase wrt RF crest & 5 &
    $^{\circ}$ \\ \hline
\end{tabular}
\end{center} \vbbelow
\end{table}

The rms bunch length remains constant along the linacs, while the
bunch fractional energy spread decreases roughly as $E_0/E$, where $E$ is the
beam energy and $E_0$ is the initial main linac beam energy. The
bunches are phased 5$^{\circ}$ off-crest to minimize their energy
spread. No BNS energy spread is included to suppress resonant
head-to-tail bunch trajectory growth as the short-range wakefield is
fairly weak. For this same reason, the focusing strength of the
quadrupole lattice in the linacs is kept fairly weak to reduce
emittance growth from quadrupole misalignments.

\subsection{System Description}

\subsubsection{RF Unit}

The main linacs are composed of RF units whose layout is illustrated
in Figure \ref{fig:RFUnit} and whose parameters are listed in Table
\ref{tab:RFUnit}. Each unit has a stand-alone RF source that powers
three contiguous cryomodules containing a total of 26 cavities (with
9, 8 and 9 cavities in each cryomodule, respectively). The RF source
includes a high-voltage modulator, a 10 MW klystron and a waveguide
system that distributes the RF power to the cavities. It also
includes the low-level RF (LLRF) system to regulate the cavity field
levels, interlock systems to protect the source components, and the
power supplies and support electronics associated with the operation
of the source. To facilitate maintenance and limit radiation
exposure, the RF source is housed mainly in a separate service
tunnel that runs parallel to the beam tunnel.

\stepcounter{figlcl}\begin{figure}[htb]  \vbabove
\includegraphics[width=\textwidth, clip]{\picturefolder RFUnitDiagram.pdf}
\vbabovecaption
\caption{RF unit layout.}
\label{fig:RFUnit}
\vbbelow
\end{figure}

\stepcounter{tablcl}\begin{table}[htb] \vbabove \caption{RF unit
parameters.} \label{tab:RFUnit}
\begin{center}
\begin{tabular}{| l | r | l |} \hline
Parameter & Value & Units \\ \hline & & \vbdlspacing \hline
  Modulator overall efficiency & 82.8 & \% \\ \hline
  Maximum klyston output power & 10 & MW \\ \hline
  Klystron efficiency & 65 & \% \\ \hline
  RF distribution system power loss & 7 & \% \\ \hline
  Number of cavities & 26 & \\ \hline
  Effective cavity length & 1.038 & m \\ \hline
  Nominal gradient with 22\% tuning overhead & 31.5 & MV/m \\ \hline
  Power limited gradient with 16\% tuning overhead & 33.0  & MV/m \\  \hline
  RF pulse power per cavity & 293.7 &  kW \\ \hline
  RF pulse length & 1.565 & ms \\ \hline
  Average RF power to 26 cavities & 59.8 & kW \\ \hline
  Average power transferred to beam & 36.9 & kW \\ \hline
\end{tabular}
\end{center} \vbbelow
\end{table}

The modulator is a conventional pulse-transformer type with a
bouncer circuit to compensate the voltage droop that occurs in the
main storage capacitor during the pulse. The modulator produces 120
kV, 130 A, 1.6 ms, 5 Hz pulses with an efficiency of  83\%,
including the charging supply and rise time losses. These high
voltage pulses power a multi-beam klystron (MBK) that amplifies
$\sim100$ W, 1.6 ms RF pulses from the LLRF system up to 10 MW. This
klystron has higher power and improved efficiency (65\% goal) relative to commercial 5 MW tubes (40-45\%). Two
waveguides transport the power from the dual MBK outputs through a
penetration to the beam tunnel where the power in each waveguide is
then split to feed half of the middle cryomodule and one end
cryomodule (see Figure \ref{fig:RFUnit}).

The distribution system is composed primarily of aluminum WR650
(6.50'' x 3.25'') waveguide components. For long runs, WR770 is
substituted to minimize distribution losses, estimated to be 7\%,
including 2\% in the circulators. Along each cryomodule, RF power is
equally distributed among the cavities through a series of
hybrid-style 4-port tap-offs, each with appropriate fractional
coupling (e.g. 1/9, 1/8, ...1/2). Between each tap-off output and
its associated cavity power coupler, there are a bend, a
semi-flexible section, a circulator, a three-stub tuner, and a
diagnostic directional coupler. The three-stub tuner allows fine
adjustment of cavity phase and can be used to adjust the cavity
$Q_{\rm ext}$, although this is mainly adjusted via motor control of
the position of the inner conductor in the cavity power coupler. The
circulator, with a load on its third port, absorbs the RF power
reflected from the cavities during filling and discharge, and so
provides protection to the klystron and isolation between cavities.

\begin{figure}[htb] \vbabove
\includegraphics[width=\textwidth, clip]{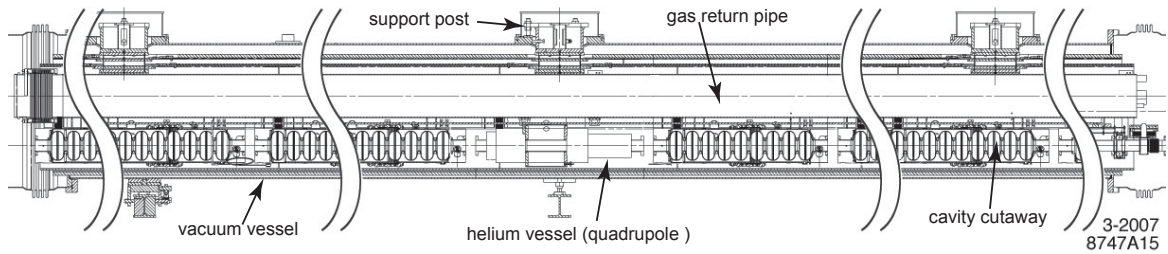}
\vbabovecaption
\caption[Side view of a cryomodule with a quadrupole magnet in the
center.] {Side view of a cryomodule with a quadrupole magnet in the
center. The figure has been compressed as indicated by the two white
gaps, so not all eight cavities are shown.}
\label{fig:CMCutaway} \vbbelow
\end{figure}

The cryomodule design is a modification of the Type-3 version developed
and used at DESY (see Figure \ref{fig:CMCutaway}). Within the
cryomodules, a 300 mm diameter He gas return pipe serves as a
strongback to support the cavities and other beam line components.
Invar rods are used to maintain the spacing between the components
when the cryomodule cools down, which requires roller-type support
fixtures. The gas return pipe itself is supported at three locations
off of the top of the outer vacuum vessel, with only the center
support fixed. The middle cryomodule in each RF unit contains eight
cavities, rather than nine, to accommodate a quad package that
includes a superconducting quadrupole magnet at the center, a cavity
BPM, and superconducting horizontal and vertical corrector magnets.
All cryomodules, whether with or without the quad package, are
12.652 m long so the active length to actual length ratio in a
9-cavity cryomodule is 73.8\%. Each also contains a 300 mm long HOM
beam absorber assembly that removes energy through the 40-80K cooling
system from beam-induced higher order modes above the cavity cutoff
frequency.

\stepcounter{figlcl}

The cavities illustrated in Figure \ref{fig:CMCutaway} are ``dressed''
in that the cells are enclosed in a titanium vessel containing
the liquid helium, a tuner system is mounted around the center to
control the cavity length, and a coaxial power coupler (not shown)
connects the cavity to the external waveguide feed.
The cavity spacing within the cryomodules is 5 3/4 $\lambda_0$ = 1.326 m,
which
%The free space between cavities is 0.290 meters, which
facilitates powering the cavities in pairs via 3 db hybrids as
an alternate distribution scheme that eliminates or reduces the number of circulators.
However, the spacing would not be significantly reduced otherwise
due to the required length of bellows between cavities and space for
flange accessibility.

To operate the cavities at 2K, they are immersed in a saturated He
II bath, and helium gas-cooled shields intercept thermal radiation
and thermal conduction at 5--8 K and at 40--80 K. The estimated
cryogenic heat loads per RF unit are listed in Table
\ref{tab:heatload}, and were obtained by scaling the TESLA TDR estimates.
Also, for each of the three cooling systems, the associated
cryoplant power is listed for both the static and dynamic
contributions from an RF unit and associated transfer line and
distribution components, including a 50\% overcapacity factor. The dynamic
2 K heat loss, attributable mainly to the RF and beam HOM losses in
the cavities, constitutes about half the total installed power.

\stepcounter{tablcl}\begin{table}[htb] \vbabove \caption{RF unit
cryogenic heat loads and installed
  AC cryogenic plant power to remove the heat.}
\label{tab:heatload}
\begin{center}
\setlength{\tabcolsep}{4pt}
\begin{tabular}{| l || c | c || c | c || c | c || c |} \hline
   & \multicolumn{2}{c ||}{40--80 K} &
     \multicolumn{2}{c ||}{5--8 K} &
     \multicolumn{2}{c ||}{2 K} & Total \\ \hline
   & Static & Dynamic & Static & Dynamic & Static & Dynamic & \\
     \hline & & & & & & & \vbdlspacing \hline
  Heat load (W) & 177.6 & 270.3 & 31.7 & 12.5 & 5.1 & 29.0 & \\
    \hline
  Installed power (kW) & 4.4 & 6.2 & 9.6 & 3.5 & 8.1 & 28.5 & 60.4 \\
  \hline
\end{tabular}
\end{center} \vbbelow
\end{table}

\subsubsection{Linac Layout}

\begin{figure}[htb] \vbabove
\includegraphics[width=\textwidth, clip]{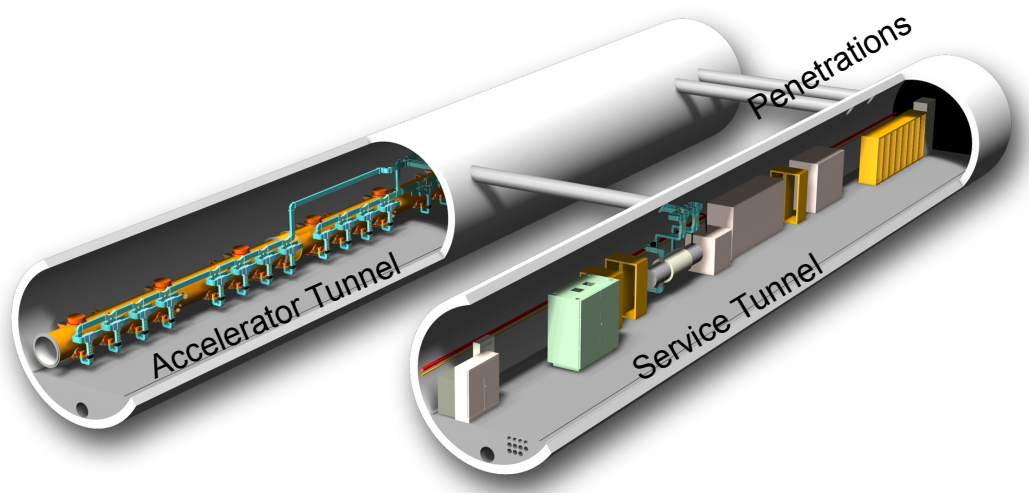}
\vbabovecaption
\caption{Cutaway view of the linac dual-tunnel configuration.}
\label{fig:TwoTunnels} \vbbelow
\end{figure}

The Main Linac components are housed in two tunnels, each of which
has an interior diameter of 4.5 meters.  The tunnels are separated
from one another by 5.0 m to 7.5 m depending on the geology at the
ILC site. As illustrated in Figure \ref{fig:TwoTunnels}, the
cryomodules occupy the beam tunnel while most of the RF system,
including modulators, klystrons, power supplies, and instrumentation
racks, are located in the service tunnel. This arrangement permits
access to the equipment in the service tunnel for maintenance,
repair, or replacement during beam operation and limits radiation
exposure to most of the electronics (except motors in or near the
cryomodules). The two tunnels are connected by three penetrations
along each RF unit: one for the waveguide, one for signal cables,
and one for power and high voltage cables. Personnel access points
between the two tunnels are located at roughly 500 meter intervals.
Rather than being ``laser straight'', the tunnels are curved in the
vertical plane, with a radius of curvature slightly smaller than
that of the Earth. This allows the beam delivery system to lie in a
plane at the center of the site, while the cryomodules nearly follow
a gravitational equipotential to simplify distribution of cryogenic
fluids.

\stepcounter{figlcl}

The positron linac contains 278 RF units, and the electron linac has
282 RF units; the four additional RF units are needed to compensate
for the beam energy lost in the undulator that is used to generate
gamma rays for positron production. The positron system section
within the electron linac is 1,257 m long and is located near the 150
GeV point (see Section \ref{sect:POSps}). Coasting sections, about 400 m long,
are included at the end of the linacs so that additional RF units can be installed as an upgrade to provide up to 3.5\% energy overhead
during 500 GeV CM operation. No additional tunnel is included for a
future upgrade to higher energies, although the site is sized
to allow expansion for 1 TeV CM operation.

The tunnels in the present sample sites are 100-150 meters underground and are
connected to the surface through vertical shafts.  Each of the main
linacs includes three shafts, roughly 5 km apart as dictated by the
cryogenic system. The upstream shafts in each linac have diameters
of 14 m to accommodate lowering cryomodules horizontally, and the
downstream shaft in each linac is 9 m in diameter, which is the
minimum size required to accommodate tunnel boring machines. At the
base of each shaft is a 14,100 cubic meter cavern for staging
installation and housing utilities and parts of the cryoplant, most of which are located on the surface.

The layout of the RF units in the main linac is not uniform, but
includes an additional 2.5 m long ``end box'' after every 4 RF units
that terminates the 2K He distribution to the upstream cavities and
restarts it from the main 2K feed line for the downstream cavities.
The linac section from one such end box to the next is called a
``cryo string.''  In a few locations, cryo-strings of three RF units
are used instead of four RF units.
Cryo-strings are arranged in groups of 10 to 16 to form a cryogenic
unit which is supported by a single cryoplant. Each cryogenic unit
also includes 2.5 m long ``service boxes'' on each end (one service
box replaces a cryo-string end box), and is separated from the next
cryogenic unit by a 7.7 m warm section that includes vacuum system
components and a laser wire to measure beam size. Accounting for
these additional sections and the quad package length, the active to
actual length ratio in the linacs is 69.7\% (excluding the undulator
section and the coasting section at the end of each linac). Table
\ref{tab:subdivision} summarizes the linac component lengths and
numbers.

\stepcounter{tablcl}\begin{table}[htb] \vbabove \caption[Subdivision
lengths and numbers in the two main linacs.] {Subdivision lengths
and numbers in the two main linacs.  Total linac lengths exclude the
length of the positron production insertion and the coasting length
at the end of each linac.} \label{tab:subdivision}
\begin{center}
\begin{tabular}{| l | c | c |} \hline
  Subdivision & Length (m) & Number \\ \hline & & \vbdlspacing \hline
  Cavities (9 cells + ends) & 1.326 & 14,560 \\ \hline
  Cryomodule (9 cavities or 8 cavities + quad) & 12.652 & 1,680 \\
    \hline
  RF unit (3 cryomodules) & 37.956 & 560 \\ \hline
  Cryo-string of 4 RF units (3 RF units) & 154.3 (116.4) & 71 (6) \\
    \hline
  Cryogenic unit with 10 to 16 strings & 1,546 to 2,472 & 10 \\
    \hline
  Electron (positron) linac & 10,917 (10,770) & 1 (1) \\ \hline
\end{tabular}
\end{center} \vbbelow
\end{table}

There are five, 4 MW-size cryoplants in each linac that also provide
cooling for the RTML and undulator region. The total cryogenic capacity of the
ILC linacs is comparable to that of the LHC. The plants are paired at
each linac shaft, one feeding downstream cryomodules and the other
upstream cryomodules, except for the downstream most shaft, where
there is only one plant that feeds upstream cryomodules. The plants
are sized with a 40\% overcapacity to account for degradation of
plant performance, variation in cooling water temperature, and
operational overhead.

Conventional water cooling towers are also located on the site
surface near each linac shaft.
Through various distribution loops, they provide 35$^{\circ}$C 'process' water that removes most of the heat generated by the RF system, and 8$^{\circ}$C 'chilled' water for heat exchangers that maintain the tunnel air temperature at 29$^{\circ}$C and cool electronics racks via closed, circulated-air systems. In each RF unit, roughly 10 kW of heat are dissipated in the racks, and another 10 kW are dissipated into the air from convection off of the RF source components.

The electrical requirements of the main linac are supplied by two high-voltage cable systems.  One of the systems supports the conventional
services, while the other supports the RF system.  Table
\ref{tab:linacpower} summarizes the combined power consumption of
the two main linacs.  Of this power, 20.5 MW is transferred to the
beams, for a net efficiency of 13.7\%.

\stepcounter{tablcl}\begin{table}[htb] \vbabove \caption{AC power
consumption of the two main linacs.} \label{tab:linacpower}
\begin{center}
\begin{tabular}{|l|c|} \hline
 System & AC Power (MW) \\ \hline & \vbdlspacing \hline
 Modulators & 81.4 \\ \hline
 Other RF system and controls & 8.4 \\ \hline
 Conventional facilities & 25.7 \\ \hline
 Cryogenic & 33.8 \\ \hline & \vbdlspacing \hline
 Total & 149.3 \\ \hline
\end{tabular}
\end{center} \vbbelow
\end{table}

\subsection{Accelerator Physics Issues}

\subsubsection{Optics}

\begin{figure}[htb] \vbabove
\begin{center}
\includegraphics[width=0.65\textwidth, clip]{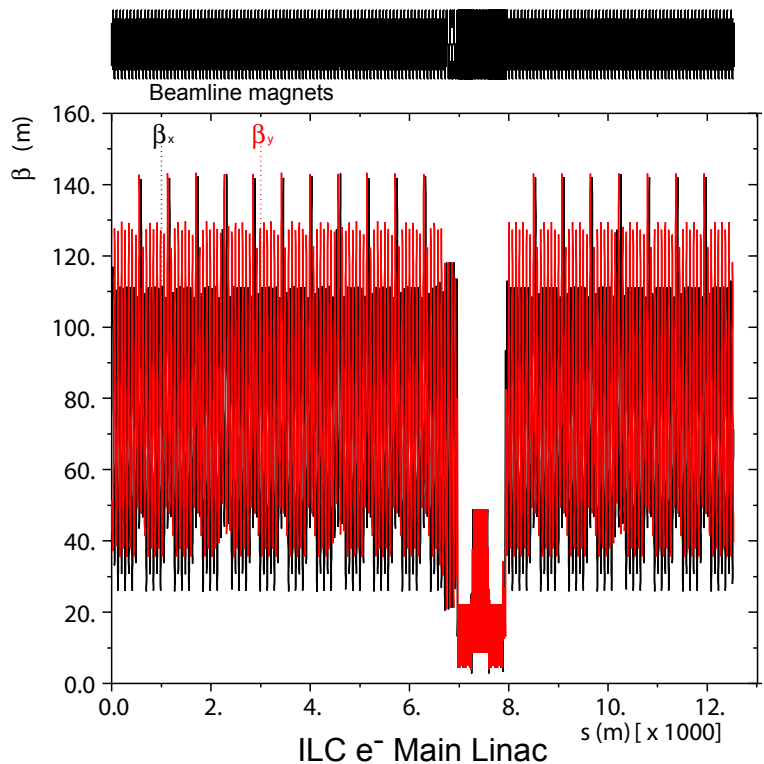}
\vbabovecaption
\caption[Beam optics functions for the electron main linac.]
{Beam optics functions for the electron main linac. The
discontinuity of the pattern around s$\sim8$ km represents the
undulator section for positron production.}
\label{fig:eminusoptics} \vbbelow\vbbelow
\end{center}
\end{figure}

The main linac lattice uses a weak focusing FODO optics, with a quad
spacing of 37.956 m, corresponding to one quad per RF unit. Each
quadrupole magnet is accompanied by horizontal and vertical dipole
correctors and a cavity BPM which operates at 1.3 GHz. Because of
the aperiodicity conditions imposed by the cryogenic system, the
lattice functions are not perfectly regular. The mean phase advance
per cell is 75$^{\circ}$ in the horizontal plane and 60$^{\circ}$ in
the vertical plane. The vertical curvature is provided by the
vertical correctors at the quadrupole locations, rather than by
dedicated bend magnets. Dispersion matching and suppression at the
beginning and end of the linac and around the undulator insertion
are achieved by supplying additional excitation to small numbers of
correctors in ``dispersion-bump" configurations. Figure
\ref{fig:eminusoptics} shows the optical functions of the electron
linac, including the undulator insertion. The functions for the
positron main linac are basically the same except that the undulator
insertion is not present.

\stepcounter{figlcl}

\subsubsection{Beam Dynamics}

A key requirement of the main linacs is that they preserves the
small emittances which are produced in the damping rings and
transported through the RTMLs. This is particularly true for the
vertical emittance, which is smaller than the horizontal emittance
by a factor of 400. The main obstacles to emittance preservation in
the linacs are transverse wakefields, betatron coupling, and
dispersion.

The short-range transverse wakefields in the ILC cavities are quite
weak compared to the wakefields typically associated with
higher-frequency RF cavities. Alignment tolerances for cavities and
cryomodules in the range of 200-300 $\mu$m RMS are expected to yield
emittance growth on the order of 2 nm (10\%) in the vertical plane.
It is possible that even this small amount of emittance dilution can
be corrected by the use of ``wake bumps'' (local orbit distortions
which excite wakefields but not other aberrations).

The long-range wakefields in the ILC cavities are potentially more
harmful given the high Q values typical in superconducting cavities.
These wakefields are mitigated through HOM damping ports on the
cavities, additional HOM absorbers in each RF unit at the location
of the quadrupole magnet package, and detuning of the HOM's at the
level of $10^{-3}$. The combination of damping and detuning reduces
the multi-bunch emittance growth to 0.3~nm (1.5\%).

Azimuthal deformations to the cavities from construction errors or
from the placement of the HOM and fundamental mode ports can cause
the HOM's to develop diagonal polarizations instead of horizontal
and vertical polarizations. Diagonally-polarized (or
``mode-rotation'') HOM's can couple beam jitter from the horizontal to
the vertical, resulting in unacceptable vertical emittance dilution.
This is mitigated in the main linacs by making the horizontal and
vertical betatron tunes highly unequal.
%, as described in Section \ref{sect:MLLayoutOptics}.
Setting the horizontal phase advance per cell to 75$^{\circ}$ and
the vertical to 60$^{\circ}$ limits emittance growth from this
effect to 0.4~nm (2\%).

Betatron coupling between the relatively large horizontal mode and
the relatively small vertical mode is driven by unwanted rotations
of the main linac quadrupole magnets. By limiting the rms rotations
of the quads to 0.3 mrad, the resulting emittance growth can be
limited to 2~nm (10\%). Most of this emittance growth can be
globally corrected by the decoupler at the start of the beam
delivery section (see Section \ref{sect:BDSDecouple}), subject to the
resolution limits of the laser wire profile monitors in the BDS.

Dispersion in the main linac is created by misaligned quadrupole
magnets and pitched RF cavities. Emittance growth from this effect
is mainly corrected through local or quasi-local steering algorithms
such as Ballistic Alignment (BA), Kick Minimization (KM), or
Dispersion Free Steering (DFS), with additional correction achieved
through local orbit distortions which produce measured amounts of
dispersion in a given phase (``dispersion bumps''). Simulations
indicate that emittance growth from dispersion can be limited to
about 5 nm (25\%) through combinations of these techniques.

The principal main linac beam diagnostic is the suite of beam
position monitors: a BPM with horizontal and vertical readout and
sub-micron single-bunch resolution is located adjacent to each
quadrupole magnet. For beam size monitoring, a single laser wire is
located in each of the warm sections between main linac cryogenics
units (about every 2.5 km). Upstream quadrupole magnets are varied
to make local measurements of the beam emittances.

The main linacs do not contain any equipment for intra-train
trajectory control.  Such trajectory control is implemented only in
the warm regions upstream and downstream of the main linacs and in
the undulator section. There are no diagnostics for measuring energy
or energy spread in the main linacs. These measurements are made
upstream and downstream of the main linacs and in the undulator
section. There are no beam abort systems in the main linacs. Machine
protection in the linac is ensured by verifying the state of the
main linac hardware (both RF and magnets) prior to beam extraction
from the RTML, and by verifying that the orbit in each damping ring
is correct.  The limiting aperture along the main linacs is the 70
mm diameter cavity iris.

\subsubsection{Operation}

Within each RF unit, a low level RF (LLRF) system monitors the
vector sum of the fields in the 26 cavities. It makes adjustments
to flatten the energy gain along the bunch train and keeps the
beam-to-rf phase constant. It compensates for perturbations
including cavity frequency variations (e.g. due to microphonics and
residual Lorentz force detuning after feed-forward piezo-electric
controller compensation), inter-pulse beam current variations, and
non-flatness of the klystron pulse. In addition to the phase and
amplitude of the klystron, this system has remote control over
individual cavity phases (through the RF distribution system),
external quality factors $Q_{\rm ext}$ (through the moveable coupler
center conductor), and resonant frequencies (through slow and
faster tuners).

The cavities are qualified at 35 MV/m or greater during initial
testing (i.e. so-called ``vertical'' tests) prior to installation in
cryomodules. This should allow them to run at 31.5 MV/m on average, installed, although the variation of sustainable gradients may be significant according to current data. Some cryomodule gradient variation within an
rf unit can be accommodated by one-time adjustments in the main feed
line power splitters and the in-line attenuators in each of the two
feed lines.

For 500 GeV operation, there is no energy overhead if the average
sustainable cavity gradient is the design value of 31.5 MV/m. With failed RF units, the ILC can only reach 500 GeV if the cavities
achieve a higher average gradient (power limited to 33 MeV/m) or if additional RF units are eventually
installed in the reserved drift region at the end of the linacs. The
beam energy is coarsely adjusted by turning on or off RF units, each
of which contributes about 0.3\% of the beam energy, and finely
adjusted by cross-phasing RF units near the end of the linacs.

\subsection{Accelerator Components}

\subsubsection{Cavities and Cryomodules}

\begin{figure}[htb] \vbabove
\includegraphics[width=\textwidth, clip]{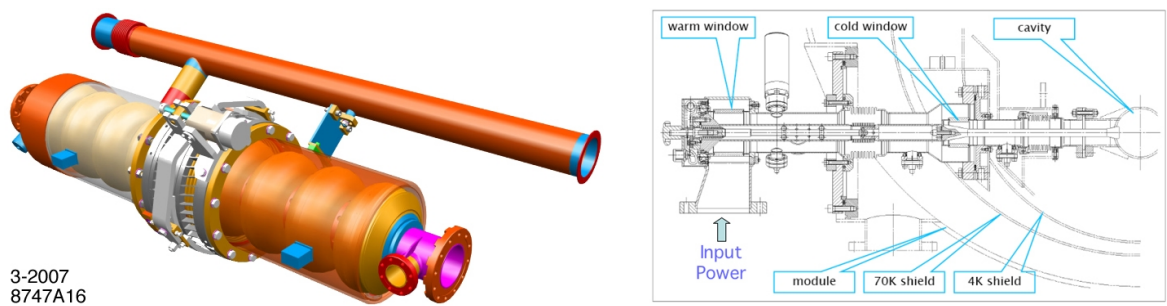}
\vbabovecaption
\caption[A partially dressed cavity and coaxial power coupler.] {Left: A partially dressed cavity including the helium
vessel, 2K He feed line and frequency tuners. Two HOM couplers and
an RF pickup (not visible) are located near the ends of the cavity.
Right: Schematic of the coaxial power coupler that attaches to the
off-axis port shown in the left figure.}
\label{fig:dressedcavity} \vbbelow\vbbelow
\end{figure}

The 1.3 GHz superconducting accelerating cavity is the fundamental
building block of the ILC main linacs.  Its parameters are listed in
Table \ref{tab:cavparams}.  A partially ``dressed'' cavity for
installation in a cryomodule is shown Figure
\ref{fig:dressedcavity}, together with the power coupler schematic.
Each cavity is qualified for installation in the main linac in a
vertical test stand; cavities which can sustain a gradient in excess
of 35 MV/m with a Q value in excess of $0.8\times10^{10}$ are then
installed in cryomodules for use in the main linac.  More
information on the construction and testing of cavities can be found
in Section \ref{sect:Cavity}.
\begin{comment}
The ``bare'' cavities are made from 2.5 mm thick, pure (> 300 RRR)
niobium sheets which are deep-drawn to form half-cells and then
electron-beam welded together. The cavities are then processed to
prepare them for high gradient operation.  This processing includes
electro-polishing, heat treatments, high-pressure rinsing and bakes.
Cavities that achieve gradients in excess of 35 MV/m and Q values in
excess of  $0.8\times10^{10}$ in a vertical test facility are then
``dressed'' by first welding on a titanium outer vessel and a 2K
helium feed pipe. A set of cavities is then assembled in a class-10
clean room, where fundamental and HOM pickups are attached, the cold
portions of the power couplers are installed and the cavities are joined
via beam pipe bellows. Outside of the clean room, magnetic shielding
(Cryoperm) and slow and fast tuners are installed, and the assembly
is suspended from a 300 mm diameter pipe that serves as both the
helium gas return line and the support structure for the 5K and 40K
radiation shielding. The insulation package is then completed and
the cavity assembly inserted into a cryomodule vacuum vessel where
it is supported from above off of three fiberglass posts. Finally,
the warm parts of the power couplers are attached and bolted to the
outer vessel.
\end{comment}

\stepcounter{figlcl}

\stepcounter{tablcl}\begin{table}[htb] \vbabove \vskip0.2in
\caption{Cavity Parameters.} \label{tab:cavparams}
\begin{center}
\begin{tabular}{| l | r | l |} \hline
  Parameter & Value & Units \\ \hline & & \vbdlspacing \hline
  Type & 9 cell, $\pi$-mode & \\ \hline
  R/Q of fundamental mode & 1036 & $\Omega$ \\ \hline
  Iris diameter & 70 & mm \\ \hline
  Cell-to-cell coupling & 1.9 & \% \\ \hline
  Average $Q_0$ & $1.0\times10^{10}$ & \\ \hline
  Average $Q_{\rm ext}$ & $3.5\times10^{6}$ & \\ \hline
  Fill time & 596 & $\mu$s \\ \hline
  Cavity resonance width & 370 & Hz \\ \hline
\end{tabular}
\end{center} \vbbelow
\end{table}

\subsubsection{Quad Package}

In addition to cavities, the center cryomodule in each RF unit
contains a 1.2 m long quad package that includes a quadrupole
magnet, combined horizontal and vertical corrector magnets, and a
cavity beam position monitor.
%
%Like the cavities, these components are supported from the 300 mm
%diameter gas return pipe as illustrated in Fig. 6 for the TTF-3 type
%cryomodule used at TTF. Unlike the TTF cryomodules, the ILC quad
%packages are located at the center of the cryomodule, the quad and
%corrector dipoles are separate magnets (instead of combined
%function) and they operate at 2K (instead of 4K).
At the low-energy end of the linac the quadrupoles and correctors
are superferric types, while cos(2$\theta$) and cos($\theta$)
superconducting magnets are used at the high-energy end of the
linac.  The maximum gradient required in the quadrupoles at the high
energy end of each linac is 60 T/m, while the maximum dipole
integrated strength required is about 0.05 T-m.  The beam position
monitor is an L-band design capable of measuring horizontal and
vertical positions with 1 micrometer resolution for a single bunch
at full charge.
All of the elements in the quad package have an aperture which
is larger than the 70 mm aperture of the superconducting cavities.
\begin{comment}
The strongest magnets are at the high energy ends of the linacs
where the quads are about 600 mm long and have a maximum gradient of
60 T/m, and the combined horizontal/vertical correctors are about
200 mm long with a maximum strength of 0.05 T-m. The required magnet
strengths are proportional to the beam energy, and likely three magnet
styles are used for this range (the quad package length remains
the same however). With the large beam aperture required (> 70 mm),
cos(2$\theta$) and cos($\theta$) type magnets are used at the downstream ends of
the linacs, and probably super-ferric types are used at the
upstream ends. To measure horizontal and vertical bunch positions,
cavity-type monitors with a relatively low frequency (1.5 GHz) are
used due to the large aperture. The goal is to individually measure
bunch positions within the train with less than 1 micron resolution
at the design bunch charge, which still affords a reasonable
resolution with the ten-times lower keep-alive bunch charge.
\end{comment}

\subsubsection{Vacuum System}

There are three independent vacuum systems along the accelerator:
the beamline system that includes the volume in the cavities and
other beamline components, the coupler system that includes the
volume between the two windows in each coupler, and the insulation
system that includes the volume within the cryomodule vacuum vessel.
The beamline system runs the length of the linacs and includes slow
valves with second-scale response times in each 154 m cryo-string
plus fast valves with ms-scale responses in the warm sections
between cryogenic units.  In the event of a major vent, these
systems will limit the length of linac which is exposed to air to one or
two cryo strings.  Finally, the coupler vacuum system is segmented
by cryomodule, and all couplers therein are pumped in common. With
this system, a leak in one of the cold windows is fairly benign.
%\end{comment}

\subsubsection{Beamline Components}

Table \ref{tab:blcomps} lists the basic beamline components and the total number of each contained in the two main linacs, excluding those in the positron production undulator region.

\stepcounter{tablcl}\begin{table}[b!] \vbabove \caption{Main Linac
Beamline Components.} \label{tab:blcomps}
\begin{center}
\begin{tabular}{|l|c|} \hline
  Component & Number (total) \\ \hline & \vbdlspacing \hline
  Cavities & 14,560 \\ \hline
  SC quadrupole magnets & 560 \\ \hline
  X-correctors & 560 \\ \hline
  Y-correctors & 560 \\ \hline
  SRF BPMs & 560 \\ \hline
  Laser wire scanners & 7 \\ \hline
\end{tabular}
\end{center} \vbbelow
\end{table}

\clearpage %\setcounter{page}{1}
\setcounter{figlcl}{0}
\setcounter{tablcl}{0}
\setcounter{secnumdepth}{4} \setcounter{subsection}{0}
\setcounter{subsubsection}{0}
\usecounter{subsubsection}

\setcounter{section}{6} \renewcommand{\picturefolder}{./bds/}

\section{Beam Delivery Systems  }

\subsection{Overview }
%Overview \\ Beam Parameters Angal-Kalinin, Seryi, Yamamoto

The ILC Beam Delivery System (BDS) is responsible for transporting
the ${\rm e^+/e^-}$ beams from the exit of the high energy linacs,
focusing them to the sizes required to meet the ILC luminosity goals
($\sigma_x^*=639$ nm, $\sigma_y^*=5.7$~nm in the nominal
parameters), bringing them into collision, and then transporting the
spent beams to the main beam dumps.
%focusing them to very small beam sizes  (few 100~nm horizontally and
%few nm vertically) at the interaction point (IP).
In addition, the
BDS must perform several critical functions:
\begin{itemize}
 %\addtolength{\topsep}{-2\baselineskip}
% \addtolength{\itemsep}{-0.6\baselineskip}
 \item measure the linac beam and match it into the final focus; \itemspace
 \item protect the beamline and detector against
mis-steered beams from the main linacs; \itemspace % and safely
%extract them to a dump;
 \item remove any large amplitude particles
(beam-halo) from the linac to minimize background
in the detectors; \itemspace
 \item measure and monitor the key physics parameters such as
energy and polarization before and after the collisions; \itemspace
% \item ensure that the extremely small
%beams collide optimally at the IP;
% \item safely extract the beams after collision to the high-power beam dumps.
\end{itemize}
The BDS must provide sufficient instrumentation, diagnostics and
feedback systems to achieve these goals.

%{\it Need to check what parameters are already included in the RDR
%Introduction and what should be shown here. }

\subsection{Beam Parameters}

Table~\ref{table_bds} shows the key BDS parameters. The IP beam
parameters are shown for the nominal parameter set at 500~GeV~CM.

\stepcounter{figlcl}\begin{figure}[htb]
\begin{center} \vbabove
%\vspace{-.2mm}
\includegraphics[width=0.95\textwidth]{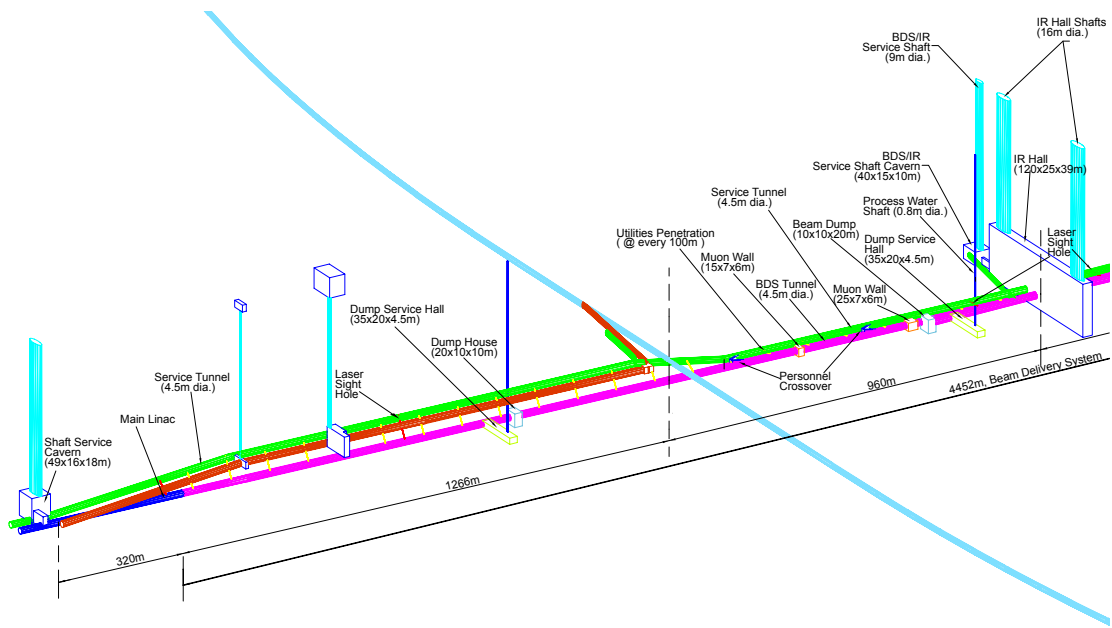}
\vbabovecaption
%\vspace{-5mm}
\caption[BDS layout, beam and service tunnels.]
{BDS layout, beam and service tunnels
(shown in magenta and green), shafts, experimental hall.}
\label{bds_layout}
\end{center} \vbbelow
%\vspace{-5mm}
\end{figure}

\subsection{System Description}

\begin{figure}[htb]
\begin{center} \vbabove
%\vspace{-.05mm}
%\includegraphics[width=130mm]{BDS/bsy}
%\includegraphics[width=130mm]{BDS/BDS_1000_ip_layout}
\includegraphics[width=0.95\textwidth]{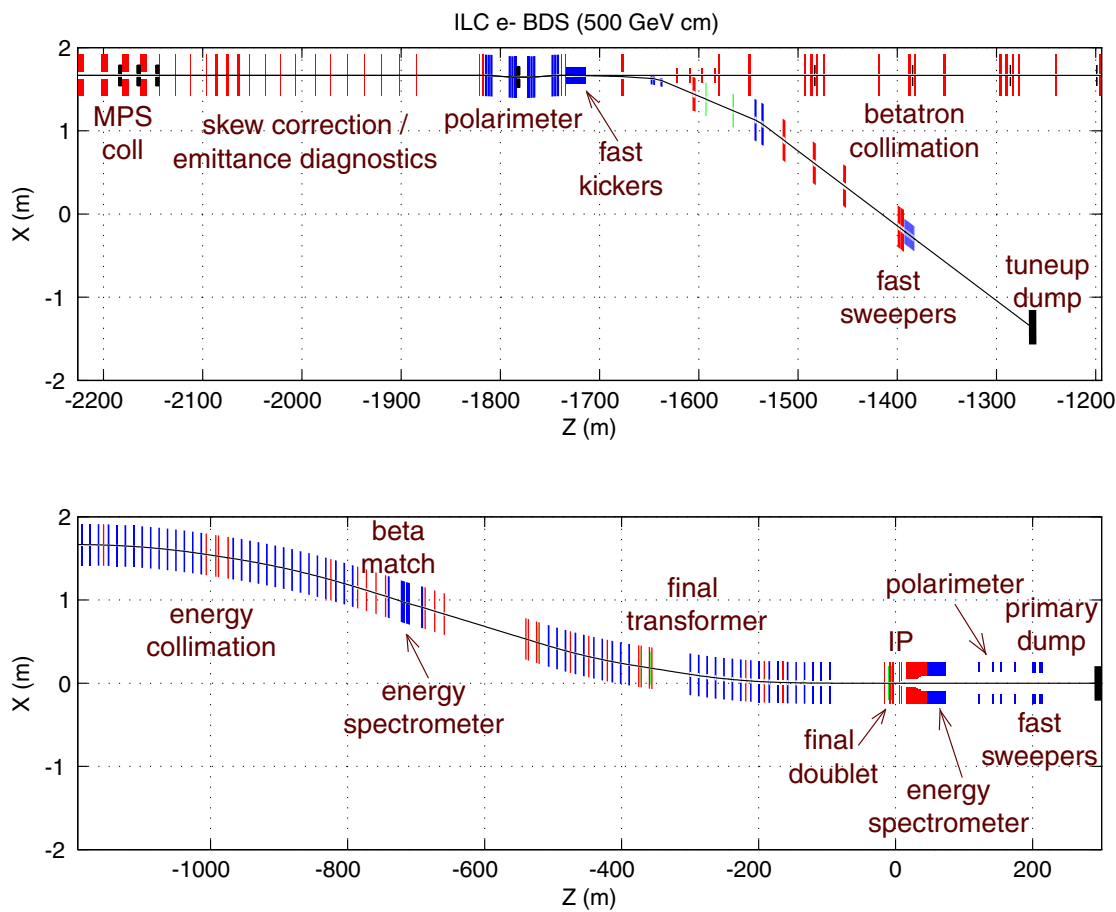}
%\vspace{-10mm}
\vbabovecaption \caption[BDS layout showing functional subsystems.]
    {BDS layout showing functional subsystems, starting
from the linac exit; X -- horizontal position of elements,
Z -- distance measured from the IP.
%{\it (SHOW EXTRACTION LINE OF
%THE OPPOSITE BEAMLINE. SHOW WHERE MUON SPOILERS ARE AND ALSO TAIL
%FOLDING OCTUPOLES.)}
}
\label{bds_functions} \end{center} \vbbelow
%\vspace{-5mm}
\end{figure}

The main subsystems of the beam delivery starting from the exit of
the main linacs are the diagnostics region, the fast extraction and
tuneup beamline, the betatron and energy collimation, the final
focus, the interaction region and the extraction line. The layout of
the beam delivery system is shown in Figures~\ref{bds_layout} and
\ref{bds_functions}.
%The length of the BDS is 4452~m from linac to
%linac and the extraction lines are 300~m long each.
The BDS is designed for 500~GeV center of mass but can be upgraded
to 1~TeV with additional magnets.

\stepcounter{tablcl}\begin{table}[htb]
  \centering \vbabove
  \caption[Key parameters of the BDS.] {Key parameters of the BDS. The range
of L$^*$, the distance from the final quadrupole to the IP, corresponds to values considered for the existing detector concepts.}
\label{table_bds}
  \begin{tabular}{| l | c | c |}
  \hline
    Parameter & Units & Value \\ \hline & & \vbdlspacing \hline
    Length (linac exit to IP distance)/side & m & 2226 \\   \hline
    Length of main (tune-up) extraction line & m & 300 (467) \\   \hline
    Max Energy/beam (with more magnets) & GeV & 250 (500) \\   \hline
    Distance from IP to first quad, L* & m & 3.5-(4.5) \\   \hline
    Crossing angle at the IP  & mrad & 14 \\   \hline
    Nominal beam size at IP, $\sigma^*$, x/y & nm & 639/5.7 \\   \hline
    Nominal beam divergence at IP, $\theta^*$, x/y & $\mu$rad & 32/14 \\   \hline
    Nominal beta-function at IP, $\beta^*$, x/y & mm & 20/0.4 \\   \hline
    Nominal bunch length, $\sigma_z$ & $\mu$m & 300 \\   \hline
    Nominal disruption parameters, x/y & & 0.17/19.4 \\   \hline
    Nominal bunch population, N & & $2\times 10^{10}$ \\   \hline
    Beam power in each beam & MW & 10.8 \\   \hline
    Preferred entrance train to train jitter & $\sigma_y$ & $<0.5$\\   \hline
    Preferred entrance bunch to bunch jitter & $\sigma_y$ & $<0.1$\\   \hline
    Typical nominal collimation aperture, x/y &  & 8--10/60 \\   \hline
    Vacuum pressure level, near/far from IP & nTorr & 1/50 \\ \hline
  \end{tabular} \vbbelow
\end{table}

There is a single collision point with a 14~mrad crossing
angle. To support future energy upgrades, the beam delivery systems
are in line with the linacs and the linacs are also oriented at a
14~mrad angle. The 14~mrad geometry provides space for separate
extraction lines and requires crab cavities to rotate the bunches
horizontally for head-on collisions. There are two detectors in a
common IR hall which alternately occupy the single collision point,
in a so-called ``push-pull'' configuration. The detectors are
pre-assembled on the surface and then lowered into the IR hall in large
subsections once
the hall is ready for occupancy.

\subsubsection{Diagnostics, Tune-up dump, Machine Protection }
% -- Blair, Mattison, Woodley, et al

%\paragraph{Introduction}

The initial part of the BDS, from the end of the main linac to the
start of the collimation system (known for historical reasons as the
Beam Switch Yard or ``BSY''), is responsible for measuring and
correcting the properties of the beam before it enters the
Collimation and Final Focus systems. In addition, errant beams must
be detected here and safely extracted in order to protect the
downstream systems.
%Figure~\ref{bsy} shows the beamline layout of the BSY.
Starting at the exit of the main linac, the system
includes the MPS collimation system, skew correction section,
emittance diagnostic section, polarimeter with energy diagnostics,
fast extraction/tuning system and beta matching section.

\stepcounter{figlcl}

\paragraph{MPS collimation}

At the exit of the main linac is a short \(90^{\circ}\) FODO
lattice, composed of large bore quadrupoles, which contains a set
of sacrificial collimators of decreasing aperture.  The purpose of
this system is to protect the 12~mm aperture BDS from any beam
which develops an extremely large trajectory in the 7~cm aperture
main linac (the effective aperture is $R/\beta^{1/2}$, which is
3--4 times smaller in the BDS than in the linac). This section
also contains kickers and cavity BPMs for inter- and intra-train
trajectory feedback.

\paragraph{Skew Correction}\label{sect:BDSDecouple}

The skew correction section contains 4 orthonormal skew
quadrupoles which provide complete and independent control of the
4 betatron coupling parameters. This scheme allows correction of
any arbitrary linearized coupled beam.

\paragraph{Emittance Diagnostics}

The emittance diagnostic section contains 4 laser wires which are
capable of measuring horizontal and vertical RMS beam sizes down to
1~$\mu$m. The wire scanners are separated by \(45^{\circ}\) in
betatron phase to allow a complete measurement of 2D transverse
phase space and determination of the projected horizontal and
vertical emittances.

\paragraph{Polarimeter and Energy Diagnostics}

Following the emittance diagnostic section is a magnetic chicane
which is used for both Compton polarimetry and beam energy
diagnostics.  At the center of the chicane is the Compton IP, a BPM
for measuring relative beam energy changes, and a sacrificial
machine protection system (MPS) energy collimator which defines the
energy acceptance of the tune-up extraction line.
The length of the chicane is set to
limit horizontal emittance growth due to synchrotron radiation to
less than 1\% with a 500~GeV/beam.
A detector for
the Compton-scattered photons from the laser wires is included in
the chicane.

\paragraph{Tune-up and Emergency Extraction System}

The BSY pulsed extraction system is used to extract beams in the
event of an intra-train MPS fault. It is also used any time when
beams are not desired in the collimation, final focus, or IR areas,
for example during commissioning of the main linacs.  The extraction
system includes both fast kickers which can rise to full strength in
the 300~ns between bunches, and pulsed bends which can rise to full
strength in the 200~ms between trains.  These are followed by a
transfer line with $\pm10\%$ momentum acceptance which transports
the beam to a full-beam-power water-filled dump. There is a
125~m drift which allows the beam size to grow to an area of 2$\pi$~mm$^{2}$%\(mm^{2}\)
~at the dump.
A set of rastering kickers sweep the beam in a 3~cm radius circle on
the dump window.
By using the nearby and upstream
BPMs in the polarimeter chicane and emittance sections, it is
possible to limit the number of errant bunches which pass into the
collimation system to 1--2.

\subsubsection{    Collimation System} % \& machine backgrounds -- 2p} –
%Jackson, Watson, Mokhov, Keller, Drozhdin, Maruyama, et al
\label{sectBDSColl}

%Dec 19 version
Particles in the beam halo produce backgrounds in the detector and
must be removed in the BDS collimation system. One of the design
requirements for the ILC BDS is that no particles are lost in the
last several hundred meters of beamline before the IP. Another
requirement is that all synchrotron radiation passes cleanly through
the IP to the extraction line. The BDS collimation must remove any
particles in the beam halo which do not satisfy these criteria.
These requirements define a system where the collimators have very
narrow gaps and the system is designed to address the resulting machine protection,
survivability and beam emittance dilution issues.

The collimation system has a betatron collimation section followed
by energy collimators. The downstream energy collimators help to
remove the degraded energy particles originating from the betatron
collimation section but not absorbed there. The betatron collimation
system has two spoiler/absorber x/y pairs located at high beta
points, providing single-stage collimation at each of the final
doublet (FD) and IP betatron phases. The energy collimation section
has a single spoiler located at the central high dispersion point
(1530~$\mu$m/\%).  All spoilers and absorbers have adjustable gaps.
Protection collimators (PC) are located throughout to provide local
protection of components and additional absorption of scattered halo
particles.

The spoilers are 0.5 to 1~X$_0$ (radiation length) thick, the
absorbers are 30~X$_0$, and the protection collimators  are
45~X$_0$. The betatron spoilers as well as the energy spoiler are
``survivable'' -- they can withstand a hit of two errant bunches of
250~GeV/beam, matching the emergency extraction design goal. With
500~GeV beam, they would survive only one bunch, and would therefore
require more effective MPS or the use of a pre-radiator scheme.

The collimation apertures required are approximately
${\sim}8-10\sigma_x$ in the x plane and ${\sim}60-80\sigma_y$ in the y
plane. These correspond to typical half-gaps of betatron spoiler of
$\sim$1~mm in the x plane and $\sim$0.5~mm in the y plane.

Wakefield calculations for the BDS spoilers and absorbers give IP
jitter amplification factors \cite{wake-1} of ${\cal A}_x=0.14$ and
${\cal A}_y=1.05$. Estimated as $\delta\varepsilon/\varepsilon=(0.4
n_{\mathrm{jitter}} {\cal A})^2$ this gives emittance dilutions of
0.08\% and 4.4\% in the x and y planes respectively, for 0.5 $\sigma$ incoming
beam jitter. Energy jitter at the collimators also
amplifies the horizontal jitter at the IP. An energy jitter of 1\%
produces a horizontal emittance growth of 2.2\%.

\paragraph{Muon suppression}

%add figure showing spoiler drawing back in

\begin{figure}[htb]
\begin{center} \vbabove
%\vspace{-.05mm}
\includegraphics[width=0.95\textwidth]{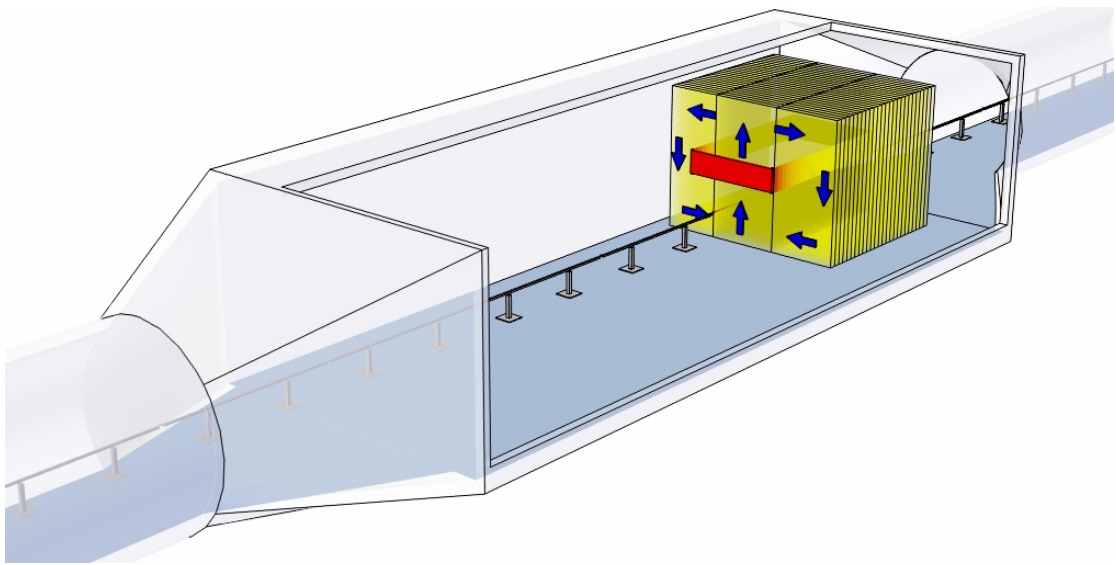}
%\vspace{-10mm}
\vbabovecaption \caption[Schematic of the 5-meter magnetized muon shield.]
{Schematic of the 5-meter magnetized muon shield
installed in a tunnel vault which is configured to accommodate possible
upgrade to 19-meter shield. The coil is shown in red, and blue arrows indicate
direction of the magnetic field in the iron.}
\label{muon_wall} \end{center} \vbbelow
%\vspace{-5mm}
\end{figure}

Electromagnetic showers created by primary beam particles in the
collimators produce penetrating muons which can easily reach the
collider hall.  The muon flux through the detector is reduced by a 5~m
long magnetized iron shield 330~m upstream of the
collision point which fills the cross-sectional area of the tunnel
and extends 0.6~m beyond the ID of the tunnel,
as shown in Figure~\ref{muon_wall}. The shield has a
magnetic field of 1.5~T, with opposite polarities in the left and
right halves of the shield such that the field at the beamline is
zero. The shield also provides radiation protection for the collider
hall during access periods when beam is present in the linac and
beam switch yard.

\stepcounter{figlcl}

\paragraph{Halo power handling}\label{sect:BDSHaloPower}

The power handling capacity of the collimation system is set by two
factors:  the ability of the collimators to absorb the incident beam
power, and the ability of the muon suppression system to reduce the
muon flux through the detector.  In the baseline design, the muon
suppression system presents the more restrictive limitation, setting
a tolerance of $1-2\times10^{-5}$ on the fraction of the beam which
is collimated in the BDS. With these losses and the 5~m wall,
the number of muons reaching
the collider hall would be a few muons per 150 bunches (a reduction of more than 10$ ^{-2} $ ).
Since the actual beam halo conditions are
somewhat uncertain, the BDS includes caverns large enough to increase
the muon shield from 5~m to 18~m and to add an additional 9~m shield
downstream.  Filling all of these caverns with magnetized muon shields
would increase the muon suppression capacity of the system to
$1\times10^{-3}$ of the beam.  The primary beam spoilers and
absorbers are water cooled and capable of absorbing $1\times10^{-3}$
of the beam continuously.

\paragraph{Tail-folding octupoles}

The final focus includes two superconducting octupole doublets.
These doublets use nonlinear focusing to reduce the amplitude of
beam halo particles while leaving the beam core untouched
\cite{octupoles}. This ``tail-folding'' would permit larger
collimation amplitudes, which in turn would dramatically reduce
the amount of beam power intercepted and the wakefields.  In the
interest of conservatism the collimation system design described
above does not take this tail folding into account in the
selection of apertures and other parameters.

\subsubsection{    Final focus }
%– Angal-Kalinin, Kuroda,Seryi
% beam size numbers need updating, also parameter table

The role of the final focus (FF) system is to demagnify the beam to
the required size ($\sim$639~nm (horz) and $\sim$5.7~nm (vert)) at the IP. The
FF optics creates a large and almost parallel beam at the entrance
to the final doublet (FD) of strong quadrupoles. Since particles of
different energies have different focal points, even a relatively
small energy spread of $\sim$0.1\% significantly dilutes the beam size,
unless adequate corrections are applied. The design of the FF is
thus mainly driven by the need to cancel the chromaticity of the FD.
The ILC FF has local chromaticity correction \cite{prlff} using
sextupoles next to the final doublets.
A bend upstream generates dispersion across the FD, which is
required for the sextupoles to cancel the chromaticity. The
dispersion at the IP is zero and the angular dispersion is about $\eta'_x{\sim}0.009$, i.e. small
enough that it does not significantly increase the beam divergence.
Half of the total horizontal chromaticity of the whole final focus
is generated upstream of the bend in order for the sextupoles to
simultaneously cancel the chromaticity and the second-order
dispersion.

\stepcounter{figlcl}\begin{figure}[htb]
%\vspace{-.05mm}
\begin{center} \vbabove
\includegraphics[width=0.995\textwidth]{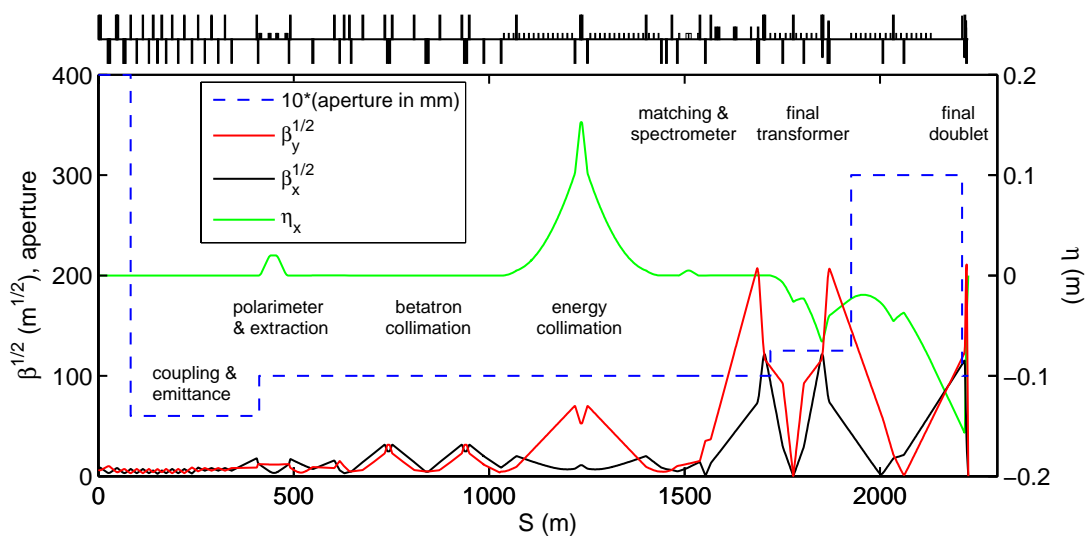}
%\vspace{-10mm}
\vbabovecaption \caption[BDS optics, subsystems and vacuum chamber aperture.]
{BDS optics, subsystems and vacuum chamber aperture; S is distance measured from the entrance. }
\label{ffoptics} \end{center} \vbbelow
%\vspace{-5mm}
\end{figure}
% def of s is reversed wrt Z-0 in fig 2.2

The horizontal and the vertical sextupoles are interleaved in this
design, so they generate third-order geometric aberrations.
Additional sextupoles upstream and in proper phases with the FD
sextupoles partially cancel the third order aberrations. The residual
higher-order aberrations are minimized further with octupoles and
decapoles. The final focus optics is shown in Figure~\ref{ffoptics}.

Synchrotron radiation from the bending magnets causes emittance
dilution, so it is important to maximize the bending radius,
especially at higher energies.  The FF includes sufficient bend
magnets for 500~GeV~CM and space for additional bend magnets which
are necessary at energies above 500~GeV~CM.  With the reserved space
filled with bends, the emittance dilution due to bends at 1~TeV~CM
is about a percent, and at 500~GeV~CM, with only every fifth bend
installed, about half of that.

In addition to the final doublet and chromaticity correction optics,
the final focus includes: an energy spectrometer (see Section
\ref{energy_meas}); additional absorbers for the small number of
halo particles which escape the collimation section; tail folding
octupoles (see Section \ref{sectBDSColl}); the crab cavities (see
Section \ref{sectCrabCav}); and additional collimators for machine
protection or synchrotron radiation masking of the detector.

\subsubsection{    IR design and integration to detector}\label{ir_design}
% – Parker et al

%\section{BDS Magnets: Final Focus}

\begin{figure}[htbp]
%\vspace{-.5mm}
\begin{center} \vbabove
\includegraphics[width=0.95\textwidth]{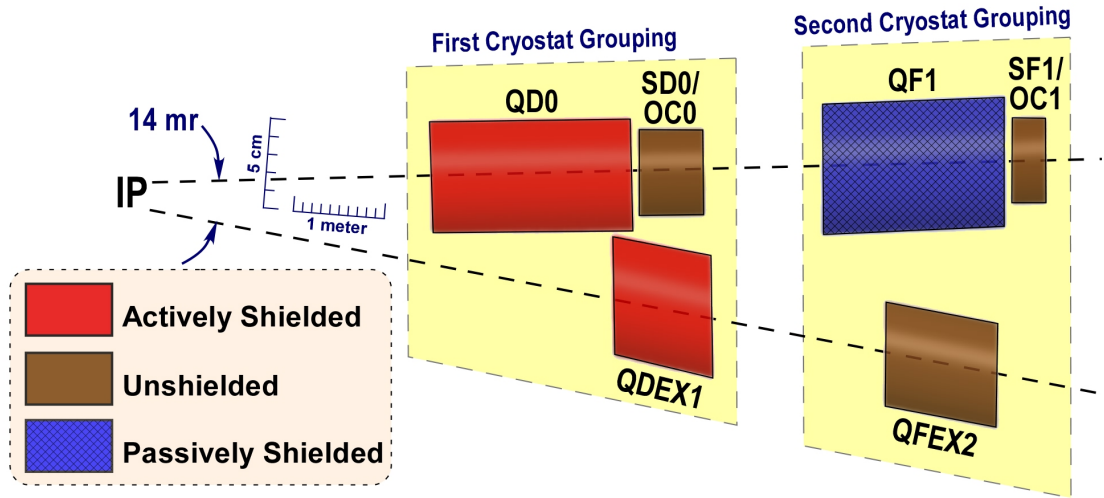}
%\vspace{-5mm}
\vbabovecaption \caption{Schematic layout of magnets in the IR.  }
\label{fd_push_pull}
\end{center} \vbbelow
%\vspace{-5mm}
\end{figure}

\begin{figure}[htbp]
%\vspace{-.5mm}
\begin{center} \vbabove
\includegraphics[width=0.99\textwidth]{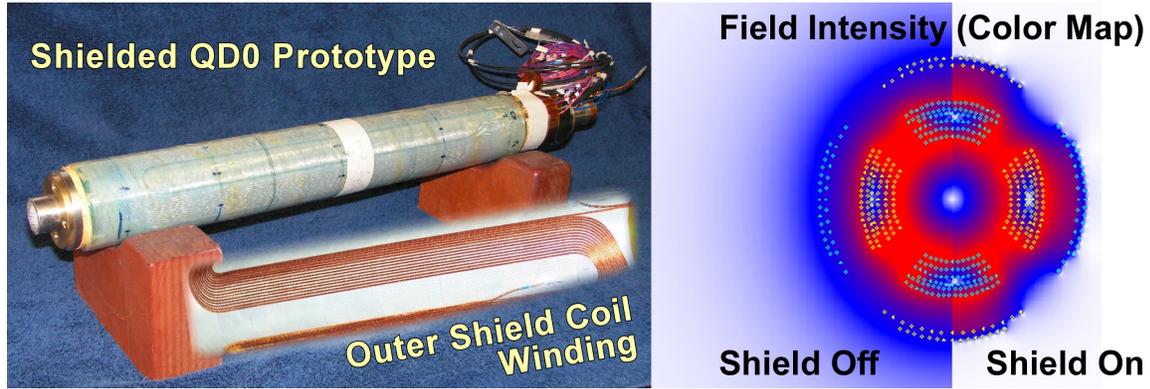}
%\vspace{-5mm}
\vbabovecaption \caption[Prototype of QD0 quadrupole and its active shield.] {Prototype of QD0 quadrupole and its active shield (left);
calculated field pattern with and without activation of the
shielding coils (right).}
\label{fd_prototype}
\end{center} \vbbelow
%\vspace{-5mm}
\end{figure}

The ILC final focus uses independently adjustable compact
superconducting magnets for the incoming and extraction beam
lines.
%\cite{ir_magnets}.
The adjustability is needed to
accommodate beam energy changes and the separate beamline allows
optics suitable for post IP beam diagnostics. The BNL direct wind
technology is used to produce closely spaced coil layers of
superconducting multi-strand cable. The design is extremely
compact and the coils are almost touching in shared cold mass
volumes.  Cooling is provided by superfluid helium at 2~K.
%, which also minimizes unwanted vibration. (explain or drop)
% add figure showing prototype coil, maybe winding techniques, BNL references
The technology has been
demonstrated by a series of short prototype multi-pole coils. The
schematic layout of magnets in the IR is shown in
Figure~\ref{fd_push_pull} and Figure~\ref{generic-ir}. The
quadrupoles closest to the IP are actually inside the detector
solenoidal field and therefore cannot have magnetic flux return
yokes; at the closest coil spacing the magnetic cross talk between
the two beam apertures is controlled by using actively shielded
coil configurations and by use of local correction coils, dipole,
skew-dipole and skew-quadrupole or skew-sextupole, as appropriate.
Figure~\ref{fd_prototype} shows the prototype of QD0 quadrupole and
illustrates the principle of active shielding.

\stepcounter{figlcl}

To facilitate a rapid, ``push-pull'' style exchange of detectors at
a shared IP, the superconducting final focus magnets are arranged
into two groups so that they can be housed in two separate cryostats
as shown in Figure \ref{fd_push_pull}, with only warm components and
vacuum valves placed in between.  The cryostat on the left in Figure
\ref{fd_push_pull} moves with the detector during switchover, while
the cryostat on the right remains fixed on the beamline.

\stepcounter{figlcl}

Additional optical elements are required in the IR to compensate the
effects of the detector solenoid field interacting with the
accelerator IR magnets.  The first is a large aperture anti-solenoid
in the endcap region to avoid luminosity loss due to beam optics
effects \cite{asol-prstab}. The second is a large aperture Detector
Integrated Dipole (DID) \cite{did-prstab} that is used to reduce
detector background at high beam energies or to minimize orbit
deflections at low beam energies.

The vertical position of the incoming beam line quadrupole field
center must be stable to order of a few tens of nanometers, in order
to stay within the capture range of the intra-train collision
feedback (see \ref{sectBDSfeedback}). This requirement is well
beyond experience at existing accelerators and is being addressed in
ongoing R\&D.

\subsubsection{    Extraction line}
% – Nosochkov et al

The ILC extraction line~\cite{extr-optics,extr-study2} has to
transport the beams from the IP to the dump with acceptable beam
losses, while providing dedicated optics for beam diagnostics. After
collision, the beam has a large angular divergence and a huge energy
spread with very low energy tails.  It is also accompanied by a high
power beamstrahlung photon beam and other secondary particles.  The
extraction line must therefore have a very large geometric and
energy acceptance to minimize beam loss.

\begin{figure}[htbp]
%\vspace{-.05mm}
\vbabove \centering
\includegraphics[width=0.9\textwidth, angle=0]{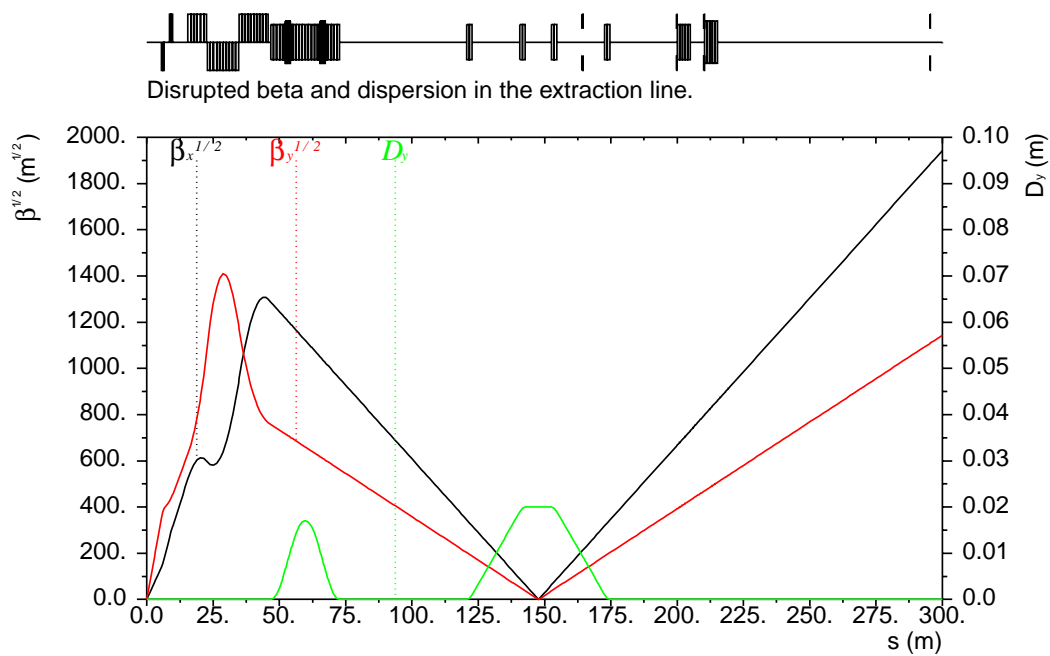}
%\vspace{-5mm}
\vbabovecaption \caption{Disrupted $\beta$-functions and dispersion in
the extraction line for the nominal 250 GeV beam.}
\label{extr-beta} \vbbelow
%\vspace{-3mm}
\end{figure}

The optics of the ILC extraction line is shown in Figure
\ref{extr-beta}.  The extraction line can transport particles with
momentum offsets of up to 60\% to the dump.  There is no net bending
in the extraction line, which allows the charged particle dump to
also act as a dump for beamstrahlung photons with angles of up to
0.75 mrad.

\stepcounter{figlcl}

The first quadrupole is a superconducting magnet 5.5 m from the IP,
as shown in Figure \ref{fd_push_pull}.  The second quadrupole is also
superconducting, with a warm section between
the cryostats for these two quadrupoles. The downstream
magnets are normal conducting, with a drift space to accommodate the
crab cavity in the adjacent beamline. The
quadrupoles are followed by two diagnostic vertical chicanes for the
energy spectrometer and Compton polarimeter, with a secondary focal
point in the center of the latter.
The horizontal angular amplification ($R_{22}$) from the IP to the
Compton IP is set to -0.5 so that the measured Compton polarization
is close to the luminosity weighted polarization at the IP.
% add ref to explain why -0.5
The lowest energy particles are
removed by a vertical collimator in the middle of the energy
chicane.  A large chromatic acceptance is achieved through the soft
D-F-D-F quadruplet system and careful optimization of the quadrupole
strengths and apertures. The two SC quadrupoles are compatible with
up to 250~GeV beam energy, and the warm quadrupoles and the chicane
bends with up to 500~GeV beam.

The diagnostic section is followed by a 100~m long drift to allow
adequate transverse separation ($>$3.5~m) between the dump and the
incoming line. It also allows the beam size to expand enough to
protect the dump window from the small undisrupted beam. A set of
rastering kickers sweep the beam in a 3~cm circle on the window to
avoid boiling the water in the dump vessel. Three protection
collimators in the 100~m drift remove particles that would hit
outside of the 15~cm radius dump window and protect the rastering
kicker magnets.

\stepcounter{figlcl}\begin{figure}[htbp]
%\vspace{-.05mm}
\vbabove
\centering
\includegraphics[width=0.75\textwidth, angle=0]{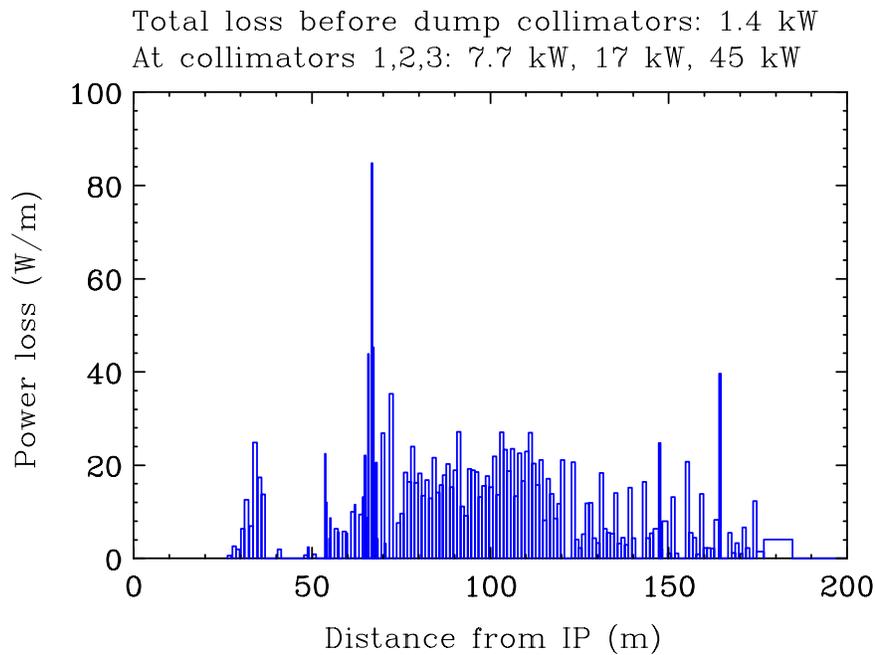}
%\vspace{-5mm}
\vbabovecaption \caption[Power loss density in the magnet region for
disrupted beam at 250 GeV] {Power loss density in the magnet region for
disrupted beam at 250 GeV, with an extreme choice of parameters.}
\label{extr-losses} \vbbelow
%\vspace{-3mm}
\end{figure}

Extraction beam loss has been simulated for realistic 250 GeV
GUINEA-PIG beam distributions \cite{gp}, with and without beam
offset at the IP. No primary particles are lost in the SC
quadrupoles, and all particles above 40\% of the nominal beam
energy are transmitted cleanly through the extraction magnets. The
total primary loss on the warm quadrupoles and bends is a few
watts, and the loss on the protection collimators is a few kW for
the nominal beam parameters. Figure~\ref{extr-losses} shows that
even for an extreme set of
parameters, with very high beamstrahllung energy loss, the radiation deposition in the magnet region is manageable.

\subsection{Accelerator Components}

The BDS accelerator components are described in the following
sections and the total counts are shown in
Table~\ref{bds_components}.

\subsubsection{   Crab cavity system }\label{sectCrabCav}
% – McIntosh, Bellantony, Li, et al

\begin{figure}[h]
%\vspace{-.5mm}
\vbabove
\centering
\includegraphics*[width=0.9\textwidth]{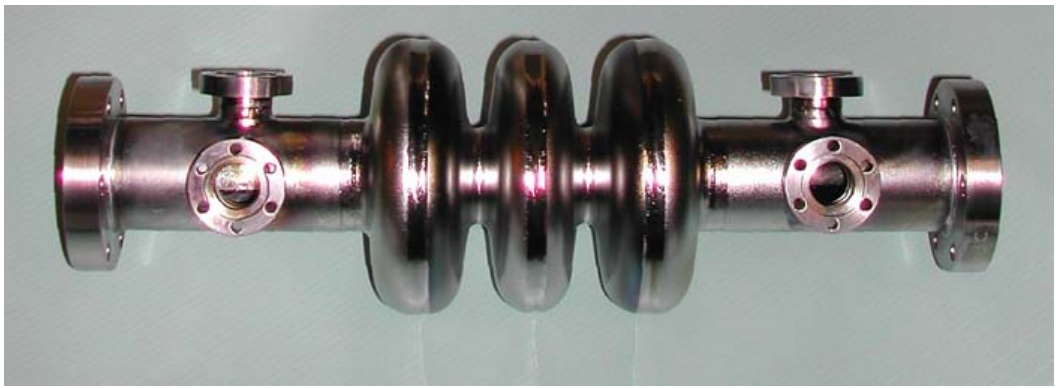}
%\vspace{-5mm}
\vbabovecaption \caption{Photo of a 3.9GHz 3-cell deflecting cavity
built at Fermilab, which achieved 7.5MV/m.}
\label{ccav_3cell} \vbbelow
%\vspace{-.5mm}
\end{figure}

With a 14~mrad crossing angle, crab cavities are required to rotate
the bunches so they collide head on. Two 3.9~GHz SC 9-cell cavities
in a 2--3~m long cryomodule are located 13.4~m from the IP. The
cavities are based on the Fermilab design for a 3.9~GHz TM$_{110}$
$\pi$ mode 13-cell cavity \cite{cc-ref4}. The three cell prototype
of this cavity is shown in Figure~\ref{ccav_3cell}. The ILC has two
9-cell versions of this design operated at 5~MV/m peak deflection.
This provides enough rotation for a 500~GeV beam and 100\%
redundancy for a 250~GeV beam.

\stepcounter{figlcl}

The most challenging specification of the crab cavity system is on
the uncorrelated phase jitter between the incoming positron and
electron cavities which must be controlled to 61~fsec to maintain
optimized collisions \cite{cc-ref5}. A proof-of-principle test of a
7 cell 1.5~GHz cavity at the JLab ERL facility \cite{cc-ref6} has
achieved a 37~fsec level of control, demonstrating feasibility.  The
higher- and lower-order modes of the cavity must be damped
effectively to limit unwanted vertical deflections at the IP, as
must the vertical polarization of the main deflecting mode.

Couplers with lower $Q_{ext}$ and greater power handling capability
are required to handle beam loading and LLRF feedback for off-axis
beam. The crab cavity needs $\sim$3~kW per cavity for about 10~msec,
with a $Q_{ext}$ of $\sim10^6$ \cite{cc-ref7,cc-new07}. The crab cavity is placed in a
cryostat with tuner, x-y and roll adjustment which provides proper
mechanical stability and microphonic rejection. The cryostat also
accommodates the beampipe of the extraction line which passes about
19~cm from the center of the cavity axis.

\subsubsection{   Feedback systems and Stability
}\label{sectBDSfeedback}
% – Burrows, White et al

%Dec 12 version

Maintaining the stability of the BDS is an essential prerequisite to
producing luminosity.  Since the beams have RMS vertical sizes of
5.7~nm at the IP, vertical offsets of about 1~nm will
noticeably reduce the luminosity.  In addition, especially for parameter sets with higher disruption, the beam-beam
interaction is so strong that the luminosity is extremely sensitive
to small variations in the longitudinal shape of the bunch caused by
short-range wakefields.  Finally, the size of the beam at the IP is
sensitive to the orbit of the beam through the final doublet quads,
the sextupoles, and other strong optical elements of the BDS.
Care must be taken to minimize thermal and mechanical disturbances,
by stabilizing the air temperature to 0.5$^\circ$C and the cooling water
to 0.1$^\circ$C, and by limiting high frequency vibrations due to local
equipment to the order of 10~nm.

Beam-based orbit feedback loops are used to maintain the size and
position of the beam at the IP.  All of the feedback loops use beam
position monitors with at least micron-level (and in some cases sub-micron) resolution to detect the
beam position, and dipole magnets or stripline kickers to deflect
the beam.  There are two basic forms of feedback in the BDS:
train-by-train feedbacks, which operate at the 5~Hz repetition rate
of the ILC, and intra-train feedbacks, which can apply a correction
to the beam between bunches of a single train.

\paragraph{Train-by-train feedback}

A train-by-train feedback with 5 correctors controls the orbit
through the sextupoles in the horizontal and vertical planes, where
the optical tolerances are tightest.  Additional correctors
throughout the BDS help reduce long-term beam size growth. The orbit
control feedback can maintain the required beam sizes at the IP over
periods from a few hours to several days depending on details of the
environment.  On longer timescales, IP dispersion and coupling knobs
need to be applied.

\paragraph{Intra-Train IP position and angle feedbacks}

The intra-train feedbacks use the signals detected on early bunches
in the train to correct the IP position and angle of subsequent
bunches.  The offset of the beams at the IP is determined by
measuring the deflections from the beam-beam interaction; this
interaction is so strong that nm-level offsets generate deflections
of tens of microradians, and thus BPMs with micron-level resolution
can be used to detect offsets at the level of a fraction of a
nanometer. Corrections are applied with a stripline kicker located
in the incoming beamline between SD0 and QF1.
The angle of the beams at the IP is determined by
measuring the beam positions at locations 90$^{\circ}$ out of phase
with the IP; at these locations the beam is relatively large so
micron resolution is sufficient to directly measure the beam
position (and hence the IP angle) to a small fraction of its RMS
size.  A stripline kicker is located at the entrance to the FF
causing a latency of about 4 bunch spacings.

The position feedback BPM is located near the IP in a region where
electromagnetic backgrounds or particle debris from the collisions
are a concern. Preliminary results from simulations and from a
test-beam experiment indicate that backgrounds are an order of
magnitude too small to cause a problem \cite{feedback-ref3}.

\begin{comment}
The sources of instability relevant to the beam position at the IP
have characteristic frequencies of Hz to kHz, not MHz.  Thus, the
intra-train feedback is orders of magnitude faster than the
instabilities and will quickly converge on stable solutions.
\end{comment}

\paragraph{Luminosity feedback}

Because the luminosity may be extremely sensitive to bunch shape, the
maximum luminosity may be achieved when the beams are slightly
offset from one another vertically, or with a slight nonzero
beam-beam deflection.  After the IP position and angle feedbacks
have converged, the luminosity feedback varies the position and
angle of one beam with respect to the other in small steps to
maximize the measured luminosity.

\paragraph{BDS Entrance Feedback ('train-straightener')}

A bunch-to-bunch correction at the end of the Linac removes
systematic transverse position offsets within the train due to
long-range wakefield kicks in the accelerating cavities. This system
consists of two kicker-BPM systems similar to those described above.
Each pair operates at a different phase to null the orbit in both
vertical degrees of freedom.
%This system may also {\it optionally}
%be used to remove pulse-to-pulse jitter caused by vibration of
%linac components.

For stripline kickers the maximum correction would be about
8--10~$\mu$m, and the BPM resolution requirements are about
200~nm.
This requires cavity BPMs that are read out
in bunch-bunch mode and processed with low-latency electronics.
The kicker-BPM separations imply
latencies of about 400~ns,
allowing feedback on every-other bunch.

\paragraph{Hardware Implementation for intra-train feedbacks}

High bandwidth, low-latency ($\sim$5~ns) signal processors for
stripline and button BPMs have been tested at the NLCTA and ATF
\cite{feedback-ref1}. The feedback processor has been prototyped
using fast state of the art FPGAs; a system prototype has been
demonstrated with a FB board latency of $\sim$70~ns
\cite{feedback-ref2}. Commercial boards that meet the latency
requirement are not available without custom firmware modification;
one such board has been tested by the FONT group and would meet the
ILC latency specification for bunch-bunch operation.

\subsubsection{Energy, Luminosity and polarization measurements}

\paragraph{Energy measurements}\label{energy_meas}

Absolute beam energy measurements are required by the ILC physics
program to set the energy scale for particle masses. An absolute
accuracy better than 200~ppm is required for the center-of-mass energy,
which implies a
requirement of 100~ppm on determination of the absolute beam energy.
The intra-train relative variation in bunch energies must be measured with a
comparable resolution. Measurements of the disrupted energy spectrum
downstream of the IP are also useful to provide direct information
about the collision process.

To achieve these requirements, there are two independent and
complementary detectors for each beam. Upstream from the IP, a
spectrometer based on the LEP-II energy spectrometer is capable of
making high-precision bunch-to-bunch relative measurements in
addition to measuring the absolute beam energy scale. A four-magnet
chicane in the instrumentation region provides a point of dispersion
which can be measured using triplets of high-precision RF BPMs.
The maximum displacement of the beam is a few millimeters
and must be measured to a precision below 100 nanometers.
Precision movers keep the beam nearly centered in the BPMs in
order to achieve this accuracy.
% ref for LEP spectrometer? and for sync stripe below

Downstream from the IP, there is a synchrotron radiation
spectrometer.  A three-magnet chicane in the extraction line
provides the necessary beam deflection, while the trajectory of the
beam in the chicane is measured using synchrotron radiation produced
in wiggler magnets imaged $\sim$70 meters downstream at a secondary
focus near the polarimeter chicane.

\paragraph{Luminosity measurements}

The ILC luminosity can be measured with a precision of $10^{-3}$ or
better by measuring the Bhabha rate in the polar-angle region from
30-90~mrad. Two detectors are located just in front of the final doublets
as shown in Figure~\ref{generic-ir}.
The {\it LumiCal} covers the range from 30-90~mrad
and the {\it BeamCal} covers the range from 5-30~mrad.
At 500~GeV center-of-mass energy, the expected rate in
the {\it LumiCal} region is $\sim$10 Bhabhas per bunch train, which is too low to
permit its use as an intra-train diagnostic for tuning and feedback.
At smaller polar angles of 5-30~mrad the rate or energy deposition
of beamstrahlung e+e- pairs can be measured for a fast luminosity
diagnostic. The expected rate in this region is 15,000 pairs (and
50~TeV energy deposition) per bunch crossing.  Furthermore, the
spatial distributions of pairs in this region can be used to
determine beam collision parameters such as transverse sizes and
bunch lengths.

\paragraph{Polarization measurements}

%Section updated December 14, 2006. Primary authors responsible for
%this section: K.C. Moffeit, K.P. Schuler, M. Woods

\begin{figure}[htbp]
%\vspace{-.05mm}
\begin{center} \vbabove
\includegraphics[width=\textwidth]{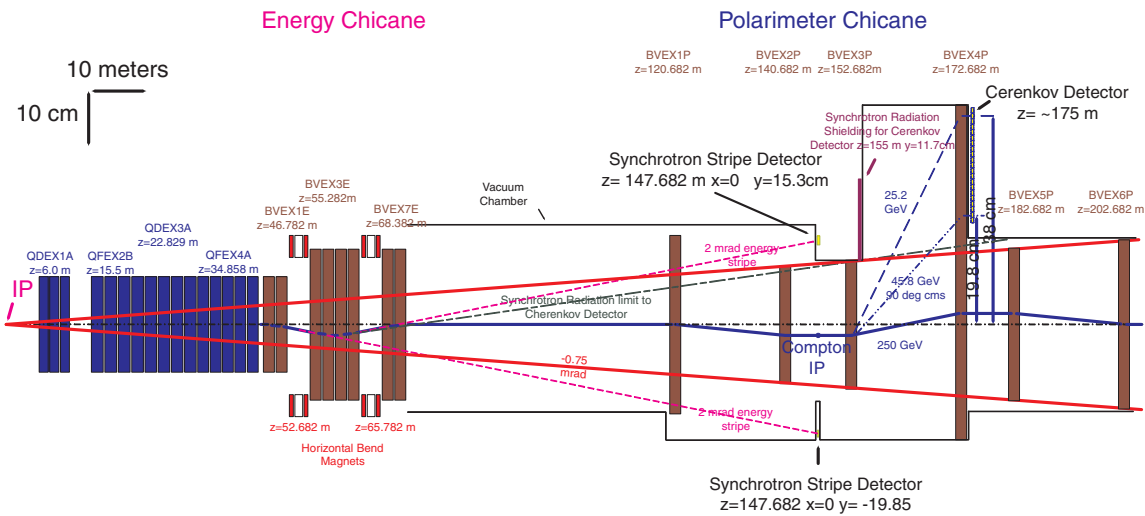}
%\vspace{-5mm}
\vbabovecaption \caption[Schematics of energy and polarimeter
chicanes in the 14~mrad extraction line.]{Schematics of energy and polarimeter
chicanes in the 14~mrad extraction line, shown in a
configuration with two additional bends at the end. Longitudinal
distances are given from the IP. Also shown is the 0.75~mrad beam
stay-clear from the IP.}
\label{downstream_diag} \end{center} \vbbelow
%\vspace{-5mm}
\end{figure}

Precise polarimetry with 0.25\% accuracy is needed to achieve the
ILC physics goals %\cite{Moortgat-Pick-hep-ph0507011}.
Compton polarimeters \cite{Moffeit-SLACPUB11322,Meyners-LCWS05}
are located both $\sim$1800~m upstream of the IP, as shown in
Figure~\ref{bds_functions}, and downstream of the IP, as shown in
Figure~\ref{downstream_diag}, to achieve the best accuracy for
polarimetry and to aid in the alignment of the spin vector.

% why is this an optional, not standard configuration – removed word optional

\stepcounter{figlcl}

The upstream polarimeter measures the undisturbed beam before
collisions. The relatively clean environment allows a laser system
that measures every single bunch in the train and a large lever arm
in analyzing power for a multi-channel polarimeter, which
facilitates internal systematic checks. The good field region of
the individual dipoles is wide enough to accommodate the lowest
expected beam energy of 45.6~GeV. The downstream polarimeter
measures the polarization of the outgoing beam after collision. The
estimated average depolarization for colliding beams is 0.3\%, and
for the outgoing beam 1\%. A schematic drawing of the extraction line is
shown in Figure~\ref{downstream_diag}.

Each polarimeter has a dedicated 4-bend chicane to facilitate
injection of the laser light and extraction of the Compton signal.
The upstream polarimeter uses a horizontal chicane to minimize
emittance growth from synchrotron radiation, while the downstream
polarimeter uses a vertical chicane to maximize analyzing power. The
systems are designed to meet the physics requirements at all
energies from the Z pole to the full energy of the ILC.

\subsubsection{Beam dumps and Collimators}\label{sect:BDSdump}
% – Ban, Markiewicz, Densham, et al

%Dec 12 version

The beam delivery system contains two tune-up dumps and two main
beam dumps.  These four dumps are all designed for a peak beam power
at nominal parameters of 17~MW at 500~GeV per beam. These dumps consist of
1.5~m diameter cylindrical stainless steel high pressure (10 bar)
water vessels with a 30~cm diameter 1~mm thick Ti window; and also
include their shielding and associated water systems.

The dumps absorb the energy of the electromagnetic shower cascade
in 6.5~m (18~$X_0$) of water followed by 1~m of water cooled Cu
plates (22~$X_0$).  Each dump incorporates a beam sweeping magnet
system to move the charged beam spot in a circular arc of 3~cm
radius during the passage of the 1~ms long bunch train.  Each dump
operates at 10 bar pressure and also incorporates a vortex-flow
system to keep the water moving across the beam at 1.0-1.5~m/s. In
normal operation with 250~GeV beam energy, the combination of the
water velocity and the beam sweepers limits the water temperature
rise during a bunch train to 40$^\circ$C.  The pressurization
raises the boiling temperature of the dump water; in the event of
a failure of the sweeper, the dump can absorb up to 250 bunches
without boiling the dump water.

The integrity of the dump window, the processing of the
radiolytically evolved hydrogen and oxygen, and containment of the
activated water are important issues for the full power dumps.  The
dump service caverns include three loop pump driven 2300 gallon per
minute heat exchanger systems, devices to remotely exchange dump
windows as periodic maintenance, catalytic H$_2$-O$_2$ recombiners,
mixed bed ion exchange columns for filtering of $^7$Be, and
sufficient storage to house the volume of tritiated water during
maintenance operations.

In addition to the main dumps, the BDS contains 16 stoppers, of
which 14 are equipped with burn-through monitors, and the
extraction lines have 6 fixed aperture high power devices composed
of 10~mm aluminum balls immersed in water. The beam delivery
system contains 32 variable aperture collimators and 32 fixed
aperture collimators.  The devices with the smallest apertures are
the 12 adjustable spoilers in the collimation system. To limit
their impedance to acceptable levels, these 0.6-1.0~$X_0$ Ti
spoilers have longitudinal Be tapers.

% component counts do not match text

\stepcounter{tablcl}\begin{table}[htb]
  \centering \vbabove
  \caption{BDS components, total counts.}\label{bds_components}
% adjust "Xpt" to squeeze table into width of text on the page
\setlength{\tabcolsep}{4pt}
\begin{tabular}{| l | c || l | c || l | c |}
\hline \multicolumn{2}{|c||}{ Magnets  } &
            \multicolumn{2}{|c||}{ Instrumentation} &
            \multicolumn{2}{|c|}{ Dumps} \\ [-6pt]
\multicolumn{2}{|c||}{} &
            \multicolumn{2}{|c||}{} &
            \multicolumn{2}{|c|}{ \& Collimators} \\   \hline & & & & & \vbdlspacing \hline
%\hline
 Warm dipoles & 190           & BPMs C-band & 262                 & Full power dumps & 4  \\   \hline
 Warm quads   & 204           & BPMs L-band & 42                  & Insertable dumps & 2  \\   \hline
 Warm sextupoles & 10         & BPMs S-band & 14                  & Adjustable collim. & 32  \\   \hline
 Warm octupoles  & 4          & BPMs stripline/button  & 120      & Fixed apert. collim. & 32  \\   \hline
 SC quads & 32                & Laser wire & 8                    & Stoppers & 14  \\   \hline
 SC sextupoles & 12            & SR transv. profile imager & 10 &        &    \\   \hline
 SC octupoles  & 14            & OTR screens & 2                   & \multicolumn{2}{|c|}{ Vacuum  }   \\   \hline
 Muon spoilers & 2        & Crab \& deflection cavities & 4               & Pumps & 3150  \\   \hline
 Anti-solenoid & 4      & Loss monit. (ion chamb., PMT) & 110 & Gauges & 28  \\   \hline
 Warm correctors & 64        & Current monitors & 10              & Gate valves & 30  \\   \hline
 SC correctors & 36        & Pick-up phase monitors & 2           & T-connections & 10  \\   \hline
 Kickers/septa & 64        & Polarimeter lasers & 3               & Switches & 30  \\
\hline
\end{tabular} \vbbelow
\end{table}

\subsubsection{  BDS Magnets}
% – Spencer, Tompkins, Sugahara, et al

The BDS has a wide variety of different magnet requirements, and the
most distinct magnet styles (67) of any ILC area, even though there
are only 636 magnets in total. Of these, 86 are superconducting
magnets clustered into 4 cryostats close to the IP, as described in
section \ref{ir_design}, and the tail-folding octupoles described
below. There are 64 pulsed magnets: 5 styles of abort kickers,
sweepers and septa. These are used to extract the beams to a fast
extraction/tuning dump and to sweep the extracted beam in a 3~cm
circle on a dump window.

The remaining 474 magnets are conventional room temperature magnets,
mostly with water-cooled hollow copper conductor coils and low
carbon steel cores. The bend magnets in the final focus have fields
of less than 0.5~kG to minimize synchrotron radiation that would
cause beam dilution; they use solid wire coils. The quadrupoles
and sextupoles have straightforward designs adequate for up
to 500~GeV beam. The extraction line magnets have large
apertures, e.g. over 90~mm and up to 272~mm, to accommodate the
disrupted beam and the photons emerging from the IP. These magnets
must fit in alongside the incoming
beamline.

The main technical issue with the BDS magnets is their
positional stability. All the incoming beamline quadrupoles and
sextupoles sit on 5 degree of freedom magnet movers with a 50~nm smallest
step size. BPMs  inserted in the
magnet bores provide data on the relative position of each
magnet with respect to the beam so that it can be moved if necessary. The absolute field
strength of the BDS magnets has a tight tolerance, requiring
power supplies with stability of a few tens of ppm. Magnet temperature changes lead to strength and position variations so the ambient temperature in the
tunnel must be controlled to within about 0.5$^\circ$C and the
cooling water to within 0.1$^\circ$C.

\paragraph{BDS Magnets: Tail Folding Octupoles}

The tail folding octupoles are the only superconducting
magnets in the BDS (other than the FD and extraction quadrupoles) and
have the smallest, 14~mm ID, clear working aperture in order to reach the highest practical operating
gradient. The magnets are energized via NbTi conductor cooled to
4.5~K. With such a small aperture, the beam
pipe must have high conductivity to minimize the impact of
wakefields. This can be achieved with a cold
aluminum beam pipe at 4.5~K or a cold stainless steel beam pipe
with a high conductivity coating. Because these magnets are
isolated in the BDS, being far from either the IP or the end of
the linac, cryocoolers are used to provide standalone cooling.

\subsubsection{Vacuum system}
% – Suetsugu, Bane, et al – 0.5p

While the aperture of the BDS vacuum chamber is defined by the
sizes of the beam, its halo and other constraints, the design
of the chambers and vacuum level are governed mainly by two
effects: resistive and geometric wakes and the need to preserve
the beam emittance; beam-gas scattering and minimization of
detector background.

\paragraph{Wakes in vacuum system}

The resistive wall (RW) wakefield of the BDS vacuum system and the
geometric wakefield of the transitions in the beam pipe may cause
emittance growth due to incoming (transverse) jitter or drift, or
due to beam pipe misalignment.  In order to limit these effects to
tolerable levels, the BDS vacuum chamber must be coated with copper,
the vacuum chambers must be aligned with an RMS accuracy of
$\sim$100~$\mu$m \cite{bane_vacuum}, and incoming beam jitter must
be limited to 0.5 $\sigma_y$ train-to-train and 0.25 $\sigma_y$
within a train, to limit the emittance growth to 1-2\%.

\paragraph{Beam-gas scattering}

%Dec 12 version

The specification for the pressure in the BDS beam pipe is driven by
detector background tolerance to beam-gas scattering.  Studies have
shown that electrons which are scattered within 200~m of the IP
can strike the beam pipe within the detector and produce intolerable
backgrounds, while electrons which scatter in the
region from 200 to 800~m from the IP are much more likely to
hit the protection collimator upstream of the final doublet and
produce far less severe detector backgrounds \cite{bg-3}.  Based on
these studies, the vacuum in the BDS is specified to be 1~nTorr
within 200~m of the IP, 10~nTorr from 200~m to 800~m from the IP,
and 50~nTorr more than 800~m from the IP.

In the extraction lines the pressure is determined by beam-gas
scattering backgrounds in the Compton Polarimeter located about
200~m from the IP.
Here the signal rates are large enough that 50~nTorr would
contribute a negligible background in the detectors.

\paragraph{Vacuum system design}

The BDS vacuum is a standard UHV system. The main beampipes are
stainless steel, copper coated to reduce the impedance, with the
{\it option} of an aluminum alloy chamber. In locations where there
is high synchrotron radiation (SR) power ($\geq$10~kW/m) (e.g. in
the chicanes or septa regions), the beampipe is copper with a
water-cooled mask to intercept the photons. The beampipes are
cleaned and baked before installation. There is no {\it in situ}
baking required
 except possibly for the long drift before the IP.

The required maximum pressure of 50~nTorr (N$_2$/CO equivalent) can
be achieved by standard ion pumps located at appropriate intervals.
The beampipe near the IP must have pressure below 1~nTorr for
background suppression, and may be baked {\it in situ} or
NEG-coated.

% comment from NP – way too much detail for a standard system

\subsubsection{IR arrangements for two detectors}
% – Angal-Kalinin, Seryi, Yamamoto

There are two detectors in a common IR hall which alternately occupy
the single IR, in a so-called ``push-pull'' configuration. The
detector hall is 120~m (long) $\times$ 25~m (wide) $\times$ 38~m (high). The
layout of the hall is compatible with surface assembly of the
detectors.
The previous layout with two 14~mrad IRs is kept as an {\it alternative
configuration}, and is about 50\% more expensive than the single IR.

\stepcounter{figlcl}\begin{figure}[htbp]
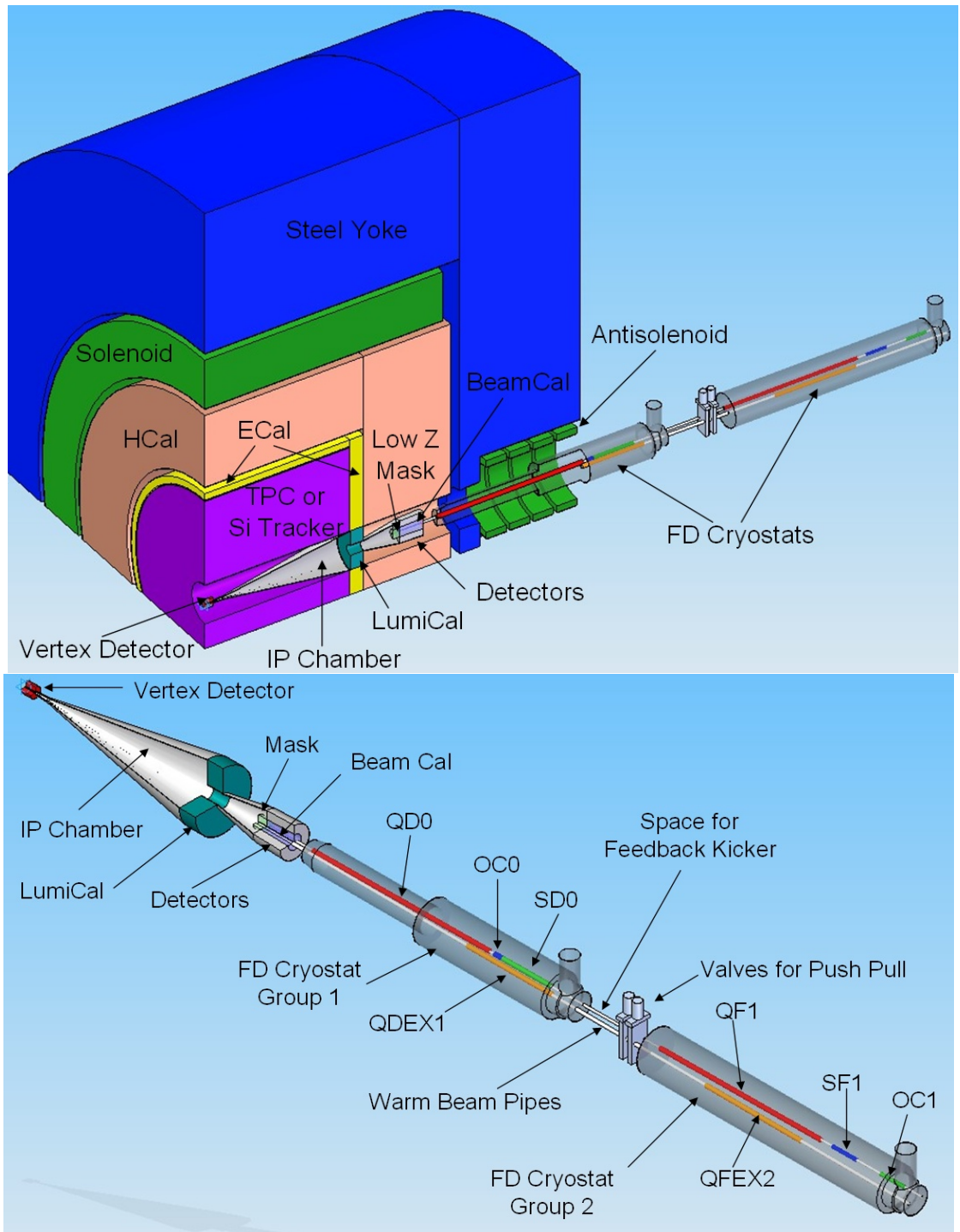

%\vspace{-.05mm}
\vbabove \centering
\includegraphics[width=0.95\textwidth, angle=0]{\picturefolder gen_detector_2}
\includegraphics[width=0.95\textwidth, angle=0]{\picturefolder gen_ir_2}
%\vspace{-5mm}
\vbabovecaption \caption[Generic detector and IR arrangements.]{Generic detector and IR arrangements, showing the location of beamline elements near the IR and their integration with the detector. }
\label{generic-ir}
\vbbelow
%\vspace{-3mm}
\end{figure}

To facilitate the exchange of detectors, there is a breakpoint in
the beam line near the edge of the detector, between the two final
doublet cryostat halves as shown in Figure~\ref{fd_push_pull} and Figure~\ref{generic-ir}. A
necessary condition for efficient push-pull operation is to avoid
disconnecting any of the systems for the detectors during the exchange. One
possible solution is to equip each detector with an adjacent
services platform which moves together with the detector. The
platforms would house the cryogenic systems, high current power
supplies for solenoids and FD, and detector electronics. All the
connections between the platform and detector would be fixed and not
disconnected during the exchange. The movable detector service
platform would have flexible connections to fixed services
(including high voltage AC, room temperature high pressure He supply
and return, data I/O, etc.), that do not need to be disconnected
during the exchange.

The FD alignment and support system is designed to
be compatible with rapid exchange, in particular, an
interferometer network between the two parts of the FD and the
walls may be needed. The push-pull arrangement of two detectors
implies specific requirements for the radiation safety design of the
detector and of the collider hall shielding. Since the off-beam axis
detector needs to be accessible during beam operation, the detectors
either need to be self shielded or there must be a
shielding wall between them.

Several technical solutions for moving the detectors are under
consideration, including rails, Hilman Rollers or air pads. A
guiding mechanism is needed to
determine the path for the detector motion and its accurate
positioning. The motion of a heavy detector (up to 14 kton) in the collider
hall produces deformations, which are estimated to be less
than a millimeter \cite{ja_deform}. The detector support system
must ensure that those deformations, as well as
possible deformations during lifting, do not affect its internal
alignment. To minimize deformations, the detector may require a support platform.
The 5~cm thick steel plates covering most of the
experimental hall area also facilitate stability and allow the use of
air-pads.

\subsubsection{Diagnostic and Correction devices}

Each quadrupole, sextupole, and octupole magnet in the incoming BDS
beamlines is placed on an x/y/roll/pitch/yaw mover, and has an associated BPM.
There are also several tens of correctors in the incoming
beamlines for 5~Hz feedback, and in the extraction lines,
where there are no movers. The BPMs in the incoming beamline are
RF-cavities, either S, C or L-band, depending on the beamline
aperture. Long chains of bends or kickers have sparsely placed BPMs.
BPMs in the extraction lines are button or strip-line design.

Additional instrumentation in the BDS includes a deflecting cavity to
measure Y-T correlation, ion chamber and PMT loss monitors,
X-synchrotron light transverse profile monitors, OTR monitors,
current monitors, pickup phase monitors, etc.

\clearpage %\setcounter{page}{1}
\setcounter{figlcl}{0}
\setcounter{tablcl}{0}
\setcounter{secnumdepth}{4} \setcounter{subsection}{0}
\setcounter{subsubsection}{0}
\usecounter{subsubsection}

\setcounter{section}{7} \renewcommand{\picturefolder}{./BeamDynamics/}

\section{Emittance Preservation and Luminosity Stabilization
% in the RTML, Main Linac and BDS
} \label{sectBeamDynamics}

\subsection{Overview}
The luminosity performance of the ILC will be affected by many issues
ranging from space charge effects at the electron gun to instabilities
in the damping rings to timing errors at the IP.  This section addresses
issues associated with the emittance preservation from the damping ring
extraction to the IP which is referred to as the Low Emittance Transport (LET).
Other accelerator physics issues are addressed in the respective subsystem
descriptions.

Static and dynamic imperfections in the LET impact the luminosity performance;
examples are the survey errors of beamline components and ground motion. Preserving the ultra-small emittances requires component alignment tolerances far beyond that which can be achieved by traditional mechanical and optical alignment techniques, hence the use of beam-based alignment and tuning techniques are essential in obtaining the design luminosity. The corresponding sensitivity to ground motion and vibration mandates the use of continuous trajectory correction feedback systems in maintaining that luminosity.
The accelerator physics group must develop the necessary procedures, specify
the required hardware and assess the potential luminosity degradations.

Estimation of the luminosity performance relies on complex simulations. Experimental verification of the
predictions of the codes is difficult, since the very small emittances are
not readily available in test facilities. Nonetheless, several of the fundamental aspects of the algorithms have been successfully tested.
Many of the emittance transport concepts were demonstrated and benchmarked
in the Stanford Linear Collider (SLC) which operated from 1987 to 1998. Beam-based alignment has been demonstrated in
SLC~\cite{c:dfs_slc}, LEP~\cite{c:dfs_lep}, and the Final Focus Test Beam~\cite{c:bba_fftb} --- a first test of the final focus which demonstrated the demagnification required for the ILC, and achieved a final spot size of 50~nm. The SLC operated with tolerances that are very similar to those required
for the ILC. The Accelerator Test Facility (ATF) Damping Ring at KEK is a low-emittance test facility that addresses some of the emittance issues, and is being extended into a Beam Delivery System test facility (ATF2). While not a full-scale test of the ILC damping ring and beam delivery system, they can test a number of aspects of low-emittance generation and preservation.

The simulation tools in use have been developed and refined over many years. Extensive studies for a superconducting linear
collider were performed for the TESLA TDR~\cite{tdr} and the ILC
Technical Review Committee~\cite{bib:ACCiltrc}, many of which are quite applicable to the present ILC and provide confidence in the design concepts.
In addition, extensive studies have been made for linear collider designs based on normal conducting acceleration at X-band (JLC/NLC) and K-band (CLIC),
which were designed to operate under more stringent beam dynamics regimes.

For the aspects of the machine performance that cannot be tested
experimentally before construction, simulations provide the only
tool. Fully integrated and realistic simulations are needed to study both the static (peak luminosity) and dynamic (integrated luminosity) behavior of the machine.

The performance of the ILC has been simulated for a variety of errors and
procedures. Design performance was achieved in essentially all of these
studies. Although the studies are not yet complete, they are not expected
to uncover major obstacles that would prevent the ILC from reaching
design performance.

\subsection{Sources of Luminosity Degradation}

The performance of the real machine
is rapidly degraded by errors in both component alignment and field quality. For example, misaligned magnets result in beam trajectory errors which cause emittance growth via chromatic effects (dispersion) or impedance effects (wakefields). The primary sources of emittance degradation considered are:

\begin{itemize}
\item Dispersion: The anomalous kicks from misaligned quadrupoles, coupled with the non-zero energy spread of the beam, cause dispersive emittance growth. \itemspace
\item X-Y Coupling: Rotated quadrupoles and vertically misaligned sextupoles (for
example) couple some fraction of the large horizontal emittance into
the small vertical emittance leading to beam emittance growth. \itemspace
\item Single-bunch wakefields: An off-axis bunch in a cavity or beampipe generates a dipole wakefield, causing a transverse deflection of the tail of the bunch with respect to the head. The wakefields are relatively weak for the SCRF accelerating cavities, and the cavity alignment tolerances correspondingly loose. \itemspace
\item Multi-bunch wakefields (Higher-Order modes): Leading bunches kick trailing bunches, which can lead to individual bunches in a train being on different trajectories. \itemspace
\item Cavity tilts: The transverse component of the accelerating field causes a transverse kick on the beam. \itemspace
\end{itemize}

\begin{comment} this is all said better above
The above error sources can be further categorized in terms of time-scale into static and dynamic effects. Static alignment errors, i.e. mechanical installation errors due to fiducialization and survey errors, must be initially corrected for by using beam-based alignment and tuning techniques. Dynamic errors, such as quadrupole vibration and slow ground motion, must be compensated for using continuous trajectory correction (feedback systems). The most critical feedback system maintains the nanometer-size beams in collision at the Interaction Point (IP).
\end{comment}

\subsection{Impact of Static Imperfections}

\subsubsection{Beam-Based Alignment and Tuning}

The beam emittance at damping ring extraction is $\gamma\epsilon_x=8~\mu$m and
$\gamma\epsilon_y=20$~nm. In a perfect machine, the emittance would be essentially the same at the interaction point. To allow for imperfections,
the ILC parameters specify a target emittance at the IP of $\gamma\epsilon_x=10~\mu$m
and $\gamma\epsilon_y=40$~nm. An emittance growth budget for the various regions
has been set at $\Delta\epsilon_y\le 4$~nm for the RTML,
$\Delta\epsilon_y\le 10$~nm for the main linac, including the positron source,
and $\Delta\epsilon_y\le 6$~nm for the BDS. These allocations may be
redistributed as the RTML budget currently appears too optimistic while the main linac budget appears generous.
Depending on the actual misalignments, the machine
performance can differ significantly. The goal for the alignment and tuning
procedures is to ensure that the emittance growth is within the budget with a likelyhood of at least 90\%.

Similar beam-based alignment and tuning procedures are applied in the different
subsystems of the LET. First, the elements are aligned in
the tunnel with high precision. When the beam is established, the corrector dipoles are used to zero the readings in the Beam Position Monitors (BPMs) (so-called one-to-one steering).
Even with a very good installation accuracy, the final emittance will be significantly above the
target; table~\ref{t:ml_tol} lists the expected main linac alignment errors together with the emittance growth resulting from each error after simple steering. The most important error source is the total BPM offset (with respect to the design ideal reference; note that an offset of a cryomodule also results in a BPM offset). Achieving the emittance goal requires beam-based alignment (BBA) to minimize the dispersive emittance growth (the dominant source of aberration).

\stepcounter{tablcl}\begin{table} \vbabove \caption[Assumed
installation errors in the main linac.] {Assumed installation errors
in the main linac, and the emittance growth for each error assuming
simple one-to-one steering. The required emittance preservation can
only be achieved using beam-based alignment of the magnets/BPMs.}
\label{t:ml_tol}
\begin{center}
\begin{tabular}{|*5{c|}}
\hline
Error & with respect to & value & $\Delta\gamma\epsilon_y$ [nm]\\
\hline  & & &  \vbdlspacing  \hline
Cavity offset & module     & $300\;{\rm \mu m}$       & $0.2$ \\ \hline
Cavity tilt   & module     & $300\;{\rm\mu radian}$   & $<0.1$ \\ \hline
BPM offset    & module     & $300\;{\rm\mu m}$        & $400$\\ \hline
Quadrupole offset & module & $300\;{\rm\mu m}$        & $0$\\ \hline
Quadrupole roll & module   & $300\;{\rm\mu radian}$   & $2.5$ \\ \hline
Module offset & perfect line &$200\;{\rm\mu m}$       & $148$ \\ \hline
Module tilt   & perfect line &$20\;{\rm\mu radian}$   & $0.7$\\
\hline
\end{tabular}
\end{center} \vbbelow
\end{table}

\begin{comment}
All BBA algorithms attempt to steer the beam in a straight dispersion-free line by effectively removing the unknown BPM offsets. Since the quadrupole offsets are responsible for perturbing the trajectory, this is equivalent to placing the magnetic centers on a straight line, either by physically moving the magnet (remote magnet movers) or by using corrector dipoles close to the quadrupoles. However, the exact details of the algorithms and their relative merits differ.
\end{comment}
All BBA algorithms attempt to steer the beam in a dispersion-free path through the centers of the quadrupoles, either by physically moving the magnets (remote magnet movers) or by using corrector dipoles close to the quadrupoles.  The exact details of the algorithms and their relative merits differ. The three most studied methods are:

\begin{itemize}
\item Dispersion Free Steering (DFS): Beam trajectories are measured for different beam energies, and the final trajectory minimizes the difference, thereby minimizing the dispersion. \itemspace
\item Kick Minimisation (KM): The BPM offsets with respect to the associated quadrupole magnetic centers are determined by varying the quadrupole strength
and monitoring the resulting downstream beam motion. This information is
used in a second step to find a solution for the beam trajectory where the
total kick from quadrupoles and correctors on the beam is minimized. \itemspace
\item Ballistic Alignment (BA): A contiguous section of quadrupoles (and in the linac the RF) is switched off and the ballistic beam
is used to determine the BPM offsets with respect to a straight line. The
quadrupoles/RF are then restored, and the beam is steered to match the established straight line. \itemspace
\end{itemize}

All BBA techniques rely on precise measurements of the BPMs to determine a near dispersive-free trajectory. The final performance of the algorithms is determined by the resolution of the monitors.

Once BBA is complete, a final beam-based tuning either minimizes the beam emittance by direct measurement of the beam size (emittance) or maximizes the luminosity. Closed-trajectory bumps or specially located and powered tuning magnets are used as orthogonal knobs to generate specific aberrations, such as dispersion or X-Y coupling. The knobs are tuned to minimize the emittance (or maximize the luminosity) by canceling the remaining aberrations in the beam.

\subsubsection{RTML before the Bunch Compressor}
The issue of static emittance growth from misalignments and errors
has been studied in some detail for the section of the RTML from the
turnaround to the launch into the bunch compressor.  The strong
focusing, strong bending, strong solenoids, and large number of
betatron wavelengths in this area can potentially lead to very
serious growth in the vertical emittance, despite the relatively low
energy spread of the beam extracted from the damping rings.

The tolerances used in the study were similar to those found at the
Final Focus Test Beam for warm, solid-core iron-dominated magnets
and are listed in table~\ref{t:misalignments}.
\stepcounter{tablcl}\begin{table} \vbabove \caption{Alignment
tolerance for RTML section up to the bunch compressors.}
\label{t:misalignments}
\begin{center}
\begin{tabular}{ | l | r | l | }
      \hline
      Misalignment  & RMS Value & Reference  \\ \hline
    & &   \vbdlspacing  \hline
      Quadrupole Misalignment (x,y) & 150 $\mu$ m & Survey Line \\ \hline
      BPM Misalignment (x,y)       & 7 $\mu$ m   & Quad Center   \\ \hline
      Quadrupole Strength Error     & 0.25\%      & Design Value  \\ \hline
      Bend Strength Error           & 0.5\%       & Design Value  \\ \hline
      Quadrupole Rotation           & 300 $\mu$ rad & Survey Line \\ \hline
      Bend Rotation                 & 300 $\mu$ rad & Survey Line \\ \hline
    \end{tabular}
\end{center} \vbbelow
\end{table}

{\bf Dispersion Correction:}  The preferred dispersion correction
method was found to be a combination of Kick Minimization (KM) and
dispersion knobs, the latter consisting of pairs of dedicated skew quadrupoles located in the turnaround, where there is non-zero horizontal dispersion. The two skew quads in a pair are separated by a $-I$ transform such that
exciting the quads with equal-and-opposite strengths causes the
resulting betatron coupling to cancel and the dispersion coupling to
add. There are two such dispersion knobs in the turnaround, which
allows correction of dispersion at each betatron phase. Simulations indicate that in the absence of
measurement errors, the combination of KM and dispersion knobs (DK)
can completely eliminate dispersion as a source of
emittance growth in this part of the RTML.  The principal remaining source of emittance dilution is betatron coupling, which typically contributes about 7.2~nm of emittance growth.

{\bf Coupling Correction:}  The coupling correction section consists
of four skew quads phased appropriately to control all four betatron
coupling parameters of the beam. The skew quads are used to minimize the vertical beam sizes as measured in the downstream emittance measurement station. The correction system is can completely eliminate the betatron coupling introduced by misalignments and errors in this section of the RTML.

In addition to the studies described above, the emittance
preservation issues in the long transfer line from the damping ring
to the turnaround have been examined.  Because of the weaker focusing, the alignment tolerances are much
looser than in the turnaround area, and emittance preservation is
relatively straightforward~\cite{c:transfer1}. One possible remaining error source in the long transfer line is the impact of time-varying stray fields, which can drive orbit oscillations.
This can be cancelled by the feed-forward system located across the turnaround. Measurements at existing laboratories~\cite{c:jf} indicate a reasonable estimate for the magnitude of time-dependent stray field is \~2~nTesla, which will not cause a problem.
% The only remaining possible issue is the emittance growth in the
%turnaround itself.

\subsubsection{Bunch Compressors}

%First studies of the bunch compressor alignment and tuning show that the
%additional emittance growth in this area can be significant (in the order of a
%few nm)\cite{c:bc}. Further study is ongoing to establish firm
%conclusions\cite{c:rtml}.

The RF in the bunch compressor introduces an energy correlation along the bunch. The long bunch from the Damping Ring (9~mm) makes the beam particularly sensitive to cavity tilts in the bunch compressor RF. The near-zero phase crossing of the bunch induces a strong transverse kick which is also correlated to the longitudinal location in the bunch (i.e. the bunch is crabbed), and therefore also strongly correlated to the induced energy spread. The resulting kick-energy correlation can effectively be compensated using downstream dispersion knobs. Wakefield-driven head-tail correlations can also be compensated the same way. As with other sections of the LET, the other primary source of emittance dilution is dispersion due to misaligned quadrupoles.

Simulation studies with RMS random quadrupole offsets of 0.3~mm, cavity offsets of 0.3~mm and cavity pitch of 0.3~mrad, indicate that
combined DFS and DK will reduce the mean emittance dilution to less than 2~nm ~\cite{c:bc}. Simulations of combined KM and DK result in several nm of residual emittance dilution. The expected 0.3~mrad RMS quadrupole roll
errors cause a modest average increase of the vertical emittance of less than 1 nm without any corrections. Although very promising, these results are preliminary and further study is required with more realistic errors~\cite{c:rtml}.

\subsubsection{Main Linac}

\stepcounter{figlcl}\begin{figure}[htb]
  \begin{center} \vbabove
       \includegraphics[width=0.95\textwidth]{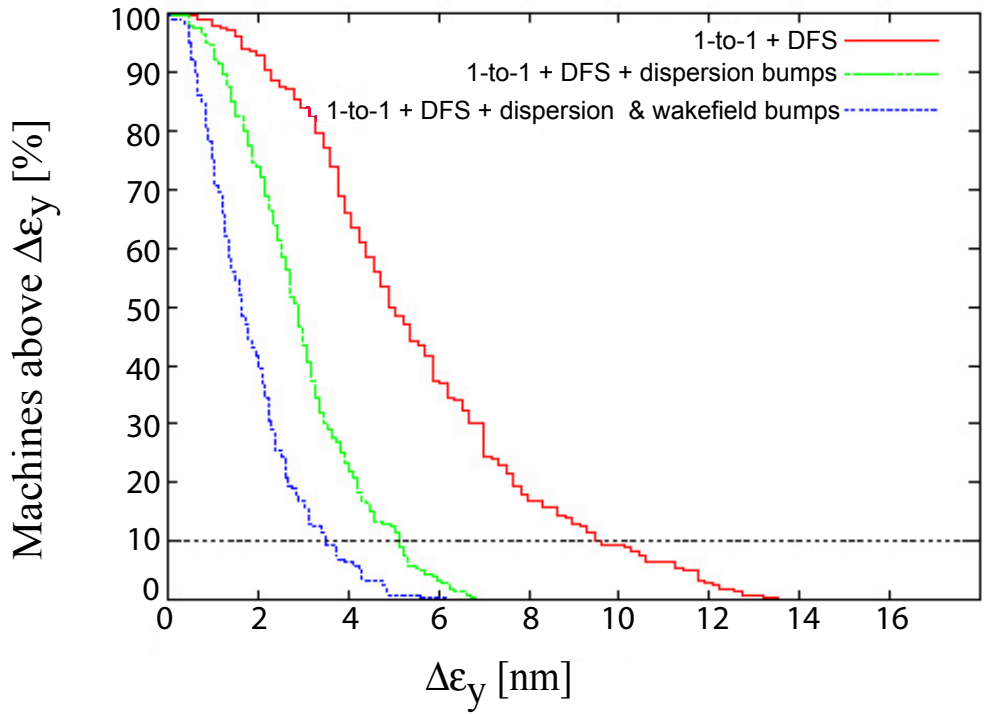}
      %\centerline{\epsfxsize=12cm\epsfbox{acc_phys_ml.eps}}
\vbabovecaption
\caption[The fraction of simulated cases staying below the emittance
growth target for the main linac.]{The fraction of simulated cases staying below the emittance
growth target for the main linac after Dispersion Free Steering
(red), followed by application of dispersion knobs (blue), followed
by wakefield knobs (green).}
\label{f:bump}  \end{center} \vbbelow
\end{figure}

Single-bunch emittance dilution in the main linac is dominated by chromatic (dispersive) effects and wakefield kicks arising from misaligned quadrupoles and cavities respectively. X-Y coupling arising from quadrupole rotation errors also adds a small contribution to the vertical emittance growth. The assumed installation errors are listed in table~\ref{t:ml_tol}. The tolerances for cavity offsets and quadrupole rolls can be achieved mechanically, but beam-based tuning is required for the quadrupole and BPM offsets.

The main linac follows the gravitational equipotential of the earth, and is therefore not laser-straight. This gentle bending in the vertical plane results in a small but non-negligible design dispersion which must be matched, and taken into consideration during beam-based alignment. A variant of dispersion free steering, dispersion matched steering (DMS), is used to attain the matched dispersion function along the lattice. This modified form of DFS requires well calibrated BPMs~\cite{c:curved3} to the level of 5-10\% with
very stable readout. The method achieves the required
performance in simulations\cite{c:curved1,c:curved3,c:curved2,c:curved4}.
Additional tuning knobs to modify the dispersion at the
beginning and end of the linac reduce the emittance growth still further to well below the budget\cite{c:bump1}, see Fig.~\ref{f:bump}. Further improvement is possible with wakefield tuning knobs.

Studies of kick minimisation have shown similar performance as DMS\cite{c:kickML}. The ballistic alignment method has not been
applied to the latest ILC lattice, however studies for TESLA showed that ballistic alignment and dispersion free steering yielded comparable results (for a laser-straight machine).

The suppression of multi-bunch wakefields (high-order modes, HOMs) have been a major part of the SCRF R\&D effort over the last decade. If left unsuppressed, the HOMs, which can have very high Q-values, would lead to unacceptable multi-bunch emittance growth. Suppression is achieved by random cavity detuning ($\sim$0.1\% spread in the HOM frequencies, expected from the manufacturing process), and by damping using HOM couplers (one per cavity) and HOM absorbers (one absorber per cryomodule for those modes above cut-off). All the modes for the baseline TESLA cavity shape have been calculated and measured at FLASH. The resulting multi-bunch emittance growth due to cavity misalignment is expected to be below 0.5~nm. If the transverse wakefield modes are rotated due to fabrication errors, they can lead to a coupling of the horizontal and vertical plane, potentially increasing the vertical emittance~\cite{c:roger}. This
effect is mitigated by using a split-tune lattice in which the
vertical and horizontal beam oscillation wavelengths are different, thus avoiding resonant coupling.

Different codes have been compared in detail for the main
linac\cite{c:MLcompare} finding excellent agreement for both tracking
and performance predictions for a specific beam-based alignment
method. This cross-benchmarking increases confidence in the results of each individual code.

\subsubsection{Undulator Section for Positron Production}
At the nominal 150~GeV point in the electron main linac, the beam passes through an undulator and emits hard photons for positron production. This insert has several potential consequences for emittance preservation which still require detailed study:

\begin{itemize}
\item Stronger focusing in the 1.2~km insert leads to additional dispersive emittance growth. This should be correctible using BBA methods.
\item The undulator increases the energy spread of the beam, which increases the dispersive emittance growth in the downstream linac. This effect is expected to be small.
\item The narrow bore vacuum vessel in the undulator is a potential source of transverse wakefields. Initial studies indicate these effects to be small.
\item The restricted bandwidth of the undulator chicane may hinder the use of DFS algorithms in the downstream linac. This problem can be alleviated by a straight-ahead bypass for tuning purposes.
\end{itemize}

Preliminary studies indicate the total emittance growth in this insertion to be small compared to the overall main linac budget. Further studies are required, however, including integration into the complete electron LET simulations.

\subsubsection{Beam Delivery System (BDS)}
Beam-based procedures have been developed to align and tune the BDS.
First, all multipoles are switched off and the quadrupoles and BPMs are
aligned. Second the multipoles are switched on and aligned. Finally,
tuning knobs are used to correct the different beam aberrations at the
interaction point. Detailed simulations have been made assuming the realistic installation alignment errors and magnet field errors given in Table~\ref{t:bdserr}.

\stepcounter{tablcl}\begin{table} \vbabove \caption[Assumed
imperfections in the BDS.] {Assumed imperfections in the BDS. The
assumed magnet strength errors are very tight, it is expected that
more realistic larger errors mainly lead to slower convergence of
the procedures.} \label{t:bdserr}
\begin{center}

\begin{tabular}{| l | c | r |}
\hline
Error&with respect to&size\\
\hline  & &   \vbdlspacing  \hline
Quad, Sext, Oct x/y transverse alignment&perfect machine&200 um\\  \hline
Quad, Sext, Oct x/y roll alignment&element axis&300 urad\\ \hline
Initial BPM alignment &magnet center&30 um\\ \hline
Strength Quads, Sexts, Octs&nominal &1e-4\\ \hline
Mover resolution (x/y)& &50 nm\\ \hline
BPM resolutions (Quads)&&1 um\\ \hline
BPM resolutions (Sexts, Octs)&&100 nm\\ \hline
Power supply resolution&&14bit\\ \hline
%FCMS: Assembly alignment200 um / 300urad
%FCMS: Relative internal magnet alignment10um / 100 urad
%FCMS: BPM-magnet initial alignment (i.e. BPM-FCMS Sext field centers)30 um
Luminosity measurement&&0.1\%\\
\hline
\end{tabular}
\end{center} \vbbelow
\end{table}

The current studies have been performed using the beam-beam interaction code GUINEA-PIG to give a realistic estimate of the luminosity, but assuming that the luminosity is measured accurately. Further studies are planned including a realistic simulation of luminosity measurement. The studies will also be crosschecked with other simulation codes. Results to date indicate that the design goals can be achieved with some overhead.

\subsection{Dynamic Effects}

The ILC relies on several different feedback systems to mitigate the
impact of dynamic imperfections on the luminosity. These feedback systems act on different timescales. The long $\sim$ 1~ms pulse length and relatively large bunch spacing ($\sim$ 300~ns) makes
it possible to use bunch-to-bunch (or intra-train) feedbacks located at critical points, the most important one being the beam-beam feedback at the interaction point which maintains the two beams in collision. Other feedback systems act from train to train (inter-train) at the 5~Hz pulse repetition rate of the machine. Over longer timescales (typically days or more) the beam may have to be invasively re-tuned.

The performance of the feedback systems is governed by the effective loop gain. A large gain (large bandwidth) is desirable to decrease the response time of the feedback; this is particularly true for the intra-train feedback, which reacts to each new pulse. A fast response time minimizes the number of initial bunches over which the feedback converges (normally a few percent effect). A low gain is desirable to reduce the amplification of high-frequency noise in the beam, and to effectively integrate away (average over) monitor resolution. The exact choice of gain is an optimization based on the noise spectrum being corrected (both in the beam and the monitors).

The number, type and location of feedback systems along the machine is also an optimization which is currently under study.

Very important sources of dynamic imperfections are ground motion and
component vibration. The ground motion depends strongly on the site
location. For the ILC-TRC study three ground motion models were
developed, all based on measurements at existing sites: Model A represents a very quiet site (deep tunnel at CERN); Model B a medium site (linac tunnel at SLAC); Model C a noisy site (shallow tunnel at DESY). A fourth model (K) was later developed based on measurements at KEK and is roughly equivalent to C. These models have been used in all subsequent simulations of the dynamic behavior of the ILC.

\subsubsection{Bunch-to-Bunch (Intra-Train) Feedback and Feedforward Systems}

The damping ring extraction kicker extracts each bunch individually.
If this kicker does not fully achieve the required reproducibility,
the beam will have bunch-to-bunch variations that cannot be removed by an
intra-pulse feedback system (effective white-noise). The feedforward system in the RTML is designed to mitigate this effect. The position jitter of each bunch
is measured before the turn-around and then corrected on that very bunch
after the turn-around.

Quadrupole vibration in the downstream bunch compressor and (predominantly) in the main linac will induce transverse beam jitter (coherent betatron oscillations). The tolerance on the amplitude of this jitter (and hence on the quadrupole vibration) from the main linac itself is relatively relaxed. Quadrupole vibration amplitudes of the order of 100~nm RMS lead to negligible pulse-to-pulse emittance growth. However the resulting oscillation (one- to two-sigma in the vertical plane) in the BDS could lead to significant emittance degradation from sources such as collimator wakefields. An intra-train feedback at the exit of the linac solves this problem. In addition, this feedback could correct any residual static HOM disturbance in the bunch train. If the main linac quadrupole vibrations are significantly less than 100~nm (e.g. 30~nm RMS, as expected for a typical quiet site), then a intra-train feedback at the exit of the linac may not be required.

Small relative offsets of the two colliding beams, in the range of
nanometers, lead to significant luminosity loss. The offsets are particularly
sensitive to transverse jitter of the quadrupoles of the final doublet. Fortunately, the strong beam-beam kick causes a large mutual deflection of the offset beams, which can be measured using BPMs just downstream of the final quadrupoles. The intra-train feedback system zeros the beam-beam kick by steering one (or both) beams using upstream fast kickers. The system typically brings the bunch trains into collision within several leading bunches (depending on the gain).
The IP fast feedback and the long bunch train also affords the possibility to optimize the luminosity within a single train, using the fast pair monitor as a luminosity signal~\cite{c:acc_opt}.

Studies performed as part of the TRC indicated that in a quiet site
(B or better) the fast beam-beam feedback and a slow orbit correction in the beam delivery system keeps the luminosity loss due to dynamic effects negligible~\cite{bib:ACCiltrc,c:nano}. In a noisy site (e.g. C) some luminosity loss occurs.
% and the orbit feedback needs to act within a few pulses~\cite{c:nano}.
%Simulations have thus concentrated on the noisy cites C and K.
%NJW: I don't understand this statement.

\subsubsection{Train-to-Train (5Hz) Feedback}

The exact layout of the train-to-train feedback has not yet been finalized and
different options are being studied. A simple but workable option is to use
a number of local feedbacks. At certain locations in the machine a few correctors are used
to steer the beam back through a few selected BPMs thus keeping the trajectory
locally fixed. These feedback systems can be used in a cascaded mode where each
of the feedback anticipates the trajectory change due to the upstream feedback
systems. Such a system was successfully implemented at SLC.

Since the system corrects only locally, a residual of the dynamic
imperfections will remain, due to deterioration of the trajectory between the feedback locations. After longer times this will require a complete
re-steering of the machine back the exact trajectory determined from the initial beam-based alignment (gold-orbit).

Other envisaged options are to perform permanent re-steering with a very low
gain; this method avoids the additional layer of steering but may be slower than local feedback. A further option is the use of a
MICADO-type correction. In this procedure all BPMs are used to determine the
beam orbit. A small number of most effective correctors is identified after
each measurement and these are used to correct the trajectory.

\subsubsection{Feedback Performance (Luminosity Stabilization)}

A complete and realistic simulation of the dynamic performance of the collider requires complex software models which can accurately model both the beam physics and the errors (e.g. ground motion and vibration). The problem is further complicated by the various time scales which must be considered, which span many orders of magnitude: performance of the fast intra-train feedbacks requires modelling of the detailed 10~MHz bunch train; fast mechanical vibrations at the \~Hz level need to be accurately modelled to test the performance of the pulse-to-pulse feedback systems; long-term slow drifts of accelerator components over many days must be studied to determine long-term stability and the mean time between invasive (re-)application of BBA. Ideally all these elements need to be integrated into a single simulation of the complete machine.

\begin{figure}
  \begin{center} \vbabove
    \includegraphics[width=0.95\textwidth]{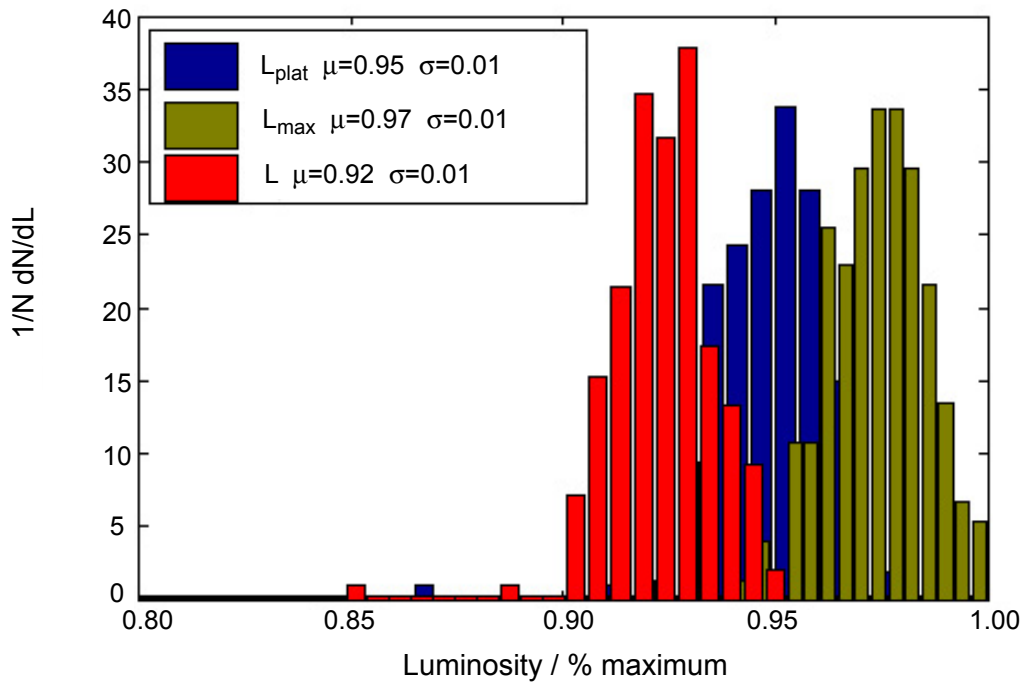}
\vbabovecaption
\caption[Example of integrated dynamic simulations.]{Example of integrated dynamic simulations, showing the
performance of the beam-beam intra-train feedback system with
realistic beams and beam jitter (simulated from the Main Linac and
BDS). The histograms show performance over 100 seeds of random
vibration motion: green - achieved luminosity for an infinitely fast
beam-beam feedback and no bunch-to-bunch variations (3\% reduction
from ideal); blue – performance including bunch-to-bunch variations
(driven by long-range wakefields in the Main Linac); red – as blue
but including a finite response time for the feedback (8\% reduction
from ideal). Taken from \cite{c:glenfdbk2,c:glenfdbk}.}
\label{f:beambeam}  \end{center} \vbbelow
\end{figure}

Progress towards such complete simulations is on-going. However,  many simulations have already been made, which have focused on individual aspects of the problem (time-scales), with varying degrees of sophistication of the feedback models. The results thus far give every indication that the ILC can achieve and maintain the desired performance. For example:

\begin{itemize}
\item Extensive simulations have been made of the performance of the fast beam-beam (and other) intra-train feedback using a model of the main linac and BDS to generate realistic bunch trains ~\cite{c:glenfdbk2,c:glenfdbk}. For realistic component vibration amplitudes, the results indicate that feedback can maintain the luminosity within a few percent of peak on a pulse-to-pulse timescale (5~Hz). See for example Figure~\ref{f:beambeam}. These results are in agreement with earlier studies~\cite{c:nano,bib:ACCiltrc}.

\item Drifts of components on the timescale of seconds to minutes have been studied \cite{c:nano,bib:ACCiltrc}. Simulations of 5~Hz operation with all ground motion models, and assuming the beams are maintained in collision by the fast IP feedback, indicate a slow degradation in luminosity. This can be mitigated by pulse-to-pulse feedback, especially in the BDS, where the tolerances are tightest. Noisy sites (model C) showed the most pronounced effect, and would place most demand on the slower feedback systems.

\item Longer term stability has been studied, assuming a variety of configurations for the slower pulse-to-pulse feedbacks. Studies of the main linac \cite{c:ds_dar1} using local distributed feedback systems indicate that the time between re-steering ranges from a few  hours  to a few days for ground motion models C and B respectively. After 10/200 days (models C/B) simple re-steering does not recover the emittance, at which point a complete re-tuning would be necessary.

\item Recent dynamic studies integrating the main linac and BDS, again based on distributed local pulse-to-pulse feedback systems (including one in the BDS) and incorporating many error sources and comparing all ground motion models have been made~\cite{c:glen5hz}. The noisy sites (models C and K) show a luminosity reduction of up to \~30\%, coming almost entirely from the BDS. Based on results from earlier simulation of other collider designs (notably TESLA), it is expected that optimization of the BDS feedback configuration, together with possible additional stabilization of critical magnets, can recover a significant fraction of the loss. By contrast, quiet sites (models A and B) show only a few percent loss for the configuration studied.
\end{itemize}

\stepcounter{figlcl}

Figure~\ref{f:beambeam} shows the
results of running 200 such simulations with differing random seeds.
The brown histogram shows the achieved luminosity for an infinitely fast
feedback and no bunch-to-bunch variations, it is 3\% below the case without
dynamic effects. The blue histogram includes the
bunch-to-bunch variations while the red one also includes the time to
convergence, leading to an average luminosity 8\% below nominal.
By optimizing the feedback gain and the intra-pulse luminosity tuning strategy,
one can hope to recover part of the 5\% additional loss.

\subsection{Remaining Issues}
Simulation tools must be developed further to fully specify the tuning algorithms, and in particular instrumentation, required to achieve and maintain the design luminosity. The cost impact should be minor, but the impact on the performance, in particular during commissioning of the machine, will be important.

The beam-based alignment and tuning procedures need to be fully specified for
all sub-systems. In particular, both the RTML and the positron source insertion in the electron linac need further detailed study. Detailed, fully-integrated and realistic studies of all the feedback systems also remains to be done.
Of particular importance is to quantify the impact of dynamic errors and equipment failure on the initial static error tuning (beam-based alignment and knob-based tuning). Preliminary studies have shown no indication of a severe problem but more study is required.

More thorough studies of the effect of the phase stability of the crab-cavities in the Interaction Region are required, and particularly their interaction with the beam-beam feedbacks. In addition the crab-cavity wakefields can potentially amplify beam jitter and lead to emittance dilution~\cite{c:jones2}. This can be avoided by ensuring that the transverse modes are not resonant with the bunch frequency.
Further study of collimator wakefields (and other general impedance issues in the BDS) is also required.
Depolarization during the beam transport from the damping ring to the IP has
been found not to be a problem\cite{c:depol1,c:jeff}.

%\end{document}

\clearpage %\setcounter{page}{1}
\setcounter{figlcl}{0}
\setcounter{tablcl}{0}
\setcounter{secnumdepth}{4} \setcounter{subsection}{0}
\setcounter{subsubsection}{0}
\usecounter{subsubsection}

\setcounter{section}{8} \renewcommand{\picturefolder}{./operations/}

\section{Availability, Commissioning and Operations}
\label{sec:Operations}
%{\bf (DRAFT: under revision)}
\subsection{Overview}

The ILC is a complex machine with hundreds of thousands of components
most of which must be tuned with exquisite precision to achieve design
luminosity.  This high luminosity must be maintained routinely in
order to deliver the required integrated luminosity.  Great care must
be taken at all stages of the design to ensure that the ILC can be
commissioned rapidly and operate efficiently with minimal downtime.
Some of the critical design issues are:

\begin{itemize}
   \item high availability components and redundancy to minimize
     downtime; \itemspace
   \item ease of commissioning; \itemspace
   \item separation of regions to allow beam in one region while
     another is in access; \itemspace
   \item Machine Protection System (MPS) to prevent the beam from
     damaging the accelerator while ensuring automated rapid recovery; \itemspace
   \item feedback systems and control procedures to maintain optimum
     performance. \itemspace
\end{itemize}

Many of these issues are mentioned elsewhere but are presented here as
an integrated package to emphasize their importance to the ILC.

\subsection{Availability}\label{sec:Availability}

\subsubsection{Importance of Availability}

The important figure of merit for the ILC is not the peak luminosity
but the {\it integrated} luminosity.  The integrated luminosity is the
average luminosity multiplied by the uptime.  Having surveyed the
uptime fraction (availability) of previous accelerators, a goal of
75\% availability has been chosen for the ILC.  This is comparable to
HEP accelerators whose average complexity is much less than that of
the ILC.  As such it should be a challenging, but achievable goal.  This
goal is made even more challenging by the fact that all ILC subsystems
must be performing well to generate luminosity.  In contrast, a
storage ring has an injector complex that can be offline between fills
without impacting performance.

Because it has more components and all systems must be working all the
time, attaining the target availability for the ILC requires
higher availability components and more redundancy than previous
accelerator designs.  High availability must be an essential part of
the design from the very beginning. A methodology is in place to
apportion the allowed downtime among various components and arrive at
availability requirements for the components.

\subsubsection{Methodology}

A simulation has been developed that calculates accelerator
availability based on a list of parts (e.g. magnet, klystron, power
supply, water pump). Input includes the numbers of each component, an
estimate of its mean time between failure (MTBF) and mean time to
repair (MTTR), and a characterization of the effect of its failure
(e.g. loss of energy
headroom, minor loss of luminosity, or ILC down). The simulation
includes extra repair time for components in the accelerator tunnel
(for radiation cool-down and to turn devices off and on), repair of
accessible devices while the accelerator is running,
repair of devices in parallel to overlap their downtimes, and
extra time to recover
the beam after repairs are completed. It also allows repairs to be
made in one region of the ILC while beam is used for accelerator
physics studies in an upstream region.

The inputs to the simulation were varied to test different machine
configurations and different MTBFs/MTTRs to develop a machine design
that had a calculated downtime of 15\%. The ILC design goal is
$>$75\% uptime, but 10\% downtime was reserved as contingency for
things that are missing from the simulation or for design errors.
The major design issues are described in the next section.

\subsubsection{High Availability Design Features}

There are some design features of the ILC that are particularly
important to achieving a high availability.  These features were
assumed in the simulation and if for some reason the ILC design is
changed so these assumptions are no longer valid, then other
improvements need to be made to maintain an adequate
availability.

{\bf RF Power Sources:} High power rf sources typically have a short
MTBF either due to faults or to component lifetime. Large linacs
ensure smooth operation by having spare units that can be switched in
quickly to replace the energy lost by the failed unit. The ILC was
assumed to have a 3\% energy overhead in each main linac. In the low
energy linac regions (5 GeV booster, bunch compressor, crab
cavities...), the fractional energy change due to a klystron failure
is very high making it impractical to replace the energy lost with a
unit in a different location. In these regions, there are hot spare
klystron/modulator units with waveguide switches that can immediately
replace the power to the same section of accelerator. One hot spare
for each low energy linac is sufficient. Klystrons and modulators are
accessible with the beam on and can be replaced with only a few minute
interruption to the beam to disconnect the waveguide.

{\bf Power supplies and electronics:} Power supplies are designed to
have a modular architecture with an extra module for redundancy. Most
electronics modules not in the accelerator tunnel are designed to be
replaceable without interrupting power to their crate. This allows
broken modules to be replaced without further impact on the beam.

{\bf Separation of regions:} There are tune up dumps and shielding
between each region of the accelerator so that one region can be run
while people are in another region. The ILC regions are injectors, DR,
main linac and BDS.

{\bf Site power:} Problems with the overall site power are allocated
only 0.5\% downtime. Present experience is that a quarter second power
dip can bring an accelerator down for 8 to 24 hours. For the ILC, one
24 hour outage would consume much of the downtime budget. This places
very stringent requirements on the reliability of the incoming power
and on-site power distribution system. The present design of the power
distribution system has some redundancy but the projected performance
needs to be reviewed and possibly a larger downtime budget
allocated.

{\bf Cryo:} The downtime budget allocated to the entire cryogenic
systems is set at 1\%. This includes all the time for which there
cannot be full power beam due to the outage, including even possible cool downs
after a warm up. With 10 large cryo plants for the main linac and 3
smaller plants for other systems, the required availability of each
plant is 99.9\% including outages due to incoming utilities
(electricity, house-air, cooling water, ventilation). This is 10-20
times better than the existing Fermilab or LEP cryo plants, where
around half of the cryo system downtime is due to the incoming
utilities. Achieving this goal requires both more reliable
utilities and more reliable cryo plants.

{\bf Positron source:} The positron target and capture section will
become too radioactive for hands-on maintenance. As the present design
does not have a spare target and capture section on the beam line, it
is vitally important that the components be designed so they can be
replaced with the use of remote handling equipment in less than a
day. There is a positron keep-alive source (KAS) that can provide low
intensity positrons to the e+ DR when the electron system in
inoperable for some reason. The intensity is high enough for BPMs to
work at their full specifications, about 10\% of the design intensity.
The KAS is expected to improve the availability by as much as 7\%.

\subsubsection{Required MTBF and MTTR Improvements}

In addition to all the specific design features described in the
previous section, many of the individual components must be designed
to have better MTBF and/or MTTR than measured in present
accelerators. Note that for all practical purposes, decreasing the
MTTR is equivalent to increasing the MTBF, so components can be
improved in whichever manner is most practical.

Table~\ref{tab:mtbf} shows the MTBFs and MTTRs needed to attain the desired
downtime goals. As engineering continues, these goals will be refined
to minimize the cost of the project while maintaining the desired availability.

\stepcounter{tablcl}\begin{table}[htb!] \vbabove \caption[Table of
MTBFs used.]{Table of the MTBFs that were used to obtain the
  desired 15\% downtime. Note that the desired MTBF is the product of
  the nominal MTBF and the improvement factor. The nominal MTBFs give
  a rough idea of what has been achieved at present accelerators. The
  third column shows the percentage downtime caused by the devices
  with the MTBF improvements given in the second column. These can be
  used to estimate the effect of not meeting one of the MTBF goals.}
\begin{center} \label{tab:mtbf}
%{\small
\setlength{\tabcolsep}{4pt}
   \begin{tabular}{| l | c | c | r |  r | } \hline
     Device         &  Needed     &   Downtime &  Nominal &  Nominal \\ [-6pt]
                    & improvement &  to these       & MTBF     & MTTR \\ [-6pt]
                    & factor      &   devices (\%)       & (hours)  & (hours)      \\  \hline  & & & & \vbdlspacing  \hline
Power supplies                  & 20 & 0.2 &     50,000 & 2 \\   \hline
Power supply controllers        & 10 & 0.6 &    100,000 & 1 \\   \hline
Flow switches                   & 10 & 0.5 &    250,000 & 1 \\   \hline
Water instrumentation near pump & 10 & 0.2 &     30,000 & 2 \\   \hline
Magnets - water cooled          &  6 & 0.4 &  3,000,000 & 8 \\   \hline
Kicker pulser                   &  5 & 0.3 &    100,000 & 2 \\   \hline
Coupler interlock sensors       &  5 & 0.2 &   1000,000 & 1 \\   \hline
Collimators and beam stoppers   &  5 & 0.3 &    100,000 & 8 \\   \hline
All electronics modules         &  3 & 1.0 &    100,000 & 1 \\   \hline
AC breakers $<$ 500 kW          &    & 0.8 &    360,000 & 2 \\   \hline
Vacuum valve controllers        &    & 1.1 &    190,000 & 2 \\   \hline
Regional MPS system             &    & 1.1 &      5,000 & 1 \\   \hline
Power supply - corrector        &    & 0.9 &    400,000 & 1 \\   \hline
Vacuum valves                   &    & 0.8 &  1,000,000 & 4 \\   \hline
Water pumps                     &    & 0.4 &    120,000 & 4 \\   \hline
Modulator                       &    & 0.4 &     50,000 & 4 \\   \hline
Klystron - linac                &    & 0.8 &     40,000 & 8 \\   \hline
Coupler interlock electronics   &    & 0.4 &  1,000,000 & 1 \\   \hline
Vacuum pumps                    &    & 0.9 & 10,000,000 & 4 \\   \hline
Controls backbone               &    & 0.8 &    300,000 & 1 \\
\hline
   \end{tabular}
%}
\end{center}
\vbbelow
\end{table}

Note that an MTBF of 1 million hours does not mean that a device must
run for 114 years without attention and without failing. Preventive
maintenance and even periodic replacement of components is allowed. It
is only failures that occur while the accelerator is running that
count towards the MTBF.

\begin{figure}[htb!] \vbabove
\begin{center}
   \includegraphics[width=11cm]{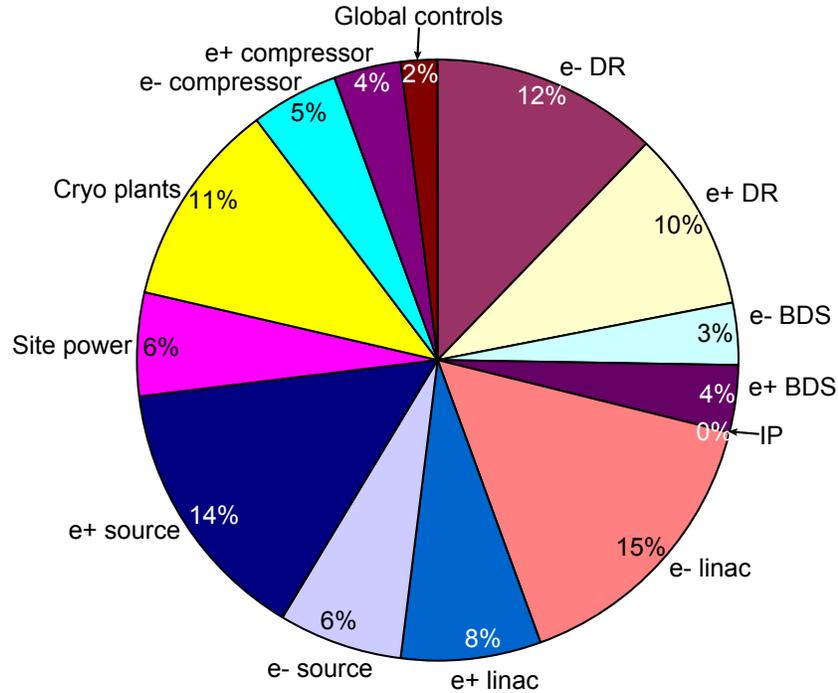}
\end{center}
\vbabovecaption
\caption{Distribution of the total downtime of 17\% among the various regions of the ILC.}
\label{fig:opsdownarea}
\vbbelow
\end{figure}

\begin{figure}[htb!] \vbabove
\begin{center}
   \includegraphics[width=11cm]{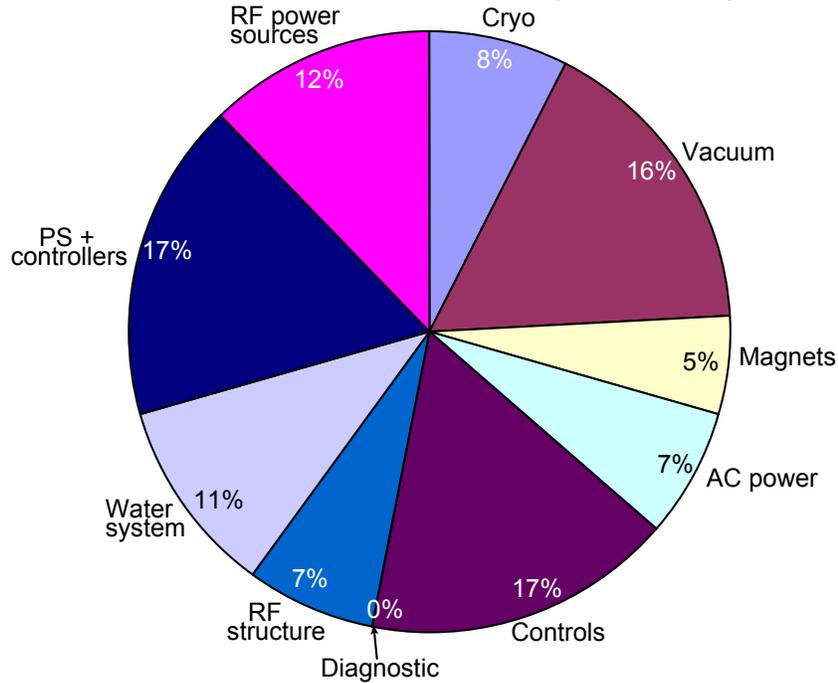}
\end{center}
\vbabovecaption
\caption[Distribution of the total downtime among the
  various systems of the ILC.]{Distribution of the total downtime among the
  various systems of the ILC. Note that the global system (site power,
  cryo plants, site-wide controls) are not shown in this chart.}
\label{fig:opsdownsys}
\vbbelow
\end{figure}

The pie chart in Figure~\ref{fig:opsdownarea} summarizes how much of the downtime is
caused by the various regions of the ILC. The chart in Figure~\ref{fig:opsdownsys} shows
which systems are causing the downtime. These charts give starting
values for how the unavailability budget is divided among the regions
and systems.

\stepcounter{figlcl}

\stepcounter{figlcl}

\subsubsection{High Availability R\&D}

Table~\ref{tab:mtbf} gives
the MTBFs and MTTRs that were
used to obtain a 15\% downtime. The desired MTBF
is the product of the nominal MTBF in column four and an improvement factor
in column two. The
nominal MTBFs give a rough idea of what has been achieved at present
accelerators. The third column shows the percentage downtime caused by
the devices after the MTBF improvements listed. These can be used to {\it estimate} the effect of not meeting
  one of the MTBF goals. Fairly large improvements are needed for
several types of hardware components. Some of these are being
addressed by ILC R\&D projects summarized here.

The factor of 20 improvement in magnet power supply MTBFs is mainly
to be accomplished with redundancy. SLAC has purchased a commercial
supply with five 1~kW regulators to feed a 4~kW load. Tests show a very
short dip in the current when one of the regulators dies. Another 20
supplies of an improved version are to be installed in ATF2 to provide
an extended test. Other improvements such as redundant controllers and
embedded diagnostic boards are planned.

Magnets need an MTBF of 18 million hours. While the average MTBF at
SLAC and FNAL was about 3 million hours, measured MTBFs range from 0.5
to 12 million hours depending on the sets of magnets used and the time
period. The magnet designers from different facilities are working
together to develop a set of
best practices that should result in magnets with MTBFs near the ILC
requirements. Serious consideration should be given to having an
analog readout for each thermal and flow interlock so that impending
problems can be fixed before affecting operations.

There is no active work on flow switches and water
instrumentation. The approach is likely to be to reduce the number of
flow switches and/or give them analog readout and add redundancy to
the other water instrumentation.

Kicker pulsers with built-in redundancy to provide high availability
have been developed. Most kicker installations also use multiple
kickers each with its own pulser, with one or more spare units to
replace a failed unit.

Electronic modules were assumed to have a factor of 3 improvement in
MTBF. The plan is to achieve this through the use of the advanced
telecommunications architecture, ATCA. This provides crates with
redundant power supplies and fans, and modules that are hot
swappable. ATCA prototype systems are being tested to learn how this
technology can best be used for the ILC. Commercial ATCA modules
provide redundant CPUs and networking with automatic fail-over, but the
ILC also needs to develop I/O boards that are sufficiently
reliable. The ability to replace a module with the crate powered
eases requirements on modules such as BPMs which degrade performance
without causing actual downtime.

\subsection{Commissioning}

This section describes initial ideas on commissioning. The actual
implementation will evolve with the schedule for construction of the conventional facilities
and the availability of early access to regions of the accelerator.  The plan needs further development during the EDR
phase.

\subsubsection{Phased Commissioning}

To minimize the time from completion of construction of the ILC to
operation for high luminosity, it is desirable to complete upstream
regions of the accelerator early.  Commissioning
can then start on these regions while construction continues downstream.
This is called {\it phased commissioning}.  In particular,
it would be beneficial to complete the electron injector and damping rings in time to allow
one or two years of commissioning while construction of the linacs and BDS
continues.

If a sufficient number of tunnel boring machines are available to
start all civil engineering projects simultaneously (i.e. the main
linacs, the beam delivery system and the damping rings), there is
about a year period available for phased commissioning. This is
because the damping ring tunnel, being shorter than the main linacs,
can be completed earlier.  If the number of boring machines is
limited, the preferred order of completion is injector, damping rings,
electron linac, positron linac and then beam delivery system.

\begin{figure}[htb!] \vbabove
\begin{center}
   \includegraphics[width=0.95\textwidth]{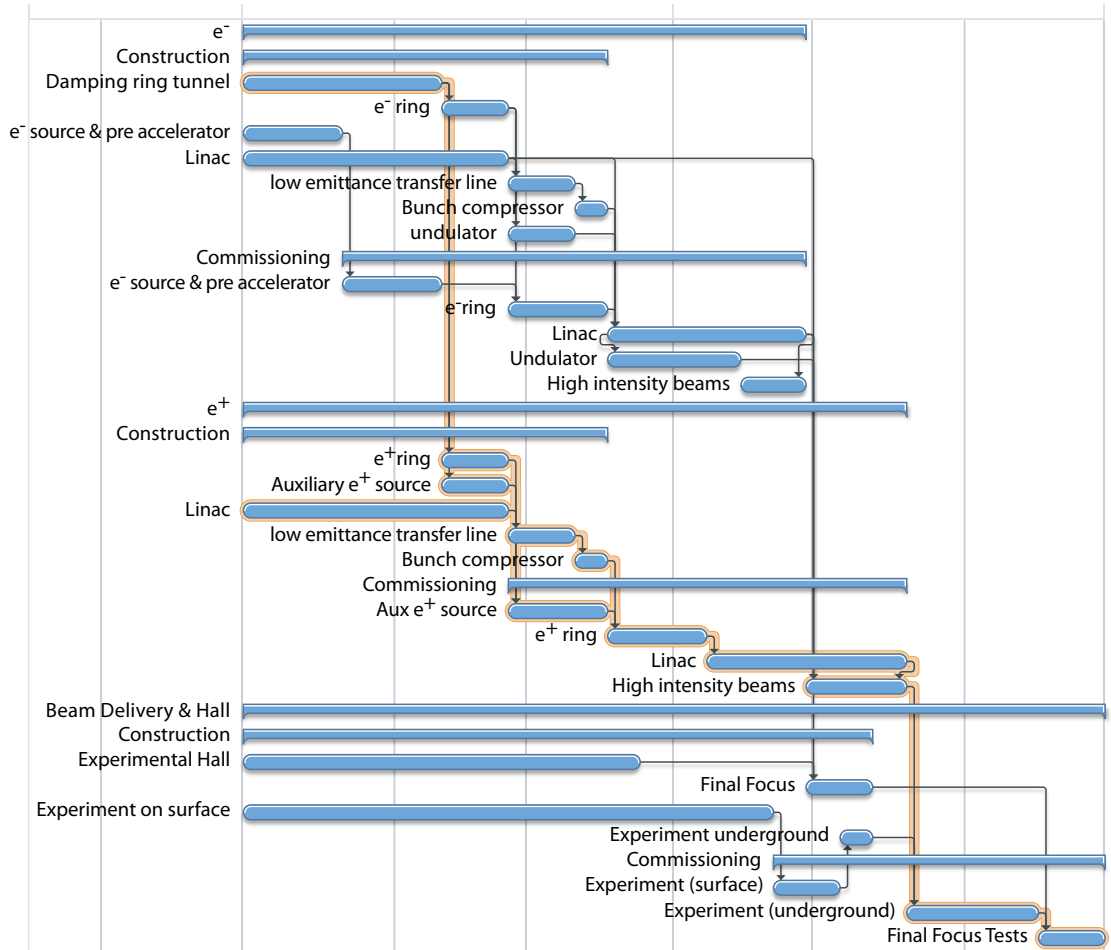}
\end{center}
\vbabovecaption
\caption[Sketch of the dependencies of the various construction
  tasks and the implications on commissioning.]{Sketch of the dependencies of the various construction
  tasks and the implications on commissioning. Construction of $e^-$,
  $e^+$ and Beam Delivery are shown and separated from the respective
  commissioning steps. Length of time lines are not to scale and
  the critical path indicated varies accordingly.}
\label{fig:commsketch}
\vbbelow
\end{figure}

A large amount of hardware validation and alignment and beam
commissioning studies are necessary to produce low emittance beams
with good stability and availability. Consequently, it is important to
allocate a sufficient amount of commissioning time at an early
stage. A major function of the DR commissioning period is to achieve
the alignment of optical components and to establish a small beam
emittance. In addition, there are beam intensity related issues that
need to be checked and high intensity beams are needed for vacuum
chamber scrubbing. The use of the damping rings obviously necessitates
functional beam source systems. Since both DRs are in the same tunnel,
a schedule optimization has to be done to determine if it is
best to install both DRs at the same time or if the e- ring should be
installed and commissioned followed by the e+ ring. The trial
construction schedule shown in Figure ~\ref{fig:commsketch} assumes
that both rings are installed together.

\stepcounter{figlcl}

Commissioning of the positron damping ring can begin with electrons
and then positrons from the positron-keep-alive source. However,
high-current commissioning must await partial commissioning of the
electron main linac up to the undulator at the 150GeV point and the e+
transport line to the DR.  The electrons that are used for producing
the positrons are dumped at the end of the main linac. Care has to
be taken to avoid interference with installation work on the beam
delivery system which may still be ongoing at that time.

The actual commissioning scenario depends on the construction
duration which is largely influenced by financial
resources. Nonetheless the general features are seen in
Figure~\ref{fig:commsketch}.

The construction of the experiment is likely to consume the largest
contiguous amount of time. It is recognized that construction of the
underground detector hall is a major undertaking which cannot be
completed several years after groundbreaking of ILC. To mitigate the
schedule impact, most of the sub-assemblies of ILC detector
facilities are built on surface and lowered later into the hall in
large pieces \cite{ops-gastal}.
%\footnote{Gastal, Martin ``Draft ILC Construction Schedule'' ; talk given at ALCPG
%meeting, September 21, 2006
%http://ilcagenda.linearcollider.org/materialdisplay.py?contribID=0$\&$amp;materialId=slides$\&$amp;confID=1144} .

\subsubsection{Electron Source and Reversible Positron Damping Ring for Commissioning}

The positron sources have a large emittance which by design nearly
fills the aperture of the positron damping ring.  This makes initial
commissioning of the $e^+$ ring with positrons challenging, since
initial construction and alignment errors cause substantial beam
losses which tend to produce false data from beam instrumentation.
Hence it is desirable to conduct some aspects of very early
commissioning of the positron damping ring with low emittance
electrons. The electron injector of the keep-alive-source is designed
so it can provide these electrons. For this commissioning with electrons, The positron ring
needs to have its magnet polarities reversed.
Reversal of the DR polarity may be allowed to take
several days as it is not done frequently.

\begin{comment}
this system was not costed for the RDR. it belongs in the EDR chapter R&D sections
\subsubsection{Automated surveying system}

The commissioning time is severely influenced by the accuracy and
efficiency of the survey and alignment. An automated system as
proposed for the linac of the XFEL could reduce the survey time of the
long straight sections by almost an order of magnitude. The
LiCAS\cite{ops-reichold}
system has a rail based laser that makes use of reference
markers along the length of the survey path.
Challenges to the automated survey are
presented by access shafts and other discontinuities in the tunnel
that make the passage of the survey train difficult. Nonetheless, it
may be possible to find technical solutions to bridge these points.
Once such solutions are demonstrated, putting such a system into the
ILC baseline should be examined.

The beam delivery and in particular the interaction point are areas
where repeated surveys are mandatory. Detector facilities weighing
thousands of tons are moved in the experimental hall and will
affect alignment. It is thus mandatory to implement a highly redundant
system of reference points in the hall that enable the survey to link
one linac alignment with the other.
\end{comment}

\subsection{Radiation Shielding and PPS Zones}\label{ssect:OPSrad}

To enable efficient operation and commissioning, the personnel
protection system (PPS) is designed to allow personnel access in one
region while beam is in another region. As an example, the main linac
beam tunnel can be in access while there is beam in the damping
ring. It is assumed that all accelerator housings could have radiation
levels that exceed the requirements for non-radiation
workers. Therefore, the radiation shielding and PPS zones described
here are designed for radiation workers.

\subsubsection{Summary of Regions' Radiation Requirements}

Maximum allowable radiation levels for radiation workers for each
region are summarized in Table~\ref{tab:radlevel}. Radiation shielding and PPS devices
must be designed to satisfy these criteria under the ILC beam-loss
scenarios.

\stepcounter{tablcl}\begin{table}[htb!] \vbabove \caption[Maximum
allowable radiation levels and doses.]{Maximum allowable radiation
levels and doses.
%The one in bold is the lowest and have been used in our design.
\newline
$~{(a)}$ Radiation Protection Instructions, DESY, June 2004.\newline
$~{(b)}$ Radiation Safety Instructions, KEK, in Japanese, June 2004.\newline
$~{(c)}$ Radiation Safety System, SLAC, April, 2006.\newline
$~{(d)}$ Fermilab Radiological Control Manual, FNAL, July, 2004.}
\begin{center}\label{tab:radlevel}
%{\small
   \setlength{\tabcolsep}{2pt}
   \begin{tabular}{| l | l | l | l | l | l |} \hline
    &   DESY$~{(a)}$  & TESLA  & KEK$~{(b)}$   & SLAC$~{(c)}$  & FNAL$~{(d)}$  \\
\hline & & & & & \vbdlspacing   \hline
Standard       & 20 mSv/yr    & 1.5 mSv/yr & 20 mSv/yr      &   & 50 mSv/yr  \\  \hline
Fertile women  &  2 mSv/month & 1.5 mSv/yr &  6 mSv/yr      &   &            \\ [-6pt]
               &              &            &  2 mSv/3months &   &            \\  \hline
Pregnant women &  1 mSv       & 1.5 mSv/yr &  1 mSv         &   & 5 mSv      \\ [-6pt]
               & /pregnancy   &            & /pregnancy     &   & /pregnancy \\ \hline & & & & &  \vbdlspacing \hline
Operating &        &            &                &   &            \\ [-6pt]
conditions &        &            &                &   &            \\ \hline & & & & & \vbdlspacing \hline
Normal         &              &            & 20 $\mu$Sv/hr  & 5 $\mu$Sv/hr   &    \\ [-6pt]
               &              &            & (1mSv/week )   & (10 mSv/year)  &    \\ \hline
Mis-steering   &              &            & 20 mSv/event   & 4 mSv/hr       &    \\ [-6pt]
               &              &            & (20 mSv/year ) &                &    \\ \hline
System failure &              &            & 20 mSv/event   & 250 mSv/hr for &    \\ [-6pt]
               &              &            & (20 mSv/year ) & max. credible  &    \\ [-6pt]
               &              &            &                & beam           &    \\ [-6pt]
               &              &            &                & (30 mSv/event) &    \\
\hline
   \end{tabular}
%}
\end{center}
\vbbelow
\end{table}

The TESLA TDR cited beam-loss scenarios for the main linac as 0.1~W/m
loss for normal operation and 100~W/m loss for 100 hours per year for
mis-steering condition.

The SLAC maximum credible beam loss condition is the full beam power
of 18~MW. Using these scenarios and the maximum allowable radiation
levels, the most stringent criteria comes from the SLAC maximum
credible beam condition. This gives a limit of 0.014~mSv/hr/kW loss
for the main linac.

The interaction region will be occupied by many experimentalists.
Hence tighter radiation design critera have been used so that
occupants do not need to have radiation worker training.
For normal operation, the IR hall radiation design limit is 0.5 microSv/h.
For the maximum credible incident the limits are 250 mSv/h and 1mSv/event.

\subsubsection{Shielding Calculation between Two Tunnels}

The linac design has a beamline and a service tunnel separated by 7.5~m. Radiation levels must be low enough in the service tunnel to allow
occupancy for repairs when beam is in the beam tunnel. Radiation dose
rates were evaluated using the Monte Carlo codes, MARS and FLUKA, and
the two tunnel configuration satisfies the radiation dose limit of
0.014~mSv/hr/kW. Here are a few selected results.

\begin{figure} \vbabove
\begin{center}
   \includegraphics[width=\textwidth]{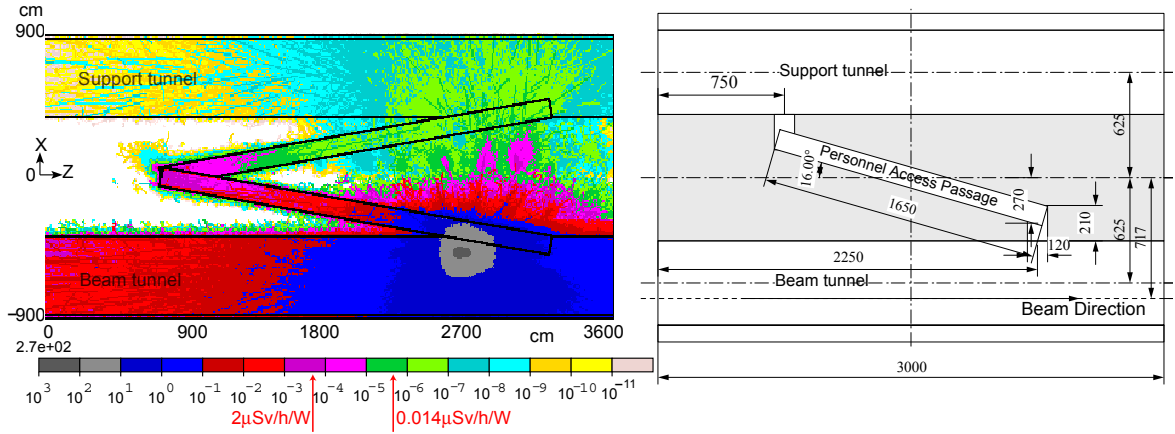}
\end{center}
\vbabovecaption
\caption{Two designs for passageways between the tunnels that give
  adequate radiation shielding.}
\label{fig:passways}
\vbbelow
\end{figure}

\begin{itemize}
  \item For sections with no penetrations between two tunnels, 4~m
   of earth provides adequate shielding. This was evaluated by
    the MARS code and the Jenkins formula with a soil density 1.9~g/cm$^3$
       and a 250~GeV electron beam incident on the worst case
    target: a thick copper cylinder 20~$X_0$ long and a radius of
    1~$X_0$. \itemspace

  \item In sections which have a penetration for waveguides or cables,
    the radiation near the penetration is above the allowed
    limit. However, the radiation level falls off rapidly with
    distance so it is sufficient to fence off the area immediately
    next to the penetration. The penetrations are located near the top
    of the tunnel, well above the personnel passage, so the fencing
    does not significantly restrict access in the service tunnel. In
    the calculations, the penetration was assumed to be a 7.5~m long
    circular hole with a diameter of 48~cm, and no shielding in the
    penetration. Suitable shielding could potentially lower the
    radiation levels further. \itemspace

  \item Personnel access passages between the two tunnels are located
    every 500~m along the main linac for emergency egress. Heavy
    movable shielding doors cannot be used because of the need for a
    fast escape route. These passages cannot have a direct
    line-of-sight or the radiation dose in the service tunnel would be
    unacceptable, but two designs adequately reduced the radiation in
    the service tunnel below the limit. These are shown in
    Figure~\ref{fig:passways} and described below. \itemspace

  \begin{enumerate}
    \item A rotated ``V'' shape passageway gave the lowest dose rate,
      about 20\% of the limit. In this simulation, the access
      passageway had a width of 1.2~m, a height of 2~m and the arms of
      the ``V'' were angled 10 degrees away from the accelerator
      tunnels. The total length of the passage was about 50~m. \itemspace

     \item Another design is a modified crank with an inclined center
       passageway. The dose rate calculated by the simulation was
       about 80\% of the limit. \itemspace
  \end{enumerate}
\end{itemize}

\stepcounter{figlcl}

\subsubsection{PPS Zones}

The personnel protection system (PPS) prevents people from being in
the accelerator tunnel when beam is on. A system of gates and
interlocks turn off the beam before allowing access to the accelerator
housing. Access to the service tunnel is not part of the PPS
system. The ILC is divided into different regions (PPS zones) with
tune up dumps and shielding to allow beam in one region while there is
access in another region. The PPS zones are the injectors, DR, main
linac and BDS. Entrance gates for PPS zones are monitored and stop the
beam when opened.

The ILC PPS zones are long and it would be burdensome to search the
full region after each permitted access. To ameliorate this problem,
they are divided into multiple search zones separated by fences with
gates that are also monitored. The search zones are up to several
hundred meters long. For example, in the linac a search zone is 500~m
long and is separated by gates midway between each cross tunnel
passageway.

Personnel access from a service area (service tunnel, shaft, detector
hall etc.) to an accelerator area is controlled by PPS gates, as is
the access from one accelerator region (PPS zone) to another
accelerator region. Fences, doors, or moving shields are used for
these gates and they have redundant gate-closed status switches for
PPS monitoring. They are locked to prevent careless access but have an
unlocking mechanism for emergencies. Information and communication
systems are provided at the gates to show the operational status and
allow communication between a person at the gate and an operator
granting permission to go through the gate.

There are personnel access passages between accelerator area and
service area at the main linac, shafts, alcoves and the detector hall
with PPS gates near each end. Since the passageways are used as
emergency exits, heavy moving doors are avoided if possible. PPS gates
between the accelerator areas and the service areas (including the
access passageway) need to restrict the flow of activated air from the
accelerator tunnel to the service area.

\begin{comment} this is all said above
For example, the accelerator tunnel of main linac has a personnel
access passage to/from the service tunnel every 500~m. Fences to
minimize searching sections are placed every 500~m, midway
between the passageway connections. Each end of each passageway has a
PPS gate. Opening a gate near the service tunnel shuts off the beam in
the main linac. The gate near the accelerator restricts the flow of
radioactive air.
\end{comment}

\subsubsection{Shielding between PPS Zones}

Shielding between PPS zones is designed to allow beam in the upstream
zone while people are in the downstream zone. The upstream beam is
deflected into a tune-up dump and there are triply redundant beam
stoppers between the beam and the accessed region to ensure the beam
does not enter the accessed region.

\subsection{Machine Protection System}

The task of the machine protection system, MPS, is to protect the
machine components from being damaged by the beam when equipment
failure or human error causes the beam to strike the vacuum envelope.
The MPS design must take into account the types of
failures that may occur and the damage they could produce.

\subsubsection{Overview}

The ILC Machine Protection System (MPS) is a collection of devices
intended to keep the beam from damaging machine components. The
nominal average beam power is 20~MW, consisting of 14,000 bunches of
2$\times10^{10}$ particles per second, and typical beam sizes near
10~$\times$~1~$\mu$m. Both the damage caused by a single bunch and
the residual radiation or heating caused by small (fractional)
losses of many bunches are important for MPS. The MPS consists of 1)
a single bunch damage mitigation system, 2) an average beam loss
limiting system, 3) a series of abort kickers and dumps, 4) a
restart ramp sequence, 5) a fault analysis recorder system, 6) a
strategy for limiting the rate with which magnetic fields (and
insertable device positions) can change, 7) a sequencing system that
provides for the appropriate level of protection depending on
machine mode or state, and 8) a protection collimator system. The
systems listed must be tightly integrated in order to minimize time
lost to aberrant beams and associated faults.

\subsubsection{Single Pulse Damage}

Single pulse damage is mitigated by systems that check the
preparedness of the machine before the high power beam passes.
Single pulse damage control is only necessary in the `damped-beam'
section of the ILC. Three basic
subsystems are involved: 1) a beam permit system that
surveys all appropriate devices before damping ring beam extraction
begins and provides a permit if each device is in the proper state 2) an abort
system that stops the remaining bunches of a train if a bunch does not
arrive at its intended destination 3) spoilers upstream of devices (typically collimators)
to expand the beam size enough that
several incident bunches do not cause damage. In
addition, some exceptional devices (damping ring RF and extraction
kickers for example) have fast monitoring systems and redundancy.

Spoilers or sacrificial collimators are placed before the bunch compressors,
in the undulator chicane, at the beginning of the BDS system and in the
collimator section of the BDS. Locations with dispersion downstream of an accelerator section have
spoilers to intercept off-energy beam caused by klystron faults or
phase errors before the beam can hit
a downstream collimator or beam pipe.  The spoilers are designed
to survive the number of incident bunches that hit
before the abort system can stop the beam.  If this design becomes problematic, the use of a {\it pilot bunch} is being kept as an option.  A
pilot bunch is one percent of nominal current and is spaced 10 $\mu$s
ahead of the start of the nominal train. If it does not arrive at its
intended destination, the beam abort system is triggered to prevent
full intensity bunches from hitting the spoiler.

Studies \cite{ops-eliasson} have shown that for many failure
scenarios such as quadrupole errors or klystron phase errors, the
beam is so defocused by the time it hits the linac aperture that it
does not cause damage.  For this reason, no spoilers or extra beam
abort kickers are included in the linac.

The beam abort system uses BPMs and current detectors to monitor the
beam trajectory and detect losses.  On a bunch by bunch basis, the
system checks for major steering errors or loss of beam.  When a
problem is detected it inhibits extraction from the damping ring and
fires all abort kickers upstream of the problem.  The abort kickers
cleanly extract the beam into dumps, protecting downstream beamlines.

In the few
milliseconds before the start of the pulse train,
the beam permit system checks the readiness
of the modulators and kicker pulsers, and the settings of
many magnets before allowing extraction of
beam from the damping rings.

\subsubsection{Average Beam Loss Limiting System}

Average beam loss is limited, throughout the ILC, by using a
combination of radiation, thermal, beam intensity and other special
sensors. This system functions in a manner similar to other
machines, such as SLC, LHC, SNS and Tevatron. If exposure limits are
exceeded at some point during the passage of the train, damping ring
extraction or source production (e$^+$/e$^-$) are stopped. For
stability, it is important to keep as much of the machine as
possible operating at a nominal power level. This is done by
segmenting it into operational MPS regions. There are 11 of these
regions, as noted in Table~\ref{tab:beampoints}. Beam rate or train
length can be limited in a downstream region while higher rate and
train lengths are maintained in upstream regions.

\stepcounter{tablcl}\begin{table}[htb!] \vbabove \caption[Beam shut
off points.]{Beam shut off points. Each of these segmentation points
is
  capable of handling the full beam power, i.e. both a kicker and dump
  are required. These systems also serve as fast abort locations for
  single bunch damage mitigation.}
   \label{tab:beampoints}
\begin{center}
   \begin{tabular}{| l | l | l | l |} \hline
    &  Region name &  Begin &  End  \\ \hline & & & \vbdlspacing \hline
     1 & e$^-$ injector & Source (gun) & $e^-$ Damping ring injection (before)  \\  \hline
     2 & e$^-$ damping ring & Ring injection & $e^-$ Ring extraction (after) \\  \hline
     3 & e$^-$ RTML & Ring extraction & $e^-$ Linac injection (before) \\  \hline
     4 & e$^-$ linac & Linac injection & Undulator (before) \\  \hline
     5 & Undulator & Undulator  & BD; $e^+$ target  \\  \hline
     6 & e$^-$ BDS & BD start & $e^-$ Main dump \\  \hline
     7 & e$^+$ target & $e^+$ target  & $e^+$ damping ring injection \\  \hline
     8 & e$^+$ damping ring & Ring injection & $e^+$ ring extraction \\  \hline
     9 & e$^+$ RTML & Ring extraction & $e^+$ linac injection \\  \hline
     10 & e$^+$ linac & Linac injection & $e^+$ BDS \\  \hline
     11 & e$^+$ BDS & $e^+$ BDS & $e^+$ main dump \\ \hline
   \end{tabular}
\end{center}
\vbbelow
\end{table}

\subsubsection{Abort Kickers and Dumps}

Abort systems are needed to protect machine components from single
bunch damage. It is expected that a single bunch impact on a vacuum
chamber will leave a small hole, roughly the diameter of the beam.
Each abort system uses a fast kicker to divert the beam onto a dump. The kicker rise time must be fast enough to produce a
guaranteed displacement of more than the beampipe radius in an inter-bunch
interval.

There are three abort systems in each RTML, one at the undulator
entrance, and one at the entrance to each BDS.

There will be many meters of fast kickers needed at each dump and
megawatts of peak power from pulsers.  R\&D is need to optimize the
system and and ensure its reliability.

\subsubsection{Restart Ramp Sequence}

Actual running experience is needed to exactly define the restart
ramp sequence.  For that reason the sequencer must be flexible
and programmable. Depending on the beam dynamics of the long trains,
it may be advisable to program short trains into a restart
sequence. There may also be single bunch, intensity dependent effects
that require an intensity ramp. In order to avoid relaxation
oscillator performance of the average beam loss MPS, the system must
be able to determine in advance if the beam loss expected at the next
stage in the ramp sequence is acceptable. Given the number of stages
and regions, the sequence controller must distribute its intentions so
that all subsidiary controls can respond appropriately and data
acquisition systems are properly aligned.

The sequence may need to generate a `benign' bunch sequence with the
nominal intensity but large emittance. The initial stages of the
sequence can be used to produce `diagnostic' pulses to be used
during commissioning, setup and testing.

\subsubsection{Fault Analysis Recorder System}

A post mortem analysis capability is required that captures the
state of the system at each trip. This must have enough information to
allow the circumstances that led to the fault to be uncovered. Data to
be recorded on each fault include: bunch by bunch trajectories,
loss monitor data, machine component states (magnets, temperature, RF,
insertable device states), control system states (timing system,
network status) and global system status (sequencer states, PPS,
electrical, water and related sensors). The fault analysis system
must automatically sort this information to find what is relevant.

\subsubsection{Rapidly Changing Fields}

In addition to the above, there are critical devices whose fields (or
positions) can change quickly, perhaps during the pulse, or (more
likely) between pulses. These devices need 1) special controls
protocols, 2) redundancy or 3) external stabilization and verification
systems.

\begin{enumerate}
\item Depending on the state of the machine, there are programmed
  (perhaps at a very low level) ramp rate limits that keep critical
  components from changing too quickly. For example, a dipole magnet
  is not allowed to change its kick by more than a small fraction
  of the aperture (few percent) between beam pulses during full power
  operation. This may have an impact on the speed of beam based
  feedbacks. Some devices, such as collimators are effectively
  frozen in position at the highest beam power level. There may be
  several different modes, basically defined by beam power, that
  indicate different ramp rate limits. \itemspace

\item There are a few critical, high power, high speed devices
  (damping ring kicker and RF, linac front end RF, bunch compressor RF
  and dump magnets) which need some level of redundancy or extra
  monitoring in order to reduce the consequence of failure. In the
  case of the extraction kicker, this is done by having a
  sequence of independent power supplies and stripline magnets that
  have minimal common mode failure mechanisms. \itemspace

\item There are several serious common mode failures in the timing and
  phase distribution system that need specially engineered
  controls. This is necessary so that, for example, the bunch
  compressor or linac common phase cannot change drastically compared
  to some previously defined reference, even if commanded to do so by
  the controls, unless the system is in the benign beam-tune-up mode. \itemspace

\end{enumerate}

\subsubsection{Sequencing System Depending on Machine State}

The ILC is divided into segments delineated by beam stoppers and
dump lines. There may be several of these in the injector system, two
beam dumps in each RTML, and 2 (or 3) in the beam delivery and
undulator system. In addition, the ring extraction system effectively
operates as a beam stopper assuming the beam can remain stored in the
ring for an indefinite period. This part of the MPS assumes that the
beam power in each of these segments can be different and reconfigures
the protection systems noted above accordingly.

\subsubsection{Protection Collimators}

The entire ILC requires protection collimators and spoilers that
effectively shadow critical components. These devices must be
engineered to withstand innumerable single pulse impacts. The number
and locations of these protection collimators are documented in the
descriptions of each accelerator region.

\subsection{Operability}

To ensure high average luminosity it is important that the ILC have
many features built in to make its operation mostly automatic and
efficient. These features include:

\begin{itemize}

  \item Accurate, reliable, robust diagnostics \itemspace

  \item Monitoring, recording, and flagging of out-of-tolerance readings
    of all parameters that can affect the beam.  Some of these
    must be checked milliseconds before each pulse train so beam
    can be aborted if there is a problem. \itemspace

  \item Beam-based feedback loops to keep the beam stable through
    disturbances like temperature changes and ground motion \itemspace

  \item Automated procedures to perform beam based alignment,
    steering, dispersion correction, etc. \itemspace

  \item Automatic recovery from MPS trips starting with a low
    intensity high emittance beam and gradually increasing to nominal
    beam parameters \itemspace

\end{itemize}

%\subsubsection{Partitioning the feedback loops}
\subsubsection{Feedback systems}

The transport of the beam through the ILC requires a large
number of feedback systems to be active to steer the beam to the
interaction point.
%Survey alignment tolerances will position components
%to on the order of 300 $mu$m, which is not
%adequate to transport the beam through the beam delivery system.
These feedback systems include measurements from various beam
position monitors, from laserwires scanning the beam profile and other
diagnostics. The feedback loops must be carefully designed to be orthogonal
and to maintain corrections that are within the device ranges.
The feedback systems must avoid trying to compensate for large
deviations of the beam due to component failure.
\begin{comment}
most of this next is garbled
%particularly demanding to ascertain that their action be independent
%and that the implied corrector settings correspond to the known
%tolerances of the components and their alignment.
%It is mandatory in the beginning and to monitor
%component malfunction to implement simple feedback loops that can be
%cross checked with the known alignment tolerances. As better
understanding is obtained the action of the feedback loop can be
extended to better optimize overall performance. Nevertheless one has
to avoid that the action of one feedback loops works against a
subsequent one with the result of erratic settings of the machine.
\end{comment}
It is hence necessary to use flexible setups for the control
loops such as provided by MATLAB tools and analysis techniques.

\setcounter{chapter}{2}

\chapter{\textsf{Technical Systems}\label{chapTS}}

\setcounter{figlcl}{0}
\setcounter{tablcl}{0}
\setcounter{secnumdepth}{4} \setcounter{subsection}{0}
\setcounter{subsubsection}{0}
\usecounter{subsubsection}

\setcounter{section}{0} \renewcommand{\picturefolder}{./magnets/}
\section{Magnet Systems}\label{sectMagnet}

\subsection{Overview}

The ILC has $\sim$80~km of beamlines which require magnets for
focusing and steering the beams. There are over 13,000 individual
magnets, of which approximately 18\% are superconducting and the
rest ``conventional'' warm iron-dominated magnets with copper coils.
About 40\% are low-current corrector magnets. Superconducting
technology is primarily used for the magnets located in the RF
cryomodules, but it is also required for the spin rotation
solenoids, damping ring wigglers, positron source undulator and beam
delivery octupoles, sextupoles and final doublet quadrupoles.

\subsection{Technical Description}

The scope of the Magnet Technical System includes the magnets and
their power systems, as well as the magnet support stands and
positioning devices needed for precise magnet alignment in the
beamlines.  Power systems include the power supplies, cabling to the
magnets, sensors and systems for local control, monitoring, and
magnet protection.  Pulsed kicker and septum magnets used for beam
injection, extraction, and/or protection are particularly
challenging. Almost all of the room temperature magnets have easily
achievable requirements. The major technical issues, challenges and
special purpose magnet systems are presented below.  Challenging
technical issues unique to particular Areas (especially BDS and DR)
are discussed in those sections.

The magnet design process starts with the Area System leaders who
specify a standard set of requirements based upon the lattice
designs, machine layout, and envisioned operating scenarios.
Conceptual magnet designs follow primarily from the specified
integrated strength, field quality, clear bore aperture, and slot
length constraints. Given the conceptual magnet design, a power
supply (PS) design is developed based on the magnet-specific
current, voltage and stability requirements, and the need to power
magnets individually or in series.

\subsection{Technical Issues and Challenges}

\subsubsection{High Availability and Low Cost}

A major criterion for ILC magnet design is to achieve very high
availability in spite of the very large number of magnets. The
``availability'' goal of the ILC is 75\% (or better) and the magnets
have been budgeted to incur no more than 0.75\% down time. The
availability ``A'' of a component is given as A = MTBF/(MTBF+MTTR),
where MTBF is Mean Time Between Failures, and MTTR is Mean Time To
Repair a magnet, turn it back on and restore the beam. Detailed
studies of magnet failures at three high energy physics labs
indicate that most failures are with conventional water-cooled
magnets, which had an MTBF ranging from about 0.5 million to 12
million hours based on tens of millions of  integrated magnet-hours.
The ILC has 6873 such magnets. With an MTTR of 16 hours, the MTBF of
each one must be longer than 18 million hours in order to achieve
the desired availability.

This reliability level should be achievable, without incurring a
significant increase in cost, by applying the assembled magnet
design, production and operation experience at existing HEP
accelerators. The approach is to apply best modern magnet
engineering practices, ensure adequate quality control of materials
and procedures during fabrication, and use established guidelines
for operating within reasonable environmental limits (such as
ambient temperature and allowed temperature rise, maintaining proper
water flow conditions, and keeping electronic components out of
radiation areas where possible). Power system electronic components
typically have much lower MTBF values of around 100,000 hours. Here,
the solution is to build in redundancy for crucial elements, and use
``smart" electronics that can detect failure and rapidly switch to
redundant units. Replacement of failed units can then be scheduled
to occur during beam downtimes.  Comprehensive failure mode and
effects analyses (FMEA) are thus viewed as an essential part of the
magnet system engineering effort.
% needs a [ref].

\subsubsection{Field Quality and Alignment}

The field quality requirements in most normal-conducting ILC magnets
are similar to those at other accelerators currently in operation,
and not particularly challenging. Higher order harmonics must be on
the order of a ``few units'' (1 unit = $10^{-4}$) of the main field
strength, and are most stringent in the Damping Rings where beam
circulates for many turns. For corrector magnets, a few tens of
units is characteristic.  In warm iron-dominated electromagnets,
these levels are achieved through careful control of pole shapes and
their positioning.  Similarly, control of coil position is important
for superconducting magnet field quality, and is achievable with
proper mechanical design and tooling.  Large room temperature
magnets have a split yoke design to reduce repair time in the
tunnel; experience shows that field quality can be maintained with
proper design and care in re-assembly.

Alignment and mechanical stability requirements in many areas are
very challenging. In the BDS, beam positions must be maintained at
sub-micron levels to collide the beams, so precision 5-axis magnet
positioning mounts, or ``movers,'' are needed for continuous
adjustment of all the quads and sextupoles in the final focus
region. For the regions where movers are not required, room
temperature and cryogenic magnet stands use a robust and precisely
adjustable design.  In some areas, pedestals are required to offset
the precision stands from the tunnel floor. Alignment tolerances on
the relative positions of Beam Position Monitors (BPMs) to
quadrupoles differ by area ($\sim$10 $\mu$m in BDS, $\sim$100 $\mu$m
in ML).  In the BDS this results in stringent temperature control
requirements locally, where geometry-sensitive cavity style BPMs are
affixed to the thermally active magnets.

\subsubsection{Superconducting Magnets}

There are 2318 superconducting (SC) magnets in a variety of
applications throughout the ILC, but fewer than 10\% of them require
high integrated field strength in limited space and about 60\% are
correcting coils wound in the same physical space as the main coils.
Most of the SC magnets are not very strong and are located in the RF
cryomodules. A package containing a focusing quadrupole (quad),
steering dipole correctors and a Beam Position Monitor is located at
the center of every third main linac cryomodule. This location makes
it challenging to maintain the quad positions during thermal cycles
and to measure and relate the quad positions to external survey
fiducials and to the BPMs used to keep the beam centered in the
quads (at the 100 micron level, over a distance of $\sim$6 meters).
The resulting magnetic center in nested dipole and quad designs may
also be affected by persistent current effects. Alternative designs
and further research are needed to understand these issues and
develop the magnet support and measurement techniques; there could
be significant advantages to moving the magnets and BPMs from the RF
cryomodule out to a separate cryostat.

The superconducting wigglers in the damping rings and the
superconducting undulators in the positron source also require great
mechanical precision; their particular challenges are described in
their respective chapters. The most challenging superconducting
magnets, those just before the interaction point, are described in
the BDS chapter. They have strong gradient fields with many layers
of correcting coils, and must fit into as small a radius possible to
not interfere with the detector. In the ILC sources, there are
superconducting solenoids for spin rotation and a few large aperture
magnets that may be either conventional or superconducting,
depending on detailed optimization of operating versus capital cost.

\subsubsection{Power Systems Design}

The non-pulsed ILC magnets operate with DC currents, with set points
that may be adjusted periodically but only slowly ($\sim$5 A/s or
less). Most magnets are individually powered to allow independent
control. Power supply stability requirements are assumed to be
comparable to performance of existing commercial units. The design
of a power system, whether for an individual or a string of magnets,
requires a conceptual magnet design which determines the required
operating current, defines the coil resistance and inductance and
cooling requirements (air or water).  The magnet position, with
respect to power supplies located in alcoves, defines the cable
length; cables are sized for the maximum operating current using two
sizes above the NEC rated minimum, to reduce voltage drop and heat
generation. The required maximum power supply voltage is then
determined by the cable and coil resistive drop at maximum current,
plus the inductive drop at the maximum ramp rate.  The supply is
sized with a 10\% margin on the power rating, to accommodate
uncertainties in magnet strength, inductance, cable lengths, etc.
The summed power ratings set requirements for AC power, air and
water conditioning in each area.

\begin{figure}[htb]
   \begin{center} \vbabove
      \includegraphics[width=13cm]{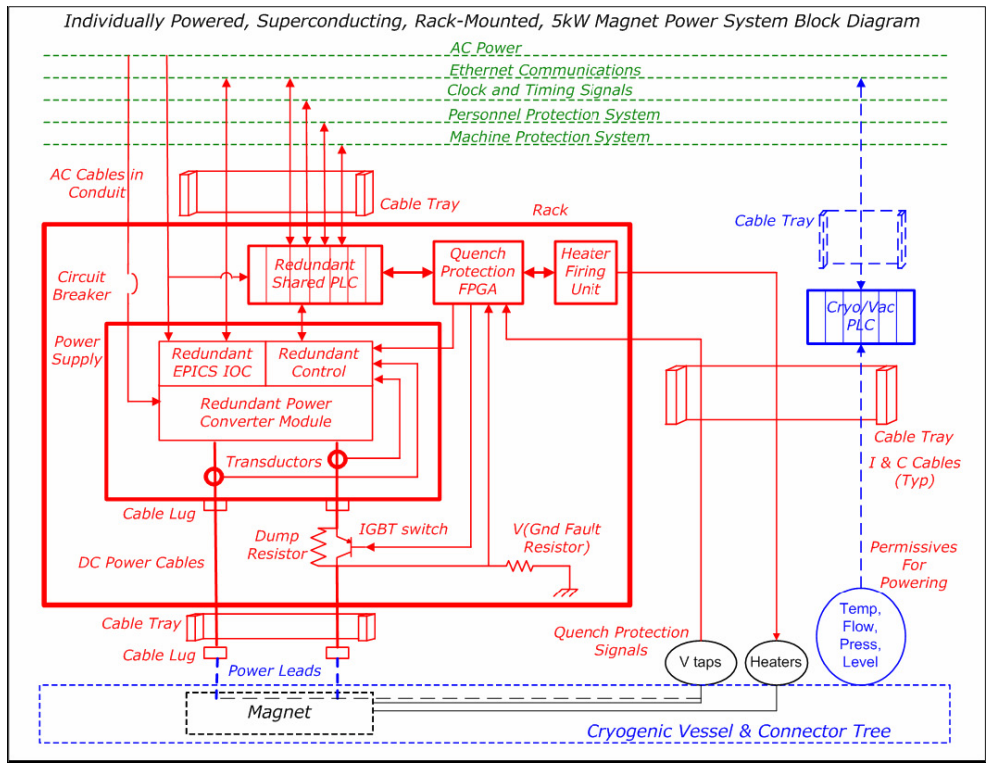}
      \vbabovecaption \caption[An example DC power system.] {An example DC power system style:
      items in red are specific to the power system,
      magnet elements are in black; relevant interfaces are shown,
      where blue and green lines are responsibility of other groups
      (global controls, cryogenics, vacuum, facilities, etc.).}
      \label{fig:MagnetPSBlockDiagram}
   \end{center} \vbbelow
\end{figure}

Power systems are classified by their size and type, and standard
models were developed for each of the various system categories:
they are distinguished by ``normal'' versus ``superconducting'',
``individually powered'' versus ``series connected'', ``rack
mounted'' versus ``free standing''.  These styles have certain
elements in common, but may differ in detail (water versus air
cooling, for example); Figure \ref{fig:MagnetPSBlockDiagram} shows
one example which contains all of the power system elements, and
illustrates interconnections between components and systems.  Each
power system provides local control and magnet protection (via PLCs
and FPGAs), and has the capability of diagnostic data capture.  The
design incorporates redundant current transductors, controllers, and
Ethernet IOCs, which are utilized for communication with machine
control, protection, and other technical systems (e.g., to obtain
cryogenic or LCW process variables for operating permissive).
Smaller rack-mount supplies can accommodate a redundant supply
within the rack, for automatic switch-over in case of a failure. The
conceptual design for the superconducting magnets is based on the
generic model shown, although the protection elements may be
simplified after detailed magnet design.

\stepcounter{figlcl}

\subsubsection{Kicker, Septum, and Pulsed Magnets}

A kicker is a device that makes fast time-dependent changes in the
beam path. A septum magnet has regions with very different magnetic
field with a material septum between them.  A kicker diverts the
beam from one side of the septum to the other, and the much higher
field of the septum diverts the beam by a much larger angle,
typically around some downstream obstacle like a quadrupole magnet.
A pulsed magnet changes its field as part of normal operation, but
less rapidly than a kicker.  The high power beam dumps have pulsed
magnets upstream to sweep the beam across the dump to avoid
localized damage.  Fast actuators in the beam feedback systems in
the damping rings and at the IP, which are also sometimes called
``kickers,'' are described in the area chapters.

There are several classes of kickers in the ILC. The damping ring
injection and extraction kickers are pulsed every few hundred ns for
single bunches during each millisecond linac pulse, and need rise
and fall times of a few ns. A damping ring abort kicker is only
fired when an abnormal beam condition is detected, to divert the
stored beam to a dump, and avoid damage to machine components.  The
rise time must be less than the ion-clearing or abort gap in the
ring filling pattern, with a pulse width of a full ring turn
($\approx$22 $\mu$s).

There are other abort kickers at several locations outside the
damping rings, with rise times of less than the time between bunches
($\approx$100 ns). When used as abort kickers, the pulse rate is
nearly zero, and the pulse width need only be long enough for the
bunches that cannot be stopped upstream.  An abort kicker can also
be used to limit the beam power downstream, by firing it after a
fraction of the beam bunches in the train have passed.  In this
application, the kicker may be fired on every linac pulse, for the
full linac pulse duration.

The ILC kickers are all stripline structures inside the vacuum
chamber, driven by pulsed power supplies.  The injection and
extraction kickers have short strips and extremely fast pulsers to
achieve fast rise and fall times.  The required total kicker
strength (kilovolt-meters) is set essentially by the beam size and
energy, with the result that a large number of stripline and pulser
units are needed for each installation.  The damping ring abort
kickers use more conventional thyratron or FET pulsers and longer
strips since the rise time can be longer, the pulse length is
moderate, and the rate is low.  The other abort kickers have relaxed
rise time requirements, but the pulse may need to be a millisecond
long at full linac rate, and higher beam energies require more
kicker field energy.  The pulser power required scales inversely
with the cube of the available length, and can be quite high.  The
beam delivery system abort kicker installations each require several
pulsers of the scale of main linac modulators.

The baseline design has a thin and a thick pulsed eddy-current
septum magnet for each damping ring kicker.  This design is inspired
by the Argonne APS septa, but R\&D is required to make a millisecond
flat top to the required tolerance.  An alternative optics design is
under consideration for injection and extraction that allows a DC
current-sheet septum of moderate current density to be used. An
abort septum must be DC, and could be a current sheet septum, or an
iron magnet with a beam-hole in its pole region (Lambertson septum).
The damping ring abort region optics, and thus the type and
parameters of its septum, are discussed in the Damping Ring chapter.
The RTML and beam delivery abort septa are DC current sheets.  The
undulator protection abort septa are dogleg bend magnets modified to
be Lambertson septa.  All of the abort dumps downstream of the
damping ring require sweeper magnets.

\subsubsection{Fabrication, Test, and Storage}

The program of fabrication and testing of ILC magnets follows a 7
year schedule, with one year of preparation, five years of
production and testing, and (overlapping) four years of
installation.  Magnet fabrication utilizes industrial suppliers
world-wide; tooling developed for ILC magnet fabrication belongs to
ILC for future use.  A large fraction of solid-wire corrector
magnets are tested by manufacturers, and all non-corrector magnets
are tested and magnetically measured at the ILC test facility.
Superconducting test stands share cryogenic resources with nearby
SRF test facilities and have both production and special measurement
areas and test systems.  The conventional magnet measurement area is
large, with multiple stands for efficient and high throughput, with
space for temporary magnet storage. Alignment and survey
capabilities are needed for all magnet styles.  In the long term,
some space could be converted for storage of tooling and spare
magnets (or coils and parts), and part of the facility could be
devoted to repair and new magnet fabrication.  Also an area remains
dedicated to making tests and measurements, and conducting R\&D for
later machine improvements.  Such a facility is necessary to ensure
the initial high quality of ILC magnets.

\subsection{Cost Estimation}

The cost estimate is based on the conceptual designs for magnets,
power systems, stands and movers described above, with additional
assumptions about estimated costs of material and labor.  Given time
and resource limitations, detailed conceptual designs were developed
for only a small number of the magnet styles. The majority of
estimates are ``engineering estimates'' based on existing designs
with similar requirements. Standardized labor rates were determined
from laboratory and industrial sources\footnote{It should be noted
that rates for different world regions have not been incorporated at
this time.  It should also be recognized that labor rates and
production hours are not necessarily uncorrelated:  the lowest labor
rates are quite often in regions with less automation and
infrastructure resulting in longer task times.}.
In order to determine the material costs, the weights of magnet and
cable materials, primarily copper and iron, have been estimated and
summed, and current world commodity prices obtained.  Similarly,
prices have been obtained for commercially available electronic
components such as power supplies, FPGAs and PLCs, controllers and
Ethernet interfaces.

In one instance, a design and a complete set of drawings was
developed for a e$^+$ Source transfer line quadrupole (a large
quantity item) and a request for quote sent to a number of magnet
vendors. The cost estimates obtained were in reasonable agreement
with an internal estimate:  the average agreed within a few percent
of the internal estimate, with a spread of $\sim$25\%. For a few
magnet systems, more detailed cost estimates were provided based on
either existing designs (Cornell wigglers) or R\&D prototypes
already in progress (Daresbury/Rutherford undulators); in a similar
fashion, Brookhaven provided detailed cost estimates for the
superconducting insertion magnets at the IR based on experience with
similar magnet designs.

Estimates of EDIA labor costs were based upon reviews of recent
large accelerator magnet and power supply projects at SLAC and
Fermilab, where the materials, fabrication and EDIA labor fractions
are well known.  The fractional distribution of EDIA among several
types of laborers, which were costed at the standardized labor
rates,  was assigned on the basis of project management experience.

\subsection{Component Counts}

The number of conventional and superconducting magnets and magnet
styles in each of the ILC Areas is shown in Table
\ref{tab:MagnetStyleCount}. There are compelling reasons to reduce
the number of magnet styles - to reduce cost and increase
maintainability and reliability - and this process is an iterative
one, that has not yet been fully optimized.

\begin{landscape}
\stepcounter{tablcl}\begin{table}
   \caption{Numbers of Conventional (Normal Conducting, NC) and Superconducting
   Magnets and Magnet Styles in ILC Areas.}
   \label{tab:MagnetStyleCount}
   \begin{center}
\setlength{\tabcolsep}{2pt}
      \begin{tabular}{| l || c | c ||
       c | c | c || c | c |
       c | c | c || c | c |
       c | c | c | c |}
% start with the top column headings
\hline
 & \multicolumn{2}{c||}{Grand Totals} &
   \multicolumn{3}{ c|| }{Sources} &
   \multicolumn{5}{c||}{Damping Rings} &
   \multicolumn{2}{ c| }{2 RTMLs} &
   \multicolumn{2}{c|}{2 Linacs} &
   \multicolumn{2}{c|}{2 BDS} \\  \cline{2-17}  \cline{2-17}
% line 2 headings
   \multicolumn{1}{|c||}{Magnet Type} &   &   &
   & e- & e+ & & e- DR & e+ DR & e- Inj/Ext & e+ Inj/Ext &
  &   &  &   &  &
   \\ \cline{5-6} \cline{8-11}
 & \# of &  \multicolumn{1}{c||}{total} & \# of
   & total & total  & \# of & total &  total & total & total &
 \# of &  total & \# of &  total & \# of &  total
   \\ [-8pt]
% line 3 headings
 & styles & qty. & styles & qty. & qty. & styles & qty. & qty. & qty. & qty. &
   styles & qty. & styles & qty. & styles & qty. \\ \hline & & & & & & & & & & & & & & & & \vbdlspacing \hline
% now for the actual data
  ~Dipole & 22 & 1356 & 6 & 25 & 157 & 2 & 126 & 126 & 8 & 8 &
    6 & 716 & 0 & 0 & 8 & 190 \\ \hline
  ~NC quad & 37 & 4165 & 13 & 76 & 871 & 4 & 747 & 747 & 76 & 76 &
    5 & 1368 & 0 & 0 & 15 & 204 \\ \hline
  ~SC quad & 16 & 715 & 3 & 16 & 51 & 0 & 0 & 0 & 0 & 0 &
    0 & 56 & 3 & 560 & 10 & 32 \\ \hline
  ~NC sextupole & 7 & 1050 & 2 & 0 & 32 & 2 & 504 & 504 & 0 & 0 &
    0 & 0 & 0 & 0 & 3 & 10 \\ \hline
  ~SC sextupole & 4 & 12 & 0 & 0 & 0 & 0 & 0 & 0 & 0 & 0 &
    0 & 0 & 0 & 0 & 4 & 12  \\ \hline
  ~NC solenoid & 3 & 50 & 3 & 12 & 38 & 0 & 0 & 0 & 0 & 0 &
    0 & 0 & 0 & 0 & 0 & 0  \\ \hline
  ~SC solenoid & 4 & 16 & 1 & 2 & 2 & 0 & 0 & 0 & 0 & 0 &
    1 & 8 & 0 & 0 & 2 & 4  \\ \hline
  ~NC corrector & 9 & 4016 & 1 & 0 & 840 & 3 & 540 & 540 & 0 & 0 &
    4 & 2032 & 0 & 0 & 1 & 64  \\ \hline
  ~SC corrector & 14 & 1374 & 0 & 32 & 102 & 0 & 0 & 0 & 0 & 0 &
    0 & 84 & 2 & 1120 & 12 & 36  \\ \hline
  ~Kickers/septa &  11 & 227 & 0 & 0 & 19 & 5 & 46 & 46 & 0 & 0 &
    1 & 52 & 0 & 0 & 5 & 64 \\ \hline
  ~SC wiggler & 1 & 160 & 0 & 0 & 0 & 1 & 80 & 80 & 0 & 0 &
    0 & 0 & 0 & 0 & 0 & 0  \\ \hline
  ~NC oct/muon spoiler & 3 & 8 & 0 & 0 & 0 & 0 & 0 & 0 & 0 & 0 &
    0 & 0 & 0 & 0 & 3 & 8  \\ \hline
  ~SC octupole & 3 & 14 & 0 & 0 & 0 & 0 & 0 & 0 & 0 & 0 &
    0 & 0 & 0 & 0 & 3 & 14  \\ \hline
  ~SC undulator & 1 & 27 & 1 & 0 & 27 & 0 & 0 & 0 & 0 & 0 &
    0 & 0 & 0 & 0 & 0 & 0  \\ \hline & & & & & & & & & & & & & & & & \vbdlspacing \hline
  ~Overall Totals & 135 & 13190 & 30 & 163 & 2139 & 17 & 2043 & 2043 & 84 &
  84 &
    17 & 4316 & 5 & 1680 & 66 & 638 \\ \hline
  ~Totals w/o Correctors & 112 & 7800 &
    \multicolumn{14}{ c }{} \\ \cline{1-3}
  \multicolumn{1}{ r ||}{Total NC} & 92 & 10872 &
    \multicolumn{14}{c }{} \\ \cline{2-3}
  \multicolumn{1}{ r ||}{Total SC} & 43 & 2318 &
    \multicolumn{14}{c }{} \\ \cline{2-3}
      \end{tabular}

   \end{center}
\end{table}
\end{landscape}

\clearpage %\setcounter{page}{1}
\setcounter{figlcl}{0}
\setcounter{tablcl}{0}
\setcounter{secnumdepth}{4} \setcounter{subsection}{0}
\setcounter{subsubsection}{0}
\usecounter{subsubsection}

\setcounter{section}{1} \renewcommand{\picturefolder}{./vacuum/}
\section{Vacuum Systems}\label{sectVacuum}

\subsection{Overview}

The ILC has over $\sim$80~km of beamlines which must be kept under vacuum
to limit the beam-gas scattering and operate the RF cavities.
Different areas of the machine
present different challenges but fortunately, there is an experience
base at existing accelerators for essentially all of the systems, to
facilitate design and costing~\cite{NLCvacuum,NLCEriksson,TESLAvacuum}. The largest and most complex are the
vacuum systems for the cryomodules containing superconducting
cavities that accelerate the beam. There are $\sim$1680
cryomodules in the main linacs,
electron and positron booster linacs and bunch compressors. There
are also single cavity cryomodules in the damping rings and beam
delivery systems. These cryogenic units require separate vacuum
systems for the beam line, the insulating vacuum and the waveguides.

Other beamlines throughout the ILC pose particular challenges. The
lifetime of the electron source photocathode requires a vacuum in
the range of a pico-Torr. The superconducting undulator for the
positron source is a warm bore chamber with a very small aperture.
Chambers for bending magnets in the damping rings and elsewhere
require antechambers and photon absorbers for the synchrotron
radiation. The presence of electron cloud in the positron damping
ring and ions in the electron damping ring can seriously impact
performance and requires mitigation. Beam-gas scattering in the beam
delivery must be limited to reduce backgrounds in the experimental
detectors. The designs for each system and costing approach are
discussed in more detail below and in reference~\cite{bib:VacTechNote}.

\subsection{Technical Issues}

\subsubsection{Linac Cryomodules}

There are $\sim$20~km of cryomodules in the main linac and another $\sim
$1.6~km of modules in the sources and bunch compressors. Each cryomodule
has separate vacuum systems for the accelerating structures, the
insulating vacuum and the transmission waveguides. The structure
vacuum vessel holds the niobium cavities and is at 2K cryogenic
temperature. This system must produce very low quantities of
particulates as these can contaminate the cavities causing field
emission and lowering the available gradient. The system must also
be able to produce ultra-high vacuum at room temperature to
eliminate the risk of residual gases condensing on the niobium
walls during cooldown. The beamline vacuum is segmented into
strings of 154.3~m. Each string has an insulating vacuum break and a
port for valves and ion pumps. Every other string has additional
valves, pumps, leak detection, and vacuum diagnostics. Each group of
4 strings (617~m) has cold vacuum isolation valves.  A
vacuum/diagnostics station is installed between every 16 strings
(2.472~km).  These stations have slow start turbo-molecular pumps,
leak detection, clean venting systems, and warm isolation valves.

\stepcounter{figlcl}\begin{figure} [t]
   \begin{center} \vbabove
      \includegraphics[width=13cm]{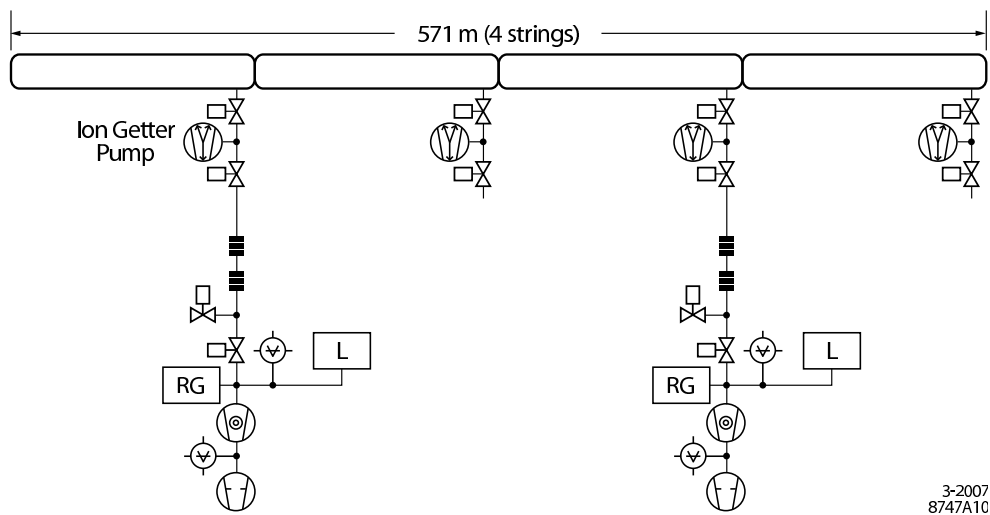}
      \vbabovecaption \caption[Beamline vacuum system.] {Beamline vacuum system -- 2 turbo-molecular pumps (TMP)
      with high sensitivity leak detector (LD)
     and residual gas analyzer (RGA), safety,
      clean venting system, slow start pumping etc.}
      \label{fig:vacuum1}
   \end{center} \vbbelow
\end{figure}

\stepcounter{figlcl}\begin{figure} [b]
   \begin{center} \vbabove
      \includegraphics[width=13cm]{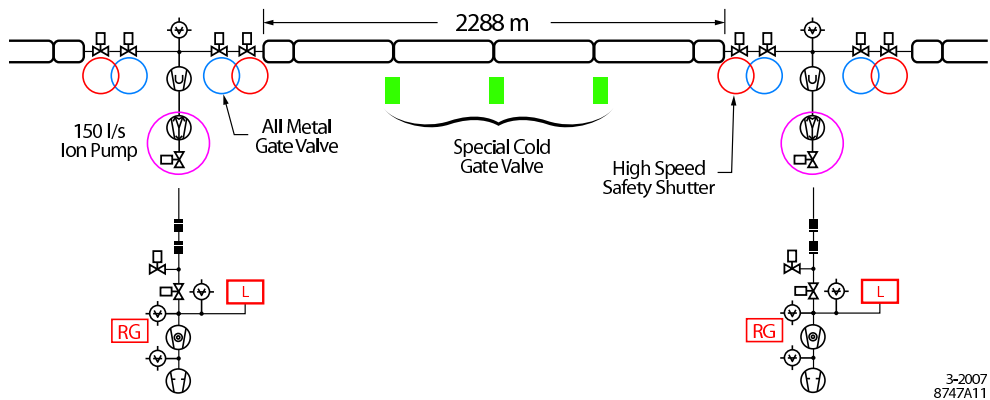}
      \vbabovecaption \caption{Beamline vacuum system gates and valves.}
      \label{fig:vacuum2}
   \end{center} \vbbelow
\end{figure}

The insulating vacuum system must maintain a typical pressure of
$\sim0.1$~mTorr, a regime where high voltage breakdown is a serious
issue. It is complicated by the pump cabling from the main system
which must pass through the insulating vacuum. The system is
segmented into 154.3~m strings consistent with the beamline vacuum.
Each string has valves, a turbomolecular pump, and bypass valves.
Every other string additionally has a leak detector and a large
screw pump.

\stepcounter{figlcl}\begin{figure} [t]
   \begin{center} \vbabove
      \includegraphics[width=13cm]{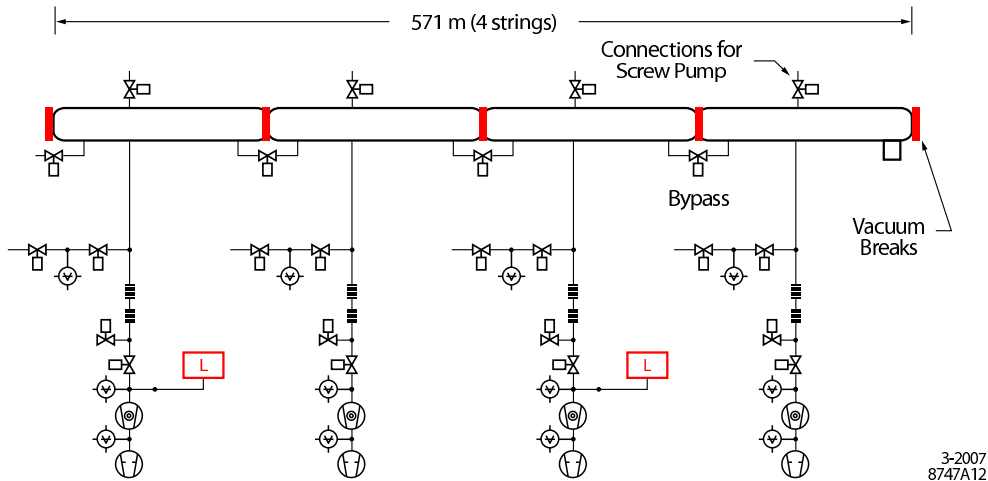}
      \vbabovecaption \caption[Insulating vacuum system.]{Insulating vacuum system -- 4 TMP pumping units:
      2 with LD (leak detector) + 2 large screw pump for fore pumping.}
      \label{fig:vacuum3}
   \end{center} \vbbelow
\end{figure}

Much of the transmission waveguide vacuum is at room temperature,
but it must transition to helium temperatures at the couplers.  In
addition, the rf power being transmitted is very high, so
multipactoring and arcing must be considered in the design.  There
is a valve for each coupler.  Every cryomodule has an ion pump and
titanium sublimation pump, and every 3 cryomodules have a
turbomolecular pump, a scroll fore pump and a leak detector.

\stepcounter{figlcl}\begin{figure} [b]
   \begin{center} \vbabove
      \includegraphics[width=13cm]{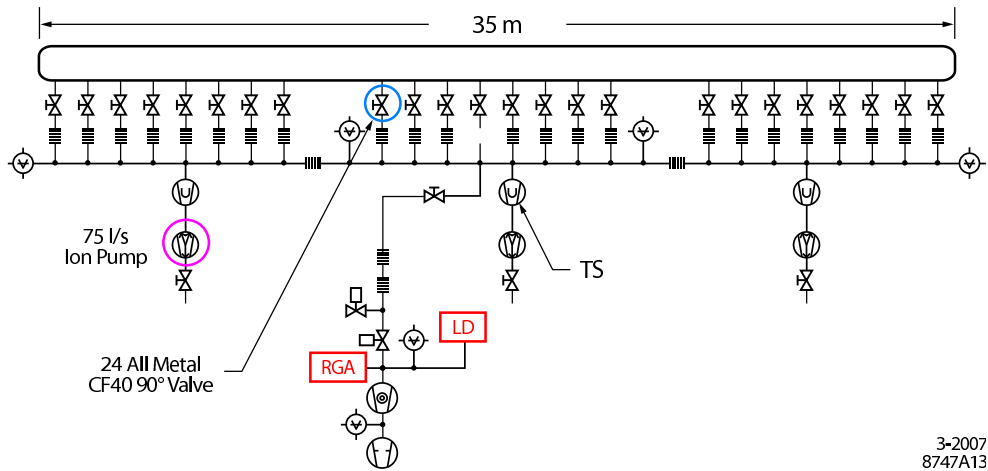}
      \vbabovecaption \caption{Waveguide and coupler vacuum system.}
      \label{fig:vacuum4}
   \end{center} \vbbelow
\end{figure}

While the cryomodule vacuum system is complex, costs can be
estimated from work done for the TESLA TDR proposal and from recent
projects such as SNS. Standard parts, were estimated from vendor
quotations and from recent large quantity procurements.

\subsubsection{Damping Ring and Beam Delivery Cryomodules}

The damping ring accelerating rf is single 650 MHz cavities in
individual cryomodules. The beam delivery also uses superconducting
crab cavities with individual cryomodules.  (See Sections \ref{sectDR} and
\ref{sectCrabCav} for a description of damping ring cryomodules and crab
cavity cryomodules.)

\subsubsection{Polarized Electron Source}

The electron source is a DC gun with a laser illuminated
photocathode similar to the electron guns at SLAC and Jefferson Lab.
To maintain photocathode lifetime, the pressure must be $<$ 3$\times
10^{-11}$~torr. This is achieved by incorporating large ion pumps
and non-evaporable getter (NEG) pumps.

\stepcounter{tablcl}\begin{table} \vbabove
   \caption[Transport lines for the ILC Electron Source System.]{Transport lines for the ILC Electron Source System.
   Vacuum specifications, beam aperture inner diameters,
   and lengths are noted. Except in the case of the accelerator
   sections, the vacuum chamber material is stainless steel.}
   \label{tab:ElectronSourceVacuum}
\setlength{\tabcolsep}{4pt}
   \begin{center}
%   {\tiny
      \begin{tabular}{|l|c|c|c|c|l|}\hline
  Beamline & Max & Aperture & Length & Number & Comments \\ [-6pt]
    & Pressure & Diameter & (m) & of & \\ [-6pt]
    & (nTorr) & (cm) & & Beamlines & \\ \hline & & & & & \vbdlspacing \hline
  Gun & $10^{-3}$ & 4 & 0.2 & 2 & Integrated into gun design \\
    \hline
  Gun & $10^{-3}$ to 0.1 & 3 & 1 & 2 & Differential pumping
   \\ [- 6pt]
  combining & & & & & needed to protect \\ [- 6pt]
  beam line & & & & & gun vacuum \\ \hline
  Transport & 1 & 4 & $\sim$15 & 1 & \\ [- 6pt]
  through   &  &  &  &  &  \\ [- 6pt]
  Bunching  &  &  &  &  &  \\ [- 6pt]
  System    &  &  &  &  &  \\ \hline
  NC beam & 10 & 4 & $\sim$17.5 & 1 & \\ [- 6pt]
  lines & & & & & \\ \hline
  SC RF & $<$1 & 7 & $\sim$273 & 1 & 8 strings (of 3) cryomodules, \\ [- 6pt]
  & & & & & adopt vacuum specification \\ [- 6pt]
  & & & & & for Main Linac\\ \hline
  Dump beam & 10 & 4 & 12 & 1 & \\ [- 6pt]
  line & & & & & \\ \hline
  ELTR & 10 & 4 & $\sim$140 & 1 & Linac to Ring beam line \\ \hline
      \end{tabular}

 %  }
   \end{center} \vbbelow
\end{table}

\subsubsection{Positron Source}

The positron source undulator and target vacuum systems are
particularly challenging.  Electrons are transported through a superconducting undulator to
produce $\gamma$-rays.  The superconducting undulator is a cold bore
chamber with a small aperture.  The $\gamma$-rays are then directed
onto a target to produce positrons.  The positron target has a very
large power load deposted into the target and nearby structures.

\stepcounter{tablcl}\begin{table} [b] \vbabove
   \caption[Transport lines for the ILC Positron System.]{Transport lines for the ILC Positron System. The reasoning
   behind the specification is noted and is subject to discussion.
   Vacuum specifications and aperture inner diameters are noted.
   Except in the case of the accelerator sections, the vacuum chamber
   material is stainless steel.}
   \label{tab:PositronSourceVacuum}
\setlength{\tabcolsep}{4pt}
   \begin{center}
      \begin{tabular}{|l|c|c|c|l|}\hline
  Beamline & Max Pressure & Aperture & Length & Comments \\
    & (nTorr) & (cm) & (m) &  \\ \hline & & & & \vbdlspacing \hline
  Chicane 1 & 50 & 2 & 300 & halo generation \\ \hline
  Undulator & 100 & 0.6 & 290 & fast ion, Daresbury \\ \hline
  Chicane 2 & 50 & 20 & 300 & halo generation \\ \hline
  Photon line & 1000 & 4.5 & 500 & \\ \hline
  Positron transport & 100 & 15 & 5,100 & \\ \hline
  NC RF & 20 & 6 and 6-4.6 & 115 & 1.27 m and 4.3 m sections \\
    \hline
  SC RF & $<$1 & 7 & 280 & 12.6 m sections \\ \hline
  Linac-to-Ring & 50 & 2 & 80 & \\ \hline
  Other & 100 & 6 & 300 & \\
  \hline
      \end{tabular}

   \end{center}  \vbbelow
\end{table}

\subsubsection{Damping Ring}

The most challenging issues for the damping ring vacuum systems are
suppression of the electron cloud in the positron damping ring and
ions in the electron damping ring. A variety of techniques are used,
including low residual pressure, low SEY coatings, and possibly
grooved chambers or clearing electrodes. Lifetime considerations
require pressures of less than 1 nTorr which is achieved with neg
coated chambers. The bend magnet vacuum pipe requires an antechamber
with a photon absorber to collect synchrotron radiation emitted.
(For details see \ref{sectDR}.)

The wiggler straight vacuum system for the ILC damping rings
consists of separate chambers for the wiggler and quadrupole
sections.
%A schematic cross-section of the wiggler chamber is shown
%Figure \ref{fig:vacuum5}.
The chamber is a machined and welded
aluminum unit designed as a warm bore insert which is mechanically
decoupled from the wiggler and cryogenic system. A NEG pumping
system \cite{bertolini} and photon absorber are incorporated in ante
chambers. Integral cooling is incorporated to minimize distortion of
the chamber and thermal load on the wiggler cryostat during NEG
regeneration. A NEG surface coating will be used on the main chamber
bore to minimize secondary electron yield \cite{benvenuti}. Clearing
electrodes will also be incorporated to reduce the electron cloud.

%\stepcounter{figlcl}\begin{figure}
%   \begin{center} \vbabove
%      \includegraphics[width=13cm]{\picturefolder vacuumfig5.jpg}
%      \vbabovecaption  \caption{ILC damping ring wiggler chamber.}
%      \label{fig:vacuum5}
%   \end{center} \vbbelow
%\end{figure}

The quadrupole chamber is welded aluminum, also incorporating NEG
coating for secondary electron yield reduction. Bellows, a BPM
assembly and an ion pump are incorporated. The quadrupole chamber is
completely shadowed by the wiggler chamber photon absorbers and does
not absorb any of the photon power from upstream wigglers.

\subsubsection{Ring to Main LINAC}

Each of the two Ring to Main Linac transport sections contains a
room temperature transport line of $\sim$15~km length,
superconducting RF sections of $\sim$0.5~km length, and additional
room temperature beamlines of $\sim$1.0~km length.  The vacuum level
in the long room temperature transport line is set by requirements
on the beam-ion interaction in the electron system to
$\sim$20~nTorr. The vacuum level in the remaining room temperature
beamlines is set by beam scattering requirements to 100~nTorr, at
which level about $1\times10^{-6}$ of the beam population is
scattered out of the acceptance.  The superconducting RF sections
have vacuum requirements and system designs which are identical to
those of the main linac, i.e., beamline and isolation vacuum
systems.  Although the RTML contains room temperature bending
sections they are not expected to need photon stops or other photon
power absorbers because the average current is low and the bending
radii in the RTML are kept large to limit emittance growth from
incoherent synchrotron radiation effects.

\subsubsection{Beam Delivery System}

The beam delivery system transport requires special attention to
limit backgrounds in the experimental detectors. In order to reduce
the residual beam-gas scattering to acceptable levels, the line
pressure near the interaction region needs to be $<$1~nTorr.  The
design is complicated by the requirement for small chamber
diameters.  The small chamber diameter and the low pressure require
close spacing of the ion pumps, bake-outs and the use of NEG coated
chambers.

\subsection{Cost Estimation}

The main parts of the vacuum systems were obtained from quotations
from vendors and from recent large quantity procurements.
``Consumables,'' such as flanges, gaskets, bolts and nuts, cables,
etc, were either not yet included or were estimated for quantity
discounts of catalog items.

\clearpage %\setcounter{page}{1}
\setcounter{figlcl}{0}
\setcounter{tablcl}{0}
\setcounter{secnumdepth}{4} \setcounter{subsection}{0}
\setcounter{subsubsection}{0}
\usecounter{subsubsection}

\setcounter{section}{2} \renewcommand{\picturefolder}{./modulator/}
\section{Modulators}\label{sectModulator}

\subsection{Overview}

The accelerating gradient for the ILC main linacs is supplied by
superconducting 1.3~GHz cavities powered by 560 10~MW RF stations,
each with a modulator, klystron and RF distribution system. Another
86 similar stations are used in the e$^+$ and e$^-$ Sources and RTML
bunch compressors. The damping ring RF power is supplied by 650~MHz
superconducting cavities powered by 1.2~MW peak power klystrons.
These are fed from a DC supply and do not have pulsed modulators.
There are also a few special purpose S-band RF stations for
instrumentation and a 3.9~GHz RF station to power the crab cavities
near the Interaction
\stepcounter{tablcl}\begin{table} [htb] \vbabove
   \caption{Modulator Specifications \& Requirements Assuming Klystron $\mu$P=3.38, Effy=65\%.}
   \label{tab:ModulatorRequirements}
   \begin{center}
      \begin{tabular}{|l|c|c|}\hline
  \multicolumn{1}{|l}{Specification} &
      \multicolumn{1}{|c}{Typical} &
      \multicolumn{1}{|c|}{Maximum} \\ \hline & & \vbdlspacing \hline
  Charger input voltage kV RMS & 7.67 & 8 \\ \hline
  Charger average power input kW & 147.9 & 161.7 \\ \hline
  Charger efficiency & 0.93 & 0.93 \\ \hline
  Charger DC output voltage = Modulator kV$_{\rm in}$ & 10.8 & 11.3 \\ \hline
  Charger DC avg output current = Modulator A$_{\rm in}$ & 13.26 & 13.26 \\
    \hline
  Charger average power output @ 5 Hz kW & 137.5 & 150.3 \\ \hline
  Modulator efficiency & 0.94 & 0.94 \\ \hline
  Modulator pulse voltage output = Pulse Transformer kV$_{\rm in}$ & 10.16 &
    10.18 \\ \hline
  Modulator pulse current output = Pulse Transformer A$_{\rm in}$ & 1560 &
    1680 \\ \hline
  Modulator average power output @ 5 Hz kW & 129.3 & 141.3 \\ \hline
  Pulse transformer step-up ratio & 12 & 12 \\ \hline
  Pulse transformer efficiency & 0.97 & 0.97 \\ \hline
  Pulse transformer voltage out = Klystron kV$_{\rm pk}$ & 115.7 & 120 \\
    \hline
  Pulse transformer current out = Klystron A$_{\rm pk}$ & 133.0 & 140 \\ \hline
  Pulse transformer average power output @ 5 Hz kW & 125.4 & 137.1
    \\ \hline
  High voltage pulse duration (70\% to 70\%) ms & 1.631 & 1.7 \\
    \hline
  High voltage rise and fall time (0 to 99\%) ms & $<$0.23 & 0.23 \\
    \hline
  High voltage flat top (99\% to 99\%) ms & 1.565 & 1.565 \\ \hline
  Pulse flatness during flat top \% & $<\pm0.5$ & $\pm0.5$ \\ \hline
  Pulse to pulse voltage fluctuation \% & $<\pm0.5$ & $\pm0.5$ \\ \hline
  Energy deposit in klystron from gun spark J & $<20$ & 20 \\ \hline
  Pulse repetition rate, Hz & 5 & 5 \\ \hline
  Klystron filament voltage V & 9 & 11 \\ \hline
  Klystron filament current A & 50 & 60 \\ \hline
      \end{tabular}
   \end{center} \vbbelow
\end{table}
\noindent
Point. This section describes only the 1.3~GHz modulators, Damping
Ring HVPS system and associated components.

\subsection{Technical Description}

The 10~MW L-Band RF power stations for the ILC are installed in the
support tunnel, spaced approximately 38 meters apart. The L-Band
Modulator baseline design was developed for the TESLA Test Facility
at DESY, and has been adopted for the European XFEL. Three FNAL
units and 5 commercial units have brought online at DESY starting in
1993. The design has a series on-off solid state switch with partial
capacitor discharge. The ILC unit varies from this design in two
minor ways: (1) A new solid state redundant switch is employed to
form the 1.7~msec output pulse, for better reliability; and (2) the
input charger will operate from a voltage of 8~kV instead of 480~V
to eliminate the AC input step-up transformer in the current design.
The modulator specifications and requirements are summarized in
Table \ref{tab:ModulatorRequirements}.

\stepcounter{figlcl}\begin{figure} [htb]
   \begin{center} \vbabove
      \includegraphics[width=\textwidth]{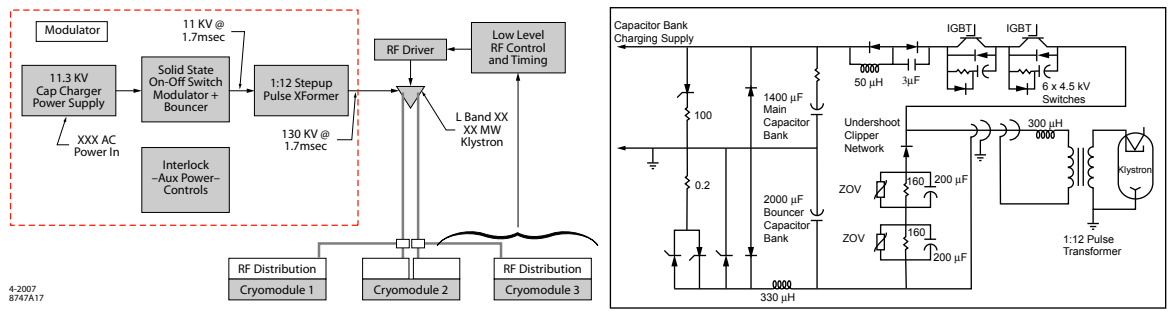}
      \vbabovecaption \caption{Modulator schematic and L-Band RF station
      block diagram (1 of 646).}
      \label{fig:ModulatorBlockSchematic}
   \end{center} \vbbelow
\end{figure}

\stepcounter{figlcl}\begin{figure} [htb]
   \begin{center} \vbabove
      \includegraphics[width=\textwidth]{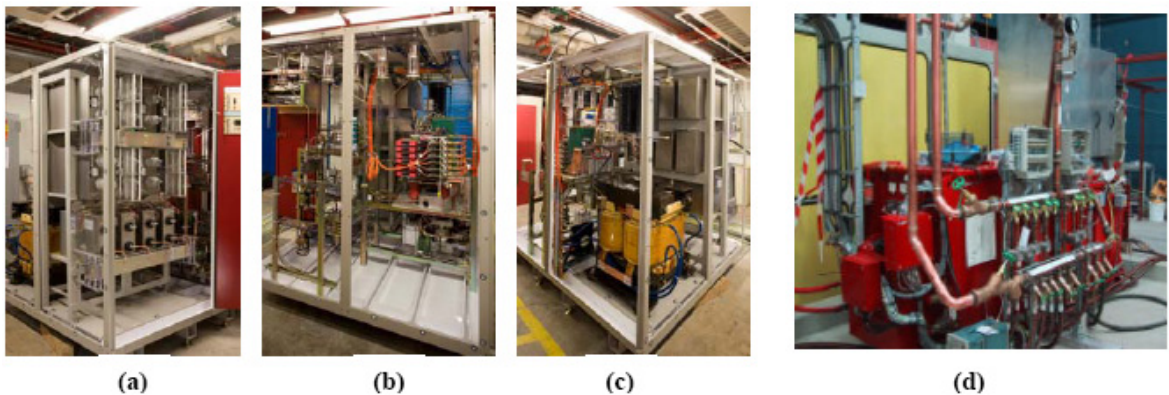}
      \vbabovecaption \caption[Modulator photos]{(a) Capacitor stack, (b) Dual IGBT switch, (c)
      Bouncer choke, (d) Pulse transformer.}
      \label{fig:ModPhotos}
   \end{center} \vbbelow
\end{figure}

The block schematic is shown in
Figure~\ref{fig:ModulatorBlockSchematic}. Photos of current
prototypes are shown in Figure~\ref{fig:ModPhotos}. Operation is
straightforward: The charger delivers a DC voltage to the storage
capacitors of approximately 11~kV. The modulator main switch is then
triggered and held closed for 1.7~msec. Capacitor current flows
through the switch to the step-up transformer input. At the same
time, an auxiliary droop compensation ``bouncer" circuit is fired to
maintain the pulse top flat to within  $\pm$0.5\% during the RF
drive period. The slightly above 10~kV drive pulse (to compensate
for Bouncer voltage) is delivered to the input of the pulse
transformer in order to produce at least 115.7~kV 133.0~A to the
klystron for rated 10~MW peak output.

\begin{figure}
   \begin{center} \vbabove
      \includegraphics[width=11cm]{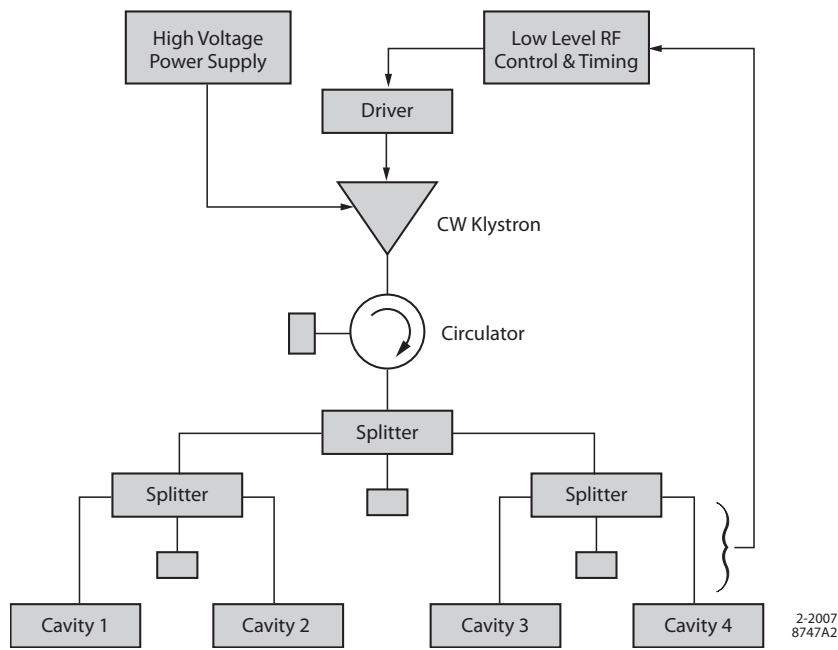}
      \vbabovecaption \caption{Damping Ring 1.2~MW RF station (1 of 20).}
      \label{fig:DRBlockSchematic}
   \end{center} \vbbelow
\end{figure}

The Damping Rings have 650~MHz CW stations using 1.2~MW peak power
klystrons, 20 in total for 2 rings. Power is supplied from a DC
supply of 2.0~MW delivering 50-75~kV at 17-10~A~DC. The RF envelope
is controlled by the low level RF and timing to maintain
stability and clearing gaps as needed. The station block diagram is
shown in Figure~\ref{fig:DRBlockSchematic}.

\stepcounter{figlcl}

\subsection{Technical Issues}

\subsubsection{L-Band}

There are no major technical issues with the L-Band modulator as
long as the entire system has sufficient overhead (redundancy) to
compensate for a failed station. To achieve an acceptable
availability, the linac energy and beam current parameters must be
chosen to provide some RF spare stations. Redundancy of internal
components such as IGBT switches and sectioning of chargers for N+1
redundancy\footnote{N+1 design segments a single unit such as a
power supply into N parallel or series smaller modules components
plus an additional spare so one module can fail without interrupting
operation. N+1 design is used in stacked or parallel power supplies,
capacitors and IGBT's. Such designs can achieve much higher overall
Availability especially if modules can be exchanged without
interrupting operation (Hot Swap capability). This is only possible
in lower voltage units.} is also important. Currently this is only
partially implemented in the prototypes.

The present design which develops the drive pulse at low voltage and
high current has larger losses than would be experienced with a
higher voltage design. This is not a major technical issue, but a
cost, size and weight issue. Installation and repair during
operations will be more difficult with multi-ton components such as
the transformer and main capacitor-switch multi-cabinet assembly.

An alternative modulator design is being investigated to address
these issues, including the possibility of significant cost
reduction. The design would reduce the overall footprint and
eliminate the step-up transformer and other oil-filled components.

\subsubsection{Damping Rings 650~MHz}

The Damping Ring stations are modeled after similar stations in
operation in Italy, Japan and the US. The power supply systems are very
well understood. The only change desired would be to make them N+1
redundant internally for higher reliability. This will be investigated and
will not have a large cost impact.

\subsection{Cost Estimation}

The L-Band modulator cost model was derived directly from the latest
FNAL design, extrapolated as needed to fit the ILC specifications. A
traditional bottom-up estimate was made and learning curves applied
to first-unit costs for an estimated manufacturing cost. Both single
and dual source factory models were examined, as well as sensitivity
to learning curve assumptions. These costs were also compared with
industrial estimates from both Europe and Japan. In general, the US
estimated cost lies between the two offshore commercial estimates.
Conservative learning curve exponents (``alphas''\footnote{``Alpha''
refers to the exponential decrease of costs with each doubling of
manufacturing volume. For details see section \ref{sect:VALvem}.})
were used for both parts and labor. Profit and factory support costs
were than applied, as well as the staging costs of preparing the
units for installation and final system checkout. These costs were
compiled in M\&S and FTE's. The factory models were documented in
detail for each Area subsystem and given to the responsible managers
for the Area rollups.

The modulator and charger costs were based on recent fabrication of
units at SLAC in partnership with LLNL. All parts were recently
purchased or fabricated at outside shops, and small additional
extrapolations were made for the total quantities.

The cost of the HV power supply for the Damping Ring CW tubes was
estimated based on recently built PEPII stations at SLAC, and
separate estimates from Italy and Japan. All estimates were in
reasonable agreement. The CW power rating needed is 25\% lower than
for PEP but there will be some additional cost for the N+1
implementation. Again a conservative learning curve was applied for
20 units.

\subsection{Table of Components}

Table \ref{tab:ModulatorCounts} shows the modulator component counts
in various Areas.

\stepcounter{tablcl}\begin{table} [h] \vbabove
   \caption{Modulator distribution by type and area.}
   \label{tab:ModulatorCounts}
\setlength{\tabcolsep}{4pt}
   \begin{center}
      \begin{tabular}{| l | c c c c c c c c c |}\hline
      Modulator type & Total & e$ ^{-} $ & e$ ^{+} $ & e$ ^{-} $ & e$ ^{+} $ & e$ ^{-} $ & e$ ^{+} $ & e$ ^{-} $ & e$ ^{+} $ \\ [-6pt]
       & & Inj & Inj & RTML & RTML & Linac & Linac & DR & DR \\
        \hline & & & & & & & & & \vbdlspacing \hline
      10 MW--1.3 GHz--5 Hz & 646 & 13 & 39 & 17 & 17 & 282 & 278 & 0 &
        0 \\ \hline
      1.2 MW--650 MHz--CW & 20 & 0 & 0 & 0 & 0 & 0 & 0 & 10 & 10 \\
        \hline
      \end{tabular}
   \end{center} \vbbelow
\end{table}

\clearpage %\setcounter{page}{1}
\setcounter{figlcl}{0}
\setcounter{tablcl}{0}
\setcounter{secnumdepth}{4} \setcounter{subsection}{0}
\setcounter{subsubsection}{0}
\usecounter{subsubsection}

\setcounter{section}{3} \renewcommand{\picturefolder}{./klystron/}
\section{Klystrons}\label{sectKlystrons}

\subsection{Overview}

The accelerating gradient for the ILC main linacs is supplied by
superconducting 1.3~GHz cavities powered by 560 10~MW RF stations,
each with a modulator, klystron and RF distribution system. Another
86 identical klystron/modulator systems are used in the e$^+$ and
e$^-$ Sources and RTML bunch compressors. The damping ring RF power
is supplied by 650 MHz superconducting cavities powered by 1.2~MW CW
klystrons. These are fed from a DC charging supply and do not have
modulators. There are also a few special purpose S-band RF stations
for instrumentation and a 3.9~GHz RF station to power the crab
cavities near the Interaction Point. This section describes the
1.3~GHz and damping ring klystrons.

\subsection{Technical Description}

\subsubsection{L-Band Klystrons}

The 10~MW L-band source in the ILC baseline design is a Multi-Beam
Klystron (MBK), chosen as a result of ten years of R\&D for TESLA and the European XFEL. The MBK is a design approach for linear beam
tubes that achieves higher efficiency by using multiple low space
charge (low perveance) beams. This
allows MBKs to operate at a lower voltage yet with a higher efficiency
than simpler single round beam klystrons, and provides a cost-effective and simplified design configuration for the ILC RF systems.

\begin{comment}
The multiple
low perveance beams in individual drift tubes, are joined together
in a single vacuum envelope with a common input and output cavity.
The gain cavities may or may not be common. The mechanical
complexity of the MBK is the fact that a single vacuum envelope
contains a large number of braze joints which scale linearly as the
number of beams is increased.  In addition, the electron gun tends
to be a problem in the multi-beam designs. Standard high convergence
gun optics is more difficult to implement with closely packed beams.
Low convergence optics and high current density cathodes produces
limited lifetimes.
\end{comment}

\stepcounter{figlcl}\begin{figure}[htb]
   \begin{center} \vbabove
      \includegraphics[width=13cm]{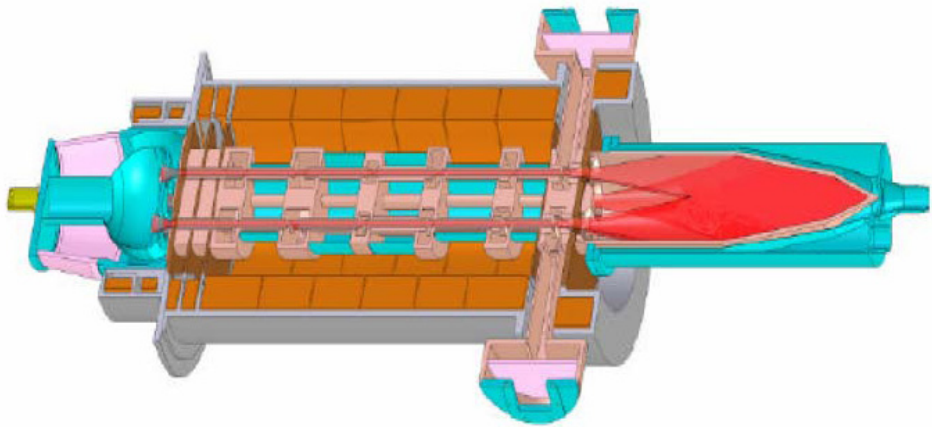}
      \vbabovecaption   \caption{Toshiba E3736 Multi-Beam Klystron.}
      \label{fig:ToshibaMBK}
   \end{center} \vbbelow
\end{figure}

\stepcounter{tablcl}\begin{table}[htb] \vbabove
   \caption{10 MW MBK parameters.}
   \label{tab:MBKpars}
   \begin{center}
      \begin{tabular}{|l|l|}\hline
  Parameter & Specification \\ \hline & \vbdlspacing  \hline
  Frequency & 1.3 GHz \\ \hline
  Peak Power Output & 10 MW \\ \hline
  RF Pulse Width & 1.565 ms \\ \hline
  Repetition Rate & 5 Hz \\ \hline
  Average Power Output & 78 kW \\ \hline
  Efficiency & 65\% \\ \hline
  Saturated Gain & $\ge$47 db \\ \hline
  Instantaneous 1 db BW & $>$3 MHz \\ \hline
  Cathode Voltage & $\le$120 kV \\ \hline
  Cathode Current & $\le$140 A \\ \hline
  Power Asymmetry & $\le$1\% \\ \hline
  Lifetime & $>$40,000 hours \\ \hline
      \end{tabular}
   \end{center} \vbbelow
\end{table}

MBK prototypes have been successfully built for the XFEL by three major
electron tube manufacturers: Thales, CPI and Toshiba.  These prototypes
were designed for essentially the same peak RF output power specifications
as required at ILC, yet with nearly twice the duty cycle as
required for the XFEL. All of these manufacturers have extensive past
experience in bringing prototype klystrons of state-of-the-art designs
into production models, and they are regarded as fully capable of
ramping up and producing the required quantities of MBKs to meet the
delivery schedule for the construction of the ILC.  A summary of the
MBK specifications is shown in Table~\ref{tab:MBKpars}.

\stepcounter{figlcl}\begin{figure}[t]
   \begin{center} \vbabove
      \includegraphics[width=\textwidth]{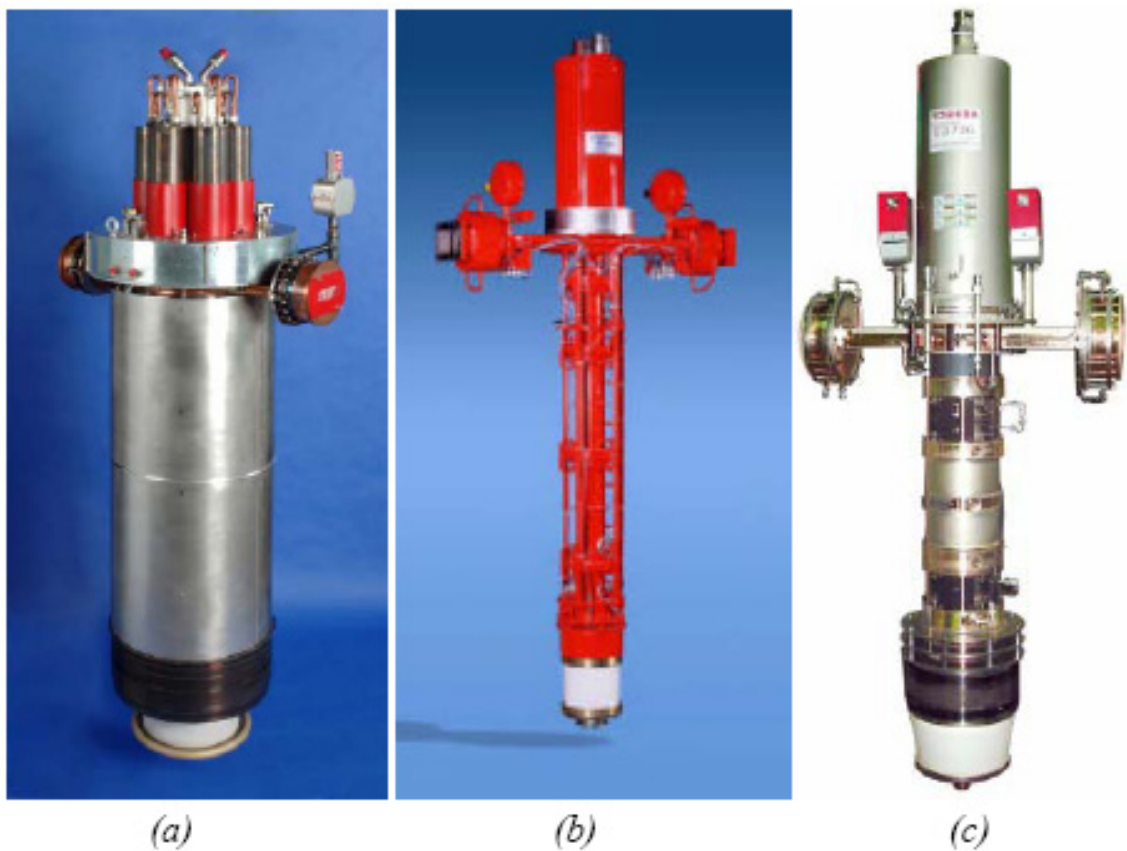}
      \vbabovecaption \caption{(a) CPI VKL-8301 (b) Thales TH1801 (c) Toshiba MBK E3736.}
      \label{fig:MBKPhoto}
   \end{center} \vbbelow
\end{figure}

\subsubsection{Damping Ring Klystrons}

The CW Klystron used in the damping rings is a frequency scaled
version of the 1.2~MW 500~MHz CW klystrons currently operating reliably at
SLAC and KEK \cite{dr14}.
Frequency scaling of klystrons is a common
practice in industry, which has a thorough understanding of the engineering
procedures to follow. Therefore, availability of the 650 MHz
klystrons is not considered to be a technical concern.

\subsection{Technical Issues}

\subsubsection{L-Band Klystrons}

\begin{comment}
The technical issues for the 10~MW MBKs are rather minor, and
consistent with the prototype stage of building state-of-the-art
microwave tubes. The RF design of the 10~MW MBK is mature and has
been developed over the last ten years. The future manufacturing
issues that might well be encountered during procurement for the ILC
should be identified by the early performance issues of the prototypes
built for the XFEL.  The initial results are strong indicators of
future success and are listed below:
\end{comment}

\begin{figure}[b]
   \begin{center} \vbabove
      \includegraphics[width=\textwidth]{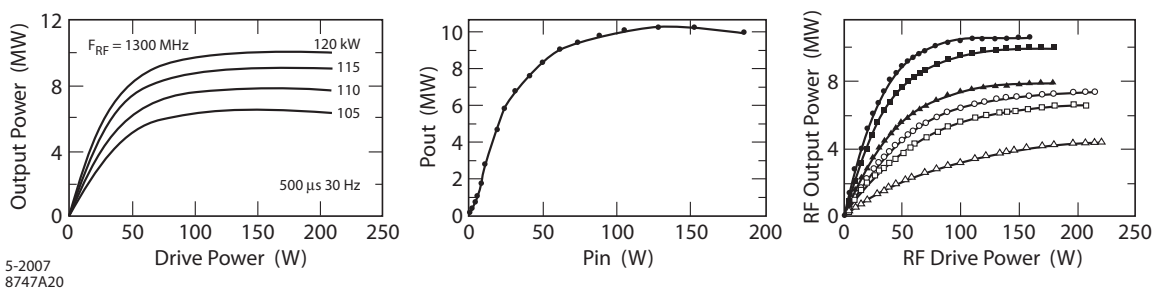}
      \vbabovecaption  \caption[Klystron test results.]{Test results for:  (a) CPI VKL-8301 at reduced          pulse width; (b) Toshiba MBK E3736 at full spec pulse width; (c)
      Thales TH1801 at reduced pulse width.}
      \label{fig:TestResults}
   \end{center} \vbbelow
\end{figure}

The RF design of the MBK klystron has matured through
several iterations of design and testing, and today
essentially all aspects of the electrical design are considered solved,
in particular, the choice of resonant frequencies to use for the cavities within
the klystron body, the beam focusing and others \cite{klys2},
\cite{klys3}, \cite{klys4}, \cite{klys5}.  Test results for all
three manufacturers are summarized in Figure \ref{fig:TestResults}.

\stepcounter{figlcl}

\begin{comment}
The Toshiba tube was reworked twice.  Once for a collector leak, and
once for gun arcing. Most recently, the Toshiba tube has been
operating in excess of 420 hrs and achieved 10.8~MW at 67\%
efficiency, Figure \ref{fig:TestResults} (b) \cite{klys2}. The CPI
tube was reworked for a waveguide leak and tested at DESY with 55\%
efficiency. Earlier it was tested at 10MW at CPI with reduced pulse
width, Figure \ref{fig:TestResults} (a) \cite{klys3}.  The output
circuit of the CPI tube was known to be less than optimal and
additional heating problems developed during further testing at DESY
at maximum saturated output power and nearly twice the ILC duty
cycle \cite{klys4}.  The Thales tube produced 10 MW at some rated
conditions Fig \ref{fig:TestResults} (c) \cite{klys5}, and may have
developed other problems due to high cathode current density
\cite{klys6}.
\end{comment}

The three most important technical issues for the MBK are lifetime,
manufacturability, and reliability. Lifetime for linear beam tubes
is dominated by cathode performance.  Both
the CPI and Toshiba MBKs have gun designs with cathode loading close to 2 A/cm$^2$. For
an M-type dispenser cathode, this low current density corresponds to
a lifetime in excess of 50,000 hours. However, this lifetime has
to be confirmed by suitable long-term operation tests.
The ``lifetime'' quoted in
Table~\ref{tab:MBKpars} is the time during which the klystron can operate at the
design performance specifications.

Construction of the MBK is inherently more complex than that of
single-beam klystrons due to the several linear
beam tubes being built into a single vacuum envelope. The number of
braze joints, the fixturing and tooling, and the processes required
to successfully construct, bakeout, and test an MBK are issues that
require attention in developing an efficient
assembly procedure that reduces the unit cost.
%, as noted by the number of vacuum leaks found in the prototypes.

For reliable performance, a robust thermal design of the output
circuit (output cavity, waveguide, and RF window) is important.
Since ILC MBK klystrons are being built for the
European XFEL, where they will operate at nearly twice the duty
cycle of the ILC, there will be significant thermal/mechanical
margin when operated for ILC specifications. The XFEL, however,
does not require operation at full power, so reliability
at 10~MW must also be demonstrated.

A remaining open issue is that the existing prototypes are vertical klystrons but a horizontal version is required for installation in the tunnel. While this is an engineering challenge, DESY is already working with the manufacturers to produce a horizontal klystron for the XFEL.

An alternate design is being developed to improve on the
manufacturability and reliability of the MBK. The Sheet Beam
Klystron (SBK) \cite{klys5} has fewer parts and processes than an
MBK.  It is focused with a periodic permanent magnet (PPM) system
and, as a result, is smaller and weighs less than an MBK.

\subsection{Cost Estimation}

The cost estimate for the MBKs was derived from cost estimates from
the manufacturers themselves, from the actual costs of the
prototypes, and from a bottoms-up factory model. The manufacturers'
estimates have inherent in them a set of assumptions that are
company specific and not transparent to an outside reviewer. These
assumptions cover the spectrum from proprietary processes to
corporate policy decisions regarding risk assessment.  The actual
costs of prototypes are useful to determine the characteristics of
possible learning curves a company may have used for quantity
discounting, and may be useful in benchmarking models such as those
used in the bottoms-up factory model.

The bottoms-up factory model used for the MBK was derived from the
model used for the NLC X-band klystron.  It is a comprehensive
factory model with explicit assumptions about variable costs such as
yield and learning curves, and fixed costs, such as up-front costs
of tooling and fixturing.  Fixed costs are more than 50\% of the
total cost during the prototype and pre-production stage of
manufacturing, and taper off to 10\% during the years of maximum
production rates. The range of estimates from all sources is well
within the risk associated with those estimates.

The cost estimate for the Damping Ring klystrons was based on actual
procurement costs for 1.2~MW klystrons already produced by industry.

\subsection{Components}
\stepcounter{tablcl}\begin{table}[h] \vbabove
   \caption{Klystron requirements by area.}
   \label{tab:KlysCount}
   \begin{center}
      \begin{tabular}{|l|c|c|c|c|c|}\hline
  Klystron & Main Linac & RTML & e$ ^{-}$ source & e$^{+}$ source & DRs \\ \hline & & & & & \vbdlspacing  \hline
  1.3 GHz & 560 & 34 & 13 & 39 & 0 \\ \hline
  650 MHz & 0 & 0 & 0 & 0 & 10 \\ \hline
      \end{tabular}
   \end{center} \vbbelow
\end{table}

\clearpage %\setcounter{page}{1}
\setcounter{figlcl}{0}
\setcounter{tablcl}{0}
\setcounter{secnumdepth}{4} \setcounter{subsection}{0}
\setcounter{subsubsection}{0}
\usecounter{subsubsection}

\setcounter{section}{4} \renewcommand{\picturefolder}{./RFDistribution/}
\section{RF Distribution}\label{sectRFDistribution}

\subsection{Overview}

The accelerating gradient for the ILC main linacs is supplied by
superconducting 1.3 GHz cavities powered by 560 10 MW RF stations,
each with a modulator, klystron and rf distribution system. Another
86 similar stations are used in the e$^+$ and e$^-$ sources and RTML
bunch compressors. The injector stations have slightly fewer
cavities (24-25) per RF unit. This section describes the baseline
design for distributing the high-power RF to the cavities.

\subsection{Technical Description}

\begin{figure}[htb]
   \begin{center} \vbabove
      \includegraphics[width=\textwidth]{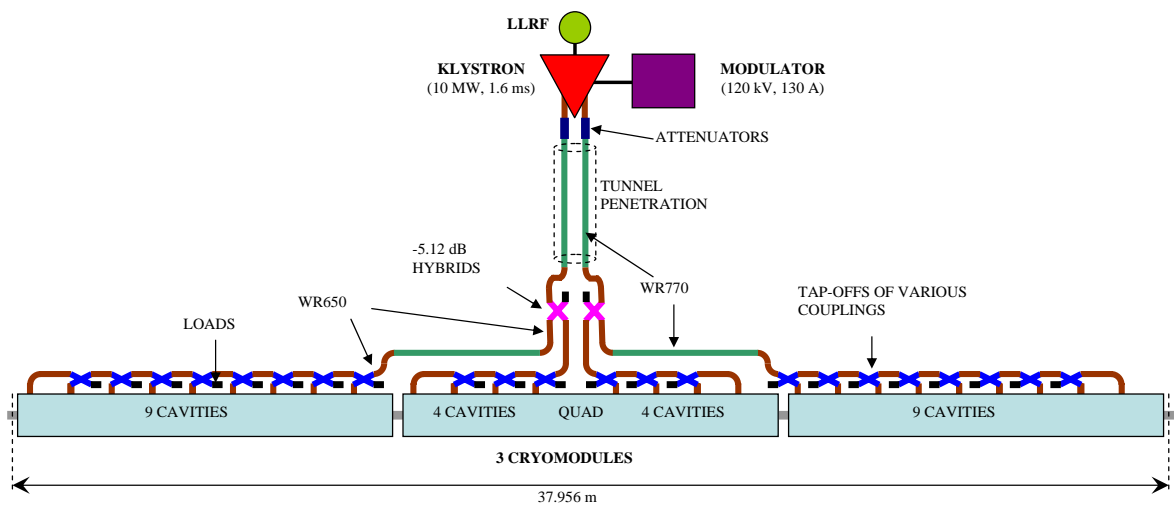}
      \vbabovecaption \caption[RF unit diagram showing the basic waveguide
        distribution layout.] {RF unit diagram showing the basic waveguide
        distribution layout between the klystron and 26 cavities
        in three cryomodules.}
      \label{fig:RFUnitDiagram}
   \end{center} \vbbelow
\end{figure}

The high-power L-band RF from each 10 MW klystron is brought to the
accelerator cavity couplers through an RF distribution system (see
Figure \ref{fig:RFUnitDiagram}). The standard linac RF unit powers
26 nine-cell superconducting cavities filling three cryomodules. The
upstream and downstream cryomodules contain nine cavities each, and
the middle one contains eight, with a superconducting quadrupole
magnet replacing the center cavity. This three cryomodule unit
occupies 37.956 m and, at the nominal 31.5 MV/m cavity gradient,
provides 846.6 MeV of acceleration (5$ ^{\circ} $ off-crest). With a 9 mA beam current, the
total power needed in the cavities is 7.62 MW, so some overhead is
included.

\stepcounter{figlcl}

The dual outputs of the klystron feed into two waveguides, each
carrying half the power, which run roughly 11 m to the linac through
a penetration between the service tunnel and the main tunnel. High-power
in-line attenuators allow more power to be sent through one arm than the other to accommodate different average gradient capabilities in the sets of cavities they feed. The
penetration emerges approximately at the center of the middle
cryomodule of the unit. Here, a hybrid splitter divides the power in
each waveguide with a 4:9 ratio (-5.12 dB). The lower power output
of each splitter feeds half the center cryomodule and the higher
power output is carried approximately 6 m to one of the outer
cryomodules.

Along each cryomodule, RF power is equally distributed among the
cavities in a linear waveguide system, passing through a series of
hybrid-style 4-port tap-offs. These tap-offs couple the appropriate
sequential fraction (1/8, 1/7, ...1/2 or 1/4,1/3,1/2) of the power
remaining in the line to all but the last cavity, which is directly
fed the remainder. The nominal power required in each cavity is
293.7 kW. Between the tap-offs, the remainder of the 1.326 m coupler
spacing is filled with modified straight waveguide sections whose
width is symmetrically tapered, with 1/4-wave transformer matching
steps, varying the guide wavelength to roughly yield the proper
inter-cavity phasing.

\begin{figure}[htb]
   \begin{center} \vbabove
      \includegraphics[width=13cm]{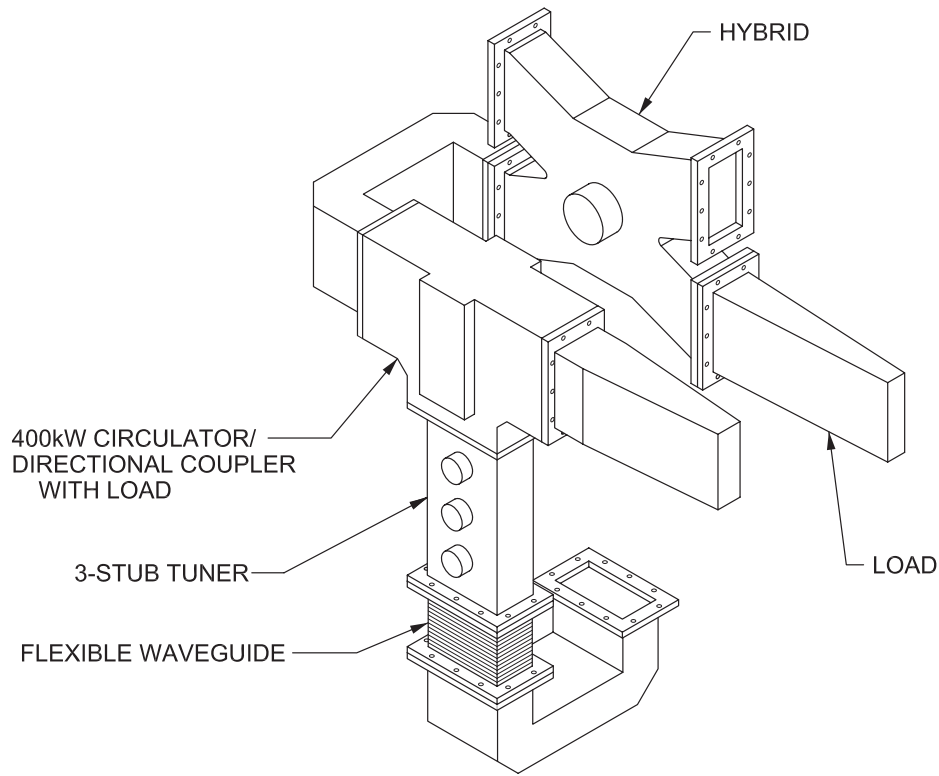}
      \vbabovecaption \caption[Waveguide circuit from tap-off hybrid to coupler
         input.] {Waveguide circuit from tap-off hybrid to coupler
         input, showing the various components (except for
         the directional coupler).}
      \label{fig:WGCircuit}
   \end{center} \vbbelow
\end{figure}

Between each tap-off output and its associated cavity coupler are a
circulator, a three-stub tuner, and a diagnostic directional coupler
(see Figure \ref{fig:WGCircuit}). The three-stub tuner allows fine
adjustment of both cavity phase and external coupling. The
circulator, with a load on its third port (thus technically an
isolator), absorbs RF power reflected from the standing-wave cavity
during filling and emitted during discharge. It provides protection
to the klystron and isolation between cavities. A couple of E-plane
waveguide U-bends are also needed to keep the system compact and
feed into the downward pointed coupler flange, and a short
semi-flexible section is included to relieve stress and ease
alignment tolerances.

\stepcounter{figlcl}

\subsection{Technical Issues}

\subsubsection{Waveguide}

The bulk of the distribution system consists of aluminum WR650
waveguide (6.50'' $\times$ 3.25'') components. This is the standard
rectangular waveguide for 1.3 GHz. Larger WR770 waveguide (7.70''
$\times$ 3.85''), which has 32\% lower attenuation, is used, with
matched transition sections, for the long runs through the
penetration and to the outer cryomodules in order to reduce system
losses.  The remaining loss, estimated at about 6.5\%, may be
further reduced by plating the inner walls of waveguide and/or
components with copper, which is 22\% less lossy.

The entire waveguide system, from the klystron window to the outer
coupler window, is pressurized with dry nitrogen to a pressure of 3
bar absolute. This prevents RF breakdown at the klystron window and
potential problems in the circulator or elsewhere. It requires
thicker-walled (0.25'') waveguide, but is more economical than
evacuating the system and also avoids multipactoring. The option of
using SF6 was considered undesirable due to safety regulations and
the risk of corrosion. Gas loss due to an open connection provides a
signal to disable RF operation as a safety measure during
installation and maintenance.

Relative phase changes along an RF unit due to temperature change
during installation, maintenance or operation are at most about
1.1$^{\circ}$ per degree Celsius. This can be easily controlled with
water cooling and insulation on some waveguide runs and components.
In addition to the water cooling required on the loads and
circulators, this water removes heat from the system that would be
more expensive to remove from the tunnel air.

As a cost-saving measure, electron-beam welding of waveguide joints
is used in place of expensive waveguide flanges and gaskets where
feasible. This is particularly useful for the penetration
waveguide, which cannot be put through in one piece, as it reduces
the effective waveguide cross-section.

\subsubsection{Tap-offs, Circulators and Tuners}

The tap-offs are compact four-port hybrids with WR650 ports of the
type used at TTF. Eight different designs are required, with various
coupling fractions: four each with 1/4, 1/3, and 1/2, and two each
with 1/9, 1/8, 1/7, 1/6, and 1/5. The 4/13 hybrids providing the 4:9
split of the power from each klystron arm may be of the same type.
Alternatively, a ``button type" hybrid with slight adjustability of
the split ratio might be used to provide added flexibility to tailor
the system for unequal cryomodule performance.

The circulators are ferrite-based, with a T-junction configuration
that provides a needed H-plane bend. The third port of this device
is matched into a load, which absorbs power propagating
backward, away from the cavity. In addition to being the most
expensive components in the distribution system, circulators
contribute the most loss (2\% out of $\sim$6.5\%). R\&D for an
alternate distribution system aims at eliminating the need for them~\cite{bib:Rfdist1}.

With three degrees of freedom, the three-stub tuner is a complicated
tool to use. It is, however, compact and well tested in TTF. It may
be desirable to replace it with an alternate phase shifter~\cite{bib:Rfdist2}, with the
movable coupler antenna providing Q$_{\rm ext}$ adjustment.

\subsection{Cost Estimation}

The cost estimate for the RF distribution system was derived from
cost estimates from component manufacturers, from the actual costs
of purchased components in small quantities, and from a bottoms-up
factory model. The estimates from waveguide component manufacturers
have inherent in them a set of company specific assumptions that are
not transparent to an outside reviewer. However, it is possible to
quantify high quantity discounts and learning curves from a
manufacturer if small quantities of a component were already
purchased. The actual costs of purchased components were used in two
different ways: 1) to calculate learning curves and quantity
discounts as mentioned above, and 2) to use in bottoms-up factory
models. The bottoms-up factory model approach developed a cost for a
single unit using information from previous experience in
constructing components such as loads and couplers. These costs
included material, labor and overhead, and fixturing costs. Once a
completed first unit cost was computed, a learning curve was applied
for high quantities.

The difference between the manufacturers' estimates and the factory
model was about 10\%. Cost estimates for some components reflected a
wide range in capacity and high-volume manufacturing experience
among the three regions.

\subsection{Components}

Table \ref{tab:RFPartsCount} gives a rough part count for the
components in the baseline RF distribution system of a single RF
unit. There are a total of 560 L-Band RF units in the main linacs
and approximately 86 more (some normal conducting) in the injectors and RTMLs.

\stepcounter{tablcl}\begin{table}[htb] \vbabove
   \caption{Component count for a single L-Band
   RF distribution system to 26 Cavities.}
  \label{tab:RFPartsCount}
   \begin{center}
      \begin{tabular}{| l | c || l | c |}\hline
  Component & \#/RF Unit & Component & \#/RF Unit \\  \hline & & & \vbdlspacing  \hline
  Circulators w/loads & 26 & H-Plane bends & 24 \\ \hline
  3-stub tuners & 26 & E-Plane bends & 4 \\ \hline
  Directional couplers & 32 & E-Plane U-bends & 52 \\ \hline
  Hybrids & 24 & Meters of WR770 & 34 \\ \hline
  Loads & 24 & Meters of WR650 & 4 \\ \hline
  Flex guides & 30 & WR650-770 trans. & 8 \\ \hline
  Phasing sections & 25 & Gaskets & 306 \\ \hline
        \end{tabular}
   \end{center} \vbbelow
\end{table}

\clearpage %\setcounter{page}{1}
\setcounter{figlcl}{0}
\setcounter{tablcl}{0}
\setcounter{secnumdepth}{4} \setcounter{subsection}{0}
\setcounter{subsubsection}{0}
\usecounter{subsubsection}

\setcounter{section}{5} \renewcommand{\picturefolder}{./cavities/}
\section{Cavities}\label{sect:Cavity}

\subsection{Overview}

The accelerating gradient in the ILC main linac is supplied by over
16,000 9-cell superconducting RF cavities, grouped into
approximately 12.6~m long cryomodules. Another $\sim$1200 9-cell cavities
provide acceleration in the sources and bunch compressors. The
baseline cavities use the TESLA design developed at DESY over the
past 10 years. The cavities are qualified at 35~MV/m gradient in a
vertical test and operated at an average gradient of 31.5~MV/m. At
these gradients, piezo-electric tuners are required to compensate
for Lorentz force detuning.

\subsection{Technical Description}

\subsubsection{Cavity Design}

\begin{figure}[htb]
   \begin{center} \vbabove
      \includegraphics[width=\textwidth]{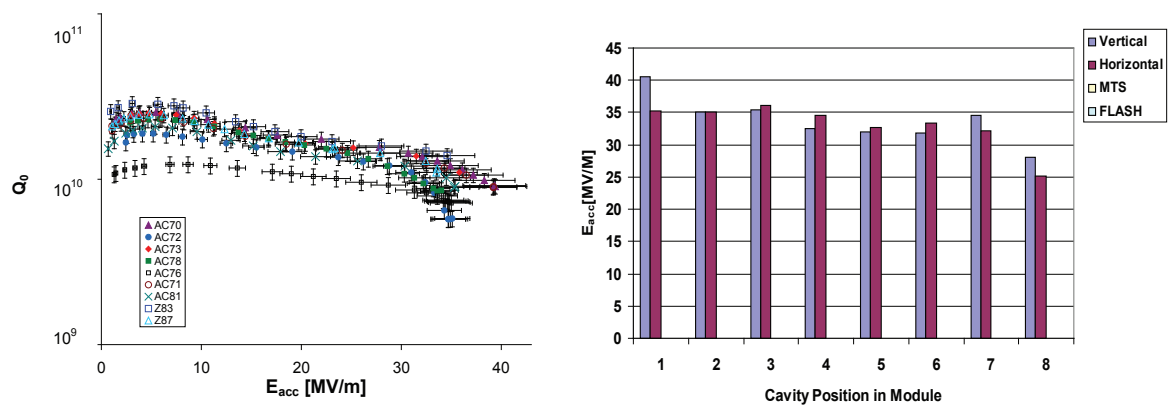}
      \vbabovecaption \caption[Q$_0$ vs. E curves for the best 9 Cell vertical
      qualification tests.]{Q$_0$ vs. E curves for the best 9 Cell vertical
      qualification tests at DESY (left) and data for a high gradient
      cryomodule assembled at DESY (right).}
      \label{fig:cavperform}
   \end{center} \vbbelow
\end{figure}

The TESLA 9-cell superconducting cavity was chosen as the baseline
design because it has achieved the highest qualification gradients
to date for multi-cell cavities, approximately within the range
required for ILC. There is significant operational experience with
these cavities and it has been demonstrated with beam that
accelerating gradients of greater then 30~MV/m are possible after
full installation in a cryomodule. Figure \ref{fig:cavperform} shows
examples of the best vertical test performance for individual cavity
structures at DESY (left) and results for a recent DESY cryomodule
assembled with the best collection of high gradient cavities
(right).

\stepcounter{figlcl}

\begin{figure}[htb]
   \begin{center} \vbabove
      \includegraphics[width=7cm]{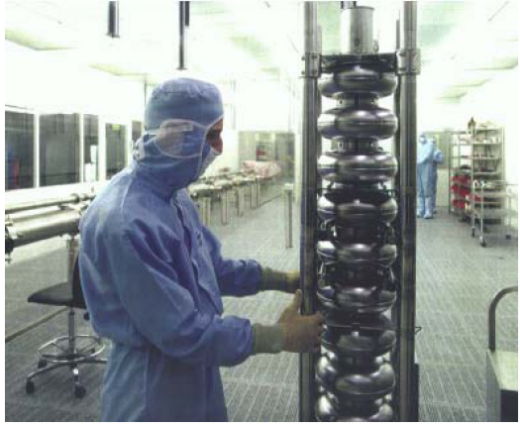}
     \vbabovecaption  \caption{A TTF cavity assembled and prepared for RF
      qualification testing.}
      \label{fig:cav9cell}
   \end{center}\vbbelow
\end{figure}

Each 9 cell cavity consists of nine accelerating cells between two
end group sections. One end group has a port for coupling RF power
from the power source into the structure, and the other end has a
port for a field sampling probe used to determine and control the
accelerating gradient. Each of these ports accepts an electric field
antenna required for qualification and operation. Each end group
also has a resonant higher order mode (HOM) coupler structure with a
probe port and small electric field antenna for extracting HOM power
and for diagnostics. In the process of building a cryomodule, these
cavity structures are cleaned, tested and placed in a helium jacket
for cooling together with additional peripheral components assembled
on them (dressing the cavity). A dressed cavity contains one 9-cell
niobium cavity structure, coarse and fine tuners for adjusting the
frequency of the structure, magnetic shielding material to minimize
the cavity losses, a variable coupling high power input antenna for
powering the cavity, an electric field sampling antenna and two
higher order mode electric field antennas. Nine of these dressed
cavities are usually connected into a string and are a subcomponent
of a superconducting cryomodule. Figure \ref{fig:cav9cell} shows a
TTF cavity undergoing clean assembly for RF qualification. The basic
design parameters for this cavity are listed in Table
\ref{tab:cavparams2}.

\stepcounter{figlcl}

\stepcounter{tablcl}\begin{table}[htb] \vbabove
 \caption{ILC 9-Cell superconducting niobium cavity design parameters.}
 \label{tab:cavparams2}
   \begin{center}
      \begin{tabular}{| l | c |} \hline
      Parameter & Value \\ \hline & \vbdlspacing \hline
      Type of accelerating structure & Standing Wave \\   \hline
      Accelerating Mode & TM$_{010}$, $\pi$ mode \\  \hline
      Fundamental Frequency & 1.300 GHz \\  \hline
      Average installed gradient & 31.5 MV/m \\  \hline
      Qualification gradient & 35.0 MV/m \\  \hline
      Installed quality factor & $\ge$1$\times10^{10}$ \\  \hline
      Quality factor during qualification & $\ge$0.8$\times10^{10}$ \\  \hline
      Active length & 1.038 m \\  \hline
      Number of cells & 9 \\  \hline
      Cell to cell coupling & 1.87\% \\  \hline
      Iris diameter & 70 mm \\  \hline
      R/Q & 1036 $\Omega$ \\  \hline
      Geometry factor & 270 $\Omega$ \\  \hline
      E$_{\rm peak}$/E$_{\rm acc}$ & 2.0 \\  \hline
      B$_{\rm peak}$/E$_{\rm acc}$ & 4.26 mT MV$^{-1}$m$^{-1}$ \\  \hline
      Tuning range & $\pm300$ kHz \\  \hline
      $\Delta$f/$\Delta$L & 315 kHz/mm \\  \hline
      Number of HOM couplers & 2 \\ \hline
      \end{tabular}
   \end{center} \vbbelow
 \end{table}

\subsubsection{Cavity Fabrication}

The fabrication of high quality superconducting cavities starts with
high quality niobium materials. The cavity accelerating cells and
end group components are fabricated from high purity, high RRR
niobium sheets. The RRR (residual resistivity ratio) value
indirectly indicates the purity of the bulk metal as well as
interstitial contamination that can affect the performance of the
superconducting properties. An RRR value of 300 is considered
desirable Table \ref{tab:ILCNb} shows the typical properties of
niobium sheets considered suitable for ILC. The transition joints to
the helium vessel are fabricated from a lower grade niobium sheet
called ``reactor grade'' with a RRR value of around 40.  The cavity
flanges and transitions to the helium vessel are made from bar or
round stock niobium alloy, typically NbTi55. The alloy is harder and
stronger, and it prevents deformation near the vacuum seals and
provides strong transition joints at the cavity connections.

\stepcounter{tablcl}\begin{table}[b] \vbabove
 \caption{Typical properties of high-RRR Niobium suitable for use in
    ILC cavities.}
 \label{tab:ILCNb}
   \begin{center}
 %     \begin{tabular}{| l | l @{\hspace{1in}} || l | l |} \hline
      \begin{tabular}{| l | c || l | c |} \hline
      Element & Impurity content & Property & Value \\ [-6pt]
       & in ppm (wt) & & \\ \hline & & & \vbdlspacing \hline
      Ta & $\le$500 & RRR & $\ge$300 \\  \hline
      W & $\le$70 & Grain size & $\approx$50 $\mu$m \\  \hline
      Ti & $\le$50 & Yield strength & $>$50 MPa \\  \hline
      Fe & $\le$30 & Tensile strength & $>$100 MPa \\  \hline
      Mo & $\le$50 & Elongation at break & 30\% \\  \hline
      Ni & $\le$30 & Vickers hardness & \\  \hline
      H & $\le$2 & HV 10 & $\le$50 \\  \hline
      N & $\le$10 & & \\  \hline
      O & $\le$10 & & \\  \hline
      C & $\le$10 & & \\ \hline
      \end{tabular}
   \end{center} \vbbelow
\end{table}

As a final quality assurance check prior to use, the cell material
is sometimes eddy-current scanned to a depth of 0.5 mm into the
surface of the sheet material. Cavity cells are traditionally formed
by deep drawing or hydro-forming methods where the sheets are
pressed into dies to form the necessary shapes. These fabrication
methods require machining of surfaces to form the weld joints. All
cavity subcomponents are joined by electron beam welding in a vacuum
chamber. This reduces the contamination at the welds and is
considered the cleanest form of joining metals together. Prior to
electron beam welding, all subcomponents are inspected, degreased
then prepared typically by mechanical polishing of surfaces to be
welded, as necessary. The welded components are degreased and
chemically etched and rinsed to remove inclusions and surface
contamination from the machining and welding steps. The completed
cavity has both internal and external chemistry to further remove
the damage layer from the fabrication steps of both welding and
handling. A smooth outer surface is necessary to provide good
thermal contact with the cryogenic bath.

In total about 150-250~$\mu$m of niobium material is removed from the
interior RF surface of the cavity through several cleaning steps.
After each of these acid etchings the cavity has a new
superconducting RF surface and can have different RF performance and
a different gradient limitation.

The two primary issues with cavity fabrication are quality assurance
on the niobium materials and on the electron beam welds. Niobium
materials must be scanned to detect and eliminate surface defects,
and then protected with care throughout the manufacturing process.
Defective material will ultimately limit the gradient performance of
a completed cavity. As with the surface defects, impurities in the
welds and heat affected zones next to welds will also limit the
gradient performance. Welds must have a smooth under-bead and form
no surface irregularities, in particular, sharp edges where the weld
puddle meets the bulk material. Defects in the equator welds will
limit the gradient by thermal quenches due to the high magnetic
fields there. Thermal mapping of quench locations suggests that they
are typically located at or near the equator region.

\subsubsection{Cavity Processing}

The current technology for preparing cavity surfaces consists of a
series of process steps~\cite{bib:cv1} that: remove niobium material
damage incurred during the fabrication process or handling; remove the
chemical residues left over from the material removal steps; remove
the hydrogen from the bulk niobium that has entered during the
chemistry steps; remove any particulate contamination that entered
during the cleaning and assembly steps; and close up the cavity to
form a hermetically sealed structure. The following steps are typical
of those used to qualify a cavity structure in a vertical RF test.

{\bf Mechanical Inspection:}
The cavity is mechanically measured with a coordinate measuring
machine to compare dimensional measurements against mechanical
tolerances identified on design drawings.

{\bf RF Inspection:}
The cavity fundamental frequency is measured. A bead is pulled
through the beam axis of the cavity to determine and record the
stored energy of each cell. The bead disturbs the fields in each
cell as it passes through which changes the frequency by an amount
equal to the stored energy in that cell.

{\bf Bulk Chemistry:} Both the internal and external surfaces of the
cavity are ultrasonically treated in hot de-ionized water to remove
grease from the handling and surface particulates that have collected
since fabrication. The cavity is then internally chemically etched
with electropolishing~\cite{bib:cv2} to remove 150-250~$\mu$m of
material. The cavity is placed horizontally into an alignment fixture,
which levels the cavity and seals the openings while allowing the
cavity to rotate. A high purity aluminum cathode rod is inserted on
the beam axis to pump cooled electrolyte into each cell of the cavity
through a series of holes in the cathode. The cathode is electrically
connected to the negative contact of a DC power supply. The cavity
itself is the anode and is typically connected on the cells to the
positive contact of the DC supply. The electrolyte is a mixture of
hydrofluoric and sulfuric acid in a 1:9 or 1:10 parts by volume
respectively. At the start of the process, the cavity is filled to the
60\% level covering the entire cathode with the cavity slowly rotating
at $\approx$1~RPM. The DC power supply provides a current density of
about 50~mA/cm$^2$ and the cavity is polished for approximately 6-7
hours for an etching depth of 150~$\mu$m.  Temperatures are monitored
during the process to control the current and etch rate which is
0.4~$\mu$m/minute at 30 degrees C. After etching, the cavity is rinsed
extensively with ultra pure water to remove any chemical residues or
chemical safety hazards.

{\bf Heat treatment:}
After bulk chemistry the cavity is cleaned and dried before
inserting into a vacuum furnace for heat treatment. The chamber is
evacuated to $\approx$10$^{-7}$ mbar and the bare cavity is heated
to 800 $^{\circ}$C and soaked at that temperature to remove any
excess hydrogen gained during the chemistry. This additional
hydrogen, if not removed, lowers the cavity Q-value due to formation
of a niobium hydride during cool-down, that adds surface losses. The
cavity is then cooled to room temperature and removed from the
furnace.

{\bf RF Tuning:}
The cavity is tuned to the correct warm frequency and the stored
energy (field flatness) in each cell equalized. The cells are
measured with a bead pull and then plastically deformed by pulling
or squeezing in a mechanical tuner. The cavity is mechanically
adjusted to correct any alignment errors that would affect tuning
for field flatness.

{\bf Final Chemistry:}
The final internal chemistry refreshes the niobium surface by
electropolishing removal of 10-30~$\mu$m of material. The processing
steps are the same as for the bulk chemistry although the processing
time is much shorter. After the standard water rinsing, additional
steps should be taken to remove any sulfur particulates from the
surface and several methods are now under evaluation.

{\bf High Pressure Rinsing:} The cavity is inserted vertically into
the high pressure rinse (HPR)~\cite{bib:cv3} system where a wand is
moved slowly through the beam axis of the cavity and the cavity is
rotated. The head of the wand has small diameter nozzles tilted at
angles through which high pressure ultra pure water is sprayed. A
water pressure to the wand of 80-100 bar produces up to 40 N of force
on the surface at impact. The wand is repeatedly moved up and down
spraying all surfaces of the cavity with water to remove surface
particulates which are attached on the cavity interior. The HPR
process is considered the most effective cleaning method to remove
surface contamination.

{\bf First Assembly:}
Assembly takes place in a Class 10 cleanroom, where the cavity has
been left open to air dry over night. Once the cavity has dried,
cleaned subcomponents are carefully attached to the cavity by hand.
Each flange connection is sealed using a diamond shaped aluminum
alloy gasket that is crushed between flange faces with high line
loading forces. High strength bolts and nuts with washers provide
the force necessary to crush the seal. All subcomponents are
assembled to the cavity except the bottom beam-line flange to allow
for the second high pressure rinse.

{\bf Second HPR:}
The second rinse is typically longer then the first rinse and is
necessary to remove any additional particulates that have entered
during the assembly steps, either from the personnel, the cleanroom
environment, or the subcomponents. The cavity is removed from the
HPR system after the rinse is completed and is moved to the Class 10
area to dry again overnight, this time with only the lower beam-line
flange open.

{\bf Second Assembly:}
The bottom beam-line flange is connected to
the cavity. It typically has an isolation vacuum valve with pump-out
port and an RF input probe to power the cavity. The cavity is now
hermetically sealed and ready for evacuation.

{\bf Cavity Evacuation:}
The cavity pump-out port is connected to
a vacuum pump and evacuated. The pump system usually has a turbo
molecular pump with a scroll type dry mechanical backing pump. The
cavity is pumped overnight and the following day tested for vacuum
leaks by spraying the cavity flange joints with helium gas and using
a residual gas analyzer on the vacuum system to detect helium.

{\bf 120$^{\circ}$C Vacuum Bake:}
To improve the high field
Q-value, the cavity is baked at 120 degrees C for 12-24 hours while
actively being pumped by the vacuum system. After being cooled to
room temperature, the cavity is ready for RF testing.

{\bf RF Qualification Testing:}
The cavity is mounted vertically into a cryogenic test stand, RF
cables are connected to the cavity probes and the stand is inserted
into a cryogenic dewar. The dewar is cooled to 4.2 K by helium gas
and liquid is collected to fill the dewar. The dewar is pumped down,
lowering the temperature to 2.0 K where the cavity is RF tested to
determine its gradient, Q-value and limitations. Once testing is
complete, the dewar is warmed up to room temperature and the stand
with cavity is removed. With existing technology and infrastructure,
this cryogenic cycle usually takes about 2 days at DESY, and about 4
to 5 days at KEK.

\subsubsection{Peripheral Components}

The DESY variable input coupler has been chosen as part of the
baseline cavity design. This coupler features two ceramic RF windows
as wells as two bellows which allow the center conductor to be
mechanically moved into or out of the cavity structure thus changing
the RF coupling of klystron power to the cavity. Further R\&D will
focus on reducing the cost of construction and adapting it to large
scale production in industry.

The ILC cavities have both a mechanical coarse tuner and a piezo
electric fine tuner. There are several viable designs for both the
mechanical and the piezo electric tuners such as the blade tuner
(mechanical). Currently no tuner has been chosen for the baseline
design, and R\&D is required to determine the reliability and
installed performance of current designs.

The ILC cavities use a DESY style helium vessel made from titanium,
which is thermally matched to the cavity material to avoid
distortion of the cavity shape during cool down.

\subsubsection{Cavity Performance Requirements}

The ILC cavities must meet specific requirements on accelerating
gradient and Q-value, both in the vertical qualification test and
after assembly in a cryomodule. For the vertical test, each cavity
must achieve 35MV/m gradient with a Q-value of $0.8\times10^{10}$ or
greater. The Q-value is a ratio of the stored energy within the cell
structure to the losses dissipated in the cell walls. Lower Q-values
increase the heat load to the cryogenic system. A cavity assembled
within a cryomodule must reach 31.5MV/m on average, with a Q-value
of $1\times10^{10}$. The installed gradient and Q-value requirements
are believed to be achievable with current fabrication and
processing techniques. The vertical test gradient requirement is
higher then that of the cryomodule in order to increase the success
rate of assembled cryomodules. The performance of a cryomodule can
be limited by additional system variability and administrative
interlocks for the protection of peripheral components as well as
from the cavity. The baseline design of ILC has been developed under
the assumption that cavities qualified at 35MV/m will meet the
requirement of 31.5MV/m on average once installed in a cryomodule,
with overheard to compensate for microphonics and for limitations
from weaker cavities powered by the same RF source.

With current fabrication and process procedures, cavities have a
large spread in gradient and Q-value performance and do not reliably
reach 35MV/m in the vertical tests. The primary issue is the
magnitude and onset of field emission, which lowers the Q-value
below specification. Field emission is typically caused by surface
contamination located in regions of high electric field. Electrons
emitted from the contamination site bombard other cavity surfaces,
increasing surface heating and surface losses, thus lowering the
cavity Q-value at that gradient. Once field emission starts, it is
typically stable and the Q-value continues to drop with increasing
gradient.

When not dominated by field emission, the high gradient performance
of a cavity is typically limited by a thermal magnetic quench of the
niobium material. Quenches can be caused by a variety of surface
defects such as bad welds, lossy oxides and imbedded materials
entering during fabrication or handling, or even by field emitted
electrons from surface contamination.

The highest priority for ILC accelerator cavity R\&D is to increase
the success rate in producing cavities that reach the required
performance. Increased quality control of the processing and
assembly steps is expected to address the field emission issues
which currently appear to dominate the limitation and variation of
cavity performance. Better control of the process variables are
being pursued through a global R\&D effort. Current emphasis is on
understanding and improving the electropolishing process. To reach
the desired gradient and Q-value, a high level of quality control
must be implemented for the preparation of material used in cavity
fabrication, throughout the fabrication of the structures, and
during the cleaning and assembly processes.

\stepcounter{figlcl}\begin{figure}[htb]
   \begin{center} \vbabove
      \includegraphics[width=12cm]{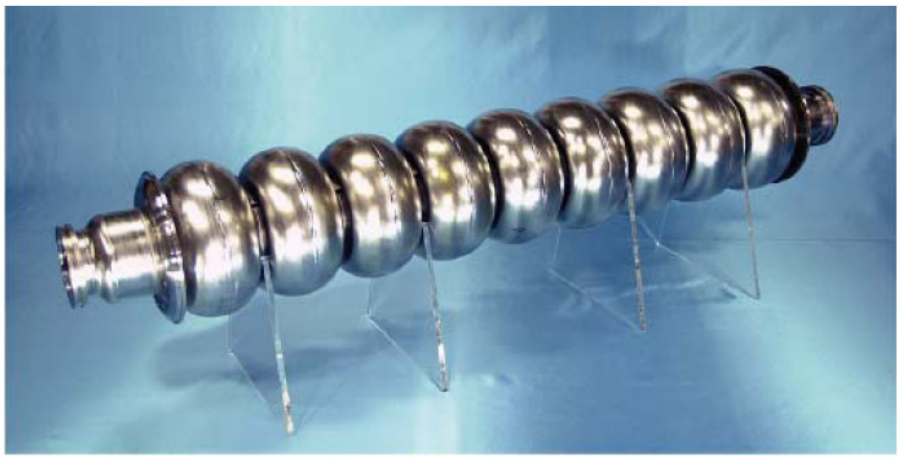}
      \vbabovecaption \caption{A low loss nine cell prototype RF structure under
        development.}
      \label{fig:lowloss}
   \end{center} \vbbelow
\end{figure}

\subsubsection{Alternative Cavity Designs}

Alternative cavity shapes and fabrication materials are being
studied that could potentially reduce the cost of fabrication or
increase the achievable gradient. If successful, either could
significantly reduce the ILC cost.

By slightly changing the shape of the cavity cell walls, it is
possible to reduce the peak magnetic flux on the walls and allow the
cavity to reach higher accelerating gradients before reaching a
critical field limit on the niobium surface and starting to quench.
New cavity shapes have been successfully tested as single cell
structures up to gradients of 50MV/m at both Cornell University, with
a reentrant shape~\cite{bib:cv5}, and at KEK with the ``Ichiro''
design~\cite{bib:cv6}.  Figure \ref{fig:lowloss}. shows a low loss nine
cell prototype RF structure. However, the cavity shape affects many
other operational parameters such as the effectiveness of higher order
mode damping, multipactoring, shunt impedance, peak electric fields,
energy dissipation, beam impedance and mechanical properties since a
different aperture size is to be adopted. These aspects must be
reoptimized.

Recent niobium material studies at Jefferson Lab have led to new
methods for cavity fabrication using either large grain or single
crystal niobium. These new materials may lead to significant cost
savings in cavity preparation as well as simplification of the
processing procedures.

\clearpage %\setcounter{page}{1}
\setcounter{figlcl}{0}
\setcounter{tablcl}{0}
\setcounter{secnumdepth}{4} \setcounter{subsection}{0}
\setcounter{subsubsection}{0}
\usecounter{subsubsection}

\setcounter{section}{6} \renewcommand{\picturefolder}{./cryomods/}

\section{Cryomodules}\label{sect:CRYMc}

\subsection{Overview}\label{ssect:CRYMo}

The accelerating gradient in the ILC main linac is supplied by over
16,000 9-cell superconducting RF cavities.  These cavities are
grouped into approximately 1,700 12.7 m long cryomodules. Each
cryomodule holds nine cavities, their supporting structure, thermal
shields, associated cryogenic piping, and the insulating vacuum
vessel.  Every third cryomodule in the main linac contains a
superconducting quadrupole/corrector/BPM package in place of the
center cavity. Another 150 cryomodules are located in the e$^+$ and
e$^-$ sources and RTML bunch compressors. Most of these are the
standard linac configuration of 9 cavities or 8 cavities plus quad.
A few have special configurations of cavities and quadrupoles.

\subsection{Technical Description}\label{ssect:CRYMtd}

The cryomodule design is a modification of the type developed and
used in the TESLA Test Facility (TTF) at DESY, with three separate
vacuum envelopes.  The ILC cryomodules contain either nine 9-cell
cavities or eight cavities plus a quadrupole package, and have a uniform
length of 12.652 m.  The cavity spacing within this modified
cryomodule is (6-1/4) $\lambda_{0}$ = 1.327 m.  This facilitates
powering the cavities in pairs via 3 db hybrids with reflection
cancelation in an alternate distribution scheme that may allow the
elimination of circulators.
%The primary differences in the ILC
%design are reduced cavity-to-cavity spacing and the relocation of
%the quadrupole/corrector/BPM package from the end to the center of
%the cryomodule, directly beneath the fixed center support.  The ILC
%cryomodule design will continue to evolve, with the goals of
%improving the design for manufacturability, ease of assembly, unit
%cost reduction, reliability and robustness.
Present day accelerators
with superconducting RF cavities typically have many separate
cryogenic supply boxes and associated warm-to-cold transitions,
which represent a significant fraction of the cost. The concept
adopted for the ILC is to significantly reduce this number by having
a single long continuous string of about 2.5km---called a cryogenic
unit---which is connected to one cryogenic supply box at the
beginning and one end box.

\stepcounter{figlcl}\begin{figure}[htb]
   \begin{center} \vbabove
      \includegraphics[width=12cm]{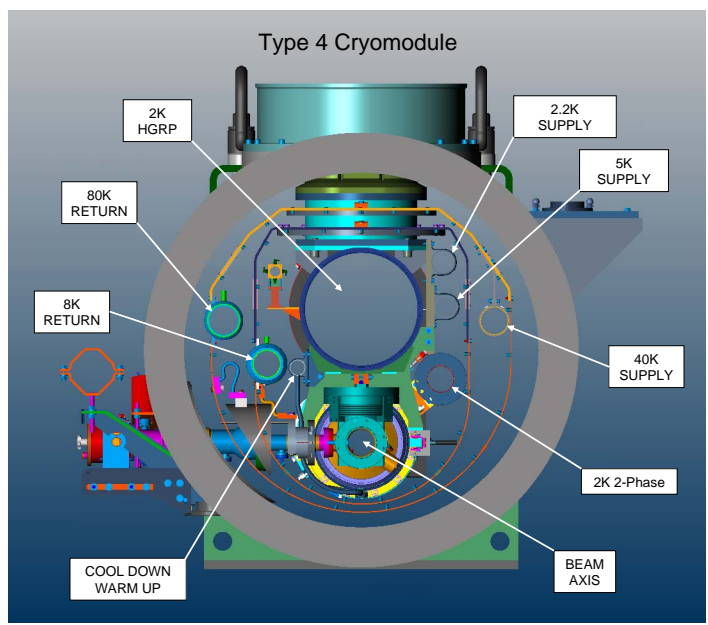}
      \vbabovecaption \caption{Representative Cryomodule Cross-Section.}
      \label{fig:CRYMcrymodcs}
   \end{center} \vbbelow
\end{figure}

%\stepcounter{figlcl}\begin{figure}[htb]
%   \begin{center}
%      \includegraphics[width=16cm]{\picturefolder cryomodlong.jpg}
%      \caption{Longitudinal View of a Cryomodule}
%      \label{fig:CRYMcryomodlong}
%   \end{center}
%\end{figure}

\subsection{Technical Issues}\label{ssect:CRYMti}

\subsubsection{The Cryomodule }\label{sssect:POSpp}

Figure \ref{fig:CRYMcrymodcs} shows a cross-section of a TTF-III
cryomodule \cite{tdr}.
 The 300~mm diameter helium gas return pipe (GRP) is the main
support structure for the string of cavities and the quadrupole
package. The GRP is supported from above by three posts which
provide the necessary thermal insulation to room temperature. The
posts are fastened to large flanges on the upper part of the vacuum
vessel by adjustable suspension brackets, allowing the axis of the
cavities and quadrupoles to be correctly aligned, independent of the
flange position.  The support system is designed to allow the GRP to
contract/expand longitudinally with respect to the vacuum vessel
during thermal cycling. The center post is fixed to the vacuum
vessel, while the two end brackets can move in the axial (z)
direction to accommodate differential shrinkage. A post consists of
a fiberglass pipe terminated by two shrink-fit stainless steel
flanges. Two additional shrink-fit aluminum flanges are provided to
allow intermediate heat flow intercept connections to the 5-8 K and
40-80 K thermal shields; the exact location of these flanges has
been optimized to minimize the heat leakage~\cite{bib:CM-Nicol}.

Each of the cavities is encased in a titanium helium vessel,
supported by the GRP by means of stainless steel brackets connected
to four titanium pads on the helium vessel itself; each bracket is
equipped with a longitudinal sliding mechanism and adjusting screws
and pushers for alignment. A mechanical, coaxial (blade) and a
piezo-electric tuner are mounted to the vessel. The inter-cavity
spacing---which accommodates RF- and HOM-couplers and a flanged
interconnecting bellows---amounts to 291 mm.
Manually operated valves required by the clean-room
assembly terminate the beam pipe at both module ends.
The valves are fitted with simple RF shields.

During cool down the two ends of the $\sim$12 m long gas return pipe
move by up to 18mm toward the center of the module. To keep the cold
input coupler head of each cavity fixed longitudinally within an
accuracy of 1 mm, each cavity is anchored to a long invar rod
attached to the longitudinal center of the gas return pipe. The beam
pipe interconnection between the cryomodules consists of a 0.38 m
long section that incorporates a Higher Order Mode (HOM) absorber, a
bellows, and a vacuum pumping port; the latter connected to a flange
in the vacuum vessel every ninth cryomodule.

The cryostat includes two aluminum radiation shields operating
in the temperature range of 5-8K and 40-80K respectively~\cite{bib:CM-Barni}. Each
shield is constructed from a stiff upper part (divided into two halves),
and multiple lower sections (according to the number of the cold
active components, e.g. cavities, magnets). The upper parts are
supported by the intermediate flanges on the fiberglass posts;
they are screwed to the center post but can axially slide on the
other two posts, to which they are still thermally connected.
The `finger welding' technique~\cite{bib:CM-Pagani} is used both to connect each
thermal shield to its properly shaped aluminum cooling pipe,
and the lower shield parts to the upper ones.

Blankets of multi-layer insulation (MLI) are placed on the outside
of the 5-8 K and the 40-80 K shields. The 5-8 K shield blanket is
made of 10 layers while the 40-80 K blanket contains 30 layers. In
addition the cavity and quadrupole helium vessels, gas return pipe
and 5-8 K pipes are wrapped with 5 layers of MLI to reduce heat
transfer in the event of a vacuum failure.

%Figure~\ref{fig:CRYMcrymodcs} shows a cross section of the type-III cryomodule
%and Figure~\ref{fig:cryomodlong}. shows a longitudinal view.

The cryostat outer vacuum vessel is constructed from carbon steel
and has a standard diameter of 38''. Adjacent vacuum vessels are
connected
 to each other by means of a cylindrical sleeve with a bellows,
which is welded to the vessels during installation. Radiation
shield bridges are also provided. In the event of accidental
spills of liquid helium from the cavity vessels, a relief valve
on the sleeve together with venting holes on the shields
prevent excessive pressure build-up in the vacuum vessel.
Wires and cables of each module are extracted from the
module using metallic sealed flanges with vacuum tight
connectors.  The insulating vacuum system is pumped
during normal operation by permanent pump stations
located at appropriate intervals. Additional pumping ports
 are available for movable pump stations, which are used
for initial pump down, and in the event of a helium leak. The RF
power coupler needs an additional vacuum system on its room
temperature side; this is provided by a common pump line for all
couplers in a module, which is equipped with an ion getter and a
titanium sublimation pump.

The following helium lines~\cite{bib:CM-Jensch} are integrated into the cryomodules:

\begin{itemize}

      \item The 2~K forward line transfers pressurized single phase
              helium through the cryomodule to the end of the cryogenic unit. \itemspace
     \item The 2~K two phase supply line (made from titanium) is
              connected to the cavity and magnet helium vessels. It supplies
              the cavities and the magnet package with liquid helium and returns
              cold gas to the 300~mm GRP at each module interconnection. \itemspace
     \item  The 2~K GRP returns the cold gas pumped off the saturated He II
              baths to the refrigeration plant. It is also a key structural
              component of the cryomodule \itemspace
     \item The 5-8~K forward and return lines. The 5K forward line is used
              to transfer the He gas to the end of the cryogenic unit. The 5-8~K return
              line directly cools the 5-8~K radiation shield and, through the shield,
              provides the heat flow intercept for the main coupler and diagnostic
              cables, and the higher-order mode (HOM) absorber located in the
              module interconnection region. \itemspace
     \item The 40-80~K forward and return lines. The 40~K forward line is
             used to transfer He gas to the cryogenic unit end and cools the high
             temperature superconductor (HTS) current leads for the quadrupole
             and correction magnets. The 40-80~K return line directly cools the
             40-80K radiation shield and the HOM absorber and, through the shield,
             provides an additional heat flow intercept for the main coupler
             and diagnostic cables. \itemspace
     \item The warm-up/cool-down line connects to the bottom of each cavity
              and magnet helium vessel. It is used during the cool down and
               warm up of the cryostat. \itemspace
\end{itemize}

The helium lines connected to the cavities and the magnets withstand
a pressure of 4 bar; all other cryogenic lines withstand a pressure
of 20 bar. The helium lines of adjacent modules are connected by
welding, as was done for the HERA superconducting magnets.
Transition joints (similar to those used in the HERA magnets) are
used for the aluminum to stainless steel transition on the thermal
shield cooling lines. The cryostat maintains the cavities and
magnets at their operating temperature of 2 K. A low static heat
load is an essential feature of the cryostat design; the total heat
load is dominated by the RF losses, and is thus principally
determined by cavity performance.
%Tables~\ref{tab:CRYMcryocomp} ,
%\ref{tab:CRYMtl2k} , \ref{tab:CRYMtl5k} , \ref{tab:CRYMtl40k} shows
%the parameters for the various cryomodules and the calculated heat
%load for a cryomodule with cavities operating at the nominal
%accelerating gradient of 31.5MV/m~\cite{bib:CRYMadolphpeter}.
Table \ref{tab:heatloadcryomod} lists the heat loads for an RF unit
scaled from the 12-cavity cryomodule heat loads calculated for TESLA
and documented in the TESLA TDR. For the scaling to the ILC, it was
assumed that the gradient is 31.5 MV/m, the cavity Q$_0$ is
$1\times10^{10}$, and the beam and RF parameters are those listed in
section \ref{sect:ML}.

\stepcounter{tablcl}\begin{table}[htb] \vbabove \caption[Heat loads
for one RF unit of 3 cryomodules with 26 cavities.]{Heat loads for
one RF unit of 3 cryomodules with 26 cavities.  All values are in
watts.} \label{tab:heatloadcryomod}
  \begin{center}
\setlength{\tabcolsep}{6pt}
    \begin{tabular}{| l || c | c || c | c || c | c |}\hline
      & \multicolumn{2}{c||}{2 K} &
        \multicolumn{2}{c||}{5-8 K} &
        \multicolumn{2}{c|}{40-80 K} \\  \cline{2-7}
      & Static & Dynamic & Static & Dynamic & Static & Dynamic \\
        \hline & & & & & & \vbdlspacing \hline
      RF Load & & 22.4 & 4.2 & & 97.5 & \\ \hline
      Supports & 1.8 & 0.0 & 7.2 & & 18.0 & \\ \hline
      Input coupler & 1.6 & 0.5 & 4.4 & 4.0 & 46.5 & 198.2 \\ \hline
      HOM coupler (cables) & 0.0 & 0.6 & 0.9 & 5.5 & 5.5 & 27.1 \\
        \hline
      HOM absorber & 0.4 & 0.0 & 9.4 & 1.6 & 9.8 & 32.6 \\ \hline
      Beam tube bellows &  & 1.1 & & & & \\ \hline
      Current leads & 0.9 & 0.9 & 1.4 & 1.4 & 12.4 & 12.4 \\ \hline
      HOM to structure & & 3.6 & & & & \\ \hline
      Coax cable (4) & 0.2 & & & & & \\ \hline
      Instrumentation taps & 0.2 & & & & & \\ \hline
      Diagnostic cable & & & 4.2 & & 7.4 & \\ \hline & & & & & & \vbdlspacing \hline
      Sum & 5.1 & 29.0 & 31.7 & 12.5 & 177.6 & 270.3 \\ \hline
    \end{tabular}
  \end{center} \vbbelow
\end{table}

Most losses occur at lower frequencies where the conductivity of the
superconducting surfaces is several orders higher than that of
normal conducting walls. Part of this power is extracted by input-
and HOM-couplers, but high frequency fields will propagate along the
structure and be reflected at normal and superconducting surfaces.
In order to reduce the losses at normal conducting surfaces at 2 K
and 4 K, the cryomodule includes a special HOM absorber that
operates at 70 K, where the cooling efficiency is much higher. The
absorber basically consists of a pipe of absorbing material mounted
in a cavity-like shielding, and integrated into the connection
between two modules. As the inner surface area of this absorber
(about 280 cm$^2$) is small compared to that of all the normal
conductors in one cryomodule, the absorber has to absorb a
significant part of all the RF power incident upon it. In field
propagation studies, which assume a gas-like behavior for photons,
it has been shown that an absorber with a reflectivity below 50\% is
sufficient. Theoretical and experimental studies have suggested that
the required absorption may be obtained with ceramics like MACOR or
with artificial dielectrics.

%The axes of the cavities must be aligned to the ideal beam axis to
%within $\pm0.5$ mm.  This has been achieved at the TESLA Test
%Facility \cite{cavalign}.
%%and quadrupole axes to within $\pm0.2$ mm.
%%
%The quadrupole axes must be aligned to within $\pm0.2$ mm of the
%ideal beam axis, and must meet a tolerance of $\pm0.1$ mrad on their
%rotation about the beam axis.
The ambient magnetic field in the
cavity region must not exceed 0.5 $\mu$T to preserve the low surface
resistance.  The magnetic field tolerance is achieved by
demagnetizing the vacuum vessel before assembly of the cryomodule,
and placing a passive shield made of Cryoperm around each cavity's
helium vessel.

\subsubsection{Quadrupole/Corrector/BPM Package}\label{sssect:CRYMqcbpt}

%Design details of the quadrupole/corrector/BPM package are provided
%elsewhere in this report and will not be discussed here.
%
The quadrupole/corrector/BPM package is discussed in Section
\ref{sect:ML}.
An important feature that must be addressed is the package
fiducialization and subsequent transfer of these features to
reproducible, external cryomodule fiducials to assure the correct
alignment of the package with respect to the cryomodule string.

\subsubsection{Damping Ring and Beam Delivery Cryomodules}\label{sssect:CRYMdrbdc}

The damping ring accelerating RF is single 650 MHz cavities in
individual cryomodules. The beam delivery also uses superconducting
crab cavities with individual cryomodules.  This system is discussed
in Section \ref{sectDR}.
%Details of systems are discussed elsewhere
%in this document and will not be presented here.

\subsubsection{Shipping of Cryomodules}\label{sssect:CRYMsc}

To date, no engineering design to facilitate the shipping of completed
cryomodules exists.  It is essential that a reliable method be developed
and incorporated into the ILC cryomodule deign.

\subsection{Cost Estimation}\label{ssect:CRYMce}

The cryomodules represent nearly one third of the total ILC project cost.
Cost studies have been conducted in all three regions , Americas, Asia and
Europe.  Much of the original effort relied on the TESLA TDR costing as
a basis for comparison.  However, independent regional studies and the
cost study for the XFEL have proved useful in improving the reliability of
the ILC cost numbers.

Significant effort has been expended to understand the cost drivers for
cryomodules. The cavities are the largest item, with over 40\% of the
cryomodule cost for cavity fabrication, processing, dressing and qualification.
The next largest items are the power couplers, the helium vessel fabrication,
the quad package and the tuners, which represent another 30\%.
It is anticipated that joint studies between ILC engineers
and designers and industrial partners utilizing design for manufacture
methodology and value engineering principles will lead to significantly
reduced cryomodule component and assembly costs.

\subsection{Table of Cryomodule Types}\label{ssect:CRYMtct}

The different cryomodule types and required quantities of each type
are listed in Table~\ref{tab:CRYOscnums}.  As can be seen in this
table, there are basically four types of cryomodules required for
the 1.3 GHz portion of the ILC.

\begin{comment}
\stepcounter{tablcl}\begin{table} \vbabove \caption{Summary of
Superconducting RF Cryomodules in the ILC. }
   \label{tab:CRYMcryonums}
   \begin{center}
      \begin{tabular}{| l |c |c | c | c | l |}
         \hline
         Location & Quantity & Multiplier & Total & Unit Breakdown & Notes \\
         \hline & & & & & \vbdlspacing \hline
         Main Linac & 936 & 2 & 1872     & 624 standard & w/quad \\
         & & & & 624 standard & w/o quad \\  \hline
         RTML & 60 & 2 & 120 & 64 standard & w/quad \\
         & & & & 56 standard & w/o quad   \\ \hline
         e- source & 21 & 1 & 21 & 11 standard & w/quad \\
         & & & & 4 special & 6 cavities + 6 quads \\
         & & & & 6 special & 8 cavities and 2 quads  \\ \hline
         e+ source & 22 & 1 & 22 & 12 standard & w/quad \\
         & & & & 4 special & 6 cavities + 6 quads \\
         & & & & 6 special & 8 cavities and 2 quads  \\ \hline
         e+ keep alive & 2 & 1 & 2 & 2 special & w/quad(s) \\ \hline
         Undulator & 13 & 1 & 13     & 9 standard & w/quad \\
         & & & & 4 standard & w/o quad \\  \hline
         Total 1.3 GHz & & & 2050 & 2050 & 1.3 GHz \\ \hline \hline
         Damping Rings & 32 & 3 & 96 & 96 special & 650 MHz \\ \hline
      \end{tabular}
   \end{center} \vbbelow
\end{table}
\end{comment}

\clearpage %\setcounter{page}{1}
\setcounter{figlcl}{0}
\setcounter{tablcl}{0}
\setcounter{secnumdepth}{4} \setcounter{subsection}{0}
\setcounter{subsubsection}{0}
\usecounter{subsubsection}

\setcounter{section}{7} \renewcommand{\picturefolder}{./cryo/}

\section{Cryogenic Systems}\label{sect:CRYOcs}

\subsection{Overview}\label{ssect:CRYOo}

With superconducting equipment throughout the ILC, cryogenic systems of extensive size and capacity will be required.  Superconducting RF cavities operating at 2~Kelvin in the main linacs are the primary accelerating structures in the ILC and comprise the largest cryogenic cooling load.  Although not as extensive, the positron and electron sources, damping rings, RTML, and beam delivery systems include a large number and variety of superconducting RF cavities.  Table~\ref{tab:CRYOscnums} summarizes the numbers of various types of superconducting RF modules in the ILC.

In addition to the RF modules listed in Table~\ref{tab:CRYOscnums}, there are a variety of superconducting (SC) magnets in the ILC.  About one third of the 1.3~GHz cryogenic modules contain SC magnets.  As part of the positron source, the electron linac includes about 150~meters of SC helical undulators in 2 to 4~meter length units.  The Damping Rings have 8~strings of superconducting wiggler magnets, and there are special SC magnets in the sources, RTML, and beam delivery system.

\stepcounter{figlcl}\begin{figure} [htb]
   \begin{center} \vbabove
      \includegraphics[width=\textwidth]{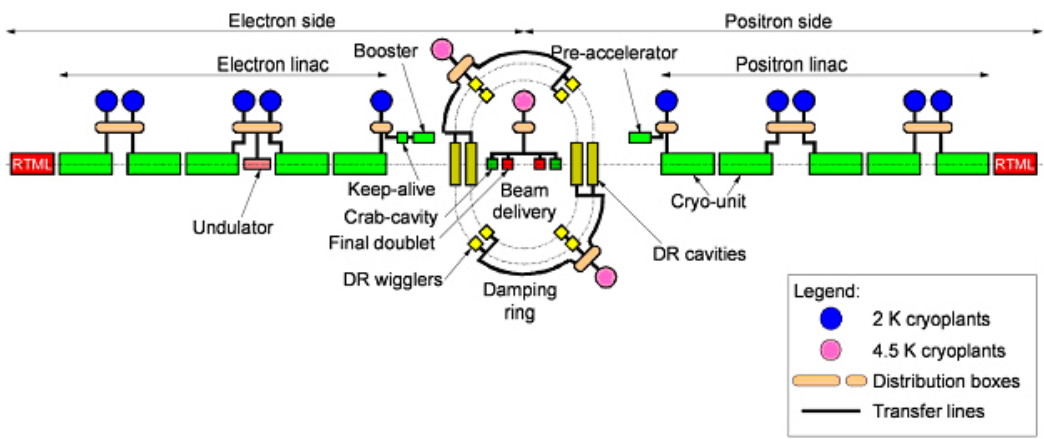}
     \vbabovecaption  \caption{The overall layout concept for the cryogenic systems.}
      \label{fig:CRYOsyslayout}
   \end{center} \vbbelow
\end{figure}

Figure~\ref{fig:CRYOsyslayout} illustrates the concept for the cryogenic system arrangement in ILC.  Ten large cryogenic plants with 2~Kelvin refrigeration cool the main linac, RTML and the electron and positron sources. Three smaller cryogenic plants with mostly 4.5~K loads cool the damping rings and beam delivery system.

\stepcounter{tablcl}\begin{table} \vbabove \caption[Superconducting
RF modules in the ILC.] {Superconducting RF modules in the ILC,
excluding the two 6-cavity energy compressor cryomodules located in
the electron and positron LTRs}
   \label{tab:CRYOscnums}
   \begin{center}
      \begin{tabular}{| l | c | c | c | c | c | c | c |}
         \hline
         Cryomodules & & & & & Total & & \\ [-6pt]
         ~~(cavities/cryomoule)& 8-C & 9-C & 8-C & 6-C &  & 1-C & 2-C \\ [-6pt]
         ~~(quads/cryomodule) & 1-Q & 0-Q & 2-Q & 6-Q &  &  &  \\ [-6pt]
          ~~(frequency, MHz)& 1300 & 1300 & 1300 & 1300 & 1300 & 650 & 3900 \\
         \hline & & & & & & & \vbdlspacing \hline
         Main Linac  e$^{-} $ & 282 & 564 & & & 846 & &  \\ \hline
         Main Linac  e$^{+} $ & 278 & 556 & & & 834 & & \\ \hline
         RTML   e$^{-} $ & 18 & 30 & & & 48 & &  \\ \hline
         RTML  e$^{+} $ & 18 & 30 & & & 48 & &  \\ \hline
          e$^{-} $ source & 24 & &  &  & 24 & & \\ \hline
          e$^{+} $ booster & 12 & & 6 & 4 & 22 & & \\ \hline
          e$^{+} $ Keep Alive & 2 & & & & 2 & & \\ \hline
          e$^{-} $ Damping Ring & & & & & & 18 & \\ \hline
          e$^{+} $ Damping Ring & & & & & & 18 & \\ \hline
         Beam Delivery System & & & & & & & 2 \\ \hline
         \hline
         Total & 634 & 1180 & 6 & 4 & 1824 & 36 & 2 \\ \hline
      \end{tabular}
   \end{center} \vbbelow
\end{table}

\subsection{Technical Issues}\label{ssect:CRYOti}

\subsubsection{Cryogenic System Definition}\label{sssect:CRYOcsd}

The ILC cryogenic systems are defined to include cryogen
distribution as well as production.  Thus, components of the
cryogenic system include the cryogenic plants, distribution
and interface boxes, transfer lines, and non-magnetic,
non-RF cold tunnel components.  Although cryomodules,
SC magnets, and production test systems also include
significant cryogenics, those are not considered in this
section of the RDR.

\subsubsection{Cryogenic Cooling Scheme for the Main Linac}\label{sssect:CRYOccsml}

Main linac cryogenic modules each containing eight
(with magnet package) or nine (without magnet package)
nine-cell niobium cavities, cold helium pipes, and thermal
shields are the dominant load to be cooled by the
cryogenic system. The magnet package, in one third
of the cryomodules, includes a superconducting quadrupole
and corrector magnets.  The ILC cryomodule design for
the 1.3~GHz RF is based on the TESLA Test Facility (TTF)
type III design~\cite{tdr} which contains all the cryogenic
pipework inside its vacuum enclosure. There are
approximately 23~km of 1.3~GHz cryomodules including
main linac, RTML, and sources.

Series architecture is mostly used in the cryogenic unit
cooling scheme. Like for the TESLA cryogenic concept,
saturated He II cools RF cavities at 2~K, and helium gas cooled
shields intercept thermal radiation and thermal conduction
at 5 - 8~K and at 40 - 80~K.  A two-phase line (liquid helium
supply and concurrent vapor return) connects to each helium
vessel and connects to the major gas return header once per
module.  A small diameter warm-up/cool-down line connects
the bottoms of the He vessels.

A subcooled helium supply line connects to the two-phase
line via a Joule-Thomson valve once per "string" (typically 12~modules).
The 5~K and 40~K heat intercepts and radiation screens are
cooled in series through an entire cryogenic unit of up to 2.5~km in length.  For the 2~K cooling of the RF cavities, a
parallel architecture is implemented with the parallel
cooling of cryo-strings resulting in operational flexibility.
Consequently, each cryogenic unit is subdivided into
about 14 to 16~cryo-strings, each of which corresponds
to the 154~meter length elementary block of the cryogenic
refrigeration system.

\begin{figure}
   \begin{center} \vbabove
      \includegraphics[width=0.95\textwidth]{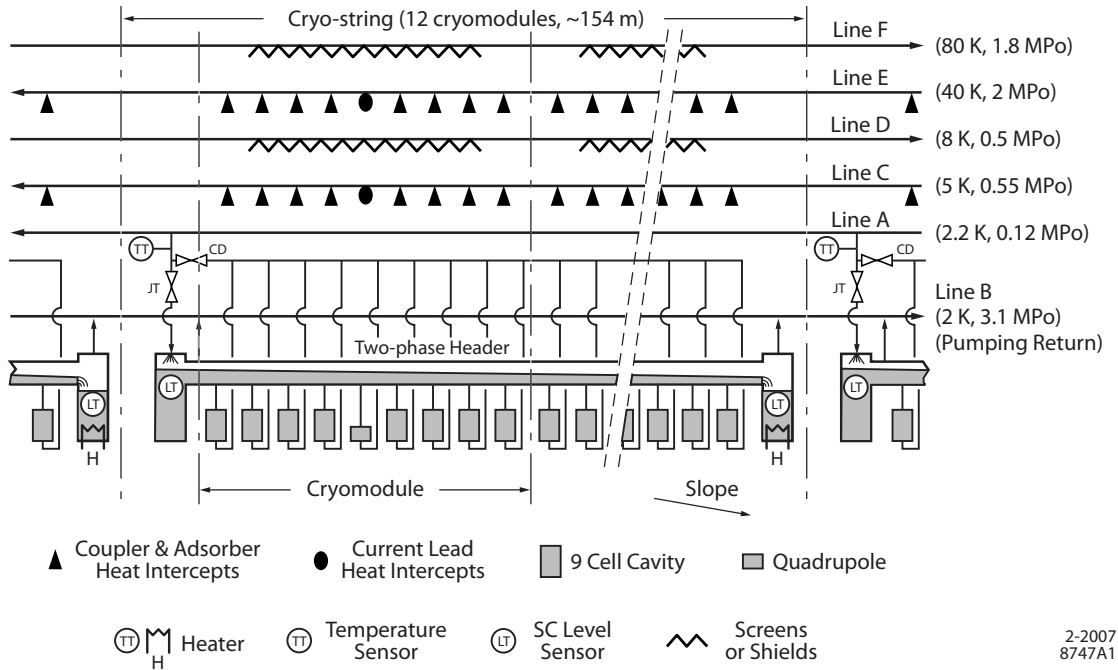}
      \vbabovecaption \caption{Cooling scheme of a cryo-string.}
      \label{fig:CRYOcooling}
   \end{center} \vbbelow
\end{figure}

Figure~\ref{fig:CRYOcooling} shows the cooling scheme
of a cryo-string, which contains 12~cryomodules. The
cavities are immersed in baths of saturated superfluid
helium gravity filled from a 2~K two-phase header.
Saturated superfluid helium is flowing all along the
two-phase header for filling the cavities and phase
separators located at both ends of the two-phase header.
The first phase separator is used to stabilize the saturated
liquid produced during the final expansion. The second
phase separator in used to recover the excess of liquid,
which is vaporized by a heater. At the interconnection of
each cryomodule, the two-phase header is connected to
the pumping return line.

\stepcounter{figlcl}

\stepcounter{figlcl}\begin{figure} [b]
   \begin{center} \vbabove
      \includegraphics[width=0.95\textwidth]{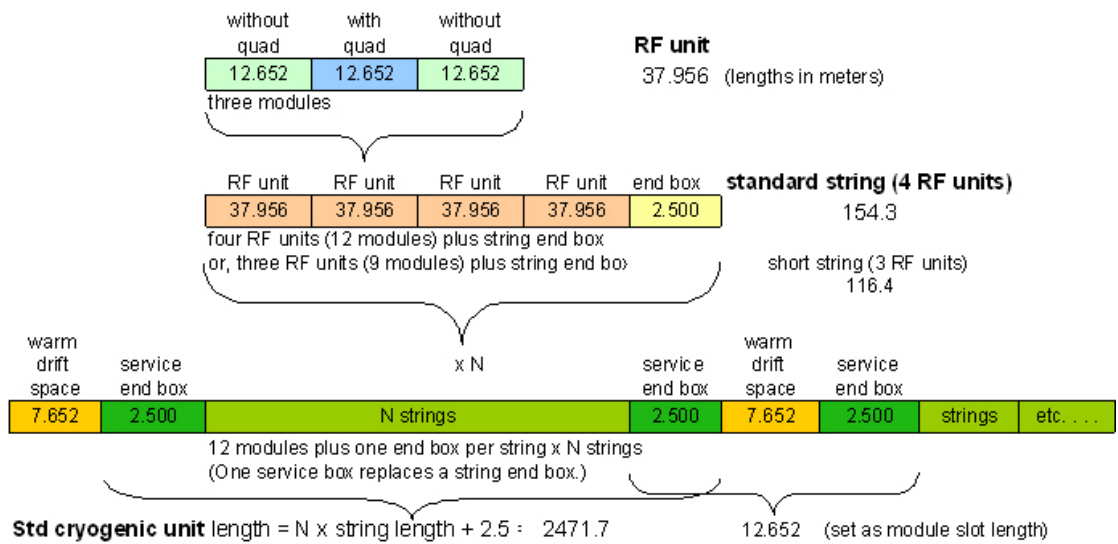}
      \vbabovecaption \caption{Lengths and typical arrangement of modules in the electron Main Linac.}
      \label{fig:CRYOlengths}
   \end{center} \vbbelow
\end{figure}

The division of the Main Linac into cryogenic units is
driven by various plant size limits and a practical size
for the low pressure return pipe.  A cryogenic plant of 25~kW
equivalent 4.5~K capacity is a practical limit due to industrial
production for heat exchanger sizes and over-the-road shipping
size restrictions.  Cryomodule piping pressure drops also
start to become rather large with more than 2.5~km distances.
Practical plant size and gas return header pressure drop limits
are reached with 192~modules in a 16-string cryogenic unit,
2.47~km long.  Five cryogenic units divide the main linac
conveniently for placing the positron source undulators at 150~GeV.
Figure~\ref{fig:CRYOlengths}, illustrates the division of the
main linac into strings and units.

\stepcounter{figlcl}\begin{figure}
   \begin{center} \vbabove
      \includegraphics[width=0.7\textwidth]{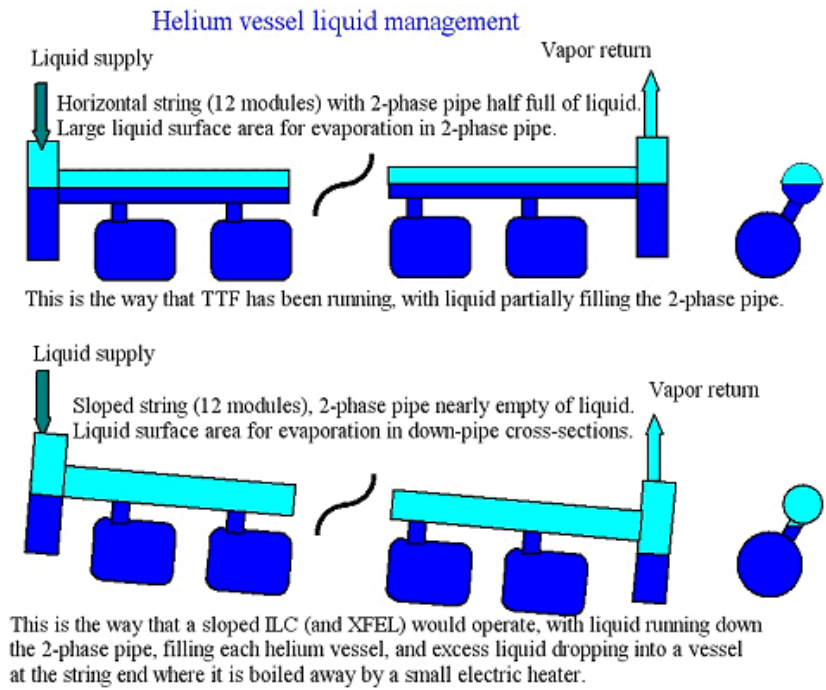}
      \vbabovecaption \caption{Two-phase helium flow for level and for sloped systems.}
      \label{fig:CRYOheliumflow}
   \end{center} \vbbelow
\end{figure}

\subsubsection{Liquid Helium Management in 1.3 GHz Modules}\label{sssect:CRYOlhm}

As the ILC site has not yet been selected, the cryogenic
system concept must accommodate different configurations
of tunnel and civil works. The tunnel may follow the earth's
curvature or be laser-straight with a maximum slope of
up to 0.6\% creating large elevation differences. To avoid
harmful instabilities, all fluid should ideally be transported
over large distances in a mono-phase state. Local
two-phase circulation of saturated liquid can be tolerated
over limited lengths, within a controlled range of vapor quality.
Figure~\ref{fig:CRYOheliumflow} illustrates two methods of liquid management
in the two-phase supply pipe for main linac cryogenic
modules, one case for a sloped system and the other for a level system.

\subsubsection{Sources, Damping Rings, and Beam Delivery Systems}\label{sssect:CRYOsdrbds}

As listed above in Table~\ref{tab:CRYOscnums}, electron
and positron sources each include just over 20~SRF
modules containing 1.3~GHz RF cavities cooled to 2~Kelvin.
The sources also include several superconducting magnets,
as well as about 150~meters of superconducting positron source undulators.
These undulators are cooled by one of the cryogenic
plants in the electron linac cryogenic system.  The
electron and positron source linacs are also cooled
from main linac cryogenic plants, as illustrated in Figure~\ref{fig:CRYOsyslayout}.

Damping ring cryogenic loads include 4.5~K superconducting
wigglers, 4.5~K 650~MHz cryomodules, associated cryogenic
distribution systems, and 70~K thermal shields for all of these.
Two cryogenic plants serve the damping rings.

The beam delivery system has one 3.9~GHz cryomodule
(containing two cavities) on each side of the interaction point,
superconducting final focus quadrupoles, and other special
superconducting magnets spaced several hundred meters
from the IR.  One cryogenic plant serves both sides of the
interaction region.  This plant could also serve the cryogenic
needs of the detectors, but that aspect of these cryogenic
systems is not considered here.

\stepcounter{tablcl}\begin{table} \vbabove \caption{Main Linac heat
loads and cryogenic plant size.}
   \label{tab:CRYOmlheatloads}
   \begin{center}
      \begin{tabular}{| l | c | c | c |}
         \hline
         & 40-80 K & 5-8 K & 2 K \\
         \hline & & & \vbdlspacing \hline
         Predicted module static heat load (W/mod) & 59.19 & 10.56 & 1.70  \\ \hline
         Predicted module dynamic heat load (W/mod) & 94.30 & 4.37 & 9.66  \\ \hline
         Modules per cryo unit & 192 & 192 & 192  \\ \hline
         Non-module heat load per cryo unit (kW) & 1.0 & 0.2 & 0.2  \\ \hline
         Total predicted heat per cryo unit (kW) & 30.47 & 3.07 & 2.38 \\ \hline
 %        Static Heat Uncertainity Factor (Fus) & 1.10 & 1.10 & 1.10 \\ \hline
 %        Dynamic Heat Uncertainity Factor (Fud) & 1.10 & 1.10 & 1.10 \\ \hline
         Efficiency (fraction Carnot) & 0.28 & 0.24 & 0.22 \\ \hline
         Efficiency  (Watts/Watt)  & 16.45 & 197.94 & 702.98 \\ \hline
         Uncertainty \& overcapacity factor (Fo)  & 1.54 & 1.54 & 1.54 \\ \hline
 %        Overall Net Cryogenic Capacity Multiplier & 1.54 & 1.54 & 1.54 \\ \hline
         Heat Load per Cryo Unit including Fo (kW) & 46.92 & 4.72 & 3.67 \\ \hline
         Installed power (kW) & 771.7 & 934.9 & 2577.6 \\ \hline
         Installed 4.5 K equivalent (kW) & 3.5 & 4.3 & 11.8 \\ \hline
         Percent of total power at each level & 18.0 & 21.8 & 60.2 \\ \hline
%     \end{tabular}
%\vspace{4pt}
%\setlength{\tabcolsep}{10pt}
%    \begin{tabular}{|  l  | c @{\hspace{11pt}} |}
%        \hline
       & & & \vbdlspacing \hline
    \multicolumn{3}{| l | }{Total operating power for one cryo unit based on        predicted heat (MW)} & 3.34 \\ \hline
      \multicolumn{3}{| l | }{Total installed power for one cryo unit (MW)} &       4.33  \\ \hline
      \multicolumn{3}{| l | }{Total installed 4.5 K equivalent power for one            cryo unit (kW)} & 19.57  \\ \hline
      \end{tabular}
   \end{center} \vbbelow
\end{table}

\subsubsection{Heat Loads and Cryogenic Plant Power}\label{sssect:CRYOhlcpp}

Table~\ref{tab:CRYOmlheatloads} shows the predicted heat load for a typical Main Linac Cryogenic Unit.  This table lists a combined uncertainty and overcapacity factor, Fo, which is a multiplier of the estimated heat loads.  The factor Fo is used to estimate a total required cryogenic plant capacity as follows.  Installed cryogenic capacity = Fo~$\times$~(Qd + Qs), where Fo is overcapacity for control, off design operation, seasonal temperature variations, and heat load uncertainty.  Qd is predicted dynamic heat load, and Qs is predicted static heat load.  Note also that cryogenic plant efficiency is assumed to be 28\% at the 40 to 80~K level and 24\% at the 5 to 8~K temperature level.  The efficiency at 2~K is only 20\%, however, due to the additional inefficiencies associated with producing refrigeration below 4.2~Kelvin.  All of these efficiencies are in accordance with recent industrial conceptual design estimates.

A similar analysis has been done for the sources, damping
rings, and beam delivery system in order to estimate size
requirements for each.  (RTML cooling is included with the main linac.)

Table~\ref{tab:CRYOdrc}, below, lists the estimated heat loads
and required cryogenic plant size for the damping rings.

\stepcounter{tablcl}\begin{table} \vbabove \caption{Damping Ring
cryogenics (per ring, two total).}
   \label{tab:CRYOdrc}
   \begin{center}
      \begin{tabular}{| l | c | c |}
         \hline
         & Units & Value \\ \hline & & \vbdlspacing \hline
        Total predicted 4.5 K heat & W & 1660 \\ \hline
        Total predicted 4.5 K liquid production (for current leads) & grams/sec & 0.80 \\ \hline
        Total predicted 70 K heat & W & 5080 \\ \hline
        Uncertainity and overcapacity (total combined) Margin & & 1.54 \\ \hline
        Installed power & MW & 1.13 \\ \hline
        Cryogenic plant capacity (converted to 4.5 K equiv) & kW & 3.45 \\ \hline
      \end{tabular}
   \end{center} \vbbelow
\end{table}

Table~\ref{tab:CRYOplantsizes} summarizes the required
capacities of the cryogenic plants for the different area systems.
The maximum required plant capacities (equivalent at 4.5~K)
are comparable with the present state of the art cryogenic
plants used in the Large Hadron Collider [3]. Total installed
power for the cryogenic system is 48~MW, with an expected
typical operating power of 37~MW.

\stepcounter{tablcl}\begin{table} [hb] \vbabove \caption[ILC
cryogenic plant sizes.] {ILC cryogenic plant sizes (sources listed
separately here, but may be combined with Main Linac).}
   \label{tab:CRYOplantsizes}
   \begin{center}
      \begin{tabular}{| l | c | c | c | c | c |}
         \hline
         & & Installed & Total & Operating & Total \\ [-6pt]
         &\# of & Plant Size & Installed & Power & Operating \\ [-6pt]
         Area & Plants & (each) & Power & (each) & Power \\ [-6pt]
        & & (MW) & (MW) & (MW) & (MW) \\ \hline & & & & & \vbdlspacing \hline
        Main Linac + RTML & 10 & 4.35 & 43.52 & 3.39 & 33.91 \\ \hline
        Sources & 2 & 0.59 & 1.18 & 0.46 & 0.92 \\ \hline
        Damping Rings & 2 & 1.26 & 2.52 & 0.88 & 1.76 \\ \hline
        BDS & 1 & 0.41 & 0.41 & 0.33 & 0.33 \\ \hline\hline
        Total & & & 47.63 & & 36.92 \\ \hline
      \end{tabular}
   \end{center} \vbbelow
\end{table}

If the tunnel is located near the surface, i.e. with depth
of access shafts smaller than 30~m, the entire cryogenic
plant can be installed above ground. If the tunnel is deep,
certain components must be installed at tunnel level
because of the hydrostatic pressure loss.

\subsubsection{Helium Inventory}\label{sssect:POShi}

\begin{figure} [t!]
   \begin{center} \vbabove
      \includegraphics[width=0.7\textwidth]{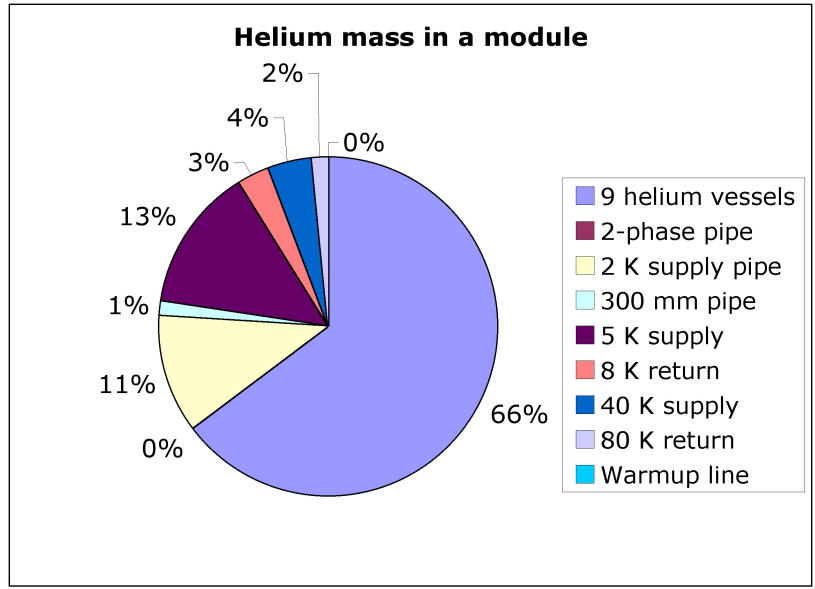}
     \vbabovecaption  \caption{Helium mass in a module.}
      \label{fig:CRYOheliumvolume}
   \end{center} \vbbelow
\end{figure}

As illustrated in Figure~\ref{fig:CRYOheliumvolume},
most of the helium inventory consists of the liquid
helium which bathes the RF cavities in the helium vessels.
The total helium inventory in ILC will be roughly equal
to that of the LHC at CERN, about 650,000 liquid liters, or
about 100~metric tons.

\stepcounter{figlcl}

\stepcounter{tablcl}\begin{table} [t!] \vbabove \caption{Main Linac
helium inventory.}
   \label{tab:CRYOmlheinventory}
\setlength{\tabcolsep}{8pt}
   \begin{center}
      \begin{tabular}{| l | c | c | c | c | c |}
         \hline
         Volumes & & Helium & &  & Inventory \\ [-6pt]
         & & (liquid liters & Tevatron & LHC & Cost \\ [-6pt]
         & & equivalent) & Equiv. &  Equiv. & (k\$) \\
        \hline & & & & & \vbdlspacing \hline
        One module & & 370 &  &  &  \\ \hline
        String & 12 modules & 4,500 & 0.1 &  & 13.4 \\ \hline
        Cryogenic unit & 14-16 strings & 68,000 & 1.1 & 0.1 & 203.6 \\ \hline
        ILC Main Linacs & 2x5 cryo units & 680,000 & 11.3 & 0.9 & 2,037 \\ \hline
      \end{tabular}
   \end{center} \vbbelow
\end{table}

\subsection{Cost Estimation}\label{ssect:CRYOce}

The cryogenic system cost estimate has been
generated based on experience in procurement of
cryogenic plants and equipment at Fermilab, CERN, DESY,
and other laboratories.

\clearpage %\setcounter{page}{1}
\setcounter{figlcl}{0}
\setcounter{tablcl}{0}
\setcounter{secnumdepth}{4} \setcounter{subsection}{0}
\setcounter{subsubsection}{0}
\usecounter{subsubsection}

\setcounter{section}{8} \renewcommand{\picturefolder}{./llrf/}
\section{Low Level RF Controls}\label{sect:LLRF}

\subsection{Overview}

The Low-Level RF system (LLRF) controls the phase and amplitude of the RF cavities used to accelerate the beam, and is essential for stable and reliable beam operation. The LLRF includes feedback and
feed-forward, exception handling and extensive built-in diagnostics
with suitable speed and accuracy. Each of the $\sim$650 L-Band RF units in the main linacs, sources and bunch compressors have a LLRF controller, as do the damping ring RF stations. LLRF also controls the crab cavities in the beam delivery and various RF diagnostic devices.

A primary challenge for the ILC LLRF is the large number of cavities driven by a single klystron. The LLRF controls the vector-sum of all
cavities as well as controlling the individual cavities. Most of the needed requirements have been demonstrated in the LLRF systems in operation at the FLASH facility at DESY \cite{bib:LLRFflash}. The DESY LLRF uses state-of-the-art technologies for digital control of the
operational parameters. Similar systems are being implemented at FNAL and KEK.

\subsection{Technical Description}

The performance requirements for the LLRF are set by the gradient desired from the cavities and by the stability required for
beam parameters such as energy and energy spread, both
bunch to bunch and pulse to pulse. There are also stringent requirements on the bunch compressor RF to set the arrival time of the beams at the IP, and on the crab cavity RF to fix the beam position at the IP.

Three issues of particular importance for the ILC LLRF are:

\begin{enumerate}

  \item Lorentz force detuning: The radiation pressure of the
    electromagnetic field during the RF pulse deforms the cavity and
    pulls it off resonance.  The static detuning ($\Delta f$) due to
    the Lorentz forces is proportional to the square of the
    accelerating field ($E_{acc}$) and is approximately 600~Hz
    \cite{bib:LLRFliepe}
    for operation at design gradient in the main linac (31.5 MV/m).

       To maximize the RF power efficiency, and to reduce the
electric fields at the cavity input coupler, it is essential to cancel the Lorentz force detuning by a fast frequency tuner (for example,
       piezoelectric actuators). \itemspace

  \item Microphonics: External mechanical vibrations can be
    transferred to the cavities via the supporting system within the
    cryostat.  Modulation of the resonant frequency due to microphonics
    is estimated to be $\sim$10~Hz rms.  This modulation is not
    correlated to the macro pulse and therefore can only be
    corrected by the feedback system. \itemspace

  \item Beam loading: The beam loading by individual bunches is about
    0.15\% at design bunch charge, which is considered acceptable.
    However, slow bunch charge
    fluctuations within the bandwidth of the RF system cause cavity
    vector disturbances that need to be controlled on the order of
    0.05\% at each station as the bunch charge fluctuations are
    correlated through the accelerator chain.  Bunch charge is
    measured in the DRs and processed by the LLRF to create a
    correction feedforward term before beam is injected into the
    linac. \itemspace

\end{enumerate}

\stepcounter{tablcl}\begin{table}[htb!] \vbabove \caption[Summary of
tolerances for phase and amplitude control.]{Summary of tolerances
for phase and amplitude control.
  These tolerances limit the average luminosity loss to $<$2\% and
  limit the increase in RMS center-of-mass energy spread to $<$10\% of
  the nominal energy spread.}
\label{tab:LLRFtols}
\begin{center}
\setlength{\tabcolsep}{4pt}
    \begin{tabular}{| l | c | c  | c | c | l |} \hline
    Location & \multicolumn{2}{| c |}{Phase (degree)} &
       \multicolumn{2}{| c |}{Amplitude (\%)} & limitation \\
    & correlated & uncorr. & correlated & uncorr. & \\
       \hline & & & & & \vbdlspacing \hline 
Bunch Compressor & 0.24& 0.48& 0.5& 1.6& timing stability at IP \\
    & & & & &  (luminosity) \\
Main Linac& 0.35& 5.6& 0.07& 1.05& energy stability $\le$0.1\% \\ \hline

    \end{tabular}
\end{center} \vbbelow
\end{table}

The RF systems in the main linacs and RTML require tight field
control on the order of up to 0.07\% for amplitude errors and
0.35$^\circ$ for the phase. Due to microphonics, the measurement of the vector sum must be calibrated to an accuracy on the order of
1\% for amplitude and 1.0$^\circ$ for phase. The phases of
crab cavities in the beam delivery system must be stabilized to
better than 0.015$^\circ$. Table~\ref{tab:LLRFtols} gives an
overview of the regulation requirements of the Main Linac and RTML
bunch compressor.

Besides field stabilization, the LLRF provides automatic
beam-based system calibration and diagnostic signals to the accelerator
control system. Exception handling is required to avoid
unnecessary beam loss and to allow for maximum operable gradient.
% ED note: this doesn't say anything.

Availability and maintainability are also critical
considerations in the LLRF system design. Although most of the LLRF system
components are located in the service tunnel, the large number of units
requires a high availability design.
Possible failure modes must be understood, their
operational impacts examined, and mitigation measures developed and implemented. Adequate redundancy such as a simple
feed-forward technique in the complex feedback scheme should be an
integral part of the system design. Built-in diagnostics for both hardware
and software are required to support preventative maintenance and increase
reliability.

\subsection{Technical Issues}
\subsubsection{Hardware Architecture }

%\subsubsection{System Principles}

\begin{figure}[htb!]
\begin{center} \vbabove
\includegraphics[width=14cm]{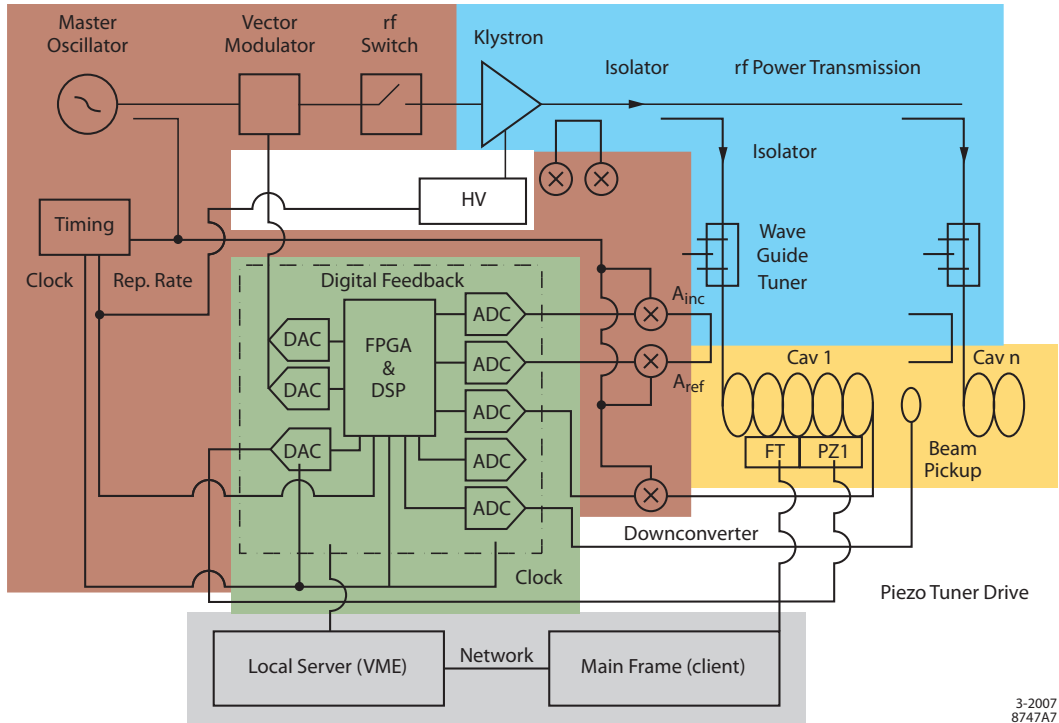}
\vbabovecaption
\caption{Typical configuration of an RF control system using digital
  feedback control.}
\label{fig:LLRFconfig}
\end{center} \vbbelow
\end{figure}

The most basic function of any LLRF control is a feedback that measures
the cavity field vector and attempts to hold it to a
desired set-point. The vector
difference between the measured field and the set-point is filtered
and amplified, then used to modulate the klystron drive and thereby
the incident power to the cavities. The forward and
reflected power signals are also processed to measure the
resonant frequencies of the ILC cavities, for automated adjustment by
slow motor-controlled tuners and fast piezoelectric actuators.
The architecture of a typical LLRF control system is shown in
Figure~\ref{fig:LLRFconfig}. The signal
from the master oscillator, brought through the RF distribution
system, is used as the RF reference.

\stepcounter{figlcl}

The LLRF has to combat numerous perturbations with various time
patterns and frequencies. Some of these perturbations recur at the machine
repetition rate (5~Hz for ILC),
like Lorentz force detuning and beam loading. An adaptive
feedforward system is used to compensate for the average repetitive
errors. The set-points for cavity fields are also implemented in a
table to accommodate the time-varying gradient and phase
during the cavity filling.

\subsubsection{Digital Technologies}

The key technologies to be used are modern Analog to Digital
Converters (ADCs), Digital to Analog Converters (DACs), as well as
powerful Field Programmable Gate Arrays (FPGAs) and Digital Signal
Processors (DSPs) for signal processing. Low latency can be realized,
with time delays
from ADC input to DAC output ranging from a few 100~ns to several $\mu$s
depending on the chosen processor and the complexity of the
algorithms. Gigabit links are used for the high speed data transfer
between the large number of analog input and output channels and the
digital processor as well as for communication between various signal
processing units. Typical parameters for the ADCs and DACs are sample
rates of 65-125~MHz and 14-bit resolution.  The signal processing
uses FPGAs with several million gates, including many
fast multipliers. More complex algorithms are implemented on slower
floating point DSPs

A down-converter module translates the 1.3~GHz RF cavity probe signal
to the Intermediate Frequency (IF) where it can be digitized and
processed further. The down converter can degrade
overall performance if not properly designed. Problems with nonlinearities,
thermal noise, phase noise and thermal stability must be addressed in order to
maintain the integrity of the detected signal from the cavity.
The up-converter module translates a digitally generated IF signal
back to the RF in a process similar to that of the down converter. The
up converter has less stability issues since it is
within the feedback loop.

A fast piezoelectric actuator
and a slow motor-driven tuner control the resonant frequency of each individual cavity.  The frequency error of the cavity is measured during and after the flattop.  This
error can be reduced by suitable excitation of the piezoelectric
actuator (fast tuner), or it can be compensated via additional RF power.  The
motor-driven tuner is only used to correct for long-term drifts.  The
station LLRF system must interface to High Level RF, beam transfer
control, machine protection, sector and global energy and phase
regulation, and the control system.  A control system IOC is built
into the LLRF system to handle parameter and data collection.

\subsubsection{Software Architecture}

%\subsubsection{System Principles}

\begin{figure}[htb!]
\begin{center} \vbabove
\includegraphics[width=14cm]{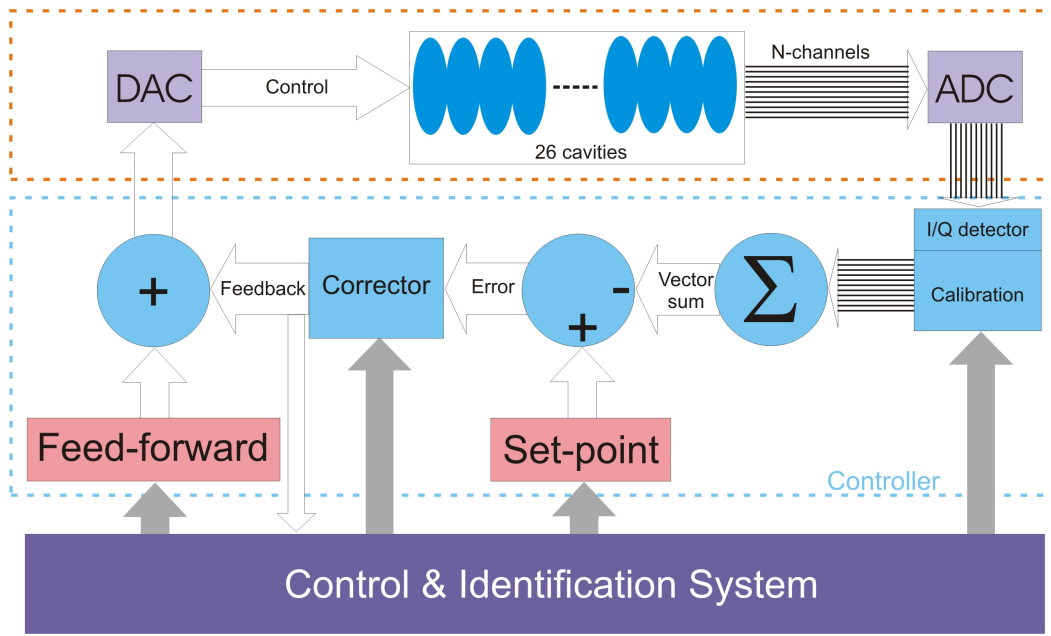}
\vbabovecaption
\caption{Basic functional diagram of the LLRF software system.}
\label{fig:LLRFsw}
\end{center} \vbbelow
\end{figure}

A major benefit of a digital RF feedback and feed-forward system is
that it supports automated operation with minimal operator
intervention. This is accomplished by deploying a number of algorithms
to maintain best field stability (i.e. lowest possible rms amplitude
and phase errors), to allow for fast trip recovery, and to support
sophisticated exception handling. Beam-based feed-forward further
improves the field stability. Figure~\ref{fig:LLRFsw} shows the basic
functional diagram of the LLRF software system.

\stepcounter{figlcl}

The software implementation of the RF control system must also support
high availability. The main requirements for the algorithms are low
latency for feedback, modularity to simplify interfacing, and support
of a high degree of automation. Important applications include
exception handling, built-in diagnostics and beam-based feedback.

subsubsection{Software Implementation}

The massive parallel processing in the FPGAs provides low latency
for the feedback algorithm. Complex algorithms requiring floating point
calculations such as adaptive feed-forward can be also
implemented.

The setting of system parameters and piezoelectric tuner control
are implemented on floating point DSP processors since the latency
requirements are not as stringent. Automated
operation can also be implemented on a middle layer server CPU since the
timing requirements are not as critical.

The distribution of the modular algorithms requires well-defined
interfaces to ensure simplicity in performing trouble shooting,
maintenance, and upgrades. Low latency links use in-house
protocols while commercial protocols are available for links
needing high bandwidth but not low latency.

System redundancy is achieved with algorithms, which calculate the
key results from multiple signal sources. It is, for example, possible to
calculate the cavity field from forward and reflected power although
the measurement error is larger. Any discrepancy between the
independently derived signals flags potential errors in hardware or
algorithms.

Data storage is provided locally on most processor boards and is
distributed to the central servers between pulses for further signal
processing. With almost 15,000 cavities to control, automation is
essential to ensure simplicity of operation and high availability.
To support automation, the front-end hardware and software must
as a minimum include the following features:
field vector measurement, loop phase and loop gain,
loaded Q and cavity detuning, beam phase and beam induced voltage,
calibration of cavity field and phase, vector-sum calibration,
calibration of forward and reflected wave, beam loading compensation
(current and phase), klystron linearization, exception detection and
handling, RMS field errors, warnings and alarms.

It is desirable to implement the algorithms as close a possible to the
LLRF station controller to reduce network traffic. However, if the algorithms and applications are implemented in middle
layer servers or as client applications, it can simplify the programming,
facilitate later upgrades and improve maintainability.

\subsection{Components}

Table~\ref{tab:LLRFcount} gives a rough parts count for the components
in the baseline LLRF system for a single RF unit in the main linac.

\stepcounter{tablcl}\begin{table}[htb!] \vbabove \caption{Rough
parts count for the components in the baseline LLRF system for a
single RF unit at the main linac.} \label{tab:LLRFcount}
\begin{center}
\begin{tabular}{| l | l | l | } \hline
  Module           & Specification              & Quantity     \\ \hline & & \vbdlspacing \hline
  Precision cable  & 1/2 Coax--low temp.coef. & 94           \\ \hline
  Down converter   & 1300MHz to IF              & 95           \\ \hline
  ADC channel      & 14 bit, 65MHz or higher    & 95           \\ \hline
  FPGA \& DSP      & State of the art           & 3 to 10 each \\ \hline
  DACs             & 16 bit, 100 MHz or higher  & 6            \\
  \hline
\end{tabular}
\end{center} \vbbelow

\end{table}
There are a total of 14,540 cavity modules in the main linacs, where
560 klystrons (i.e. 560 RF units) provide the drive power for 26
cavities each. The e$^{-}$ source, e$^{+}$ source, RTMLs have 11, 39
and 36 RF units, respectively. The e$^{-}$ and e$^{+}$ damping rings
have 10 klystrons driving 36 superconducting cavity modules in
total.  Each of these cavity modules has three signals monitored by
the LLRF, a cavity field probe, and a forward and reflected power
signal.  Each signal is routed in temperature-stabilized coaxial
cable.

\clearpage %\setcounter{page}{1}
\setcounter{figlcl}{0}
\setcounter{tablcl}{0}
\setcounter{secnumdepth}{4} \setcounter{subsection}{0}
\setcounter{subsubsection}{0}
\usecounter{subsubsection}

\setcounter{section}{9} \renewcommand{\picturefolder}{./instrum/}

\section{Instrumentation}\label{sect:RUMi}

\subsection{Overview}\label{ssect:RUMo}

To deliver high luminosity, the ILC must produce very low emittance
beams in the damping rings, preserve that low emittance through more
than 20 kilometers of beam transport, bunch compression and
acceleration, to finally focus the beams to a few nanometers at the
collision point. This requires extensive beam instrumentation with
requirements at or often beyond the current state-of-the-art. Most
of the beam instrumentation in the linac and beam delivery requires
single-pass, bunch-by-bunch signal processing and data acquisition.
The damping ring requires turn-by-turn or multi-turn measurements
similar to modern storage rings. Beam
instrumentation is a critical component of:

\begin{itemize}
\item     diagnostic systems characterizing machine
             performance,  beam properties and collision parameters, \itemspace
\item     beam-based feedbacks, \itemspace
\item     machine protection system. \itemspace
\end{itemize}

The beam position monitors (BPM), beam profile monitoring
systems and feedbacks are particularly challenging, and
include devices based on RF cavities or lasers. In many
cases, individual devices have been built that satisfy the
minimal requirements, but these must be integrated into large,
highly reliable systems to achieve the required levels of beam
monitoring and control.

\subsection{Technical Description}\label{ssect:RUMtd}

Instrumentation includes all direct {\it beam} monitors, e.g.
beam position, profile, bunch length and bunch charge monitors,
as well as beam feedbacks, but not general machine infrastructure
monitoring systems such as RF control and protection interlocks, temperature
and pressure monitors, flow meters, etc. Near the interaction point (IP),
there is also specialized beam instrumentation, e.g. luminosity
and background monitors, energy spectrometers and polarimeters,
that is not within the scope of the instrumentation technical system.

In both physical and cost terms, the largest instrumentation
systems are the beam position monitors (BPMs)
and the laser-based beam profile monitors ({\it laser-wires}).
The BPM systems
consist of $\sim$4500 beam pickups of two basic types, i.e.
resonant cavity-sytle and broadband button (or stripline) style,
with associated analog front-end electronics, digital signal
processing, and related infrastructure such as cables,
power-supplies, racks, crates, etc., distributed along the beamlines.
The laser-wires include 68 laser/beam Interaction Points, fed by 17 lasers with 29 Compton
gamma detectors.

\subsubsection{Beam Position Monitors}\label{sssect:RUMbpom}

\stepcounter{figlcl}\begin{figure}[htbp]
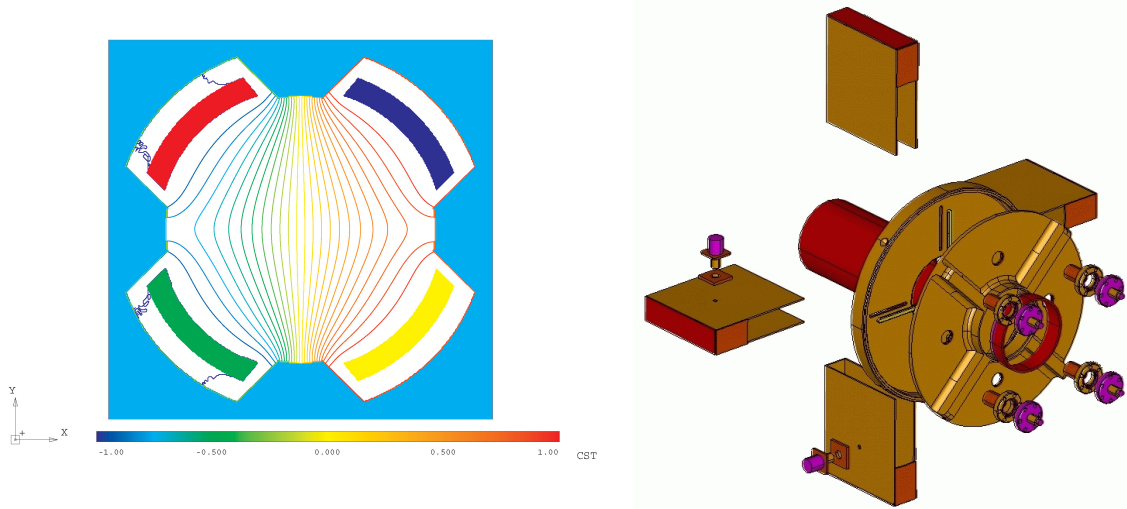

   \begin{center} \vbabove
      \begin{minipage}[c]{0.53\textwidth}
      \resizebox*{\textwidth}{!}{\includegraphics{\picturefolder hor2d.jpg}}
      \end{minipage}\hfill
      \begin{minipage}[c]{0.43\textwidth}
\resizebox*{\textwidth}{!}{\includegraphics{\picturefolder BPM_exploded.jpg}}
      \end{minipage}
     \vbabovecaption  \caption[Broadband and resonant BPM pickups.] {Broadband (left: hor.\ equipotentials of the ATF
      damping ring button BPM) and resonant BPM pickups (right: exploded view
      of the L-Band cavity BPM).}
      \label{fig:RUMbpom}
   \end{center} \vbbelow
\end{figure}

The beam position monitor systems in the ILC accelerator complex are
the most essential and most extensive beam instrumentation tool.
Four different types of beam position monitors (BPMs) are used
throughout the ILC. Broadband BPMs of {\it stripline} or {\it
button} style (Fig.~\ref{fig:RUMbpom}, left) are used for
applications requiring medium    or lower resolution,
$\sim$10-30~$\mu$m RMS (single bunch). Button pickups  are used in
the gun region, in the damping rings and in other space critical
areas. Stripline pickups are used in most warm sections of the
sources and in the BDS. Cavity BPMs are used for higher resolution
applications, where few- or sub-micron RMS single-bunch resolution
is required (see Table~\ref{tab:RUMcounts}). Three different basic
styles, C-Band, S-Band and L-Band,       are used according to the
needs of different beam pipe apertures. A ``cold'' version of the
L-Band cavity BPM is used in the cryostats of the Main Linacs, RTMLs
and Sources (Fig.~\ref{fig:RUMbpom}, right). ``Warm'' cavity BPMs of
all styles are used thoughout the ILC accelerator complex downstream
of the damping rings.

Except for the damping rings, all BPM systems are designed to be able to provide the
beam position of each bunch in the macropulse (bunch-by-bunch).
This requires a measurement or integration time smaller than the
bunch-by-bunch time spacing (369~ns, nominal) for all BPM system
components.
The damping ring BPMs have to time resolve the beam position on a
turn-by-turn basis (t$_{\rm rev}$~$\sim$20~$\mu$s) or measure in a
narrow-band (BW~$\sim$1~kHz) averaging mode.
A common
set of readout, timing and auxiliary hardware and software is used for all
BPMs, apart from the RF analog signal processing front-end section.
This minimizes cost, and simplifies commissioning, maintenance and troubleshooting. A beam position monitor consists of:

\begin{itemize}
\item     A pickup detector, which detects the beam's electromagnetic field and
            converts it to an electrical signal, usually in the range of RF
            or microwave frequencies. \itemspace
\item     A set of analog and digital read-out electronics, which
            processes the pickup signals to extract the required beam
            displacement information. \itemspace
\item     Trigger and timing hardware to time-resolve position data
             for individual bunches or turns. \itemspace
\item     A system for calibration and self-diagnosis tests. \itemspace
\item     Digital data acquisition and control hardware and software,
             including a control system interface. \itemspace
\item     Auxiliary systems and components (racks, crates, power supplies, cables, etc.). \itemspace
\end{itemize}

There are a variety of R\&D activities for ILC BPMs at the
laboratories, mostly including university collaboration. Warm cavity
BPMs studied under ILC-like beam conditions (nanoBPM collaboration)
at the KEK Accelerator Test Facility (ATF) have achieved a
single-bunch position resolution of $\sim$20~nm. The ATF damping
ring is also developing high resolution BPMs based on a digital
receiver readout system~\cite{nanoBPM}. The DESY FLASH linac has a
variety of button and stripline-BPMs, and uses RF-signals from the
HOM-couplers of the accelerating structures for beam position and
alignment studies (HOM collaboration) ~\cite{HOM}. S-Band cavity
BPMs tested in SLAC ``End Station A'' (ESA) achieved a single bunch
resolution well below 1~$\mu$m~\cite{S-Band-cavity}. A ``cold''
L-Band cavity-BPM for use in the cryomodule is being built at FNAL
using a read-out digitizer based on the high availability ATCA
standard ~\cite{L-band-BPM}.

\subsubsection{Beam Profile Monitors}\label{sssect:RUMbprm}

\stepcounter{figlcl}\begin{figure}[htbp]
   \begin{center} \vbabove
      \includegraphics[width=\textwidth]{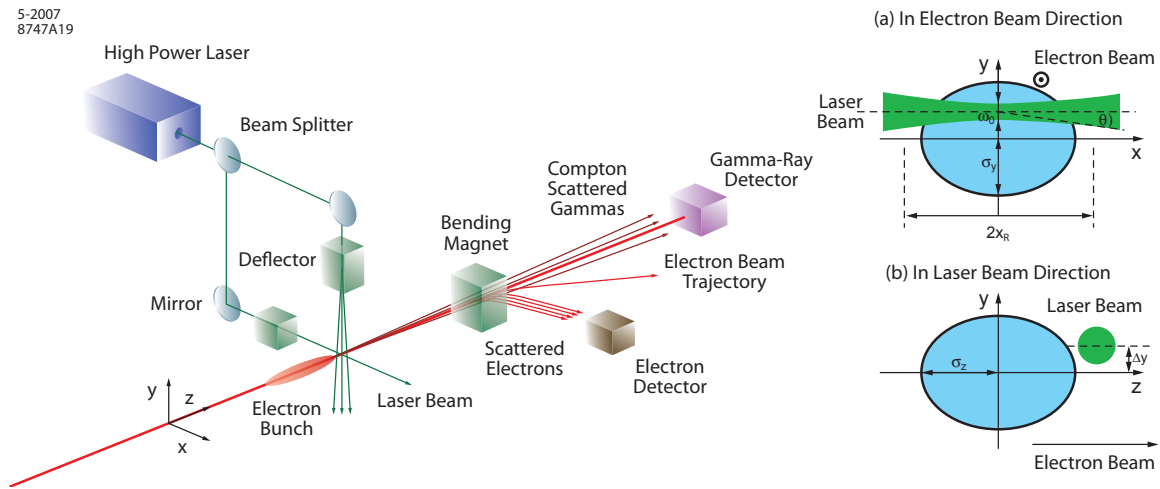}
      \vbabovecaption \caption{Schematic of a laser-wire beam profile monitor.}
       \label{fig:RUMbprm}
  \end{center} \vbbelow
\end{figure}

A variety of beam profile monitors are used throughout the ILC.
Conventional wire scanners are used for beam transverse
emittance measurements in upstream low-energy sections, i.e.
the electron and positron sources. However, in the damping rings
and downstream areas of the machine the low emittance beam would destroy
any conventional wire
scanner. In these areas laser-wires must be used for any measurements of the
beam's transverse dimensions.

The laser-wire (Fig.~\ref{fig:RUMbprm}) operates by scanning a finely focussed beam of
laser light across the electron/positron bunches. The resulting rate of Compton
scattered photons is measured in a downstream detector, as a function of
relative position of laser and beam. The laser-wire is a relatively
non-invasive device and can be used to measure the beam properties
continuously during ILC operations. Prototype laser-wire systems are
being developed at PETRA~\cite{PETRAlaserwire} and ATF~\cite{ATFlaserwire}~\cite{ATFlaserwire2}.
In the latter case the key R\&D
challenge is to push the spatial resolution to the micron level, as required for ILC.

Other optical beam monitors are used to analyze transverse
and longitudinal beam parameters, and beam energy.
There are OTR (optical transition radiation) and OTRI (OTR interferometer)
 screen monitors for beam emittance and energy measurements in the sources,
RTML and BDS. Screen monitors are also used in other ways (e.g. YaG, slits, etc.)
X-ray synchrotron light monitors are used for transverse and
longitudinal beam imaging in the positron source, damping rings,
RTML and BDS. In the damping rings they can image the 3D
parameters of a bunch on a turn-by-turn basis (as done at LEP).
Other optical-based beam monitor systems, currently not in the RDR
baseline, may be required (e.g. bunch length measurements based on
elecro-optical sampling (EOS),
optical diffraction radiation (ODR) monitors, interferometers, etc.).
As these are lower cost
single system installations, they would not affect significantly
the overall instrumentation costs and requirements.

\subsubsection{Bunch Length Monitors}\label{sssect:RUMblm}

The electron and
positron sources and RTMLs have Deflecting Mode Cavity (DMC) or {\it LOLA}~\cite{lola} structures, based on
normal-conducting technology to measure bunch length and
longitudinal charge distribution.
A pair of DMCs based on superconducting technology are located near
the crab-cavity bunch rotation system just upstream of the IP. Streak
cameras are used for beam imaging in the damping rings.

\subsubsection{Beam Current Monitors}\label{ssect:RUMbcm}

There are a variety of beam current monitors used to measure the bunch charge, including toroids, wall current monitors (WCM), Faraday cups and DC Current Transformers (DCCT). The WCM and Faraday cups are located in the sources, and the DCCTs in the damping rings. These monitors measure the charge
of every bunch in the macro pulse. Like the BPMs, the measurement time
has to be $<$~369 ns to time resolve the charge of individual bunches.
Monitors with higher bandwidth are required in the damping rings where
the bunch-to-bunch spacing is 6~ns. Synchronized bunch
charge measurements also quantify the injection/ejection efficiency
to/from the damping rings, and are used to detect beam losses as part of the machine protection system (MPS). For luminosity
monitoring, a high precision bunch charge measurement is required
in both Beam Delivery Systems. All of these devices are commonly available and require little or no R\&D.

Toroids are the simplest and most
reliable detector for bunch charge measurements, with medium to high bandwidth
(100$\ldots$1000~MHz), and a cut-off frequency as low as 10 Hz.
Toroids for accelerator applications are offered by several
smaller companies but are also developed in-house at some laboratories (eg. DESY, CERN).

Faraday cups are used in the electron source
at the end of the low-energy spectrometer and in the gun region. As they
physically collect the particles, they have a very high sensitivity
and can also be used for dark current investigations. The bandwidth
is sufficient to resolve the charge of individual
bunches.

The wall current monitor (WCM) is a broadband beam current / bunch
charge monitor which offers very high bandwidth (typically 5-10~GHz).
It is used in the electron and positron sources as an excellent source of bunch timing signals and as a diagnostic
for issues in the timing and trigger distribution system, e.g. filled
neighbor buckets (parasitic bunches), time-of-flight measurements, etc.

A DCCT monitor in each damping
ring measures the DC beam current component with high resolution.
The system can also serve for diagnostic purposes and machine
development studies, e.g. beam lifetime studies.

\subsubsection{Beam Phase Monitors }\label{sssect:RUMbphm}

Beam phase monitors are used in the electron and positron sources,
RTMLs and BDS. The precise measurement of the phase or time of the
bunch center (or the average of all bunches) with respect to the
1.3~GHz RF-drive signal, is crucial for successful ILC operation
(see Section~\ref{sect:LLRF}). The beam phase can be used to
diagnose numerous machine performance issues, such as unwanted
signal content generated in the different sections of the RF sources
and distribution (noise, jitter, wrong set points, problems in
feedback, feed-forward or state machine systems), problems in the
related auxiliary systems (water cooling, power distribution), in
the accelerating cavities (slow tuners, Lorenz force compensation),
and finally issues driven by the beam itself (beam loading,
wakefields). The resolution requirements for a beam phase monitor
ranges from 0.1-0.01$^{\circ}$ of 1.3~GHz (equivalent to
$\sim$200-20~fs). In many instances, an average beam phase
measurement is sufficient, but in some cases, a bunch-by-bunch beam
phase gives additional, valuable information. Two or more broadband
detectors can provide a time-of-flight (TOF) measurement of
particular interest in the bunch compressors. DESY is currently
developing two beam phase measurement methods, relevant for ILC.

\begin{itemize}
\item     A broadband, bunch-by-bunch beam phase
            and TOF measurement system is based on an electrical pickup
            (similar to a button BPM) read-out by an optical Terahertz sampler.
          Beam tests show a bunch-by-bunch
            resolution of 30~fs RMS. This method is used in several locations
           in the sources, RTML and BDS areas of the ILC. \itemspace
\item     A broadband read-out (oscilloscope based) of the
            HOM signals is used for a high resolution (0.08 degree RMS,
            equivalent 170~fs) measurement of the beam phase, by comparing
            the signal of the RF-driven fundamental TM$ _{010} $ mode (1.3~GHz) with
            the beam-driven first higher monopole mode TM$_{011}$ mode. This
technique is used in the RF cryomodules to measure the average beam phase of all bunches. \itemspace
\end{itemize}

\subsubsection{Beam Loss Monitors}\label{sssect:RUMblom}

 Two types of beam loss monitors are used throughout the entire machine complex.
Long ion-chambers (LION) run along the tunnel sections and
photo-multiplier tube (PMT) based beam loss monitors are attached to
scintillation paddles or aluminum foils. Both systems are used for
machine commissioning and for the machine protection system (MPS). A
reliable detection of low beam losses $<$0.01 \% of the total beam
intensity is required, along with good calibration and linearity.

\subsubsection{Beam Feedback Systems }\label{sssect:RUMbfs}

Beam based feedback systems stabilize the beam current, energy and
trajectory throughout the machine. There are slow, pulse-to-pulse
(5~Hz), and bunch-to-bunch ('intra-train') feedbacks. Only
beam-based feedback systems are discussed here, all of which employ
instrumentation such as beam position monitors (BPMs) and fast
kickers. Other feedback and feed-forward systems (including non-beam
based), such as adaptive LLRF control loops, cavity temperature
control, etc.\ are covered elsewhere. A partial list of feedback
loops is given in Table~\ref{tab:instrfdbk}.

\stepcounter{tablcl}\begin{table} [htb] \vbabove
 \caption{Partial list of feedback loops.}
   \label{tab:instrfdbk}
   \begin{center}
      \begin{tabular}{| l | c |}
      \hline
      \multicolumn{2}{| l |}{Damping Ring} \\ \hline & \vbdlspacing \hline
     ~~~~Injection and extraction trajectory control & 5~Hz \\ \hline
     ~~~~Dynamic orbit control & 10-20~KHz \\ \hline
     ~~~~Bunch-by-bunch transverse feedback & \\ \hline \multicolumn{2}{|c|}{}  \vbdlspacing \hline
      \multicolumn{2}{| l |}{Ring to Main Linac} \\ \hline & \vbdlspacing \hline
     ~~~~Pre- and post-turnaround emittance correction & 5~Hz \\ \hline
     ~~~~Turnaround trajectory feed-forward & bunch-by-bunch \\ \hline
     ~~~~Beam energy at bunch compressor & two stages \\ \hline \multicolumn{2}{|c|}{} \vbdlspacing  \hline
      \multicolumn{2}{| l |}{Main Linac} \\ \hline & \vbdlspacing \hline
     ~~~~Trajectory Feedback (several cascaded loops) & 5~Hz \\ \hline
     ~~~~Dispersion measurement and control & \\ \hline
     ~~~~Beam energy (several cascaded sections) & 5~Hz \\ \hline
     ~~~~End of linac trajectory control & bunch-by-bunch \\ \hline \multicolumn{2}{|c|}{} \vbdlspacing \hline
      \multicolumn{2}{| l |}{Positron Source} \\ \hline & \vbdlspacing \hline
     ~~~~Beam energy at undulator & 5~Hz \\ \hline \multicolumn{2}{|c|}{} \vbdlspacing \hline
      \multicolumn{2}{| l |}{Beam Delivery System} \\ \hline & \vbdlspacing \hline
     ~~~~Trajectory feedback & 5~Hz \\ \hline
     ~~~~Interaction Point collision feedbacks & 5~Hz and bunch-by-bunch \\ \hline
      \end{tabular}
   \end{center}
\vbbelow
\end{table}

Damping ring orbit stability requirements are similar to those for
existing storage rings such as B~factories and synchrotron light sources.
Orbit feedback based on a
response matrix method takes position measurements from multiple
BPMs around the ring, and corrects the orbit with multiple
distributed correctors, using algorithms and
technology that are well established.

A turnaround in the Ring to Main Linac (RTML) allows bunch-by-bunch trajectory measurements to be fed forward
over a shorter path length to two
fast correctors/kickers per plane, separated by 90 degree phase advance.
Processing time is critical as the turnaround length
is only 170~m, which allows less than 0.5~$\mu$sec to measure, process,
and apply the kick angle correction.

All trajectory feedback, except the RTML feed-forward, has the same basic elements, the same algorithm,
and similar or identical hardware. The algorithm is based on response matrices, but most of the trajectory correction loops operate
synchronously at the 5~Hz ILC pulse rate. BPM measurements are processed locally, and read by the middle-ware layer of
the control system, which then calculates corrector magnet
settings for the subsequent ILC pulse, and distributes the corrector
setpoints synchronously.

Several cascaded feedback loops provide position and energy control in the
sources, bunch compressor, main linac, and beam delivery system.
In addition to the trajectory feedback, two BPMs in each section are used to
measure beam energy and provide local feedback using klystron
phase/amplitude control. There is a 5~Hz BDS trajectory feedback
system that may be cascaded with the linac 5~Hz systems, and/or
augmented with feed-forward information from upstream in the machine
(i.e. from the linacs and/or the damping rings). In addition, there is a
5~Hz interaction-point (IP) feedback system. All of these systems
will use similar hardware software based on state space analysis and adaptive
feedback algorithms.

For collision optimization, and luminosity stabilization, there is
an intra-train (bunch-to-bunch) feedback system at the IP. A BPM
sensor  several meters downstream of the IP measures the
position of the outgoing bunches, and a kicker several meters
upstream of the IP corrects the incoming bunches. Such a system
can {\it lock in} within $\sim$100 bunch crossings to achieve roughly 80\%
of the luminosity attainable if the beams were in perfect collision.
Additional upstream BPM-kicker sets provide angle correction.
An intra-train position/angle scan(s) is used to optimize a
bunch-by-bunch luminosity signal from the detector.
Inputs to the feedbacks from additional diagnostics such as beam
charge, transverse size, and bunch length monitors allow adaptive
gain control as collision conditions change.

\subsection{Technical Issues}\label{ssect:RUMti}

\subsubsection{ Feedback Hardware}\label{sssect:RUMfh}

The relatively low correction rates and the distributed nature of
many of the monitors and actuators make it possible to implement
the 5~Hz feedback in the integrated controls infrastructure without
requiring dedicated hardware and interfaces. Dedicated local
systems are required for intra-bunch feedback systems that must
operate at the bunch rate of $\sim$3~MHz, such as the RTML
turnaround trajectory feed-forward control, and intra-bunch trajectory
control at the IP. In addition, a fast synchronous infrastructure will
allow implementation of delayed bunch-to-bunch feedback/feed-forward
along the length of the linac.

Modern storage rings have refined orbit
correction systems to the level likely required for the ILC damping ring.
Ongoing advances in digital processor performance and fast high
performance analog to digital conversion chips has allowed the
conversion from the analog to digital domains to be performed
much earlier in the signal chain. Most challenging are systematic
effects in beam position monitoring when required resolutions are
at or below the few micron level.

Fast intra-bunch trajectory control for the IP is presently being
developed by the FONT collaboration, with the latest implementation being
(FONT-4)~\cite{FONT4} aiming to demonstrate feedback with 100~ns latency in the electronics, and stabilization at $\mu$m level.

\subsubsection{Layout }\label{sssect:RUMl}

\begin{figure}[htb]
   \begin{center} \vbabove
      \includegraphics[width=0.95\textwidth]{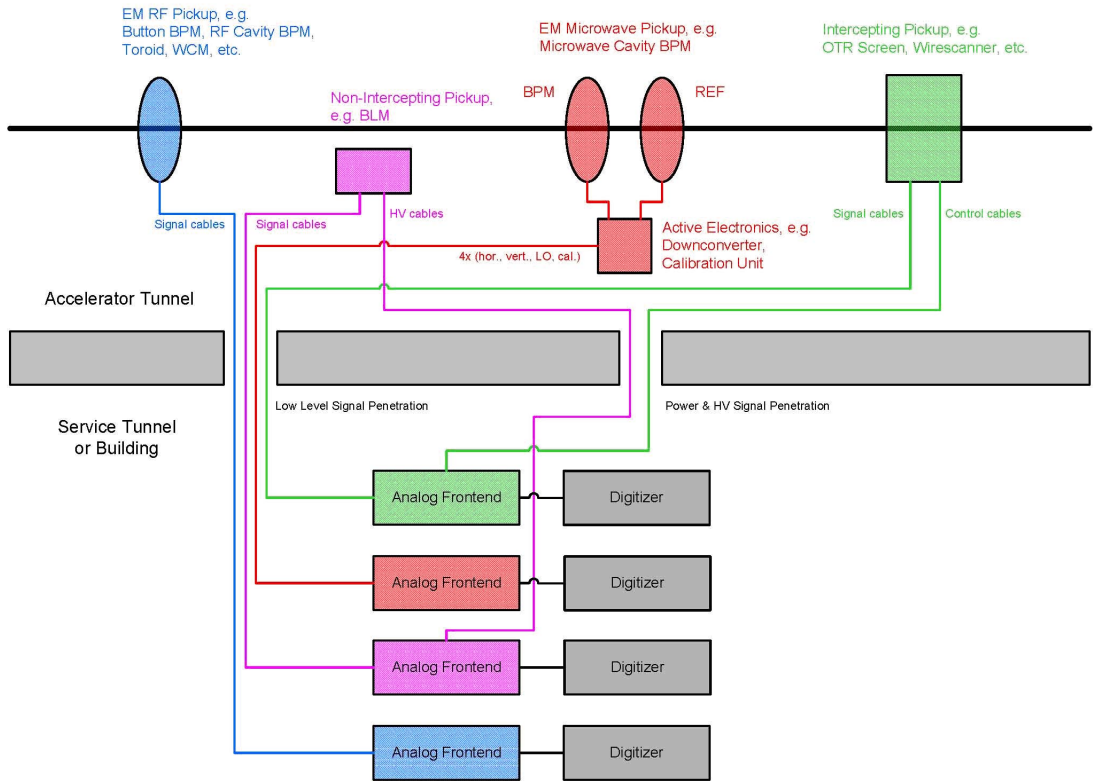}
      \vbabovecaption \caption{Generalised Schematic of Beam Pickups and Read-Out Systems.}
      \label{fig:RUMbeampickup}
   \end{center} \vbbelow
\end{figure}

A generalized schematic of an Instrumentation system is shown
in Fig.~\ref{fig:RUMbeampickup}. While the pickup monitors of the beam instruments
are located in the accelerator tunnel (in most cases they are part
of the vacuum system), the read-out electronics are typically installed
in an accessible service tunnel or in service buildings. Pickup stations
and electronics are connected by cables through the penetrations
between the parallel tunnels. Most of the signal processing is done in
the digital domain, if applicable. Standardized, common hardware
(e.g. ATCA, VME) is used over the entire system complex. Data
management, collection and distribution are part of the Control
system. Auxiliary systems for trigger and clock signals (timing),
AC power and cooling and the infrastructure for racks, crates, cabling, etc. are required.

\stepcounter{figlcl}

\subsection{Cost Estimation Methodology}\label{ssect:RUMcem}

For beam monitors the Instrumentation cost estimation covers:
\begin{itemize}

\item     all pickup stations, as part of the vacuum system; \itemspace
\item     scintillators, PMTs, laser systems, calibration systems; \itemspace
\item     RF systems and infrastructure for the DMC-based bunch length monitors; \itemspace
\item     associated motors, switches, and mechanical set-up; \itemspace
\item     signal and control cables, connectors, patch-cables, etc.; \itemspace
\item     dedicated read-out electronics (analog \& digital), control
            units, local timing electronics, calibration electronics, local
             software and firmware. \itemspace
\end{itemize}

Except for special cases, e.g. certain feedback systems,
data acquisition infrastructure is covered by the control
system cost estimation. Controls includes global trigger and
clock signals, global electronics infrastructure (racks, crates,
power supplies, cabling), global communication and data
acquisition hardware, firmware and software.

For costing purposes, instrumentation was classified into 17
different {\it systems}. Core cost and manpower information was
estimated for each individual component of an instrumentation
system and its subcomponents, including the cost reductions
due to volume or/and technology advances. Counts for each type
of instrumentation were supplied by the Area Systems. No spares
were included. Counts of control racks required for data acquisition
were generated form the above data. Labor information
(in person years) was estimated separately for Prototyping, Testing and Installation. The Installation labor was then incorporated into the
Installation estimate and not included in Instrumentation.

\subsection{Table of Components}\label{ssect:RUMtc}

\begin{landscape}
\stepcounter{tablcl}\begin{table} \caption[Counts of Beam
Instrumentation System Installations.] {Counts of Beam
Instrumentation System Installations in the
              ILC Accelerator Complex (along with some basic requirements).
              (The BPMs and the laser-wires are the cost-drivers) }
   \label{tab:RUMcounts}
   \begin{center}
      \begin{tabular}{| l |c |c | c | c | c | c |}
         \hline
         INSTRUMENT & \multicolumn{6}{| c |}{ AREA}  \\
         requirements & e$^{-}$ & e$^{+}$ & DR & RTML & ML & BDS \\
         (e.g. resolution) & source &  source & & & &  \\
         \hline\hline
         Button/stripline BPM & 69 & 400 & 2 $\times$ 747 & & & 120 \\
         resolution ($\mu$m) & 10-30 & 10-30 & $<$0.5 & & & $<$100\\ \hline
         C-Band Cavity BPM (warm) &   & 109 &  & 2 $\times$ 649 & & 262 \\
         resolution ($\mu$m) &  & $<$0.1-0.5 & & $<$0.1-0.5  & & $<$0.1-0.5 \\ \hline
         S-Band Cavity BPM (warm) &  &  &  & & & 14 \\
         resolution ($\mu$m) &  &  &  & & & $<$ 0.1-0.5  \\ \hline
         L-Band Cavity BPM (warm) &  &  &  & 2 $\times$ 27 & & 42 \\
         resolution ($\mu$m) &  &  &  & $<$1-5 & & $<$1-5 \\ \hline
         L-Band Cavity BPM (cold) &  &  &  & 2 $\times$ 28 &2 $\times$ 280 & \\
         resolution ($\mu$m) &  &  &  & $\sim$0.5-2 & $\sim$0.5-2 & \\ \hline
         Laser-wire IP & 8 & 20 & 2 $\times$ 1 & 2 $\times$ 12 &2 $\times$ 3 & 8 \\
         resolution ($\mu$m) & $<$0.5-5 & $<$0.5-5 & $<$0.5-5 & $<$0.5-5 & $<$0.5-5 & $<$0.5-5 \\ \hline
          Wirescanner & 12 & 8 & & & & \\ \hline
         Optical Monitors & 6 & 17 & 2 $\times$ 2 & 2 $\times$ 8 & & 11\\ \hline
          DMC  & 3  &  4 &  & 2 $\times$ 2 &  &  2 (cold) \\
          resolution $\Delta$E $\sim$0.1\% / s$ _{z}\sim$100 $\mu$m &  &  &  &  &  & \\ \hline
          Beam Current Monitors & 7 & 11 & 2 $\times$ 1 & 2 $ \times $ 2 & 2 $ \times $ 3 &  10 \\ \hline
          Beam Phase Monitor & 4 & 2 & & 2 $\times$ 3 & & 2 \\ \hline
         BLM (PMT/IC) & 60/2 & 400/20 & 2 $\times$ 40/4 &  2 $\times$ 75/2 & 2 $ \times $ 325/10 & 100/10 \\ \hline
         Feedback System & 5 & 10 & 2 $\times$ 2 & 2 $\times$ 1 & 2 $\times$ 10 & 12 \\ \hline
      \end{tabular}
   \end{center}
\end{table}

\end{landscape}

\clearpage %\setcounter{page}{1}
\setcounter{figlcl}{0}
\setcounter{tablcl}{0}
\setcounter{secnumdepth}{4} \setcounter{subsection}{0}
\setcounter{subsubsection}{0}
\usecounter{subsubsection}

\setcounter{section}{10} \renewcommand{\picturefolder}{./dumps/}

\section{Dumps, Collimators, and Stoppers}\label{sect:DUMdc}
%{\bf (DRAFT: under revision)}

\subsection{Overview}

%The ILC has 26 beam dumps, 85 fixed aperture collimators, 113
%variable aperture collimators and 25 beam stoppers.

The ILC requires a total of 26 beam dumps, each of which must be
capable of absorbing its rated beam power indefinitely without
failing.  Most of these dumps are used primarily during personnel
access, during invasive beam tuning, or as locations where the beam
can be extracted in the event of a machine protection system (MPS)
fault.  There are also 2 main beam dumps near the interaction point
and 1 photon dump in the positron source which are
used during normal luminosity delivery.  Almost all of the dumps
require water cooling.

In addition to the dumps, there are 25 beam stoppers in the ILC.
These stoppers are never intended to see beam during normal operation, but
are only used as backups to other devices and/or systems which are
expected to contain the beam power.  The stoppers are thus designed
as sacrificial devices, which are expected to be damaged if struck by the
beam. Their failure then cause a beam abort.  Stoppers are used as part of the
Personnel Protection System (PPS) as well as MPS.

The ILC collimators are required to absorb a fixed
fraction of the beam power indefinitely without failing.  In
general, this fraction is between 0.1\% and a few percent.  The
collimators are used to reduce detector backgrounds, to protect
downstream devices and apertures from damage, and to limit radiation
deposition and activation to specific regions of the beamline,
which can then be shielded
locally.  The ILC has 113 collimators with adjustable apertures and
85 collimators with fixed apertures.

\subsection{Technical Description}

The design of each beam dump, collimator or stopper is determined by
the peak incident power, power density, beam energy, and particle
type.  Electron and positron beam dumps and collimation devices that
absorb from 0-25 W of power can be made of uncooled metal; this
category of devices includes the abort dumps in the damping rings,
which are only used in the event of a hardware failure in the rings
themselves, and the faraday cups at each electron source.
Devices which are required to absorb from 25 W to 40 kW can also be
made of metal, with peripheral cooling that is provided by the
facility's low conductivity water (LCW) system; this category of
devices includes the low-power tune-up dumps in the BDS, and the
full power dumps at low-energy (100-400 MeV) locations in the
electron and positron sources.
For beam power in the range of 40 kW to 600 kW, the dump contains aluminum balls
immersed in water; this category of devices includes
the tune-up dumps at the 5 GeV end of the electron and positron
sources, the tune-up dumps in the RTML, and the tune-up dump in the
positron production undulator hall.
Beam power above 600 kW requires water as the absorbing medium; this
category of devices includes the main beam dumps and the tune-up
dumps in the BDS.  The photon dump downstream of the positron
production undulator is also a pure-water dump.

The ILC collimation system includes devices with fixed apertures and
devices which are adjustable, either in one plane or in two.  There
are 6 fixed-aperture collimators in the post-collision extraction
lines which require water-cooled aluminum balls as their primary
absorber; the remainder of the collimators are solid metal with
peripheral cooling.  In many locations, a thin (0.6-1.0 $X_0$)
collimator (or ``spoiler'') is placed in front of a thick
($>20\;X_0$) collimator (or ``absorber''); if the primary beam
leaves the collimation acceptance, the spoiler expands the beam size
via multiple Coulomb scattering to reduce the power density on the
absorber.  This approach is used to improve the survivability of
collimators in some locations, most notably the collimators
downstream of the damping ring and upstream of the final focus.

Beam stoppers that are part of the ILC Personnel Protection System
(PPS) are low power devices that self-destruct when struck by the beam,
such that the loss of beamline vacuum causes the beam to be shut
off; they are inserted into the
beam path during access periods as insurance against the failure of
the primary beamline components that protect the area under access.
Beam stoppers that are part of the ILC Machine Protection System
(MPS) have fixed or adjustable apertures; if the beam violates the
defined aperture their burn through monitors protect the remainder
of the beamline by spoiling the vacuum and shutting down the beam.

For dumps and absorbers that bring water into direct contact with
ionizing radiation, underground plumbing must be provided to safely
remove or contain the radiolytically evolved gases or isotopes while
providing adequate cooling.  All dumps and collimators require local
steel and concrete shielding to protect equipment and personnel from
residual radiation from the activated devices. If
the site chosen for the ILC tunnels is not dry, additional shielding
to protect ground water from tritium activation will be required.

\subsection{Technical Issues}

\subsubsection{18 MW Beam Dumps}

\begin{figure} [htb] \vbabove
   \begin{center}
      \includegraphics[width=0.95\textwidth]{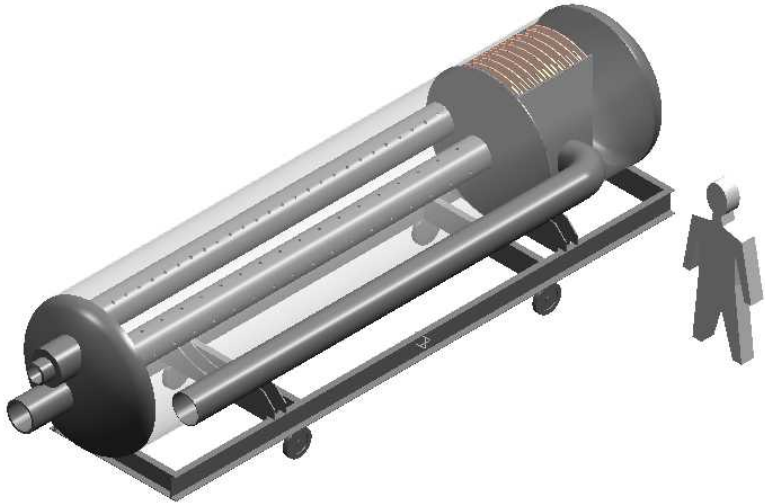}
\vbabovecaption
      \caption{Schematic of the 18MW water dump.}
      \label{fig:main_dump}
   \end{center}
\vbbelow
\end{figure}

The four linac tune-up and main beam dumps are sized for a peak
power at nominal 1~TeV beam parameters of 18~MW.  These dumps
(Figure~\ref{fig:main_dump}) consist
of 1.5~m diameter cylindrical stainless steel high pressure water
vessels with a 30~cm-diameter 1~mm-thick Ti window; they, their
shielding and associated water systems represent most of the cost of the
Beam Delivery System dumps and collimators.  The design
is based on the SLAC 2.2~MW water dump \cite{bib:Walz1965}\cite{bib:Walz1967}
that has been used without
problems for over 40 years.

\stepcounter{figlcl}

The dumps absorb the energy of the electromagnetic shower cascade in
6.5~m (18~$X_0$) of water followed by 1~m of water cooled Cu plates
(22~$X_0$).  Each dump incorporates a beam sweeping magnet system to
move the charged beam spot in a circular arc of 3~cm radius during
the passage of the 1~ms long bunch train.  Each dump operates at 10~bar
pressure and also incorporates a vortex-flow system to keep the
water moving across the beam at 1.0-1.5~m/s. In normal operation
with 250~GeV beam energy, the combination of the water velocity and
the beam sweepers limits the water temperature rise during a bunch
train to 40$^\circ$C.  The pressurization raises the boiling
temperature of the dump water; in the event of a failure of the
sweeper, the dump can absorb up to 250 bunches without boiling the
dump water.  The power which is absorbed in the dump is finally
removed by a heat exchanger system with a capacity of 2300 gallons
per minute.

The integrity of the dump body and dump window, the management of
radionuclides, the processing of the radiolytically evolved hydrogen
and oxygen, and containment of the activated water are important
issues for the 18~MW dumps.

\paragraph{Mechanical Failure of Dump or Dump Window}

The main vessel is welded using low carbon stainless steel
(316L) and all welds radiographed to ensure quality; the 10
atmosphere radioactive water cooling system is closed but
communicates with the atmosphere via a small diameter tube from the
gas space on top of the surge tank to avoid it being classified as a
nuclear pressure vessel.  Several materials are under consideration
for use in the dump window: 316L stainless, Ti-6Al-4V, and Inconel
(A601,718,X750).  All of these materials have been extensively used
in nuclear reactors; their mechanical properties, thermal
properties, and reaction to radiation damage have been thoroughly
studied.  As described above, the bunches in each train are
swept in a circle to further reduce the thermal stress and radiation
damage to the dump windows; the windows also have additional
water cooling from multiple water jets in a separate cooling loop
from the main vessel. Each dump incorporates a remote controlled
mechanism for exchanging the highly activated windows on a regular
schedule driven by integrated specific dose, along with local
temporary storage for all tritiated water.  As a final backup to
guarantee environmental safety in the event of a failure of the dump
body or dump window, the dump enclosure is air tight and
incorporates adequate sump volume and air drying capacity to prevent
the release of tritiated water even in the case of catastrophic dump
failure.  Since a failure of the window could create a catastrophic
water-to-vacuum leak with highly radioactive tritated water, a
pre-window, with peripheral and gas cooling, isolates the
beamline vacuum system and provide secondary containment. Storage
space for a damaged dump and a removable cavern wall are provided
for dump replacement.

\paragraph{Water Activation Products}

Activation products are primarily the result of photo-spallation on
$^{16}$O, primarily $^{15}$O, $^{13}$N, $^{11}$C, $^7$Be and $^3$H
(tritium). The first three radionuclides have short half lives and
decay after $\sim3$ hours.  $^7$Be is removed from
the system by filtering it out in a mixed bed ion exchange column
located in the dump support cavern. Tritium, a $\sim20$ keV
emitter with a half life of 12.3 years builds up in the water to
some equilibrium level; the tritium is contained by the
integrity of the dump system and the backup measures described in
the preceding section.

\paragraph{Radiolysis and Hydrogen and Oxygen Evolution}

Hydrogen is produced via the reaction H$_2$O $\rightarrow$
H$_2$+H$_2$O$_2$ at the rate of 0.3~l/MW-s, or 5.4~l/s at 18 MW beam
power. The lower explosive limit (LEL) of hydrogen in air is
$\sim4$\%. Experience at SLAC \cite{bib:Walz-Hydrogen}
indicates that a catalyst consisting
of a high-nickel stainless steel ribbon coated with platinum and
palladium, in the form of a 46~cm diameter 6.4~cm thick mat, will
reduce the H$_2$ concentration to the 25\% of the LEL in one pass.
Other types of higher density catalyst are also available.  The
gases released in a surge tank are heated to 65$^{\circ}$~C and are
pumped through the catalyst, which does not need replacement or
servicing.

\paragraph{Shielding and Protection of Site Ground Water}

Assuming a dry rock site, as in the baseline configuration, 50~cm of
iron and 150~cm of concrete shielding are needed between the dump
and other areas of the tunnel enclosure to protect equipment from
radiation damage. If the chosen site is not dry, the area
surrounding the dump must be enveloped by an additional 2~m thick
envelope of concrete to prevent tritium production in the ground
water.

\subsubsection{Undulator Photon Dump}

The dump that absorbs non-interacting undulator photons from the
positron production target must absorb 300~kW continuously.  The
photon energy spectrum spans the range 0-140~MeV, with an average
energy of 10~MeV; 300~kW corresponds to $1.9\times10^{17}$
photons/sec.  The photons are transported 500~m to the rapidly
rotating 1.4~mm Ti positron production target and then 150~m to a
stationary dump.  The important issues are the energy density and
temperature rise in the dump window and in the body of the dump
absorber.  The cross section of the photons is such that aluminum
balls cannot be used despite the relatively low total power; the
primary absorber in this case must be water.  With the current
undulator-dump separation the power density on a 1~mm Ti window is
0.5~kW/cm$^2$ and the resultant temperature rise after the passage
of one bunch train is 425$^{\circ}$~C. This is to be compared with a
limit of 2 kW/cm$^2$ and a fracture temperature of 700$^{\circ}$~C.
In the core of the beam the rise in the water temperature would be
190$^{\circ}$~C. With this geometry a compact (10~cm diameter by 100~cm
long) pressurized (12~bar) water vessel and Ti window, with a
radioactive water processing system, is required. Lengthening the
target to dump distance by several hundred meters would result in a
less technically challenging and less expensive system, but with the
added expense of boring a longer hole for the undulator photon
transport.

\subsubsection{Aluminum Ball Dumps}

The water-cooled aluminum ball dump \cite{bib:Walz1969}
consists of a 40~cm diameter by
250~cm long stainless vessel which is filled with 10~mm aluminum
balls and water. The water is circulating with a flow rate of
approximately 30~gallons per minute.  The dump is backed up by a
short length of peripherally-cooled solid copper.  The aluminum ball
dumps have technical issues which are qualitatively similar to some
of those of the all-water main dumps:  generation of hydrogen and
oxygen, activation of the water, and local shielding.  Because of
the much lower power levels and the use of aluminum as the main
absorbers, all of these issues are much less severe.
\begin{comment}
The water-cooled aluminum ball dump consists of a 40 cm diameter by
250 cm long stainless vessel filled with 10 mm diameter aluminum
balls, with a water flow rate of approximately 30 gallons per
minute; it will be backed by a short length of peripherally cooled
solid copper. The dump will need to be shielded from the access
passageway by 10 cm of steel and 40 cm of concrete.  A service
cavern that houses a heat exchanger, pumps and a system to treat the
water for hydrogen, oxygen and $^7$Be is required.
\end{comment}

\subsubsection{Stoppers and Collimators}

The stoppers and collimators are largely based on well-understood
designs in regular use at accelerator laboratories all over the
world.  The technical issues in these devices are not considered
important, with two exceptions.

The first exception is the collimators in the extraction lines,
which use water-cooled aluminum balls to absorb the beam power. This
system has similar issues to the main dumps in terms of activation
and risks of water-to-vacuum leaks, although on a much smaller
scale.  These collimators share the radioactive water system of
the nearby main dumps.

The other exception is limiting the deleterious effects of
wakefields in the collimators, in particular the geometric wakes of
the short spoilers and the resistive-wall wakes of the long
absorbers.  The wakes are limited by the use of copper coatings on
all surfaces in the vacuum system, and by longitudinal tapering of
the apertures to limit geometric wakes.

\subsection{Cost Estimation}

The systems that put water into direct contact with the beam
dominate the cost estimate of this technical system.  For the main
18~MW dumps, the cost estimate is based on industrial studies
 \cite{bib:Fichtner_Dump} \cite{bib:Framatome_Dump}
by two
German companies expert in nuclear reactor technology.
Their estimates have been examined by the staff responsible for the
ISIS neutron spallation target and adjusted,  for
example, to add the costs of the remote controlled window replacement
system and air drying systems.  For the aluminum ball
dumps that do not operate at high pressure, the cost of the 2006
ISIS target cooling system was used as the basis of estimate.

Items with peripheral cooling supplied by the tunnel low
conductivity water (LCW) system have only mechanical design and
construction costs.  Whether for collimators or solid dumps, these
costs are estimated based on the production costs of similar devices
in use at SLAC.

\subsection{Table of Components}

\stepcounter{tablcl}\begin{table}[h] \vbabove
  \begin{center}
\setlength{\tabcolsep}{4pt}
\caption{Dump types and locations.}
    \begin{tabular}{| l | c | l |} \hline
      Item & \# & Locations \\ \hline  \multicolumn{3}{|c|}{} \vbdlspacing  \hline
      \multicolumn{3}{|c|}{Beam Dumps} \\ \hline & & \vbdlspacing  \hline
      10 MW 10 atm water & 4 & Ends of linacs and BDS dumplines \\ \hline
      300 kW undulator photon & 1 & Behind positron production target \\ \hline
      250 kW aluminum ball & 9 & DR injectors (2) \\ [-6pt]
        & & RTML (6) \\ [-6pt]
        & & positron production undulator chicane (1) \\ \hline
      Fixed 10 kW solid metal, & 6 & 100 MeV points in e- sources (2), \\ [-6pt]
      peripherally cooled & & 114 and 400 MeV points in e+ sources (4) \\ \hline
      Insertable low power tuning dumps & 2 & Final focus \\ \hline
      Faraday cups & 2 & Electron guns \\ \hline
      Uncooled aluminum blocks & 2 & DR abort dumps \\  \hline  \multicolumn{3}{|c|}{} \vbdlspacing  \hline
      \multicolumn{3}{|c|}{Adjustable Aperture Collimators}
        \\ \hline & &  \vbdlspacing  \hline
      Short 2 jaw (H,V) tapered & 60 & RTML (36) \\ [-6pt]
        uncooled beam spoilers & & BDS Collimation (24) \\ \hline
      Long 2 jaw (H,V) cooled & 43 & BDS Collimation (32) \\ [-6pt]
        beam absorbers & & BDS FF SR masks (4) \\ [-6pt]
        & & Electron sources (2) \\ [-6pt]
        & & Positron 5 GeV point (5) \\ \hline
      Short 2 jaw uncooled collimator & 10 & Positron sources  \\ \hline  \multicolumn{3}{|c|}{} \vbdlspacing  \hline
      \multicolumn{3}{|c|}{Fixed Aperture Protection Collimators}
        \\ \hline & & \vbdlspacing  \hline
        30 cm cooled solid metal & 74 & RTML (52) \\ [-6pt]
          with circular aperture & & BDS (22) \\ \hline
        High power water cooled & 6 & BDS extraction lines \\ [-6pt]
          aluminum balls & & \\ \hline
        Single jaw cooled device & 2 & BDS collimation \\ \hline
        Uncooled block with rectangular aperture & 2 & BDS Crab Cavities \\ \hline
        Photon collimator & 1 & Undulator / positron source \\
          \hline  \multicolumn{3}{|c|}{} \vbdlspacing \hline
      \multicolumn{3}{|c|}{Beam Stops with Burn Through Monitors}
        \\ \hline & & \vbdlspacing \hline
      PPS stoppers & 14 & Positron source (2) \\ [-6pt]
                             & & RTML (6) \\ [-6pt]
                             & & BDS(6) \\ \hline
      Fixed aperture MPS & 9 & Positron source (3) \\ [-6pt]
        Stoppers & & BDS (6) \\ \hline
      Variable aperture MPS & 2 & BDS Tuneup Dump Line \\ \hline
    \end{tabular}
  \end{center} \vbbelow
\end{table}

\clearpage %\setcounter{page}{1}
\setcounter{figlcl}{0}
\setcounter{tablcl}{0}
\setcounter{secnumdepth}{4} \setcounter{subsection}{0}
\setcounter{subsubsection}{0}
\usecounter{subsubsection}

\setcounter{section}{11} \renewcommand{\picturefolder}{./controls/}

\section{Control System }\label{Controls}

%{\bf (DRAFT: under revision)}

\subsection{Overview}

Rapid advances in electronics and computing
technology in recent decades have had a profound effect on the
performance and implementation of accelerator control systems. These advances will continue through the time of ILC construction, when network and
computing capabilities will far surpass that of equipment available today.  Nevertheless, a machine of the scope of an
ILC presents some unique control system challenges independent of
technology, and it is important to begin the process of determining functional
requirements for the ILC control system.

This chapter discusses the control system requirements for the ILC, and describes a functional and physical model for the
system. In several places implementation details are described, but
this has been done largely as a means to describe representative
technologies, and in particular, to establish a
costing model. Regardless of the final technology implementation, the control system model described in this chapter contains a number of architectural
choices that are likely to survive.

\subsection{Requirements and Technical Challenges}

The broad-scope functional requirements of the ILC control system are
largely similar to those of other modern accelerator control systems,
including control and monitoring of accelerator technical systems,
remote diagnostics, troubleshooting, data archiving, machine
configuration, and timing and synchronization. However, several
features of the ILC accelerator push implementation beyond the
present state of the art.  These are
described below.

\subsubsection{Scalability}

The ILC has an order of magnitude more technical system devices
than other accelerators to date. The primary challenges of scalability
in relation to existing accelerator control systems are the physical
distances across the accelerator, the large number of components
and number of network connections, and the implied network
bandwidth. Real-time access to control system parameters must be
available throughout the site, and by remote access.
These challenges are also present in the commercial domain,
notably in telecommunication applications, and lessons learned there are
almost certainly applicable to the ILC control system.

\subsubsection{High Availability}

Requirements for high availability drive many aspects of the ILC
control system design and implementation. These requirements were
derived from accelerator-wide availability simulations. The control
system as a whole is allocated a 2500 hour MTBF and 5 hour MTTR (15
hours downtime per year). This translates to control system
availability between 99\% and 99.9\% (2-nines and 3-nines). A
detailed analysis of how control system availability relates to beam
availability is complicated. However, a coarse analysis shows that
if the control system comprises some 1200 controls shelves
(electronics crates), then each shelf must be capable of providing
99.999\% (5-nines) availability. Such availability is routinely
implemented in modern telecom switches and computer servers, but has
not been a requirement of present accelerator control systems.

\subsubsection{Support extensive automation and beam-based feedback}

A very complex series of operations is required to produce the beams
and deliver them to the collision point with the required
emittance. The control system must provide functionality to automate
this process. This includes both getting beam through the entire chain and
also tune-up procedures to maximize the luminosity. Beam-based feedback
loops are required to compensate for instabilities and time-dependent
drifts in order to maintain stable performance. Inter-pulse feedback
should be supported in the control system
architecture to minimize development of custom hardware and
communication links. The automation architecture should have some
built-in flexibility so procedures can easily be changed and feedback
loops added or modified as needed. Automation and feedback
procedures should incorporate online accelerator models where
appropriate.

\subsubsection{Synchronous Control System Operation}

The ILC is a pulsed machine operating at a nominal rate of 5~Hz.
Sequences of timing events must be distributed throughout the complex
to trigger various devices to get beam through the accelerator
chain. These events are also used to trigger acquisition of
beam instrumentation and other hardware diagnostic information so that
all data across the machine can be properly correlated for each pulse.

\subsubsection{Precision RF Phase Reference Distribution}

The control system must generate and distribute RF phase references
and timing fiducials with stability and precision consistent with the
RF system requirements.

\subsubsection{Standards and Standardization, Quality Assurance}

A critical aspect of implementing a high availability control system will be the use of consistent (``best'') work practices and a level of quality assurance process that is unprecedented in the accelerator controls environment. Additional technical solutions to HA will rely on this foundation of work practices and quality assurance processes. Commercial standards should
be used wherever they can meet the requirements, for such things as
hardware packaging and communication networks.

The control system must specify standard interfaces between internal
components and to all other systems. This makes integration,
testing, and software development easier and more reliable.
Standard interfaces allow parts of the system to be more easily
upgraded if required for either improved performance or to replace
obsolete technologies.

\subsubsection{Requirements on Technical Equipment}

Technical equipment comprises field hardware such as power supply
controllers, vacuum equipment, beam instrumentation, and motion
control devices. These systems are the responsibility of the technical
groups. However, they must interface to the control system in a
coherent way to allow equipment to be accessed via a common interface
for application programming, data archiving, and alarms. In order to
meet the very stringent requirements for overall system reliability,
as well as provide for more efficient R\&D and long-term maintenance,
standards must be applied to the technical equipment for packaging, field
bus, communication protocol, cabling, and power distribution.

\subsubsection{Diagnostic Interlock Layer}

A Diagnostic Interlock Layer complements normal self-protection
mechanisms built into technical equipment. The DIL utilizes
information from diagnostic functions within the technical equipment
to monitor the health of the equipment and identify anomalous behavior
indicative of impending problems. Where possible, corrective action
is taken, such as pre-emptive load balancing with redundant
spares, to avert or postpone the fault before internal protective
mechanisms trip off the equipment.

\subsection{Impact of Requirements on the Control System Model}

In order to meet the  high availability requirements of the ILC, a
rigorous failure mode analysis must be carried out in order to
identify the significant contributors to control system downtime. Once
identified, many well-known techniques can be brought to bear at
different levels in the system, as well as system wide, and at
different time scales (i.e. bunch-to-bunch, macro pulse, process
control) to increase availability. The techniques begin with
relatively straightforward, inexpensive practices that can have a
substantial impact on availability. A careful evaluation and selection
of individual components such as connectors, processors, and chassis
are crucial. Administrative practices such as QA, agile development
methodology, and strict configuration management must also be
applied. Other techniques are much more complex and expensive, such as
component redundancy with automatic detection and failover \cite{ctrl1}. The
control system must be based on new standards for next-generation
instrumentation that
\begin{enumerate}
   \item are modular in both hardware and software for ease in repair
     and upgrade; \itemspace
   \item include inherent redundancy at internal module, module
     assembly, and system levels; \itemspace
   \item include modern high-speed, serial, inter-module
     communications with robust noise-immune protocols; and \itemspace
   \item include highly intelligent diagnostics and board-management
     subsystems that can predict impending failure and invoke evasive
     strategies. \itemspace
\end{enumerate}
The Control System Model incorporates these principles through the
selection of the front-end electronics packaging standard and
component redundancy.

In addition to its intrinsic availability, the control system is
responsible at the system level for adapting to failures in other
technical systems. For example, the feedback system is responsible for
reconfiguring a response matrix due to the loss of a corrector, or
switching on a spare RF unit to replace a failed station.

Scalability requirements are met through a multi-tier hierarchy of
network switches that allow for the flexible formation of virtual
local area networks (VLANs) as necessary to segment network
traffic. Control system name-servers and gateways are utilized
extensively to minimize broadcast traffic and network
connections. These software components manage the otherwise
exponential growth of connections when many clients must communicate
with many distributed control points.

Automation and flexible pulse-to-pulse feedback algorithms are
implemented by a coordinated set of software services that work
together through global coordination and distributed execution. The
distributed execution is synchronized with the machine pulse rate via
the timing event system which can produce software interrupts where
needed. The network backbone accommodates the distribution of any
sensor value to any feedback computation node. This distribution can
be optimized to allow for efficient local as well as global
feedback.

%----------------------
\subsection{Control System Model}

\begin{figure}[!htb]
\begin{center} \vbabove
   \includegraphics[width=\textwidth]{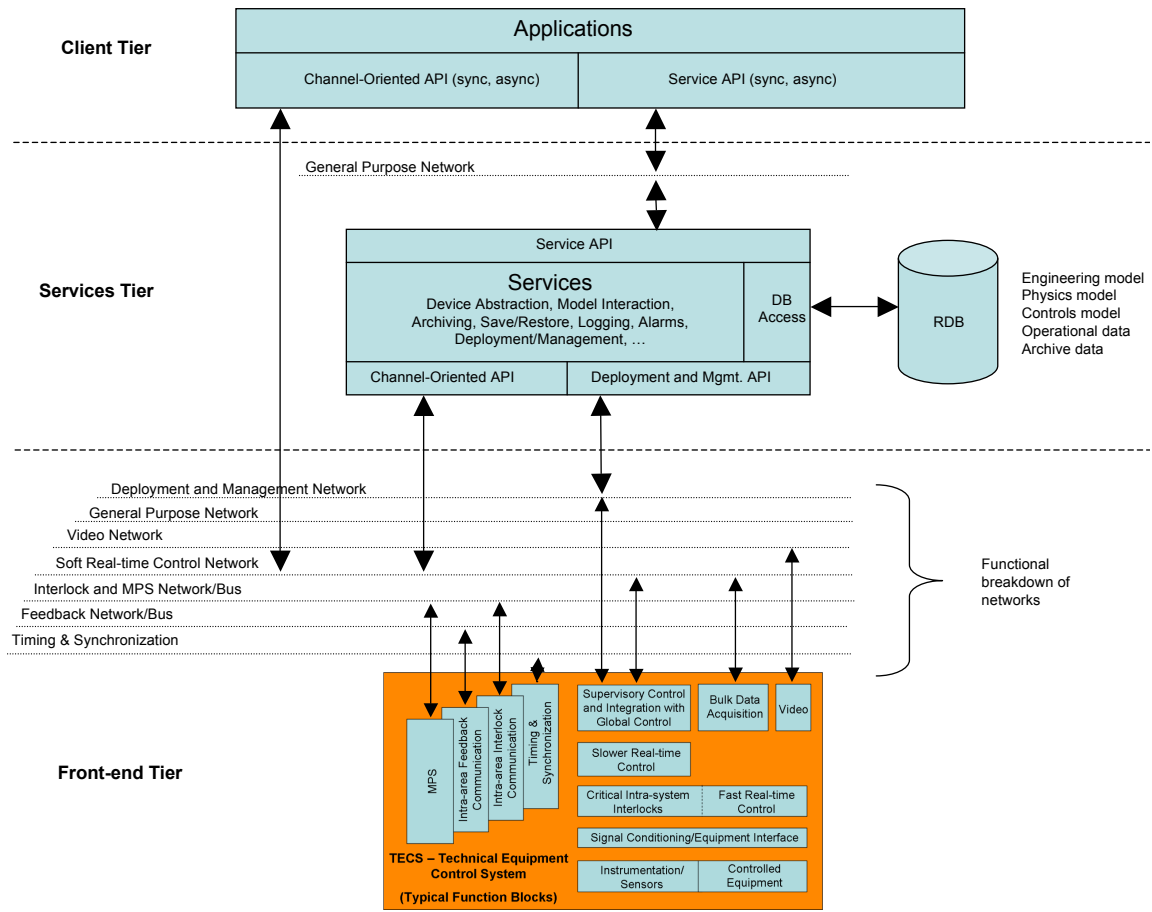}
\vbabovecaption \caption{Control system functional model.}
\label{fig:CtrlFuncModel}
\end{center} \vbbelow
\end{figure}

\begin{figure}[!htb]
\begin{center} \vbabove
   \includegraphics[width=\textwidth]{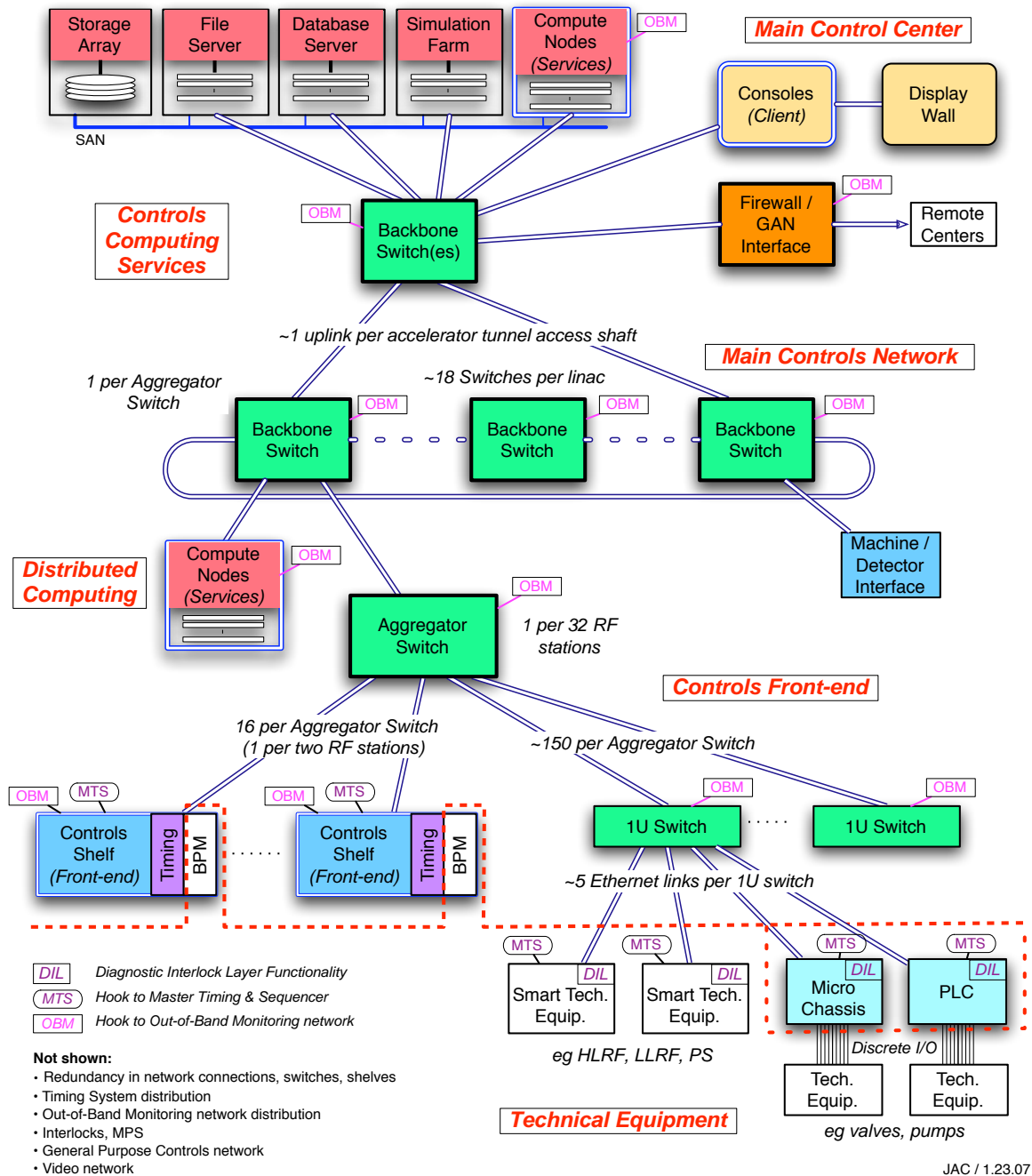}
\vbabovecaption \caption{Control system physical model.}
\label{fig:CtrlPhysModel}
\end{center} \vbbelow
\end{figure}

The model of the ILC control system is presented here from both
functional and physical perspectives. This model has served
as a basis for the
cost estimate, as well as to document that the control
system requirements have been satisfied.
Functionally, the control system architecture is
separated into three tiers, as shown in
Figure~\ref{fig:CtrlFuncModel}. Communication within and between these
tiers is provided by a set of network functions. A physical
realization, as applied to the Main Linac, is shown in
Figure~\ref{fig:CtrlPhysModel}. The remainder of the chapter describes
the functional and physical models in more detail.

\stepcounter{figlcl}

\stepcounter{figlcl}

\subsubsection{Functional Model}

The control system model is functionally composed of three distinct
tiers, as shown in Figure~\ref{fig:CtrlFuncModel}. The 3-tier model
includes a middle tier that implements significant portions of the
logic functionality through software services that would otherwise
reside in the client tier of a 2-tier system \cite{ctrl2}.  The three tiers are
described in more detail below:

{\bf Client Tier}: Provides applications with which people directly
interact. Applications range from engineering-oriented control
consoles to high-level physics control applications to system
configuration management applications. Engineer-oriented consoles are
focused on the operation of the underlying accelerator
equipment. High-level physics applications require a blend of
services that combine data from the front-end tier and supporting data
from the relational database in the context of high-level device
abstractions (e.g., magnets, BPMs).

{\bf Services Tier}: Provides services that coordinate many activities
while providing a well-defined set of public interfaces
(non-graphical). Device abstractions such as magnets and BPMs that
incorporate engineering, physics, and control models are represented
in this tier. This makes it possible to relate high-level machine
parameters with low-level equipment settings in a standard way.
For example, a parameter save/restore service can prevent two clients
from simultaneously attempting to restore a common subset of
operational parameters. This centralization of control provides many
benefits in terms of coordination, conflict avoidance, security, and
optimization.

{\bf Front-end Tier}: Provides access to the field I/O and underlying
dedicated fast feedback systems. This tier is configured and managed
by the services tier, but can run autonomously. For example, the
services tier may configure a feedback loop in the front-end tier, but
the loop itself runs without direct involvement. The primary
abstraction in this tier is a channel, or process variable, roughly
equivalent to a single I/O point.

\subsubsection{Physical Model}

The ILC control system must reliably interact with more than 100,000
technical system devices that could collectively amount to several
million scalar and vector Process Variables (PVs) distributed across the
many kilometers of beam lines and facilities at the ILC
site. Information must be processed and distributed on a variety of
timescales from microseconds to several seconds. The overall
philosophy is to develop an architecture that can meet the
requirements, while leveraging the cost savings and rapid evolutionary
advancements of commercial off-the-shelf (COTS) components.

\paragraph{Main Control Center}

The accelerator control room contains consoles, servers, displays, and
associated equipment to support operations of the ILC accelerator from
a single location. Operators and technical staff run the
accelerator and interact with technical equipment through Client Tier
applications that run in the Main Control Center.

\paragraph{Controls Computing Services}

Conventional computing services dedicated to the control system
include storage arrays, file servers, and compute nodes. A separate
simulation farm is anticipated for offline control system modeling and
simulation, and for potentially performing model-reference comparisons
to dynamically detect off-normal conditions. Enterprise-grade
relational databases act as a central repository for
machine-oriented data such as physics parameters; device descriptions;
control system settings; machine models; installed components; signal
lists, and their relationships with one another.

\paragraph{Controls Networks and Distributed Computing}

\subparagraph{Main Controls Network}

Data collection, issuing and acting on setpoints, and pulse-to-pulse
feedback algorithms are all synchronized to the pulse repetition
rate. The controls network must therefore be designed to ensure
adequate response and determinism to support this pulse-to-pulse
synchronous operation, which in turn requires prescribing
compliance criteria for any device attached to this
network. Additionally, large data sources must be prudently managed to
avoid network saturation.

For example, in the Main Linac, waveform capture from the LLRF systems
likely dominates linac network traffic. Full-bandwidth raw
waveforms from individual RF stations could be required for post-event
analysis and therefore must be captured on every pulse. However, only
summary data is required for archiving and performance
verification. By grouping multiple RF stations together (notionally
into groups of 32), full-bandwidth waveforms can be locally captured
and temporarily stored, with only summary data sent
on.

Dedicated compute nodes associated with each backbone network switch
run localized control system services for monitoring, data
reduction, and implementing feedback algorithms.

\subparagraph{Other Physical Networks}

To accommodate communication functions that are not compatible with
the Main Controls Network, several other physical networks are
envisioned, namely: a {\it General-purpose controls network} for general
controls network access, including wireless access and controls
network access to non-compliant devices; an {\it Out-of-band monitoring
network}: to provide independent means to access and configure all
Network switches and Controls Shelves; a {\it Video network} to distribute
video data streams facility wide.
A {\it Technical Equipment Interlock Network} provides a means to distribute interlock signals. Functionally, this has similarities with the Machine Protection System described elsewhere. Technical equipment may report equipment or sensor status for use by other systems or utilize status information provided by other technical systems.

Based on initial assessments, commodity-computing equipment
(e.g. 10-GB redundant Ethernet) is adequate to meet the requirements
for all the networks.

\paragraph{Controls Front-end}

The control system model front-end comprises the following three main
elements:

{\bf 1U Switch}: Aggregates the many Ethernet controlled devices in a
rack or neighborhood of racks. Some of these devices speak the
controls protocol natively, while others have proprietary
protocols that must be interfaced to the control system. It is assumed
these 1U switches reside in many of the technical equipment
racks.

{\bf Controls Shelf}: Consists of an electronics chassis, power
supplies, shelf manager, backplane switch cards, CPUs, timing cards,
and instrumentation cards (mainly BPMs). The Controls Shelf serves
several purposes: (1) hosts controls protocol gateways, reverse
gateways, and name servers to manage the connections required for
clients to acquire controls data; (2) runs the core control system
software for managing the various Ethernet device communication
protocols, including managing any instrumentation (BPM) cards in the
same shelf; (3) performs data reduction, for example, so that
full-bandwidth RF/BPM waveforms need not be sent northbound in the
control system. The control system physical model references the
commercial standard AdvancedTCA (ATCA) for the Controls Shelves. This
is a specification that has been developed for the telecommunications
industry \cite{ctrl3}, and has applicability for the ILC control system in part
because of its  high availability feature set.

{\bf Aggregation Switch}: Aggregates network connections from the 1U
switches and Controls shelves and allows flexible formation of virtual
local area networks (VLANs) as needed.

\paragraph{Technical Equipment Interface}

It has been common practice at accelerator facilities for the control
system to accommodate a wide variety of interfaces and protocols,
leaving the choice of interface largely up to the technical system
groups. The large scale of the ILC accelerator facility means that
following this same approach would almost certainly make the controls
task unmanageable, so the approach must be to
specify a limited number of interface options.  For the purpose of
the conceptual design and for the costing exercise, two interface
standards were chosen: a Controls-shelf compliant electronics module
for special sensor signals and specific beam instrumentation
applications such as BPM electronics; a controls compliant redundant
network for all {\it smart} technical systems. While not explicitly part
of the control system model, it is assumed that discrete analog and
digital I/O can be provided through micro-controller chassis' or
PLCs.

In addition to conventional interfaces for controls purposes, the
control system provides functionality for remote configuration
management of technical equipment for micro- controllers, PLCs,
application oriented FPGAs, etc.

\subsubsection{Pulse-to-Pulse (5~Hz) Feedback Architecture}

Many of the beam-based feedback algorithms required for ILC apply
corrections at the relatively low machine pulse rate (nominally
5~Hz). This low correction rate and the distributed nature of many of
the monitors and actuators make it desirable to use the integrated
controls infrastructure for these feedback systems.

Using the integrated control system architecture to implement the
feedback algorithms offers many advantages, including:

\begin{itemize}
\item Simpler implementation, since dedicated interfaces are not
  required for equipment involved in feedback loops. \itemspace

\item Higher equipment reliability, since there are fewer components
  and interfaces. \itemspace

\item Greater flexibility, since all equipment is inherently
  available for feedback control, rather than limited to predefined
  equipment. \itemspace

\item Simplified addition of ad hoc or un-anticipated feedback loops
  with the same inherent functionality and tools. This could
  significantly enhance the commissioning process and operation of the
  ILC. \itemspace
\end{itemize}

Referring to Figure~\ref{fig:CtrlPhysModel}, feedback algorithms are
implemented as services running in both distributed and centralized
compute nodes. Design and implementation of feedback algorithms is
enhanced through high-level applications such as Matlab \cite{ctrl4}
integrated into the Services Tier shown in Figure~\ref{fig:CtrlFuncModel}.

Implementing feedback at the machine pulse rate demands synchronous
activity of all involved devices and places stringent compliance
criteria on technical equipment, control system compute nodes, and the
main controls network.

\subsection{Remote Access / Remote Control}

It is becoming commonplace for accelerator-based user facilities to
provide means for technical experts to remotely access machine
parameters for troubleshooting and machine tuning purposes. This
requirement for remote access is more critical for the ILC because
of the likelihood that expert personnel are distributed worldwide.

\subsection{Timing and RF Phase Reference}

Precision timing is needed throughout the machine to control RF phase
and time-sampling beam instrumentation \cite{ctrl5}. The timing system
emulates the architecture of the control system, with a centrally
located, dual-redundant source distributed via redundant fiber signals
to all machine sector nodes for further local distribution. Timing is
phase-locked to the RF system.

\subsubsection{RF Phase Reference Generator}

The RF phase reference generator is based on dual phase-locked
frequency sources for redundancy. It includes fiducial generation
(nominally at 5~Hz) and line lock. The macro-pulse fiducial is
encoded on the distributed phase reference by a momentary phase
shift of the reference signal. Failure of the primary frequency
source can be detected and cause an automatic failover to the backup
source.

\subsubsection{Timing and RF Phase Reference Distribution}

\begin{figure}[!htb]
\begin{center} \vbabove
   \includegraphics[width=14cm]{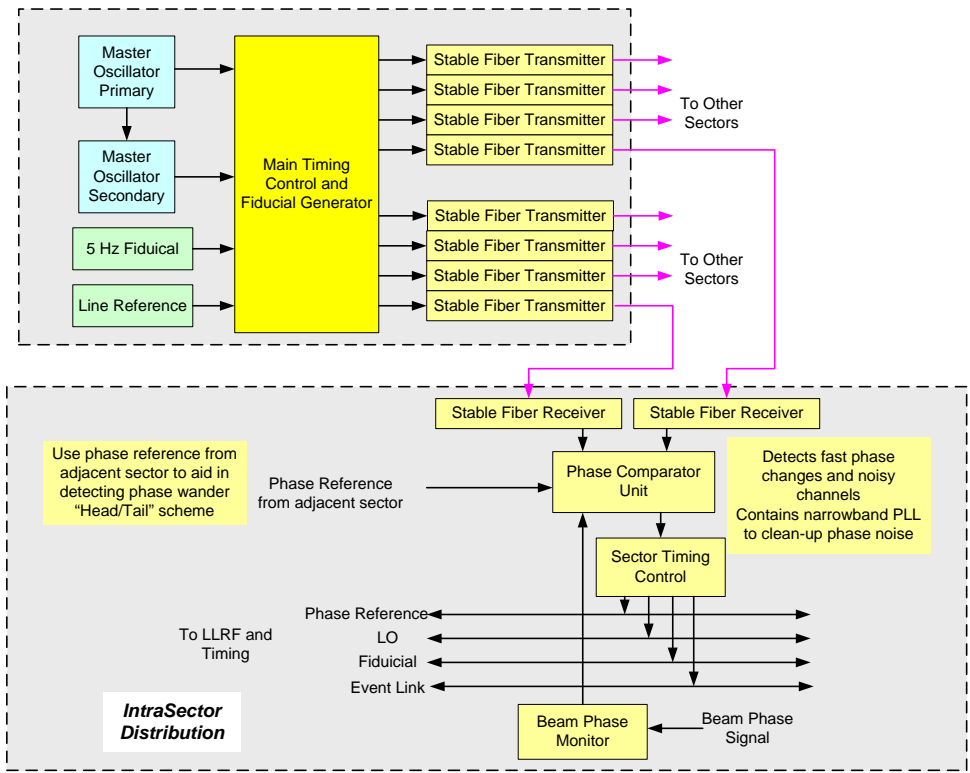}
\vbabovecaption \caption{Timing system overview showing redundant phase reference
  distribution and local intra-sector timing distribution.}
\label{fig:CtrlTimingSystem}
\end{center} \vbbelow
\end{figure}

The phase reference is distributed via dual redundant active phase
stabilized links. Figure~\ref{fig:CtrlTimingSystem} shows an overview of
dual redundant phase reference transmission and local, intra-sector
distribution.

\stepcounter{figlcl}

The Phase Comparator unit detects failures in the primary phase
reference link and automatically fails over to the secondary
link. Both the Phase Comparator unit and the Sector Timing Control
units are fault tolerant. A local DRO or VCXO is phase-locked to
the phase reference to develop a low phase noise local reference for
distribution within an RF sector of the main linac.

\begin{figure}[!htb]
\begin{center} \vbabove
   \includegraphics[width=\textwidth]{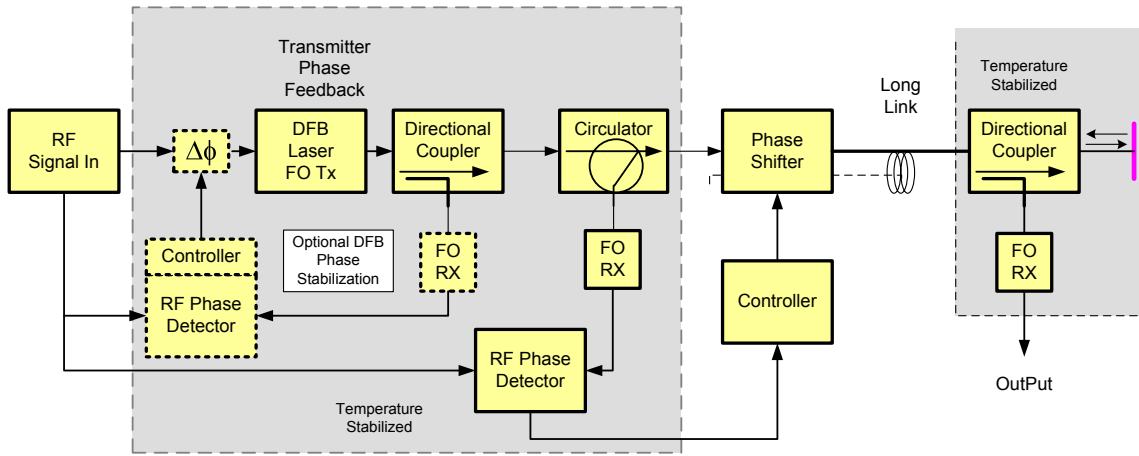}
\vbabovecaption \caption{Phase stabilized reference link.}
\label{fig:CtrlRefLink}
\end{center} \vbbelow
\end{figure}

Figure~\ref{fig:CtrlRefLink} shows a block diagram of a single active
phase-stabilized link. A portion of the optical signal is reflected at
the receiving end. The phase of the reflected optical signal is
compared with the phase of the frequency source. The resulting error
signal controls the temperature of the shorter series section of fiber
to compensate for environmentally induced phase shifts \cite{ctrl6}.

\stepcounter{figlcl}

\subsubsection{Timing and Sequence Generator}

An event stream is distributed via dual redundant links in a star
configuration. The system automatically fails over to the redundant link
upon detection of a failure. The event system provides a means for
generating global and local sequences, synchronizing software
processing to timing events, and generating synchronous time stamps.

\subsection{Beam-based Feedback}

Beam-based dynamical feedback control is essential for meeting the
high performance and luminosity needs of the ILC.
Feedback systems stabilize the electron and positron trajectories
throughout the machine, correct for emittance variations, and provide
measurement and correction of dispersion in the Main Linac. Two
timescales of beam-based feedback are anticipated, namely
pulse-to-pulse feedback at the 5~Hz nominal pulse repetition rate, and
intra-train feedback that operates within the macropulse
containing $\sim$3000 bunches spaced at $\sim$300ns intervals.

\subsubsection{Architecture for Intra-Bunch Feedback Systems}

Unlike pulse-to-pulse feedback, which is implemented through the
control system, dedicated systems are required for intra-bunch
feedback. These must operate at the bunch rate of $\sim$3~MHz, and include
the RTML turnaround trajectory feed-forward control and intra-bunch
trajectory control at the IP. Orbit feedback in the damping ring is synchronized to the damping ring revolution frequency.

Local input/output processors acquire beam position, cavity
fields, beam current, and other local beam parameters at the full
3~MHz bunch rate and distribute that information to a fast synchronous
network.  Local interconnections with the low-level RF systems provide
opportunities for local feedback loops at the full 3~MHz bunch
rate. Dedicated processing crates provide both dedicated
real-time bunch-to-bunch control, such as RF cavity fields, and
dispersion-free steering, while additional uncommitted crates could
provide feedback systems to be implemented as required.

\subsubsection{Hardware Implementation }

Most of the feedback processing requirements described in this section
can be met using commercial hardware, including dynamic orbit control
in the damping ring. Custom hardware solutions are required in cases
where low latency or unique capabilities are required, such as for the
RTML turnaround trajectory feed-forward and the IP intra-bunch
trajectory feedback. High availability solutions are implemented as
appropriate, using the same standards and approach as for other
instrumentation and control system equipment.

\subsection{Information Technology (IT) Computing Infrastructure}

The ILC requires an Information Technology infrastructure. For the
purposes of the RDR, this infrastructure is costed assuming that it
resides at the ILC site.  Equivalent functionality can be achieved
by outsourcing many of the required services, but it is expected that
the cost is similar. There is a central computing
building to house the machines and network infrastructure, a network
internal to the laboratory, a connection to the wide area network,
computer hardware and software for business computing, computing tools
for engineering support (excluding civil engineering), basic services
(web, email, file servers, databases, backups, help desk, accounts),
and computer security that complies with regulations and allows for
secure access to and dissemination of information.

\subsection{Cost Estimation, Bases of Estimates}

An inherent assumption is that the control system hardware model can
be implemented largely using COTS equipment.

Manpower estimates were developed top down, using assumptions about
the level of effort required to implement a control system for ILC,
and were compared with levels of effort from recent accelerator
projects. It is assumed that the ILC control system software framework
is founded on an existing framework, rather than developing a new
framework from the ground up. Assumptions were made on the level of
extra effort needed to implement  high availability control system
hardware and software.

Materials and Services cost estimates were derived from a bottom-up
assessment of the controls requirements from each accelerator area and
technical system. Costs for computing infrastructure (servers,
networking, storage) were based on current commodity computing vendor
prices, with an inherent assumption that technology advances will
bring commodity computing to the level of performance required for the
ILC by the time of project construction. Estimates for RF phase reference
distribution were developed from a reference design and vendor
quotes. Estimates for ATCA front-end electronics were based on
technically comparable components in other electronics platforms since
equivalent components are not yet available (or at least not in
quantity) for ATCA.

The IT infrastructure estimates were based on actual costs for
building and running IT infrastructure at Fermilab, assuming that
an ILC laboratory requires equivalent functionality at
approximately the same scale.

\subsection{Table of Components}

The following table shows a snapshot of the counts of the major
control system elements.

\stepcounter{tablcl}\begin{table}[!htb] \vbabove \caption{Snapshot
counts of the major control system elements.} \label{tab:CtrlCount}
\begin{center}
\setlength{\tabcolsep}{4pt}
\begin{tabular}{| l | p{6.5cm} | c |} \hline
%\begin{tabular}{lcr} \hline
Component      & Description & Quantity \\ \hline & & \vbdlspacing \hline
1U Switch      & Initial aggregator of network connections from technical systems  & 8356 \\  \hline
Controls Shelf & Standard chassis for front-end processing and instrumentation cards & 1195 \\  \hline
Aggregator Switch & High-density connection aggregator for 2 sectors of equipment & 71 \\  \hline
Fiber Channel RAID Disks & Controls computing high performance disk storage & 350 Terabytes \\  \hline
Tape Library   & Automated tape system for backup \& retrieval, plus front-end disk cache & 1 \\  \hline
Controls CPU   & Controls computing CPUs (other than real-time front-end processors) & 452 \\  \hline
Database CPU   & CPUs for running development, staging, and production databases & 30 \\  \hline
Controls Backbone Switch & Backbone networking switch for controls network & 126 \\  \hline
General Purpose Backbone Switch & Backbone networking switch for general purpose network & 126 \\  \hline
Monitoring Backbone Switch & Backbone networking switch for monitoring (SNMP, IPMI) network & 126 \\  \hline
Video Backbone Switch & Backbone networking switch for video distribution network & 126 \\  \hline
Phase Ref. Link & Redundant fiber transmission of 1.3-GHz phase reference & 68 \\  \hline
Phase Comparator & Phase comparison of dual phase references and adjacent sector & 68 \\  \hline
Sector Phase Ref. Timing Control & Local sector receiver of phase ref. and fiducial & 68 \\  \hline
Event System Link & Fiber link for event code distribution & 68 \\  \hline
Local Timing Card & Controls shelf timing receiver and intra-shelf timing distribution & 1134 \\  \hline
Controls Rack & Standard rack populated with one to three controls shelves & 753 \\  \hline
LLRF Controls Station & Two racks per station for signal processing and motor/piezo drives & 668 \\
\hline
\end{tabular}
\end{center} \vbbelow
\end{table}

\setcounter{chapter}{3}

\chapter{\textsf{Conventional Facilities and Siting}\label{chapCFS}}

\setcounter{figlcl}{0}
\setcounter{tablcl}{0}
\setcounter{secnumdepth}{4} \setcounter{subsection}{0}
\setcounter{subsubsection}{0}
\usecounter{subsubsection}

\setcounter{section}{0} \renewcommand{\picturefolder}{./cfs/}
\setcounter{figlcl}{0}
\setcounter{tablcl}{0}
\setcounter{secnumdepth}{4} \setcounter{subsection}{0}
\setcounter{subsubsection}{0}
\usecounter{subsubsection}

\section{Overview}\label{sect:CFSo}

This section provides an overview of the ILC Conventional
Facilities and Siting (CFS) which has been adopted as
the basis of the RDR cost estimate.  A more detailed
description can be found in~\cite{bib:cfs1}. In the absence of a specific
ILC site, three reference sites -- one in each region -- have
been developed in parallel by the CFS Group. The
reference sites (described in  Chapter \ref{chapSS}) are all
deep-tunnel sites, but have varying geologies and topographical
 constraints. An evaluation of an optimized shallow site
(either a shallow tunnel or \lq cut and cover\rq) was beyond the
scope of the current RDR activities, but will be done in the
near future. While the focus of the CFS design work has
been on the 31~km long 500~GeV machine, the sites are
required to support the footprint of the 1 TeV upgrade, both
in terms of space and available infrastructure (e.g. power).

The CFS Sample Site designs were generated using criteria provided by each of the ILC Area Systems.  Overall tunnel lengths were specifically determined by the machine parameters.  However, the size of tunnels, shafts, underground caverns and surface buildings, as well as the related CFS systems, have been developed to accommodate specific equipment installation, maintenance and personnel access and egress requirements.  For all these systems, the original criteria have been iterated in order to minimize overall costs while meeting the requirements of the present state of the ILC design. Specific examples include the reduction of Service and Beam tunnel diameters, the number and size of shafts, electrical power and process cooling loads. Further documentation can be found in references~\cite{bib:cfs2} to~\cite{bib:cfs16}.

\stepcounter{figlcl}\begin{figure}[htb]
   \begin{center} \vbabove
      \includegraphics[width=\textwidth]{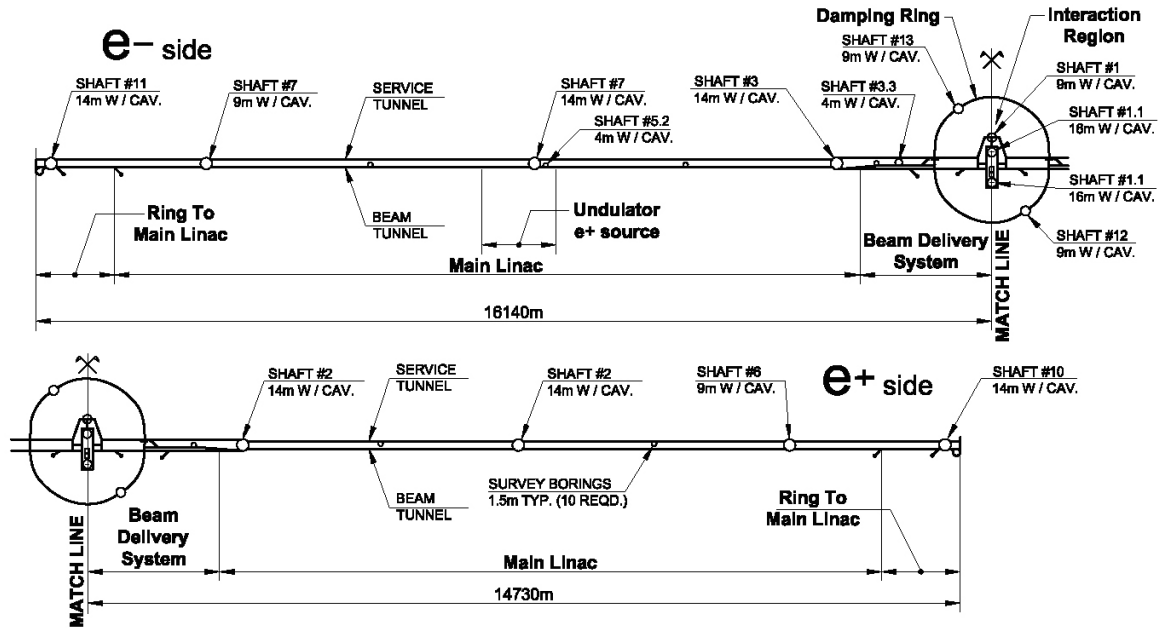}
      \vbabovecaption
      \caption{Layout of the civil construction, indicating the position of shafts and caverns.}
      \label{fig:CFScclayout}
   \end{center} \vbbelow
\end{figure}

Figure \ref{fig:CFScclayout} indicates the basic scope
of the civil construction of the ILC layout:

\begin{itemize}

\item    Two parallel 31 km long 4.5m diameter underground
tunnels house the main accelerators and the Beam Delivery
Systems (Beam Tunnel), and their associated support hardware
(Service Tunnel, containing klystrons, modulators, power supplies,
controls and instrumentation electronics etc.). The tunnels
are generally separated horizontally by $\sim$11 m (center-to-center),
and are connected via small diameter penetrations every 12 m
supporting cables, waveguides etc. Personnel access
connection tunnels (primarily for safety egress) are located every 500 m. \itemspace

\item    A total of 13 shafts along the length of the machine
provide access to underground caverns linking to the tunnels.
They primarily support the large cryogenics plants required
for the superconducting linacs. \itemspace

\item    A single collider hall at the Interaction Region (IR)
is large enough to support two physics detectors in a
push-pull configuration. \itemspace

\item    A single 5 m inner-diameter $\sim$7 km approximately
circular tunnel located around the central IR region
and $\sim$10 m above the BDS elevation houses both the
electron and positron Damping Rings in a stacked configuration.  \itemspace

\item    Several additional tunnels and service shafts house
the electron and positron sources and injector linacs
(injection into the Damping Ring), and connect the
damping ring to the main accelerator housing. \itemspace

\end{itemize}

Civil Engineering, Electrical, and Process Cooling Water
comprise greater than 90\% of the total cost of the CFS.
The Civil Engineering portion of the project is almost two
thirds of the total CFS cost, with the Underground Facilities
equating to 75\% of Civil Engineering.  The more than
72 km of tunnel is the single largest cost element.
Although formal value engineering has not yet been
accomplished, the designs have been reiterated with
the project team several times to develop a cost efficient,
workable design.

%------------------------

\clearpage %\setcounter{page}{1}
\setcounter{figlcl}{0}
\setcounter{tablcl}{0}
\setcounter{secnumdepth}{4} \setcounter{subsection}{0}
\setcounter{subsubsection}{0}
\usecounter{subsubsection}
 
\section{Civil Engineering and Layout}\label{sect:CFScel}

\subsection{Main Accelerator Housing}\label{ssect:CFSmah}

The largest underground structures are the two parallel 4.5 m
diameter tunnels, which effectively run for the entire length of the
machine footprint ($\sim$31 km).  One tunnel (the Beam Tunnel)
contains the beamline components (SCRF accelerator cryomodules,
magnets, vacuum systems etc.) The second so-called ``Service''
Tunnel houses the entire support infrastructure: RF power sources
(klystrons, modulators, pulse transformers); dc magnet power
supplies; radiation-sensitive instrumentation and controls
(electronics). Unlike the Beam Tunnel, the Service Tunnel is
designed to be accessed during beam operation, allowing in-situ
repairs and adjustment of equipment during running.

\begin{figure}[htb]
   \begin{center} \vbabove
      \includegraphics[width=\textwidth]{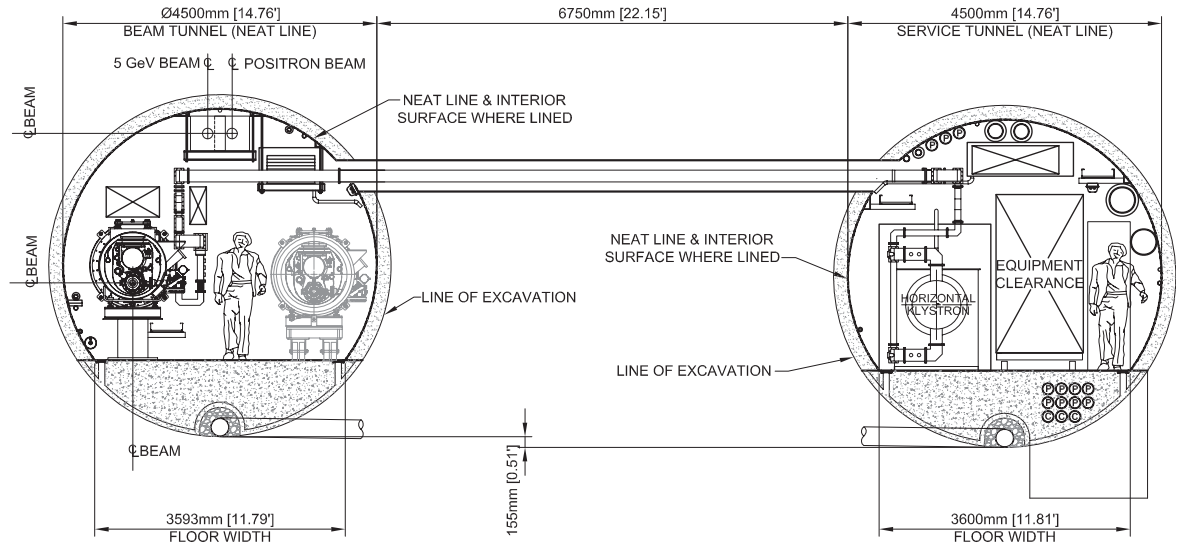}
      \vbabovecaption
      \caption{Cross-section of the Main Linac
                  housing (Beam Tunnel, left) and Service Tunnel, showing the
                  connecting waveguide penetration.}
      \label{fig:CFSmlhousingcc}
   \end{center} \vbbelow
\end{figure}

Figure~\ref{fig:CFSmlhousingcc} shows a cross-section of the Main
Linac twin-tunnel, with the Beam Tunnel on the left. The 4.5 m inner
diameter accommodates the cryomodules and RF distribution
(waveguides), at the same time as allowing space for cryomodule
installation (or removal), while maintaining a minimum ``clear
passage'' for emergency egress (see Figure~\ref{fig:CFSmlhousingcc}
left). The Cryomodules and other floor standing components are
placed on short stands mounted to a concrete floor. The beam is
centered 1.1 meters above the floor and 0.8 meters away from the
wall, which is considered sufficient to allow for cryomodule
installation (welding) and the installation of the RF waveguides.
Space needed for the survey lines of sight has also been considered.
The outer positioning of the cryomodules allows for clear access to
the egress passageways connecting the two tunnels, spaced at 500
meter intervals (not shown).

\stepcounter{figlcl}

The lateral separation of the tunnels is $\sim$11 m (center to center).
The $\sim$7 m rock and concrete separation between the Service
Tunnel and Beam Tunnel is required for structural reasons
and to provide the required radiation shielding mass allowing
workers to enter the Service Tunnel while the accelerator is
operating.  Penetrations between tunnels have been sized
and configured to provide the required radiation shielding.

The regions of superconducting RF (accelerator) dominate
the length of the main accelerating housing. These sections
are made up of many consecutive identical RF units. An
RF unit is approximately 38 m long (three cryomodules),
and is supplied from the Service Tunnel by three cross
penetrations at intervals of approximately 12 m: one for
the RF waveguides, and two additional ones for cables
and signals. The main RF unit components housed in
the Service Tunnel and their approximate space
requirements are given in Table~\ref{tab:CFSmstequip}. For
the \lq warm\rq ~sections of the Ring-to-Main-Linac (RTML)
as well as the Beam Delivery System (BDS), the Service
Tunnel accommodates the many independent magnet
power supplies, as well as electronics for controls and
instrumentation.

\stepcounter{tablcl}\begin{table} \vbabove \caption{Main Service
Tunnel equipment for a single RF unit. }
   \label{tab:CFSmstequip}
   \begin{center}
\setlength{\tabcolsep}{6pt}
      \begin{tabular}{| l | c | l |}
         \hline
         ~Item Name & Size (Meters) & Comments  \\ \hline & & \vbdlspacing \hline
         ~Klystron & 1. $ \times $ 3.38 &   \\ \hline
         ~Pulse transformer & 1.34 $ \times $ 1.25 & \\ \hline
         ~Modulator & 1 $ \times $ 4.27 &  \\ \hline
         ~Electronic racks & 9 – 0.80 $ \times $ 1.1 & Self Contained w/ integral cooling  \\ \hline
         ~LCW \& CW skids & 1.22 $ \times $ 2.06 &  \\ \hline
         ~RF transformer & 1.353 $ \times $ 1.499   & Plus 800 Amp Panel   \\ \hline
         ~Charging supply transformer & 1.22 $ \times $ 2.44 &  \\ \hline
         ~Conventional transformer & 1.575 $ \times $ 1.245 & Plus 800 Amp Panel   \\ \hline
         ~Emergency transformer & 1.575 $ \times $ 1.245 &   \\ \hline
      \end{tabular}
   \end{center} \vbbelow
\end{table}

In addition to the RTML, Main Linac and BDS
beamline components, the Beam Tunnel also
houses the 5 GeV low-emittance transport line
(part of the RTML) which transports the beam
from the central Damping Rings to the far ends
of the machine. The RTML \lq turnarounds\rq ~at the ends
of the machine are housed in a 4.5 m diameter looped
tunnel with an average bending radius of $\sim$30 m in the
 horizontal plane. The length of each loop is
approximately 140 m. On the electron linac side,
a third beamline from the undulator-based positron
source (nominal 150 GeV point) is required to transport
the 400 MeV positrons from the source to the Damping
Rings.  Both the long 5 GeV low-emittance and the
400 MeV positron transport lines are supported from
the Beam Tunnel ceiling, and are positioned towards
the center of the tunnel to allow for installation and
replacement without removing a cryomodule. Power
and cooling services for these elements are provided
from equipment in the Service Tunnel.

The BDS and RTML bunch compressor tunnels (a total of $\sim$5.3~km
and $\sim$2~km, respectively) lie in a plane, while the Main Linac
tunnels (47.8~km) and associated beamline components, including the
long RTML transfer lines, follow the Earth's curvature.

\begin{figure}[htb]
   \begin{center} \vbabove
      \includegraphics[width=\textwidth]{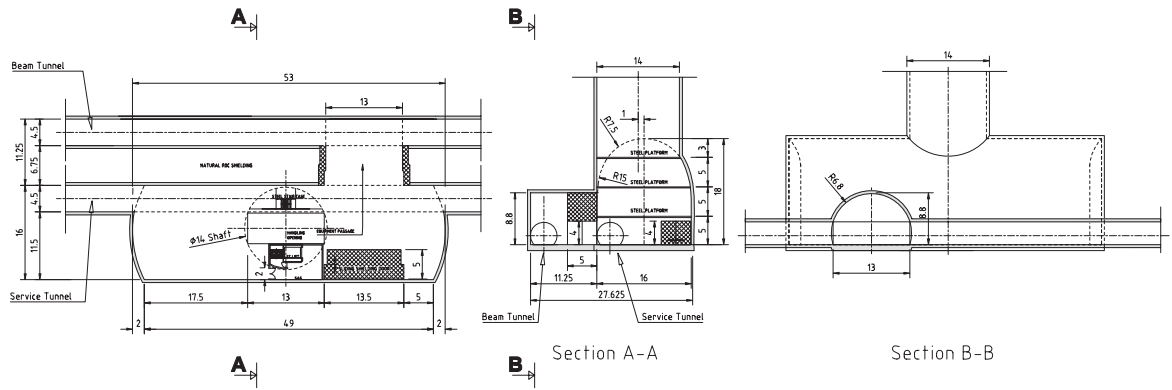}
      \vbabovecaption
      \caption{Example of a 9 m shaft with underground cavern,
      Service and Beam Tunnels (European Sample Site).}
      \label{fig:CFS9meterss}
   \end{center} \vbbelow
\end{figure}

\begin{figure}[htb]
   \begin{center} \vbabove
      \includegraphics[width=0.6\textwidth]{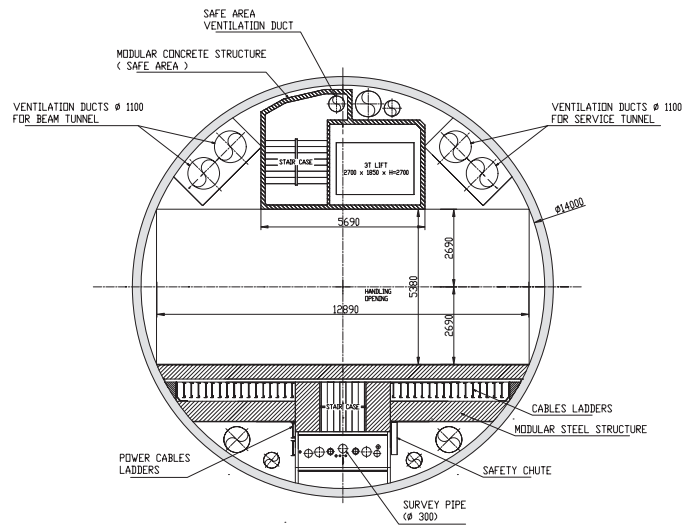}
      \vbdlspacing
      \caption{Detailed view of 14 meter shaft.}
      \label{fig:CFSmlhousingsection}
   \end{center} \vbbelow
\end{figure}

The large cryogenic plants (see Section 3.8), required
primarily for the SCRF RF cryomodules, are housed in
eight underground caverns connected to the surface via
shafts (four per side, spaced approximately 5000 m apart):
shaft nos. 2, 3, 4, 5, 10 and 11 are 14 m in diameter,
while shaft nos. 6 and 7 are 9 m diameter (see Fig.~\ref{fig:CFScclayout}
for shaft locations). Figures~\ref{fig:CFS9meterss} , \ref{fig:CFSmlhousingsection} shows a
schematic of a typical 9 m shaft and cavern. In addition
to housing the cryogenic plants, these shafts are also
used for: installation of machine components (including
cryomodules at the 14 m shafts); normal and safety
egress from the tunnels; and for supporting all services
such as cooling water, power etc. The 14 m shafts are
also used to lower, assemble and prepare the Tunnel
Boring Machines (TBM) which are used extensively
for tunnel excavation.

\stepcounter{figlcl}

\stepcounter{figlcl}

Temperature neutral air is routed through the tunnel from
the shafts, no additional heating or cooling is required in
the Beam Tunnel.  Where needed there is dehumidification
equipment installed to maintain humidity levels below
the dew point.  Seepage is directed to a drain and routed
to the sumps located at the shaft caverns.

A special underground cavern is required to service
the undulator-based positron source, located at the
nominal 150 GeV point in the Main Electron Linac.
A 4 m vertical shaft is provided for removal and
installation of \lq hot\rq ~targets.

\subsection{Central Injectors}\label{ssect:CFSci}

\begin{figure}[htb]
   \begin{center} \vbabove
      \includegraphics[width=\textwidth]{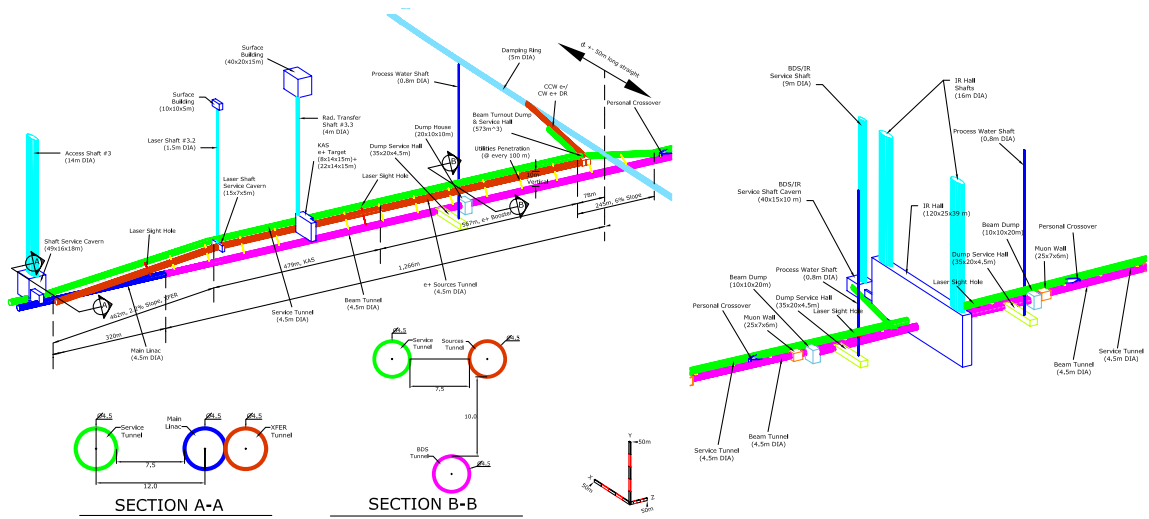}
      \vbabovecaption
      \caption{Layout of the Central Injector Complex (electron side).}
      \label{fig:CFSciclayout}
   \end{center} \vbbelow
\end{figure}

The central injector systems include: the 6.7 km
circumference Damping Rings; the polarized
electron source; the positron Keep Alive Source (KAS);
and the electron and positron 5 GeV SCRF injector linacs.
Figure~\ref{fig:CFSciclayout} shows the primary tunnel and shaft arrangements.

\stepcounter{figlcl}

\begin{figure}[htb]
   \begin{center} \vbabove
      \includegraphics[width=0.8\textwidth]{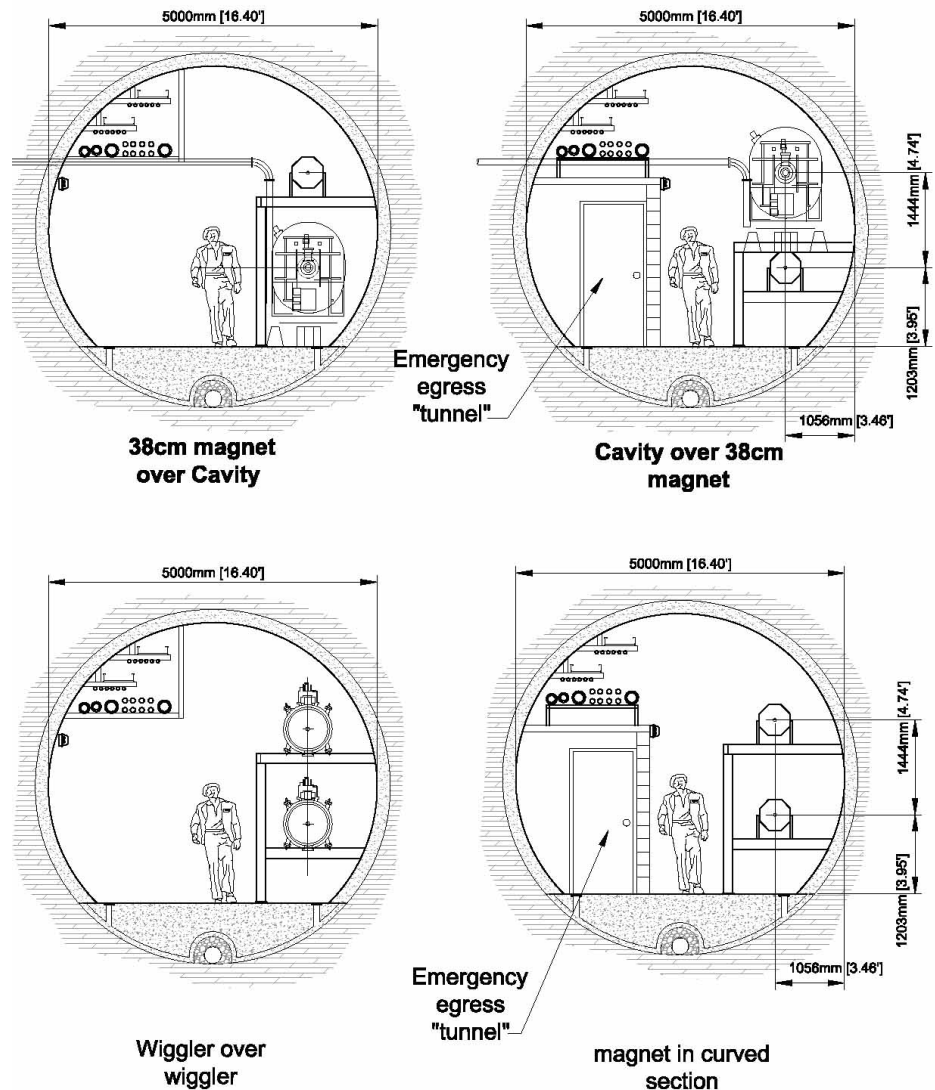}
     \vbabovecaption
      \caption{Cross-sections of the 5 m diameter Damping Ring tunnel
                  showing vertical stacked rings at several locations.}
      \label{fig:CFSdrsections}
   \end{center} \vbbelow
\end{figure}

The electron and positron Damping Rings are housed in a single
5 m inner-diameter quasi-circular tunnel with a total circumference
of 6704 m. The tunnel is located in the horizontal plane,
approximately 10 m above the plane of the BDS. The ring is
made up of six arc sections, two long straight sections for
injection and extraction and four short straight sections containing
the superconducting damping wigglers and RF. The DR tunnel is
connected to the injection tunnels from the sources in the middle
of the long straight sections (see Fig.~\ref{fig:CFSciclayout}). The tunnel has 6 alcoves
in total, located in the middle of the straight sections. Two main
alcoves are accessed via two 9 m diameter shafts, and are used
to house the cryogenic plants and RF power sources for the wigglers
and RF cavities (and are also used for installation).  The four smaller
alcoves in the remaining straight sections are not connected to
the surface, but two of them are vertically connected to the BDS
portion of the Service Tunnel 10 m below to allow personnel access.
As there is no separate service tunnel for the DR, all service and
support equipment are housed in the two shaft caverns and the
four smaller alcoves. A cross-section of the Damping Ring tunnel is
shown in Fig.~\ref{fig:CFSdrsections}; the 5 m inner diameter is required to house the
two rings (vertically stacked), and the emergency egress
passage, while allowing enough space for component installation.

\stepcounter{figlcl}

The electron and positron 5 GeV injector linacs are each housed in 4.5 m diameter tunnels, and share the main Service Tunnel with
the BDS. The sources also make use of the 14 m diameter shafts
 located directly at the end of each Main Linac, where the
connecting tunnel to the Damping Rings has a 2\% slope
to accommodate the 10 m vertical offset between the Damping
Ring and Main Accelerator Housing. The KAS source requires
an underground cavern similar to the positron production vault
in the electron Main Linac, again with a 4 m diameter vertical
shaft for removal and installation of the hot target.

\subsection{Interaction Region and BDS}\label{ssect:CFSirb}

\stepcounter{figlcl}\begin{figure}[htb]
   \begin{center} \vbabove
      \includegraphics[width=\textwidth]{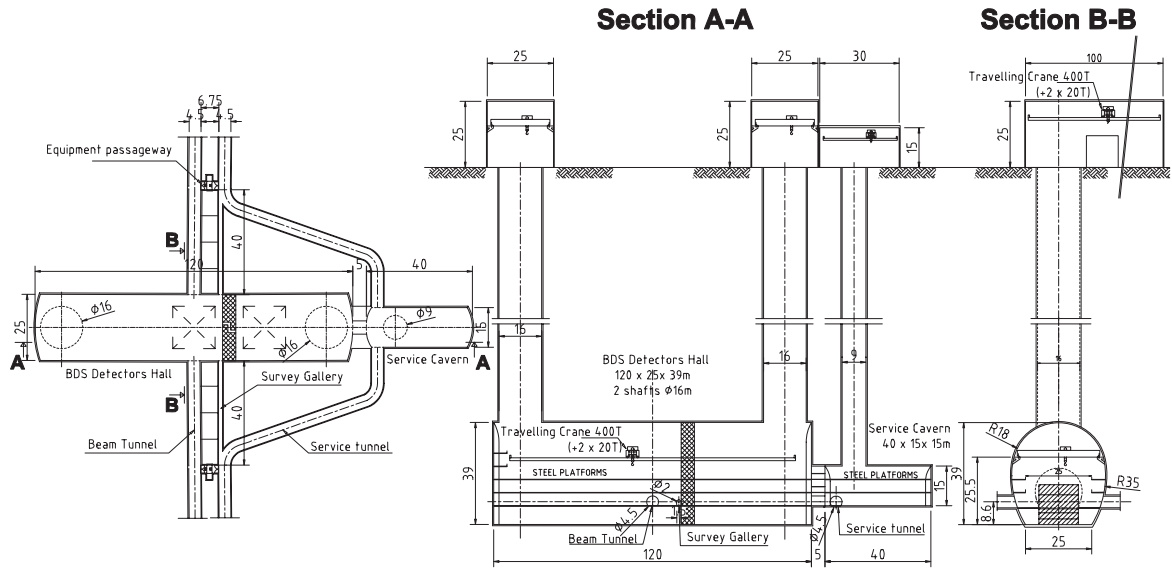}
      \vbabovecaption
      \caption{Schematic of the Physics Detector Hall, showing BDS service cavern arrangement.}
      \label{fig:CFSirsections}
   \end{center} \vbbelow
\end{figure}

The Physics Detector Hall is the largest cavern in the project. It is sized
to accommodate two Physics Detectors in a "Push Pull" type
configuration assuming surface assembly of each detector. The hall
is connected to the surface assembly buildings via two 16 m diameter
shafts, one for each Detector. It is also connected to the Beam Tunnel,
to a service cavern through a passageway, and to the survey galleries.
The floor slab is thick enough to accommodate the weight of the two
Detectors and the weight of the movable shielding wall (2 pieces) in
between the Detectors. The walls of the hall are equipped with 3 to 5
levels of steel platforms to be used for services and access at various
levels to the Detectors. The hall also has beams and rails for one 400
ton crane and two 20 ton cranes, assuming the surface
pre-assembled Detector elements weigh 400 tons at most.

An additional service cavern for the BDS is located next to the Detector Hall
(see Fig.~\ref{fig:CFSirsections}). It houses all the equipment needed for the running of the
Detectors and ancillary facilities, which need to the shielded from radiation or the magnetic field of the detectors.
It has two steel platforms as intermediate floors for equipment, and is
connected to the Detector Hall through a shielded passage for personnel
and goods, and to the BDS service tunnel on both sides. The service
cavern is accessible from the surface via a 9 m diameter vertical shaft,
which supports all services, houses a safe staircase and lift for
personnel and equipment, and leaves space for lowering all components
to be installed in the service cavern.

There are four full-powered Beam Dump facilities in the BDS System,
two on each side. For each one there is a cavern which houses the
high-pressure water dump itself, and a service cavern located
$\sim$30~m away to house all electrical, control and cooling equipment.

The inner diameter of the BDS beam tunnels are locally
enlarged at four locations (two per side) to house the large
magnetized toroids (so-called Muon spoilers) for reducing the
muon background to the experimental hall.

\subsection{Surface Buildings}\label{ssect:CFSsb}

A total of 96, 140 and 133 buildings are
foreseen for the Americas, Asian and European sites, respectively. The type, number and dimensions of the buildings include only those surface facilities
required for construction, installation and operation of the project,
taking into account the specifics of each of the three sample
sites. For instance, additional infrastructure such
as seminar rooms, guest-houses, restaurants, administrative
facilities etc. are assumed to be supplied by a nearby
(host) laboratory, and are not included in the cost estimate.
The Asian sample site does not have a nearby laboratory
and that estimate does include such central campus facilities.

Types of surface buildings considered included: surface equipment
buildings, including cooling towers and pump stations; shaft head
buildings; storage areas; local workshops and assembly areas, local
technical offices etc. The majority of these buildings are
concentrated at the ``central campus'' and specifically at the
Interaction Region. The remaining buildings are located at the
shafts positioned along the Main Accelerator Housing. Concrete
construction with acoustic absorbent material is used for buildings
which contain ``noisy'' equipment; the remaining buildings have
steel structures and insulated steel ``sandwich'' type panels for
roof and wall cladding on concrete foundations. In all cases, the
design of the buildings takes into account the local climatic loads,
seismic load (according to local standards) and fixed or moving
loads linked to the use of that building. The requirements for each
building type have been considered in making the cost estimate.
Overhead cranes, gantries, elevators and other lifting gear with
appropriate capacity are included where necessary in surface and
underground structures. A detailed breakdown of the surface
buildings can be found in~\cite{bib:cfs1}.

\subsection{Site Development}\label{ssect:CFSsd}

For the areas where surface buildings are located (central campus,
shaft positions), the following items have been included in the cost
estimate:

\begin{itemize}

\item    fences and gates; \itemspace
\item    roads and car parks within fences and from fence to existing road network; \itemspace
\item    pedestrian walkways; \itemspace
\item    lighting for the above and around buildings including buried electrical connections; \itemspace
\item    all necessary drains along roads, car parks, including sumps,
            water treatment facilities and connections to existing mains; \itemspace
\item    all needed water supply pipes, tanks and connection to existing water supply network; \itemspace
\item    landscaping and planting of trees, bushes, seeding of grass as required; \itemspace
\item    spoil dumps (where applicable) created close to the building areas, including landscaping. \itemspace

\end{itemize}

All temporary facilities needed for the construction works as well as
the necessary site preparation before start of the works are also
included in the cost estimate.

\subsection{Regional Variants}\label{ssect:CFSrv}

Both the Americas and European sites are similar deep tunnel sites
and both utilize vertical shafts for access as described in the sections
above. These shafts are respectively 135-100 m deep.
The Asian site is somewhat different, in that it is located along the
side of a mountain. With the exception of the two central shafts for
servicing the Detector Hall at the IR, long almost horizontal access
tunnels are used instead of vertical shafts. The lengths of these
access tunnels range from 700 m to 2000 m. Other variants which
are due to construction methods depending on local geology are covered
in Chapter~\ref{chapSS}.

%------------------------

\clearpage %\setcounter{page}{1}
\setcounter{figlcl}{0}
\setcounter{tablcl}{0}
\setcounter{secnumdepth}{4} \setcounter{subsection}{0}
\setcounter{subsubsection}{0}
\usecounter{subsubsection}

\section{A.C. Power Distribution}\label{sect:CFSacpd}

Electrical power is categorized by three major systems:

\begin{itemize}

\item     RF power (modulators); \itemspace
\item     conventional power (normal conducting magnet power
             supplies, cryogenic plants, electronic racks, surface
             water plant systems and infrastructure components); \itemspace
\item     emergency power provided by back-up generators
            (emergency lighting, sump pumps and ventilation
             systems for sub-surface enclosures).  \itemspace

\end{itemize}

The power requirements are dominated by the RF system
(modulators) located in the Service Tunnel along the length of the
Main Linac. Table~\ref{tab:CFSpowerloads} gives an overview of the estimated \it nominal \rm
\footnote{\it Nominal \rm electrical power requirements have been developed
(as much as practical) as \it continuous power \rm, sometimes denoted
as \it wall power \rm.  \it Installed \rm power may be 75-100\% higher.}
power consumption for 500 GeV center-of-mass operations,
broken down by system area and load types. The cost estimate
is based on a total nominal power requirement of 216.3 MW.
The additional required power for a potential upgrade to 1 TeV
centre-of-mass is not included in the current power load tabulation.

\stepcounter{tablcl}\begin{table} [hb] \vbabove \caption{Estimated
nominal power loads (MW) for 500 GeV centre-of-mass operation. }
   \label{tab:CFSpowerloads}
   \begin{center}
\setlength{\tabcolsep}{6pt}
      \begin{tabular}{| l | c | c | c | c | c | c | c |}
         \hline
      & & \multicolumn{4}{| c | }{Conventional Power}& & \\ \cline{3-6}
         ~Area & RF & Conv & NC & Water & Cryo & Emer & Total \\ [-6pt]
         ~System & Power & & Magnets & Systems & & Power & (by area)  \\ \hline & & & & & & & \vbdlspacing \hline
         ~Sources e$^{-}$ & 1.05 & 1.19 & 0.73 & 1.27 & 0.46 & 0.06 & 4.76 \\ \hline
         ~Sources e$^{+}$ & 4.11 & 7.32 & 8.90 & 1.27 & 0.46 & 0.21 & 22.27 \\ \hline
         ~DR & 14.0 & 1.71 & 7.92 & 0.66 & 1.76 & 0.23 & 26.29 \\ \hline
         ~RTML & 7.14 & 3.78 & 4.74 & 1.34 & 0.0 & 0.15 & 17.14 \\ \hline
         ~Main Linac & 75.72 & 13.54 & 0.78 & 9.86 & 33.0 & 0.4 & 134.21 \\ \hline
         ~BDS & 0.0 & 1.11 & 2.57 & 3.51 & 0.33 & 0.20 & 7.72 \\ \hline
         ~Dumps & 0.0 & 3.83 & 0.0 & 0.0 & 0.0 & 0.12 & 3.95 \\ \hline
         ~Totals (by system)  & 102.0 & 32.5 & 25.6 & 17.9 & 36.9 & 1.4 & 216.3 \\ \hline
      \end{tabular}
   \end{center} \vbbelow
\end{table}

High voltage (HV) connections to the utility system varies by
region, ranging from 275kV (Asia), 345kV (America) and 400kV (Europe).
All regions provide for a main substation located at or near the
Interaction Region/Central Damping Ring for connection to
the utility's high voltage transmission system. Standards for high voltage transmission, and medium (MV) and low voltage (LV) distribution vary across regions; consequently the approach to distributing
the power to the machine components is slightly different for the
three sample sites. However, the salient features remain the same:

\begin{itemize}

\item     Connection to the utility's HV transmission system via a main             substation
             located at the central campus; \itemspace
\item     HV transmission voltage is transformed to medium voltage
        (MV; 34-69 kV) for distribution across the site to remote shafts            (access points); \itemspace
\item     From the remote shaft locations, power is further transformed and             distributed
            to the Service Tunnel. For the Main Accelerator Housing,
            this implies a distribution of approximately $\pm$2.5 km
        (in both directions) from the shaft locations.
            Medium voltage is distributed directly to RF stations               (modulators).
            Low-Voltage transformers located along
            the Service Tunnel tap-off the MV distribution system to provide            power to the LV systems and components. \itemspace

\end{itemize}

An optimized engineering solution for the power distribution is heavily influenced by site selection, including: availability and location of utility's substations; regional voltage standards and regional safety regulations. The design work for this report was developed globally by identifying site-dependent and site-independent infrastructure requirements, the former being developed by the regions and the latter being based on a European estimate.
Details can be found in \cite{bib:cfs1}. In the following sections, the European solution is presented. Important regional variations are briefly described in Section~\ref{ssect:CFSmti}.

\subsection{System Configuration}\label{ssect:CFSnc}

Voltage levels selected for the MV distribution systems are 66/6 kV, 36 kV and 69/34 kV for the Asian, European and Americas regions, respectively. LV distribution systems for all regions are in the magnitude of hundreds of volts. Standardized switchboards powered from the LV transformers are used to locally distribute LV power.

All HV and MV substations –- including the one at the central campus –- are provided with a bus-tie-bus configuration (RF bus system and conventional bus system).  The HV and MV protection systems are based on numerical relays with facilities for programming automated sequences, and for recording network perturbations; thus allowing every major electrical system to be monitored by a Supervisory Control and Data Acquisition system (SCADA)

\subsection{Distribution for the Main Accelerator Housing}\label{ssect:CFSdmah}

Two MV cable lines are routed along those sections of
Service Tunnel containing SCRF cryomodules (Main Linac and RTML):

\begin{itemize}

\item     One MV system provides power to the RF (Modulator) system, with a ring        main unit (RMU) installed at every RF unit ($\sim$38 m) connected       directly to the RMU. \itemspace
\item     The second MV system provides power for conventional services, with a         RMU and  500 kVA transformer located at every fourth RF unit            ($\sim$152 m). A LV switchboard is powered from the 500 kVA             transformer and located near the transformer in the Service Tunnel.         The switchboard supplies LV power to $\sim$152 m of both the Beam       and Service Tunnels. \itemspace

\end{itemize}

In those sections of the Main Accelerator Housing where there are no
RF units (warm sections of RTML and the BDS), only the conventional
services MV system components are installed. The same 152~m module
structure is used in these areas. For the BDS sections, individual
transformers are rated 1000~kVA (each) in lieu of 500~kVA due to the
higher density of the load (larger number of normal conducting
magnets)

LV power is supplied to the Beam Tunnel
via the connecting penetrations, spaced approximately 12 m apart
(12 per 152 m distribution unit).

\subsection{Distribution for the Central Injectors}\label{ssect:CFSdci}

Power for the central injectors (damping rings, sources and injector linacs)
is derived from the main substation located at the central campus.

The SCRF is responsible for about two-thirds of the total power
requirement for the Damping Rings, the remaining third being the
normal conducting magnets and superconducting wigglers.
Due to the restricted tunnel cross section, the MV systems are installed partly on the surface and partly in the tunnel alcoves
(see Section~\ref{ssect:CFSci}). The Damping Ring tunnel is supplied from
a MV loop system originating at one of the two Damping Ring service shafts. A distribution substation is installed on the
surface, fed directly from the central area via a MV system.
The surface equipment at the second service shaft completes the closed loop. A LV transformer provides power in the shaft base alcoves. RF and other large power consumers are fed via RMUs and the dedicated MV system at the surface.

For the remaining four subterranean alcoves located in the
Damping Ring straight sections, an underground substation
is powered utilizing RMUs. The LV distribution in the Damping Ring tunnel uses multiple LV switchboards. The LV in the tunnel provides general power to lighting, outlets, and possibly minor machine system loads. Switchboards are generally located in the alcoves together with the substation and RMU.

The sources are essentially concentrated in short tunnels and caverns.
The equipment for each of the sources is fed from a short MV
system with RMUs. Dedicated LV transformers are installed
for the RF for the source capture sections and SCRF injector
linacs. The remaining part of the electrical load is powered from the conventional power distribution system.

\subsection{Interaction Region}\label{ssect:CFSir}

The power requirements of the detectors are currently not known.
The current design is estimated based on a detector load requirement of 3~MVA, scaled down from CMS. A MV cable system, RMUs, LV transformers and switchboards has been reserved for the detectors.

\subsection{Emergency Supply Systems}\label{ssect:CFSess}

The emergency supply system is based on stand-by diesel generator systems. Each generator set supplies a protected substation, which is normally supplied by the utility power. During a utility power interruption, the diesel engines start automatically and transfer the critical load when ready. On return of utility power, the diesel generator systems synchronize to the utility power system  and the load is re-transferred back to the utility power system, after which the diesel generator systems shut down.

Due to voltage drop considerations, the generator output voltage must be transformed up to a MV level. The Main Linacs and the RTML zones are equipped with a MV system with RMUs at regular spacing.  The Damping Ring tunnel is also equipped with a MV system, originating at one access shaft with a RMU in each alcove or cavern.  Each MV system is completed by exiting the adjacent access shaft to the surface.  Each of the RMUs feeds the critical load through a LV transformer and switchboard. Any critical system which cannot accept any power interruption is provided with an  Uninterruptible Power Supply (UPS) system, or no-break systems.

\subsection{Miscellaneous Technical Issues}\label{ssect:CFSmti}

\it Power quality considerations \rm A separation of pulsed and
non-pulsed systems may be needed to avoid interference between
certain loads. Reactive power compensation and harmonics filtering
may also be needed, depending on the non-linear load and the dynamic
behaviour of the load, especially the RF (Modulator) system.

\subsection{Regional Variations}\label{ssect:CFSrvs}

Table~\ref{tab:CFSvlevels} gives an overview of the various voltages assumed for the different regions.

\stepcounter{tablcl}\begin{table}  \vbabove \caption[Various voltage
levels utilized by regions.] {Various voltage levels utilized by
regions. Note that there
      are two levels of HV distribution utilized for the Americas and Asian sample sites }
   \label{tab:CFSvlevels}
   \begin{center}
      \begin{tabular}{| l | c | c | c |}
         \hline
         Voltages & Europe & America & Asia \\ \hline & & & \vbdlspacing  \hline
         Transmission & 400 kV & 345 kV & 275 kV \\ \hline
         Distribution & 36 kV & 69 kV & 66 kV \\ \hline
         Distribution & 36 kV & 34 kV & 6.6 kV \\ \hline
         Distribution & 400/230 V & 480/277 V & 400/200 V \\ \hline
      \end{tabular}
   \end{center} \vbbelow
\end{table}

\bf European Sample Site: \rm The description of the power distribution
given above is primarily that adopted for the European site (and the
cost estimate). The utility voltages are 400~kV, with the MV levels set to 36~kV. LV levels are typically 400~V (three phase) and 220-240~V (single phase).

\bf Asian Sample Site: \rm The distribution of power is slightly different
to that documented above. The utility voltage is 275~kV, and is transformed to 66~kV and distributed via the Service Tunnel to the secondary substations located in each access base caverns. Secondary substations transform the voltage to 6.6~kV which is then distributed to local LV transformers. A LV system of 400~V (three-phase) and 100-200~V (single-phase) is supplied via local transformers from the 6.6~kV system. The system applies to power transport and distribution in the entire underground areas.

\bf Americas Sample Site: \rm The Americas distribution also varies
slightly in that the utility voltage of 345~kV is first transformed to 69~kV
at the master substation.  The 69~kV is routed through the tunnel to
each shaft, and then up the shaft to where it is stepped down to
the medium distribution voltage (34.5~kV).

\subsection{Information Network}\label{ssect:CFSin}

Site-wide communications are in general supported via
a fiber-optic based LAN system. For the underground areas,
local LAN racks are located in the tunnel at an interval of
approximately 200~m, which serve as the primary connection
point to the end equipment (via electrical cables). From
here the signals are sent to sub-center LAN racks located
in the Shaft Bases, and finally to the main control center.

The LAN supports the following equipment:

\begin{itemize}

\item    General digital data transfer. \itemspace
\item    Telephone system:  1,800 cordless lines and 240$\sim$360 fixed
            lines are assumed.  Cordless telephones are supported in
            the underground areas via IP transmitters. \itemspace
\item    Public address system (including safety address system):
           for the underground areas, speakers are mounted every 10 m of tunnel. \itemspace
\item    Security CCTV and other video monitoring where needed
           (both surface buildings and underground areas).\itemspace
\item    Fire alarms, smoke detectors etc. \itemspace

\end{itemize}

In the case of the critical safety-related systems, emergency back-up
power is supplied from the standby generator in case of power failure.

%------------------------

\clearpage %\setcounter{page}{1}
\setcounter{figlcl}{0}
\setcounter{tablcl}{0}
\setcounter{secnumdepth}{4} \setcounter{subsection}{0}
\setcounter{subsubsection}{0}
\usecounter{subsubsection}

\section{Air Treatment Equipment}\label{sect:CFSate}

\begin{figure}[t!]
   \begin{center} \vbabove
      \includegraphics[width=\textwidth]{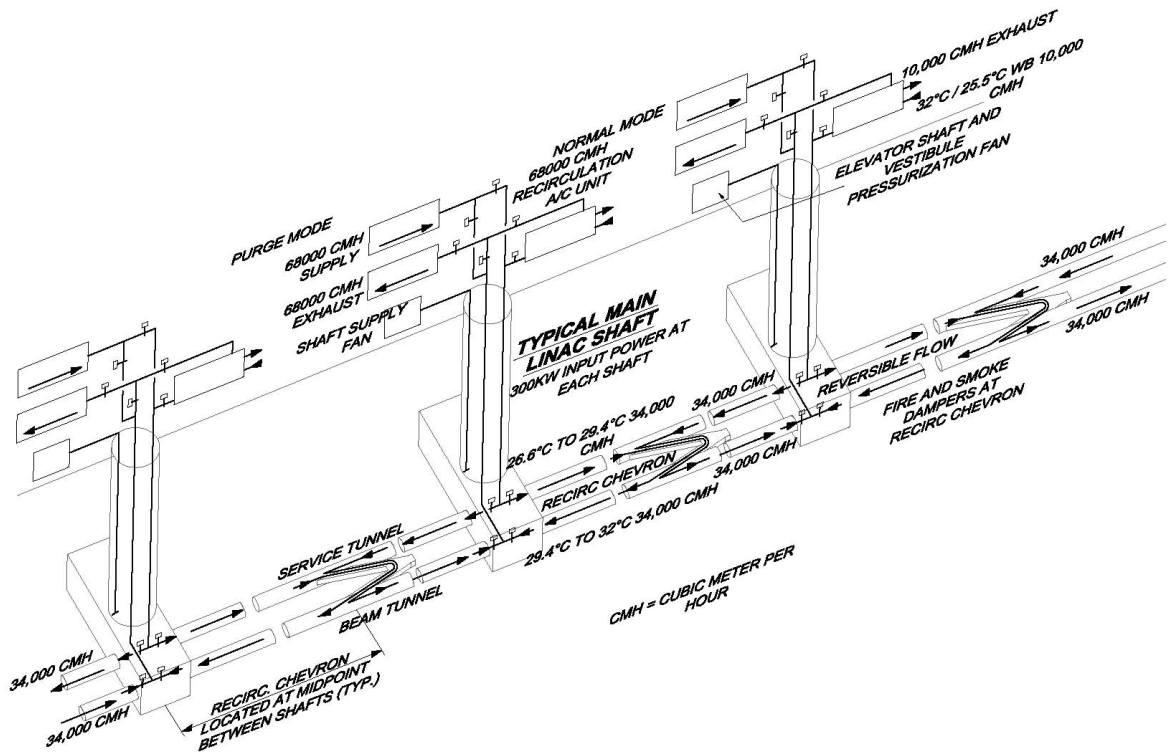}
     \vbabovecaption
      \caption{Air treatment concept for the Main Accelerator housing.}
      \label{fig:CFSairtreatment}
   \end{center} \vbbelow
\end{figure}

Figure~\ref{fig:CFSairtreatment} shows the air-flow and treatment for a typical section
of Main Linac. Conditioned ventilation airflow of 68,000 m$^{3}$/hr is
ducted down every major shaft and routed into the Service Tunnel
at the base of the shaft or Access Cavern in both directions.
Air flows through the Service Tunnel to the midpoint between
the major shafts ($\sim$2.5 km) where it is directed through a
protected air passageway into the Beam Tunnel and returned back to the shaft area.
The return air is ducted to the surface where 15\%
(10,200 m$^{3} $/hr) of the stale air is vented to the outdoors,
and an equivalent amount of fresh conditioned outside
air is mixed back in with the remaining circulated air.
While at the surface, the air is cooled, dehumidified and/or
heated as needed to achieve a neutral dry condition approximately
24-27$^{\circ} $C dry bulb and 40\% relative humidity. The supply air is
then ducted back to the Service Tunnel. This air flow pattern
requires evaluation by radiation safety personnel.

The conditioning units are located on the surface and reject tunnel
heat and moisture to the ambient air.  Air is supplied to the tunnel at a
flow rate of approximately 27 m/min; this provides one complete
air exchange every 6 hours in the entire tunnel volume. Additional
non-conditioning exhaust and supply fans are provided at each shaft
to double and/or reverse the airflow during emergency operation.
Common ducts are used for both systems separated by configuration
control dampers. Elevator shafts and exit vestibules are provided
with separate air systems for control and pressurization during
emergency operation.

\stepcounter{tablcl}\begin{table} [hb] \vbabove \caption{HVAC
requirements. }
   \label{tab:CFShvacreq}
\setlength{\tabcolsep}{6pt}
   \begin{center}
      \begin{tabular}{| l | c | c | c | c |}
         \hline
         ~Location & Temperature & Dewpoint & RH & Air Flow \\ [-6pt]
         & (drybulb) & & & \\ \hline & & & & \vbdlspacing \hline
         ~e- Source & 29$ ^{\circ} $C &  $<$13$ ^{\circ} $C & $<$35\% & 27 m/min \\ \hline
         ~Damping Ring & 40$ ^{\circ} $C & $<$13$ ^{\circ} $C & $<$20\% & 27 m/min \\ \hline
         ~Main accelerator service tunnel & 29$ ^{\circ} $C & $<$13$ ^{\circ} $C & $<$35\% & 27 m/min \\ \hline
         ~Main Linac beam tunnel (not contr.) & $>$30$ ^{\circ} $C & $<$13$ ^{\circ} $C & $<$35\% & 27 m/min \\ \hline
         ~BDS beam tunnel & 29-32$ ^{\circ} $C & $<$13$ ^{\circ} $C & $<$35\% & 27 m/min \\ \hline
         ~IR hall & 29-32$ ^{\circ} $C & $<$13$ ^{\circ} $C & $<$35\% & 27 m/min \\ \hline
      \end{tabular}
   \end{center} \vbbelow
\end{table}

\stepcounter{figlcl}

\subsection{Controls}\label{ssect:CFSc}

The temperature and humidity in the Service Tunnel are primarily
set by regional standards for allowing personnel to be in the tunnel
at moderate work levels with no required rest periods.  The
requirements for the tunnels in each of the system areas
are summarised in Table~\ref{tab:CFShvacreq}. In general, air temperature in the
Service Tunnel is controlled at 27-29$^\circ$C using chilled water Fan
Coil Units (FCU), as described in Section~\ref{ssect:CFSsd}. In the Main Linac
sections, the FCUs are located at every RF unit.

Temperature control in the Main Linac Beam Tunnel is not provided
because of the relatively low heat loads.  The humidity level is
maintained by the air circulation from the Service Tunnel and
by packaged dehumidification units located approximately every
100 meters. Beam Tunnel temperatures in the BDS are maintained
locally at 40-43$^\circ$C  by FCUs.

%------------------------

\clearpage %\setcounter{page}{1}
\setcounter{figlcl}{0}
\setcounter{tablcl}{0}
\setcounter{secnumdepth}{4} \setcounter{subsection}{0}
\setcounter{subsubsection}{0}
\usecounter{subsubsection}

\section{Process Cooling Water}\label{sect:CFSpcw}

Cooling water is required as a heat rejection medium for
technical components such as the water cooled RF components,
water cooled magnets, and water dumps in the BDS.
The majority – if not all – of these require low conductivity/deionized
water (LCW). Further study is needed to establish water quality
requirements. The following descriptions present a reference
solution that is generally applicable for all regions, ignoring minor regional site differences.

\subsection{Heat Loads}\label{ssect:CFShl}

Table~\ref{tab:CFSsumheatload} summarizes the estimated heat-loads broken down
by Area Systems. Of the total load of $\sim$182 MW, over half is
attributed to the Main Linac. Table~\ref{tab:CFSrfheatload} lists the heat loads for the
Main Linac RF unit.

\stepcounter{tablcl}\begin{table} [h] \vbabove \caption{Summary of
heat loads broken down by Area System. }
   \label{tab:CFSsumheatload}
   \begin{center}
      \begin{tabular}{| l | c | c | c |}
         \hline
         Area System & LCW & Chilled Water & Total \\ [-6pt]
        & (MW) & (MW) & (MW) \\ \hline & & & \vbdlspacing \hline
        Sources e$^{-}$ & 2.880 & 1.420 & 4.300 \\ \hline
        Sources  e$^{+}$ & 17.480 & 5.330 & 22.810 \\ \hline
        DR   e$^{-}$ & 8.838 & 0.924 & 9.762 \\ \hline
        DR   e$^{+}$ & 8.838 & 0.924 & 9.762 \\ \hline
        RTML & 9.254 & 1.335 & 10.589 \\ \hline
        Main Linac & 56.000 & 21.056 & 77.056 \\ \hline
        BDS & 10.290 & 0.982 & 11.272 \\ \hline
        Dumps & 36.000 & 0.000 & 36.000 \\ \hline
       \multicolumn{3}{| l |}{Total Heat Load (MW)} & 182 \\ \hline
      \end{tabular}
   \end{center} \vbbelow
\end{table}

\stepcounter{tablcl}\begin{table}  \vbabove \caption{Typical Main
Linac RF component heat loads. }
   \label{tab:CFSrfheatload}
   \begin{center}
\setlength{\tabcolsep}{4pt}
      \begin{tabular}{| l | c | c | c | c | c |}
         \hline
~Components & Tunnel & Total & Average & To Water & To Air    \\ [-6pt]
 &  & (KW) & (kW) & (KW) & (KW)    \\ \hline & & & & & \vbdlspacing \hline
~RF Charging Supply 34.5 KV AC-8 KV DC & service & 4.0 & 4.0 & 2.8 &
1.2    \\ \hline ~Switching power supply 4kV 50kW & service & 7.5 &
7.5 & 4.5 & 3.0    \\ \hline ~Modulator & service & 7.5 & 7.5 & 4.5
& 3.0    \\ \hline ~Pulse transformer & service & 1.0 & 1.0 & 0.7 &
0.3    \\ \hline ~Klystron socket tank / gun & service & 1.0 & 1.0 &
0.8 & 0.2    \\ \hline ~Klystron focusing coil (solenoid ) & service
& 4.0 & 4.0 & 3.6 & 0.4    \\ \hline ~Klystron collector/
body/windows & service & 58.9 & 47.2 & 45.8 & 1.4    \\ \hline
~Relay racks (instrument racks) & service & 10.0 & 10.0 & 0.0 & -1.5
\\ \hline ~Circulators, attenuators \& dummy load & beam & 42.3 &
34.0 & 32.3 & 1.7    \\ \hline ~Waveguide & beam & 3.9 & 3.9 & 3.5 &
0.4    \\ \hline ~Subtotal Main Linac RF unit (KW) &  &  & 120 &  &
\\ \hline
      \end{tabular}
   \end{center} \vbbelow
\end{table}

\subsection{System Description}\label{ssect:CFSsysd}

There are two water cooling systems; Process Water and
Chilled Water. Chilled Water is used for water cooled racks
in each RF area and for fan coils that remove the heat
rejected to the tunnel air. The Process Water handles the
water cooled technical components. The scope of the Process
Water cooling included in conventional facilities includes the
surface cooling towers, pumps, controls, cavern heat exchangers,
skids and piping headers, and distribution and valving up to the
water cooled components. Final hose connections to each
water-cooled technical component are included in the relevant
Technical System. The tower system for the cryogenics is
considered part of the Cryogenics Technical System.

All water systems are closed loop. The cooling tower type is
a closed circuit cooler similar to a dry cooler. The tower works
dry by releasing the heat directly to the air during most of the
year. During hot periods in the summer seasons, the towers/coolers
are wetted with water in order to guarantee the supply temperature.
This setup minimizes and conserves water and treatment chemicals
and associated cost, as compared to typical open type towers.
The closed circuit coolers also minimize plume from the tower.
The make-up water to the system and tower is supplied from
a well with proper water treatment, from each surface water plant.

 \stepcounter{figlcl}\begin{figure}[htb]
   \begin{center} \vbabove
      \includegraphics[width=0.6\textwidth]{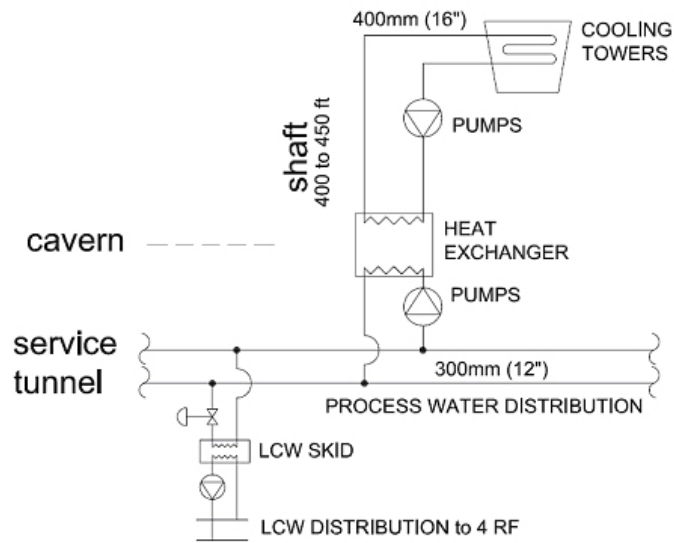}
      \vbabovecaption
      \caption{Process water system at shaft 7 plant.}
      \label{fig:CFSprocesswater}
   \end{center} \vbbelow
\end{figure}

Figure~\ref{fig:CFSprocesswater} shows a schematic of a typical
Process Water Plant. The Process water system has three closed water loops:

\begin{itemize}\itemspace

\item     The first is a water/glycol mixture loop from the surface
             cooling tower to the cavern heat exchanger at 29.4$  ^{\circ}$C supply temperature. \itemspace
\item     The second is a process water loop from the cavern heat
             exchanger at 32.2$  ^{\circ}$C supply temperature to the LCW skid
             in the Service Tunnel. The heat exchanger in the cavern is
             needed to offset the effect of the static head on system
              pressure due to the tunnel depth. \itemspace
\item     The third is the demineralized/LCW water from the skid at 35$  ^{\circ}$C supply temperature. \itemspace

\end{itemize}

The supply water temperature has a tolerance of $\pm$0.5$^{\circ}$C.
The basis for pipe sizing and costing is for a $\Delta$T $\sim$11$^{\circ}$C
water system. The return pipe is thermally insulated to
reduce the heating of the tunnel air.  This setup is applicable
for the Main Linac; a variation of this scheme is used for other
areas. In the case of the Process Water supply to the large
BDS (main) beam dumps, a near surface buried piping
distribution from a surface plant at the IR location is fed into
each dump cavern hall through individual drilled shafts.
(The cooling system for the main high-powered beam dumps is discussed elsewhere.)

\begin{figure}[htb]
   \begin{center} \vbabove
      \includegraphics[width=0.7\textwidth]{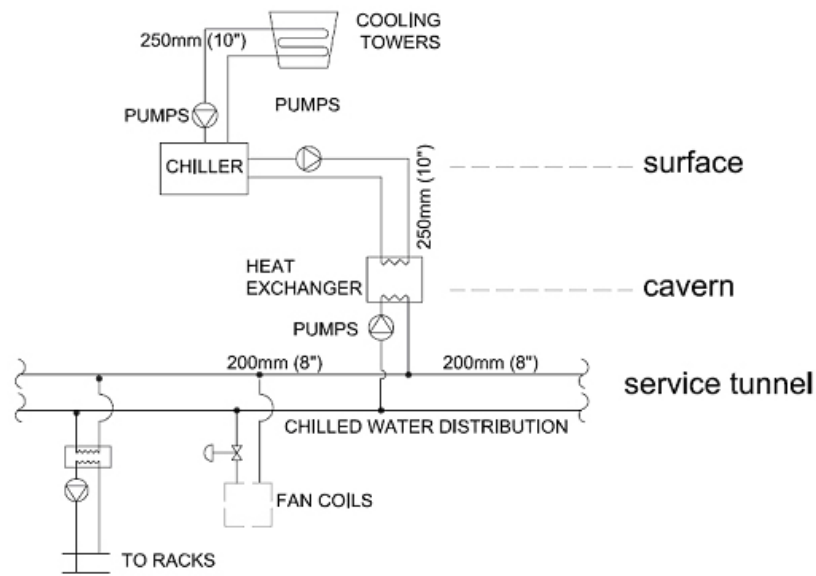}
      \vbabovecaption
      \caption{Chilled water system at shaft 7 plant.}
      \label{fig:CFSchilledwater}
   \end{center} \vbbelow
\end{figure}

The Chilled Water system provides $\sim$6.6$^{\circ}$C supply
temperature water to fan coils and to the water skid for racks.
The water skid, in turn, regulates and provides the proper temperature
(above dew point) to the water cooled RF racks. The major
components for this system are the same as for the Process
Water except for the addition of Chillers on the surface.
All chilled water piping is thermally insulated (see Fig.~\ref{fig:CFSchilledwater}).

 \stepcounter{figlcl}

\subsection{Locations and Distribution}\label{ssect:CFSld}

The main distribution of the Process Water system follows the
major shaft and cryogenic distribution locations. There are twelve
surface water plants. For the reference solution, the distribution is
simplified to minimize the number of area systems served by
each water plant (considered consistent with the current estimate).
Only the Main Linac RF system has been considered in any detail.
Estimates for other areas have been scaled from the Main Linac
model based on their respective loads.

One Low Conductivity Water (LCW) skid is used for cooling all
the water cooled technical components for every four Main Linac
RF units. Each LCW skid includes one stainless steel centrifugal
pump (with no standby), one plate heat exchanger, controls,
stainless steel expansion tanks, and miscellaneous fittings and accessories.

Similar to the LCW skid, one chilled water skid for racks is provided for
every 4 RF units. This skid is a commercially available package coolant
modulating unit typically used in data center rack applications. Each
skid includes a multi-stage centrifugal pump, brazed plate heat exchanger,
3-way control valve, expansion tank, flow switch and integrated controls.

%------------------------

\clearpage %\setcounter{page}{1}
\setcounter{figlcl}{0}
\setcounter{tablcl}{0}
\setcounter{secnumdepth}{4} \setcounter{subsection}{0}
\setcounter{subsubsection}{0}
\usecounter{subsubsection}
 
\section{Safety Systems}\label{sect:CFSss}

\subsection{Radiation safety}
The radiation safety systems are described in Section~\ref{ssect:OPSrad}

\subsection{Fire Safety }\label{ssect:CFSfs}

Because there are no existing laws and standards in
any region which directly and comprehensively stipulate
the safety measures for a facility like ILC, the currently
planned safety measures are based on examples
of existing accelerator tunnels and the regulations for
buildings and underground structures of various types.
The final plan will be subject to the approval of the competent
authority that has jurisdiction over the selected site.

\begin{figure}[htb]
   \begin{center} \vbabove
      \includegraphics[width=\textwidth]{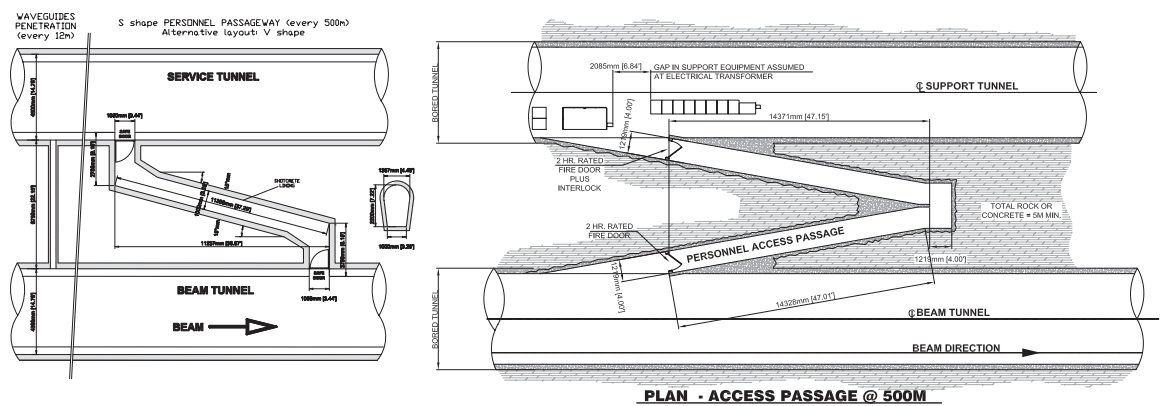}
      \vbabovecaption
      \caption[Examples of the personnel cross-connection passages
between the Service and Beam Tunnels.] {Examples of the personnel cross-connection passages
between the Service and Beam Tunnels (left Asia and Europe, right Americas).
The geometry of the passage is designed to reduce the radiation levels in the
Service Tunnel to acceptable levels.}
      \label{fig:CFSccpassage}
   \end{center} \vbbelow
\end{figure}

Evacuation of the underground Service Tunnel is the primary
concern, due to the relatively high level of cables. In the event
of a fire in the Service Tunnel, personnel can escape to the
safety of the Beam Tunnel via the Beam-Service Tunnel personnel
cross-overs, located every 500 m (see Figure~\ref{fig:CFSccpassage}). Egress to
the surface is only possible at the shafts, located every 5 km.
Assuming a walking speed of 1 m/s, 500 m between emergency
egress points is considered acceptable ($\sim$8 minutes maximum).
During beam operations, triggering of a fire alarm will immediately de-energize the machine, making it safe for personnel to enter the Beam Tunnel.

During access periods or installation, when personnel are
present in the Beam Tunnel, emergency egress can be
either to the Beam or Service Tunnel, depending on the location of the fire.

 \stepcounter{figlcl}

For the Damping Rings (during installation and maintenance), emergency
egress is to a separate safety enclosure behind a fire wall within the tunnel (see Section~\ref{ssect:CFSci}).

In all cases, personnel can either safely escape to the surface via
the nearest shaft, or remain in a fire-safe area until the emergency services
respond.

Smoke detectors are installed in all underground tunnels and
halls at intervals of 30 m. Manual alarms (buttons) are located at
intervals of 100 m. Alarm bells are also installed at intervals of
100 m. A smoke exhaust fan of 60,000 m$^3$/hr is installed at
each Shaft Access Building on the surface. 1.2 m$^2$ ducts
are installed in the Access Shafts/Tunnels to connect the fans
and the Access Hall.

No emergency smoke exhaust system is installed in the tunnels.
Instead, movement of the smoke is retarded by 1.5 m high walls mounted to the top of the tunnels at intervals of 50 m.
Simulations using software developed and widely used in
Japan indicate this solution is more effective than a mechanical exhaust system

Provisions for the required emergency fixtures are also included in the current estimate:
 \begin{itemize}

\item     emergency lighting located every 8 m \itemspace
\item     illuminated exit signs installed above every exit
            door in the underground spaces \itemspace
\item     illuminated exit direction signs installed at intervals
             of 20 meters in the underground spaces \itemspace
\item     portable chemical powder fire extinguishers (3.0 kg)
            placed every 30 m in the Service Tunnel, Beam Tunnel
            and for Asia Site the Access ramps \itemspace
\item     large size fire extinguishing equipment (30 kg) located every
             200 meters in the Access Halls and the Experimental Hall. \itemspace

\end{itemize}

Sprinklers, hydrants and water curtains have not been
specified to avoid possible water damage to the machines.

\subsection{Safety Access Control}\label{ssect:CFSsac}

Access control equipment such as a card lock is installed at the
entrances to the radiation control areas as required by the radiation safety plan.

\subsection{Safety for Helium}\label{ssect:CFSsfh}

The helium supply system is equipped with an oxygen meter
which activates an alarm and stops the gas supply in case of
oxygen deficiency. Air in the Beam Tunnel is automatically pressurized.

%------------------------

\clearpage %\setcounter{page}{1}
\setcounter{figlcl}{0}
\setcounter{tablcl}{0}
\setcounter{secnumdepth}{4} \setcounter{subsection}{0}
\setcounter{subsubsection}{0}
\usecounter{subsubsection}
 
\section{Survey and Alignment}\label{sect:CFSsaa}

Survey and alignment covers a very broad spectrum of activities,
starting from the conceptual design of the project, through the
commissioning of the machines, to the end of operations.
The cost estimate developed covers the work necessary until
successful completion of the machine installation. It includes
equipment needed for the tasks to be performed, and equipment
for a dedicated calibration facility and workshops. It also includes
the staff that undertake the field work, and the temporary
manpower for the workshops.  Full time, regular staff is considered
to be mainly dedicated to organizational, management, quality
control, and special alignment tasks. The cost estimate is mostly
based on scaling the equivalent costs of the LHC to the ILC scope.

\subsection{Calibration Facility}\label{ssect:CFScaf}

A 100 m long calibration facility is needed for the calibration of all
the metrological instruments. The facility is housed in a climate
controlled and stable building. Due to the range limit of current day
commercial interferometers – against which the instruments are to
be compared –  the facility has been restricted to 100 m.
A mechanical and an electronic workshop are also needed
during the preparation phase and throughout the entire project.
They are used for prototyping, calibration, and maintenance
of the metrological instruments.

\subsection{Geodesy and Networks}\label{ssect:CFSgan}

A geodetic reference frame is established for use across the
whole site, together with appropriate projections for mapping
and any local 3D reference frames appropriate for guaranteeing
a coherent geometry between the different beam lines and
other parts of the project. An equipotential surface in the form
of a geoid model is also established and determined to the
precision dictated by the most stringent alignment tolerances of the ILC.

The geodetic reference frame consists of a reference network
of approximately 80 monuments that cover the site. These
monuments are measured at least twice, by GPS for horizontal
coordinates, and by direct leveling for determining the elevations.
The first determination is used for the infrastructure and civil
engineering tasks. The second, and more precise determination,
is used for the transfer of coordinates to the underground networks
prior to the alignment of the beam components.  A geodetic
reference network is also installed in the tunnel and in the
experimental cavern.  For costing purposes it is assumed that
the reference points in the tunnel are sealed in the floor and/or
wall (depending on the tunnel construction) every 50 m. In the
experimental cavern, the reference points are mostly wall brackets.
The underground networks are connected to the surface by
metrological measurements through vertical shafts.
The distance between two consecutive shafts does exceed 2.5 km in most cases.

\subsection{Civil Engineering Phase}\label{ssect:CFScep}

The layout points which define the tunnel locations and shapes
are calculated according to the beam lines in the local 3D
reference frame. The tunnel axes are controlled as needed
during the tunnel construction. All tunnels, including profiles,
are measured in 3D using laser scanner techniques when
the tunnels are completed. The same process is applied
to the experiment cavern(s) and other underground structures.
The buildings and surface infrastructure are also measured
and the as-built coordinates are stored in a geographical
information system (GIS).

\subsection{Fiducialization}\label{ssect:CFSfid}

Systematic geometrical measurements are performed on
all beamline elements to be aligned prior to their installation
in the tunnels. The alignment of elements installed on
 common girders or in cryomodules is first performed,
and the fiducial targets used for the alignment in the
tunnels are then installed on the girders (cryomodules)
and all individually positioned elements. The positional
relation between the external markers and the defining
centerlines of the elements are then measured. For this
report, an estimated 10,000 magnetic elements were
assumed to need referencing.  It is also assumed that
most corrector magnets do not need fiducialization. This
number does not account for instrumentation, collimators,
or other special beam elements.

\subsection{Installation and Alignment}\label{ssect:CFSiaa}

The trajectories of all the beamlines are defined in the local 3D
reference frame which covers the entire site. The location of
reference markers at the ends of each beam line element to
be aligned are defined in this reference system, together with
the roll angle giving a full 6 degrees of freedom description of
element location and orientation.  Likewise the position of all
geodetic reference points is determined in this reference frame.

Prior to installation, the beamlines and the positions of the
elements are marked out on the floors of the tunnels.
These marks are used for installing the services, and the
element supports. The supports of the elements are then
aligned to their theoretical position to ensure that the
elements can be aligned whilst remaining within the adjustment range of the supports.

After installation of services such as LCW and cable trays,
the tunnels are scanned with a laser scanner. The point clouds
are then processed, and the results inserted into a CAD model.
A comparison with theoretical models is used by the integration
team to help identify any non-conformity and prevent interference
with the subsequent installation of components.  The current
requirements for the one sigma tolerances on the relative alignment
of elements or assemblies are given in Table~\ref{tab:CFScompalignment}.

\stepcounter{tablcl}\begin{table}  \vbabove \caption{Component
alignment tolerances. }
   \label{tab:CFScompalignment}
   \begin{center}
      \begin{tabular}{| l | l | l | }
         \hline
Area & Type & Tolerance \\ \hline & & \vbdlspacing \hline
Sources, Damping  & Offset & 150 $ \mu $m (horizontal and vertical), \\ [-6pt]
Rings and RTML &  & over a distance of 100 m. \\
& Roll & 100 $ \mu $rad \\ \hline
Main Linac & Offset & 200 $  \mu$m (horizontal and vertical), \\ [-6pt]
(cryomodules) & & over a distance of 200 m. \\
 & Pitch & 20 $  \mu$rad \\
 & Roll &  \\ \hline
BDS & Offset & 150 $  \mu$m (horizontal and vertical),  \\ [-6pt]
& & over a distance of 150 m around the IR. \\ \hline
      \end{tabular}
   \end{center} \vbbelow
\end{table}

The components are aligned in two steps:

\begin{itemize}

\item   A first alignment is performed to allow connection of the vacuum
          pipes or interconnection of the various devices. This is done using
          the underground geodetic network as reference. \itemspace
\item   After all major installation activities are complete in each
          beamline section, a final alignment, or so-called smoothing,
          is performed directly from component to component in order
            to guarantee their relative positions over long distances. \itemspace
\end{itemize}

To reach and maintain the positioning tolerances of the final doublets
in the BDS IR, a 150 m straight reference line is set up as close as
possible to the beam components. This line, consisting of a laser or
stretched wire and hydrostatic levels, is housed in a dedicated gallery
built parallel to the beam tunnel, and goes through the experimental cavern. This allows for the geometrical connection between the beam lines and the detector.

\subsection{Information Systems}\label{ssect:CFSinsys}

The theoretical positions of all the components to be aligned on
the beam lines is managed in a dedicated database. This database
is also used for managing all the geodetic and alignment measurements
and the instrument calibrations.  All measurement data are captured
and stored electronically and subsequently transferred to the database.
Pre-processing of the measurements are carried out in the database and
then dedicated software for data analysis is used to calculate the best
fit position of the elements and components. These results are also
stored in the database where they can be accessed for further
post-processing, analysis and presentation. A geographic information system
(GIS) is set up for managing all location data.

%------------------------

\clearpage %\setcounter{page}{1}
\setcounter{figlcl}{0}
\setcounter{tablcl}{0}
\setcounter{secnumdepth}{4} \setcounter{subsection}{0}
\setcounter{subsubsection}{0}
\usecounter{subsubsection}
 
\section{CFS Cost Methodology}\label{sect:CFSccm}

The cost for the ILC CFS has been developed internationally
with teams in each of the three regions (Americas, Asia and Europe).
These teams have worked closely together to optimize the CFS
design, based on the requirements supplied by the Area and Technical Systems.

To make use of the available resources for the design and cost work,
a detailed WBS for the project was produced, containing up to
5 levels of detail. This WBS was then broken down into site-dependent
and site-independent sections. For the site-dependent estimates, the
CFS group established a set of uniform definitions for underground
construction unit costs. This ensured consistency across all three regions.
Estimates for each unit cost were independently produced by experts
and consultant engineering firms in each of the three regions
(the Civil Engineering falls into this category, for example) and
then used to develop each site-dependent design. The remaining
site-independent parts were then divided up amongst the regional
teams to produce single estimates as follows:

\medskip
\begin{center}
\begin{tabular}{lr}
Civil Construction & Regionally developed \\
Electrical: site-dependent & Regionally developed \\
Electrical: site-independent & European estimate used \\
Air treatment facilities & Americas estimate used \\
Process cooling water & Americas estimate used \\
Handling equipment & European estimate used \\
Safety systems & Asian estimate used \\
Survey and Alignment systems & European estimate used \\
\end{tabular}
\end{center}
\medskip

Cost estimates in all three regions were developed using the
same criteria and drawings. information was drawn from consultant
engineers, historical data from other accelerator or similar projects,
 industry standard cost estimating guides, and where applicable the
scaling of costs from similar systems.  In all cases, the estimates
reflect a median value for the work based on the criteria provided to
date and the pre-conceptual level of design maturity.  There are
no factors for contingency contained in any of the CFS costs
estimates.  Costs for activities that take place prior to the construction
start are explicitly not included in the estimate.  Some examples of
such costs are A/E Services before the start of construction,
development costs for geotechnical and environmental investigation,
land acquisition costs and cost incurred for compliance with local
governmental statutes and regulations.  These costs cannot be
accurately identified until a specific site selection is made.

\clearpage %\setcounter{page}{1}
\setcounter{figlcl}{0}
\setcounter{tablcl}{0}
\setcounter{secnumdepth}{4} \setcounter{subsection}{0}
\setcounter{subsubsection}{0}
\usecounter{subsubsection}

\setcounter{section}{8} \renewcommand{\picturefolder}{./install/}

\section{Installation Plan}\label{sectINS}

\subsection{Overview}

\begin{figure}[htbp]
   \begin{center} \vbabove
      \includegraphics[width=\textwidth]{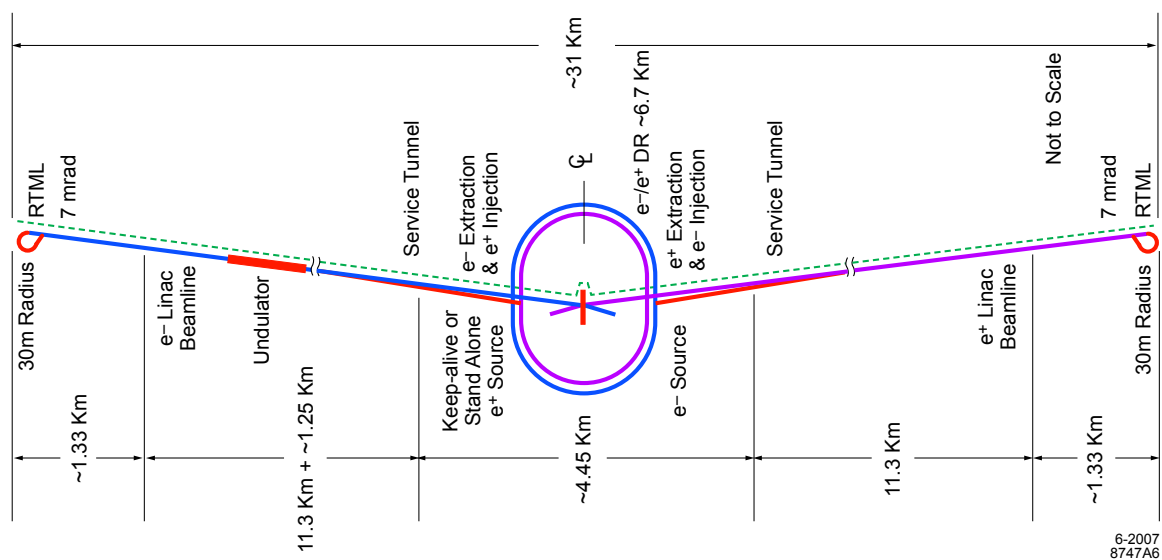}
      \vbabovecaption \caption{Schematic layout of ILC.}
      \label{fig:instLayout}
   \end{center} \vbbelow
\end{figure}

The baseline ILC covers a large geographical area over 30 kilometers
long that includes a complex network of $\sim$72~km of underground
tunnels at a depth of approximately 100~m. An overall schematic
layout of the ILC is shown in Figure~\ref{fig:instLayout}. These
tunnels house most of the technical equipment needed to operate the
accelerator. There are $\sim$2,000 cryomodules, over 13,000 magnets
and approximately 650 high level RF stations to be installed. These
and other technical components are described in the Area and
Technical System sections of this report. The schedule for
construction of the ILC is assumed to be 7 years as described in
Section \ref{ConstSchedule}. This section describes the model that
was developed and costed for the installation of all components on
an appropriate schedule.

\stepcounter{figlcl}

\subsection{Scope}

The installation plan covers all activities required to prepare,
coordinate, integrate, and execute a detailed plan for the complete
installation of the ILC components as well as associated site-wide
logistics. It includes all labor, incidental materials and equipment required to
receive, transport, situate, affix, accurately position, interconnect,
integrate, and checkout all components and hardware from a central
storage or subassembly facility to their operational location within
the tunnels. The premise is that installation recieves fully tested assemblies certified for in-tunnel installation. It does not include component fabrication, assembly, quality control or commissioning. It also does not include the basic tunnel utilities provided by conventional facilities, such as ventilation, air conditioning, fire prevention, high voltage electrical and low-conductivity water distribution.

\subsection{Methodology}

For the RDR, the goal was to understand and define the scope of
installation work sufficiently to develop a reasonable model for a
first stage of planning and costing. The model was based on a work
breakdown structure (WBS) that listed all of the activities required
for installation of the technical systems, including the management,
planning, and engineering support.

The installation WBS was broken down into two major level-of-effort
categories, General Installation and Area Systems Installation.
General Installation included all common activities and preparations
and associated logistics on the surface. Area System Installation
included all efforts required for complete installation of the
components underground. General Installation was further broken down
into logistics management, engineering support, equipment, vehicles,
shipping-receiving, warehousing, and transportation. Area System
Installation covered the six machine areas, electron source,
positron source, damping rings, RTML, main linac and beam delivery.
Each element of the WBS for both General and Area System was then
extended two levels of effort further and populated with required
labor as well as incidental material and equipment costs, as
described below. Table~\ref{tab:instWBS} shows the top-level
installation WBS.

\stepcounter{tablcl}\begin{table} [htb] \vbabove
 \caption{Top-level WBS installation. }
   \label{tab:instWBS}
   \begin{center}
      \begin{tabular}{| c |c |c | c | c || l |}
         \hline
         \multicolumn{5}{| c ||}{WBS} & Component \\
         \hline & & & & & \vbdlspacing  \hline
         1 & 7 & 3 &  &  & Installation \\ \hline
         1 & 7 & 3 & 1  &    & General Installation \\ \hline
           &     &    &    & 1 & Logistics management \\ \hline
          &     &    &    & 2 & Engineering support \\ \hline
          &     &    &    & 3 & Equipment \\ \hline
          &     &    &    & 4 & Vehicles \\ \hline
          &     &    &    & 5 & Shipping \& receiving \\ \hline
          &     &    &    & 6 & Warehousing \\ \hline
          &     &    &    & 7 &Surface transport \\ \hline & & & & & \vbdlspacing \hline
        1  &  7   &  3  &   2  &  & Area System Installation \\ \hline & & & & & \vbdlspacing \hline
          &     &    &    & 1 & Sources e$ ^{-} $ area installation\\ \hline
        &     &    &    & 2 & Sources e$ ^{+} $ area installation\\ \hline
        &     &    &    & 3 & Damping Rings area installation\\ \hline
        &     &    &    & 4 & RTML area installation\\ \hline
        &     &    &    & 5 & Main Linac area installation\\ \hline
        &     &    &    & 6 & Beam Delivery area installation\\ \hline
      \end{tabular}
   \end{center}
\vbbelow
\end{table}

The installation cost estimate for the ``Cold'' Linear Collider from
the 2003 US Technology Options Study was used as a starting point
for developing the ILC WBS. This was adjusted for the differences
between that design and the ILC as well as for lessons learned from
other projects. Available information from the WBSs developed for
NLC and TESLA was incorporated wherever possible, as was pertinent
material from similar installation projects at APS, FNAL Main
Injector, KEKB, LHC, PEP-II, SLC and SPEAR-III, as well as
installation plans for SSC and the European XFEL. The scope,
complexity and salient features of these other machines were
compared with the ILC.

To populate the WBS, a comprehensive list of
components was compiled and interfaces and boundaries with the
technical systems carefully defined. As an example of such cost/scope definition, it was assumed for magnet installation that fully tested and measured magnets complete with supports, anchor bolts, and other required materials were delivered to a surface staging area, along with any special instrumentation or handling equipment. The installation group transported the device to the proper location, arranged for alignment, installed the device and instrumentation, and connected it to the local power, water and cryogenic systems. Details of which group supplied the cables, hoses or fittings were explicitly specified, as were testing responsibilities.

The estimates for labor and equipment required to install the components came from a wide variety of sources. For conventional components, like beampipes and magnets, the technical systems provided estimates, based on experience with other projects. Visits to CERN and DESY provided data on installation of
cryomodules, LHC magnets and the CMS detector as well as the
opportunity to observe actual installation procedures. RSMeans 2006 cost data (North America's leading supplier of
construction cost information) was used in estimating total work-hours
needed for installing equivalent size/weight equipment under similar
conditions. Since the main linac is a major cost driver, the installation of cryomodules and RF sources was modeled in detail. This is described in the next section. For other systems where there was not an appropriate experience base, the estimates were scaled from similar installation tasks based on an assessment of complexity.

The resulting estimates were subjected to a variety of cross-checks and reviewed for completeness and appropriateness by technical and area system leaders. The estimates were compared with individual estimates from other sources, and with the actual manpower used for the installation of recent accelerator projects. The labor estimate for the particular cryomodule installation tasks was also independently calculated by a second engineering team, and the results were in agreement to within 13\%. An additional check was that the overall installation costs were 7\% of the total level of effort. This is consistent with the estimate from the 2003 Options Study where the installation effort was 7\% and with the installation costs for other projects studied which also averaged 7\%.

\begin{figure}[htbp!]
   \begin{center} \vbabove
      \includegraphics[width=\textwidth]{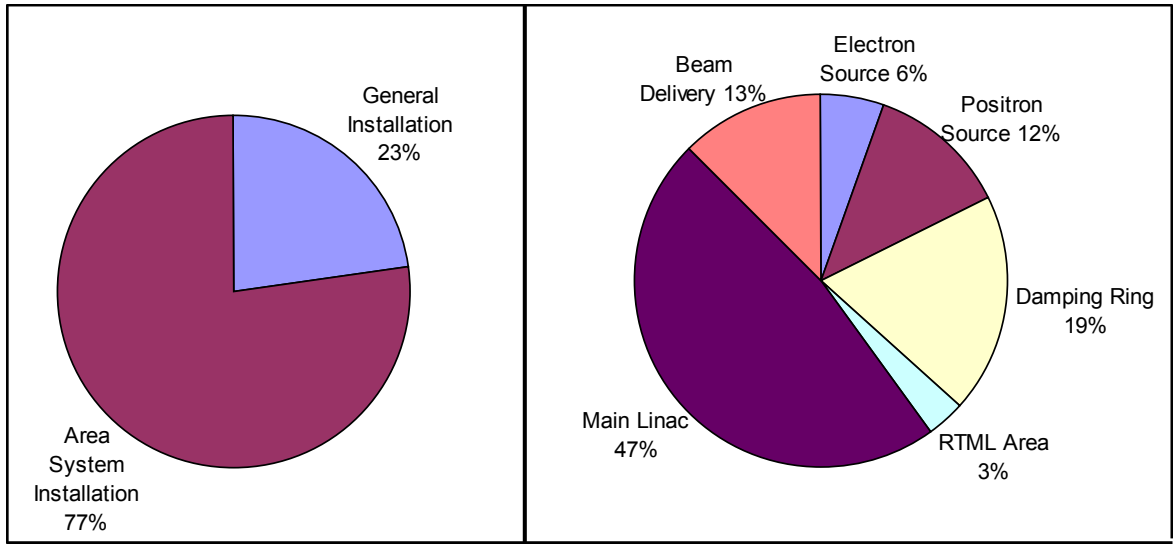}
      \vbabovecaption \caption{Distribution of effort between General and Area Systems and between Areas.}
      \label{fig:instWBS}
   \end{center} \vbbelow
\end{figure}

Figure~\ref{fig:instWBS} shows the distribution of installation effort between General and Area Systems and between the various Area Systems, where the Main Linac accounts for almost half of the effort.

\stepcounter{figlcl}

\subsection{Model of Main Linac Installation}

At this stage of the ILC design, it is too early for a complete model of the entire installation sequence. The Main Linac cryomodules and rf sources represent a major installation effort so a bottoms-up model for their installation was developed. The model was derived from that in the TESLA TDR.
Installation was assumed to take place over
a period of 3 years with half a year ramp up time. Labor productivity
was taken to be 75\%, or 6 hours per shift, given transport
distances and handling difficulty.

\begin{figure}[htbp!]
   \begin{center} \vbabove
      \includegraphics[width=0.6\textwidth]{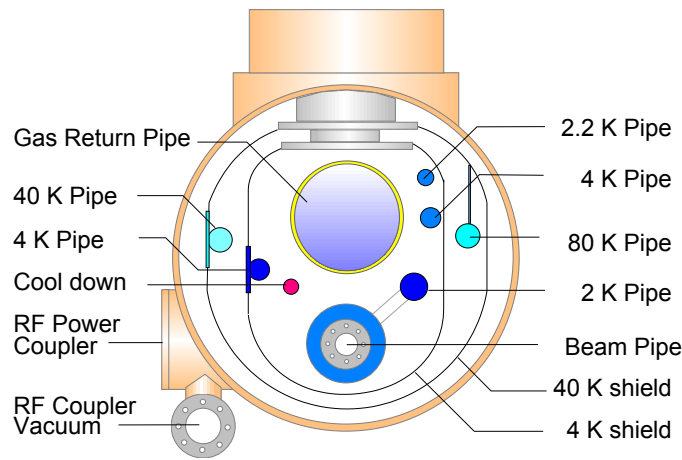}
      \vbabovecaption \caption{Schematic of the cryomodule showing multiple connections. }
      \label{fig:instModel}
   \end{center} \vbbelow
\end{figure}

Before starting installation, the section of the main linac beam and support
tunnels must be completely ready for joint occupancy, along with one large
and two small associated access shafts. The installation sequence was first to fix the cryomodule supports, then to move the cryomodules from the access shafts, install the cryomodule and complete the cryogenic, vacuum and rf connections. Figure~\ref{fig:instModel} is a schematic of the cryomodule indicating the number of connections to be made.
The installation rate was three cryomodules (one RF unit) and associated services per day for each crew. The model
included the number and size of equipment, distances to installation,
speed of transportation and estimates of number of staff and hours for
each task.

\stepcounter{figlcl}

\begin{figure}[htbp!]
   \begin{center} \vbabove
      \includegraphics[width=\textwidth]{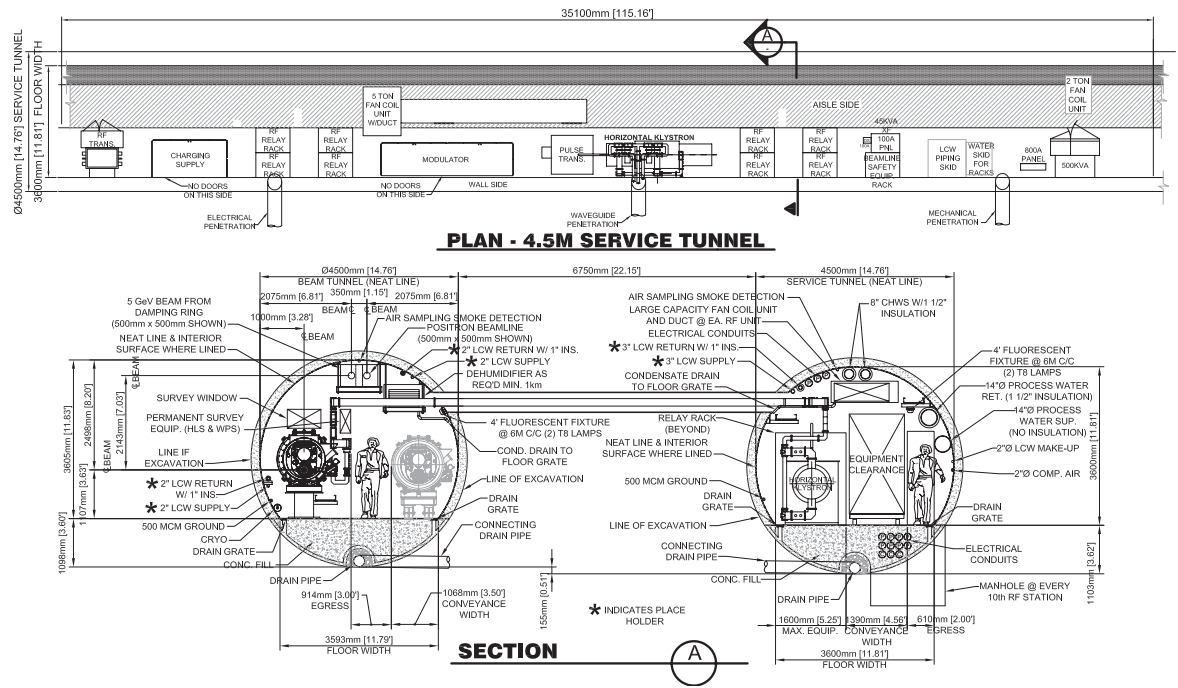} %\vspace{30pt}
      \vbabovecaption \caption[Plan view of service tunnel components in Main Linac.] {Plan view of service tunnel components in Main Linac
        (upper). Cross section of two Main Linac tunnels (lower).}
      \label{fig:instTunnelXsec}
   \end{center} \vbbelow
\end{figure}

The study concluded that a total of 72 person-days are required to install 3 cryomodules. This labor includes engineers and technicians from
a variety of specialties. The 3 cryomodules account for only about 20\% of the
effort to assemble an entire RF unit (and everything in this 38~m length of
beam and support tunnels) so the installation estimate for an entire RF unit was taken to be 5 times the cryomodule estimate. The other
components include, klystrons, modulators, control racks, cable trays,
control cables and RF waveguides. Such a
section of the tunnels and components are shown in Figure
\ref{fig:instTunnelXsec}.

\stepcounter{figlcl}

\subsection{Modelling}\label{sect:STAi}

\begin{figure}[htbp!]
   \begin{center} \vbabove
      \includegraphics[width=\textwidth]{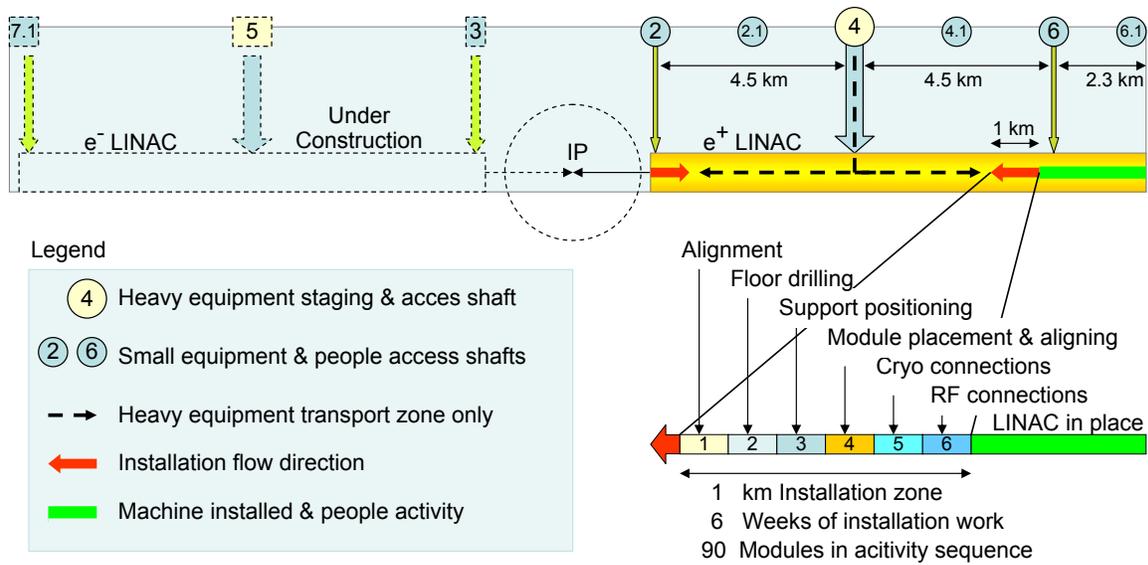}
      \vbabovecaption \caption{Installation model for main linac components
              in underground segment.}
      \label{fig:instSeq}
   \end{center} \vbbelow
\end{figure}

Installation planning of the large and complex ILC machine requires
the creation of 3-D computer models of all the major components as
well as the underground facilities. To create a cost effective, timely
and safe installation plan, certain facility conditions must be
assumed to exist prior to installation. Some examples include the
availability of utilities, communication systems, above ground
warehousing and equipment staging areas. Below ground, the personnel
access rules, including safety and emergency considerations, must be
defined and the schedule of equipment and tunnel availability must be
known. Once these and the details of the technical components are
known, a very general model, both in time and 3-D space, can be
developed as is shown below for the main linac (see Figure~\ref{fig:instSeq}).

\stepcounter{figlcl}

Here the 72 man crew is working in a (moving) 1 km section of the
tunnels at the 3 cryomodule per day rate, showing the different
activities which spread over a 6 week time span. Two crews are working
independently starting at shafts 2 and 6 and working towards shaft
4. Similar activities and crews will be working in other sections of
the linac tunnels when they become available. This is also true for
the central complex of injectors and damping rings.

\begin{figure}[htbp!]
   \begin{center} \vbabove
      \includegraphics[width=0.8\textwidth]{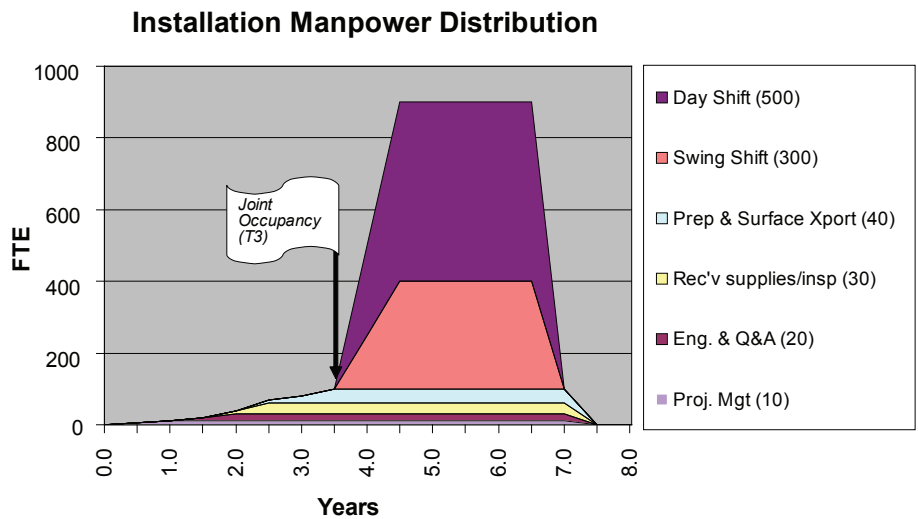}
      \vbabovecaption \caption{Required human resources versus time for
            the installation effort.}
      \label{fig:instManpower}
   \end{center} \vbbelow
\end{figure}

The RDR estimate assumed a 3 year installation schedule, a six month
period of ramp up and on the job training, and a 75\% efficiency. In tunnel activities are concentrated on day shift, with transport and staging on swing shift.
Figure~\ref{fig:instManpower}, shows a model of multishift manpower versus time, indicating the total manpower
necessary to fit all of the installation activities into that 3.5 year period. During the peak 3-year period, there are over 500 people on day shift  and another 300 on swing shift in various parts of the tunnel. There are also about 100 people involved in surface logistics.

\stepcounter{figlcl}

In the absence of a detailed fabrication plan for the major machine
components, a very top level installation schedule was developed to
integrate with a 7 year construction schedule. This will continue to
be refined as more information on fabrication schedules become
available.
\begin{comment}
Figure~\ref{fig:instTrialPlan} is a first attempt in setting-up a platform
for this activity, and shows the tunnels before and during
installation.

\stepcounter{figlcl}\begin{figure}[t!]
   \begin{center} \vbabove
      \includegraphics[width=\textwidth]{\picturefolder instTrialPlan.png}
      \vbabovecaption \caption{Example of a platform to be used for planning installation.}
      \label{fig:instTrialPlan}
   \end{center} \vbbelow
\end{figure}
\end{comment}

\setcounter{chapter}{4}

\chapter{\textsf{Sample Sites}\label{chapSS}}

\setcounter{figlcl}{0}
\setcounter{tablcl}{0}
\setcounter{secnumdepth}{4} \setcounter{subsection}{0}
\setcounter{subsubsection}{0}
\usecounter{subsubsection}

\setcounter{section}{0} \renewcommand{\picturefolder}{./sites/}

\section{Introduction}

    For this reference design, three `sample' sites for the ILC were evaluated. Each site was required to be able to accommodate all the conventional facilities for the 500 GeV CM machine; in addition, the sites needed to have the sufficient length to support an upgrade of the machine to 1 TeV CM, assuming the baseline main linac gradient.
    There were two reasons for the use of three sample sites for this reference design:

\begin{itemize}
\item This procedure demonstrates that each region can provide at least one satisfactory site for the ILC. This is important, since it shows that any of the regions has the potential to be a host for the project. \itemspace
\item The cost of, and technical constraints on, the project could depend strongly on the site characteristics. Since the actual site is not yet known, it is important to assess a range of sites with a diverse set of site characteristics, to provide confidence that when the actual site is chosen, it will not present unexpected technical difficulties or major surprises in cost. \itemspace
\end{itemize}

In addition to the three sample sites presented, a second European
sample site near DESY in Hamburg, Germany, has also been developed.
This site is significantly different from the other sites, both in
geology and depth ($\sim$25 m deep), and requires further study.

The Joint Institute for Nuclear Research has also submitted a
proposal to site the ILC in the neighborhood of Dubna, Russian
Federation.

The three sites reported in detail here are all 'deep-tunnel'
solutions. The DESY and Dubna sites are both examples of 'shallow
sites'. A more complete study of a shallow site -- either a shallow
tunnel or a cut-and-cover site -- will be made in the future as part
of the Engineering and Design phase.

\clearpage

\setcounter{figlcl}{0}
\setcounter{tablcl}{0}
\setcounter{secnumdepth}{4} \setcounter{subsection}{0}
\setcounter{subsubsection}{0}
\usecounter{subsubsection}
 \setcounter{section}{1}

\section{Americas Site}\label{sect:SITamericas}

\subsection{Location}\label{ssect:SITaml}

The Americas sample site lies in Northern Illinois near the
existing Fermi National Accelerator Laboratory. The site
provides a range of locations to position the ILC in a north-south
orientation. The range is bounded on the east by the Fermilab
site, and extends some 30 km to the west. For the purpose
of this document and the RDR estimate, a site alignment
that is roughly centered on Fermilab was selected. While
this site is more developed than an alignment to the west,
there is a reasonable construction path and the location
benefits more directly from the existing Fermilab site and facilities.

While the routing requires the tunnel to pass below residential
areas, the shafts can be located in non-residential areas.
It is highly possible that no homes will be physically affected
by this project. Roughly one quarter of the alignment is on
Fermilab property, including the ILC central campus and IR.
The Fermilab site is located approximately thirty-five miles
west of downtown Chicago. The area surrounding Fermilab is
comprised of residences, research facilities, light industry,
commercial areas, and farmland. Higher population densities
are found to the east with more rural and farm communities
to the west. The towns and villages around Fermilab vary in
population size from ten thousand to over one hundred thousand
people. The surrounding communities have established
schools, hospitals, infrastructure support functions and local governments.

The Fermilab site borders on a local railroad line with a
railroad hub located within four kilometers to the south.
Major roads connect Fermilab to the Illinois toll road system
within two miles of its gates. Access to O'Hare International
Airport and Midway Airport are via highways with travel times
to these airports of less than one hour. Steel mills and other
heavy industry are located both in Illinois and in neighboring states.

\subsection{Land Features}\label{ssect:SITamlf}

The existing surface of northern Illinois is primarily flat, with
surface elevations ranging from 200 meters to 275 meters
above sea level. Much of the eastern half of northern Illinois is
developed with Chicago suburban communities and municipalities
including many commercial, residential and industrial complexes.
Underdeveloped areas are currently used for agriculture. Major water
bodies include Lake Michigan located approximately 65
kilometers east of Fermilab, the Illinois River approximately
30 kilometers southeast of Fermilab and the Fox River
3 kilometers west of Fermilab. An intricate highway system
extends throughout the northeastern Illinois area.

The 2751 hectare (6800-acre) Fermilab site is also relatively
flat with less than 15 meters of fall from northwest to southeast.
Approximately one-third of the Fermilab site is developed with
various high-energy physics accelerator complexes or related
experimental areas. The remaining two thirds are equally split
between leased agricultural uses and open space including prairies,
wetlands and recreational areas. A series of paved roadways
exist throughout Fermilab.

\subsection{Climate}\label{ssect:SITamc}

The climate is typical of the Midwestern United States which
has four distinct seasons, and a wide variety of types and
amounts of precipitation with moderate variations between
monthly and seasonal average values. In summer, temperatures
ordinarily reach anywhere between 26$^{\circ}$C to 33$^{\circ}$C  and humidity
is moderate. Overnight temperatures in summer are usually
around 17$^{\circ}$C. Yearly precipitation averages 920 mm. Winter
temperature averages -2$^{\circ}$C  during the daytime, and -10$^{\circ}$C  at night. Temperatures can be expected to drop below -18$^{\circ}$C  on
15 days throughout the winter season.

\subsection{Geology}\label{ssect:SITamg}

\stepcounter{figlcl}\begin{figure}[bhtp]
   \begin{center} \vbabove
      \includegraphics[width=0.9\textwidth]{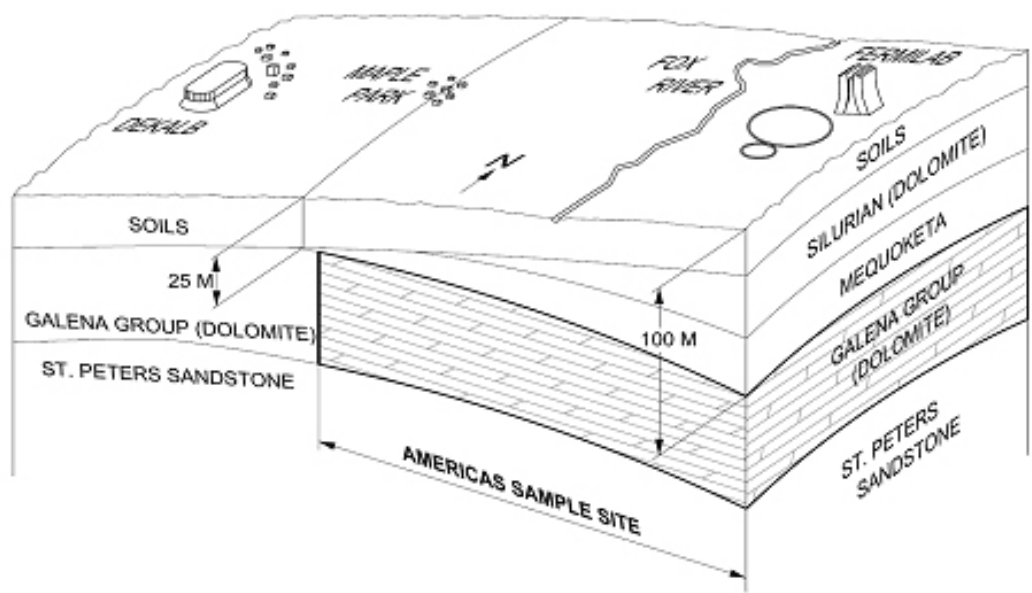}
      \vbabovecaption \caption{Geology of the Americas Sample Site.}
      \label{fig:SITssg}
   \end{center} \vbbelow
\end{figure}

The tunnels are located in the Galena Platteville layer (Figure~\ref{fig:SITssg}),
which is characterized as a fine to medium grained dolomite,
that is cherty. The Maquoketa shales overlaying the dolomite
have a low hydraulic conductivity that will act as a hydrogeologic
barrier between upper overburden aquifers and the dolomite.
At the proposed siting, the Galena Platteville varies from 100
to 125 meters in thickness, gently rising in datum elevation
from the south to the north. The Galena is covered by 15 to
30 meters of shale, 15 to 25 meters of Silurian dolomite
which in turn, is overlaid by 15 to 45 meters of overburden.
The upper Silurian dolomite found at the Fermilab site disappears
for alignments further to the west. These geologic conditions
should provide a relatively dry tunnel, both during construction
and during operations, but it is expected that some grouting will
be required. The Galena is the most structurally sound rock in
the area and, in general, should not require any extraordinary
rock support methods.

\subsection{Power Distribution System}\label{ssect:SITampd}

Electric power to the Northern Illinois area is provided by
Exelon Generation with access to approximately 35,000 MW
of electricity . Electrical power is generated by fossil fuel,
hydroelectric, wind and nuclear power generating systems
and distributed in Northern Illinois.

\subsection{Construction Methods}\label{ssect:SITamcm}

Conventional un-shielded tunnel boring machines are used for the
tunnels. No temporary support is required, permanent support can be
pattern spaced rock bolts or dowels.  Production rate is anticipated
to be 30 m/day. Caverns are excavated using drill and blast methods.
Temporary supports are required for the largest spans, permanent
support is provided by rock bolts. Production rate for medium to
large size caverns where mechanized equipment can be employed is
estimated at 1,200 cubic meters per week. Shaft overburden is excavated using
standard earth excavators and muck boxes, supported by ring beams
and timber lagging, keyed into the underlying rock. Excavation
through the limestone and shale to the final depth uses conventional
Drill \& Blast methods. Support is provided by resin encapsulated
rockbolts and the shaft is reinforced and concrete lined.

\stepcounter{figlcl}\begin{figure}[htb]
   \begin{center} \vbabove
      \includegraphics[width=\textwidth]{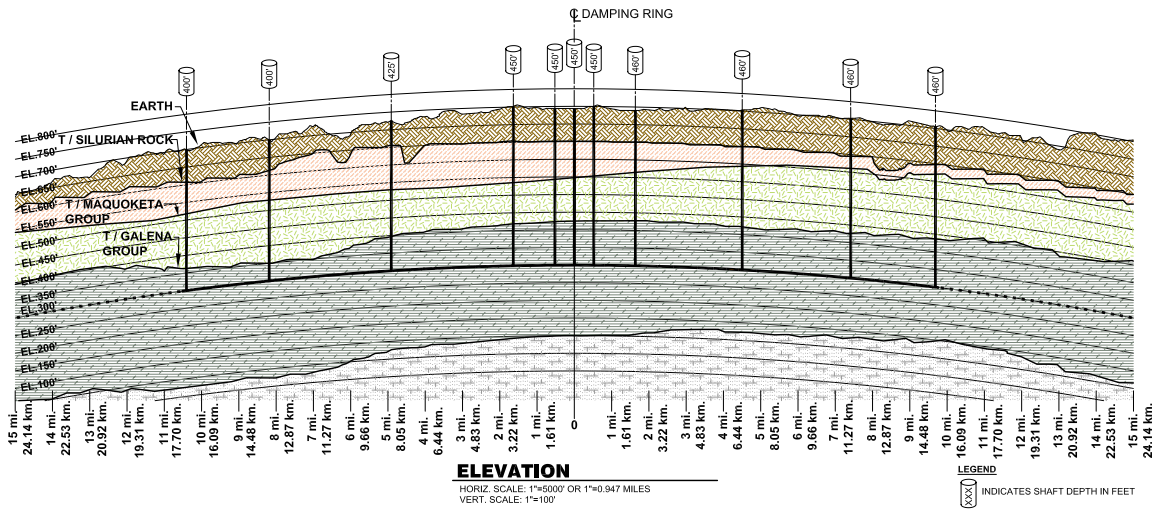}
      \vbabovecaption \caption{Longitudinal profile of the Americas Site in Northern Illinois.}
      \label{fig:SITaev}
   \end{center} \vbbelow
\end{figure}

\clearpage

\setcounter{figlcl}{0}
\setcounter{tablcl}{0}
\setcounter{secnumdepth}{4} \setcounter{subsection}{0}
\setcounter{subsubsection}{0}
\usecounter{subsubsection}
 \setcounter{section}{2}
\section{Asian Site}\label{SITasian}

\subsection{Location}\label{ssect:SITasl}

A set of about 50 criteria have been used over the past decade
to evaluate a large number of ILC candidate sites in Japan.
Out of these candidates, a sample site was selected for the
RDR with an endorsement by the ILC-Asia group at its 4th
meeting in November 2005. It satisfies the following criteria,
some overlapping with the criteria matrix developed by the CFS
Global Group:

\begin{itemize}

\item   Firm and uniform geology to ensure stable beam
           operation at the interaction region. \itemspace
\item   Sufficient length to accommodate straight tunnels spanning over 50 km. \itemspace
\item   Absence of any known, active faults in the neighborhood. \itemspace
\item   Absence of epicenters of any known earthquakes
           exceeding M6 within 50 km from anywhere in the site since AD1500. \itemspace
\item   Uniform altitude of the terrain so that the ILC tunnel
            depth is less than 600 m throughout. \itemspace
\item   Availability of sufficient electrical power for ILC operation. \itemspace
\item   Existence of a practical construction plan for the
           underground tunnels and caverns. \itemspace
\item   Suitable environment, in terms of climate and access,
           for smooth operation. \itemspace

\end{itemize}

The Asian site is located in a moderate plateau area (low mountains)
in uniform solid rock. It is within 10 to 20 km of cities which provide
a living environment with reasonable quality of life. The neighboring cities
are connected to an international airport within several hours by
both bullet train and highway.

\subsection{Land Features }\label{ssect:SITaslf}

The site surface is dominated by woods and is partly
occupied by an agricultural area which is crossed by
occasional local paved roads. Only a few local residences
exist along the tunnel route. There are no major high-ways
or streets with heavy traffic and no large river systems
which cross the tunnel route. Hence, very few sources
of natural or human-made vibrations exist. An adequate
flat surface area is available to accommodate surface facilities.
Existing local roads can be utilized as access routes to
entrances of the tunnel.

\subsection{Climate}\label{ssect:SITasc}

The climate is mild. There is snowfall in winter but
only for a short period. It is not too hot in summer.
There is no recorded history of major typhoons.

\stepcounter{figlcl}\begin{figure}[bhtp]
   \begin{center} \vbabove
      \includegraphics[width=\textwidth]{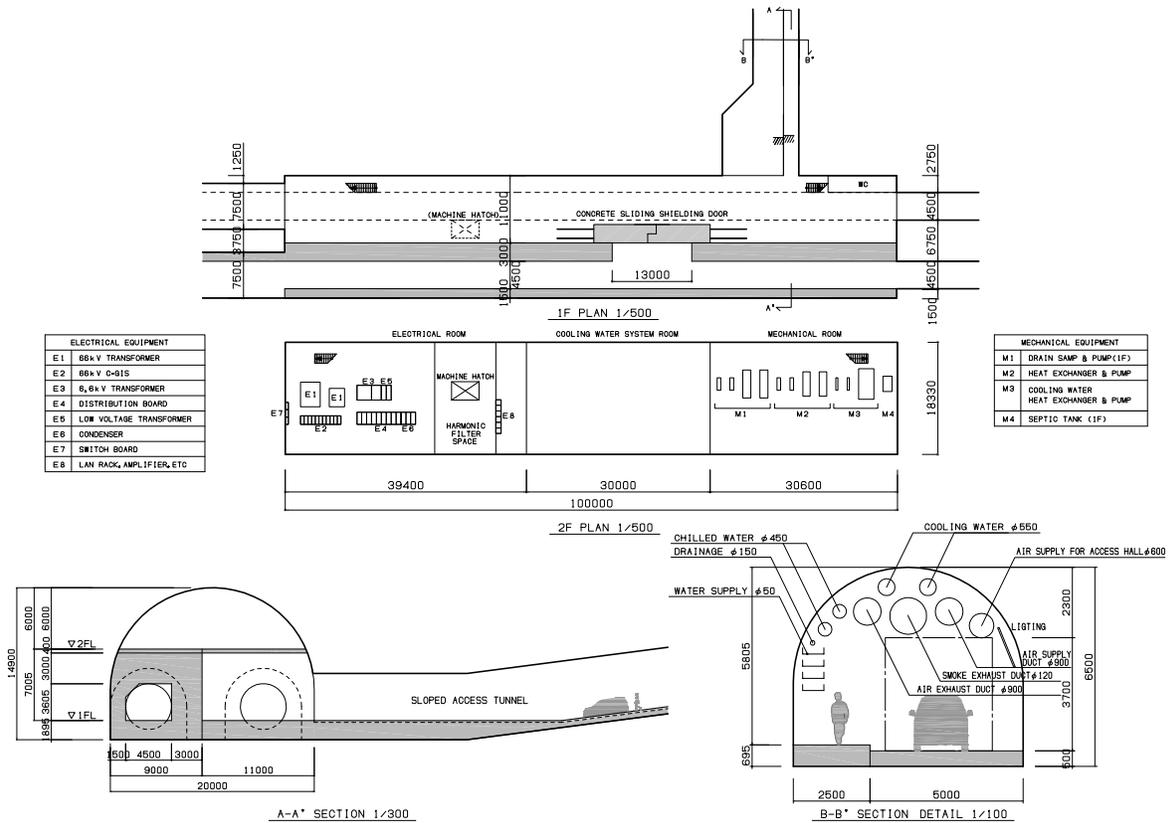}
      \vbabovecaption \caption{Detail of an access ramp for the Asian Sample Site.} \label{fig:SITasar}
   \end{center} \vbbelow
\end{figure}

\subsection{Geology and Tunnel Structure}\label{ssect:SITasgts}

The ~31 km ILC tunnels for the first project phase can
be constructed within solid hard rock. In the second project
phase, when the tunnels are extended to ~50 km, one
side of the main linac tunnel will pass through an area
with sedimentary rock, but this geology is also solid.
The depth of the tunnels, which will be built in a low
mountainous part of the site, is in the range between
40 m and 600 m. Most of the access to the tunnel is
provided by sloped ramps (Figure~\ref{fig:SITasar}). An exception is
the access to the interaction region which has a vertical
shaft approximately 112 m deep.

Past experience with Japanese construction projects
indicates that the uniform granite has sufficient strength
that the tunnels and caverns do not require reinforcement
 by rock bolts or concrete lining. Shotcrete is used to cover
the inner surfaces of the tunnels. Excavation of very large
caverns, such as the experimental hall, may require
reinforcement by rock bolts.

\subsection{Power Distribution System }\label{ssect:SITaspds}

The site is located in the neighborhood of an existing 275 kV power
grid. It is considered to be reasonably straightforward to supply
the power of 240 MW required for the 500 GeV ILC. Power failures in
Japan are very rare, and even if they occur, the system average
interruption duration index (SAIDI) \footnote{System average
interruption duration index = sum of customer interruption durations
normalized by the total number of customers served}  has been only
13 minutes, according to the statistics of the Ministry of Economy,
Trade and Industry of Japan.

\subsection{Construction Methods }\label{ssect:SITascm}

The geology is uniform hard granite below 20 m of softer
topsoil and weathered rock. The access shafts are sloped
tunnels excavated by NATM (New Austrian Tunneling Method),
except for the IR hall. These tunnels match the mountainous
geography and allow vehicle transport of personnel and materials.
They are 7.5 m x 7.0 m to accommodate access for the TBM.
From the surface to a depth of 20 m, the tunnel is reinforced by
rock bolts, a 15-20 cm thick shotcrete liner and steel supports.
In the granite, the tunnel is reinforced by rock bolts and 5 cm thick
shotcrete.

The IR vertical shafts are excavated by drill and blast, with
metal supports and a concrete lining. Caverns are excavated
by NATM. The top of the arch is excavated by advancing top
drift method with drill and blast. Reinforcement is by rock bolt,
pre-stressed bolt and sprayed concrete 20 mm thick with a
metal mesh, overlaid by a 1.5 m thick cast concrete liner on
the arch. The lower part of the cavern is excavated by drill and blast.
After reinforcement in the same method as the top, the side wall
is finished with 1.0 m thick concrete, and the concrete floor cast
2.0 m thick. Passageways are excavated manually and finished
 with sprayed mortar and pre-mixed fiber 20 mm thick.

\stepcounter{figlcl}\begin{figure}[htb]
   \begin{center} \vbabove
      \includegraphics[width=\textwidth]{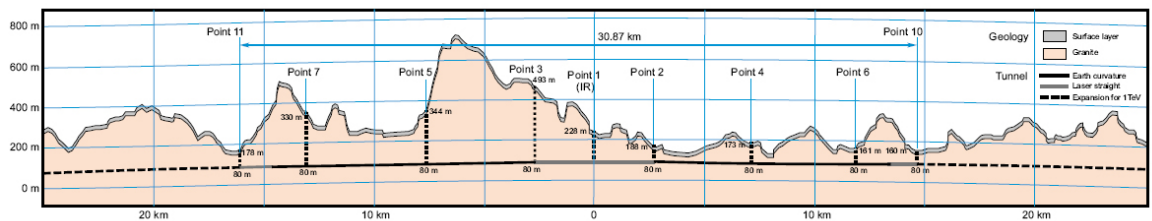}
      \vbabovecaption \caption{Longitudinal profile of the Asian Sample Site in Japan.} \label{fig:SITasev}
   \end{center} \vbbelow
\end{figure}

\clearpage

\setcounter{figlcl}{0}
\setcounter{tablcl}{0}
\setcounter{secnumdepth}{4} \setcounter{subsection}{0}
\setcounter{subsubsection}{0}
\usecounter{subsubsection}
 \setcounter{section}{3}
\section{European Site}\label{sect:SITeurope}

\subsection{Location}\label{ssect:SITeul}

The European site for the ILC is located in the north-western
part of the Geneva region near the existing CERN laboratory.
The area is fairly well populated; the more than 30 km long
path of the accelerator crosses the border between France
and Switzerland three times and passes under several villages.
The region around the accelerator path is mainly covered with
agricultural lands and some forests. There are some biologically
protected zones and historical places or memorials in the area
but the site does not affect national parks.

The proposed site meets all the main requirements of the ILC
Project. Colliders have been in operation in this area for
more than three decades, including the new Large Hadron
Collider (LHC) that will start operating soon. The geological
characteristics allow construction of tunnels for the accelerator
and its support equipment in a stable rock formation with
little seismic activity at a depth of 80 - 110 meters.

CERN and the Geneva area have at their disposal all necessary
infrastructure to accommodate specialists for the period of the
accelerator construction, to store and assemble the equipment,
and to provide for the project-production support during manufacturing
of the special-purpose equipment. Due to the importance of Geneva
as headquarters of many international organizations and to the
existing colliders at CERN, all necessary modern network and
information infrastructure is available.

The international airport of Geneva is situated only 5 km away
from CERN and is served by Swiss Rail and connected to the
European railway network. The highway connecting Switzerland
and France (Northern Europe to Southern Europe) passes nearby.
The access roads to CERN are suitable for all necessary transportation
to deliver the equipment of the accelerator itself and its technical systems.

The governments of France and Switzerland have existing
agreements concerning the support of particle accelerators in
Geneva area, which make it very likely that the land for the
accelerator location could be made available free of charge,
as they did for previous CERN projects.

\subsection{Land Features}\label{ssect:SITeulf}

The proposed location of the accelerator is situated within the
Swiss midlands embedded between the high mountain chains
of the Alps and the lower mountain chain of the Jura. CERN is
situated at the feet of the Jura mountain chain in a plain slightly
inclined towards the lake of Geneva. The surface was shaped by
the Rhone glacier which extended once from the Alps to the valley
of the Rhone. The water of the area flows to the Mediterranean Sea.
The absolute altitude of the surface ranges from 430 to 500 m
with respect to sea level.

\subsection{Climate}\label{ssect:SITeuc}

The climate is warm-continental. The mean temperature of
the air of the coldest month (January) is -0.2$^{\circ}$C. The mean
temperature of the air of the warmest month (July) is +18.4$^{\circ}$C.
The mean annual rainfall is 928 mm. Snow usually falls in
the months of December to February. On the whole, the
climate in the vicinity of Geneva is considered to be quite comfortable.

\subsection{Geology}\label{ssect:SITeug}

Most of the proposed path of the ILC
is situated within the Molasse, an impermeable sedimentary
rock of the Swiss midlands laying over the Jurassic Bedrock.
The path crosses a fault at the valley of the Allondon river which is
situated South-West
of Geneva and filled with sands and gravels. In this valley, the tunnels
are built below the groundwater level. For the 1 TeV extension of the project, the tunnel will cross a second valley at Gland, situated North-East of Geneva, and will just enter some Jurassic limestone.

\stepcounter{figlcl}\begin{figure}[htb]
   \begin{center} \vbabove
      \includegraphics[width=\textwidth]{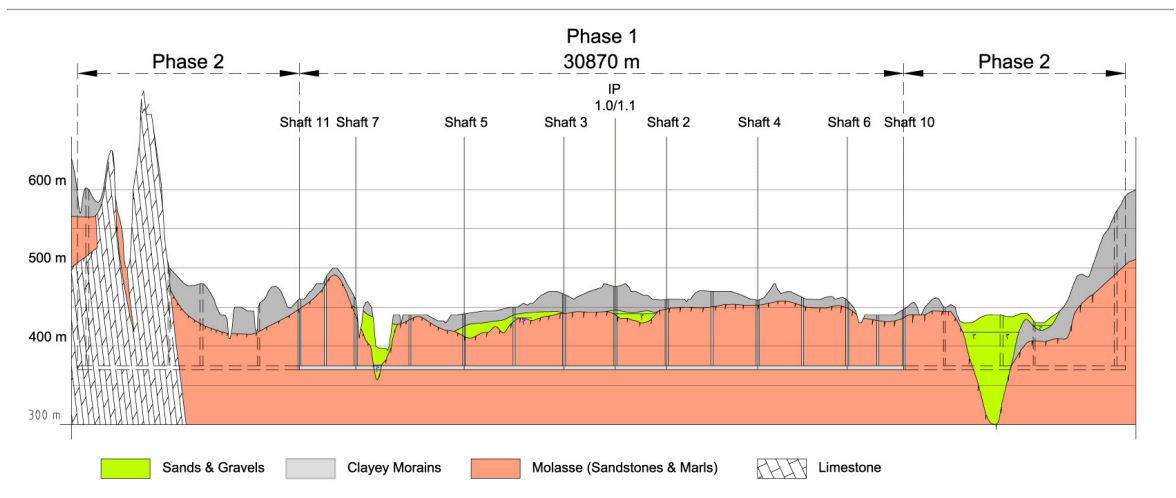}
      \vbabovecaption \caption{ Longitudinal profile of the European Sample Site near CERN.} \label{fig:SITeuev}
   \end{center} \vbbelow
\end{figure}

The alignment of the ILC accelerator is placed at a level of 370 m
in the Molasse (Figure~\ref{fig:SITeuev}). This makes it possible to excavate the
tunnels with shielded tunnel boring machines (TBM-S) with a high
penetration rate and simultaneous placement of precast concrete
segments. For the crossing of the Allondon and Gland valleys, the shielded
tunnel boring machines must be replaced by hydro mix-shield
machines (SM) able to tunnel in "closed mode" through the sands and gravels
below groundwater level and to work in "open mode" as a normal
tunnel boring machine in the Molasse.

\subsection{Power Distribution System}\label{ssect:SITeupds}

The European sample site provides sufficient electrical power
for the accelerator complex. A nearby 400 kV substation of the
French grid will serve as connection point. The availability of the
network is considered adequate for the LHC and is thus likely
to also be sufficient for the ILC.

\subsection{Construction Methods}\label{ssect:SITeucm}

The upper parts of the shafts lie in dry moraines, with
total thickness ranging from 0 to 50 m, depending on the situation.
Traditional means can be used to excavate down to sound rock, except
in water bearing sands and gravels where it will be necessary to use other techniques such as diaphragm walling to allow safe excavation of the shafts.
Once in the rock (sandstone) the shafts and
caverns are excavated with the use of rock breakers and road
headers, with blasting as a possible exceptional resort. After the temporary
lining (rock bolts, mesh and shotcrete) is in place, the walls
and vaults are sealed with waterproof membranes and covered
with cast in-situ reinforced concrete.

Shielded Tunnel Boring Machines (TBM-S) with a prefabricated
concrete segment lining are used for the long tunnels. An
average daily advance of 25 m/day is assumed. The concrete
tunnel floors are then cast in-situ. Short tunnel sections (less than 300 m)
and passageways are excavated with road headers or small
size rock breakers, then shotcreted. The penetrations between tunnels are excavated with small diameter boring machines, anchored in one of the two tunnels.

\clearpage

\setcounter{figlcl}{0}
\setcounter{tablcl}{0}
\setcounter{secnumdepth}{4} \setcounter{subsection}{0}
\setcounter{subsubsection}{0}
\usecounter{subsubsection}
 \setcounter{section}{4}

\section{Summary}\label{sect:SITsummary}

Although the three sample sites have differences, they all meet the ILC design requirements and at comparable cost. Table \ref{tab:sitesummary} compares some of the salient features.

\stepcounter{tablcl}\begin{table} [h] \vbabove
   \caption{Summary of notable features of the sample sites and
     construction methodology.}
   \label{tab:sitesummary}
\begin{center} \begin{flushleft}
\setlength{\tabcolsep}{3pt}
   \begin{tabular}{| l | l | l | l |} \hline
     ~Subject & Americas Region & Asian Region & European Region \\ \hline & & & \vbdlspacing \hline
     ~Sample site & Northern Illinois -- & Japan & Geneva Area -- \\ [-6pt]
     ~location & near FNAL. &  & near CERN \\ \hline
     ~Land features & $200\sim240$m  & $120\sim680$ m  & $430\sim480$ m  \\ [-6pt]
                         &  above sea level &  above sea level &  above sea level \\ \hline
     ~Geology & Dolomite & Granite (sedimentary  & Molasse (sedimentary  \\  [-6pt]
                  &               & rock in phase-2 extension) & rock / sandstone) \\  \hline
     ~Tunnel depth  & $100\sim150$m & $40\sim600$ m & $95\sim145$m  \\  [-6pt]
     ~from surface   &                        &                        &  (except 1 valley 30 m) \\  \hline
     ~Access paths & 13 shafts & 10 sloped tunnels (7.5m & 13 shafts  \\ [-6pt]
     ~to underground  & 9m, 14m, 16m diam &  $\times$ 7m $\times$  $700\sim2000$m)  & 9m, 14m, 16m diam \\ [-6pt]
     ~caverns &  $100\sim135$ m deep & and 3 shafts (for IR) &  $100\sim135 $m deep \\ \hline
     ~Tunnel & TBM & TBM & TBM \\ [-6pt]
     ~construction & & & \\ \hline
     ~Tunnel lining & 20\% of length  & 100\% of length & 100\% of length precast \\ [-6pt]
                      &  shotcreted         &  shotcreted      & concrete segments \\ \hline
     ~Average tunnel  & 30m/day/TBM & 16m/day/TBM  & 25m/day/TBM  \\ [-6pt]
     ~excavation speed & (boring) &  (boring + surface work) &  (boring) \\ \hline
     ~Number  & 9 & 15 (6 out of 9 accesses  & 9 \\  [-6pt]
      ~of TBMs &  & have two TBMs starting  &  \\  [-6pt]
      &  & in opposite directions) &  \\  \hline
     ~Cavern  & Drill and blast & Drill and blast& Road breaker  \\ [-6pt]
     ~construction &  & (NATM) &  /header \\ \hline
     ~Shaft  & Earth excavation  & Drill and blast  & Road breaker/header  \\ [-6pt]
    ~construction &  / Drill and blast &  (step by step method) & (Moroccan method) \\ \hline
     ~New surface & 92 & 166 & 120 \\ [-6pt]
     ~buildings & & & \\ \hline
     ~Distribution & 69/34 kV & 66/6.6kV & 36kV \\ [-6pt]
     ~voltage & & & \\ \hline
   \end{tabular}
\end{flushleft}
\end{center} \vbbelow
\end{table}

\setcounter{chapter}{5}

\chapter{\textsf{Value Estimate }\label{chapCS}}

\setcounter{figlcl}{0}
\setcounter{tablcl}{0}
\setcounter{secnumdepth}{4} \setcounter{subsection}{0}
\setcounter{subsubsection}{0}
\usecounter{subsubsection}

\setcounter{section}{0} \renewcommand{\picturefolder}{./costs/}
\section{Value Estimating Methodology}\label{sect:VALvem}

\subsection{Introduction}

The ILC is an international scientific project to be funded by a
collaboration of countries or regions around the world, each of which
have different traditions and conventions for planning and estimating
the cost of large projects. In order to equitably divide up
contributions among the collaborators, one must develop a project
estimate that is independent of any particular accounting system but
compatible with all of them. The ``VALUE'' methodology has become the
standard for such international projects. It was adopted by ITER
(the international thermonuclear experimental reactor project) and by the
LHC experiments, among others. Value is a particularly convenient
concept for dealing with ``in-kind'' contributions, for which
manufacturing costs and labor rates can vary widely between
collaborators. Conversion of the value estimate to various national
costing practices can only be done by each
participating nation.

The ILC estimate consists of two important parts: VALUE (in terms of
currency units) for items provided and LABOR (in terms of person-hours
or person-years), which may be provided by the collaborating
laboratories and institutions, or may be purchased from industrial
firms. This is similar to what has been traditionally used for
European project proposals. The value of a component is defined as the
lowest reasonable estimate of the procurement cost in adequate
quality, based on production costs in a major industrial nation.
It is expressed in
2007 currency units (not escalated to the years in which the funds are
projected to be spent) and does not include R\&D, pre- or
post-construction or operating costs, taxes or
contingency. It is effectively the barest cost estimate that would be
used by any of the funding agencies. Individual regions can then add
to the base value any other items usually included in their own
estimating system.

In this context, LABOR is defined as ``explicit'' labor, which may be
provided by the collaborating laboratories and institutions, or may be
purchased from industrial firms.  This to be distinguished from a
company's ``implicit'' labor associated with the industrial
production of components and contained (hidden) within the purchase
price.  The implicit labor is included in the VALUE part of this
estimate.

The ILC VALUE plus LABOR estimate is the basis on which contributions
are apportioned among the collaborators. Each participant makes an
agreement with the ILC management to provide a certain value of
components and services. They are then responsible for providing the
contracted items, independently of what they actually cost.

\subsection{Scope of Estimate}

The estimate is for a 500 GeV center-of-mass machine but includes some
items sized for 1 TeV to enable a later energy upgrade, such as
the beam dumps and the length of the Beam Delivery tunnel. The ILC
estimate does not include the cost of the detectors. They are assumed
to be funded by a separate agreement between the collaborating
institutes, in the way the LEP and LHC detectors were built. The
estimate does include civil engineering work for the detectors,
e.g. assembly buildings, underground experimental halls, shafts, etc.

The estimate covers all aspects of construction, including tooling-up
industry, final engineering designs and construction management. The
estimate specifically does not include costs for any of the
engineering, design, or preparation activities that can be
accomplished before construction start. It does not include Research and
Development, proof-of-principle or prototype systems tests,
pre-construction (e.g. architectural engineering, conceptual and
construction drawings, component and system designs and preparation of
bid packages), commissioning, operation, decommissioning, land or
underground easement acquisition costs. It also does not include items
which are treated differently from region to region such as
taxes, escalation, or contingency. Table~\ref{tab:inout} summarizes the
items that are included in, or excluded from, the value and labor
estimate.

The estimate assumes a seven-year construction period. The estimate
for a given item covers the cost from the day the project obtains
funding until that item is installed, tested, and ready for
commissioning. Commissioning in one area may overlap with construction
elsewhere. The construction period ends when the last component has
been installed and tested.

\stepcounter{tablcl}\begin{table} \vbabove
   \caption{Summary of the items that are included in, or excluded
     from  the value and labor estimate.}
   \label{tab:inout}
\begin{center}
   \begin{tabular}{| p{60mm} | p{60mm} |} \hline
   Included & Excluded \\ \hline & \vbdlspacing \hline
    Construction of a 500 GeV machine,
   including items sized to enable a later energy upgrade &
   \\ \hline
   Tooling-up industry, final
   engineering designs and construction management &
   Engineering, design, or preparation activities that
   can be accomplished before construction start, such as,
   research \& development, proof-of-principle or prototype
   systems tests, pre-construction  \\ \hline
Construction of all conventional facilities, including the tunnels, surface buildings, access shafts and others &
Surface land acquisition or underground easement acquisition costs \\ \hline
Construction of the detector assembly building, underground experimental halls and detector access shafts &
Experimental detectors \\ \hline
 & Commissioning, operations, decommissioning \\ \hline
Explicit labor, including that for management and administrative personnel. &
Taxes, contingency and escalation \\
\hline
   \end{tabular}
\end{center} \vbbelow
\end{table}

\subsection{Estimating Approach}

The ILC estimate was developed by the RDR matrix of Area, Technical
and Global System leaders working with the Cost Engineers. The Area
Systems Leaders (AS) defined the requirements for their accelerator
systems. The Technical (TS) and Global System (GS) Leaders provided
the estimated value and explicit labor per component unit. Specialized
components such as the polarized electron gun were estimated by the
Area Systems themselves. The AS leaders then compiled the estimate for
their areas. The estimates were iterated to optimize cost and
performance.

The cost estimates were prepared using a Work Breakdown Structure
(WBS) where each item included a description, basis of estimate,
quantity required, materials and services estimate and implicit and
explicit labor. These could then be summed to produce to an estimated
total cost for the component, system, or section of the machine. There
were 351 active WBS elements, where each element represented a roll-up
of further detailed estimating information provided by the systems
leaders.  An example of the lower level of detail for one of these
WBS elements provided by the
Conventional Facilities and Siting group is presented in the Appendix.

\subsubsection{General Guidelines}

The ILC estimate is given as the sum of VALUE (in currency units) and
explicit LABOR (in person-hours).

Guidelines and Instructions for performing, preparing, and presenting
the cost estimate are available at
%\cite{costguide1,costguide2}. include links explicitly

http://www-ilcdcb.fnal.gov/RDR\_costing\_guidelines.pdf

http://www-ilcdcb.fnal.gov/RDR\_Cost\_Estimating\_Instructions\_23may06.pdf

Estimates are quoted as median or 50\%-50\% estimates, where, if a
given item were to be independently purchased many times, taking the
lowest world-wide bid each time, half of the purchases would be below
the median estimate and half above.

\subsubsection{Currency Rates and Raw Materials}

Component estimates from all three regions were converted to a common cost
basis, the ILC Unit, where one ILC Unit is set equal to
\$1 U.S. (January 2007 value) The conversion rates used were:

\begin{center}
1 ILC Unit = 1 US 2007\$ (= 0.83 Euro = 117 Yen)
\end{center}

These currency
exchange rates are an average of the exchange rates over the five years
2003 through 2007.  The value estimates were developed during 2006 and
then adjusted to January, 2007 using the official regional cost
escalation indices.

Electricity and raw materials such as niobium, steel or copper are
assigned fixed prices as of January 1, 2007, as summarized in
Table~\ref{tab:rawmaterial}.

\stepcounter{tablcl}\begin{table}[htb!] \vbabove \caption{Assumed
prices for electricity and representative raw materials.}
\label{tab:rawmaterial}
\begin{center}
\begin{tabular}{| l | p{80mm} |} \hline
  Resource & Jan 1, 2007 Price \\ \hline & \vbdlspacing \hline
  Electricity: & \$0.10 per kWh (including supply cost) \\ \hline
  Copper: & \$8 per kilogram \\ \hline
  Black steel: & \$0.6  per kilogram
  (up to three times higher price for stainless and magnet steel) \\ \hline
  Niobium: & \$70 per kilogram \\  \hline
\end{tabular}
\end{center} \vbbelow
\end{table}

\subsubsection{Contingency and Risk}

The ILC estimate does not contain contingency. Contingency is
a quantitative measure of risk -- the final number is set higher than
the initial estimate to allow for unexpected or uncontrollable factors
that may raise the ultimate price. The ILC project will avoid any future
cost increases through R\&D, industrial studies,
vendor pre-qualifications, and competitive, global calls for tender.
The level of uncertainty in the current estimate is summarized below.
\begin{comment}
note to editors:  this just points toward wherever Ewan's Technical Risk writeup goes, either in this cost chapter or in the Engineering Design chapter.  I'll write this as if in the ED chapter.
\end{comment}
A preliminary technical risk register has been compiled and is discussed in section~\ref{ssect:EDRrand} on Critical R\&D in the Engineering Design Report Phase.

\subsection{Component Estimates}

Three different classes of items were identified and approached differently.

\begin{itemize}

   \item {\it Site specific}: The costs for many aspects of
     conventional facilities are site specific and there are separate
     estimates for sample sites in all three regions: Asia, Europe,
     and the Americas. These costs are driven by real considerations,
     e.g. different geology and landscape, availability of electrical
     power and cooling water, etc. Site dependent costs due to
     formalities (such as local codes and ordinances) are not
     included. Common items such as internal power distribution, water
     and air handling, which are essentially identical across regions
     although the implementation details differ, have a single
     estimate. The sample sites have different
     geologies. Nevertheless, they use similar tunnel-boring machine
     technologies and the value estimates are very close.
     Because a site has not yet been chosen, the ILC
     value estimate is taken as the average of the three site-dependent
     estimates. Individual estimates for each of the three sites
     are also provided. \itemspace

   \item {\it High technology}: Items such as cavities, cryomodules,
     and rf power sources, where there is interest in developing
     expertise in all three regions, have been estimated separately for
     manufacture by each region. Costs are provided for the total
     number of components along with parameters to specify the cost of
     less than the total number. The European estimate for the
     cavities and cryomodules is used for the ILC value as it is the
     most mature, in terms of R\&D and industrial studies. Estimates
     from the other regions provide a crosscheck. \itemspace

   \item {\it Conventional}: Estimates for components, such as
     conventional magnets and controls, which can be produced by many
     manufacturers in all regions, are based on a world-wide call for
     tender. \itemspace

\begin{comment}
note to editors:  the Cost Engineers would like to see the next sentence on production in a major industrial nation removed
They are the lowest reasonable cost available world-wide,
     for production in a major industrial nation.
\end{comment}
\end{itemize}

Component estimates include the manufacturer's implicit labor, EDIA
(engineering, design, inspection, and administration), quality
control/assurance, and technical testing. A single supplier is assumed
to be responsible for one deliverable, even though in practice,
multiple suppliers may be chosen to reduce risk. The estimates quoted
for mass-produced technical systems were generated either by detailed
bottom-up industrial studies for the quantities required, or by
assuming a learning curve explicitly in an in-house engineering
estimate.  The basis of estimate and cost estimating methodology for
each set of components are discussed in the individual Area System,
Global System, and Technical System sections for this report.

\subsection{Explicit Labor}

Explicit labor is estimated separately from component costs, and is
given in person-hours. It may be provided by the ILC collaborators as
in-kind contributions, drawn from existing laboratories with their own
personnel and budgets, or may be purchased from industrial firms.  To
convert person-hours to person-years, it was assumed that laboratory
staff works an average of 1,700 hours per year. Only three classes of
manpower are used: engineer/scientist, technical staff, and
administrative staff.

%------------------------

\clearpage %\setcounter{page}{1}
\setcounter{figlcl}{0}
\setcounter{tablcl}{0}
\setcounter{secnumdepth}{4} \setcounter{subsection}{0}
\setcounter{subsubsection}{0}
\usecounter{subsubsection}
 
\section{Estimate for Construction of ILC}

\subsection{Value Estimate}

The value and explicit labor estimates are current as of February 1, 2007, and   will be updated in the final report. The preliminary value estimate presented here is for the cost of the ILC in its present design and at the present level of engineering and industrialization. The estimate contains three elements:
\begin{itemize}

\item 1.83 Billion  (ILC Units) for site-dependent costs, such as the costs for tunneling in a specific region \itemspace
\item 4.79 Billion (ILC Units) for shared value of the high technology and conventional components \itemspace
\item 14,200 person-years for the required supporting manpower (=24 million person-hours) \itemspace
\end{itemize}

For this value estimate:  1 ILC Unit = 1 US 2007\$ (= 0.83 Euro = 117 Yen)
\medskip

A common estimate was used
for all
non site-specific technical components, regardless of region.
The three regional site-specific estimates were based on local costs for civil
engineering and the primary high voltage electrical power connections,
feeds, substations and primary cooling water systems.  All three site-dependent estimates
are within a few percent of the average.

There are many possible models for dividing the responsibilities among
the collaborating regions. The numbers below present one possible
model where the estimates are divided into site-specific and shared
parts.  In this model, the host region is expected to provide the
site-specific parts, because of the size, complexity, and specific
nature of these elements.  The site-specific elements include all the
civil engineering (tunnels, shafts, underground halls and caverns,
surface buildings, and site development work); the primary high-voltage electrical power equipment, main substations, medium voltage
distribution, and transmission lines; and the primary water cooling
towers, primary pumping stations, and piping.  Responsibilities for
the other parts of the conventional facilities: low-voltage electrical
power distribution, emergency power, communications, HVAC, plumbing,
fire suppression, secondary water-cooling systems, elevators, cranes,
hoists, safety systems, and survey and alignment, along with the other
technical components, could be shared between the host and non-host
regions.  Such a model may be summarized as shown in
Table~\ref{tab:division}.

\stepcounter{tablcl}\begin{table}[htb!] \vbabove \caption{Possible
division of responsibilities for the 3 sample sites (ILC Units).}
\label{tab:division}
\begin{center}
\begin{tabular}{| ll | c | c | c |}\hline
    & Region   & Site-Specific & Shared    & Total \\ \hline & & & & \vbdlspacing \hline
    & Asia     &  1.75 B     &  4.78 B &  6.53 B \\ \hline
    & Americas   &  1.89 B     &  4.79 B &  6.68 B \\ \hline
    & Europe   &  1.85 B     &  4.79 B &  6.64 B \\ \hline
and & Average    &  1.83 B     &  4.79 B &  6.62 B \\ \hline\hline
\multicolumn{5}{| l |}{~~~~plus 14 K person-years of explicit labor}  \\
\multicolumn{5}{| l |}{~~~~or 24 M person-hours \@ 1,700 hours/year}  \\ \hline
\end{tabular}
\end{center} \vbbelow
\end{table}

\begin{figure}[htb!]
\begin{center} \vbabove
   \includegraphics[width=\textwidth]{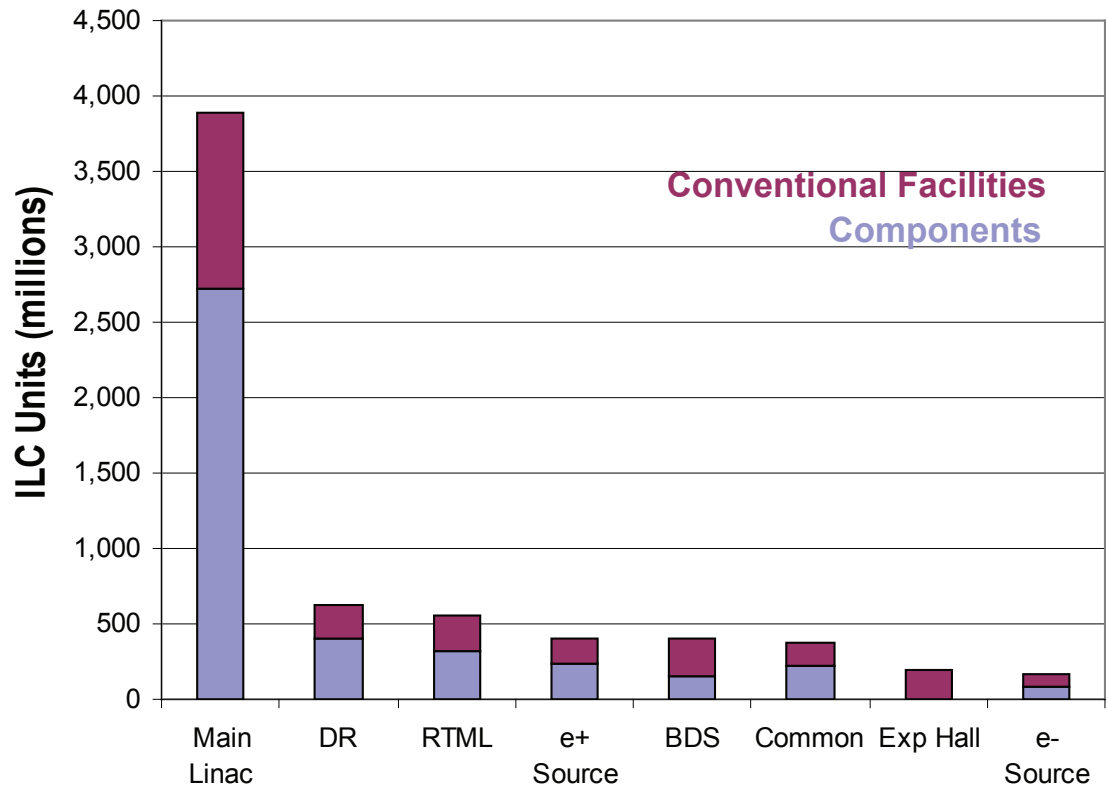}
\end{center}
\vbabovecaption \caption[Distribution of the ILC value estimate by area system and common infrastructure.]{Distribution of the ILC value estimate by area system and common infrastructure, in ILC Units.  The estimate for the experimental detectors for
  particle physics is not included.  (The Conventional Facilities estimates have been averaged over the three regional site estimates. )}
\label{fig:values}
\end{figure}

The value estimates broken down by Area System are shown separately
for both the conventional facilities and the components in
Figure~\ref{fig:values} and Table~\ref{tab:area}. Common refers to
infrastructure elements such as computing infrastructure, high-voltage
transmission lines and main substation, common control system, general
installation equipment, site-wide alignment monuments, temporary
construction utilities, soil borings and site characterization, safety
systems and communications.

The component value estimates for each of the Area (Accelerator)
Systems include their respective RF sources and cryomodules,
cryogenics, magnets and power supplies, vacuum system, beam stops
and collimators, controls, Low Level RF, instrumentation,
installation, etc.. The superconducting RF components represent
about 69\% of the estimate for all non-CF\&S components.

\stepcounter{figlcl} \vbbelow

\stepcounter{tablcl}\begin{table}[htb!] \vbabove
\caption[Distribution of the ILC value estimate by area system and
common infrastructure.]{Distribution of the ILC Value Estimate by
area system and common infrastructure, in ILC Units.  The estimate
for the experimental detectors for
  particle physics is not included.  (The Conventional Facilities estimates have been averaged over the three regional site estimates. ) }
\label{tab:area}
\begin{center}
\begin{tabular}{| l | c | c | c |} \hline
             &       &            & Conventional \\ [-6pt]
Area - M ILC Units & Total & Components & Facilities \\ \hline & & &
\vbdlspacing \hline%
 Main Linac   &   3,894 & 2,723 & 1,172 \\ \hline%
DR           &     630 &   398 &   231 \\ \hline %
RTML & 554 &   320 &   234 \\ \hline%
e$^+$ source &     398 &   232& 166\\ \hline%
 BDS          &     408 &   157 &   252 \\ \hline%
  Common&     369 &   229 &    140 \\ \hline %
  Exp Hall     & 200 &     0 & 200 \\ \hline%
  e$^-$ source &     165 &    87 &    78\\ \hline \hline%
Sum          &   6,618 & 4,146 & 2,472 \\ \hline%
\end{tabular}
\end{center} \vbbelow
\end{table}
\begin{comment}
14june07:  Wilhelm proposed a compromise removing "probably" from …estimate is probably no larger than +25\% above...  This is acceptable to me, so this now goes to editor Nan.
\end{comment}

Initial cursory analysis of the uncertainties in the individual estimates from the Technical Systems indicates that the RMS for the current RDR value estimate for the presented baseline design is likely to be in the $\sigma = \pm10-15\%$ range, and that the $95^{th}$ percentile for this estimate is no larger than +25\% above the mean.

\subsection{Explicit Labor Estimate}

The explicit labor for the Global Systems, Technical Systems, and
specific specialty items for Electron Source, Positron Source, Damping
Rings, and Ring to Main Linac, include the scientific, engineering,
and technical staff needed to plan, execute, and manage those elements
including specification, design, procurement oversight, vendor
liaison, quality assurance, acceptance testing, integration,
installation oversight, and preliminary check-out of the installed
systems.

\stepcounter{figlcl}\begin{figure}[b!]
\begin{center} \vbabove
 \includegraphics[width=\textwidth]{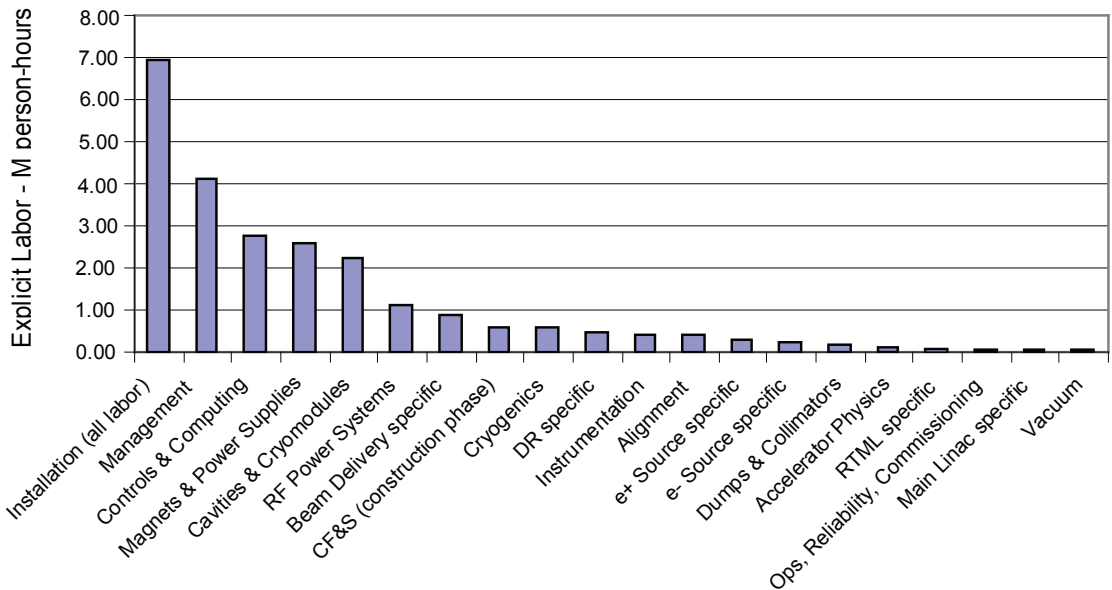}
\end{center}
\vbabovecaption \caption[Explicit labor.] { Explicit labor, which may be supplied by collaborating
  laboratories or institutions, listed by Global, Technical, and some
  Area-specific Systems.}
\label{fig:labor}\vbbelow
\end{figure}

\stepcounter{tablcl}\begin{table}[htb!] \vbabove \caption[Explicit
labor.]{Explicit labor, which may be supplied by collaborating
  laboratories or institutions, listed by Global, Technical, and some
  Area-specific Systems.}
\label{tab:labor}
\begin{center}
\begin{tabular}{| l | c |} \hline
Explicit labor &  M person-hours \\ \hline & \vbdlspacing \hline
Installation (all labor)      & 6.91 \\   \hline
Management                    & 4.09 \\  \hline
Controls \& computing         & 2.76 \\  \hline
Magnets \& Power Supplies    & 2.60 \\ \hline
Cavities \& cryomodules       & 2.23 \\  \hline
RF Power Systems              & 1.12 \\  \hline
Beam Delivery System specific & 0.89 \\ \hline
CF\&S (construction phase)   & 0.60 \\ \hline
Cryogenics                    & 0.56 \\  \hline
DR specific                   & 0.46 \\  \hline
Instrumentation               & 0.44 \\  \hline
Alignment                & 0.42 \\ \hline
$e^+$ Source specific         & 0.31 \\  \hline
$e^-$ Source specific        & 0.24 \\ \hline
Dumps \& Collimators         & 0.19 \\ \hline
Accelerator Physics      & 0.11 \\ \hline
RTML specific            & 0.09 \\ \hline
Ops, Reliability, Commissioning & 0.07 \\ \hline
Main Linac specific      & 0.06 \\ \hline
Vacuum                        & 0.05 \\  \hline
Sum                           & 24.19 \\ \hline
\end{tabular}
\end{center} \vbbelow
\end{table}

Installation is the largest fraction of explicit labor, about
29\%. Management is the second largest fraction at about 17\%.
At this stage of the ILC design, it is too early for a complete analysis of installation requirements. Instead, the RDR estimate was based on scaled information from a variety of sources, including the actual manpower used for the installation of recent accelerator projects. There was also a bottoms-up study for installation of the cryomodules for the Main Linac done by two separate engineering teams, with comparable results. The estimates were reviewed by experts and crosschecked for reasonability.

In the present estimate, the installation task is characterized almost
exclusively as explicit labor, with minimum costs for
material-handling equipment. This is on the assumption that much of the
installation and system check-out labor at the ILC site can be
contributed by the staffs of collaborating institutions or
laboratories. The validity of this assumption depends on the
availability of the necessary skilled manpower and local labor
regulations. Because of the size of the project, it is likely that
many tasks like electrical and plumbing work will need to be
outsourced to industry. Trade-offs and translations are likely between
using in-house labor and external contracts. It is estimated that a
minimum of 10\% of the installation task must be management and
supervision by in-house manpower.

The management model is similar to that of the construction phase of the
Superconducting Super Collider (SSC), but without central computing staff
which are included elsewhere. The  management personnel is estimated to be half as large as in the SSC model.  The ILC staff
consists of 345 persons, divided as shown in
Table~\ref{tab:management}.
Personnel for the Area, Global, and Technical System groups are not
included in the Project Management Division.

\stepcounter{tablcl}\begin{table}[htb!] \vbabove
\caption{Composition of the management structure at ILC.}
\label{tab:management}
\begin{center}
\begin{tabular}{| l | p{70mm} |} \hline
Unit & Responsibilities \\ \hline & \vbdlspacing \hline
Directorate (30): & Director's Office, Planning, ES\&H Oversight, Legal, External Affairs, Education, International Coordination, Technology Transfer; \\ \hline
Management Division (13): & Quality Assurance, ES\&H; \\ \hline
Laboratory Technical Services (125): & Facilities Services, Engineering Support,
Material and Logistical Services, Laboratory Fabrication Shops, Staff Services; \\ \hline
Administrative Services (94): & Personnel, Finance, Procurement, Minority Affairs; \\ \hline
Project Management Division (83): &  Management,  Administrative,
Project Management Division Office. \\ \hline
\end{tabular}
\end{center} \vbbelow
\end{table}

This explicit labor estimate is very preliminary.  Producing a more realistic explicit labor estimate will be a priority in the Engineering Design phase.

It is the practice in some regions to apply general and administrative overheads
to purchases and labor for projects. These overheads are applied as a multiplier on the underlying LABOR and VALUE, and cover the costs of the
behind-the-scenes support personnel. In this estimate, such personnel are explicitly enumerated as labor under Directorate, Management Division, Laboratory
Technical Service, and Administrative Services in Table~\ref{tab:management}.  Therefore, the overheads are included as additional explicit
LABOR, rather than as a multiplier on VALUE.

This explicit labor corresponds to 35\% scientists and engineers, 14\% administrative personnel, 27\% technical staff, and 24\% installation staff which could be either institutional or laboratory labor or contract labor or some combination.

\subsection{Operating Cost}

Operating costs are not included in the estimate for the construction
project, but a very preliminary estimate is given. It is also to be
noted that spare components (those stored in warehouses and not the
installed redundant components), although fabricated along with the
installed components, are assumed to be financed through operating
funds, and are not considered part of the construction projects.
Major factors in the operating cost include personnel costs,
electrical power, maintenance and repairs, helium and nitrogen
consumables, and components that have a limited life expectancy and
need continuous replacement or refurbishment, like klystrons.

The model assumes 9 months of machine operations per year at full
power of about 227~MW, corresponding to 500~GeV at design luminosity, plus 3
months standby at reduced power (25~MW) with the superconducting
cavities maintained at 4.5K, which is above their operating
temperature. At the current electrical power rate of \$0.1 per
kW-hr, the operating costs for these materials and services are
estimated to be approximately 150-270~M\$ per year in 2007 Dollars.
The continuing operations and
administrative staff is expected to be comparable to that at existing facilities
(not including support of the scientific
program).

Commissioning activities and operating costs are anticipated to
gradually increase over the fourth through seventh years of construction from zero up to the full level of long-term operations at
the end of the 7 year construction phase.

%%%------------------------------

\clearpage %\setcounter{page}{1}
\setcounter{figlcl}{0}
\setcounter{tablcl}{0}
\setcounter{secnumdepth}{4} \setcounter{subsection}{0}
\setcounter{subsubsection}{0}
\usecounter{subsubsection}

\section{Schedule}\label{ConstSchedule}

\subsection{Example Construction Schedule}

A detailed schedule for realization of the ILC depends on a variety of
factors and milestones including: completion of crucial R\&D,
completion and review of the conceptual design of the machine (RDR)
and detectors, and endorsement of the RDR and cost by international
funding agencies so that critical R\&D can be funded to
completion. Site specific engineering and civil designs require
international agreements on site selection which allow land
acquisition, environmental assessments, etc. In addition agreements on
cost sharing and spending profiles are required to plan the industrial
production of components and the preparation of construction
contracts.

In the absence of much of this information an attempt was made to
construct a technically limited schedule for the construction of ILC
assuming that these items have been completed prior to a physical
construction start (T0).
The RDR cost estimate has been based on this schedule.

\subsection{Conventional Facilities Schedule}

Conventional facilities include Civil Engineering of above and below
ground structures, electrical infrastructure, cooling and ventilation,
and buildings.  In what follows, an assumption was made that a site
was selected several years prior to the start of construction and that
funding was available such that Architectural \& Engineering (A\&E) firms can be retained to design
the conventional facilities and prepare bid packages prior to the
start of construction.  In the absence of other financial constraints,
the construction schedule for conventional facilities is dominated by
the required underground construction.  ILC requires about 72 km of
underground tunnel construction for the Main Linac, Beam Delivery and
Damping Ring systems.  For the purpose of this section it was assumed
that the tunnel is deep, at a depth of $\sim$100 m and located in dry rock
such that standard tunnel boring machines can be employed. (Several
possible ILC sites have different local underground
conditions. However, these are believed to alter the conclusions in
the section in only a minor fashion).

\stepcounter{figlcl}\begin{figure}[htb!]
\begin{center} \vbabove
   \includegraphics[width=\textwidth]{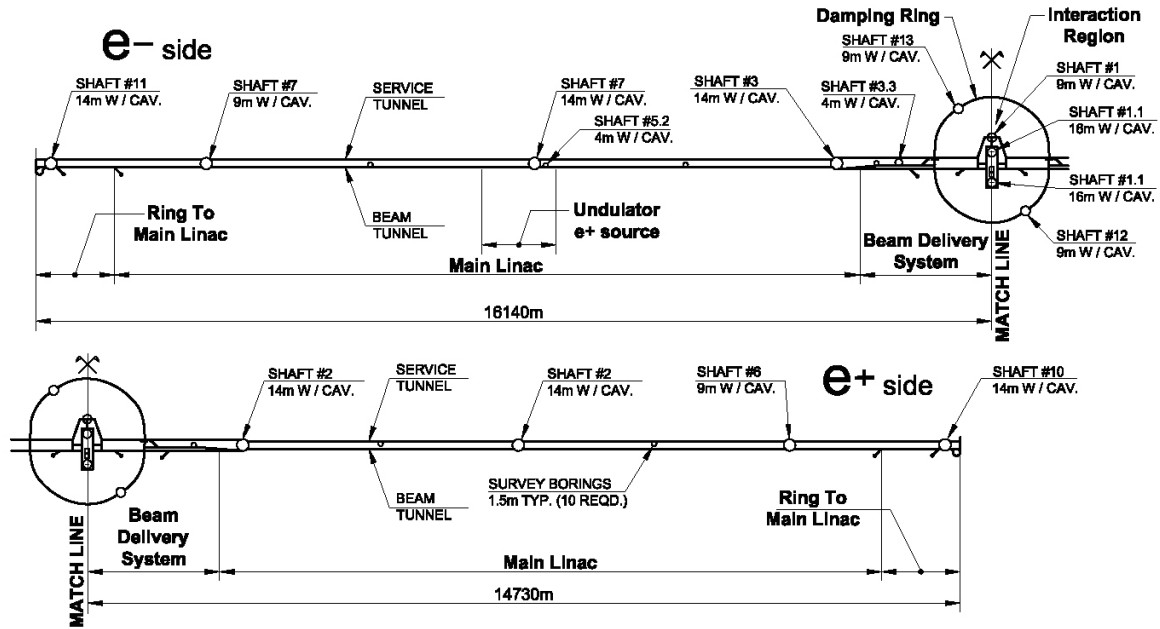}
\end{center}
\vbabovecaption \caption{ Schematic layout of the ILC. }
\label{fig:skdLayout}\vbbelow
\end{figure}

\begin{figure}[htb!]
\begin{center} \vbabove
   \includegraphics[width=14cm]{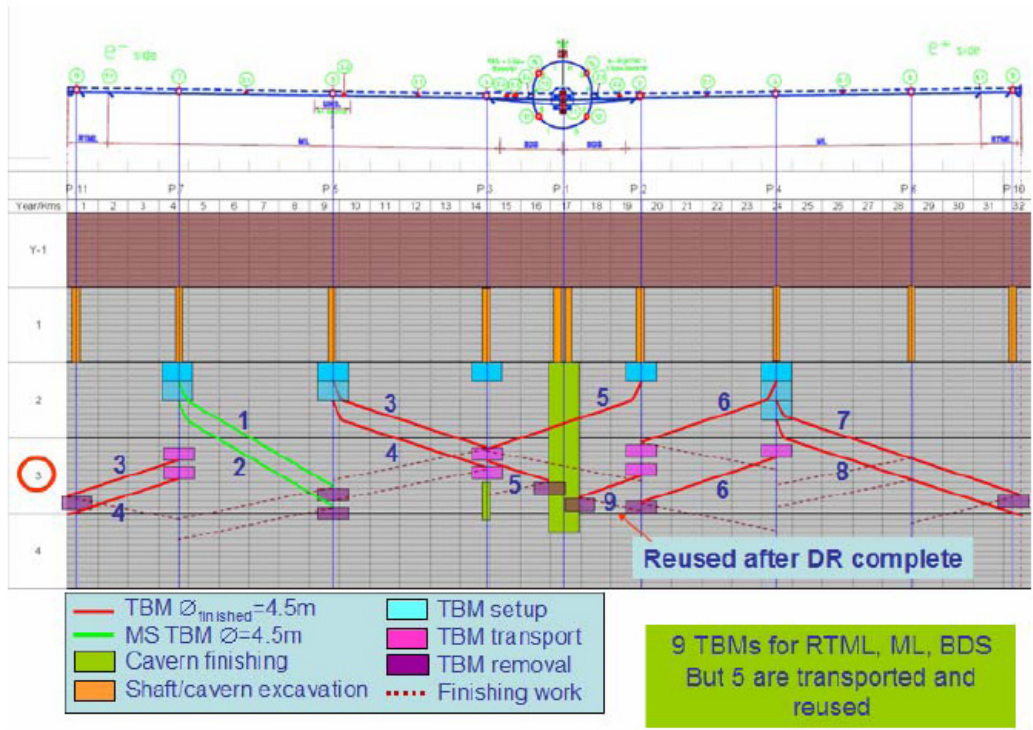}
\end{center}
\vbabovecaption \caption[Schematic of an example of an ILC civil construction plan using TBMs.]{Schematic of an example of an ILC civil construction plan using TBMs. Note that TBM \#9 is first used to excavate the tunnel for the
  damping ring (not shown).  Five Tunnel Boring machines must be
  transported in this plan. (Analysis and figure is based upon LEP and
       LHC experience at CERN.)}
\label{fig:skdConstruction} \vbbelow
\end{figure}

The layout of the ILC is shown in Figure~\ref{fig:skdLayout}.
A possible construction schedule is shown in Figure~\ref{fig:skdConstruction}
\cite{ops-gastal}, where it is
assumed that all shaft and underground construction can start
simultaneously or optimally (i.e. no funding limits) and that at least 9 tunnel
boring machines (TBM) of suitable diameter are available and employed
simultaneously. Assuming 1 year for the shaft construction (based on
LEP/LHC experience), 3 months for TBM setup, and 25 m/day boring
speed/TBM, then the actual underground construction time for the ML
and Damping Ring is about 3.5 years from ground breaking to beneficial
occupancy.

For the purposes of this schedule it was assumed that the finished
tunnel can be outfitted with services at the following rate:\\

\begin{center}
\begin{tabular}{lll}
  1 & Installation of cable trays and pipes supports &   4 weeks/km \\
  2 & Installation of cooling pipes   & 3 weeks/km \\
  3 & Installation of cables + connection  & 3 weeks/km \\
  4 & Installation of electrical equipment (transformers, switch gear) & 4 weeks/km \\
\end{tabular}\\
\end{center}

These rates are based on experience at existing facilities, not independent analysis.
It is assumed that these teams do not overlap in the tunnel at the
same time, but that a sufficient quantity of trained personnel are
available to form teams that can work in parallel at all available
locations. The installation of services require about 1 year such that
the tunnel is ready to accept technical components after about 3.5 to
4.5 years from the start of construction. (Some areas might be
available for component installation a few months sooner, but the
start of installation in these areas could disrupt the installation of
services. As a result, this was not considered in the modeling.)

\stepcounter{figlcl}

About one additional year is required to finish the underground
detector enclosure so that detector installation can proceed about 4.5
years after the start of civil construction. It is assumed that
detector assembly buildings and detector construction start at the
earliest opportunity.
Most of the detector assembly is assumed to take place above
ground following the general scheme adopted for the CMS detector at
CERN. This scheme allows detector construction and commissioning to
occur in parallel with the underground construction.  In the case of
CMS the detector assembly and commissioning took 6 years. In the
absence of more detailed information, we assume the same schedule for
the ILC detectors.

\subsection{Technical Component Schedule}

It is assumed that the high volume technical components required for
the Main Linac and Damping Ring are produced by industry. Components
in this category include SCRF cavities, cryomodules, modulators,
klystrons, SC and conventional magnets, cryogenic refrigerators,
transfer line, cables, piping, etc.  The production of
even such complex components as klystrons, modulators and
conventional magnets are well within the capability of
industry. The sequence would probably involve industrial
pre-series production by several vendors followed by tender for
production quantities of components. The number of vendors and the
region of production will largely be determined by decisions
concerning ``in-kind'' contributions from the regions participating in
the project. The required cryogenic
plants are sufficiently similar to those recently acquired for the LHC
that they can be procured from industry. Provided funding
is available, these components do not determine a
technically driven ILC schedule.

\stepcounter{figlcl}\begin{figure}[htb!]
\begin{center} \vbabove
   \includegraphics[width=14cm]{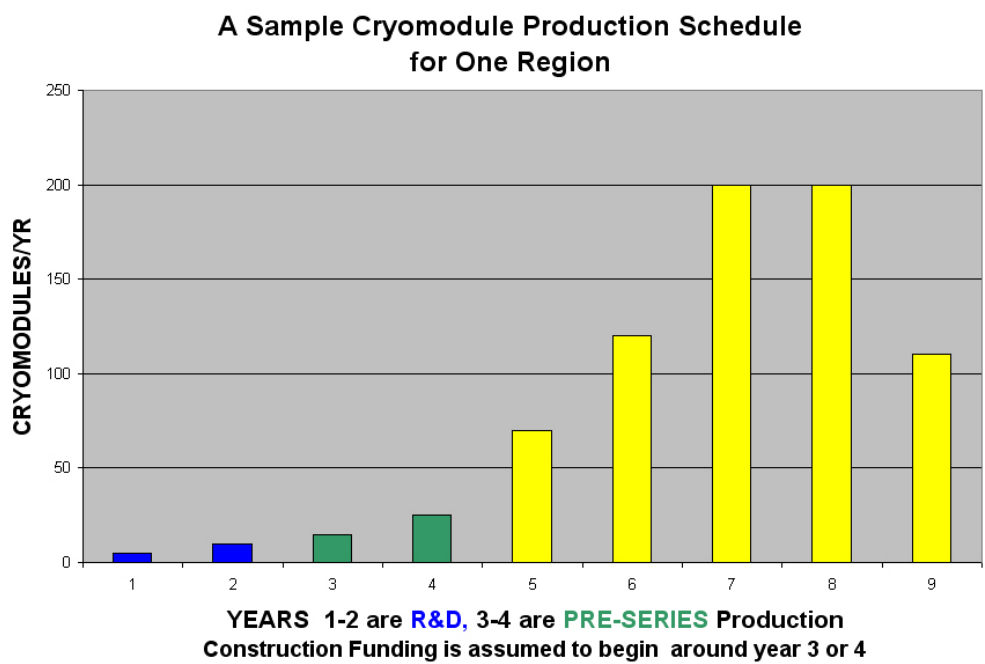}
\end{center}
\vbabovecaption \caption[A possible model schedule for cryomodule production.] { A possible model schedule for cryomodule production shows
    1/3 of the required ILC cryomodules produced in one of three regions. R\&D and   pre-series devices lead to 5 years of series production (yellow). The position  and magnitude of the peak of series production will vary with changes to the    available construction and test infrastructure. }
\label{fig:skdCryomod}\vbbelow
\end{figure}

The SCRF cavities and cryomodules are the most
technically challenging components and require the largest industrial
infrastructure and technical ramp up. The overall cost of the
cryomodules and associated infrastructure is likely to exceed
any one region's production capacity. Regional interest in SCRF
technology is also high. Both considerations suggest a model in
which three regions of the world provide these cryomodules in equal
quantity. Figure~\ref{fig:skdCryomod} shows one possible model for
the ramp up of cryomodule production in one of three regions. Note
that the five year production schedule shown in
Figure~\ref{fig:skdCryomod} assumes funding is available prior to
construction start so that infrastructure with long lead times can be
purchased early. Different regions could have earlier start times and a
flatter production schedule.

\subsection{Technical Component Installation Schedule}

The installation process follows the civil construction model with
parallel ongoing activities in separate areas but planned to minimize
interferences with other teams. The general schedule plan is based on
experience with the LHC. Transportation of components from surface
holding areas into their rough location in the underground tunnels
takes place during the evening shift. This minimizes interference with
all other activities above and below ground. These components are
installed, aligned, interconnected etc. during the day shift.

An example of this schedule for the main linac cryomodules is shown in
section~\ref{sect:STAi}.
This is accomplished with specialized crews which are appropriately
trained and have all the required support from the technical
systems. This manpower versus time profile for the linac installation
is also shown in section~\ref{sect:STAi}.
As with the civil construction schedule, installation in the central
DR/INJ complex takes place in parallel and is 6 months to 1 year ahead
of the main linac schedule.

\subsection{Example Funding Profile}

With the assumptions described above and with the value estimates, one
can model a construction schedule with its required funding
profile. This was done for a seven year construction project which is
consistent with the construction, manufacturing and installation
schedules.
The civil construction of the underground facilities is concentrated
in the first four years and the high technology cryomodules are spread
throughout the seven years. The remaining civil construction and
technical component manufacture and installation are spread throughout
years three to seven.

This plausible plan shows the need for early funding of the two cost
drivers, civil construction and the production of cavities and
cryomodules. The civil construction schedule drives the overall
schedule and therefore this funding, assumed to be mainly from the
host country, is on the critical path. The more global distribution of
funding of other systems allows flexibility in optimizing construction
and installation. This funding profile is, of course, model dependent
but it shows that there are no unusual or unattainable requirements in
a seven year construction schedule. Operations funding would begin
gradually starting in year five and would be at full operating level
in year eight.

\stepcounter{figlcl}\begin{figure}[hb]
\begin{center} \vbabove
   \includegraphics[width=12cm]{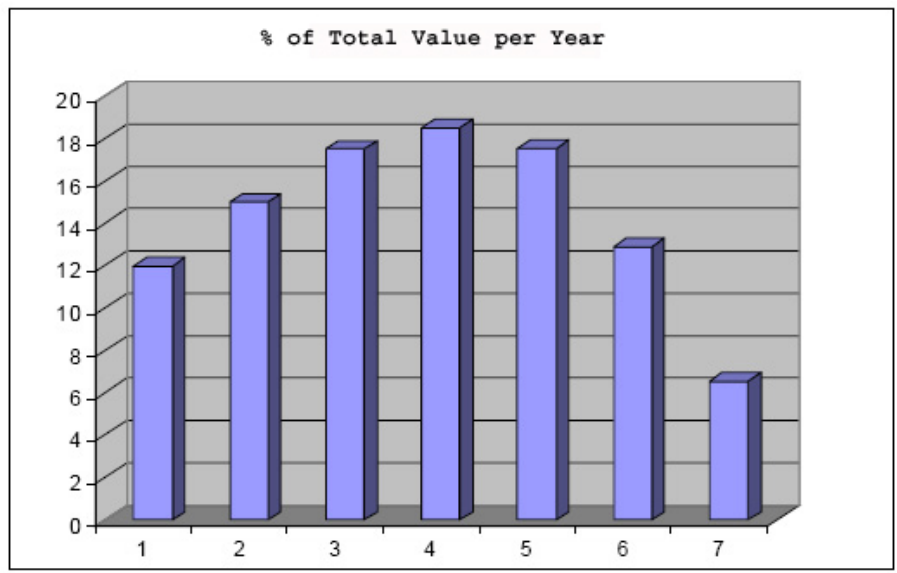}
\end{center}
\vbabovecaption \caption{A funding profile for a model seven year construction schedule.}
\label{fig:skdFunding}\vbbelow
\end{figure}
\pagebreak
{\bf Appendix:}
Example of the next level of WBS detail for Conventional Facilities
and Siting for Civil Engineering for the Main Linac Area System for
the Americas site.
%{\sf\small
\setlength{\tabcolsep}{6pt} \vbabove
 % \begin{center}
\stepcounter{tablcl}\begin{longtable}{| p{20mm} | p{90mm} | r |
p{20mm} |} \caption[WBS detail for Conventional Facilities and
Siting.] {WBS detail for Conventional Facilities and Siting for
Civil
  Engineering for the Main Linac Area System for the Americas site.}
\label{tab:costWBS1}
%\setlength{\tabcolsep}{2pt}
%   \begin{center}
%{\sf\small
%      \begin{tabular}{| l | l | r | l |}  \hline
\\ \hline
WBS \#       & WBS Title  &  Quantity  &  Unit   \\ \hline & & & \vbdlspacing \hline \endfirsthead

 \multicolumn{4}{c}{{\tablename} \thetable{} -- continued} \\ \hline
 WBS \#        &  WBS Title &  Value  &  Unit   \\ \hline & & & \vbdlspacing \hline \endhead

%This is the footer for all pages except the last page of the table...
  \multicolumn{4}{c}{{continued on next page \ldots}} \\
\endfoot

%This is the footer for the last page of the table...
  \hline
\endlastfoot

1.7 & Conventional Facilities & & \\ \hline%
1.7.1               &  Civil Engineering  &     &      \\ \hline%
1.7.1.1         &  Engineering, study work and documentation  &     &      \\ \hline%
1.7.1.1.1  &       In-house Engineering  &    &    man-hr    \\ \hline%
                           &  In-house Engineering  &  4\%  &   \%    \\ \hline%
1.7.1.1.2  &       Outsourced Consultancy Services  &     &      \\ \hline%
                           &  Outsourced Engineering  &  6\%  &   \%    \\ \hline%
1.7.1.2        &  Underground Facilities  &     &      \\ \hline%
1.7.1.2.1  &       Shafts  &     &      \\ \hline%
                          &  e$^-$ ML 14m dia. Shafts @ Points 5,3 (2x425 vert ft)  &  259  &   vert m    \\ \hline%
                          &  e$^-$ ML 9m dia. Shaft @ Point 7 (1x425 vert ft)  &  130  &   vert m    \\ \hline%
                          &  e$^-$ ML 1500mm dia. Survey Shafts @ Points 3.1, 5.1 (2x425 vert ft)  &  259  &   vert m
\\  \hline%
        &  e$^+$ ML 14m dia. Shafts @ Points 2, 4 (2x425 vert ft)  &  259  &   vert m    \\ \hline%
        &  e$^+$ ML 9m dia. Shaft @ Point 6 (1x425 vert ft)  &  130  &   vert m    \\ \hline%
        &  e$^+$ ML 1500mm dia. Survey Shafts @ Points 2.1, 4.1 (2x425 vert ft)  &  259  &   vert m
\\ \hline     & & & \\ \hline%
       &  Surface Grouting of Points 2-5 14m dia. Shafts (4x425 vert ft)  &  4  &   ea.    \\ \hline%
       &  Surface Grouting of Points 6-7 9m dia. Shafts (2x425 vert ft)  &  2  &   ea.    \\ \hline%
       &  Surface Grouting of Points 2.1, 3.1, 4.1, 5.1 Survey Shafts (4x425 vert ft)  &  4  &   ea.    \\ \hline%
       &  Points 2,3,4,5,6,7 - 14\&9m dia. Shafts, finishing (stairs, conc. wall, elev.\#2)  &  777  &   vert m
     \\ \hline%
       &  ML Underground Potable Water (1/2 of Points 2 \& 3)  &  1  &   ea.    \\ \hline%
       &  ML Underground Potable Water (Points 4,5,6,7)  &  4  &   ea.    \\ \hline%
       &  ML Underground Sanitary Sewer (1/2 of Points 2 \& 3)  &  1  &   ea.   \\ \hline%
       &  ML Underground Sanitary Sewer (Points 4,5,6,7)  &  4  &   ea.    \\ \hline%
 1.7.1.2.2  &       Tunnels  &     &      \\ \hline%
       &  e$^-$ ML 4.5m dia. Beam Tunnel, TBM Excavation (37,162 lin ft)  &  11,327  &   lin m    \\ \hline%
       &  e$^-$ ML 4.5m dia. Service Tunnel, TBM Excavation (37,162 lin ft)  &  11,327  &   lin m    \\ \hline%
       &  e$^-$ ML 4.5m dia. Tunnels, Conc. Inv. (74,324 lin ft)  &  22,654  &   lin m   \\ \hline     & & & \\ \hline%
       &  e$^+$ ML 4.5m dia. Beam Tunnel, TBM Excavation (36,660 lin ft)  &  11,174  &   lin m    \\ \hline%
       &  e$^+$ ML 4.5m dia. Service Tunnel, TBM Excavation (36,660 lin ft)  &  11,174  &   lin m    \\ \hline%
       &  e$^+$ ML 4.5m dia. Tunnels, Conc. Inv. (73,320 lin ft)  &  22,348  &   lin m  \\ \hline     & & & \\ \hline%
       &  Provide Tunnel Construction Water Treatment Plant  &  4  &   ea.    \\ \hline%
       &  Maintain and Operate Tunnel Construction Water Treatment Plant  &  4  &   ea.    \\ \hline%
       &  Treatment of Tunnel Construction Water  &  4  &   ea.    \\ \hline%
 1.7.1.2.3  &       Halls  &     &      \\ \hline%
 1.7.1.2.4  &       Caverns  &     &      \\ \hline%
       &  e$^-$ ML Shaft Base Caverns D\&B Excavation @ Points 3,5,7 (3x20,056 CY)  &  46,003  &   m$ ^{3} $    \\ \hline%
       &  e$^-$ ML Points 3,5,7 D\&B Exc. for Shield Doors (in Base Caverns) (3x959 CY)  &  2,199  &   m$ ^{3} $   \\ \hline%
       &  e$^-$ ML Beam Dump Cavern D\&B Excavation @ Point 3 (3,034 CY)  &  2,320  &   m$  ^{3}$    \\ \hline%
       &  e$^+$ ML Shaft Base Caverns D\&B Excavation @ Points 2,4,6 (3x20,056 CY)  &  46,003  &   m$ ^{3} $    \\ \hline%
        &  e$^+$ ML Points 2,4,6 D\&B Exc. for Shield Doors (in Base Caverns) (3x959 CY)  &  2,199  &   m$ ^{3} $    \\ \hline%
        &  e$^+$ ML Beam Dump Cavern D\&B Excavation @ Point 2 (3,034 CY)  &  2,320  &   m$ ^{3} $    \\ \hline%
        &  Shield Doors @ Base Caverns @ Points 2-7  &  6  &   ea.    \\ \hline%
1.7.1       &  CIVIL ENGINEERING (continued)  &     &      \\ \hline%
     1.7.1.2.5  &       Miscellaneous works  &     &      \\ \hline%
        &  e$^-$ ML Personnel Crossovers, D\&B Excavation (23 X 295.5 CY)  &  5,196  &   m$ ^{3} $    \\ \hline%
        &  e$^-$ ML Waveguides, Drill Excavation (968)  &  968  &   ea.    \\ \hline%
       &  e$^+$ ML Personnel Crossovers, D\&B Excavation (23 X 295.5 CY)  &  5,196  &   m$ ^{3} $    \\ \hline%
       &  e$^+$ ML Waveguides, Drill Excavation (968)  &  968  &   ea.    \\ \hline%
     1.7.1.3      &  Surface Structures   &     &      \\ \hline%
    1.7.1.3.1  &       Central Lab Buildings  &     &      \\ \hline%
     1.7.1.3.2  &       Detector Assembly Buildings  &     &      \\ \hline%
     1.7.1.3.3  &       Office Buildings  &     &      \\ \hline%
       &  Points 4-7 Office Buildings (4 x 3,750 sq ft)  &  1,396  &   sq m    \\ \hline%
     1.7.1.3.4  &       Service Buildings   &     &      \\ \hline%
        &  Points 2-7 Electrical Service Buildings (6 x 1,500 sq ft)  &  836  &   sq m    \\ \hline%
        &  Points 2-7 Cooling Towers \& Pump Stations Bldgs. (6 x 7,500 sq ft)  &  4,181  &   sq m    \\ \hline%
      &  Points 2-7 Cooling Ventilation Buildings (6 x 2,500 sq ft)  &  1,394  &   sq m    \\ \hline%
  1.7.1.3.5  &       Cryo- Equipment Buildings  &     &      \\ \hline%
       &  Points 2-7 Cryo - Warm Compressor Building (6 x 4,500 sq ft)  &  2,508  &   sq m    \\ \hline%
      &  Points 2-7 Cryo - Surface Cold Box Building (6 x 6,250 sq ft)  &  3,484  &   sq m    \\ \hline%
  1.7.1.3.6  &       Control Buildings  &     &      \\ \hline%
   1.7.1.3.7  &       Workshops  &     &      \\ \hline%
     &  Points 4-7 Workshop Bldg. - Machine \& Detector (4 x 11,250 sq ft)  &  4,181  &   sq m    \\ \hline%
   1.7.1.3.8  &       Site Access Control Buildings  &     &      \\ \hline%
     &  Points 4-7 Site Access Buildings (4 x 750 sq ft)  &  279  &   sq m    \\ \hline%
   1.7.1.3.9  &       Shaft Access Buildings  &     &      \\ \hline%
     &  Points 2-7 Shaft Access Buildings (6 x 9,375 sq ft)  &  5,226  &   sq m    \\ \hline%
   1.7.1.3.10  &       Miscellaneous Buildings  &     &      \\ \hline%
   1.7.1.3.11  &       User Facilities  &     &      \\ \hline%
  1.7.1.4     &  Site Development  &     &      \\ \hline%
    1.7.1.4.1  &       Off-site Site work  &     &      \\ \hline%
  1.7.1.4.2  &       Network of Monuments  &     &      \\ \hline%
    1.7.1.4.3  &       Construction Support  &     &      \\ \hline%
     1.7.1.4.4  &       Site Preparation  &     &      \\ \hline%
       &  Points 2 - 7, Clearing, Grubbing, and Initial Site Preparation (6 sites)  &  6  &   ea.    \\ \hline%
   1.7.1.4.5  &       Utility Distribution   &     &      \\ \hline%
       &  Points 2 - 7, Utility Corridors (Gas, DWS, San., Storm, Elec., Comm.)  &  6  &   ea.    \\ \hline%
     &  Points 2 - 7, Septic Field / Tank or Sanitary Sewer  &  6  &   ea.    \\ \hline%
       &  Points 2 - 7, Wells or DWS  &  6  &   ea.    \\ \hline%
      &  Points 4 - 7, Elevated Water Tank  &  4  &   ea.    \\ \hline%
       &  Points 4 - 7, Water Pump House  &  4  &   ea.    \\ \hline%
    1.7.1.4.6  &       Road, Sidewalks \& Parking Areas  &     &      \\ \hline%
       &  Points 2 - 7, Service Roads (6 sites x 1250 lin ft / site)  &  2,286  &   lin m    \\ \hline%
        &  Points 2 - 7, Paved Areas (6 sites x 8750 sy / site)  &  43,896  &   sq m    \\ \hline%
      &  Points 2 - 7, Flatwork (6 sites x 2,500 sq ft / site)  &  1,394  &   sq m    \\ \hline%
     1.7.1.4.7  &       Landscaping  &     &      \\ \hline%
       &  Points 2 - 7, Landscaping  &  6  &   ea.    \\ \hline%
       &  Points 4 - 7, Security Fencing (4 sites x 5,000 lin ft / site)  &  6,097  &   lin m    \\ \hline%
     1.7.1.4.8  &       Environmental  &     &      \\ \hline%
      &  Points 2 - 7, Sediment \& Erosion Control (6 sites)  &  6  &   ea.    \\ \hline%
    1.7.1.4.9  &       Miscellaneous Site Works  &     &
%      \end{tabular}
%}
\vbbelow
%   \end{center}
\end{longtable}
%\end{center}
%}

\setcounter{chapter}{6}

\chapter{\textsf{The Engineering Design Phase}\label{chapEDR}}
\setcounter{figlcl}{0}
\setcounter{tablcl}{0}
\setcounter{secnumdepth}{4} \setcounter{subsection}{0}
\setcounter{subsubsection}{0}
\usecounter{subsubsection}

\setcounter{section}{0} \renewcommand{\picturefolder}{./EDR/}
\setcounter{figlcl}{0}
\setcounter{tablcl}{0}
\setcounter{secnumdepth}{4} \setcounter{subsection}{0}
\setcounter{subsubsection}{0}
\usecounter{subsubsection}

\section{The Scope of the Engineering Design Phase}\label{sect:EDRscope}

The completion of the RDR is an important milestone on the way to ILC approval. The
next phase of ILC development must produce an engineering design of the project in
sufficient technical detail that approval from all involved governments can be
sought, and so that the ILC can begin construction soon after that approval is obtained.
The general plan is that the GDE will deliver an ILC Engineering Design Report
(EDR) in 2010, to demonstrate that the project can be built within the specified
budget and that it can deliver the required performance.

A fundamental management principle of the Engineering Design phase will be
the containment of the current RDR Value estimate. Areas of potential
cost-reduction via good engineering practises have been clearly identified in the
RDR. Together with the risk-mitigating prioritised R\&D program, these areas will be
the focus for the EDR.

The primary goal of the Engineering Design phase is to complete and document a
fully integrated engineering design of the accelerator. This design must satisfy
the energy, luminosity, and availability goals outlined in the ILC RDR, and include a
more complete and accurate value estimate. Specific requirements include:

\begin{itemize}

\item demonstrate through the ILC R\&D program that all major accelerator
components can be engineered to meet the required ILC performance specifications;
\itemspace
\item provide an overall design such that machine construction could start within two
to three years if the project is approved and funded; \itemspace
\item mitigate technical risks by providing viable documented fallback solutions
with estimates of their costs;  \itemspace
\item contain a detailed project execution plan including an achievable project
schedule and plan for competitive industrialization of high-volume components
across the regions;  \itemspace
\item limit options, where technical decisions are not yet final, to focus R\&D
and industrialization efforts on these issues; \itemspace
\item design the conventional construction and site-specific infrastructure in
enough detail to provide the information needed to allow potential host regions to estimate
the technical and financial risks of hosting the machine, including local
impact, required host infrastructure, and surface and underground footprints;
\itemspace
\item provide a complete value cost estimate for the machine, except for the details
not yet completed in the site-specific designs, which includes a funding
profile consistent with the project schedule proposed. \itemspace

\end{itemize}

A key component of the Engineering Design phase will be the increasing
direct involvement of industries across the regions. Industrialisation is a critical
issue for cost-effective production of the key technologies, and will also play
an important role in understanding how individual countries can contribute in-kind to
the construction project. This must be achieved on a truly world-wide basis,
including potential industrial bases which may not yet have been considered or
fully engaged.

The GDE is committed to achieving the above goals as a global project, building on
the success of the RDR. The GDE must also ensure that the internal momentum
is maintained and foster continued growth in the enthusiasm and commitment of
the international ILC community.

\clearpage %\setcounter{page}{1}
\setcounter{figlcl}{0}
\setcounter{tablcl}{0}
\setcounter{secnumdepth}{4} \setcounter{subsection}{0}
\setcounter{subsubsection}{0}
\usecounter{subsubsection}

\section{From RDR to EDR: Cost drivers and Technical Risk}\label{sect:EDR-RDR}

\subsection{The Importance of the RDR for the EDR Planning}\label{ssect:EDRplanning}

The RDR and its associated value estimate forms a solid basis from which to
prioritize and efficiently direct the GDE's efforts for the Engineering Design phase.
The fundamental assumptions on which the EDR planning will be organized are that:

\begin{itemize}

\item the RDR conceptual design is sound and complete, although the
overall engineering design remains immature; \itemspace
\item the remaining identified technical risks can be successfully mitigated via
a realistic and prioritised R\&D program during the next two years; \itemspace
\item the current estimate for the value is valid at the $<$30\% level; \itemspace
\item the value estimate can be maintained and possibly reduced by focused
and prioritized engineering program, including application of ``value engineering''
(an assumption that was independently noted by the ILCSC/FALC International
Cost Review in their report \cite{bib:EDR2}).
\itemspace

\end{itemize}

The RDR provides a design and a value estimate that is parametric in nature,
and allows us to clearly identify the cost drivers and the technical risks;
this information is critical in prioritizing both engineering and R\&D, given the
limited resources available to the GDE.

The primary cost drivers are the Superconducting RF (SCRF) linac technology
and the Conventional Facilities and Siting (CFS), which
together account for approximately 70\% of the ILC value estimate. These two
areas will correspondingly be a major focus during the Engineering Design phase.

The identification of technical risk, together with its impact and mitigation, is a
critical planning concept which the GDE has begun during the RDR phase. The
GDE Global R\&D Board (RDB) produced a prioritized list of the critical R\&D
activities that are required to mitigate many of the technical risks in the RDR
design (see section 7.2.2). A second and quasi-independent Risk
Assessment process has begun to assist management in planning a path from
the RDR through the development of the EDR and on into construction
and commissioning. The goal of this assessment is to evaluate the RDR design
for technical risks, estimate the degree of risk and define the strategy and impact
of mitigating these risks. The impact includes both the direct cost and the effects
on other ILC systems. These data have been used to begin the formation of a
risk register.

A first evaluation of the register shows (as expected) that the prioritized R\&D
plans from the RDB are well correlated with the relative value of risks and
costs. However there are also important issues that require engineering or
prototyping rather than R\&D programs, although the boundary between
engineering and R\&D cannot be sharply defined. The impact of suggested
mitigation strategies on other area or global systems is identified but requires
more development and study. In addition, areas have been identified where
a re-evaluation of the choice of the basic machine parameters can impact and
reduce the risks and associated mitigation costs.

The Risk Assessment process is at an early stage of development and will
continue with analysis and data updates throughout the Engineering Design
phase as the R\&D results become available.

One important aspect not currently included in this analysis is the impact of the
{\it design alternatives} to the current baseline, which after suitable R\&D may
provide either increased performance or a cost saving (or both). Support for R\&D
on the more promising of these alternative designs is a key part of the
Engineering Design phase. The R\&D on these supported alternative solutions will
have clear scheduled milestones and agreed-upon criteria for acceptance as the
baseline solution to be included in the EDR.

\subsection{Critical R\&D in the EDR Phase}\label{ssect:EDRrand}

The purpose of the R\&D for the ILC is to establish that:

\begin{itemize}
\item the various technologies chosen to achieve the required performance for the
machine are viable~; \itemspace
\item the cost of the technology has been minimised~; \itemspace
\item the chosen path provides sufficient operational flexibility to maximise
the chances of successful operation even when unforeseen constraints arise that affect the working point of the machine. \itemspace
\end{itemize}

All these R\&D goals can eventually be expressed in {\it equivalent cost}. For example,
the cost of a design oversight that severely limits operation of the ILC may be
a significant fraction of the total cost of the machine. In other cases, R\&D
that improves the luminosity of the machine by a certain factor can be compared to
the corresponding savings in running time.

The purpose of R\&D for the ILC is hence to reduce risk, i.e. extra cost, delays
or compromised performance. The concept of risk mitigation is one that drives
both engineering and R\&D. This approach -- implemented in a systematic manner
-- is currently being pursued both for R\&D and engineering for the ILC. It is expected
in the near future to yield quantitative assessments of the benefit of R\&D.

The ILC R\&D is currently funded almost exclusively through national (via
national laboratories and universities) and regional programs (e.g. via the
European Commission). The lack of centralized funding has been an issue
in coordinating the global R\&D, and will likely remain so during the
Engineering Design phase. At its inception, the GDE formed the Global R\&D
Board (RDB) to monitor the international R\&D activities. The RDB has since
made significant progress in identifying the critical-path R\&D, cataloguing the
global programs and available resources, and in several of the more critical
cases, achieving consensus on an R\&D program that makes most efficient use
of world-wide resources. The preeminent example here is the work supported in
all three regions on high-gradient cavities. Within the limited resources available, it
is critical that this type of activity be maintained throughout the Engineering
Design phase.

The RDB has addressed and prioritized the risk-mitigating ILC R\&D. Starting with
the highest priorities, task forces have been established to provide a
realistic cooperative international plan, which takes into account the constraints
of projected resources and the EDR timeline. At the time of writing, the top six
priority so-called ``S'' task forces\footnote{The letter ``S'' was chosen as a successor to the ``R'' requirements used in the second TRC Report}   have been formed; they are listed in
Table~\ref{tab:EDRss}. Additional task forces will be convened in the future. A full report on
the status of the S0 through S5 task-forces can be found in~\cite{bib:EDR1}. A general goal
for each task force was to produce a realistic R\&D plan to achieve required goals
within the EDR time-scale.

\stepcounter{tablcl}\begin{table} \vbabove \caption{Existing and
planned R\&D ``S'' task forces as of writing.}
   \label{tab:EDRss}
\begin{center}
\begin{tabular}{| l | l | l |}  \hline \hline
{\bf S0} & Cavity gradient & Establishment and demonstration of cavity surface  \vspace{-4pt} \\
 &  & preparation procedures which routinely (yield$ \geq $80\%)  \vspace{-4pt} \\
&  &  produce gradients of 35 MV/m with a $ Q_{0}  = 10^{10} $ in  \vspace{-4pt} \\
 &  & a vertical low-powered test. \\ \hline
{\bf S1} & Cryomodule gradient & Demonstration of a 8 cavity cryomodule operating  \vspace{-4pt} \\
 &  & at an average accelerating gradient of 31.5 MV/m  \vspace{-4pt} \\
 & &  at $ Q_{0}  = 10^{10} $, including fast tuner operation etc.. \\ \hline

{\bf S2} & Module string test & Determination of the needs, size and nature of a  \vspace{-4pt} \\
 &  &  module string test (ILC linac systems test) \\ \hline

{\bf S3} & Damping Rings & electron cloud \\
 & & fast injection/extraction kickers \\
& & lattice design \\
& & low-emittance tuning \\
& & impedance-driven single-bunch effects \\
& &  ion effects \\ \hline
{\bf S4} & Beam Delivery System & integrated IR design, including push-pull \\
& &    IR superconducting magnets \\
& &    crab-cavity system \\
& &    critical (novel) diagnostics \\
& &   intra-train feedback systems \\
& &    high-powered beam dump system \\
& &    collimator design and wakefield performance \\
& &    stabilization issues etc. \\ \hline
{\bf S5} & Positron Source & superconducting helical undulator \\
& &   photon target \\
& &   capture section (optical matching device and warm \vspace{-4pt} \\
& &   RF acceleration) \\
& &  remote handling. \\ \hline
\it{S6} & \it{Controls} & \it{in planning} \\ \hline
\it{S7} & \it{RF Power Source} & \it{in planning} \\ \hline
      \end{tabular}
   \end{center} \vbbelow
\end{table}

Beyond "S5" are additional aspects of the design, for example RF power
source, controls, etc. These areas are challenges in their own right and
certainly require R\&D. However, they are not currently considered high
risk-mitigation priorities on the time-scale of the EDR.

\subsection{The Importance of Alternative Designs}\label{ssect:EDRad}

The focus of the design work during the RDR phase was on the Baseline
Configuration, which was essentially established at the Snowmass Workshop
in August 2005, and formally adopted at the Frascati Workshop the
following November. As part of that process, a list of viable {\it alternatives} to the
baseline choices was also created. An alternative design or solution was defined
as being:
~\\ ~\\
\indent{{\it A technology or concept which may provide a significant cost reduction, increase
in performance (or both), but which will not be mature enough to be
considered baseline by the end of the RDR phase.}}
~\\ ~\\
The identified alternative designs were also formally included in the
Baseline Configuration Document (BCD), but they are not included in the RDR,
which is focused on the ILC configuration used to produce the value estimate.

Implicit in the definition of a supported alternative design is that on-going R\&D
may eventually bring an alternative to a state mature enough that it can be
considered as a replacement for the baseline solution. In evaluating an
alternative solution as viable, the time scale involved becomes relevant. If an
alternative will not reach a critical maturity on a time scale that is commensurate
with cost-effective implementation, it must be discarded or set aside for
consideration as a possible later upgrade. Unfortunately, the exact time scale for
this is not well-defined: the true time scale may go beyond the (technically
driven) Engineering Design phase due to political and financial reasons. It may
be prudent to maintain non-baseline R\&D beyond the EDR phase if
the cost/performance benefits merit it. How much of this can and should be supported
will become clearer as the Engineering Design phase evolves towards 2010.
Defining criteria for accepting the alternative designs as baseline is an important
task for the planning of the Engineering Design phase milestones; monitoring
the progress of the alternatives will be a key task throughout the process.

Examples of high-level alternative solutions are:

\begin{itemize}

\item {\bf Novel cavity shapes:} so-called low-loss structures or re-entrant designs
are currently being investigated. They offer the potential of higher gradients by
reducing the peak magnetic field on the cavity surface, or alternatively higher
Q0. Several R\&D programs are being pursued, notably at KEK and Cornell.
\itemspace
\item {\bf RF Power source:} the baseline RF power source and distribution is
considered mature and relatively low risk. However on-going R\&D at SLAC into
novel concepts could yield a significant cost reduction for this system. Of
these activities, the Marx modulator is the most mature and promising, and could
well be the first test case of the adoption of an alternative design. Other R\&D
activities are on sheet-beam klystrons and lower-cost RF distribution systems,
both of which are less mature, but have the advantage of being 'drop-in
compatible' solutions that could in principle be adopted at a late stage in the
design. \itemspace
\item {\bf IR solutions:} 2 mrad and ``head-on'' crossing-angle geometries are
being considered as alternatives to the current 14 mrad baseline. \itemspace
\item {\bf Compton-based positron source:} an independent source based on laser
Compton scattering is being investigated by an international collaboration. \itemspace

\end{itemize}

In addition to the alternatives described in the BCD, there are many technical
choices still to be made between competing technologies, where no clear
baseline emerged during the RDR phase. Examples of these include development
of the fast kicker systems for the Damping Ring (an S3 priority), and mechanical
tuner designs for the SCRF cavities.

The criteria used to monitor the progress of an alternative design during
the Engineering Design phase must be chosen carefully. Time scales for
evaluation must be based on the impact on the design work as it evolves.
The examples above of the alternative IR solutions and the Compton-based
positron source have a very large impact on the overall layout of the machine and
the associated CFS (a cost driver); such decisions will become increasingly
difficult to cost-effectively implement at a mature stage of the design. On the
other hand, the choice of cavity shape could be made relatively late, if the cost
benefits were considered adequate.

\clearpage %\setcounter{page}{1}
\setcounter{figlcl}{0}
\setcounter{tablcl}{0}
\setcounter{secnumdepth}{4} \setcounter{subsection}{0}
\setcounter{subsubsection}{0}
\usecounter{subsubsection}

\section{Restructuring the GDE: Project Management for the EDR Phase}\label{sect:EDRPM}

The scope of the EDR necessitates a robust management and an
appropriate organization, with resources sufficient to accomplish its aims.
It is essential that the current management structure of the GDE be adapted to
the needs of the Engineering Design phase. Particularly in the area of
project management, it is clear that substantial changes from the RDR
management structure are required.

The Engineering Design phase organization must have clear lines of authority
and responsibility and must effectively connect tasks with human and
financial resources (often from multiple sources across the regions). The
organization must include transparent mechanisms to establish and
communicate high-level goals and objectives, receive technical and political advice,
set priorities, manage change, resolve conflict and fill voids of human or
financial resources. All of this must be accomplished while maintaining a
strong international collaboration in the absence, at least initially, of
centralized funding.

As of writing, the exact details of the new project structure are being developed,
and will be completed and in-place by fall 2007. The following therefore reflects
a snapshot of an evolving picture which is under discussion.

\subsection{Top-Level Project-Management Structure}\label{ssect:EDRPMS}

\begin{figure}[htb]
   \begin{center} \vbabove
      \includegraphics[width=\textwidth]{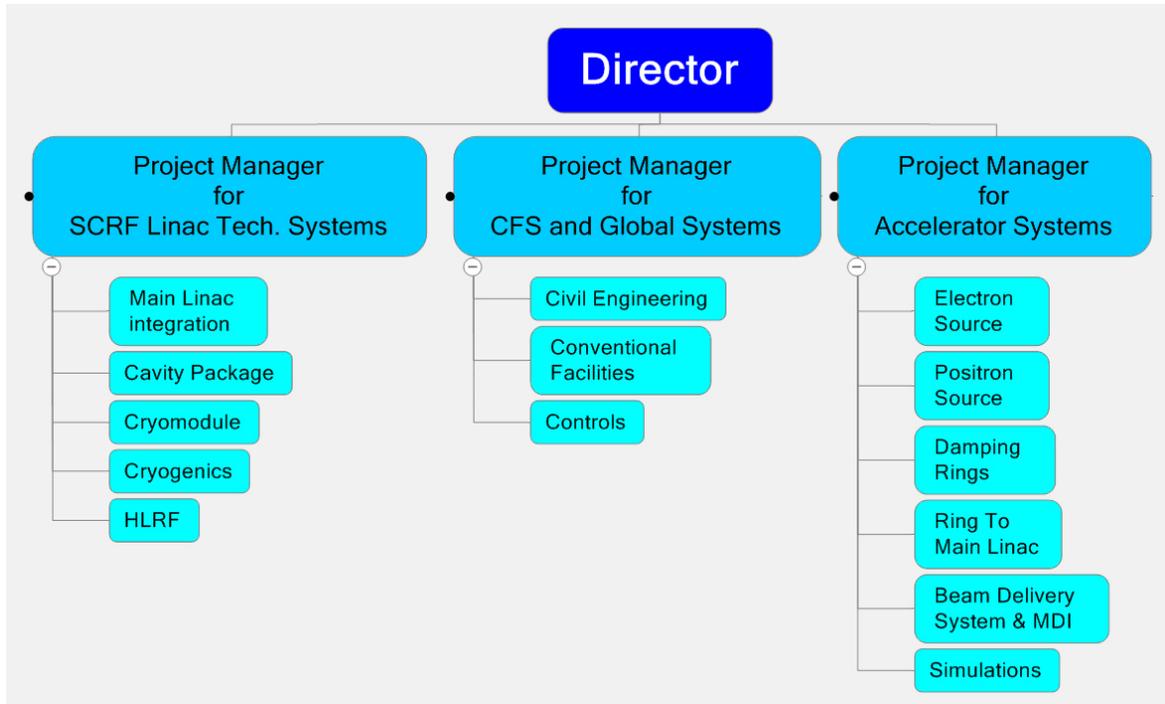}
      \vbabovecaption
      \caption[Basic proposed Project Management structure for the EDR phase.]{Basic proposed Project Management structure for the EDR phase.
The org chart indicates the top three levels of management: level-1 Director;
level-2 Project Managers; level-3 System Managers.}
      \label{fig:EDRPMS}
   \end{center} \vbbelow
\end{figure}

The RDR Management Board will effectively be replaced by a Project
Management Team consisting of three Project Managers (one from each region)
with distinct and clear responsibilities. The currently proposed structure is shown
in Figure~\ref{fig:EDRPMS}.

 \stepcounter{figlcl}

The Project Managers will report directly to the Director. The division of
responsibility reflects the primary cost drivers (SCRF technology and CFS).
Accelerator Systems is responsible for the injectors, damping rings,
bunch compressors and beam delivery systems.

The top-level management of the GDE will remain an Executive Committee similar
to the current one, chaired by the Director.

\begin{figure}[htb]
   \begin{center} \vbabove
      \includegraphics[width=0.6\textwidth]{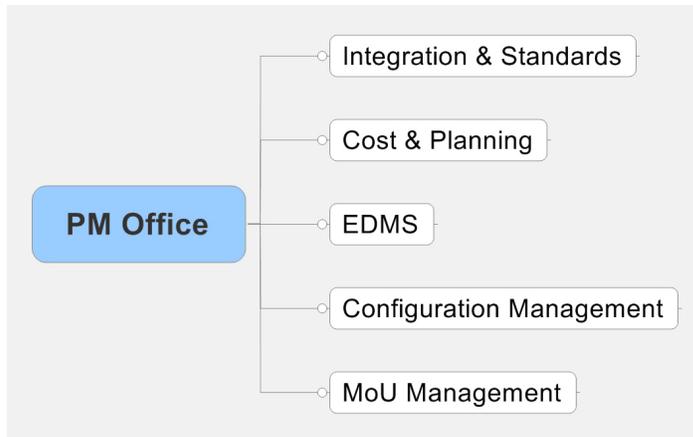}
      \vbabovecaption
      \caption{Primary central functions of the Project Management Office.}
      \label{fig:EDRPMO}
   \end{center} \vbbelow
\end{figure}

The three Project Managers will lead a Project Management Office. One
Project Manager will act as chair, and will have the final authority over
project management decisions. The project management office has a
regionally-balanced staff and supports several important central functions, depicted
in Figure~\ref{fig:EDRPMO}. (As of writing, the exact definition and structure of these functions is
still being discussed.)

 \stepcounter{figlcl}

\subsection{Work Packages}\label{ssect:EDRWP}

The technical work of the Engineering Design phase itself will be organized in
Work Packages (WPs). Each Level-3 System in Figure~\ref{fig:EDRPMS} will manage a
collection of WPs representing the Work Breakdown Structure (WBS) of that part of the
project. The WPs should reflect the critical engineering and R\&D milestones of
the EDR phase, and should have well-defined scope (deliverables), such that they
are suitable for distribution across geographically separated resources.

The formation of consortia of all types to deliver the WPs will be encouraged by
the GDE. Close consultation will take place with laboratory directors and other
senior figures to ensure that the WPs are optimally defined. Flexibility
and responsiveness will be necessary as the Engineering Design phase progresses.
In particular, it is necessary to ensure as far as possible that partners joining
the project at whatever stage can be assigned responsibilities appropriate to
their resources, competences and aspirations.

The engineering for the ILC design, and the R\&D program, will need to be
closely integrated in the work-package structure. Clear milestones and
technology choices will need to be defined to meet the EDR schedule.

The exact structure of the WBS and associated WP definitions is a process which
will evolve with time. The Project Managers together with the Level-3
(system) managers are primarily responsible for identifying the key milestones
and deliverables. Negotiations with individual institutes should start shortly
thereafter. While the new Project Management structure is being set-up, an
interim task-force was commissioned at the Beijing Workshop (February 2007)
to produce a straw man WBS and possible high-level WP definitions. This task
force will report to the Executive Committee in August 2007.

\subsection{Resources, responsibilities and organizational Issues}\label{ssect:EDRrroi}

Until such time as the GDE activities are centrally funded, the GDE must continue
to seek its resources indirectly via the institutes (funded entities) which form
the collaboration. Responsibilities for delivering a WP or part thereof must be
formally agreed upon between the GDE Project Management and the
corresponding institute via MoU. The institutes are then responsible for obtaining
the necessary resources for the task from their funding sources (agencies).
The process by which the WPs are defined, and the allocation of institutes to carry
out those WPs through MoUs, must be an open and transparent process allowing
all interested parties to make a proposal to carry out the work and to understand
and accept the criteria used in decision making.

 \stepcounter{figlcl}\begin{figure}[htb]
   \begin{center} \vbabove
      \includegraphics[width=\textwidth]{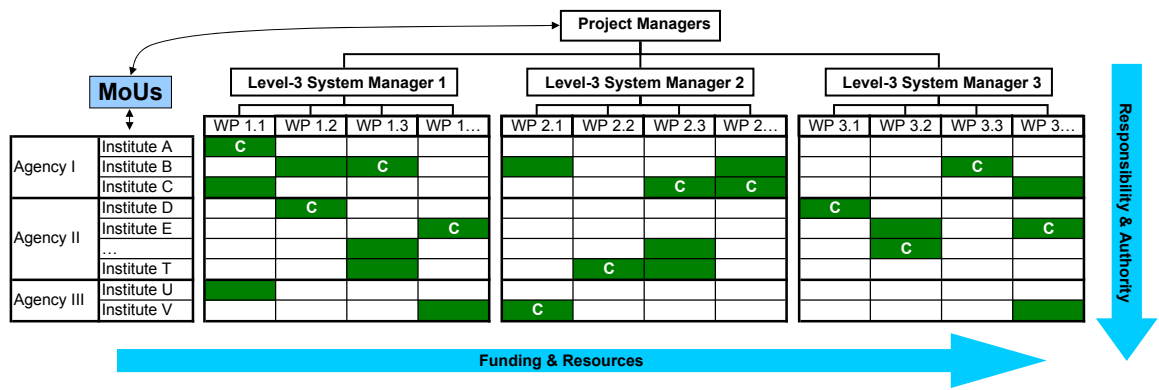}
      \vbabovecaption
      \caption[Managing a non-centrally funded project.]{Managing a non-centrally funded project. The green-filled boxes indicate
a commitment from an institute to deliver part of a WP. MoUs facilitate the
desired (and necessary) connection between the Project Management (authority
and responsibility) and the institutions (funding and resources).  The 'C' indicates
a coordinating role in a WP (an individual in an institute). Note that each WP has
only one coordinator.}
      \label{fig:EDRNCFP}
   \end{center} \vbbelow
\end{figure}

Figure~\ref{fig:EDRNCFP} represents a typical situation facing the EDR Project Management.
A top-down project management structure must have responsibility for achieving
the project goals, and requires authority over the resources to do so (represented
by the columns of Figure Figure~\ref{fig:EDRNCFP}). However, these resources are supplied
and funded via the collaborating institutes, and so the Project Management has no
direct line-authority over them (the rows of the matrix). In addition, a single institute
is likely to take on responsibilities for several parts of many Work Packages
(indicated in Figure Figure~\ref{fig:EDRNCFP} by the green highlighting).  The MoUs will be
critically important in establishing the correct and suitable level of authority for
the Project Management. Each MoU will need to be tailored to suit (a) the specifics
of the scope of the WP commitment and (b) the individual institute's management
and funding situation. Note that the many different funding agencies involved and
the associated differences in the way each works adds an additional complication
to the problem.

\subsection{Future Resource Requirements}\label{ssect:EDRfrr}

To estimate the resources required, we have looked at two international projects,
each with similarities and differences to the ILC, in order to try to estimate the
FTE years necessary to produce their equivalent of the EDR phase. These
projects were ITER, an international project approved in all three regions, and XFEL,
a predominantly but not exclusively European project recently approved. A fairly
exact estimate of the effort required to produce the XFEL TDR was possible; it is
more difficult to estimate such a directly comparable figure from ITER. However,
the information available, together with the extrapolation of estimates already made
by some GDE area systems to the whole project, leads to a similar conclusion,
viz. that an increase in the global ILC effort by approximately a factor of two to three
in manpower is required to complete an EDR of the scope of that of ITER or XFEL on the
required timescale.

\clearpage %\setcounter{page}{1}
\setcounter{figlcl}{0}
\setcounter{tablcl}{0}
\setcounter{secnumdepth}{4} \setcounter{subsection}{0}
\setcounter{subsubsection}{0}
\usecounter{subsubsection}

\section{Concluding Remarks}\label{sect:EDRcr}

The GDE remains committed to the technically driven schedule of supplying the
EDR in 2010, making start of construction possible as early as 2012. The critical
path and cost drivers have been clearly identified during the RDR phase, and
they define the priorities for the next three years of the Engineering Design phase.
The R\&D program will be fine-tuned to mitigate the remaining identified technical
risks of the design. A key element of the engineering activity will be the formation of
a qualified competitive industrial base in each region for the SCRF linac technology.
An equally critical focus will be on the civil construction and conventional facilities
-- the second primary cost driver -- where an early site selection would clearly
be advantageous; hence it is critical that the political site-selection process begin
in parallel to the technical EDR activity. This will also necessitate a movement to
a more direct involvement of the funding agencies in the governance of the ILC,
taking over functions and oversight currently performed by ICFA through the ILCSC.

Finally, the GDE remains committed to completing these challenging goals as a
truly international organization, by building on and consolidating the
successful collaboration that produced the RDR. The support of the world-wide
funding agencies is critical in this endeavour. The GDE -- together with the leaders
of the particle physics community -- will continue to work with the regional
funding agencies and governments to make this project a reality in the early part of
the next decade.

\clearpage

\backmatter

\clearpage
%\addcontentsline{toc}{chapter}{List of Figures}
\listoffigures %

\clearpage
%\addcontentsline{toc}{chapter}{List of Tables}
\listoftables

\typeout{Lastpage = \arabic{page}}
\end{document}